\documentclass[10pt,a4paper,research]{unityresearch}

\usepackage{amsmath,amsfonts,bm}

\def\eqref#1{equation~\ref{#1}}

\def\1{\bm{1}}

\DeclareMathAlphabet{\mathsfit}{\encodingdefault}{\sfdefault}{m}{sl}
\SetMathAlphabet{\mathsfit}{bold}{\encodingdefault}{\sfdefault}{bx}{n}

\usepackage{hyperref}
\usepackage{url}
\usepackage{booktabs}
\usepackage{multirow}
\usepackage{xcolor}
\usepackage{pifont}
\usepackage{graphicx}
\usepackage{amssymb}
\usepackage{wrapfig}
\usepackage{float}
\usepackage{array}
\usepackage{amsmath}
\usepackage{amssymb}
\usepackage{tabularx}
\usepackage{array}
\usepackage{longtable}
\usepackage{siunitx}
\usepackage{collcell}
\usepackage{tikz}

\ifcsname UBsplitStyleLoaded\endcsname\else
\expandafter\gdef\csname UBsplitStyleLoaded\endcsname{}

\definecolor{zRoyal}{HTML}{1C3B73}
\definecolor{zChampion}{HTML}{2286C1}
\definecolor{zGold}{HTML}{EDC870}
\definecolor{zGoldDeep}{HTML}{9A7412}
\definecolor{zParch}{HTML}{F6F4ED}
\definecolor{zParchAlt}{HTML}{F0F4EC}
\definecolor{zText}{HTML}{494B4B}
\definecolor{zShrine}{HTML}{F25C28}
\definecolor{zSheikah}{HTML}{56C7C8}
\definecolor{zZonai}{HTML}{207567}
\definecolor{zPlain}{HTML}{4B6645}
\definecolor{zRock}{HTML}{6E7676}
\definecolor{zRockLite}{HTML}{939A9A}
\definecolor{zDawn}{HTML}{7C7BA4}

\newcommand{\UBCaseStudyBlock}[4]{%
  \par\noindent
  {\small #2\par}
  \medskip
  \begin{center}
    \centering
    \includegraphics[
      width=\linewidth,
      height=0.72\textheight,
      keepaspectratio
    ]{#1}
    \par\smallskip
    \refstepcounter{figure}
    \label{#4}
    {\small\figurename~\thefigure: #3\par}
  \end{center}%
}

\newcommand{\tabcite}[2]{%
  \nocite{#2}\,\mbox{\hyperlink{cite.#2}{\textnormal{[#1]}}}%
}

\definecolor{zAblGain}{HTML}{207567}
\definecolor{zAblLoss}{HTML}{9E2A2B}
\definecolor{zAblFlat}{HTML}{6E7676}

\newcommand{\abdelta}[4]{%
  #1\hspace{0.03em}%
  \raisebox{0.52ex}{%
    \scalebox{0.42}[0.52]{\textcolor{#2}{\ensuremath{#3\!#4}}}%
  }%
}
\newcommand{\abup}[2]{\abdelta{#1}{zAblGain}{\uparrow}{#2}}
\newcommand{\abdown}[2]{\abdelta{#1}{zAblLoss}{\downarrow}{#2}}

\newcommand{\absetting}[2]{%
  \multirow{3}{*}{\shortstack[c]{\textbf{#1}\\[-0.35ex]{\scriptsize #2}}}%
}

\definecolor{zFailure}{HTML}{9E2A2B}
\newcommand{\failmax}[1]{\textcolor{zFailure}{\bfseries #1}}

\providecommand{\ubband}[2]{%
  \begingroup\setlength{\fboxsep}{4pt}%
  \colorbox{#1}{\parbox{\dimexpr\linewidth-2\fboxsep\relax}{#2}}%
  \endgroup}

\providecommand{\ubidentity}[4]{%
  \ubband{white}{%
  \noindent\textcolor{#1!72!black}{\bfseries\footnotesize #3}\hfill%
  \textcolor{zRock}{\scriptsize #4}\par\vspace{2pt}%
  \textcolor{#1!55}{\rule{\linewidth}{0.65pt}}}}

\providecommand{\ubcategory}[3]{%
  \ubband{white}{%
  \noindent\textcolor{#1!68!black}{\bfseries #2}\hfill%
  \textcolor{zRock}{\scriptsize #3}\par\vspace{1.5pt}%
  \textcolor{#1!38}{\rule{\linewidth}{0.45pt}}}}

\providecommand{\ubbadge}[2]{%
  \textcolor{#1!70!black}{\scriptsize\bfseries #2}}

\providecommand{\ubtaskhead}[6]{%
  \ubband{white}{%
    \noindent\textcolor{#1!67!black}{\ttfamily\footnotesize\bfseries #2}%
    \hfill\ubbadge{#4}{#5}\hspace{4pt}%
    \textcolor{zRock}{\scriptsize #6 tests}\par
    \vspace{1.2pt}\noindent\textcolor{zText}{\bfseries #3}\par\vspace{1.5pt}%
    \textcolor{#1!28}{\rule{\linewidth}{0.3pt}}}}

\providecommand{\ubfield}[2]{%
  \textcolor{#1!68!black}{\scriptsize\bfseries #2}}

\providecommand{\ubbullet}[2]{%
  \noindent\hangindent=1.15em\hangafter=1%
  \textcolor{#1!72!black}{\textbullet}\hspace{0.48em}#2}

\providecommand{\ubtestbadge}[1]{%
  \textcolor{zZonai!72!black}{\scriptsize\bfseries Test~#1}}

\providecommand{\ubscreenshotdir}{figures/tasks}
\providecommand{\ubscreenshotheight}{2.7cm}
\providecommand{\ubshot}[4]{
  \begin{minipage}[t]{0.485\linewidth}\centering
  \setlength{\fboxsep}{0pt}\setlength{\fboxrule}{0.55pt}%
  \textcolor{#4!48}{\fbox{%
  \IfFileExists{\ubscreenshotdir/#1/#2.png}%
    {\includegraphics[width=\dimexpr\linewidth-2\fboxrule\relax,
      height=\ubscreenshotheight,keepaspectratio]{\ubscreenshotdir/#1/#2.png}}%
    {\colorbox{white}{\parbox[c][\ubscreenshotheight][c]
      {\dimexpr\linewidth-2\fboxrule\relax}{%
      \centering\scriptsize\ttfamily\textcolor{zRockLite}{#1/#2.png}}}}}}%
  \\[2pt]{\scriptsize\textcolor{#4!65!black}{\bfseries #3}}\end{minipage}}

\providecommand{\ubshots}[2]{%
  \ubshot{#1}{template}{Initial project}{#2}\hfill
  \ubshot{#1}{completed}{Completed project}{#2}}

\fi

\definecolor{darkgreen}{RGB}{0,110,60}
\definecolor{darkred}{RGB}{160,20,30}
\newcommand{\yes}{\textcolor{darkgreen}{\ding{51}}}
\newcommand{\no}{\textcolor{darkred}{\ding{55}}}
\definecolor{chestnut}{cmyk}{0, 0.7808, 0.4429, 0.1412}

\providecommand{\ubharness}{}
\renewcommand{\ubharness}[3]{\multirow{#1}{*}{\shortstack[l]{#2\\#3}}}

\title{UnityBench-Mini3D: Evaluating LLM Agents for 3D Game Development in Unity}
\shorttitle{UnityBench-Mini3D}
\shortauthor{Yijun Zhou et al.}

\author{
Yijun Zhou$^{1,\ast}$ \and
Minghao Shao$^{1,2,\mathsection,\ast,\dagger}$ \and
Ke Zhang$^{1}$ \and
Renkai Yuan$^{1}$ \and
Hongbo Yuan$^{1}$ \and
Bin Hu$^{1}$ \and
Bozhao Qian$^{1}$ \and
Hongxia Shen$^{1}$ \and
Yanhong Liu$^{1,3,\mathsection}$ \and
Yuting Chen$^{1}$ \and
Yuntao Shi$^{1,4,\mathsection}$ \and
Zhongxuan Xu$^{1,5,\mathsection}$ \and
Shuo Mao$^{1}$ \and
Haijia Zhou$^{1}$ \and
Zhendong Li$^{1}$ \and
Tao Zhang$^{1}$ \and
Chenggang Li$^{1}$ \and
Yuxuan Wang$^{1,6,\mathsection}$ \and
Zinan Zheng$^{1,7,\mathsection}$ \and
Calvin Zhang$^{1,\mathsection}$ \and
Jia Li$^{7}$ \and
Heng Chang$^{1,\dagger}$ \and
Jie Feng$^{1,\dagger}$
}

\affiliations{%
$^{1}$Unity China \quad
$^{2}$New York University \quad
$^{3}$University of Chinese Academy of Sciences \quad
$^{4}$Xiamen University\\
$^{5}$Columbia University \quad
$^{6}$Peking University \quad
$^{7}$The Hong Kong University of Science and Technology
}

\correspondence{
$^\ast$Equal contribution. \\
$^\dagger$Corresponding authors:
\href{mailto:shao.minghao@nyu.edu}{shao.minghao@nyu.edu},
\href{mailto:heng.chang@unity.cn}{heng.chang@unity.cn},
\href{mailto:jie.feng@unity.cn}{jie.feng@unity.cn}
}

\paperdate{November 2026}
\paperversion{Version 1.0}
\division{AI Hub}
\codeurl{https://github.com/Unity-China/UnityBench-Mini3D}

\abstracttext{%
Game development stresses coding agents in ways general software engineering does not: correctness depends on how engine APIs, serialized scene state, and runtime behavior interact, not on source code alone. Existing benchmarks mainly target platforms such as Godot and heavily rely on LLM judges or coarse task-level outcomes. We present \texttt{UnityBench-Mini3D}, a benchmark of 77 project-level Unity tasks organized around seven expert-defined skill categories rather than tutorials or genres, with three difficulty tiers. Each task pairs a complete Unity project with natural-language requirements and is evaluated using 1,178 held-out, human-authored milestone tests covering scene structure, component configuration, and runtime behavior. This in-engine evaluation provides deterministic partial credit, localizes failures, and distinguishes partial implementations from complete solutions without relying on subjective judges. Across 21 models and 9 coding harnesses, the average pass rate reaches 80.9\% on easy tasks but falls to 35.2\% on hard tasks, with just 57.7\% overall. UnityBench-Mini3D provides a reproducible, diagnostic testbed distinguishing systems across difficulty levels.
}
\keywords{Game Development, Agentic Systems, Large Lanauge Models, Benchmarks}

\begin{document}
\maketitle

\begingroup
\renewcommand{\thefootnote}{\ensuremath{\mathsection}}
\footnotetext[1]{
\textbf{Minghao Shao} is a Visiting Scientist.
\textbf{Zhongxuan Xu} and \textbf{Calvin Zhang} are Student Researchers.
\textbf{Yanhong Liu}, \textbf{Yuntao Shi}, \textbf{Yuxuan Wang}, and
\textbf{Zinan Zheng} are Interns.
}
\endgroup

\section{Introduction}
\paragraph{Motivation.}
Game development is a systems-engineering discipline rather than a single programming activity. Even a small 3D game requires developers to organize scenes, implement gameplay logic, configure physical interactions, coordinate animation state machines, build responsive user interfaces, author particles and shaders, and often extend the editor with custom tools. These activities draw on distinct skills yet interact through shared project state: a small change to a collider, serialized reference, scene hierarchy, or animation transition can affect systems far from the edited asset. Consequently, correctness cannot be established from source code or compilation alone. It depends on the structure and configuration of the project as well as its runtime behavior: a project may compile and launch while containing a miswired reference, a character that clips through the floor, or an interface that never updates~\citep{la2026gameenginebench}.

Most game-development work takes place inside a game engine. Engines such as Godot, Unity, and Unreal maintain asset databases, serialization systems, and editor environments in which much of a project's state is created and persisted. Game development therefore means operating within a stateful software environment rather than editing a collection of source files. In parallel, large language models have become capable of software-development tasks~\citep{chen2021evaluating,jimenez2024swe}, and a growing body of work shows that they can support parts of the game-development workflow while demonstrating competence with engine APIs and conventions~\citep{chen2023gamegpt,yin2026autoue,jiang2026opengame}. These capabilities are now reaching production tooling~\citep{unitymanual,unreal58}. Alongside them, benchmarks have emerged to evaluate agents on tasks grounded in real engines, using engine execution, automated tests, or judgments over runnable artifacts~\citep{chi2026gamedevbench,la2026gameenginebench,luo2026gamecraft,sun2026jamer}.

Despite this progress, existing benchmarks leave three important gaps. First, engine coverage is narrow: most benchmarks use Godot~\citep{chi2026gamedevbench,luo2026gamecraft,sun2026jamer}, few target Unreal~\citep{la2026gameenginebench}, and Unity coverage remains limited despite its widespread commercial use; MAGE covers only single-file C\# generation for 26 hand-crafted 2D mini-games rather than editing complete Unity projects~\citep{liu2026mage}. Second, tasks are rarely organized around how game development is actually practiced. Many are adapted from tutorials~\citep{chi2026gamedevbench}, creating a risk that reference implementations already appear in pretraining data, while others group tasks by game genre or by topic clusters derived from public repositories~\citep{luo2026gamecraft,zhang2026v}, categories that describe what a game is rather than the development skill required to build it. Third, and most importantly, evaluation remains coarse. Many benchmarks rely on LLM judges, introducing subjectivity and limiting reproducibility~\citep{jiang2026opengame,luo2026gamecraft,zhang2026v}, while deterministic alternatives reduce an entire task to a single pass/fail outcome or an aggregate similarity score~\citep{chi2026gamedevbench,sun2026jamer}. An agent that implements a jump but never lands the character may then score the same as one whose scene does not load at all, and a solution that hard-codes an expected value or modifies the test itself may still be counted as successful. Together, these limitations motivate a Unity-based benchmark whose tasks reflect the structure of real development work and whose outcomes are verified finely enough to explain not only whether an agent succeeds, but what it builds correctly and where it fails.

\paragraph{Contribution.}
To address these gaps, we present UnityBench-Mini3D, a manually curated benchmark for evaluating LLM agents on 3D game development in Unity. Its taxonomy is grounded in expert-defined development skills rather than tutorial content or ad hoc topic groupings, and it comprises 77 project-level tasks spanning seven common development skills and three difficulty tiers. Each task is provided as a complete Unity project with a written specification and is evaluated by 1,178 held-out, human-authored milestone tests executed within Unity. By decomposing each task into independently verifiable checkpoints, these milestones provide deterministic requirement-level partial credit, identify which requirements are satisfied or missed, and distinguish partial implementations from complete solutions.

Our contributions are threefold: (1) We introduce UnityBench-Mini3D, an open-access benchmark along with an evaluation framework. Evaluation is performed in a containerized environment with the engine version, project dependencies, and testing infrastructure fully pinned, ensuring that results remain reproducible, comparable across research groups, and stable over time. (2) We systematically evaluate 21 frontier models across 9 coding harnesses, covering both general-purpose and Unity-oriented agents, and report task performance alongside cost and wall-clock time. This establishes a broad baseline for agent capability in project-level Unity development and separates the contributions of the underlying model and the harness. (3) Our evaluation, analysis, and case studies characterize what distinguishes game development from general software engineering, identify the bottlenecks that limit the transfer of conventional coding agents to this setting, and point to concrete directions for designing future game development agents and training models.

\section{Related Works}

\paragraph{LLMs for Game Development}
LLMs are increasingly used in game development. Early systems assigned planning, implementation, and review to specialized agents~\citep{chen2023gamegpt}, while LLMR enabled natural-language creation and modification of interactive Unity worlds~\citep{de2024llmr}. Newer systems cover broader workflows: UniGen integrates planning, code generation, component binding, scene construction, and debugging in Unity~\citep{yang202690}; AutoUE uses Unreal Editor tools to place assets, generate gameplay code, and automate playtesting~\citep{yin2026autoue}; and OpenGame refines browser games through generated feedback~\citep{jiang2026opengame}. Although these systems produce runnable games, their evaluations remain tied to specific pipelines and task families rather than standardized project-level assessments across development skills. Commercial engines expose similar capabilities: Unity provides a beta in-editor assistant and an official MCP server~\citep{unitymanual}, while Unreal Engine 5.8 offers an experimental first-party MCP plugin~\citep{unreal58}. As agents modify persistent engine state, evaluation should move beyond source-level checks to reproducible verification of project state and runtime behavior.

\paragraph{Benchmarks for Game Development}
A growing number of benchmarks and evaluation suites assess models and agents on creating or modifying games. They span several development environments: GameDevBench, GameCraft-Bench, and JamBench use Godot~\citep{chi2026gamedevbench, luo2026gamecraft, sun2026jamer}; GameEngineBench and AutoUE's PlayGen-20 suite operate in Unreal Engine~\citep{la2026gameenginebench, yin2026autoue}; and V-GameGym and WebGameBench use Pygame and browser-native runtimes, respectively~\citep{zhang2026v, zhang2026webgamebench}. Their tasks are constructed in different ways, including adaptations from web and video tutorials~\citep{chi2026gamedevbench}, organization by game genre~\citep{luo2026gamecraft}, and topic clusters mined from public repositories~\citep{zhang2026v}. Evaluation strategies likewise vary.
One family uses rubric-guided LLM or VLM judges over screenshots, videos, or replayed gameplay~\citep{luo2026gamecraft, jiang2026opengame, zhang2026v}, an approach suited to open-ended artifacts such as complete games.
\begin{wraptable}{r}{0.54\textwidth}
\centering
\caption{Game development benchmark comparison.
\textbf{Src.:} task source;
\textbf{Tk./Ts.:} task and test counts;
\textbf{D./P.:} judge-free deterministic evaluation and partial scoring;
\textbf{Cat.:} clustered (\textit{C}), genre-based (\textit{G}),
skill-based (\textit{S}), or pattern-based (\textit{P}).}
\label{tab:related}
\scriptsize
\setlength{\tabcolsep}{0.8pt}
\renewcommand{\arraystretch}{0.92}

\begin{tabular}{@{}llll r@{\,/\,}r c@{\,/\,}c c@{}}
\toprule
\textbf{Benchmark} & \textbf{Engine} & \textbf{Src.} & \textbf{Scope} &
\multicolumn{2}{c}{\textbf{Tk./Ts.}} &
\multicolumn{2}{c}{\textbf{D./P.}} &
\textbf{Cat.} \\
\midrule
V-GameGym\tabcite{17}{zhang2026v}
 & Pygame  & Repos.    & File  & 2{,}219 & --      & \no  & \no  & 100 (C) \\
WebGame\tabcite{18}{zhang2026webgamebench}
 & Browser & Web/user  & Proj.  & 111     & --      & \no  & \no  & 7 (G) \\
GameDev\tabcite{4}{chi2026gamedevbench}
 & Godot   & Tutorials & Proj.  & 333     & --      & \yes & \no  & 4 (S) \\
GameCraft\tabcite{12}{luo2026gamecraft}
 & Godot   & Human     & Proj.  & 140     & --      & \no  & \yes & 15 (G) \\
JamBench\tabcite{13}{sun2026jamer}
 & Godot   & Jams      & Proj. & 300     & --      & \yes & \no  & -- \\
GameLogic\tabcite{1}{che2026gamelogicbench}
 & Godot   & Repos./Games     & Proj.  & 72      & 1{,}451 & \yes & \no  & 7 (S) \\
GameEngine\tabcite{10}{la2026gameenginebench}
 & Unreal  & Repos.    & Proj.  & 110     & --      & \no  & \no  & -- \\
MAGE\tabcite{11}{liu2026mage}
 & Unity   & GPC Corpus       & File  & 26      & --      & \yes & \no  & 26 (P) \\
\midrule
\textbf{Mini3D (Ours)}
 & \textbf{Unity} & \textbf{Human} & \textbf{Proj.}
 & \textbf{77} & \textbf{1{,}178}
 & \yes & \yes & \textbf{7 (S)} \\
\bottomrule
\end{tabular}
\end{wraptable}
Another emphasizes repeatable execution-based evaluation, using task-specific engine tests that yield binary task-level outcomes~\citep{chi2026gamedevbench} or project-level compilation, structural, and behavioral measures~\citep{sun2026jamer}. Other execution-grounded approaches include tick-level state assertions for runtime game logic~\citep{che2026gamelogicbench} and the decomposition of specifications into assertable keypoints that are tested through runtime-state injection and bounded interactions~\citep{jia2026gamegen}. Within Unity, MAGE evaluates generated executable scenes along compile success, runtime success, structural fidelity, and mechanism adherence~\citep{liu2026mage}. UnityBench-Mini3D also focuses on Unity but differs by decomposing each task into human-authored milestones evaluated directly at engine runtime, exposing partial completion at the level of intended requirements.

\section{UnityBench-Mini3D}
UnityBench-Mini3D evaluates agents through Unity project final-state testing, without LLM judges or content similarity. Figure~\ref{fig:main} details the workflow, Appendix~\ref{app:benchmark-profiling} profiles the whole benchmark.
\begin{table}[htbp]
\centering
\caption{UnityBench-Mini3D task taxonomy, example tasks, and required skills. The bottom row shows one representative task from each core category (except for Miscellaneous).}
\label{tab:categories}
\fontsize{6.2pt}{6.4pt}\selectfont
\setlength{\tabcolsep}{1.4pt}
\renewcommand{\arraystretch}{1.05}
\setlength{\aboverulesep}{0.22ex}
\setlength{\belowrulesep}{0.26ex}

\begin{tabularx}{\textwidth}{@{}
>{\hsize=0.48\hsize\linewidth=\hsize\raggedright\arraybackslash}X
c
c
>{\hsize=1.22\hsize\linewidth=\hsize\raggedright\arraybackslash}X
>{\hsize=1.30\hsize\linewidth=\hsize\raggedright\arraybackslash}X
@{}}
\toprule
\textbf{Category}
& \textbf{\# Tasks}
& \textbf{\# Tests}
& \textbf{Skills exercised}
& \textbf{Primary requirement} \\
\midrule

Scene construction
& 19 & 365
& Transforms, hierarchy, prefabs,\newline
  terrain, cameras, NavMesh
& Assemble and organize scene content under spatial, hierarchy,
  and navigation constraints. \\

Animation
& 12 & 179
& Animation clips, Animator, events, layers, Avatar Masks, IK,
  Playables
& Configure animation assets and controllers\newline
  to produce coordinated runtime motion and\newline
  character poses. \\

Particles \& shaders
& 12 & 155
& Particle System, VFX Graph, Shader Graph, shaders, materials
& Construct and configure visual effects, and implement or adapt
  shaders and materials to the required behavior and rendering
  pipeline. \\

Physics
& 11 & 168
& Rigidbodies, colliders, physics materials, joints, WheelCollider,
  Cloth, procedural meshes
& Implement and configure physical motion,\newline
  collisions, and interactions under specified\newline
  runtime constraints. \\

User interface
& 9 & 111
& UGUI, Canvas, RectTransform, UI Toolkit, Sprite Editor,
  RenderTexture
& Prepare UI assets and implement layouts, input handling, and
  visual feedback. \\

Editor operations
& 11 & 142
& Hierarchy organization, AssetDatabase, importers, Undo, project
  settings, ScriptableObject, audio integration
& Organize and configure scenes, assets, and project settings,
  and integrate functionality while preserving existing references
  and behavior. \\

Miscellaneous
& 3 & 58
& Other engine features, including light cookies,\newline
  LODGroup, and Timeline.
& Address additional engine workflows, including light\newline
  projection, model level-of-detail configuration, and\newline
  timeline-based sequencing. \\

\midrule
\textbf{Total}
& \textbf{77}
& \textbf{1{,}178}
& \multicolumn{2}{r@{}}{
    \textbf{Easy: 25 \quad Medium: 29 \quad Hard: 23}
} \\
\bottomrule
\end{tabularx}

\includegraphics[width=\textwidth]{figures/task\_examples.pdf}
\end{table}

\subsection{Benchmark Construction}
\paragraph{Curation and Validation}
Six Unity experts with over two years' experience defined six core categories for small-scale 3D games and planned over ten tasks per category. Designers proposed target difficulties to balance the three levels. Tasks drew on routine Unity workflows, tutorials, and commercial-game interaction patterns. Examples include classifying road-segment GameObjects in a scene hierarchy (Task 0004), applying automatic, content-aware, and grid-based sprite slicing (Task 0010), and aligning remote third-person weapon presentation with first-person gameplay state (Task 0079). Each candidate included an initial challenge template, specification, golden reference, and milestone tests show in Appendix~\ref{app:chalspec} and ~\ref{app:chaltest}. In each review cycle, designers shared each task's inspiration, practical need, and validation logic, while peers examined its relevance, instruction clarity, and validation soundness. Five developers with LLM and AI-agent experience joined the six authors to review candidates and issue pass/fail decisions with revision suggestions. Failed candidates were revised and resubmitted until all eleven approved them. Before task evaluation, we ensured its golden reference scored 100\% and its unmodified initial challenge template scored 0\%.

\paragraph{Taxonomy and Coverage}
We organize the 77 tasks by the primary Unity development skill required to solve them rather than by game genre or application domain. Each task is assigned to the category that best represents the Unity subsystems and workflows it exercises, making category-level performance interpretable as development capability rather than game theme. The taxonomy covers seven categories: scene construction, animation, particles and shaders, physics, user interface, editor operations, and a miscellaneous category for tasks not captured by the other six. Table~\ref{tab:categories} summarizes the task and milestone-test counts, skills exercised, and primary requirements of each category, while Appendix~\ref{app:challist} lists all 77 tasks together with their categories and test counts.

\subsection{Benchmark Structure}
Each task consists of a valid initial Unity project paired with a natural-language specification. The project contains the scenes, assets, scripts, and settings relevant to the task and can be opened immediately, but does not yet satisfy the requested outcome. The specification contains four fields: \emph{Goal}, \emph{Context}, \emph{Constraints}, and \emph{Resources}. These fields define the target outcome, describe the initial project state, state the requirements to satisfy or preserve, and identify relevant Unity resources. Each task is also annotated with a unique identifier, primary skill category, difficulty level, and number of milestone tests. Starting from an existing project rather than an empty repository requires modifying interconnected project state while preserving functionality already in place.

\begin{figure}[t]
    \centering
    \includegraphics[width=\linewidth]{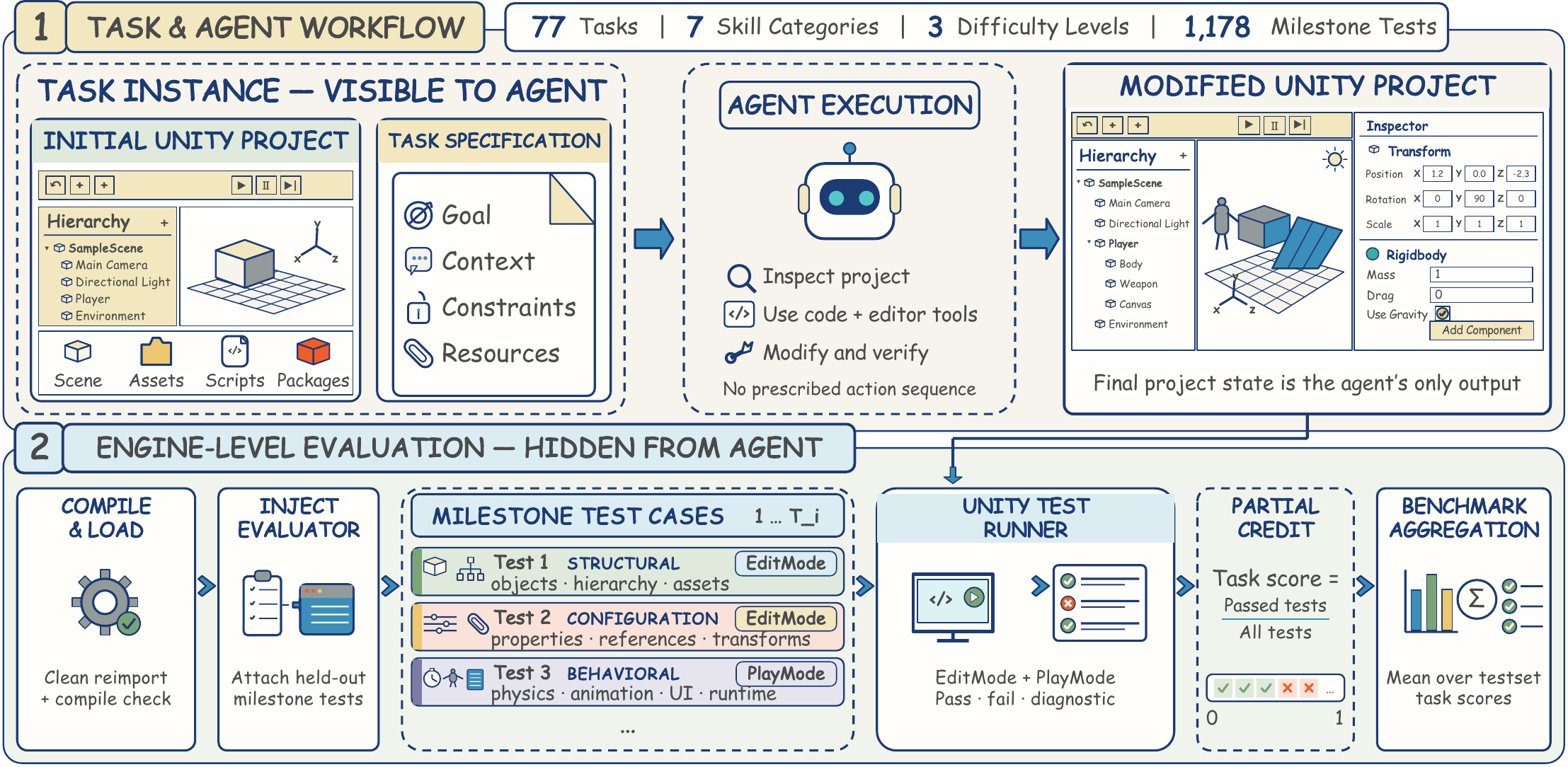}
    \caption{\textbf{UnityBench-Mini3D overview.} Agents build fully functional projects from initial challenge templates, with final project states evaluated by milestone tests.}
    \label{fig:main}
\end{figure}

\subsection{Evaluation Process}
\label{subsec:scoring}
As shown in Figure~\ref{fig:main}, our workflow separates agent execution from held-out engine-level evaluation. The agent receives only the initial project and task specification and may inspect, modify, and verify the project using the file, code, and Unity Editor tools provided by its harness. We prescribe no action sequence or implementation strategy; the final project state is the agent's only output. The milestone evaluator remains hidden until execution ends and is then injected into the project.

Evaluation begins from this final state. We reset Unity's generated library state and perform a clean reimport and compilation check. If compilation succeeds, the evaluator is injected and all milestone tests are run through the Unity Test Runner, loading relevant scenes as needed. EditMode tests inspect hierarchies, components, serialized properties, transforms, and asset references, whereas PlayMode tests assess runtime behavior, including physics, animation, and interface responses. Each test returns an independent pass/fail result and a diagnostic message.

For task \(i\) with \(T_i\) milestone tests, its score is \(s_i = \frac{1}{T_i}\sum_{j=1}^{T_i}\mathbb{I}[\text{test }j\text{ passes}]\). The overall benchmark score is \(S = \frac{1}{N}\sum_{i=1}^{N}s_i\), where \(N\) is the number of tasks. This provides requirement-level partial credit while weighting each task equally. See Appendix~\ref{app:formula-based-validation} details formula-based metrics and task-specific tolerances. We retain all test outcomes and separately record agent status (completed, timed out, or crashed) and evaluation status (completed, compilation error, or editor launch failure), distinguishing failed implementations from runs that could not be evaluated.

\section{Evaluation}

\subsection{Experiment Setup}
\label{sec:experiment}
\paragraph{Model and Harness Selection.}
We evaluate 21 proprietary and open-weight models across nine coding-agent harnesses, covering major model families and capability tiers. Models are evaluated in their provider-native harnesses whenever available, since agentic performance can depend on the integration between a model and its harness. To examine harness effects, we also evaluate a shared subset of models using two third-party systems with distinct designs. \texttt{TJ Codely} is Unity-specific and provides dedicated tools for interacting with the editor, whereas \texttt{Anomaly OpenCode} is an open-source, general-purpose coding harness. This design enables comparisons among provider-native, domain-specific, and general-purpose harnesses while helping distinguish the effects of model choice from harness design.

\paragraph{Configuration and Environment}
\label{sec:config}
We cap each run at one hour of wall-clock time, providing a common budget across harnesses with different loop structures. We conduct one attempt per task and report pass@1, avoiding best-of-\(k\) selection and the cost of repeatedly executing long-running Unity tasks. We report milestone pass rates overall and separately by difficulty and task category, together with wall-clock time and estimated cost. Costs are calculated from input and output token usage using reference prices from OpenRouter. Each challenge runs in an isolated environment with the Unity Editor and all project dependencies preinstalled without external internet access except for the model calls required for inference, preventing state from carrying across tasks. These environments execute in parallel on a shared server cluster with fixed resource limits. Harnesses are accessed through subscriptions, and used as distributed with fixed version. Unlike provider-native harnesses, Codely and OpenCode are standalone harnesses without first-party model services. We therefore connect them to models through API keys. Beyond supplying these credentials and selecting the highest available reasoning effort when supported, we retain each harness's default configuration.

\subsection{Evaluation Results}

\paragraph{Performance and Efficiency}
Table~\ref{tab:performance}(a) summarizes the performance of 32 combinations of models and harnesses on UnityBench-Mini3D. The strongest configurations achieve overall scores of about 70\%, and 25 of 32 exceed 50\%. Despite the inclusion of the easy split, 20 configurations remain between 50\% and 70\%, indicating substantial headroom. Models are also clearly stratified within a fixed harness. On Codely, Claude Opus 5 exceeds Seed 2.1 Pro by 44.35pp, or a factor of 2.82. On OpenCode, Kimi K3 outperforms DeepSeek V4 Flash by 60.06\% in relative terms. Hunyuan 3 and MiniMax M3 remain near 50\%, below the next tier, whereas configurations above 60\% are compressed within a range of roughly 11.20pp.

Comparisons of the same model across harnesses reveal two patterns. First, most models evaluated under multiple harnesses perform best in their provider's own harness. This pattern holds for GPT 5.6 Sol, Claude Opus 5, Kimi K3, Grok 4.6, and both DeepSeek models, with Qwen 3.8 Max as the sole exception. This may reflect closer alignment between provider harnesses and their models' preferred prompt formats, tool schemas, and iteration patterns. Second, Codely, tailored for Unity, consistently outperforms the general coding harness OpenCode for every shared model. The gain is largest for DeepSeek V4 Flash at 23.01pp, compared with 10.50pp for GLM 5.2 and 4.24pp for Kimi K3. These differences are comparable to those between models. DeepSeek V4 Flash spans 26.27pp across three harnesses, exceeding the entire spread of the leading group. Thus, when model performance is otherwise similar, harness choice can matter as much as model choice.

\definecolor{ubMarkRed}{HTML}{D94801}
\definecolor{ubMarkBlue}{HTML}{006FB9}
\definecolor{ubMarkCyan}{HTML}{008FA8}
\definecolor{ubMarkGold}{HTML}{A66E00}
\definecolor{ubMarkGreen}{HTML}{007A5E}
\definecolor{ubMarkPurple}{HTML}{7B3F98}
\definecolor{ubMarkMagenta}{HTML}{C2185B}
\definecolor{ubMarkCharcoal}{HTML}{374151}

\newcommand{\ubvectorshape}[2]{%
  \tikz[baseline=-0.48ex,x=0.9ex,y=0.9ex]{%
    \path[use as bounding box] (-0.45,-0.45) rectangle (0.45,0.45);%
    \begin{scope}[draw=#1,fill=none,line width=0.36pt,line join=round,line cap=round]
      #2%
    \end{scope}%
  }%
}
\newcommand{\ubmodelmark}[2]{%
  \nobreak\hspace{0.16em}%
  \makebox[1.62em][c]{%
    {\tiny\textcolor{#1}{[}}%
    \ubvectorshape{#1}{#2}%
    {\tiny\textcolor{#1}{]}}%
  }%
}
\newcommand{\ubOpusMark}{%
  \ubmodelmark{ubMarkRed}{\draw (0,0) circle (0.34);}%
}
\newcommand{\ubSolMark}{%
  \ubmodelmark{ubMarkBlue}{\draw (0,0.39)--(0.39,0)--(0,-0.39)--(-0.39,0)--cycle;}%
}
\newcommand{\ubKimiMark}{%
  \ubmodelmark{ubMarkCyan}{\draw (-0.39,-0.33)--(0.39,-0.33)--(0,0.39)--cycle;}%
}
\newcommand{\ubDSProMark}{%
  \ubmodelmark{ubMarkGold}{\draw (-0.34,-0.34) rectangle (0.34,0.34);}%
}
\newcommand{\ubDSFlashMark}{%
  \ubmodelmark{ubMarkGreen}{\draw (-0.39,0.33)--(0.39,0.33)--(0,-0.39)--cycle;}%
}
\newcommand{\ubGLMMark}{%
  \ubmodelmark{ubMarkPurple}{%
    \draw (90:0.40)--(54:0.17)--(18:0.40)--(-18:0.17)--(-54:0.40)--%
          (-90:0.17)--(-126:0.40)--(-162:0.17)--(162:0.40)--(126:0.17)--cycle;%
  }%
}
\newcommand{\ubQwenMark}{%
  \ubmodelmark{ubMarkMagenta}{%
    \draw (90:0.40)--(18:0.40)--(-54:0.40)--(-126:0.40)--(162:0.40)--cycle;%
  }%
}
\newcommand{\ubGrokMark}{%
  \ubmodelmark{ubMarkCharcoal}{%
    \draw (0:0.40)--(60:0.40)--(120:0.40)--(180:0.40)--(240:0.40)--(300:0.40)--cycle;%
  }%
}

\newcommand{\ubbest}[1]{\textcolor{zChampion}{\bfseries #1}}
\newcommand{\ubsecond}[1]{\textcolor{zGoldDeep}{\bfseries #1}}
\newcommand{\ublowbest}[1]{\textcolor{ubMarkGreen}{\bfseries #1}}
\newcommand{\ublowsecond}[1]{\textcolor{ubMarkPurple}{\bfseries #1}}

\begin{table*}[t]
\centering\scriptsize
\setlength{\tabcolsep}{2.6pt}
\renewcommand{\arraystretch}{0.92}
\setlength{\aboverulesep}{0.18ex}
\setlength{\belowrulesep}{0.22ex}
\setlength{\abovecaptionskip}{2pt}
\setlength{\belowcaptionskip}{1pt}
\caption{Performance of normal coding and CUA harnesses on UnityBench-Mini3D. O.a. is the overall pass rate and C.\ and T.\ denote average per-task API cost and wall-clock time.}
\label{tab:performance}

\begin{tabular}{@{}ll ccccccc c ccc cc@{}}
\toprule
& & \multicolumn{7}{c}{\textbf{Task category (\%)}} &
& \multicolumn{3}{c}{\textbf{Difficulty (\%)}} &
\multicolumn{2}{c}{\textbf{Efficiency}} \\
\cmidrule(lr){3-9}\cmidrule(lr){11-13}\cmidrule(lr){14-15}
\textbf{Harness} & \textbf{Model} & \textbf{UI} & \textbf{Anm.} & \textbf{Edtr.} & \textbf{Scn.} &
\textbf{P.\&S.} & \textbf{Phys.} & \textbf{Misc.} & \textbf{O.a.\,(\%)} &
\textbf{Easy} & \textbf{Med.} & \textbf{Hard} & \textbf{C.\,(\$)} & \textbf{T.\,(m)} \\
\midrule
\multicolumn{15}{c}{\textbf{(a). Coding Harnesses}} \\
\midrule
\ubharness{11}{TJ}{Codely}
 & Claude Opus 5\ubOpusMark & 62.67 & 69.48 & \ubbest{88.53} & 49.03 & 87.00 & 70.08 & 58.80 & 68.76 & \ubsecond{94.54} & 64.19 & 46.49 & 49.98 & 31.09 \\
 & GPT 5.6 Sol\ubSolMark & 45.28 & 60.59 & 67.68 & 50.81 & 83.06 & 72.15 & 54.82 & 62.33 & 84.57 & 59.18 & 42.13 & 19.14 & 38.67 \\
 & Kimi K3\ubKimiMark & 59.46 & 69.99 & 72.74 & 52.56 & 82.53 & 69.73 & 55.38 & 66.20 & 88.32 & 68.60 & 39.13 & 17.29 & 37.37 \\
 & DeepSeek V4 Pro\ubDSProMark & 62.87 & 61.00 & 75.50 & 52.22 & 73.82 & 62.37 & 55.10 & 63.09 & 90.47 & 64.36 & 31.72 & 8.25 & 34.69 \\
 & DeepSeek V4 Flash\ubDSFlashMark & 58.03 & 76.23 & 70.89 & 37.29 & 79.52 & 65.25 & 51.77 & 61.72 & 89.19 & 59.90 & 34.16 & 4.03 & 35.13 \\
 & MiniMax M3 & 59.80 & 57.14 & 59.80 & 38.30 & 60.13 & 58.66 & 16.44 & 52.28 & 77.62 & 47.55 & 30.70 & 5.28 & 23.53 \\
 & GLM 5.2\ubGLMMark & 63.85 & 66.11 & 77.41 & 42.32 & 76.41 & 70.98 & 29.17 & 62.45 & 84.10 & 63.35 & 37.80 & 12.27 & 30.37 \\
 & Qwen 3.8 Max\ubQwenMark & \ubbest{74.34} & 66.40 & 78.41 & 52.10 & 77.06 & 67.24 & 50.71 & 66.68 & 92.50 & 59.04 & 48.26 & 13.28 & 33.87 \\
 & Grok 4.6\ubGrokMark & 60.81 & 62.56 & 79.53 & 45.82 & 81.43 & 64.85 & 62.14 & 63.90 & \ubbest{94.63} & 58.04 & 37.89 & 7.08 & 24.62 \\
 & Hunyuan 3 & 42.15 & 53.94 & 66.86 & 33.03 & 66.26 & 53.15 & 53.48 & 51.04 & 76.89 & 51.33 & 22.56 & \ublowbest{0.81} & 26.46 \\
 & Seed 2.1 Pro & 18.30 & 27.92 & 27.51 & 5.63 & 38.80 & 45.86 & 0.00 & 24.41 & 40.71 & 22.58 & 9.00 & 1.44 & 14.59 \\
\midrule
\ubharness{4}{Anomaly}{OpenCode}
 & Kimi K3\ubKimiMark & 48.69 & 62.83 & 59.68 & 56.35 & 79.36 & 65.71 & 58.80 & 61.96 & 84.63 & 67.66 & 30.13 & 10.29 & 38.59 \\
 & GPT 5.6 Sol\ubSolMark & 49.46 & 67.20 & 75.00 & 50.26 & 71.74 & 60.24 & 54.73 & 61.29 & 82.27 & 59.96 & 40.16 & 20.46 & 33.58 \\
 & GLM 5.2\ubGLMMark & 50.86 & 49.12 & 65.76 & 36.52 & 61.01 & 58.06 & 55.10 & 51.95 & 82.42 & 45.48 & 27.00 & 7.49 & 29.87 \\
 & DeepSeek V4 Flash\ubDSFlashMark & 36.93 & 23.56 & 50.38 & 29.54 & 46.33 & 49.93 & 48.43 & 38.71 & 61.15 & 31.92 & 22.90 & \ublowsecond{1.09} & 10.90 \\
\midrule
\ubharness{4}{OpenAI}{Codex}
 & GPT 5.6 Sol\ubSolMark & 65.80 & 74.65 & 72.57 & \ubsecond{61.69} & \ubbest{90.04} & 78.74 & 58.80 & \ubbest{72.49} & 90.25 & 66.97 & \ubbest{60.13} & 24.88 & 19.09 \\
 & GPT 5.6 Terra & \ubsecond{72.85} & 76.41 & 83.17 & 49.79 & 76.51 & 65.52 & 55.01 & 68.02 & 90.62 & 65.41 & 46.75 & 14.06 & 20.78 \\
 & GPT 5.6 Luna & 60.28 & 77.38 & \ubsecond{85.05} & 47.19 & 74.22 & 67.50 & 58.43 & 66.38 & 88.54 & 65.95 & 42.85 & 1.94 & 20.42 \\
 & GPT 5.5 & 67.72 & 76.05 & 76.61 & 48.30 & 78.34 & 64.18 & 55.58 & 66.17 & 91.21 & 63.01 & 42.94 & 22.59 & 13.24 \\
\midrule
\ubharness{2}{Anthropic}{Claude Code}
 & Claude Opus 5\ubOpusMark & 71.07 & 69.69 & 73.48 & \ubbest{67.11} & 80.35 & 74.06 & \ubsecond{62.22} & \ubsecond{71.75} & 85.06 & \ubbest{74.34} & \ubsecond{54.02} & 49.81 & 33.99 \\
 & Claude Sonnet 5 & 47.86 & 57.00 & 55.76 & 46.98 & 62.95 & 56.07 & 55.10 & 54.00 & 79.56 & 49.84 & 31.47 & 17.00 & 31.59 \\
\midrule
\ubharness{2}{SpaceXAI}{Grok Build}
 & Grok 4.6\ubGrokMark & 66.11 & \ubsecond{78.78} & 67.78 & 60.53 & 83.04 & 75.36 & 58.80 & 70.62 & 91.56 & 66.09 & 53.57 & 11.63 & 19.22 \\
 & Grok 4.5 & 48.87 & \ubbest{81.07} & 80.81 & 58.08 & 79.94 & \ubbest{84.60} & 62.14 & 71.19 & 91.49 & \ubsecond{70.94} & 49.43 & 7.99 & 14.23 \\
\midrule
\ubharness{2}{Google}{Antigravity}
 & Gemini 3.1 Pro & 18.69 & 20.89 & 23.25 & 9.58 & 40.68 & 46.34 & 29.63 & 25.24 & 42.95 & 26.89 & 3.90 & 1.81 & \ublowbest{4.82} \\
 & Gemini 3.7 Flash & 31.26 & 26.28 & 22.84 & 14.28 & 64.55 & 37.06 & 33.33 & 31.19 & 60.69 & 23.94 & 8.25 & 6.48 & \ublowsecond{4.89} \\
\midrule
\ubharness{3}{Alibaba}{Qwen Code}
 & Qwen 3.8 Max\ubQwenMark & 55.17 & 59.79 & 72.94 & 56.91 & 81.07 & 57.24 & 32.59 & 62.31 & 93.65 & 56.34 & 35.76 & 13.19 & 41.10 \\
 & Qwen 3.7 Max & 20.12 & 58.35 & 61.31 & 39.21 & 48.07 & 60.83 & 44.73 & 47.80 & 64.02 & 51.86 & 25.06 & 5.74 & 13.28 \\
 & Qwen 3.7 Plus & 14.47 & 41.69 & 46.47 & 29.61 & 54.03 & 51.07 & 25.93 & 38.86 & 66.18 & 31.58 & 18.35 & 1.16 & 15.62 \\
\midrule
\ubharness{2}{Moonshot}{Kimi Code}
 & Kimi K3\ubKimiMark & 65.57 & 78.34 & 63.86 & 50.52 & 77.58 & 73.63 & 58.43 & 66.35 & 89.78 & 65.53 & 41.90 & 11.06 & 29.79 \\
 & Kimi K2.7 Code & 22.75 & 37.49 & 51.06 & 36.25 & 51.24 & 50.36 & 40.46 & 41.50 & 62.80 & 41.76 & 18.01 & 4.29 & 12.00 \\
\midrule
\ubharness{2}{DeepSeek}{DS Harness}
 & DeepSeek V4 Flash\ubDSFlashMark & 61.34 & 62.22 & 65.85 & 52.73 & 80.64 & 73.15 & \ubbest{68.83} & 64.98 & 84.31 & 63.26 & 46.16 & 4.98 & 30.79 \\
 & DeepSeek V4 Pro\ubDSProMark & 64.39 & 70.80 & 74.35 & 58.35 & \ubsecond{87.37} & \ubsecond{79.12} & 51.40 & 70.50 & 90.45 & 70.91 & 48.31 & 10.30 & 38.37 \\
\midrule
\multicolumn{15}{c}{\textbf{(b). CUA Harnesses}} \\
\midrule
\ubharness{2}{Simular}{Agent S}
 & Kimi K3 & 5.42 & 1.98 & 4.66 & 4.32 & 4.17 & 12.05 & 0.00 & 5.04 & 3.15 & 5.40 & 6.65 & -- & 7.31 \\
 & GPT 5.6 Sol & 22.24 & 7.53 & 29.96 & 18.76 & 23.48 & 40.76 & 0.00 & 22.16 & 31.85 & 23.92 & 9.43 & -- & 24.04 \\
\midrule
\ubharness{2}{ByteDance}{UI-TARS}
 & Kimi K3 & 5.42 & 10.31 & 3.52 & 3.81 & 15.13 & 12.05 & 25.93 & 8.77 & 16.86 & 3.73 & 6.34 & 2.02 & 48.11 \\
 & GPT 5.6 Sol & 19.99 & 13.25 & 25.91 & 9.58 & 28.07 & 57.86 & 0.00 & 23.11 & 38.50 & 21.59 & 8.28 & 3.74 & 44.34 \\
\bottomrule
\end{tabular}
\end{table*}

Cost varies far more than accuracy. Average cost per task differs by roughly 60x across the 32 coding configurations, while their scores differ by no more than 3x. Holding both provider and harness fixed, GPT 5.6 Sol exceeds GPT 5.6 Luna by only 6.11pp under Codex at nearly thirteen times the cost. Similarly, Claude Opus 5 performs within 4pp of Kimi K3 and Qwen 3.8 Max on Codely while costing up to four times as much. Small performance gains can therefore carry a substantial premium, making cost an independent dimension of model selection. Wall clock time varies less. Although runs are capped at one hour, most configurations take about half an hour per task. The two Antigravity configurations finish much faster but rank near the bottom. In contrast, Grok 4.5 with Grok Build combines a relatively short runtime with a score of 71.19\%, only 0.56pp below the best Anthropic result, at less than one sixth of the cost. Efficiency therefore depends on both runtime and cost, a distinction revisited in Section~\ref{sec:ablation}.

\paragraph{Tool Usage}
\label{sec:tooluse}
We collected the tools available in each harness and manually sorted them into seven functional categories. The full mapping is provided in Appendix~\ref{app:tools}. Figure~\ref{fig:toolusage} shows that harnesses use tools very differently, while the patterns become more similar when calls are combined for each model. For the eight models tested with multiple harnesses, engine interaction accounts for between 26\% and 37\% of all calls. This range is much narrower than the 13\% to 51\% range across harnesses, suggesting that harness design has a stronger influence on tool use than model identity. Codely provides the clearest example. With dedicated MCP engine operations, 51.44\% of Codely's calls involve engine interaction, more than twice the share of any other harness. This suggests that models use the engine directly when a suitable interface is available. Codex shows the opposite pattern. GPT 5.6 Sol frequently uses inspection tools when its calls are combined across harnesses, but it makes no inspection calls under Codex, where shell commands account for 54.21\% of all calls.

These tool use patterns are not closely related to performance. Grok Build spends over half of its calls reading and searching before changing the project, yet its strongest configuration differs from Codex's by only about 1pp in Table~\ref{tab:performance}. The total number of calls also provides little information about performance. DS Harness averages 119.77 calls per task, over 50\% above the harness average, while Kimi K3 performs as well as models making nearly twice as many calls with only 56.02. One pattern is shared across all configurations and differs from normal human practice. Although game development is highly visual, agents seldom inspect what they have built. Four harnesses never capture a rendered frame, and visual tools remain below 8.00\% of calls in any harness.

\begin{figure}[t]
    \centering
    \includegraphics[width=\linewidth]{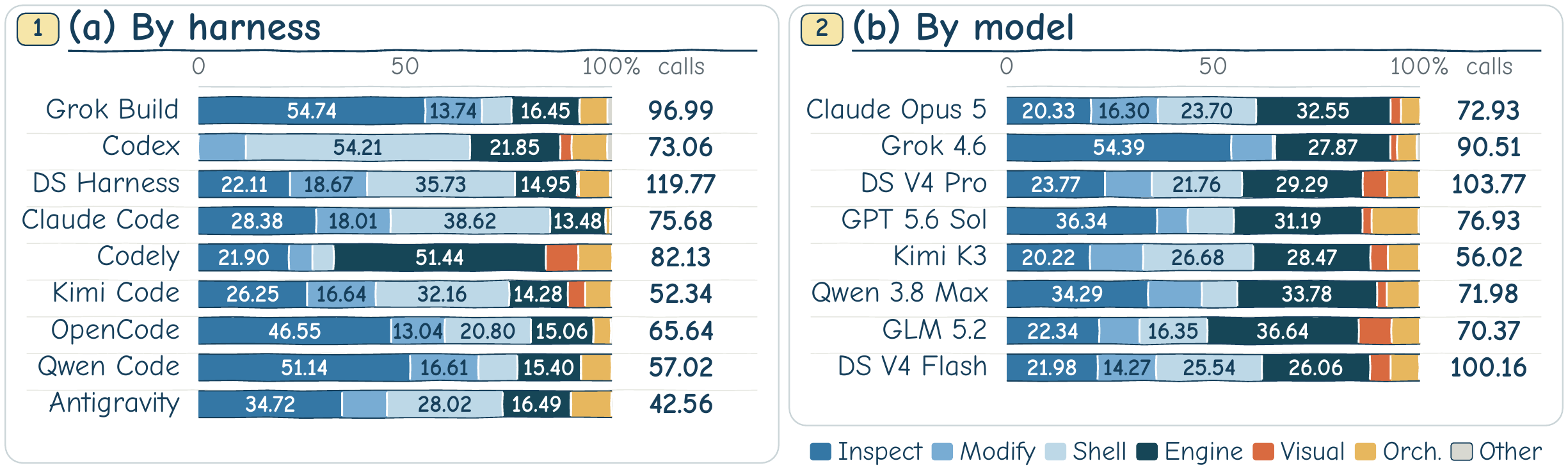}
    \caption{\textbf{Tool usage.} Panel (b) includes only models evaluated with at least two harnesses; values on the right report average calls per task.}
    \label{fig:toolusage}
\end{figure}

\paragraph{Category}
Table~\ref{tab:performance}(a) compares performance across the seven skill categories in UnityBench-Mini3D. Across all configurations, the particles and shaders category has the highest average score at 71.1\%, followed by editor tooling at 65.4\% and physics at 63.4\%. Scene construction has the lowest score at 44.3\%, below miscellaneous tasks at 48.6\%. This ranking appears to reflect task structure. Categories with higher scores often require changes to a single script or component setting, whereas lower scoring categories require agents to arrange many connected objects correctly. Scene construction remains difficult for every configuration. Its 67.11\% ceiling is the lowest among categories, while four others exceed 80\%. Dedicated engine tools do not fully solve this problem, as Codely still trails several provider harnesses on scene construction despite having dedicated Unity engine tools. Even weaker configurations can perform well in the higher scoring categories. Qwen 3.7 Plus scores only 14.47\% on user interface and 25.93\% on miscellaneous tasks, but exceeds 50\% on both physics and particles and shaders. Performance within a category can also change greatly across harnesses. Qwen 3.8 Max scores 74.34\% on user interface with Codely but 55.17\% with Qwen Code. DeepSeek V4 Flash scores 76.23\% on animation with Codely but 23.56\% with OpenCode, a difference of 52.67pp and about twice its overall score range across harnesses.

\paragraph{Difficulty}
\label{sec:difficulty}
The difficulty splits show the expected decline: average pass rate falls from 80.85\% on easy to 55.56\% on medium and 35.21\% on hard. Appendix~\ref{app:benchmark-calibration} provides difficulty diagnostics for each task. Of the 32 configurations, 22 exceed 80\% on easy, suggesting that individual skills are within reach but hard to combine at scale. Medium and hard tasks separate systems better: only 3 exceed 50\% on hard, and only GPT 5.6 Sol with Codex exceeds 60\%. Harness effects grow with difficulty. Moving GPT 5.6 Sol from Codely to Codex raises the easy score by 5.68pp but the hard score by 18.00pp, from 42.13\% to 60.13\%. Grok 4.6 is similar: its easy scores are nearly identical under Codely and Grok Build, but its hard scores differ by 15.68pp. Codely accounts for three of the five largest easy-hard gaps, showing its performance degrades sharply with increasing difficulty. The two strongest configurations, Claude Opus 5 with Claude Code and GPT 5.6 Sol with Codex, decline the least as difficulty rises. Claude Opus 5 loses only about 10.72pp from easy to medium, compared with over 20pp for most configurations. Model and harness improvements therefore appear most clearly on the medium and hard splits.

\paragraph{CUA Harnesses Performance}
Human game developers work mainly inside the GUI Editor rather than through the command line, so we further evaluate two Computer Use Agents (CUA) in Table~\ref{tab:performance}(b), Simular Agent S and ByteDance UI-TARS, which operate the editor through screenshots, mouse, and keyboard actions. Relative to each model's best coding-harness result, the CUA variants retain at most about one third of the score. In absolute score, CUA results peak at 38.50\% on the easy split and 9.43\% on the hard split, so individual editor operations are already unreliable and their errors compound over longer tasks within the same one-hour budget. Operating a GUI also appears to be a capability of its own rather than a slower form of coding: Kimi K3 matches or exceeds GPT 5.6 Sol under both coding harnesses they share, yet under UI-TARS reaches only 37.95\% of GPT 5.6 Sol's pass rate. Together with the Codex ablation in Section~\ref{sec:ablation}, where GPT 5.6 Sol remains close to its baseline score when computer use is added on top of code and shell access, these results suggest that the interface suited to human developers is not yet the one suited to agents, and that GUI control currently complements rather than replaces programmatic access.

\begin{table*}[t]
\centering
\caption{\textbf{Failure analysis.} Panel (b) includes only models evaluated with at least two harnesses. C1: missing or incomplete implementation; C2: spatial or numerical inaccuracy; C3: runtime behavior; C4: scene structure or asset wiring; C5: rendering or appearance.}
\label{tab:failure}
\scriptsize
\renewcommand{\arraystretch}{0.86}
\setlength{\aboverulesep}{0.18ex}
\setlength{\belowrulesep}{0.22ex}
\setlength{\tabcolsep}{1.6pt}

\begin{minipage}[t]{0.495\linewidth}
\centering
\textbf{(a) By harness}\\[0pt]
\begin{tabular*}{\linewidth}{@{}l@{\hspace{4pt}}r@{\hspace{3.5pt}}r@{\hspace{5pt}\extracolsep{\fill}}rrrrr@{}}
\toprule
& \multicolumn{2}{c}{Run-level (\%)} & \multicolumn{5}{c}{Failure mix (\%)} \\
\cmidrule(lr){2-3}\cmidrule(lr){4-8}
\textbf{Harness} & \textbf{TO} & \textbf{CE} & \textbf{C1} & \textbf{C2} & \textbf{C3} & \textbf{C4} & \textbf{C5} \\
\midrule
Grok Build    & 0.00 & 0.00 & 17.77 & \failmax{36.53} & 14.18 & 22.92 & 8.60 \\
OpenAI Codex  & 3.57 & 0.32 & 20.80 & 34.44 & 19.58 & 14.92 & 10.26 \\
DS Harness    & \failmax{21.43} & 1.30 & 23.28 & 28.92 & 20.66 & 16.53 & \failmax{10.61} \\
Claude Code   & 16.88 & 1.30 & 21.27 & 24.64 & \failmax{20.88} & 23.74 & 9.47 \\
Unity Codely  & 19.24 & 2.24 & 32.60 & 24.27 & 17.40 & 17.09 & 8.64 \\
Kimi Code     & 12.99 & 0.65 & 25.26 & 22.14 & 17.22 & \failmax{25.64} & 9.74 \\
OpenCode      & 18.83 & 1.95 & 24.21 & 25.18 & 18.29 & 21.79 & 10.53 \\
Qwen Code     & 11.69 & 2.16 & 32.94 & 19.50 & 20.31 & 18.27 & 8.98 \\
Antigravity   & 0.00 & \failmax{12.99} & \failmax{42.54} & 9.94 & 13.78 & 25.23 & 8.51 \\
\bottomrule
\end{tabular*}
\end{minipage}\hfill
\begin{minipage}[t]{0.495\linewidth}
\centering
\textbf{(b) By model}\\[0pt]
\begin{tabular*}{\linewidth}{@{}l@{\hspace{4pt}}r@{\hspace{3.5pt}}r@{\hspace{5pt}\extracolsep{\fill}}rrrrr@{}}
\toprule
& \multicolumn{2}{c}{Run-level (\%)} & \multicolumn{5}{c}{Failure mix (\%)} \\
\cmidrule(lr){2-3}\cmidrule(lr){4-8}
\textbf{Model} & \textbf{TO} & \textbf{CE} & \textbf{C1} & \textbf{C2} & \textbf{C3} & \textbf{C4} & \textbf{C5} \\
\midrule
Claude Opus 5
& 22.08 & 1.30 & 24.58 & 27.96 & 18.59 & 18.12 & \failmax{10.75} \\[0.72pt]
Grok 4.6
& 3.25 & 0.00 & 29.30 & 24.46 & 16.94 & \failmax{20.76} & 8.54 \\[0.72pt]
DeepSeek V4 Pro
& 27.27 & 1.30 & 24.93 & 26.73 & \failmax{20.91} & 16.90 & 10.53 \\[0.72pt]
GPT 5.6 Sol
& 18.18 & 1.30 & 24.75 & \failmax{31.13} & 16.47 & 17.96 & 9.69 \\[0.72pt]
Kimi K3
& \failmax{29.00} & 0.87 & \failmax{31.69} & 20.43 & 20.02 & 19.02 & 8.84 \\[0.72pt]
Qwen 3.8 Max
& 27.27 & 2.60 & 26.81 & 29.70 & 20.23 & 13.14 & 10.12 \\[0.72pt]
GLM 5.2
& 14.94 & 1.30 & 28.06 & 24.31 & 18.97 & 18.58 & 10.08 \\[0.72pt]
DeepSeek V4 Flash
& 13.85 & \failmax{3.03} & 23.97 & 29.86 & 18.28 & 18.01 & 9.88 \\
\bottomrule
\end{tabular*}
\end{minipage}

\end{table*}

\subsection{Failure Analysis}
\label{sec:failure}
We use expert-defined failure categories. Table~\ref{tab:failure} groups each non-passing milestones into five categories, showing what aggregate scores can hide: two configurations may fail equally often for different reasons. This contrast is clearest between Grok Build and Antigravity, which lie at opposite ends of the harness results. Grok Build places two configurations among the top five in Table~\ref{tab:performance}(a), has only 17.77\% of its non-passing milestones in missing or incomplete implementation, but 36.53\% in spatial or numerical inaccuracies. Antigravity, which ranks near the bottom, shows the reverse pattern at 42.54\% and 9.94\%. This suggests that weaker configurations more often leave work missing or incomplete, while stronger ones create the required elements but place or size them incorrectly. Failure patterns also vary more across harnesses than across models, with C1 spanning 24.77pp across harnesses but only 7.72pp across models. Further breakdowns appear in Appendix~\ref{app:behavioral-diagnostics}.

Results at the run level challenge two common assumptions. Compilation is no longer the main bottleneck. Apart from Antigravity at 12.99\%, compilation failure rates remain at or below 3.03\% and account for only 2.27\% of all runs. What the code does matters more than whether it compiles. Timeouts also reveal little about capability. Grok Build in Table~\ref{tab:performance}(a) does not have a single timeout, while DS Harness has the highest timeout rate at 21.43\% but still produces DeepSeek V4 Pro, one of the strongest configurations at 70.50\%. Qwen Code times out on only 11.69\% of runs while producing some of the weakest results. Reaching the one hour limit often means that an agent is still working, while stopping early does not imply efficiency. We also observe rare apparent false-positive model guardrail triggers that cause failures by prematurely terminating otherwise valid runs, as illustrated in Appendix~\ref{app:security-guardrail}. Rendering failures remain between 8.51\% and 10.75\% among non-passing milestones and show no clear link to visual inspection. At the harness level, the three lowest rates belong to Antigravity and Grok Build, which never capture a frame, and Codely, which uses screenshots most often. The limited benefit of multimodal input in Section~\ref{sec:ablation} therefore appears to come from the nature of the tasks rather than weak visual tools.

\subsection{Ablation Study}
\label{sec:ablation}

\begin{table}[t]
\centering
\caption{\textbf{Ablation study.} Results cover multimodal input, engine tools, and CUA; $\uparrow$/$\downarrow$ and colors indicate improvement or degradation relative to the corresponding baseline.}
\label{tab:ablation}

\footnotesize
\setlength{\tabcolsep}{0.45pt}
\renewcommand{\arraystretch}{1.08}
\resizebox{\textwidth}{!}{%
\begin{tabular}{@{}l@{\hspace{2pt}}l@{\hspace{2pt}}l@{\hspace{3.5pt}}*{13}{c}@{}}
\toprule
\multicolumn{3}{c}{\textbf{Configuration}} &
\multicolumn{7}{c}{\textbf{Task category (\%)}} &
&
\multicolumn{3}{c}{\textbf{Difficulty (\%)}} &
\multicolumn{2}{c}{\textbf{Efficiency}} \\
\cmidrule(lr){1-3}\cmidrule(lr){4-10}\cmidrule(lr){12-14}\cmidrule(lr){15-16}
\textbf{Setting} & \textbf{Harness} & \textbf{Model} &
\textbf{UI} & \textbf{Anm.} & \textbf{Edtr.} & \textbf{Scn.} &
\textbf{P.\&S.} & \textbf{Phys.} & \textbf{Misc.} & \textbf{O.a.\,(\%)} &
\textbf{Easy} & \textbf{Med.} & \textbf{Hard} &
\textbf{C.\,(\$)} & \textbf{T.\,(m)} \\
\midrule

\absetting{w/o}{MM}
& OpenCode & Kimi K3
& \abup{60.46}{11.77} & \abup{69.32}{6.49} & \abup{65.30}{5.62} & \abdown{52.91}{3.44} & \abup{83.16}{3.80} & \abdown{62.72}{2.99} & \abdown{31.99}{26.81} & \abup{63.42}{1.46} & \abup{91.51}{6.88} & \abdown{64.83}{2.83} & \abup{31.11}{0.98} & \abdown{13.64}{3.35} & \abup{34.78}{3.81} \\
& Codex & GPT 5.6 Sol
& \abdown{65.19}{0.61} & \abdown{73.38}{1.27} & \abup{88.08}{15.51} & \abup{61.99}{0.30} & \abdown{87.14}{2.90} & \abdown{74.66}{4.08} & \abdown{58.43}{0.37} & \abup{73.46}{0.97} & \abup{92.31}{2.06} & \abup{67.33}{0.36} & \abup{60.69}{0.56} & \abup{18.46}{6.42} & \abup{14.90}{4.19} \\
& Codely & Kimi K3
& \abup{65.24}{5.78} & \abup{72.24}{2.25} & \abup{78.35}{5.61} & \abdown{50.57}{1.99} & \abup{83.16}{0.63} & \abup{73.02}{3.29} & \abdown{55.10}{0.28} & \abup{68.09}{1.89} & \abup{94.10}{5.78} & \abdown{68.32}{0.28} & \abup{39.53}{0.40} & \abup{16.64}{0.65} & \abup{34.37}{3.00} \\

\midrule
\absetting{w/o}{MCP}
& \multirow{3}{*}{Codely} & GPT 5.6 Sol
& \abup{51.32}{6.04} & \abup{62.55}{1.96} & \abup{73.37}{5.69} & \abdown{44.01}{6.80} & \abdown{58.33}{24.73} & \abdown{64.45}{7.70} & \abdown{29.63}{25.19} & \abdown{56.54}{5.79} & \abdown{80.42}{4.15} & \abdown{53.40}{5.78} & \abdown{34.54}{7.59} & \abdown{25.44}{6.30} & \abup{26.42}{12.25} \\
& & GLM 5.2
& \abdown{56.84}{7.01} & \abdown{55.44}{10.67} & \abdown{71.07}{6.34} & \abdown{33.26}{9.06} & \abdown{67.13}{9.28} & \abdown{62.20}{8.78} & \abup{51.77}{22.60} & \abdown{55.01}{7.44} & \abdown{82.66}{1.44} & \abdown{51.60}{11.75} & \abdown{29.26}{8.54} & \abup{11.31}{0.96} & \abup{24.64}{5.73} \\
& & Kimi K3
& \abup{69.03}{9.57} & \abdown{66.14}{3.85} & \abdown{62.02}{10.72} & \abdown{46.45}{6.11} & \abdown{78.61}{3.92} & \abup{78.45}{8.72} & \abdown{52.14}{3.24} & \abdown{64.19}{2.01} & \abdown{84.36}{3.96} & \abdown{65.46}{3.14} & \abup{40.65}{1.52} & \abup{10.75}{6.54} & \abup{32.70}{4.67} \\

\midrule
\absetting{force}{CUA}
& \multirow{3}{*}{Codex} & GPT 5.6 Sol
& \abup{77.01}{11.21} & \abup{75.48}{0.83} & \abup{76.77}{4.20} & \abdown{59.75}{1.94} & \abdown{84.07}{5.97} & \abdown{63.25}{15.49} & \abup{62.14}{3.34} & \abdown{71.03}{1.46} & \abdown{90.24}{0.01} & \abdown{63.71}{3.26} & \abdown{59.40}{0.73} & \abdown{38.90}{14.02} & \abdown{27.56}{8.47} \\
& & GPT 5.6 Terra
& \abdown{55.83}{17.02} & \abdown{63.37}{13.04} & \abdown{53.12}{30.05} & \abdown{44.78}{5.01} & \abdown{63.85}{12.66} & \abdown{65.15}{0.37} & \abdown{25.47}{29.54} & \abdown{55.29}{12.73} & \abdown{67.17}{23.45} & \abdown{59.25}{6.16} & \abdown{37.37}{9.38} & \abdown{26.44}{12.38} & \abdown{24.22}{3.44} \\
& & GPT 5.6 Luna
& \abdown{49.04}{11.24} & \abdown{46.58}{30.80} & \abdown{64.08}{20.97} & \abdown{31.20}{15.99} & \abdown{52.46}{21.76} & \abdown{59.69}{7.81} & \abdown{6.67}{51.76} & \abdown{46.80}{19.58} & \abdown{59.49}{29.05} & \abdown{44.66}{21.29} & \abdown{35.73}{7.12} & \abdown{4.46}{2.52} & \abdown{33.04}{12.62} \\

\bottomrule
\end{tabular}
}
\end{table}

\paragraph{Impact of Multimodality}
Table~\ref{tab:ablation} supports the observation in Section~\ref{sec:tooluse}: agents rarely use visual tools even when screenshots are available. Scores rise slightly in all three reported no-MM runs: GPT 5.6 Sol with Codex, Kimi K3 with Codely, and Kimi K3 with OpenCode gain 0.97, 1.89, and 1.46pp, respectively. The archived Codex batch may also disable MCP, so we do not treat it as a clean single-factor ablation. All three also finish 3.00 to 4.19 minutes sooner per task, while the Codex configuration costs 26\% less. Two factors may explain this. First, visual input may be redundant because relevant Unity scene state is stored as text. Agents can inspect task requirements and textual project state without rendering a frame, while image tokens may consume context needed for planning and tool calls. Second, models may struggle to recover precise spatial placement and object hierarchies from images, consistent with scene construction being the weakest category in Table~\ref{tab:performance}. Overall, the clean comparisons suggest that multimodal input is not the main factor limiting performance in these configurations. Appendix~\ref{app:case-multimodality} presents contrasting multimodal case studies.

\paragraph{Engine Tools}
Removing Codely's dedicated engine tools and forcing editor access through the shell lowers overall scores for every model tested. Scores fall by 5.79, 7.44, and 2.01pp for GPT 5.6 Sol, GLM 5.2, and Kimi K3, respectively. Average loss is greatest on medium tasks at 6.89pp, versus 3.18pp on easy and 4.87pp on hard. This suggests engine tools matter most at intermediate complexity: easy tasks remain tractable via the shell, while hard tasks remain constrained by planning. If tools merely made actions more convenient, removing them should lengthen runs. Instead, runs end 4.67 to 12.25 minutes earlier while all three scores decline, and GPT 5.6 Sol costs 33\% more. Earlier termination with lower scores suggests that engine tools support continued progress rather than merely faster execution. Appendix~\ref{app:case-engine-specific} presents an illustrative case study.

\paragraph{Impact of CUA}
Codex provides CUA tools that all three models must use in this ablation. Rather than consistently improving performance, CUA widens the gap between models. The overall-score range across the three GPT configurations grows from 6.11pp at baseline to 24.23pp with CUA. GPT 5.6 Sol stays near baseline, declining by 1.46pp, while GPT 5.6 Terra and Luna decline by 12.73 and 19.58pp. GPT 5.6 Sol improves by 11.21pp on user-interface tasks, suggesting that visual grounding helps when targets are directly visible on the rendered screen. No other result improves consistently. All three configurations become slower and more expensive without an overall gain. The weaker models decline most on the easy split, where interface navigation adds overhead to tasks already inexpensive to solve directly. For these configurations, GUI control therefore adds another interaction channel but does not improve their ability to solve tasks.

\section{Conclusion, Future Work and Broader Impact}
We introduced UnityBench-Mini3D, an open-access benchmark of 77 project-level Unity challenges spanning seven development skills, with human-authored milestone tests that evaluate final project states directly in the engine. We evaluate 21 models across nine coding harnesses and further compare two CUA harnesses, showing that success depends not only on the underlying model, but also on how the harness enables it to interact with the development environment. Dedicated engine tools improve performance, whereas simply adding more tool calls, longer runtimes, multimodal input, or GUI control does not consistently help. The main bottleneck lies not in producing compilable code, but in constructing coherent scenes with accurate spatial relationships. For future work, we will continue adding new challenges as models and harnesses evolve, with a particular focus on medium and hard tasks. We will introduce new categories, such as multiplayer development, optimization, and post processing, while improving the balance across existing categories. Regularly adding new challenges will also help reduce the risk of data contamination, and we will provide recommended subsets for evaluating agents in different game development domains. By revealing the gap between conventional coding and project-level game development through fine-grained milestone feedback, UnityBench-Mini3D can guide model and harness development and may ultimately make Unity prototyping more accessible to independent developers, educators, and non-expert creators.

\bibliographystyle{plainnat}
\bibliography{references}

\clearpage
\appendix

\section{UnityBench-Mini3D Challenge Scenes}
\label{app:challist}

\providecommand{\ubshots}[2]{}
\renewcommand{\ubshots}[2]{%
  \noindent
  \begin{minipage}[t]{0.487\linewidth}
    \centering
    \includegraphics[width=\linewidth]{figures/appendix/screenshot/initial-screenshots-640p/#1.jpg}%
    \par\vspace{2pt}{\scriptsize\bfseries Initial Challenge Template}
  \end{minipage}\hfill
  \begin{minipage}[t]{0.487\linewidth}
    \centering
    \includegraphics[width=\linewidth]{figures/appendix/screenshot/golden-screenshots-640p/#1.jpg}%
    \par\vspace{2pt}{\scriptsize\bfseries Completed Golden Reference}
  \end{minipage}%
}

\begingroup
\setlength{\tabcolsep}{0pt}
\renewcommand{\arraystretch}{1}
\setlength{\LTcapwidth}{14.4cm}
\footnotesize\color{zText}
\begin{longtable}{@{}p{14.4cm}@{}}
\caption{\textbf{Challenge list.} The 77 UnityBench-Mini3D tasks, grouped by skill category. Each entry identifies the initial project and the corresponding completed reference project; difficulty and held-out milestone-test count are shown in the header.}\label{tab:challenge-list}\\
\ubidentity{zRoyal}{white}{CHALLENGE LIST}{77 tasks \textperiodcentered\ 7 skill categories}\\[5pt]
\endfirsthead
\multicolumn{1}{@{}l@{}}{\textcolor{zRock}{\itshape Table \thetable\ continued from previous page.}}\\[3pt]
\ubidentity{zRoyal}{white}{CHALLENGE LIST}{continued}\\[5pt]
\endhead
\multicolumn{1}{r@{}}{\textcolor{zRock}{\itshape Continued on next page.}}\\
\endfoot
\endlastfoot
\noalign{\vskip 3pt}
\ubcategory{zChampion}{User Interface}{9 tasks \textperiodcentered\ 111 milestone tests}\\*[5pt]
\begin{minipage}[t]{\linewidth}
\ubtaskhead{zChampion}{0025-FPSCompassHUDCalibration}{Compass HUD Direction Calibration}{zZonai}{Easy}{6}
\vspace{4pt}
\ubshots{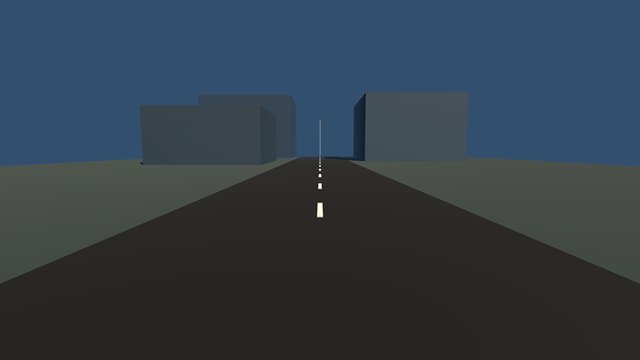}{zChampion}
\end{minipage}\\[6pt]
\begin{minipage}[t]{\linewidth}
\ubtaskhead{zChampion}{0050-UIToolkitSnake}{Snake Game Implemented with Unity UI Toolkit}{zZonai}{Easy}{23}
\vspace{4pt}
\ubshots{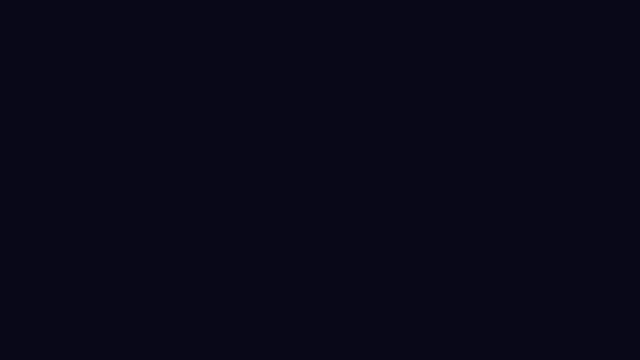}{zChampion}
\end{minipage}\\[6pt]
\begin{minipage}[t]{\linewidth}
\ubtaskhead{zChampion}{0063-MinimapRealization}{Minimap UI Implementation}{zZonai}{Easy}{7}
\vspace{4pt}
\ubshots{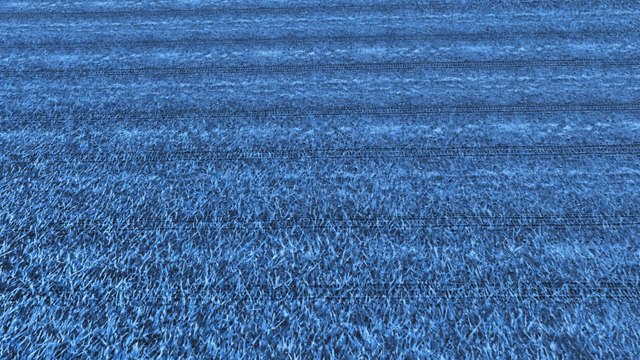}{zChampion}
\end{minipage}\\[6pt]
\begin{minipage}[t]{\linewidth}
\ubtaskhead{zChampion}{0010-SplitTheUI}{UI Sprite Sheet Slicing}{zGoldDeep}{Medium}{11}
\vspace{4pt}
\ubshots{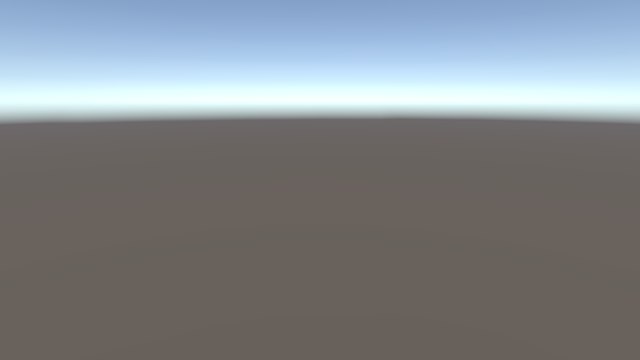}{zChampion}
\end{minipage}\\[6pt]
\begin{minipage}[t]{\linewidth}
\ubtaskhead{zChampion}{0023-UIRedoBench}{GoldenBench UI Recreation}{zGoldDeep}{Medium}{9}
\vspace{4pt}
\ubshots{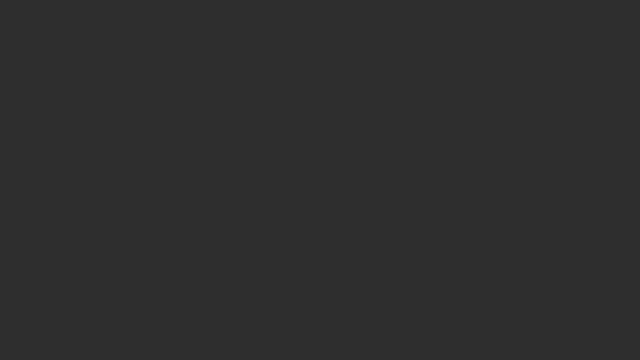}{zChampion}
\end{minipage}\\[6pt]
\begin{minipage}[t]{\linewidth}
\ubtaskhead{zChampion}{0049-UIDocument}{Recreate T-Rex Runner with UI Toolkit}{zGoldDeep}{Medium}{18}
\vspace{4pt}
\ubshots{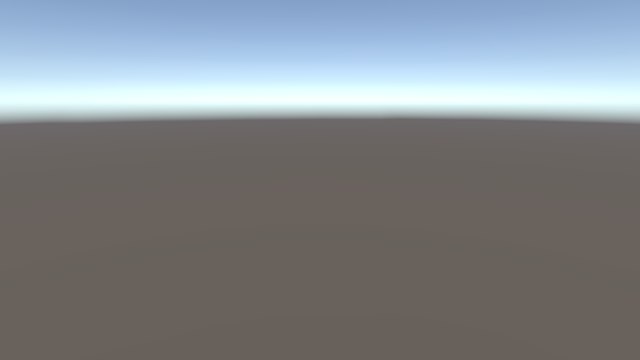}{zChampion}
\end{minipage}\\[6pt]
\begin{minipage}[t]{\linewidth}
\ubtaskhead{zChampion}{0015-ShootAheadUguiMobileHud}{ShootAhead Mobile HUD Refactor}{zShrine}{Hard}{9}
\vspace{4pt}
\ubshots{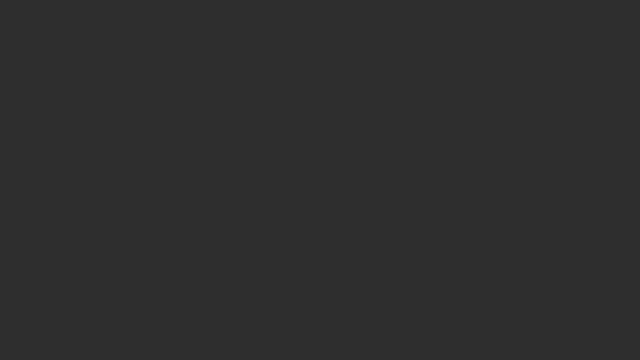}{zChampion}
\end{minipage}\\[6pt]
\begin{minipage}[t]{\linewidth}
\ubtaskhead{zChampion}{0062-UIToolkitBubbleDragon}{Build an Endless UI Toolkit Bubble Shooter}{zShrine}{Hard}{19}
\vspace{4pt}
\ubshots{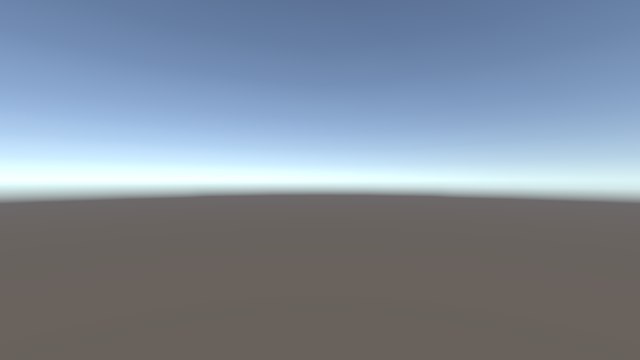}{zChampion}
\end{minipage}\\[6pt]
\begin{minipage}[t]{\linewidth}
\ubtaskhead{zChampion}{0078-JetHUD}{Build a First-Person F-22 Cockpit HUD}{zShrine}{Hard}{9}
\vspace{4pt}
\ubshots{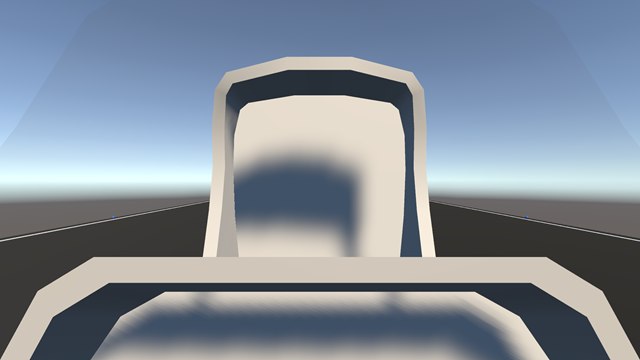}{zChampion}
\end{minipage}\\[6pt]
\noalign{\vskip 3pt}
\ubcategory{zDawn}{Animation \& Animator}{12 tasks \textperiodcentered\ 179 milestone tests}\\*[5pt]
\begin{minipage}[t]{\linewidth}
\ubtaskhead{zDawn}{0000-AutoDoor}{Build an Automatic Opening and Closing Door Scene (AutoDoor)}{zZonai}{Easy}{29}
\vspace{4pt}
\ubshots{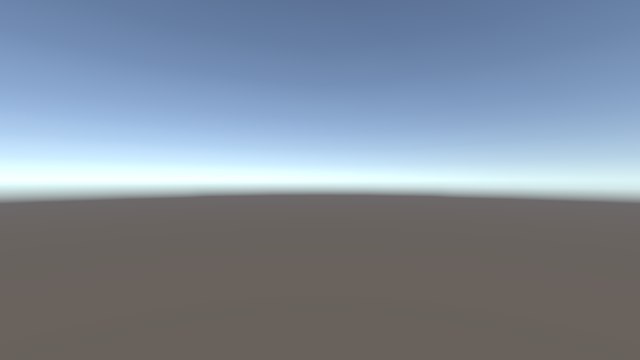}{zDawn}
\end{minipage}\\[6pt]
\begin{minipage}[t]{\linewidth}
\ubtaskhead{zDawn}{0006-TankTrackAnimation}{Tank Track Animation}{zZonai}{Easy}{2}
\vspace{4pt}
\ubshots{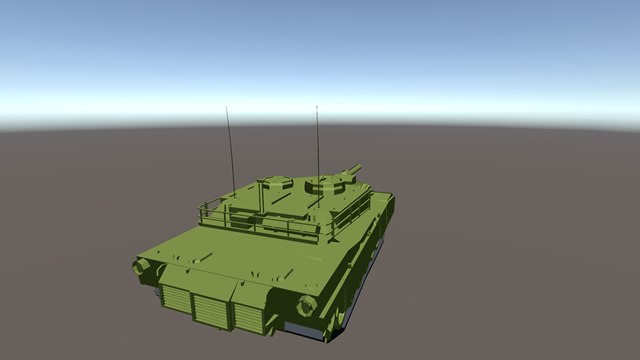}{zDawn}
\end{minipage}\\[6pt]
\begin{minipage}[t]{\linewidth}
\ubtaskhead{zDawn}{0008-SwordsmanAttack}{Swordsman Attack}{zZonai}{Easy}{4}
\vspace{4pt}
\ubshots{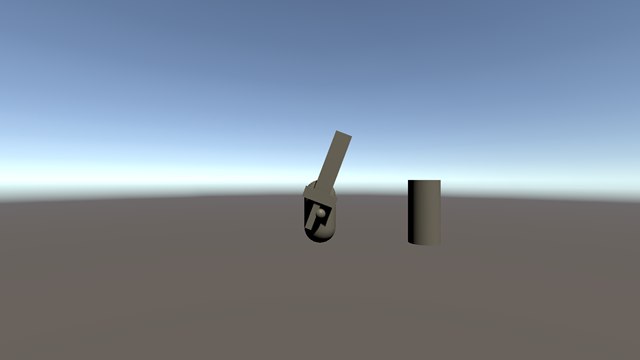}{zDawn}
\end{minipage}\\[6pt]
\begin{minipage}[t]{\linewidth}
\ubtaskhead{zDawn}{0009-AnimationEvents}{Animation Events}{zGoldDeep}{Medium}{9}
\vspace{4pt}
\ubshots{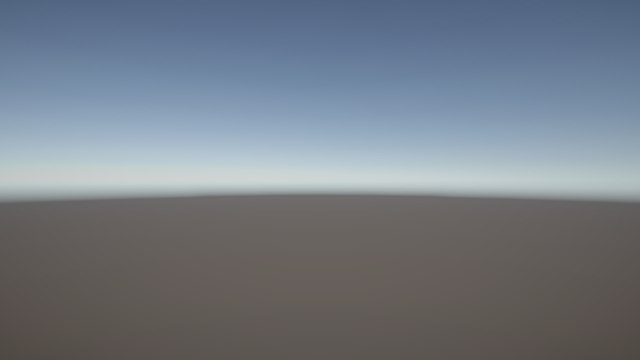}{zDawn}
\end{minipage}\\[6pt]
\begin{minipage}[t]{\linewidth}
\ubtaskhead{zDawn}{0029-AnimatorLayer}{Layered Character Animation}{zGoldDeep}{Medium}{16}
\vspace{4pt}
\ubshots{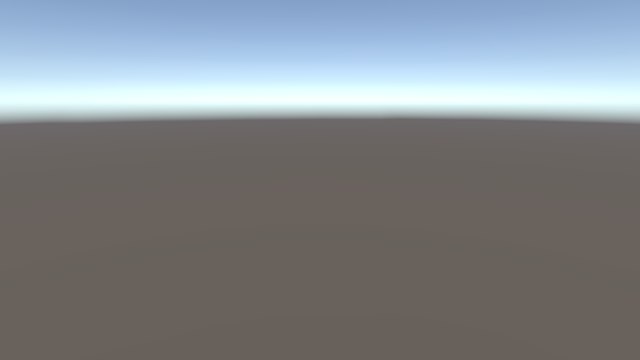}{zDawn}
\end{minipage}\\[6pt]
\begin{minipage}[t]{\linewidth}
\ubtaskhead{zDawn}{0035-HumanoidIK}{Humanoid IK}{zGoldDeep}{Medium}{13}
\vspace{4pt}
\ubshots{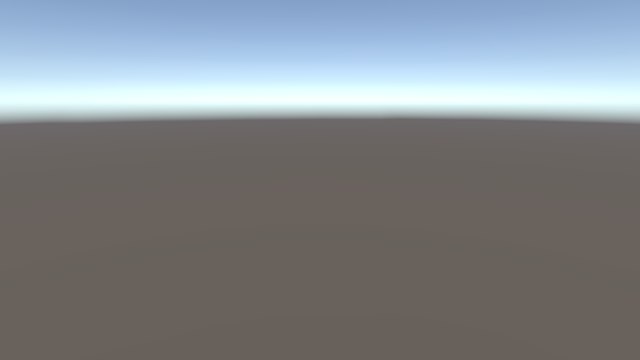}{zDawn}
\end{minipage}\\[6pt]
\begin{minipage}[t]{\linewidth}
\ubtaskhead{zDawn}{0041-ExcavatorAnimationClipDigDump}{Excavator AnimationClip Dig and Dump Task}{zGoldDeep}{Medium}{9}
\vspace{4pt}
\ubshots{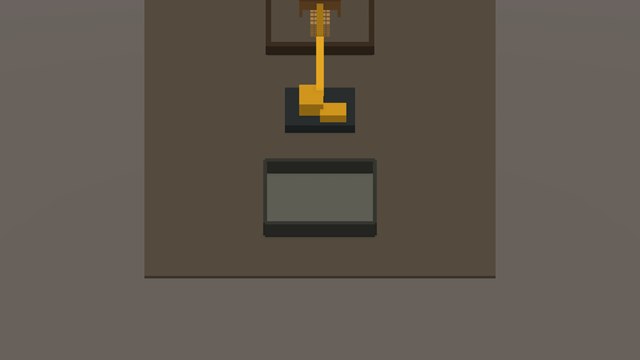}{zDawn}
\end{minipage}\\[6pt]
\begin{minipage}[t]{\linewidth}
\ubtaskhead{zDawn}{0046-AnimationScriptPlayable}{Animation Script Playable}{zGoldDeep}{Medium}{17}
\vspace{4pt}
\ubshots{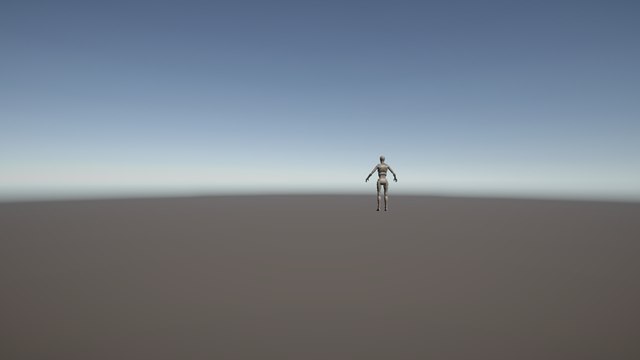}{zDawn}
\end{minipage}\\[6pt]
\begin{minipage}[t]{\linewidth}
\ubtaskhead{zDawn}{0055-ArcherAnimatorController}{Archer Animator Blend Tree Controller}{zGoldDeep}{Medium}{30}
\vspace{4pt}
\ubshots{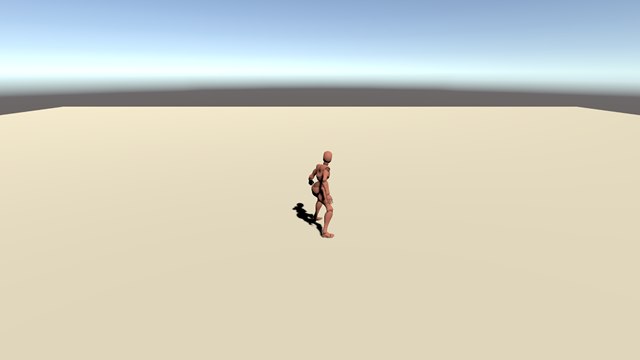}{zDawn}
\end{minipage}\\[6pt]
\begin{minipage}[t]{\linewidth}
\ubtaskhead{zDawn}{0065-ShootAheadSplitBodyAim}{ShootAhead Split Body Aim}{zGoldDeep}{Medium}{12}
\vspace{4pt}
\ubshots{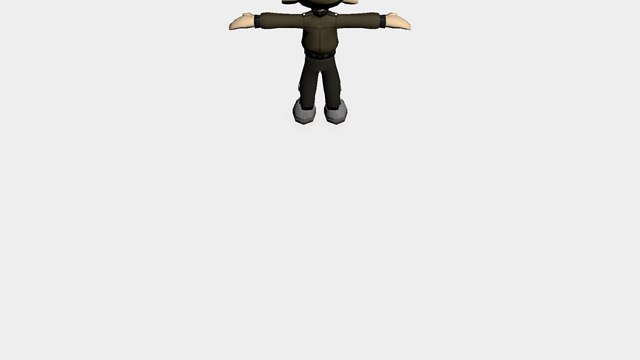}{zDawn}
\end{minipage}\\[6pt]
\begin{minipage}[t]{\linewidth}
\ubtaskhead{zDawn}{0051-AnimationClassification}{Animation Classification}{zShrine}{Hard}{20}
\vspace{4pt}
\ubshots{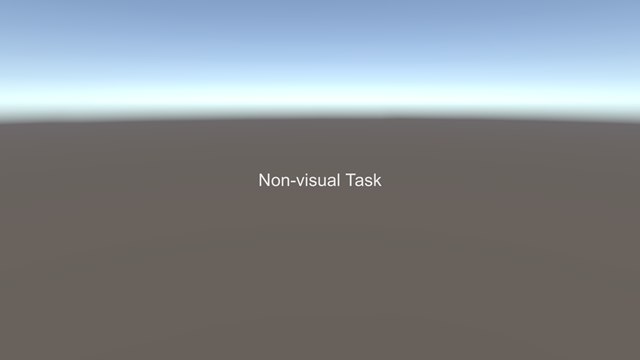}{zDawn}
\end{minipage}\\[6pt]
\begin{minipage}[t]{\linewidth}
\ubtaskhead{zDawn}{0079-FPSRemoteWeaponAnimationSync}{Synchronize Remote SCAR Weapon Presentation}{zShrine}{Hard}{18}
\vspace{4pt}
\ubshots{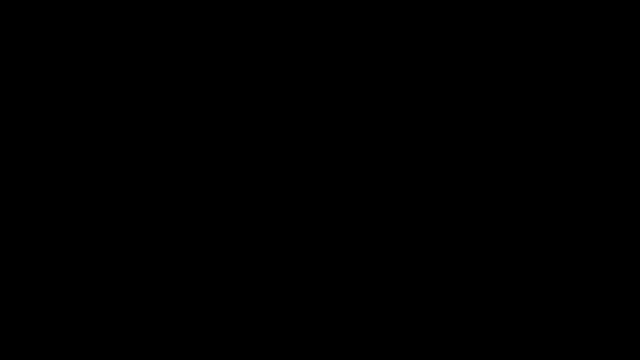}{zDawn}
\end{minipage}\\[6pt]
\noalign{\vskip 3pt}
\ubcategory{zRoyal}{Editor Operations}{11 tasks \textperiodcentered\ 142 milestone tests}\\*[5pt]
\begin{minipage}[t]{\linewidth}
\ubtaskhead{zRoyal}{0004-IslandBenchRoadClassification}{Island Bench Road Group Cleanup}{zZonai}{Easy}{9}
\vspace{4pt}
\ubshots{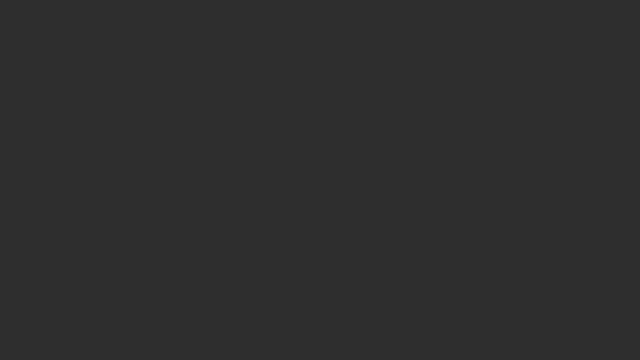}{zRoyal}
\end{minipage}\\[6pt]
\begin{minipage}[t]{\linewidth}
\ubtaskhead{zRoyal}{0013-M249MachineGun}{M249}{zZonai}{Easy}{15}
\vspace{4pt}
\ubshots{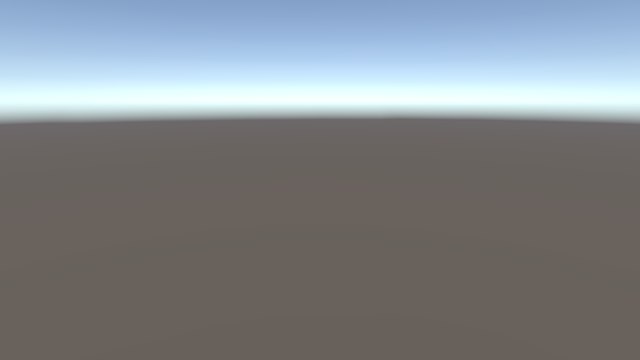}{zRoyal}
\end{minipage}\\[6pt]
\begin{minipage}[t]{\linewidth}
\ubtaskhead{zRoyal}{0014-ShootAheadAssetPruning}{ShootAhead Asset Pruning}{zZonai}{Easy}{15}
\vspace{4pt}
\ubshots{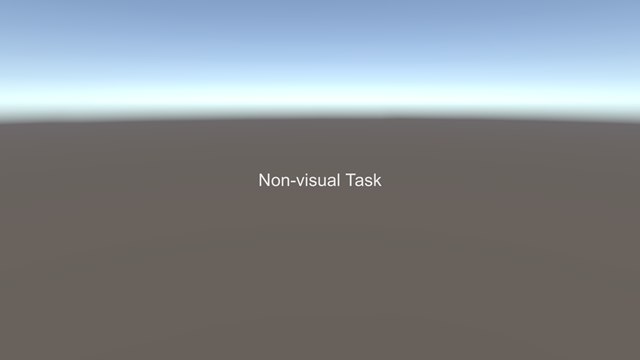}{zRoyal}
\end{minipage}\\[6pt]
\begin{minipage}[t]{\linewidth}
\ubtaskhead{zRoyal}{0027-VolumeBoundsOrganizer}{Automated Scene Grouping and Renaming Based on Spatial Bounding-Box Features}{zZonai}{Easy}{16}
\vspace{4pt}
\ubshots{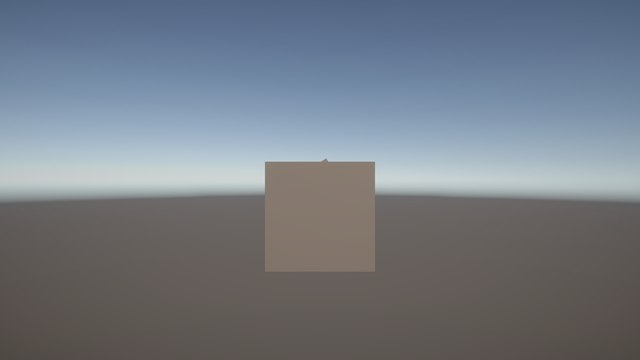}{zRoyal}
\end{minipage}\\[6pt]
\begin{minipage}[t]{\linewidth}
\ubtaskhead{zRoyal}{0028-ImportExportPackages}{Import / Export Package Asset Organization}{zZonai}{Easy}{7}
\vspace{4pt}
\ubshots{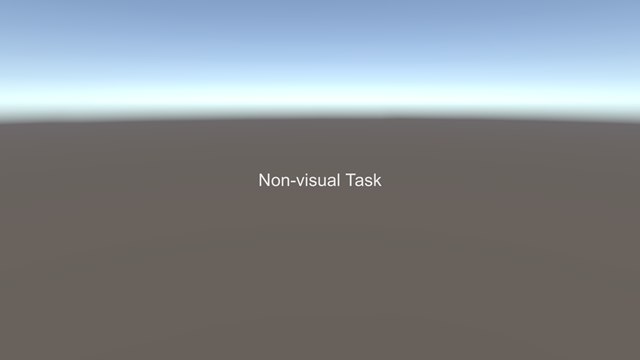}{zRoyal}
\end{minipage}\\[6pt]
\begin{minipage}[t]{\linewidth}
\ubtaskhead{zRoyal}{0042-PrefabSharedMaterialIsolation}{Prefab Asset Modification, Loose Object Replacement, and Variant Construction}{zZonai}{Easy}{16}
\vspace{4pt}
\ubshots{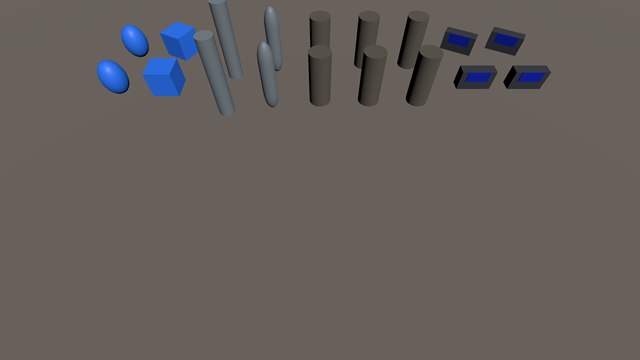}{zRoyal}
\end{minipage}\\[6pt]
\begin{minipage}[t]{\linewidth}
\ubtaskhead{zRoyal}{0052-URPPipelineConfiguration}{URP Pipeline Configuration}{zZonai}{Easy}{8}
\vspace{4pt}
\ubshots{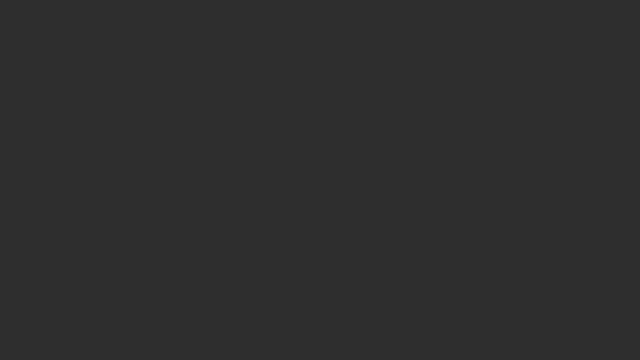}{zRoyal}
\end{minipage}\\[6pt]
\begin{minipage}[t]{\linewidth}
\ubtaskhead{zRoyal}{0016-TexturePipeline}{Automated Texture Import Pipeline Based on Semantic Routing and Low-Level Memory Defense}{zGoldDeep}{Medium}{8}
\vspace{4pt}
\ubshots{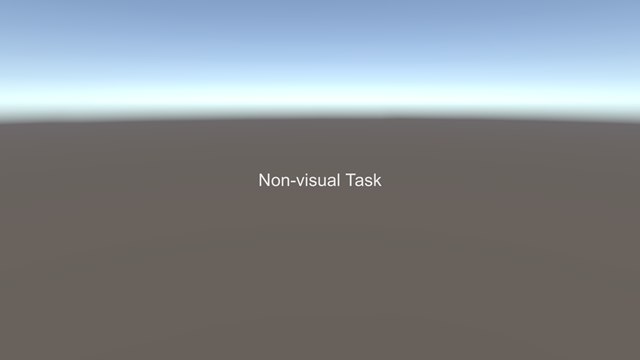}{zRoyal}
\end{minipage}\\[6pt]
\begin{minipage}[t]{\linewidth}
\ubtaskhead{zRoyal}{0018-FindGameObjectsInComplexScene}{Find GameObjects in a Complex Scene}{zGoldDeep}{Medium}{10}
\vspace{4pt}
\ubshots{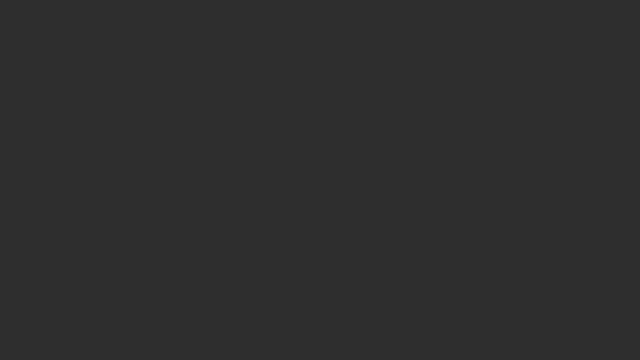}{zRoyal}
\end{minipage}\\[6pt]
\begin{minipage}[t]{\linewidth}
\ubtaskhead{zRoyal}{0060-GunShotSound}{Build a Production Gunshot Audio System}{zShrine}{Hard}{29}
\vspace{4pt}
\ubshots{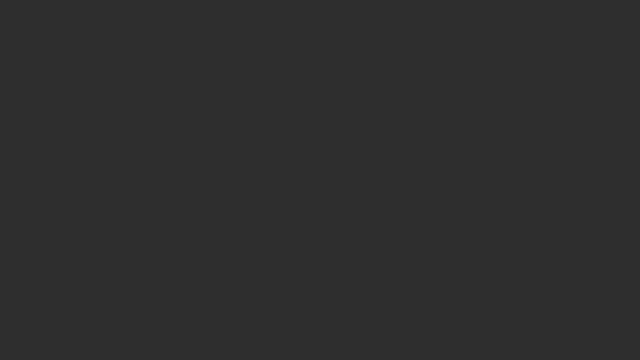}{zRoyal}
\end{minipage}\\[6pt]
\begin{minipage}[t]{\linewidth}
\ubtaskhead{zRoyal}{0072-CastleDestruction}{Classify Castle Structures into Hierarchy Groups}{zShrine}{Hard}{9}
\vspace{4pt}
\ubshots{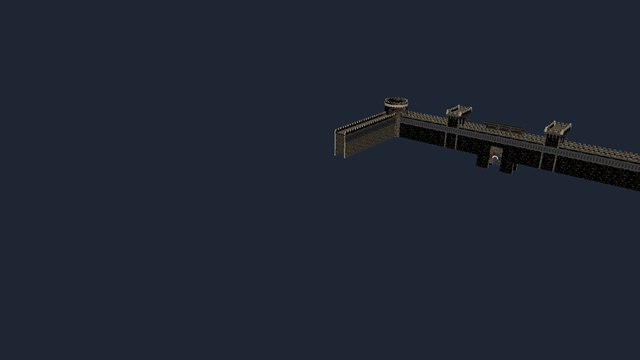}{zRoyal}
\end{minipage}\\[6pt]
\noalign{\vskip 3pt}
\ubcategory{zPlain}{Scene \& Level Construction}{19 tasks \textperiodcentered\ 365 milestone tests}\\*[5pt]
\begin{minipage}[t]{\linewidth}
\ubtaskhead{zPlain}{0001-ObjectIdentificationPrefabSorting}{Object Identification and Categorization}{zZonai}{Easy}{33}
\vspace{4pt}
\ubshots{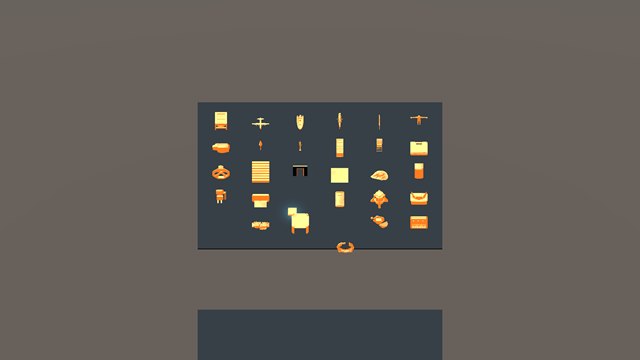}{zPlain}
\end{minipage}\\[6pt]
\begin{minipage}[t]{\linewidth}
\ubtaskhead{zPlain}{0021-ShootAheadStreetLayout}{Street Intersection Assembly}{zZonai}{Easy}{10}
\vspace{4pt}
\ubshots{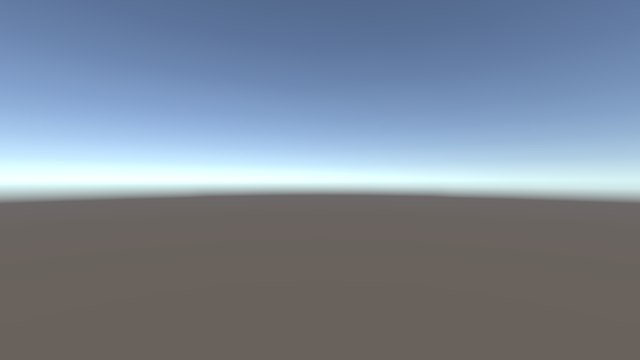}{zPlain}
\end{minipage}\\[6pt]
\begin{minipage}[t]{\linewidth}
\ubtaskhead{zPlain}{0040-ObjectIdentificationPrefabSorting}{Object Identification and Categorization}{zZonai}{Easy}{33}
\vspace{4pt}
\ubshots{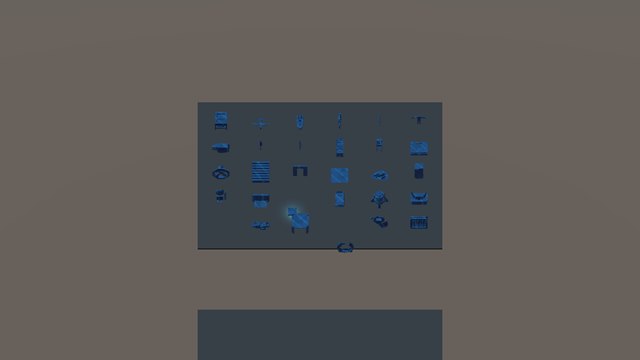}{zPlain}
\end{minipage}\\[6pt]
\begin{minipage}[t]{\linewidth}
\ubtaskhead{zPlain}{0005-BuildHouseReachability}{Paper-Thin Multistory House and Floor Reachability}{zGoldDeep}{Medium}{4}
\vspace{4pt}
\ubshots{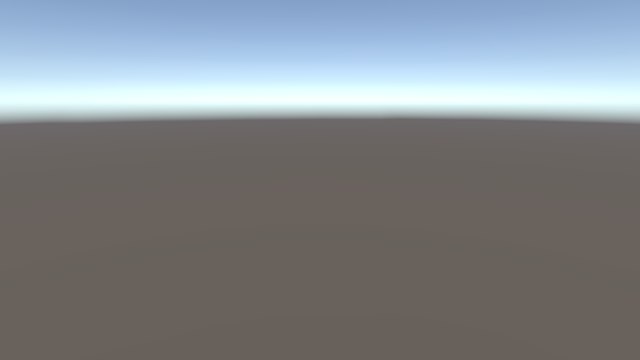}{zPlain}
\end{minipage}\\[6pt]
\begin{minipage}[t]{\linewidth}
\ubtaskhead{zPlain}{0020-InteriorFurniturePlacement}{Interior Furniture Placement}{zGoldDeep}{Medium}{39}
\vspace{4pt}
\ubshots{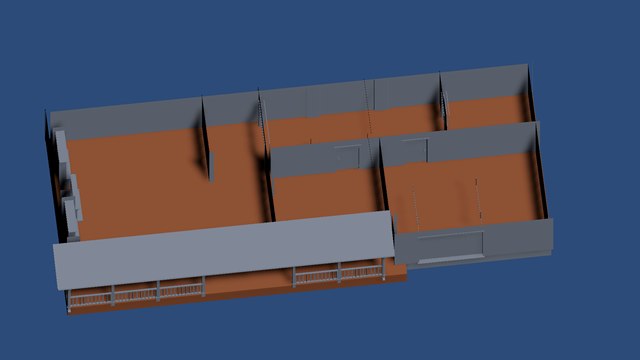}{zPlain}
\end{minipage}\\[6pt]
\begin{minipage}[t]{\linewidth}
\ubtaskhead{zPlain}{0022-ShootAheadReferenceStreet}{Reference Image Street Recreation}{zGoldDeep}{Medium}{6}
\vspace{4pt}
\ubshots{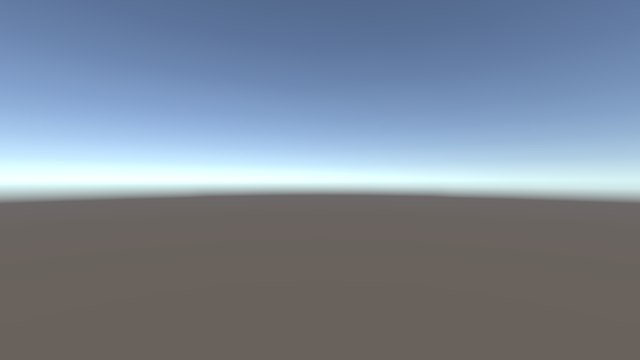}{zPlain}
\end{minipage}\\[6pt]
\begin{minipage}[t]{\linewidth}
\ubtaskhead{zPlain}{0034-SciFiRoomVariants}{Sci-Fi Room Size Variants}{zGoldDeep}{Medium}{6}
\vspace{4pt}
\ubshots{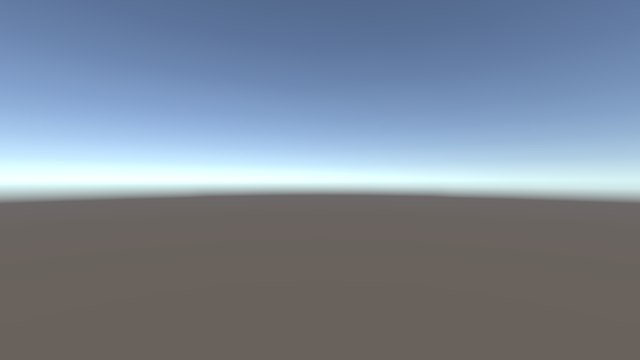}{zPlain}
\end{minipage}\\[6pt]
\begin{minipage}[t]{\linewidth}
\ubtaskhead{zPlain}{0036-NavMeshAgentOffMeshLink}{NavMeshAgent OffMeshLink Multi-Room Navigation}{zGoldDeep}{Medium}{10}
\vspace{4pt}
\ubshots{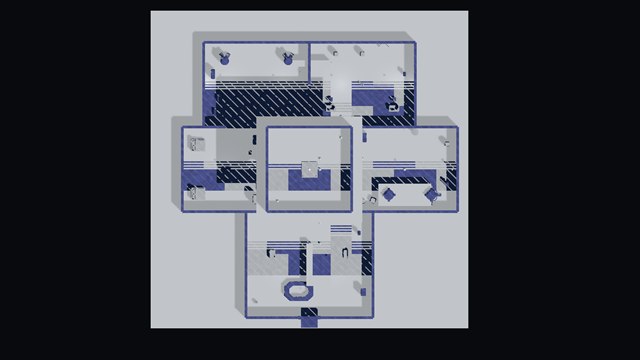}{zPlain}
\end{minipage}\\[6pt]
\begin{minipage}[t]{\linewidth}
\ubtaskhead{zPlain}{0043-MazeObstacleParenting}{11x11 Unique-Path Maze Construction}{zGoldDeep}{Medium}{10}
\vspace{4pt}
\ubshots{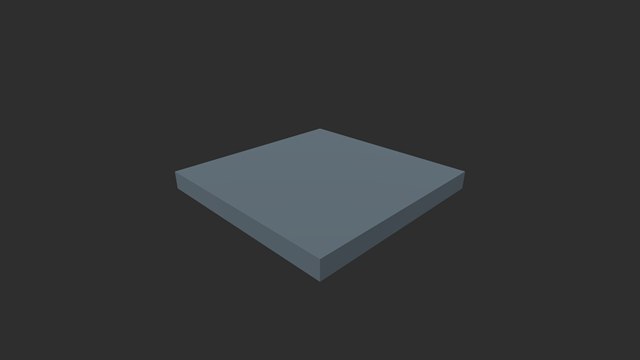}{zPlain}
\end{minipage}\\[6pt]
\begin{minipage}[t]{\linewidth}
\ubtaskhead{zPlain}{0069-TrenchAssembly}{Trench Asset Assembly}{zGoldDeep}{Medium}{16}
\vspace{4pt}
\ubshots{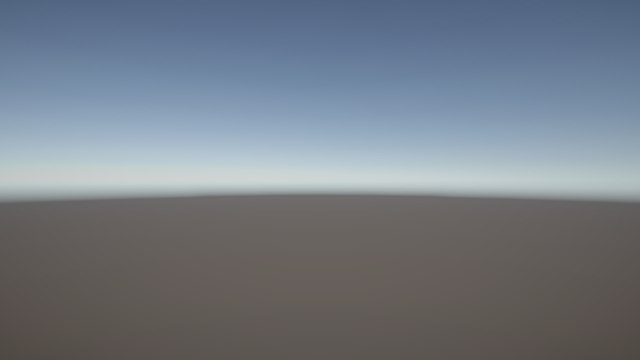}{zPlain}
\end{minipage}\\[6pt]
\begin{minipage}[t]{\linewidth}
\ubtaskhead{zPlain}{0066-Shop3DLetters}{Shop 3D Letter Signs}{zShrine}{Hard}{44}
\vspace{4pt}
\ubshots{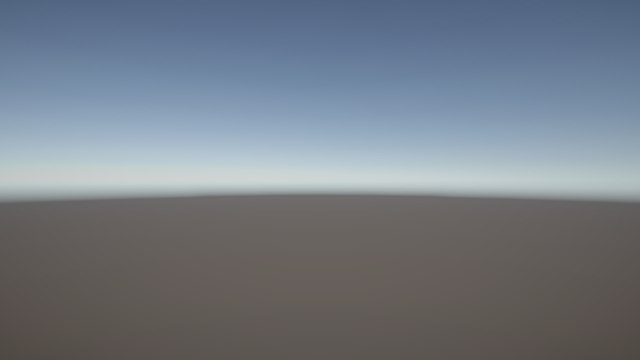}{zPlain}
\end{minipage}\\[6pt]
\begin{minipage}[t]{\linewidth}
\ubtaskhead{zPlain}{0067-TerrainRiverWaterAssembly}{Terrain River and Lake Water Surface Assembly}{zShrine}{Hard}{14}
\vspace{4pt}
\ubshots{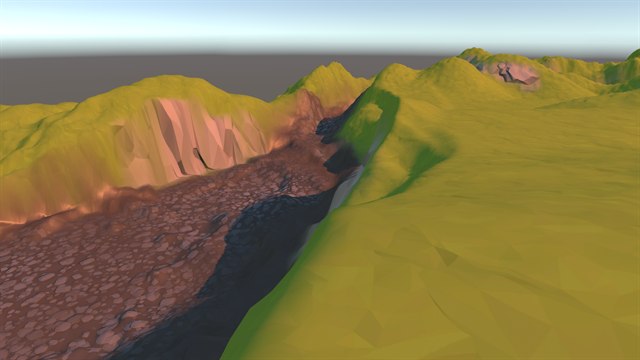}{zPlain}
\end{minipage}\\[6pt]
\begin{minipage}[t]{\linewidth}
\ubtaskhead{zPlain}{0068-ShanghaiCircuitTerrainPaint}{Shanghai Circuit Terrain Painting, White Lines, and Turn Props}{zShrine}{Hard}{28}
\vspace{4pt}
\ubshots{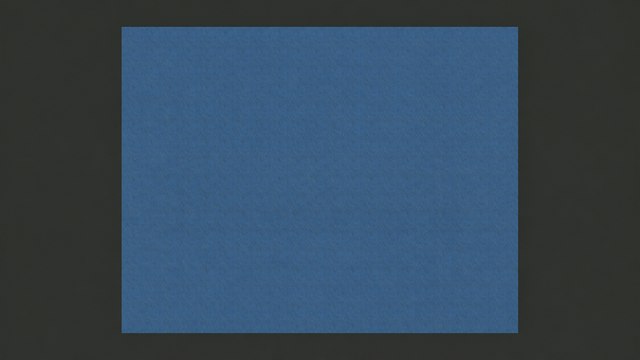}{zPlain}
\end{minipage}\\[6pt]
\begin{minipage}[t]{\linewidth}
\ubtaskhead{zPlain}{0070-CameraPlacement}{Place Indoor and Outdoor Observation Cameras}{zShrine}{Hard}{11}
\vspace{4pt}
\ubshots{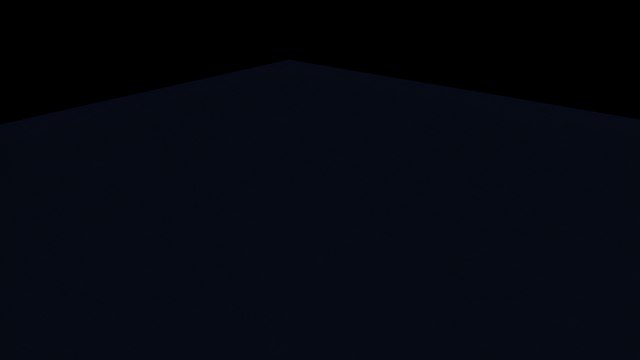}{zPlain}
\end{minipage}\\[6pt]
\begin{minipage}[t]{\linewidth}
\ubtaskhead{zPlain}{0071-PlantsDecoration}{Decorate Terrain with Natural Trees and Riverside Rocks}{zShrine}{Hard}{24}
\vspace{4pt}
\ubshots{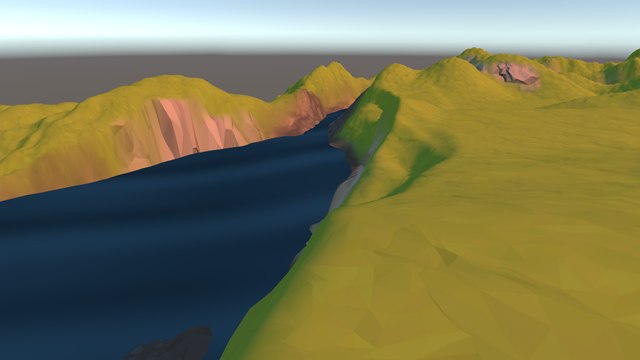}{zPlain}
\end{minipage}\\[6pt]
\begin{minipage}[t]{\linewidth}
\ubtaskhead{zPlain}{0073-CastleConstruction}{Connect Castle Circulation with Supplied Stair Prefabs}{zShrine}{Hard}{9}
\vspace{4pt}
\ubshots{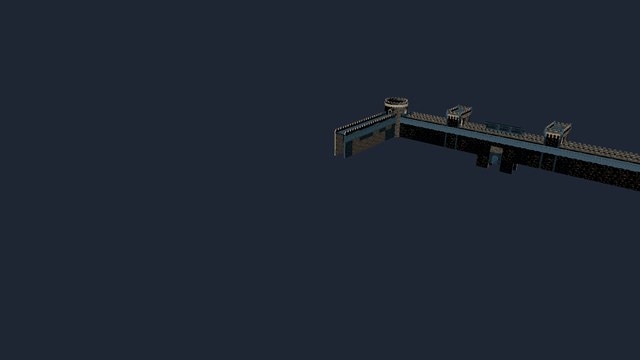}{zPlain}
\end{minipage}\\[6pt]
\begin{minipage}[t]{\linewidth}
\ubtaskhead{zPlain}{0075-CrystalFragment}{Restore the Crystal Fragment Assembly}{zShrine}{Hard}{14}
\vspace{4pt}
\ubshots{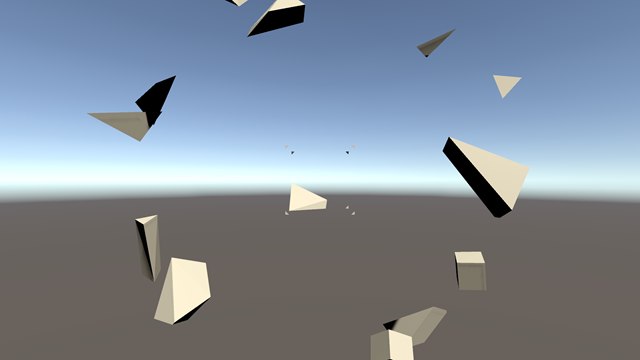}{zPlain}
\end{minipage}\\[6pt]
\begin{minipage}[t]{\linewidth}
\ubtaskhead{zPlain}{0076-BuildHouseFromDesign}{Build a White-Box House from a Floor Plan}{zShrine}{Hard}{23}
\vspace{4pt}
\ubshots{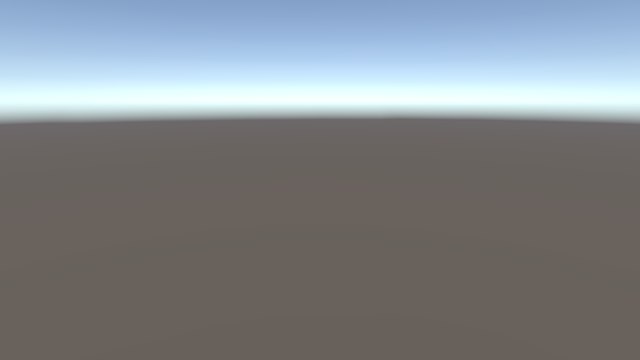}{zPlain}
\end{minipage}\\[6pt]
\begin{minipage}[t]{\linewidth}
\ubtaskhead{zPlain}{0080-UrbanRoads}{Add Correct Lane-Direction Arrows to Urban Intersections}{zShrine}{Hard}{31}
\vspace{4pt}
\ubshots{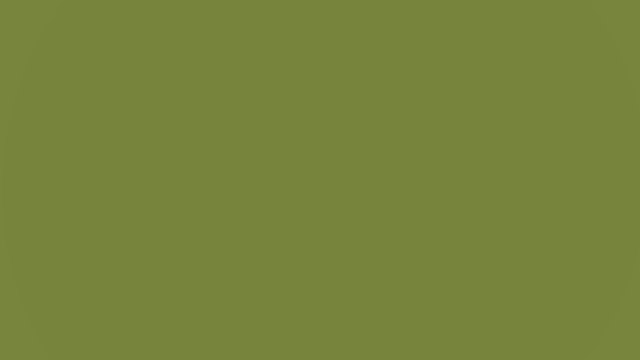}{zPlain}
\end{minipage}\\[6pt]
\noalign{\vskip 3pt}
\ubcategory{zShrine}{Particles \& Shaders}{12 tasks \textperiodcentered\ 155 milestone tests}\\*[5pt]
\begin{minipage}[t]{\linewidth}
\ubtaskhead{zShrine}{0007-SkyboxBlackHole}{Shadertoy Black Hole Skybox}{zZonai}{Easy}{26}
\vspace{4pt}
\ubshots{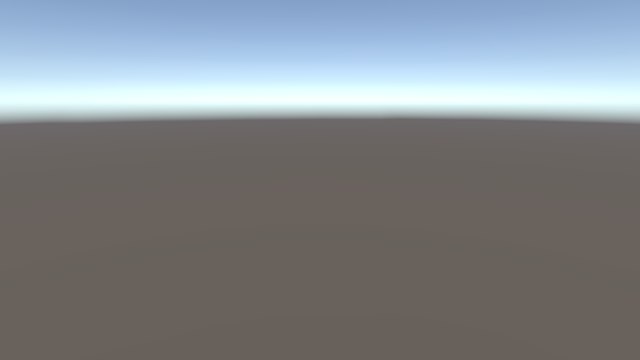}{zShrine}
\end{minipage}\\[6pt]
\begin{minipage}[t]{\linewidth}
\ubtaskhead{zShrine}{0030-Fireworks-TrailingSmoke}{Firework Trailing Smoke}{zZonai}{Easy}{13}
\vspace{4pt}
\ubshots{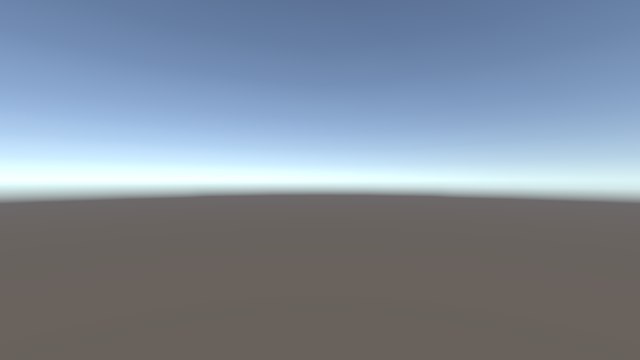}{zShrine}
\end{minipage}\\[6pt]
\begin{minipage}[t]{\linewidth}
\ubtaskhead{zShrine}{0031-Fireworks-TrailingSparks}{Firework Trailing Sparks}{zZonai}{Easy}{15}
\vspace{4pt}
\ubshots{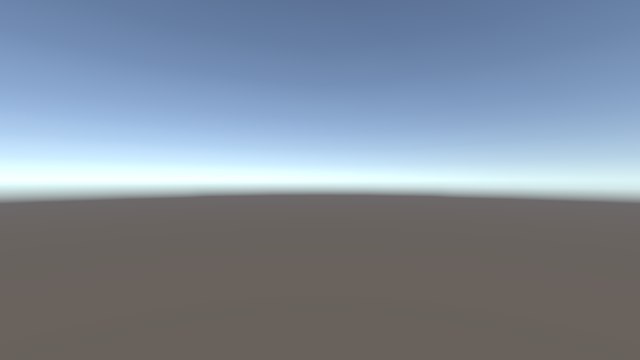}{zShrine}
\end{minipage}\\[6pt]
\begin{minipage}[t]{\linewidth}
\ubtaskhead{zShrine}{0032-Fireworks-HeadHalo}{Firework Head Halo}{zZonai}{Easy}{13}
\vspace{4pt}
\ubshots{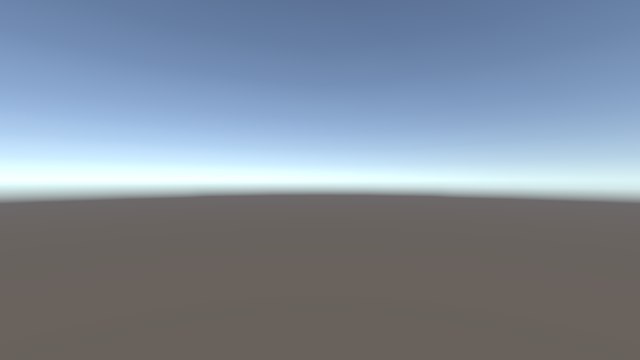}{zShrine}
\end{minipage}\\[6pt]
\begin{minipage}[t]{\linewidth}
\ubtaskhead{zShrine}{0033-Fireworks-ExplosionEffect}{Firework Explosion Effect}{zZonai}{Easy}{19}
\vspace{4pt}
\ubshots{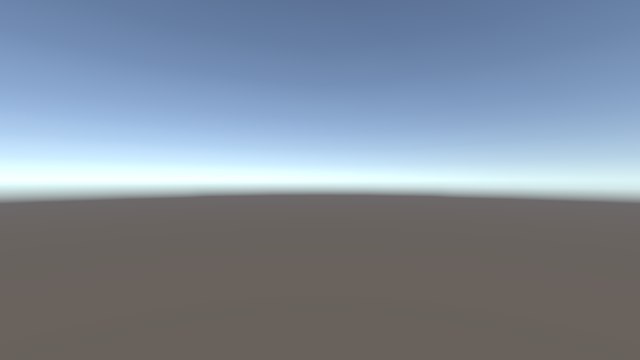}{zShrine}
\end{minipage}\\[6pt]
\begin{minipage}[t]{\linewidth}
\ubtaskhead{zShrine}{0024-CoastDunesPineUrpMaterials}{Pine Material URP Conversion}{zGoldDeep}{Medium}{5}
\vspace{4pt}
\ubshots{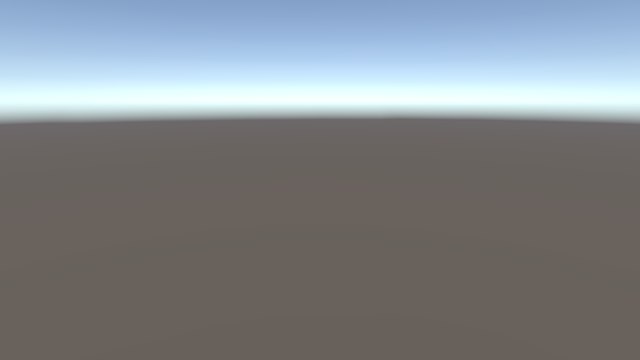}{zShrine}
\end{minipage}\\[6pt]
\begin{minipage}[t]{\linewidth}
\ubtaskhead{zShrine}{0037-ClassificationOfParticleSystem}{Particle Effect Classification}{zGoldDeep}{Medium}{10}
\vspace{4pt}
\ubshots{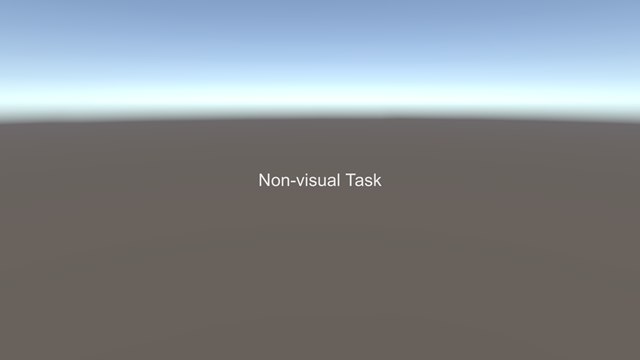}{zShrine}
\end{minipage}\\[6pt]
\begin{minipage}[t]{\linewidth}
\ubtaskhead{zShrine}{0047-DynamicDissolve}{Dynamic Dissolve}{zGoldDeep}{Medium}{10}
\vspace{4pt}
\ubshots{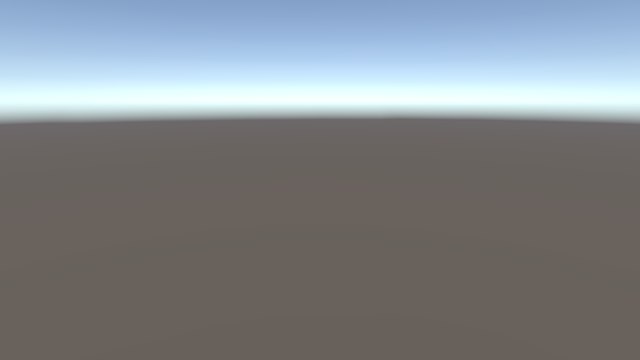}{zShrine}
\end{minipage}\\[6pt]
\begin{minipage}[t]{\linewidth}
\ubtaskhead{zShrine}{0059-RainEmitter}{Particle Collision Splash Task}{zGoldDeep}{Medium}{10}
\vspace{4pt}
\ubshots{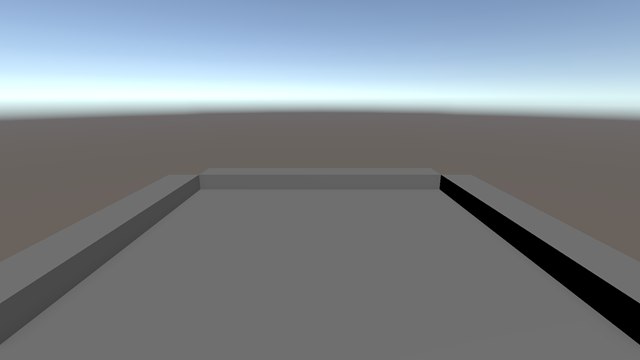}{zShrine}
\end{minipage}\\[6pt]
\begin{minipage}[t]{\linewidth}
\ubtaskhead{zShrine}{0012-Smrvfx}{Restore Skinned-Mesh VFX Graph Particles and Trails}{zShrine}{Hard}{13}
\vspace{4pt}
\ubshots{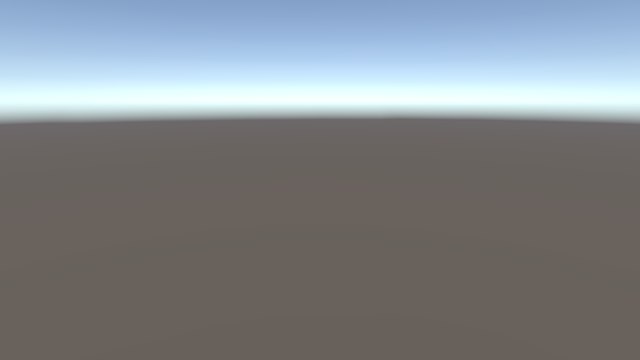}{zShrine}
\end{minipage}\\[6pt]
\begin{minipage}[t]{\linewidth}
\ubtaskhead{zShrine}{0038-DynamicRainEffect}{Create a Dynamic Rain Effect}{zShrine}{Hard}{8}
\vspace{4pt}
\ubshots{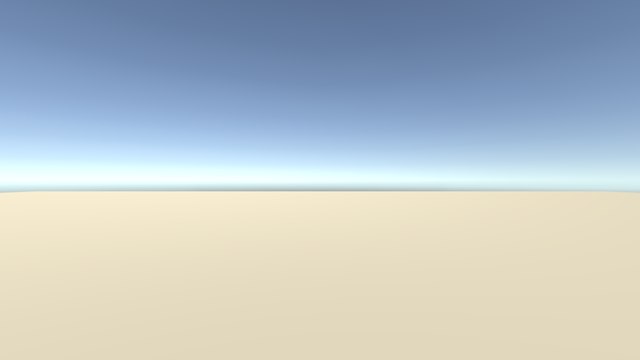}{zShrine}
\end{minipage}\\[6pt]
\begin{minipage}[t]{\linewidth}
\ubtaskhead{zShrine}{0045-InteractableLeaves}{Create an Interactable Falling Leaves Effect}{zShrine}{Hard}{13}
\vspace{4pt}
\ubshots{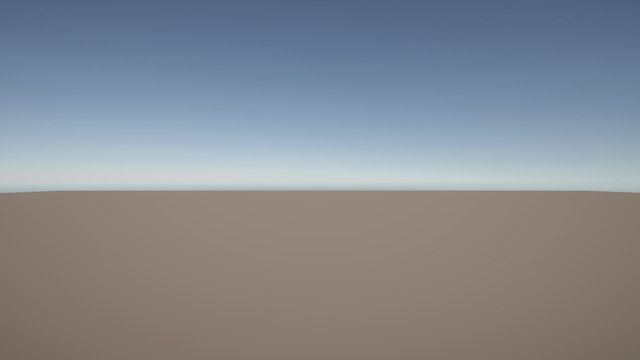}{zShrine}
\end{minipage}\\[6pt]
\noalign{\vskip 3pt}
\ubcategory{zZonai}{Physics \& Collision}{11 tasks \textperiodcentered\ 168 milestone tests}\\*[5pt]
\begin{minipage}[t]{\linewidth}
\ubtaskhead{zZonai}{0002-CubeMove}{Build a Rigidbody-Driven Cube Movement Scene}{zZonai}{Easy}{20}
\vspace{4pt}
\ubshots{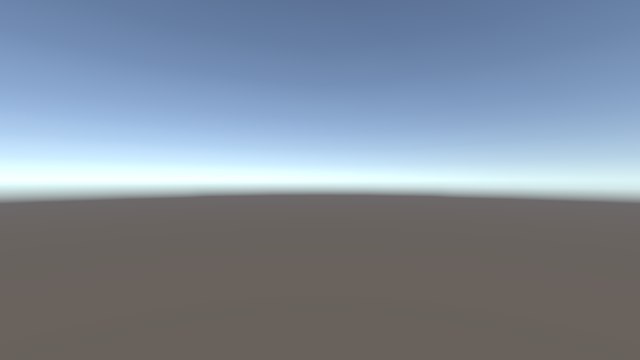}{zZonai}
\end{minipage}\\[6pt]
\begin{minipage}[t]{\linewidth}
\ubtaskhead{zZonai}{0054-ProceduralSkinnedCloth}{Procedural Skinned Cloth}{zZonai}{Easy}{10}
\vspace{4pt}
\ubshots{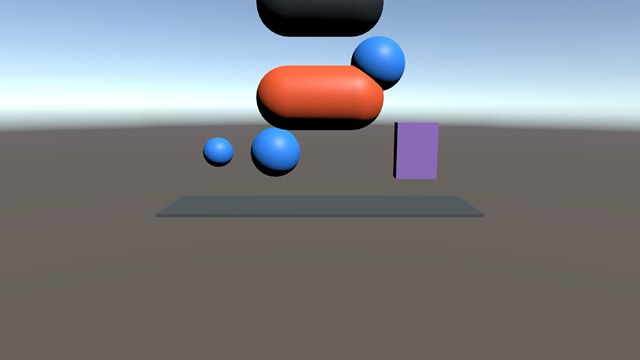}{zZonai}
\end{minipage}\\[6pt]
\begin{minipage}[t]{\linewidth}
\ubtaskhead{zZonai}{0057-VoxelWormsTerrain}{Voxel Worms Terrain}{zZonai}{Easy}{12}
\vspace{4pt}
\ubshots{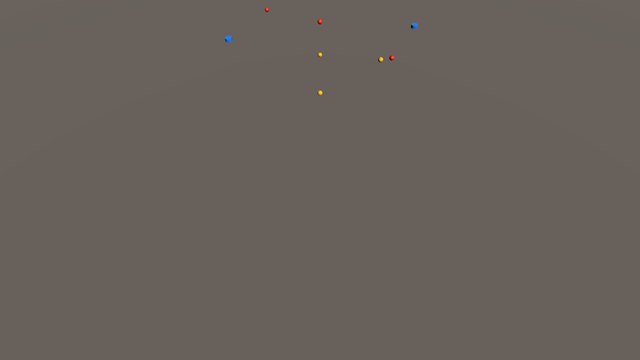}{zZonai}
\end{minipage}\\[6pt]
\begin{minipage}[t]{\linewidth}
\ubtaskhead{zZonai}{0003-CatapultExplosion}{Catapult Explosion}{zGoldDeep}{Medium}{11}
\vspace{4pt}
\ubshots{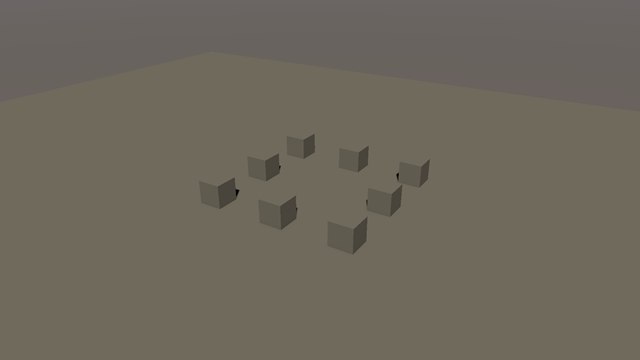}{zZonai}
\end{minipage}\\[6pt]
\begin{minipage}[t]{\linewidth}
\ubtaskhead{zZonai}{0026-ShootAheadVehicleWheelPhysics}{Vehicle Control with WheelColliders}{zGoldDeep}{Medium}{8}
\vspace{4pt}
\ubshots{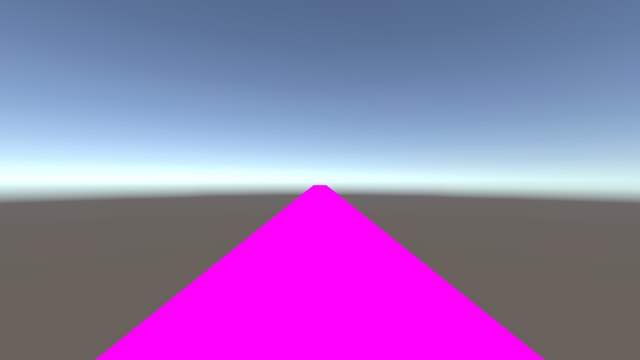}{zZonai}
\end{minipage}\\[6pt]
\begin{minipage}[t]{\linewidth}
\ubtaskhead{zZonai}{0039-RollingBallCheckpointCourse}{Inclined Rolling Ball Checkpoint Guidance}{zGoldDeep}{Medium}{15}
\vspace{4pt}
\ubshots{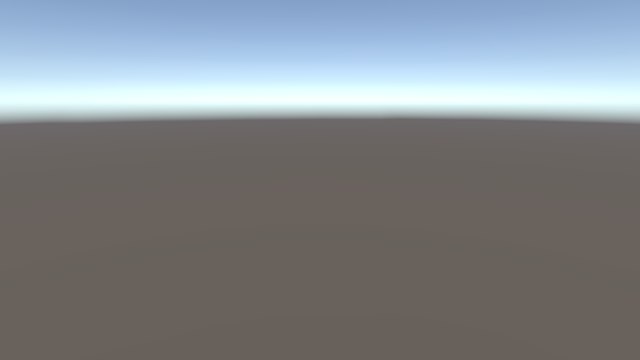}{zZonai}
\end{minipage}\\[6pt]
\begin{minipage}[t]{\linewidth}
\ubtaskhead{zZonai}{0048-RoadsideJointPhysics}{Roadside Joint Physics}{zGoldDeep}{Medium}{12}
\vspace{4pt}
\ubshots{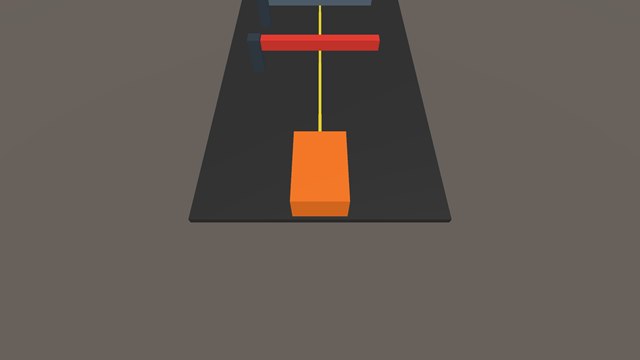}{zZonai}
\end{minipage}\\[6pt]
\begin{minipage}[t]{\linewidth}
\ubtaskhead{zZonai}{0056-RigidbodyCharacterControllerParity}{Rigidbody Character Controller Parity}{zGoldDeep}{Medium}{15}
\vspace{4pt}
\ubshots{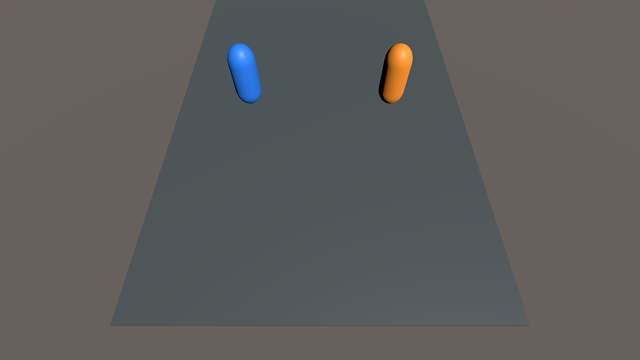}{zZonai}
\end{minipage}\\[6pt]
\begin{minipage}[t]{\linewidth}
\ubtaskhead{zZonai}{0044-ColliderWrappingChallenge}{Collider Wrapping Challenge}{zShrine}{Hard}{26}
\vspace{4pt}
\ubshots{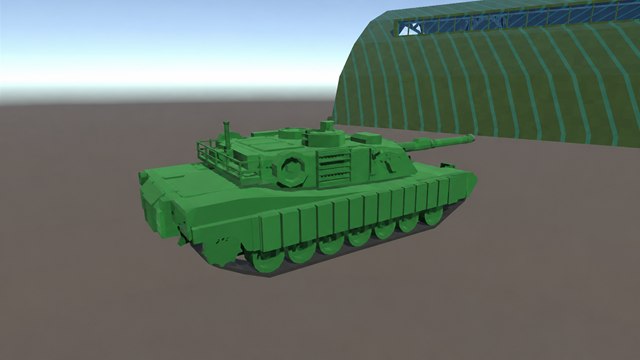}{zZonai}
\end{minipage}\\[6pt]
\begin{minipage}[t]{\linewidth}
\ubtaskhead{zZonai}{0053-PhysicsMaterial}{PhysicMaterial Tuning}{zShrine}{Hard}{8}
\vspace{4pt}
\ubshots{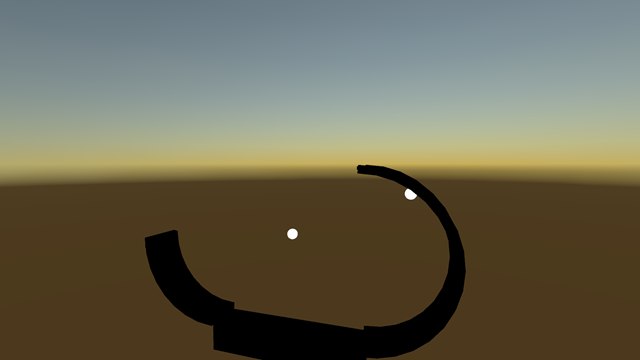}{zZonai}
\end{minipage}\\[6pt]
\begin{minipage}[t]{\linewidth}
\ubtaskhead{zZonai}{0077-TakeOffAndLanding}{Implement Physical Fixed-Wing Takeoff and Landing}{zShrine}{Hard}{31}
\vspace{4pt}
\ubshots{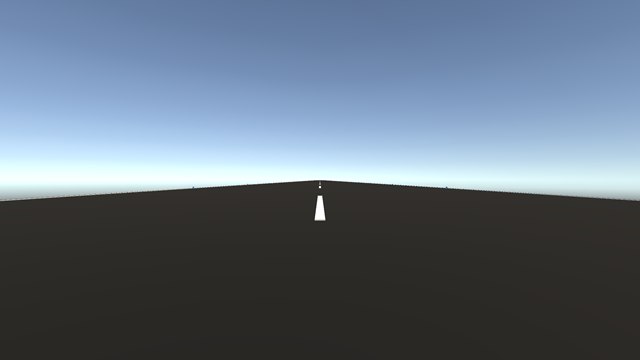}{zZonai}
\end{minipage}\\[6pt]
\noalign{\vskip 3pt}
\ubcategory{zRock}{Miscellaneous}{3 tasks \textperiodcentered\ 58 milestone tests}\\*[5pt]
\begin{minipage}[t]{\linewidth}
\ubtaskhead{zRock}{0011-CookiesApply}{Light Cookie Application}{zZonai}{Easy}{9}
\vspace{4pt}
\ubshots{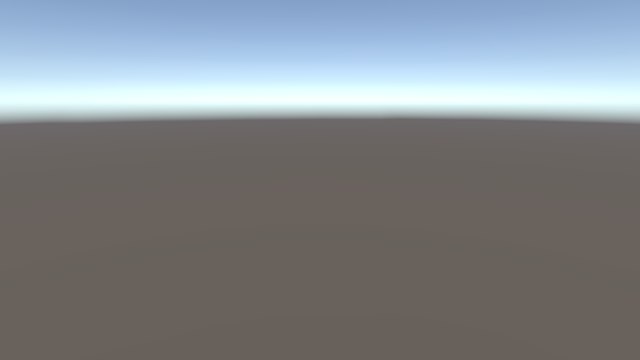}{zRock}
\end{minipage}\\[6pt]
\begin{minipage}[t]{\linewidth}
\ubtaskhead{zRock}{0017-CreateLOD}{LOD Asset Creation}{zGoldDeep}{Medium}{39}
\vspace{4pt}
\ubshots{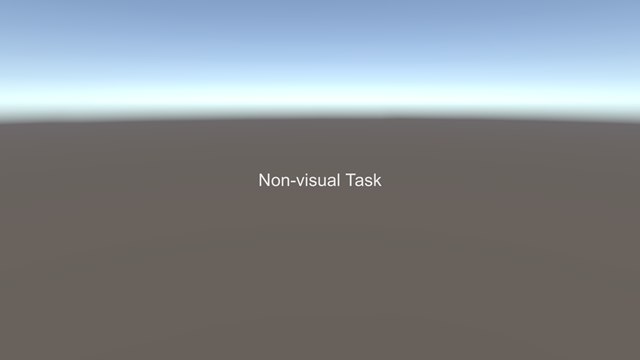}{zRock}
\end{minipage}\\[6pt]
\begin{minipage}[t]{\linewidth}
\ubtaskhead{zRock}{0019-AutoTimelineGenerator}{Timeline Cinematic Auto-Generation}{zShrine}{Hard}{10}
\vspace{4pt}
\ubshots{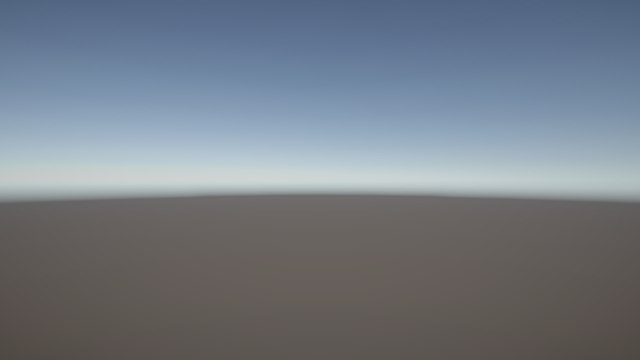}{zRock}
\end{minipage}\\[6pt]
\end{longtable}
\endgroup

\section{Challenge Specifications}
\label{app:chalspec}
\begingroup
\setlength{\tabcolsep}{0pt}
\renewcommand{\arraystretch}{1.04}
\setlength{\LTcapwidth}{14.4cm}
\footnotesize\color{zText}
\begin{longtable}{@{}>{\raggedright\arraybackslash}p{1.72cm}@{\hspace{8pt}}>{\raggedright\arraybackslash}p{12.41cm}@{}}
\caption{\textbf{Task specifications.} Agent-visible natural-language specifications for all 77 tasks. Each specification separates the requested goal, initial context, 780 explicit constraints, and relevant Unity resources.}\label{tab:specifications}\\
\multicolumn{2}{@{}p{14.4cm}@{}}{\ubidentity{zGold}{zRoyal}{SPECIFICATIONS}{goal \textperiodcentered\ context \textperiodcentered\ constraints \textperiodcentered\ resources}}\\[5pt]
\endfirsthead
\multicolumn{2}{@{}p{14.4cm}@{}}{%
  \textcolor{zRock}{\itshape Table \thetable\ continued from previous page.}}\\[3pt]
\multicolumn{2}{@{}p{14.4cm}@{}}{\textcolor{zRockLite!58}{\rule{\linewidth}{0.45pt}}}\\[3pt]
\endhead
\multicolumn{2}{r@{}}{\textcolor{zRock}{\itshape Continued on next page.}}\\
\endfoot
\endlastfoot
\noalign{\vskip 3pt}
\multicolumn{2}{@{}p{14.4cm}@{}}{\ubcategory{zChampion}{User Interface}{9 tasks \textperiodcentered\ 111 milestone tests}}\\*[5pt]
\multicolumn{2}{@{}p{14.4cm}@{}}{\ubtaskhead{zChampion}{0025-FPSCompassHUDCalibration}{Compass HUD Direction Calibration}{zZonai}{Easy}{6}}\\*[4pt]
\ubfield{zChampion}{Goal} & Fix the direction display of the compass HUD in the provided CompassHUD.cs. \\
\ubfield{zDawn}{Context} & CompassHUD already contains a compass viewport, repeating texture, and scrolling logic, but its direction mapping is incorrect. \\
\ubfield{zShrine}{Constraints} & \ubbullet{zShrine}{Only modify the direction calculation and the NorthCalibrationDegrees calibration value in CompassHUD.cs. Do not modify the compass UI creation logic, texture repetition method, stripRoot.anchoredPosition displacement formula, or smooth scrolling logic.} \\
 & \ubbullet{zShrine}{When the player's view faces world Z+, the compass center line should align with N.} \\
 & \ubbullet{zShrine}{When the player's view faces world X+, the compass center line should align with E.} \\
 & \ubbullet{zShrine}{When the player's view faces world X-, the compass center line should align with W.} \\
 & \ubbullet{zShrine}{Solve the direction mapping issue with the smallest and clearest script change possible, while preserving smooth compass scrolling.} \\*
\ubfield{zZonai}{Resources} & \textcolor{zZonai!82!black}{UGUI · RectTransform · Texture · MonoBehaviour} \\
\noalign{\vskip 3pt}
\multicolumn{2}{@{}p{14.4cm}@{}}{\textcolor{zRockLite!48}{\rule{\linewidth}{0.35pt}}}\\[4pt]
\multicolumn{2}{@{}p{14.4cm}@{}}{\ubtaskhead{zChampion}{0050-UIToolkitSnake}{Snake Game Implemented with Unity UI Toolkit}{zZonai}{Easy}{23}}\\*[4pt]
\ubfield{zChampion}{Goal} & Use Unity UI Toolkit to create a Snake game. The overall style should feel cool and futuristic, with a clear interface, smooth interaction, and adjustable game speed. The project should resolve all error reports. \\
\ubfield{zDawn}{Context} & The project starts without the required futuristic Snake game UI Toolkit implementation. \\
\ubfield{zShrine}{Constraints} & \ubbullet{zShrine}{Implement the game interface using Unity UI Toolkit.} \\
 & \ubbullet{zShrine}{Do not use a UGUI Canvas.} \\
 & \ubbullet{zShrine}{Build the main game interface with UIDocument, UXML, and USS.} \\
 & \ubbullet{zShrine}{The default scene should directly contain an editable UIDocument; do not rely entirely on runtime-generated UI.} \\
 & \ubbullet{zShrine}{Game logic may be controlled with C\# scripts, but the UI structure, styles, and main containers should be placed in UXML/USS.} \\
 & \ubbullet{zShrine}{The player controls a snake moving on a grid map.} \\
 & \ubbullet{zShrine}{The snake continuously moves in its current direction.} \\
 & \ubbullet{zShrine}{The player can control direction with arrow keys or WASD.} \\
 & \ubbullet{zShrine}{When the snake eats food, its body grows and the score increases.} \\
 & \ubbullet{zShrine}{The game ends when the snake hits a wall or itself.} \\
 & \ubbullet{zShrine}{A Game Over panel is displayed after the game ends.} \\
 & \ubbullet{zShrine}{The player can click a button to restart the game.} \\
 & \ubbullet{zShrine}{Before the game starts, there should be an idle state. The game starts after clicking the start button or pressing a key.} \\
 & \ubbullet{zShrine}{The game interface must provide a speed control feature.} \\
 & \ubbullet{zShrine}{Speed control uses a UI Toolkit Slider.} \\
 & \ubbullet{zShrine}{The player can adjust speed before the game starts.} \\
 & \ubbullet{zShrine}{Speed adjustment is also allowed during gameplay, and it should immediately affect the snake's movement interval.} \\
 & \ubbullet{zShrine}{Use the suggested speed tiers: Slow at about 0.22 seconds per move, Normal at about 0.15 seconds, Fast at about 0.09 seconds, and Extreme at about 0.06 seconds.} \\
 & \ubbullet{zShrine}{The UI must display the current speed tier or specific speed value.} \\
 & \ubbullet{zShrine}{Speed adjustment must not break game state, such as causing teleporting, duplicate movement, or logic disorder.} \\
 & \ubbullet{zShrine}{The overall style should be cool and futuristic.} \\
 & \ubbullet{zShrine}{A dark background is recommended, such as near-black, dark blue, or dark purple.} \\
 & \ubbullet{zShrine}{Use neon colors as accents, such as cyan, blue, purple, and green.} \\
 & \ubbullet{zShrine}{The map area should include a tech grid, glowing border, or scanline effect.} \\
 & \ubbullet{zShrine}{The snake body should have a glow effect or gradient tech texture.} \\
 & \ubbullet{zShrine}{Food should be clearly distinguishable from the snake, such as a highlighted block.} \\
 & \ubbullet{zShrine}{UI buttons, score, and speed controls should keep a tech feel without hurting readability.} \\
 & \ubbullet{zShrine}{The Game Over panel should stand out and provide a clear restart action.} \\
 & \ubbullet{zShrine}{Include a game title, current-score display, high-score display, speed-control area, game grid, start button, restart button or Game Over panel, and concise arrow-key/WASD control hints.} \\*
\ubfield{zZonai}{Resources} & \textcolor{zZonai!82!black}{UI Toolkit · UIDocument · UXML · USS} \\
\noalign{\vskip 3pt}
\multicolumn{2}{@{}p{14.4cm}@{}}{\textcolor{zRockLite!48}{\rule{\linewidth}{0.35pt}}}\\[4pt]
\multicolumn{2}{@{}p{14.4cm}@{}}{\ubtaskhead{zChampion}{0063-MinimapRealization}{Minimap UI Implementation}{zZonai}{Easy}{7}}\\*[4pt]
\ubfield{zChampion}{Goal} & Implement a circular minimap UI in the top-right corner of the project. \\
\ubfield{zDawn}{Context} & The project already contains minimap UI control logic and scene assets. \\
\ubfield{zShrine}{Constraints} & \ubbullet{zShrine}{Display a circular minimap in the top-right corner of the game view.} \\
 & \ubbullet{zShrine}{The minimap content is a real top-down view of the scene.} \\
 & \ubbullet{zShrine}{The player is always displayed at the center of the minimap.} \\
 & \ubbullet{zShrine}{The player is represented by a solid triangular arrow, and the arrow direction must match the player's current facing direction.} \\
 & \ubbullet{zShrine}{The minimap should accurately reflect the scene content near the current position.} \\
 & \ubbullet{zShrine}{Use only existing assets in the current project, and only modify the MinimapUI control UI logic.} \\
 & \ubbullet{zShrine}{The minimap camera is fixed to world north. The player arrow rotates, while the map itself remains stable and does not rotate.} \\
 & \ubbullet{zShrine}{The minimap UI must not create new GameObjects at runtime with new GameObject; runtime behavior may only update positions, rotations, and texture references of existing objects.} \\*
\ubfield{zZonai}{Resources} & \textcolor{zZonai!82!black}{UGUI · Camera · RenderTexture · RectMask2D} \\
\noalign{\vskip 3pt}
\multicolumn{2}{@{}p{14.4cm}@{}}{\textcolor{zRockLite!48}{\rule{\linewidth}{0.35pt}}}\\[4pt]
\multicolumn{2}{@{}p{14.4cm}@{}}{\ubtaskhead{zChampion}{0010-SplitTheUI}{UI Sprite Sheet Slicing}{zGoldDeep}{Medium}{11}}\\*[4pt]
\ubfield{zChampion}{Goal} & Correctly slice all 5 Sprite images. \\
\ubfield{zDawn}{Context} & Assets/Textures contains five Sprite images that require different slicing strategies. \\
\ubfield{zShrine}{Constraints} & \ubbullet{zShrine}{Sprite1: use Automatic slicing.} \\
 & \ubbullet{zShrine}{Sprite2: use Automatic slicing.} \\
 & \ubbullet{zShrine}{Sprite3: slice according to the image content. Each slice should have a consistent size and should reasonably and orderly separate each sub-image in the source image.} \\
 & \ubbullet{zShrine}{Sprite4: use Grid By Cell Size slicing. Inspect the image content yourself and choose the most appropriate Pixel Size.} \\
 & \ubbullet{zShrine}{Sprite5: delete all existing sliced child Sprites and keep only the root image.} \\
 & \ubbullet{zShrine}{Each slice must contain solid pixels, not only transparent background.} \\*
\ubfield{zZonai}{Resources} & \textcolor{zZonai!82!black}{Sprite Editor · Texture Importer · AssetDatabase} \\
\noalign{\vskip 3pt}
\multicolumn{2}{@{}p{14.4cm}@{}}{\textcolor{zRockLite!48}{\rule{\linewidth}{0.35pt}}}\\[4pt]
\multicolumn{2}{@{}p{14.4cm}@{}}{\ubtaskhead{zChampion}{0023-UIRedoBench}{GoldenBench UI Recreation}{zGoldDeep}{Medium}{9}}\\*[4pt]
\ubfield{zChampion}{Goal} & Open Assets/0023UIRedoBench/Scene/0023UIRedoBench.scene. Refer to Assets/0023UIRedoBench/Reference/Reference.jpg, and use the assets under Assets/0023UIRedoBench/ArtResource/Sprite/ to recreate the character selection screen under Panel\_CharacterSelect. \\
\ubfield{zDawn}{Context} & Assets/0023UIRedoBench/Scene/0023UIRedoBench.scene provides a character-selection panel, a reference image, and Sprite assets. \\
\ubfield{zShrine}{Constraints} & \ubbullet{zShrine}{Recreate the character-selection screen under Panel\_CharacterSelect in Assets/0023UIRedoBench/Scene/0023UIRedoBench.scene using the supplied Reference.jpg and Sprite assets.} \\
 & \ubbullet{zShrine}{Use a 1920x1080 landscape layout as the design baseline and display visual assets through UGUI Image or Button components.} \\
 & \ubbullet{zShrine}{GameObject names should match the asset names used, without file extensions.} \\
 & \ubbullet{zShrine}{The Button\_Chara series is special. Name them Button\_Chara\_1, Button\_Chara\_2, Button\_Chara\_3, Button\_Chara\_4, and Button\_Chara\_5.} \\
 & \ubbullet{zShrine}{The three entry buttons on the left should open the corresponding Panel\_Status, Panel\_Profile, and Panel\_Skin panels respectively.} \\
 & \ubbullet{zShrine}{In the character selection area, the first two character buttons should be selectable, while the last three character buttons should remain non-clickable.} \\
 & \ubbullet{zShrine}{The currently selected selectable character must show a selected visual state. When switching selection, the other selectable character should return to the unselected visual state.} \\*
\ubfield{zZonai}{Resources} & \textcolor{zZonai!82!black}{UGUI · Canvas · RectTransform · Sprite} \\
\noalign{\vskip 3pt}
\multicolumn{2}{@{}p{14.4cm}@{}}{\textcolor{zRockLite!48}{\rule{\linewidth}{0.35pt}}}\\[4pt]
\multicolumn{2}{@{}p{14.4cm}@{}}{\ubtaskhead{zChampion}{0049-UIDocument}{Recreate T-Rex Runner with UI Toolkit}{zGoldDeep}{Medium}{18}}\\*[4pt]
\ubfield{zChampion}{Goal} & In the Unity 0049-UIDocument project, recreate the open-source frontend game wayou/t-rex-runner using UI Toolkit. Reference materials have already been placed under Assets/t-rex-runner-gh-pages; prioritize referencing index.css, index.js, and the sprite assets there. \\
\ubfield{zDawn}{Context} & The project provides the t-rex-runner frontend source and Sprite assets under Assets/t-rex-runner-gh-pages. \\
\ubfield{zShrine}{Constraints} & \ubbullet{zShrine}{Implement the interface using UI Toolkit; do not use a UGUI Canvas.} \\
 & \ubbullet{zShrine}{Convert key layout, sizing, color, and display rules from the frontend index.css into USS.} \\
 & \ubbullet{zShrine}{Use UXML to convert the original frontend canvas/layer structure into an editable VisualElement hierarchy.} \\
 & \ubbullet{zShrine}{The default scene must directly contain an editable UIDocument.} \\
 & \ubbullet{zShrine}{Required initialization environment should be saved directly in the scene, UXML, USS, PanelSettings, or resource assets.} \\
 & \ubbullet{zShrine}{The Game view background should completely fill the view; it must not show only a white strip at the top.} \\
 & \ubbullet{zShrine}{Game logic should preserve the core behavior of the reference implementation: waiting, starting, jumping, ducking, moving ground, obstacle generation, collision, score, Game Over, and restart.} \\
 & \ubbullet{zShrine}{All visible game elements based on the original sprite sheet, such as the dinosaur, ground, clouds, obstacles, score characters, Game Over, restart button, and similar elements, must be driven by UI Toolkit Image or backgroundImage. Do not use generateVisualContent, self-drawn meshes, or other custom drawing methods invisible to tests to render these key elements.} \\*
\ubfield{zZonai}{Resources} & \textcolor{zZonai!82!black}{UI Toolkit · UIDocument · UXML · USS} \\
\noalign{\vskip 3pt}
\multicolumn{2}{@{}p{14.4cm}@{}}{\textcolor{zRockLite!48}{\rule{\linewidth}{0.35pt}}}\\[4pt]
\multicolumn{2}{@{}p{14.4cm}@{}}{\ubtaskhead{zChampion}{0015-ShootAheadUguiMobileHud}{ShootAhead Mobile HUD Refactor}{zShrine}{Hard}{9}}\\*[4pt]
\ubfield{zChampion}{Goal} & Make the mobile combat interface a stable, maintainable UGUI HUD that adapts to different screen sizes. \\
\ubfield{zDawn}{Context} & The ShootAhead combat flow still contains an immediate-mode mobile HUD that must be replaced with UGUI. \\
\ubfield{zShrine}{Constraints} & \ubbullet{zShrine}{Locate the in-combat mobile HUD content that is still implemented with immediate-mode UI.} \\
 & \ubbullet{zShrine}{Replace the relevant interface with UI elements under a Canvas.} \\
 & \ubbullet{zShrine}{Keep existing combat, interaction, and mobile control behavior working.} \\
 & \ubbullet{zShrine}{The UI must adapt to screen size and must not extend beyond screen boundaries.} \\
 & \ubbullet{zShrine}{UI elements must keep reasonable spacing and must not overlap each other.} \\
 & \ubbullet{zShrine}{Existing gameplay flows such as inventory, pickup, vehicles, and weapon shooting must continue to work normally.} \\*
\ubfield{zZonai}{Resources} & \textcolor{zZonai!82!black}{UGUI · Canvas · Input System · MonoBehaviour} \\
\noalign{\vskip 3pt}
\multicolumn{2}{@{}p{14.4cm}@{}}{\textcolor{zRockLite!48}{\rule{\linewidth}{0.35pt}}}\\[4pt]
\multicolumn{2}{@{}p{14.4cm}@{}}{\ubtaskhead{zChampion}{0062-UIToolkitBubbleDragon}{Build an Endless UI Toolkit Bubble Shooter}{zShrine}{Hard}{19}}\\*[4pt]
\ubfield{zChampion}{Goal} & Create a responsive, editor-authored UI Toolkit bubble-shooter game with aiming, wall bounces, grid attachment, matching, drops, pressure, endless difficulty, scoring, pause, and game-over flows. \\
\ubfield{zDawn}{Context} & The project provides a Unity scene and UI Toolkit packages, but the complete editor-placed game interface, styling, gameplay controller, and endless bubble-shooter behavior must be implemented. \\
\ubfield{zShrine}{Constraints} & \ubbullet{zShrine}{Create a complete endless Bubble Shooter game in the editor with all errors resolved, using an editor-placed UIDocument with assigned VisualTreeAsset and PanelSettings rather than constructing the complete UI only at runtime.} \\
 & \ubbullet{zShrine}{Present a central play board with bubble array, trajectory, visible wall boundaries, danger line, launcher, loaded bubble, and next bubble.} \\
 & \ubbullet{zShrine}{Keep the next bubble inside the play area to the right of the cannon and keep the cannon behind the danger line.} \\
 & \ubbullet{zShrine}{Place score, high score, difficulty, pressure, and pause access in a framed right sidebar with inset white cards and safe padding.} \\
 & \ubbullet{zShrine}{Support mouse aiming and firing plus A/D or arrow-key aiming and Space firing.} \\
 & \ubbullet{zShrine}{Keep cannon rotation, shot direction, and trajectory preview aligned around the same pivot.} \\
 & \ubbullet{zShrine}{Shot bubbles must travel straight, bounce from aligned side walls, and attach to a reasonable nearest grid position.} \\
 & \ubbullet{zShrine}{Clear connected groups of at least three same-color bubbles and drop bubbles disconnected from the top.} \\
 & \ubbullet{zShrine}{Accumulate pressure on non-clearing shots and move the board downward without changing honeycomb row parity.} \\
 & \ubbullet{zShrine}{End the round at the danger line and provide pause, continue, restart, main-interface, and game-over flows.} \\
 & \ubbullet{zShrine}{Track score, high score, survival progress, combos, pressure, and persistent best performance.} \\
 & \ubbullet{zShrine}{Ramp endless difficulty smoothly by color count, pressure threshold, row frequency, or board density.} \\
 & \ubbullet{zShrine}{Provide at least seven visually distinct bright bubble colors—red, yellow, blue, green, purple, orange, and cyan—and keep the overall presentation bright, saturated, and readable.} \\
 & \ubbullet{zShrine}{Provide clear animation and feedback for shots, attachment, elimination, falling bubbles, score changes, difficulty increases, approaching danger, and game over.} \\
 & \ubbullet{zShrine}{Maintain a responsive layout across common desktop resolutions, keeping the board centered, firing area operable, visible walls aligned with bounce boundaries, and sidebar content inside its cards and frame.} \\*
\ubfield{zZonai}{Resources} & \textcolor{zZonai!82!black}{Scene · UI Toolkit · UIDocument · UXML · USS · MonoBehaviour · Input System} \\
\noalign{\vskip 3pt}
\multicolumn{2}{@{}p{14.4cm}@{}}{\textcolor{zRockLite!48}{\rule{\linewidth}{0.35pt}}}\\[4pt]
\multicolumn{2}{@{}p{14.4cm}@{}}{\ubtaskhead{zChampion}{0078-JetHUD}{Build a First-Person F-22 Cockpit HUD}{zShrine}{Hard}{9}}\\*[4pt]
\ubfield{zChampion}{Goal} & Reconstruct a transparent, cockpit-screen HUD whose speed, altitude, pitch ladder, world horizon, heading tape, and delayed impact point respond to live aircraft state. \\
\ubfield{zDawn}{Context} & Assets/Scenes/0078.scene provides the aircraft, FirstPerson viewpoint, cockpit Screen, StaticHUD.png, DynamicHUD.png, and Share Tech Mono font resources. \\
\ubfield{zShrine}{Constraints} & \ubbullet{zShrine}{Render the transparent HUD on the existing cockpit Screen rather than directly across the game window.} \\
 & \ubbullet{zShrine}{Preserve the Screen transform and keep lighting, shadows, and colliders from affecting the display.} \\
 & \ubbullet{zShrine}{Match the supplied StaticHUD.png and DynamicHUD.png references in position, proportion, line work, spacing, and light-green style while keeping their black regions transparent.} \\
 & \ubbullet{zShrine}{Use the supplied Share Tech Mono font for all dynamic text.} \\
 & \ubbullet{zShrine}{Show the HUD only in first-person view and hide it completely in third-person view.} \\
 & \ubbullet{zShrine}{Drive airspeed in km/h from Rigidbody speed, starting at a zero reading, with 30 km/h tick spacing and synchronized moving labels and value box.} \\
 & \ubbullet{zShrine}{Drive altitude in metres from world height with 30 metre tick spacing and moving labels and value box.} \\
 & \ubbullet{zShrine}{Render a -60 to +60 degree pitch ladder at 15 degree intervals with positive solid lines, negative dashed lines, and no zero line.} \\
 & \ubbullet{zShrine}{Move the pitch ladder continuously with pitch and roll while keeping inverted labels readable.} \\
 & \ubbullet{zShrine}{Render a separate dashed true world horizon through the FirstPerson viewpoint.} \\
 & \ubbullet{zShrine}{Scroll heading marks continuously in 30 degree increments and wrap correctly through north.} \\
 & \ubbullet{zShrine}{Predict the delayed impact point from roughly 0.2 seconds of aim history, 1000 m/s projectile speed, 200 m target distance, aircraft velocity, and gravity, then project it into the current aiming view.} \\
 & \ubbullet{zShrine}{Drive every dynamic value and position from live runtime flight state rather than hard-coding the example values, tick positions, or impact marker shown in the references.} \\
 & \ubbullet{zShrine}{Synchronize dynamic content with the interpolated aircraft attitude and FirstPerson view without visible roll misalignment, jitter, or lag.} \\
 & \ubbullet{zShrine}{Do not change flight controls, automatic routing, or camera motion to fake HUD data.} \\
 & \ubbullet{zShrine}{Preserve AircraftPhysicsRoot, FirstPerson, Screen, and the supplied HUD reference assets.} \\
 & \ubbullet{zShrine}{Remove the old central HUD, duplicate reticles, debug text, and any temporary diagnostic content before submission.} \\
 & \ubbullet{zShrine}{Confirm that the project compiles and that the complete HUD is directly inspectable in first-person Play Mode.} \\*
\ubfield{zZonai}{Resources} & \textcolor{zZonai!82!black}{Scene · UGUI · Camera · RenderTexture · Material · Rigidbody · MonoBehaviour · Transform} \\
\noalign{\vskip 3pt}
\multicolumn{2}{@{}p{14.4cm}@{}}{\textcolor{zRockLite!48}{\rule{\linewidth}{0.35pt}}}\\[4pt]
\noalign{\vskip 3pt}
\multicolumn{2}{@{}p{14.4cm}@{}}{\ubcategory{zDawn}{Animation \& Animator}{12 tasks \textperiodcentered\ 179 milestone tests}}\\*[5pt]
\multicolumn{2}{@{}p{14.4cm}@{}}{\ubtaskhead{zDawn}{0000-AutoDoor}{Build an Automatic Opening and Closing Door Scene (AutoDoor)}{zZonai}{Easy}{29}}\\*[4pt]
\ubfield{zChampion}{Goal} & Complete an automatic opening and closing door demonstration scene in Assets/Scenes/AutoDoor.scene. \\
\ubfield{zDawn}{Context} & The project starts with Assets/Scenes/SampleScene.scene containing a Main Camera and a Directional Light; Assets/Scenes/AutoDoor.scene and its required door setup must be created. \\
\ubfield{zShrine}{Constraints} & \ubbullet{zShrine}{Complete and save the result in Assets/Scenes/AutoDoor.scene.} \\
 & \ubbullet{zShrine}{The scene must contain GameObjects named Floor, DoorFrame, DoorFrame\_Left, DoorFrame\_Right, DoorFrame\_Top, DoorRoot, Door, and MainCamera.} \\
 & \ubbullet{zShrine}{DoorFrame\_Left, DoorFrame\_Right, and DoorFrame\_Top must be under the DoorFrame hierarchy.} \\
 & \ubbullet{zShrine}{Door must be under the DoorRoot hierarchy.} \\
 & \ubbullet{zShrine}{Floor must use a Plane.} \\
 & \ubbullet{zShrine}{DoorFrame\_Left, DoorFrame\_Right, and DoorFrame\_Top must use Cubes.} \\
 & \ubbullet{zShrine}{Door must use a Cube.} \\
 & \ubbullet{zShrine}{You must use an Animator. Legacy Animation components are not accepted.} \\
 & \ubbullet{zShrine}{After the scene starts running, the door must automatically start playing the open/close animation within 3 seconds.} \\
 & \ubbullet{zShrine}{The door must loop through "closed -\textgreater{} open -\textgreater{} closed".} \\
 & \ubbullet{zShrine}{The door must open by rotating around the Y axis. Do not simulate the door opening by translating it.} \\
 & \ubbullet{zShrine}{A full open/close cycle must not exceed 6 seconds.} \\
 & \ubbullet{zShrine}{The opening and closing motion must be continuous and smooth, and must include both opening and closing phases.} \\
 & \ubbullet{zShrine}{Door must have a clear local X or Z offset relative to DoorRoot, so that DoorRoot acts as the hinge pivot instead of the door rotating around its own center.} \\
 & \ubbullet{zShrine}{MainCamera must be able to see most of the door.} \\*
\ubfield{zZonai}{Resources} & \textcolor{zZonai!82!black}{Scene · GameObject · Animation · Animator · Camera} \\
\noalign{\vskip 3pt}
\multicolumn{2}{@{}p{14.4cm}@{}}{\textcolor{zRockLite!48}{\rule{\linewidth}{0.35pt}}}\\[4pt]
\multicolumn{2}{@{}p{14.4cm}@{}}{\ubtaskhead{zDawn}{0006-TankTrackAnimation}{Tank Track Animation}{zZonai}{Easy}{2}}\\*[4pt]
\ubfield{zChampion}{Goal} & When the tank enters the running state, both the left and right tracks and the road wheels must show continuous motion, creating the visual impression that the tank is driving. \\
\ubfield{zDawn}{Context} & Assets/Scenes/TankTrackAnimatorRuntime.scene already contains the real ShootAhead Tank prefab. \\
\ubfield{zShrine}{Constraints} & \ubbullet{zShrine}{Create Assets/Scripts/Visuals/TankTrackAnimator.cs and complete the effect in Assets/Scenes/TankTrackAnimatorRuntime.scene using the existing Assets/Arts/Tank.prefab instance.} \\
 & \ubbullet{zShrine}{TankTrackAnimator must automatically identify the relevant track and road-wheel child objects on the current tank.} \\
 & \ubbullet{zShrine}{When IsAnimating is true, Advance(float deltaTime) must continuously animate both tracks moving forward.} \\
 & \ubbullet{zShrine}{When IsAnimating is true, the road wheels on both sides must rotate continuously in a direction consistent with track movement.} \\
 & \ubbullet{zShrine}{Expose configurable parameters or public fields for speed, direction, and related motion behavior.} \\
 & \ubbullet{zShrine}{Do not replace or delete the real Tank prefab in the scene.} \\
 & \ubbullet{zShrine}{Name the script class TankTrackAnimator, and attach it to the Tank root GameObject in the scene.} \\
 & \ubbullet{zShrine}{Do not fake the result by setting a one-time static final state. The animation must advance over time.} \\
 & \ubbullet{zShrine}{Do not rely on a fixed frame rate.} \\*
\ubfield{zZonai}{Resources} & \textcolor{zZonai!82!black}{Animation · Transform · MonoBehaviour · Prefab} \\
\noalign{\vskip 3pt}
\multicolumn{2}{@{}p{14.4cm}@{}}{\textcolor{zRockLite!48}{\rule{\linewidth}{0.35pt}}}\\[4pt]
\multicolumn{2}{@{}p{14.4cm}@{}}{\ubtaskhead{zDawn}{0008-SwordsmanAttack}{Swordsman Attack}{zZonai}{Easy}{4}}\\*[4pt]
\ubfield{zChampion}{Goal} & The scene Assets/Scenes/HitBox.scene contains a swordsman, his sword, and a wooden stake in front of him. Assets/Task02/Attack.anim is currently an empty animation. Modify this animation so the swordsman performs one diagonal sword slash. \\
\ubfield{zDawn}{Context} & Assets/Scenes/HitBox.scene contains a swordsman, sword, and wooden stake, while Assets/Task02/Attack.anim starts empty. \\
\ubfield{zShrine}{Constraints} & \ubbullet{zShrine}{Create the sword attack motion in Assets/Task02/Attack.anim.} \\
 & \ubbullet{zShrine}{Make the sword motion follow a clear diagonal slashing trajectory. It must not stay still, only translate along a single axis, or instantly jump to the target position.} \\
 & \ubbullet{zShrine}{Open the attack hit-detection window during the middle of the slash, and close it before the slash ends.} \\
 & \ubbullet{zShrine}{While the attack hit-detection window is open, the sword's attack collider must be able to touch the wooden stake.} \\
 & \ubbullet{zShrine}{Keep the swordsman, wooden stake, and existing control-script structure usable.} \\
 & \ubbullet{zShrine}{Use animation events to call EnableAttackWindow during the middle of the slash and DisableAttackWindow before the slash ends.} \\*
\ubfield{zZonai}{Resources} & \textcolor{zZonai!82!black}{AnimationClip · Animation Event · Animator · Collider} \\
\noalign{\vskip 3pt}
\multicolumn{2}{@{}p{14.4cm}@{}}{\textcolor{zRockLite!48}{\rule{\linewidth}{0.35pt}}}\\[4pt]
\multicolumn{2}{@{}p{14.4cm}@{}}{\ubtaskhead{zDawn}{0009-AnimationEvents}{Animation Events}{zGoldDeep}{Medium}{9}}\\*[4pt]
\ubfield{zChampion}{Goal} & After entering Play Mode, the character should play its animation normally, and the configured animation events should be triggered at the corresponding points in time during playback. \\
\ubfield{zDawn}{Context} & Assets/Scenes/AnimationEvent.scene provides a humanoid model and a walking-animation FBX asset. \\
\ubfield{zShrine}{Constraints} & \ubbullet{zShrine}{Use Assets/Resources/Armature.fbx and Assets/Resources/Locomotion--Walk\_N.anim.fbx in Assets/Scenes/AnimationEvent.scene.} \\
 & \ubbullet{zShrine}{Add StartEvent at normalized time 0 with integer parameter 1.} \\
 & \ubbullet{zShrine}{Add HalfEvent at normalized time 0.5 with float parameter 1.2 and string parameter Hello World!.} \\
 & \ubbullet{zShrine}{Implement AnimationEventTestClass with public int intParam initialized to 0, public float floatParam initialized to 0f, and public string stringParam initialized to an empty string.} \\
 & \ubbullet{zShrine}{Implement callbacks for StartEvent and HalfEvent that assign the supplied parameters to the corresponding typed fields.} \\
 & \ubbullet{zShrine}{Instantiate the humanoid character with a parent GameObject named Armature and play the required walk animation through an Animator.} \\
 & \ubbullet{zShrine}{Set the Animator culling mode to Always Animate.} \\
 & \ubbullet{zShrine}{Do not modify the variable values in Update to fake animation event triggering.} \\
 & \ubbullet{zShrine}{The script name and field names inside it must match the requirements exactly.} \\
 & \ubbullet{zShrine}{The parameter values passed by the events must match the requirements exactly.} \\
 & \ubbullet{zShrine}{Do not rename any existing assets in the project.} \\
 & \ubbullet{zShrine}{The animation playback speed must be 1.0f.} \\*
\ubfield{zZonai}{Resources} & \textcolor{zZonai!82!black}{AnimationClip · Animation Event · Animator · MonoBehaviour} \\
\noalign{\vskip 3pt}
\multicolumn{2}{@{}p{14.4cm}@{}}{\textcolor{zRockLite!48}{\rule{\linewidth}{0.35pt}}}\\[4pt]
\multicolumn{2}{@{}p{14.4cm}@{}}{\ubtaskhead{zDawn}{0029-AnimatorLayer}{Layered Character Animation}{zGoldDeep}{Medium}{16}}\\*[4pt]
\ubfield{zChampion}{Goal} & Read the project content, add the character to the TestScene scene, and configure its animation. Use the simplest reasonable method to switch the character between a full-body animation state and simultaneous body-part animation playback. The character model, walking animation, and aiming animation already exist. \\
\ubfield{zDawn}{Context} & TestScene provides a character together with walking and weapon-aiming animation assets. \\
\ubfield{zShrine}{Constraints} & \ubbullet{zShrine}{TestScene must contain the character with a root GameObject named Character at position (0,0,0), rotation (0,0,0), and scale (1,1,1). Other than the root name, do not modify child GameObject properties. Set the Animator to AlwaysAnimate, AnimatePhysics, and root motion enabled.} \\
 & \ubbullet{zShrine}{After starting the TestScene scene, the character automatically plays the full-body walking animation.} \\
 & \ubbullet{zShrine}{After pressing the E key, the character can switch so the upper body plays the gun-aiming animation while the lower body continues playing the walking animation. After switching, the upper-body gun-aiming animation must start from the beginning and include a transition.} \\
 & \ubbullet{zShrine}{Pressing E again switches back to the full-body walking animation state.} \\
 & \ubbullet{zShrine}{When switching with E, the lower-body walking animation playback progress and speed must not be affected.} \\
 & \ubbullet{zShrine}{Keyboard input detection must use UnityEngine.InputSystem.} \\
 & \ubbullet{zShrine}{When the character is in the full-body walking animation state, there must not be a state where upper-body and lower-body animation progress are inconsistent.} \\
 & \ubbullet{zShrine}{Use only animation assets that already exist in the project. Do not generate extra animation clips, and do not rename or move the original animation assets.} \\*
\ubfield{zZonai}{Resources} & \textcolor{zZonai!82!black}{Animation · Animator · Avatar Mask · Input System} \\
\noalign{\vskip 3pt}
\multicolumn{2}{@{}p{14.4cm}@{}}{\textcolor{zRockLite!48}{\rule{\linewidth}{0.35pt}}}\\[4pt]
\multicolumn{2}{@{}p{14.4cm}@{}}{\ubtaskhead{zDawn}{0035-HumanoidIK}{Humanoid IK}{zGoldDeep}{Medium}{13}}\\*[4pt]
\ubfield{zChampion}{Goal} & Read the project content and implement IK for the character in the TestScene scene, allowing specified bones to be modified while the character plays animation. \\
\ubfield{zDawn}{Context} & TestScene already contains a character, four IK targets, a sphere, a cube, and a Ground plane. \\
\ubfield{zShrine}{Constraints} & \ubbullet{zShrine}{The scene already has four IK targets. Implement the IK logic for these four targets. The world position and rotation of the bones they bind to must be fully controlled by their respective targets. Do not change the target names.} \\
 & \ubbullet{zShrine}{The scene contains a sphere. When it is close enough to be within the character's reachable range, the character should use the left hand to grab the sphere and keep following the sphere as it moves. When the sphere is out of range, stop following it and return control to the IK target.} \\
 & \ubbullet{zShrine}{The scene contains a cube. Make the character's head always look at the cube.} \\
 & \ubbullet{zShrine}{The scene contains a plane whose Layer is Ground. When the ground is within reach of the character's right foot, the right foot should conform to the ground. The right foot must only detect in the (0, -1, 0) direction. When following the ground, only its Y coordinate and rotation should be adjusted; its XZ-plane position is determined by the IK target. When out of range, stop following the ground, and the Y coordinate and rotation should again be controlled by the IK target.} \\
 & \ubbullet{zShrine}{The follow logic above has higher priority than IK targets. Do not modify the scene name, the character root Transform, or the Animator component. Do not modify any data on the sphere, cube, or plane.} \\*
\ubfield{zZonai}{Resources} & \textcolor{zZonai!82!black}{Animator · Humanoid IK · MonoBehaviour} \\
\noalign{\vskip 3pt}
\multicolumn{2}{@{}p{14.4cm}@{}}{\textcolor{zRockLite!48}{\rule{\linewidth}{0.35pt}}}\\[4pt]
\multicolumn{2}{@{}p{14.4cm}@{}}{\ubtaskhead{zDawn}{0041-ExcavatorAnimationClipDigDump}{Excavator AnimationClip Dig and Dump Task}{zGoldDeep}{Medium}{9}}\\*[4pt]
\ubfield{zChampion}{Goal} & Create Dig and Dump AnimationClips for the excavator, animating its mechanical parts and using the bucket collider to scoop and unload the provided cube. \\
\ubfield{zDawn}{Context} & Assets/0041/Scene/0041.scene provides an excavator rig, a pit, a cube, and an unloading platform. \\
\ubfield{zShrine}{Constraints} & \ubbullet{zShrine}{Complete the excavator animation task in Assets/0041/Scene/0041.scene using the existing Excavator\_Rig and its Excavator\_Turntable/BoomPivot/ArmPivot/BucketPivot mechanical hierarchy.} \\
 & \ubbullet{zShrine}{Use the existing Bucket\_Capture\_Volume and Bucket\_Cargo\_Anchor helper GameObjects to support genuine physics-based scooping, carrying, and dumping.} \\
 & \ubbullet{zShrine}{Animate the ExcavationCube\_-prefixed cubes from the front ExcavationPit into the receiving workbench behind the excavator; scoring is based on the transported quantity.} \\
 & \ubbullet{zShrine}{Create Assets/0041/Animations/Dig.anim, making the mechanical arm and bucket dig cubes out of the pit.} \\
 & \ubbullet{zShrine}{Create Assets/0041/Animations/Dump.anim, making the mechanical arm and bucket dump cubes onto the receiving workbench behind the excavator.} \\
 & \ubbullet{zShrine}{Both animation clips will be played on Excavator\_Rig by a fixed program.} \\
 & \ubbullet{zShrine}{The fixed program will first play Dig.anim, then rotate Excavator\_Turntable backward by 180 degrees through physics stepping, and then play Dump.anim.} \\
 & \ubbullet{zShrine}{You do not need to rotate Excavator\_Turntable in the animation. Focus on animating BoomPivot, ArmPivot, BucketPivot, and the bucket.} \\
 & \ubbullet{zShrine}{Do not move, delete, or rename the pit, the rear workbench, the cubes, Bucket\_Capture\_Volume, or Bucket\_Cargo\_Anchor.} \\
 & \ubbullet{zShrine}{Do not use scripts to directly move cubes, and do not directly place cubes to bypass the digging and dumping animation process.} \\*
\ubfield{zZonai}{Resources} & \textcolor{zZonai!82!black}{AnimationClip · Animator · Collider · PhysX} \\
\noalign{\vskip 3pt}
\multicolumn{2}{@{}p{14.4cm}@{}}{\textcolor{zRockLite!48}{\rule{\linewidth}{0.35pt}}}\\[4pt]
\multicolumn{2}{@{}p{14.4cm}@{}}{\ubtaskhead{zDawn}{0046-AnimationScriptPlayable}{Animation Script Playable}{zGoldDeep}{Medium}{17}}\\*[4pt]
\ubfield{zChampion}{Goal} & Read the project content and implement animation effects for the character in the TestScene scene. Use the Animator component and C\# script to play animations, but the Animator component does not need a runtime animator controller. \\
\ubfield{zDawn}{Context} & TestScene already contains a character and a LocomotionAnimationScriptPlayable script component. \\
\ubfield{zShrine}{Constraints} & \ubbullet{zShrine}{The character model already exists in TestScene, and a script component already exists on the parent GameObject. Modify LocomotionAnimationScriptPlayable.cs to complete the remaining requirements without changing the parent GameObject name, Animator component, or any animation asset.} \\
 & \ubbullet{zShrine}{When the LocomotionAnimationScriptPlayable component is disabled, the character stops playing animation. When the component is enabled again, it starts the current state animation from the beginning.} \\
 & \ubbullet{zShrine}{The character has four animation states: idle, walk, run, and inair. The default state is idle. The idle, walk, and run states can switch between each other using the float variable speed, with thresholds 0.1f for walk and 5f for run. These three states can all switch to and from inair using the bool variable isGrounded. State blending is controlled by transitionDuration; if transitionDuration is less than or equal to 0, the transition completes immediately.} \\
 & \ubbullet{zShrine}{The script must expose and implement basic animation control fields: speed defaulting to 0f, isGrounded, and transitionDuration. isGrounded defaults to true, and transitionDuration defaults to 0f.} \\
 & \ubbullet{zShrine}{Root motion must be manually controlled. The script must expose and implement root motion control fields: applyRootMotion, defaulting to true, to toggle RootMotion; removeInAirVerticalRootMotion, defaulting to false, to toggle vertical RootMotion removal in the inair state; and inAirRootMotionMultiplier, defaulting to 1f, to control the scale of RootMotion in the inair state.} \\
 & \ubbullet{zShrine}{Animation must be manually controlled. The script must expose and implement control fields: hipYOffset, defaulting to 0f, which adds a Y-axis offset to the character's hips bone; and maxForwardLeanDegrees, defaulting to 0f, which controls the character spine's forward lean. The spine bone should add forward lean on top of the currently playing animation curve. When speed is 0 there is no forward lean; when speed reaches its maximum, forward lean also reaches its maximum, with smooth interpolation during the increase.} \\*
\ubfield{zZonai}{Resources} & \textcolor{zZonai!82!black}{Playables · AnimationScriptPlayable · Animator · MonoBehaviour} \\
\noalign{\vskip 3pt}
\multicolumn{2}{@{}p{14.4cm}@{}}{\textcolor{zRockLite!48}{\rule{\linewidth}{0.35pt}}}\\[4pt]
\multicolumn{2}{@{}p{14.4cm}@{}}{\ubtaskhead{zDawn}{0055-ArcherAnimatorController}{Archer Animator Blend Tree Controller}{zGoldDeep}{Medium}{30}}\\*[4pt]
\ubfield{zChampion}{Goal} & Complete XBot archer character control and Animator configuration in the scene Assets/Scene/0055.scene: the character can move, run, crouch, aim, dive forward, and play a fall animation while falling. \\
\ubfield{zDawn}{Context} & Assets/Scene/0055.scene provides XBot and the required archer animation assets. \\
\ubfield{zShrine}{Constraints} & \ubbullet{zShrine}{Copy or extract all required clips as standalone .anim assets under Assets/Art/AnimClip: Standing Idle 01; Standing Walk Forward, Back, Left, and Right; Standing Sprint Forward; Standing Run Left and Right; Crouch Idle 01; Crouch Walk Forward, Back, Left, and Right; Standing Aim Walk Forward, Back, Left, and Right; Fall A Loop; and Standing Dive Forward.} \\
 & \ubbullet{zShrine}{If the aiming loop causes upper-body jitter, create an additional stable idle-aiming .anim clip.} \\
 & \ubbullet{zShrine}{Add a character control script and CharacterController to the XBot in the scene.} \\
 & \ubbullet{zShrine}{XBot's Animator uses Assets/Archer.controller.} \\
 & \ubbullet{zShrine}{The Animator must have a valid Avatar; create and save an in-project Avatar asset if necessary.} \\
 & \ubbullet{zShrine}{Add Float parameters MoveX, MoveY, MoveSpeed, and AnimationSpeed; Bool parameters IsMoving, IsRunning, IsAiming, IsCrouching, IsGrounded, and IsFalling; and a Trigger parameter named Dive.} \\
 & \ubbullet{zShrine}{Do not retain a Jump parameter, jump state, or any transitions that depend on Jump.} \\
 & \ubbullet{zShrine}{WASD controls movement.} \\
 & \ubbullet{zShrine}{LeftControl / RightControl controls crouching.} \\
 & \ubbullet{zShrine}{LeftShift / RightShift controls running.} \\
 & \ubbullet{zShrine}{Left mouse button enters aiming.} \\
 & \ubbullet{zShrine}{Space triggers Standing Dive Forward as a forward dive, no longer a jump.} \\
 & \ubbullet{zShrine}{If camera follow is implemented, avoid obvious jitter.} \\
 & \ubbullet{zShrine}{Configure all of the following Base Layer locomotion behaviors.} \\
 & \ubbullet{zShrine}{Create Standing Locomotion: a 2D BlendTree using MoveX / MoveY. Include idle, forward/back/left/right walking, forward sprint, and left/right running.} \\
 & \ubbullet{zShrine}{Create Crouch Locomotion: a 2D BlendTree using MoveX / MoveY. Include crouch idle and forward/back/left/right crouch movement.} \\
 & \ubbullet{zShrine}{Shift + W + A and Shift + W + D should blend to diagonal running close to forward sprint, not fall back to walking or strafing. Left and right diagonal running should be symmetrical and natural.} \\
 & \ubbullet{zShrine}{Diagonal running can be slightly faster, left/right walking should be slightly slower, and diagonal walking should not drag.} \\
 & \ubbullet{zShrine}{Configure all of the following Dive and Fall behaviors.} \\
 & \ubbullet{zShrine}{Space triggers a one-shot Standing Dive Forward, and it cannot be triggered again before the animation ends.} \\
 & \ubbullet{zShrine}{Dive enters from locomotion in a controlled way and returns to locomotion after it ends; do not use the old jump animation.} \\
 & \ubbullet{zShrine}{Fall uses Fall A Loop and should enter only when truly airborne and falling.} \\
 & \ubbullet{zShrine}{Fall should not be inserted during flat-ground idle, flat-ground movement, or near the end of a roll.} \\
 & \ubbullet{zShrine}{Do not use Any State to bypass locomotion and enter Dive or Fall directly.} \\
 & \ubbullet{zShrine}{Configure all of the following upper-body aiming behaviors.} \\
 & \ubbullet{zShrine}{Create an Upper Body Aim layer, enabled by left mouse button.} \\
 & \ubbullet{zShrine}{This layer only affects the upper body; the lower body continues to use the standing/walking animation from the Base Layer.} \\
 & \ubbullet{zShrine}{Use an AvatarMask that includes spine, head, arms, and hands, and excludes hips, legs, feet, and toes.} \\
 & \ubbullet{zShrine}{The aiming pose should be stable and not continuously shake. Layer weight should change smoothly when entering and exiting.} \\
 & \ubbullet{zShrine}{Save Assets/Scene/0055.scene, Assets/Archer.controller, the control scripts, and every newly added or modified asset.} \\*
\ubfield{zZonai}{Resources} & \textcolor{zZonai!82!black}{Animator · AnimatorController · Blend Tree · Input System} \\
\noalign{\vskip 3pt}
\multicolumn{2}{@{}p{14.4cm}@{}}{\textcolor{zRockLite!48}{\rule{\linewidth}{0.35pt}}}\\[4pt]
\multicolumn{2}{@{}p{14.4cm}@{}}{\ubtaskhead{zDawn}{0065-ShootAheadSplitBodyAim}{ShootAhead Split Body Aim}{zGoldDeep}{Medium}{12}}\\*[4pt]
\ubfield{zChampion}{Goal} & This project is a minimal URP slice of ShootAhead character animation logic. It preserves the real character2 character art, skeleton, AnimatorController, and base animations. Complete the player's split-body aiming animation logic so the character simultaneously satisfies movement semantics and aiming semantics in a top-down shooting scene. \\
\ubfield{zDawn}{Context} & The minimal ShootAhead URP project preserves character2, its skeleton and AnimatorController, and an AimMarker. \\
\ubfield{zShrine}{Constraints} & \ubbullet{zShrine}{Complete and save the player components, scripts, AnimatorController, AvatarMask, and prefab or scene structure changes in Edit Mode; do not create or adjust the required setup only temporarily at runtime.} \\
 & \ubbullet{zShrine}{When the player is not aiming, the player should move according to movement input direction, and the character root/lower body should naturally face the movement direction.} \\
 & \ubbullet{zShrine}{When the player aims while standing, the character should face TopDownAimController.AimWorldPosition.} \\
 & \ubbullet{zShrine}{When the player moves while aiming, preserve lower-body movement semantics while making the upper body, muzzle, and line of sight face the aim point.} \\
 & \ubbullet{zShrine}{When movement direction clearly diverges from aim direction, trigger backward-movement semantics, such as setting the MoveBackward Animator parameter or an equivalent state, instead of making the character turn their back to the aim point.} \\
 & \ubbullet{zShrine}{Upper-body facing must be implemented through Spine-related bones, an upper-body AvatarMask/Animator Layer, or an equivalent character bone chain. Simply rotating the crosshair, camera, or muzzle empty object is not accepted as a substitute for upper-body facing.} \\
 & \ubbullet{zShrine}{Use the main scene Assets/Bench0065/Scenes/Bench0065.scene with player root GameObject Bench0065\_Player and aim marker Bench0065\_AimMarker.} \\
 & \ubbullet{zShrine}{Preserve Assets/Arts/Character/character2.prefab, Assets/Arts/Character/Player\_Gaming.controller, and Assets/Arts/Character/Player\_Gaming\_UpperBody.mask while completing the required logic.} \\*
\ubfield{zZonai}{Resources} & \textcolor{zZonai!82!black}{Animator · AnimatorController · Avatar Mask · Prefab} \\
\noalign{\vskip 3pt}
\multicolumn{2}{@{}p{14.4cm}@{}}{\textcolor{zRockLite!48}{\rule{\linewidth}{0.35pt}}}\\[4pt]
\multicolumn{2}{@{}p{14.4cm}@{}}{\ubtaskhead{zDawn}{0051-AnimationClassification}{Animation Classification}{zShrine}{Hard}{20}}\\*[4pt]
\ubfield{zChampion}{Goal} & In the TestScene scene, move each character root node that matches a description under the corresponding answer node. Each character in the scene plays one animation clip. Read and analyze the animation data, then move the character root node under the node whose name describes the matching animation effect. Each node corresponds to exactly one answer. \\
\ubfield{zDawn}{Context} & Each character in TestScene plays one animation clip, and the nodes under Answer describe the available categories. \\
\ubfield{zShrine}{Constraints} & \ubbullet{zShrine}{In TestScene, classify each character's animation clip by moving the character's parent GameObject under the matching description GameObject below Answer.} \\
 & \ubbullet{zShrine}{Every character GameObject has exactly one description GameObject that fully matches its animation.} \\
 & \ubbullet{zShrine}{Only change the scene hierarchy needed to reparent character parent GameObjects; do not modify any other GameObjects, components, or assets.} \\*
\ubfield{zZonai}{Resources} & \textcolor{zZonai!82!black}{AnimationClip · Animator · Hierarchy} \\
\noalign{\vskip 3pt}
\multicolumn{2}{@{}p{14.4cm}@{}}{\textcolor{zRockLite!48}{\rule{\linewidth}{0.35pt}}}\\[4pt]
\multicolumn{2}{@{}p{14.4cm}@{}}{\ubtaskhead{zDawn}{0079-FPSRemoteWeaponAnimationSync}{Synchronize Remote SCAR Weapon Presentation}{zShrine}{Hard}{18}}\\*[4pt]
\ubfield{zChampion}{Goal} & Complete the third-person SCAR humanoid rig, weapon binding, grip targets, semantic animation playback, aim pose, hands, fingers, and recoil synchronization while preserving first-person behavior. \\
\ubfield{zDawn}{Context} & The FPSFramework slice contains a working first-person SCAR and an incomplete third-person player visual, SCAR binding, Humanoid setup, and runtime presentation code under Assets/MultiplayerVisuals. \\
\ubfield{zShrine}{Constraints} & \ubbullet{zShrine}{Configure the third-person character as a valid Humanoid with complete arm, hand, and finger mappings.} \\
 & \ubbullet{zShrine}{Configure a visible third-person SCAR binding with correct muzzle direction.} \\
 & \ubbullet{zShrine}{Define calibrated position and orientation targets for both palms.} \\
 & \ubbullet{zShrine}{Drive the remote Animator, upper-body aim, weapon pose, palms, and fingers from semantic snapshots.} \\
 & \ubbullet{zShrine}{Keep the third-person SCAR muzzle consistent with first-person aim while idle, aiming, and firing.} \\
 & \ubbullet{zShrine}{Allow authored reload deviation but return the weapon cleanly to the aim direction afterward.} \\
 & \ubbullet{zShrine}{Make both palms and fingers wrap the SCAR grips naturally without visible hand or finger penetration.} \\
 & \ubbullet{zShrine}{Play authored reload and fire actions remotely without restarting repeated snapshots every frame.} \\
 & \ubbullet{zShrine}{Preserve first-person SCAR behavior and add non-zero look or camera recoil for a shot.} \\
 & \ubbullet{zShrine}{Keep arm, wrist, hand, and finger bones finite, plausible, and free of flips or discontinuous jumps.} \\
 & \ubbullet{zShrine}{Do not replace the first-person SCAR, remove authored animations, or disable the remote body, hands, fingers, weapon, recoil, or action behavior.} \\*
\ubfield{zZonai}{Resources} & \textcolor{zZonai!82!black}{Prefab · Animator · Animation · Humanoid IK · Avatar Mask · MonoBehaviour · Transform · Model Importer} \\
\noalign{\vskip 3pt}
\multicolumn{2}{@{}p{14.4cm}@{}}{\textcolor{zRockLite!48}{\rule{\linewidth}{0.35pt}}}\\[4pt]
\noalign{\vskip 3pt}
\multicolumn{2}{@{}p{14.4cm}@{}}{\ubcategory{zRoyal}{Editor Operations}{11 tasks \textperiodcentered\ 142 milestone tests}}\\*[5pt]
\multicolumn{2}{@{}p{14.4cm}@{}}{\ubtaskhead{zRoyal}{0004-IslandBenchRoadClassification}{Island Bench Road Group Cleanup}{zZonai}{Easy}{9}}\\*[4pt]
\ubfield{zChampion}{Goal} & A set of straight road segments is still placed under the Unsorted group. Based on each road segment's spatial position, connection relationship, and the existing named road groups, move these straight road segments under the corresponding road groups. \\
\ubfield{zDawn}{Context} & Road/Unsorted in the Island\_Bench scene contains unclassified straight road segments. The unusual material appearance in the scene is an intentional distractor, not an initial-environment defect, and does not affect the road-classification task. \\
\ubfield{zShrine}{Constraints} & \ubbullet{zShrine}{Positive world X points west, and positive Z points south.} \\
 & \ubbullet{zShrine}{The map contains two small islands connected by an east-west bridge and a south-north bridge.} \\
 & \ubbullet{zShrine}{The west island is in the northwest part of the map and contains Factory Rd. and Bridge Rd.(W).} \\
 & \ubbullet{zShrine}{The east island is in the east and south parts of the map and contains SShore Ave., EShore Ave, Connection Rd., and Bridge Rd.(E).} \\
 & \ubbullet{zShrine}{Factory Rd. is north of the west-island intersection where the two bridges meet, and the road is slightly curved.} \\
 & \ubbullet{zShrine}{Bridge Rd.(W) is west of the west-island intersection where the two bridges meet, and the road has a right-angle turn.} \\
 & \ubbullet{zShrine}{SShore Ave. is at the end of the south-north bridge on the east island. It has a right-angle turn, intersects Connection Rd., and then extends north.} \\
 & \ubbullet{zShrine}{EShore Ave is at the end of the east-west bridge on the east island, and it runs straight north-south.} \\
 & \ubbullet{zShrine}{The eastern section of Connection Rd. is parallel to EShore Ave. Its southern section intersects SShore Ave. and EShore Ave, and after meeting SShore Ave. it extends west.} \\
 & \ubbullet{zShrine}{Bridge Rd.(E) is perpendicular to the eastern section of Connection Rd., intersects it to the north, and also intersects EShore Ave.} \\
 & \ubbullet{zShrine}{Road segments before each bridge meets a road belong to the bridge.} \\
 & \ubbullet{zShrine}{Preserve the existing groups Bridge Rd.(E), Bridge Rd.(W), Connection Rd., EShore Ave, EW Bridge, Factory Rd., SN Bridge, SShore Ave., and Unsorted with their exact names.} \\
 & \ubbullet{zShrine}{Open Assets/Scenes/Island\_Bench.scene.} \\
 & \ubbullet{zShrine}{Inspect the Road\_Straight* road segments under Unsorted.} \\
 & \ubbullet{zShrine}{Based on the road directions and connection relationships in the scene, move the straight road segments belonging to each road under the corresponding road group.} \\
 & \ubbullet{zShrine}{Keep the Unsorted group itself, but do not leave straight road segments that need classification directly under Unsorted.} \\
 & \ubbullet{zShrine}{Do not delete or rename existing road groups.} \\
 & \ubbullet{zShrine}{Do not delete road segments.} \\
 & \ubbullet{zShrine}{Do not bypass the classification requirement by renaming road segments.} \\
 & \ubbullet{zShrine}{Only adjust road-segment parent-child relationships; leave model positions, rotations, and scales unchanged.} \\*
\ubfield{zZonai}{Resources} & \textcolor{zZonai!82!black}{Scene · Hierarchy · Transform} \\
\noalign{\vskip 3pt}
\multicolumn{2}{@{}p{14.4cm}@{}}{\textcolor{zRockLite!48}{\rule{\linewidth}{0.35pt}}}\\[4pt]
\multicolumn{2}{@{}p{14.4cm}@{}}{\ubtaskhead{zRoyal}{0013-M249MachineGun}{M249}{zZonai}{Easy}{15}}\\*[4pt]
\ubfield{zChampion}{Goal} & Make the M249 behave like the existing weapons in the project: it should be recognized by the data system as an item, carried by the player, and used correctly by the shooting system. \\
\ubfield{zDawn}{Context} & The existing ShootAhead project already contains item, weapon, inventory, pickup, shop, and shooting flows. \\
\ubfield{zShrine}{Constraints} & \ubbullet{zShrine}{The M249 uses 7.62 ammo.} \\
 & \ubbullet{zShrine}{The M249 magazine capacity is 100.} \\
 & \ubbullet{zShrine}{Integrate the M249 using the existing AssaultRifle type in the project.} \\
 & \ubbullet{zShrine}{Follow the existing implementation patterns for weapons, items, and gameplay flow in the project.} \\
 & \ubbullet{zShrine}{Do not break existing weapons, inventory, pickup, shop, or Raid flows.} \\*
\ubfield{zZonai}{Resources} & \textcolor{zZonai!82!black}{ScriptableObject · Gameplay Data · Inventory · Audio · AssetDatabase} \\
\noalign{\vskip 3pt}
\multicolumn{2}{@{}p{14.4cm}@{}}{\textcolor{zRockLite!48}{\rule{\linewidth}{0.35pt}}}\\[4pt]
\multicolumn{2}{@{}p{14.4cm}@{}}{\ubtaskhead{zRoyal}{0014-ShootAheadAssetPruning}{ShootAhead Asset Pruning}{zZonai}{Easy}{15}}\\*[4pt]
\ubfield{zChampion}{Goal} & Keep the scenes, scripts, data, runtime resources, and direct dependencies required by the build. Delete examples, demos, unused art assets, and unrelated resources that will not enter the current build flow. \\
\ubfield{zDawn}{Context} & The ShootAhead project contains enabled Build Settings scenes together with many potentially redundant assets. \\
\ubfield{zShrine}{Constraints} & \ubbullet{zShrine}{Use the enabled scenes in Build Settings as the source of truth.} \\
 & \ubbullet{zShrine}{Keep the project able to open and compile normally.} \\
 & \ubbullet{zShrine}{Keep runtime data, scripts, materials, textures, audio, and prefab references required by the current game flow valid.} \\
 & \ubbullet{zShrine}{Reduce the size of the Assets directory as much as possible.} \\*
\ubfield{zZonai}{Resources} & \textcolor{zZonai!82!black}{Editor Build Settings · AssetDatabase · Dependency Analysis} \\
\noalign{\vskip 3pt}
\multicolumn{2}{@{}p{14.4cm}@{}}{\textcolor{zRockLite!48}{\rule{\linewidth}{0.35pt}}}\\[4pt]
\multicolumn{2}{@{}p{14.4cm}@{}}{\ubtaskhead{zRoyal}{0027-VolumeBoundsOrganizer}{Automated Scene Grouping and Renaming Based on Spatial Bounding-Box Features}{zZonai}{Easy}{16}}\\*[4pt]
\ubfield{zChampion}{Goal} & Write or complete the editor automation script SceneOrganizer.cs in the DevTools.Automation namespace. It must implement automatic spatial-feature recognition and hierarchy reorganization for unordered scene objects. \\
\ubfield{zDawn}{Context} & The active scene root contains many flat, unorganized static objects imported from DCC tools. \\
\ubfield{zShrine}{Constraints} & \ubbullet{zShrine}{Scan all GameObjects under the current active scene root that both have MeshRenderer and MeshFilter components and currently have no parent object. Only objects that have both components are valid scan targets. Objects missing either component must be strictly skipped. Objects outside the root hierarchy must also be strictly skipped even if they have both MeshRenderer and MeshFilter; do not capture them beyond the allowed scope.} \\
 & \ubbullet{zShrine}{To defend against world-space axis-aligned bounding box (AABB) "projection expansion" errors caused by object rotation, volume calculation must use the object's local oriented bounding box (OBB) features. The formula is strictly: Volume = abs(size.x * scale.x) * abs(size.y * scale.y) * abs(size.z * scale.z), where size is MeshFilter.sharedMesh.bounds.size, and scale is transform.lossyScale. Take the absolute value of each component to prevent negative scale from producing negative volume. Objects whose calculated volume is zero, including scale 0 on any axis, should still be classified according to the Volume \textless{} 1.0 rule. If sharedMesh is null, do not calculate volume and skip the object directly.} \\
 & \ubbullet{zShrine}{Based on the calculated volume, rename each object and place it under the specified root-level parent GameObject, creating that parent when it does not exist.} \\
 & \ubbullet{zShrine}{If Volume \textless{} 1.0, rename to Prop\_OriginalName and place under Environment\_Props.} \\
 & \ubbullet{zShrine}{If 1.0 \textless{}= Volume \textless{} 10.0, rename to Furn\_OriginalName and place under Environment\_Furniture.} \\
 & \ubbullet{zShrine}{If Volume \textgreater{}= 10.0, rename to Struct\_OriginalName and place under Environment\_Structures.} \\
 & \ubbullet{zShrine}{When renaming, if the object name already starts with Prop\_, Furn\_, or Struct\_, first strip the existing prefix before applying the correct prefix, so duplicate prefixes such as Prop\_Prop\_OriginalName are never produced. If an existing prefix does not match the prefix indicated by the volume calculation, correct the prefix and move the object to the corresponding group.} \\
 & \ubbullet{zShrine}{ExecuteOrganization(Scene) must return an int representing the number of valid objects that successfully completed the full processing flow: volume calculation, name decision, renaming, and parent reassignment. Skipped objects, such as missing sharedMesh, non-root objects, or objects missing required components, do not count. Note: once an object passes validation and completes the flow, it must count even if its name and parent do not actually change, for example if it already has a compliant prefix. Since already-grouped objects are no longer root objects on repeated execution, repeated execution should return 0.} \\
 & \ubbullet{zShrine}{Provide a [MenuItem("Tools/Scene/Organize by Bounding Box Volume")] entry method named OrganizeActiveSceneMenu(). This method is responsible for calling ExecuteOrganization and managing the surrounding Undo group with Undo.IncrementCurrentGroup, Undo.SetCurrentGroupName, and exception rollback protection.} \\
 & \ubbullet{zShrine}{The entire reorganization operation must support one-click undo. Correctly use Undo.IncrementCurrentGroup to merge the operation group, Undo.RegisterCreatedObjectUndo for newly created parent GameObjects, Undo.RecordObject before renaming, and Undo.SetTransformParent when modifying hierarchy. Pressing Ctrl+Z must roll back perfectly without loss. All operations must be merged into a single Undo group so one undo fully reverts the operation.} \\
 & \ubbullet{zShrine}{Repeatedly running this tool on the same scene must not produce duplicate prefixes such as Prop\_Prop\_OriginalName, repeatedly migrate objects between groups, or nest grouping folders inside themselves. For objects already in the correct group with a compliant prefix, repeated execution should leave their state unchanged.} \\
 & \ubbullet{zShrine}{Volume boundary checks use strict comparison with no extra tolerance: Volume \textless{} 1.0 goes to Props, 1.0 \textless{}= Volume \textless{} 10.0 goes to Furniture, and Volume \textgreater{}= 10.0 goes to Structures. Do not offset with a floating-point epsilon.} \\
 & \ubbullet{zShrine}{Implement the automation as elegantly, robustly, and readably as possible. Handle edge cases such as missing sharedMesh and zero volume so the script runs stably in any complex unordered scene.} \\*
\ubfield{zZonai}{Resources} & \textcolor{zZonai!82!black}{UnityEditor · AssetDatabase · Undo · Mesh Bounds · Transform} \\
\noalign{\vskip 3pt}
\multicolumn{2}{@{}p{14.4cm}@{}}{\textcolor{zRockLite!48}{\rule{\linewidth}{0.35pt}}}\\[4pt]
\multicolumn{2}{@{}p{14.4cm}@{}}{\ubtaskhead{zRoyal}{0028-ImportExportPackages}{Import / Export Package Asset Organization}{zZonai}{Easy}{7}}\\*[4pt]
\ubfield{zChampion}{Goal} & Complete one asset package export and one asset package import organization pass in the current Unity project. All assets are under Assets/Bench0028. \\
\ubfield{zDawn}{Context} & Assets/Bench0028 provides ShootAhead assets to export and a unitypackage whose contents must be imported and organized. \\
\ubfield{zShrine}{Constraints} & \ubbullet{zShrine}{Export Assets/Bench0028/Exports/Bench0028\_SelectedPrefabs.unitypackage containing exactly SM\_Bld\_House\_02, SM\_Bld\_SmallBuilding\_03, SM\_Generic\_Mountains\_Grass\_02, SM\_Veh\_Bike\_01, and SM\_Wep\_Rifle\_Assault\_05 from their supplied Assets/Bench0028/Arts paths.} \\
 & \ubbullet{zShrine}{Do not export any other extra prefabs.} \\
 & \ubbullet{zShrine}{Include the models, materials, textures, and other dependency assets that these 5 prefabs actually need.} \\
 & \ubbullet{zShrine}{From Assets/Bench0028/Arts/0028Import.unitypackage, import only Cabinet 03, Bin 01, Coffee Table 01, Computer Desk 01, Corner Pillar 01, and their required dependencies.} \\
 & \ubbullet{zShrine}{Place the five imported target prefabs at Assets/Bench0028/Imported/Prefabs with their package names unchanged.} \\
 & \ubbullet{zShrine}{These prefabs must load normally and must not have missing Mesh or material references.} \\
 & \ubbullet{zShrine}{Dependency assets required by these prefabs must also be organized under Assets/Bench0028/Imported.} \\
 & \ubbullet{zShrine}{Do not leave the 5 target prefabs under the original package path Assets/Vendor/CUBE - Sci Fi Interiors Pack/Prefabs.} \\
 & \ubbullet{zShrine}{Do not place any extra prefabs in Assets/Bench0028/Imported/Prefabs.} \\*
\ubfield{zZonai}{Resources} & \textcolor{zZonai!82!black}{Package Import · Package Export · AssetDatabase · Prefab} \\
\noalign{\vskip 3pt}
\multicolumn{2}{@{}p{14.4cm}@{}}{\textcolor{zRockLite!48}{\rule{\linewidth}{0.35pt}}}\\[4pt]
\multicolumn{2}{@{}p{14.4cm}@{}}{\ubtaskhead{zRoyal}{0042-PrefabSharedMaterialIsolation}{Prefab Asset Modification, Loose Object Replacement, and Variant Construction}{zZonai}{Easy}{16}}\\*[4pt]
\ubfield{zChampion}{Goal} & Modify shared prefab assets safely, replace loose scene objects with prefabs, and build the required prefab variants while preserving instance links and material isolation. \\
\ubfield{zDawn}{Context} & Bench0042 provides a shared-material experiment, loose scene objects to replace, and scenes for building prefab variants. \\
\ubfield{zShrine}{Constraints} & \ubbullet{zShrine}{Change the Visual of HazardCrystal.prefab to a red warning emissive appearance; keep GuideCrystal.prefab unchanged.} \\
 & \ubbullet{zShrine}{Change the Pipe of WarningPipe.prefab to a yellow warning appearance; keep UtilityPipe.prefab unchanged.} \\
 & \ubbullet{zShrine}{Change the Liquid of PoisonBarrel.prefab to a green toxic liquid appearance; keep WaterBarrel.prefab and FuelBarrel.prefab unchanged.} \\
 & \ubbullet{zShrine}{Change the Screen of AlarmConsole.prefab to a red alarm emissive appearance; keep ControlConsole.prefab unchanged.} \\
 & \ubbullet{zShrine}{Do not only modify instances in SharedMaterialLab.scene; the correct result must come from the prefab asset.} \\
 & \ubbullet{zShrine}{Save any new material resources used by the modifications under Assets/Bench0042/Materials.} \\
 & \ubbullet{zShrine}{Do not unpack prefab instances in the scene.} \\
 & \ubbullet{zShrine}{Do not replace meshes, delete renderers, change renderer child object names, or change material slot order.} \\
 & \ubbullet{zShrine}{Save the prefab assets and the scene.} \\
 & \ubbullet{zShrine}{Red warning: base color close to (1.0, 0.05, 0.03), with emissive color clearly red as well.} \\
 & \ubbullet{zShrine}{Yellow warning: base color close to (1.0, 0.78, 0.05), with emissive color clearly yellow as well.} \\
 & \ubbullet{zShrine}{Green toxic liquid: base color close to (0.10, 0.85, 0.20), with emissive color clearly green as well.} \\
 & \ubbullet{zShrine}{Create a new prefab asset: Assets/Bench0042/Prefabs/WarningBeacon.prefab.} \\
 & \ubbullet{zShrine}{WarningBeacon.prefab must contain four direct child objects: Base, Lens, Antenna, and Trigger.} \\
 & \ubbullet{zShrine}{Replace all 12 ordinary LooseBeacon\_XX objects under ReplacementTargets with instances of WarningBeacon.prefab.} \\
 & \ubbullet{zShrine}{After replacement, each instance must preserve the original object's world position, world rotation, world scale, and parent.} \\
 & \ubbullet{zShrine}{All 12 replacement results must come from the same WarningBeacon.prefab; do not create 12 different prefabs.} \\
 & \ubbullet{zShrine}{Do not unpack the replaced prefab instances.} \\
 & \ubbullet{zShrine}{Do not leave behind the original ordinary LooseBeacon\_XX objects.} \\
 & \ubbullet{zShrine}{Save WarningBeacon.prefab and LooseReplacementLab.scene.} \\
 & \ubbullet{zShrine}{LockedSecurityDoor.prefab: a prefab variant based on SecurityDoorBase.prefab; Panel and AccessLight have a red warning emissive appearance; Panel.localScale is close to (1.22, 1.68, 0.18); place 2 instances under VariantTargets at world positions (-3.2, 0, 0) and (-1.35, 0, 0).} \\
 & \ubbullet{zShrine}{SurveyScannerCart.prefab: a prefab variant based on UtilityCartBase.prefab; attach ScannerModule.prefab as a direct child named ScannerModule, with localPosition close to (0, 0.78, 0.08) and localScale close to (0.85, 0.85, 0.85); place 2 instances under VariantTargets at world positions (1.25, 0, 0) and (4.05, 0, 0), with the second instance rotated 25 degrees around the Y axis.} \\*
\ubfield{zZonai}{Resources} & \textcolor{zZonai!82!black}{Prefab · Prefab Variant · Material · AssetDatabase} \\
\noalign{\vskip 3pt}
\multicolumn{2}{@{}p{14.4cm}@{}}{\textcolor{zRockLite!48}{\rule{\linewidth}{0.35pt}}}\\[4pt]
\multicolumn{2}{@{}p{14.4cm}@{}}{\ubtaskhead{zRoyal}{0052-URPPipelineConfiguration}{URP Pipeline Configuration}{zZonai}{Easy}{8}}\\*[4pt]
\ubfield{zChampion}{Goal} & Configure this project to use the Universal Render Pipeline and convert the sample materials. \\
\ubfield{zDawn}{Context} & The project contains Built-in Render Pipeline Standard-material samples and no completed URP setup. \\
\ubfield{zShrine}{Constraints} & \ubbullet{zShrine}{Install the Universal Render Pipeline package for this Unity version.} \\
 & \ubbullet{zShrine}{Create four quality levels named exactly Low, Medium, High, and ExtraHigh.} \\
 & \ubbullet{zShrine}{Create and assign a separate URP Render Pipeline Asset for each quality level, with settings that form a reasonable progression rather than four identical copies.} \\
 & \ubbullet{zShrine}{Set the default/active project quality level to Medium.} \\
 & \ubbullet{zShrine}{Make Medium the pipeline used by the project by default.} \\
 & \ubbullet{zShrine}{Convert every material in Assets/Bench0052/Materials to the shader Universal Render Pipeline/Lit.} \\
 & \ubbullet{zShrine}{Preserve each material's visible base color and alpha while converting.} \\*
\ubfield{zZonai}{Resources} & \textcolor{zZonai!82!black}{Package Manager · URP · Render Pipeline Asset · Quality Settings · Material · Shader} \\
\noalign{\vskip 3pt}
\multicolumn{2}{@{}p{14.4cm}@{}}{\textcolor{zRockLite!48}{\rule{\linewidth}{0.35pt}}}\\[4pt]
\multicolumn{2}{@{}p{14.4cm}@{}}{\ubtaskhead{zRoyal}{0016-TexturePipeline}{Automated Texture Import Pipeline Based on Semantic Routing and Low-Level Memory Defense}{zGoldDeep}{Medium}{8}}\\*[4pt]
\ubfield{zChampion}{Goal} & Write the editor automation script AutoTextureImporter.cs in this project. It must inherit from AssetPostprocessor and take control of the import lifecycle for all textures in the project. \\
\ubfield{zDawn}{Context} & The project contains textures with varied formats, misleading paths or names, and NPOT dimensions. \\
\ubfield{zShrine}{Constraints} & \ubbullet{zShrine}{Create AutoTextureImporter.cs as an AssetPostprocessor that controls the first-import texture pipeline.} \\
 & \ubbullet{zShrine}{Override GetVersion() and return an integer greater than 0 so script-version changes invalidate the engine's cached import results.} \\
 & \ubbullet{zShrine}{Non-destructive behavior: It is strictly forbidden to overwrite parameter settings on old textures that already exist in the project and have been manually modified by artists. This script may only configure assets on a brand-new first import.} \\
 & \ubbullet{zShrine}{Disable Read/Write Enabled by default for all new textures.} \\
 & \ubbullet{zShrine}{All non-UI 3D textures must set Non-Power of 2 handling to scale to the nearest power of two (ToNearest).} \\
 & \ubbullet{zShrine}{Detection rule: An asset is considered a UI asset only when its full path strictly starts with Assets/TestEnv/UI/. Do not be fooled by other path names that merely contain "UI".} \\
 & \ubbullet{zShrine}{UI handling: Set it to Sprite (2D and UI), and force MipMap off.} \\
 & \ubbullet{zShrine}{Atlas extraction: Dynamically extract the name of the UI asset's direct parent folder and assign it to Sprite Packing Tag.} \\
 & \ubbullet{zShrine}{For non-UI semantic routing, remove the file extension and compare suffixes case-insensitively.} \\
 & \ubbullet{zShrine}{Ends with \_Albedo: set to Default type and enable sRGB.} \\
 & \ubbullet{zShrine}{Ends with \_MaskMap: set to Default type and disable sRGB (Linear space).} \\
 & \ubbullet{zShrine}{Ends with \_Normal: set to Normal map type. Important: strictly forbid the engine from converting it as a grayscale image. Do not enable Create from Grayscale.} \\
 & \ubbullet{zShrine}{Fallback logic: If the asset does not match any of the naming suffixes above (a 3D model texture with no suffix or an unrecognized name), set it to Default type and force disable sRGB so it stays in Linear space and avoids material overexposure.} \\
 & \ubbullet{zShrine}{Build corresponding TextureImporterPlatformSettings for iOS and Android separately. Enable overrides (overridden = true) and limit Max Size uniformly to 2048.} \\
 & \ubbullet{zShrine}{Use native engine APIs to detect whether the source file contains an alpha channel, then select platform formats as follows.} \\
 & \ubbullet{zShrine}{UI textures: always use ASTC\_4x4 on both platforms.} \\
 & \ubbullet{zShrine}{3D textures (iOS): use ASTC\_6x6 when there is no Alpha; downgrade to ASTC\_4x4 when Alpha is present to prevent color leakage.} \\
 & \ubbullet{zShrine}{3D textures (Android): use ASTC\_8x8 when there is no Alpha; downgrade to ASTC\_5x5 when Alpha is present.} \\*
\ubfield{zZonai}{Resources} & \textcolor{zZonai!82!black}{AssetPostprocessor · Texture Importer · AssetDatabase · Platform Settings} \\
\noalign{\vskip 3pt}
\multicolumn{2}{@{}p{14.4cm}@{}}{\textcolor{zRockLite!48}{\rule{\linewidth}{0.35pt}}}\\[4pt]
\multicolumn{2}{@{}p{14.4cm}@{}}{\ubtaskhead{zRoyal}{0018-FindGameObjectsInComplexScene}{Find GameObjects in a Complex Scene}{zGoldDeep}{Medium}{10}}\\*[4pt]
\ubfield{zChampion}{Goal} & Read the project structure, search the scene for GameObjects that match the requirements, and move the found objects under the corresponding child nodes of the scene's Answer node. \\
\ubfield{zDawn}{Context} & TestScene contains many tables, lights, and other objects, plus Answer/Q1-Q10 nodes for submission. \\
\ubfield{zShrine}{Constraints} & \ubbullet{zShrine}{Find a lit chandelier that is rotated 45 degrees around the Y axis.} \\
 & \ubbullet{zShrine}{Find an unlit floor lamp that is scaled by 2x on the XZ plane, with no scaling on the Y axis.} \\
 & \ubbullet{zShrine}{Find a long table whose layer is Ignore Raycast.} \\
 & \ubbullet{zShrine}{Find an unlit desk lamp that is marked Static.} \\
 & \ubbullet{zShrine}{Find a short table with exactly one lit floor lamp within 1f behind it.} \\
 & \ubbullet{zShrine}{Find a lit chandelier with no object directly below it and exactly one lit chandelier within 1f behind it.} \\
 & \ubbullet{zShrine}{Find a floor lamp with a table within 1f in front of it, where the table's Z-axis direction has equal angles to the world X-axis and world Z-axis directions.} \\
 & \ubbullet{zShrine}{Find a table with two desk lamps placed on it.} \\
 & \ubbullet{zShrine}{Find a desk lamp that has a chandelier, a floor lamp, and a table within 2f. Its Z axis is opposite to the Z axes of the chandelier and floor lamp, and forms a 90-degree angle with the table's Z axis.} \\
 & \ubbullet{zShrine}{Find the desk lamp on the lowest level of the building that is farthest to the southeast. The scene's positive Z axis points north.} \\
 & \ubbullet{zShrine}{Reparent the ten unique matches in order under Answer/Q1 through Answer/Q10.} \\
 & \ubbullet{zShrine}{Do not reuse a GameObject, change object names or Inspector properties, or modify anything other than hierarchy parent-child relationships.} \\*
\ubfield{zZonai}{Resources} & \textcolor{zZonai!82!black}{Scene · Hierarchy · Transform · MonoBehaviour} \\
\noalign{\vskip 3pt}
\multicolumn{2}{@{}p{14.4cm}@{}}{\textcolor{zRockLite!48}{\rule{\linewidth}{0.35pt}}}\\[4pt]
\multicolumn{2}{@{}p{14.4cm}@{}}{\ubtaskhead{zRoyal}{0060-GunShotSound}{Build a Production Gunshot Audio System}{zShrine}{Hard}{29}}\\*[4pt]
\ubfield{zChampion}{Goal} & Create a data-driven, pooled, mixer-routed gunshot system for AK-74, M4, and M249 profiles with automatic fire, suppression, distance layers, environment tails, casings, occlusion, debugging, and stress testing. \\
\ubfield{zDawn}{Context} & The project supplies the Modern Weapons ALL-IN-ONE Sound System clip library; the 0060 scene, profiles, mixer, runtime architecture, debug panel, and audit tooling must be created without adding or replacing audio. \\
\ubfield{zShrine}{Constraints} & \ubbullet{zShrine}{Use only the supplied library under Assets/Modern Weapons ALL-IN-ONE Sound System/Sounds; do not delete, rename, duplicate, or replace it, and do not import additional audio or third-party code.} \\
 & \ubbullet{zShrine}{Create Assets/0060/Scenes/GunShotAudioLab.scene, enable it in Build Settings, and ensure it runs without Console errors or missing references.} \\
 & \ubbullet{zShrine}{Create Assets/0060/Audio/Mixers/GunAudio.mixer with Master, Gunshots/Near, Gunshots/Distant, Gunshots/Tails, Mechanical, and Casings groups plus Outdoor, Indoor, and Corridor snapshots.} \\
 & \ubbullet{zShrine}{Create independent AK74, M4, and M249 ScriptableObject profiles using only their corresponding supplied weapon clips.} \\
 & \ubbullet{zShrine}{Each profile must provide near, distant, automatic-loop, stop, suppressed, environment-tail, cadence, randomization, cooldown, and concurrency data.} \\
 & \ubbullet{zShrine}{Create all required named system, emitter, listener, zone, marker, wall, and debug-panel scene anchors exactly as listed in the Prompt.} \\
 & \ubbullet{zShrine}{Use exactly one enabled AudioListener; place the Near, Mid, and Far markers approximately 2 m, 35 m, and 100 m from PlayerGunEmitter\_0060, and place a collidable wall physically between the player emitter and Listener\_Occluded\_0060.} \\
 & \ubbullet{zShrine}{Provide collidable Concrete, Metal, and Wood surface examples.} \\
 & \ubbullet{zShrine}{Support at least three independent emitters and the specified player controls for fire, suppressor, weapon, listener distance, environment, and stress test.} \\
 & \ubbullet{zShrine}{Automatic fire must use a safe start-loop-stop lifecycle with at most one main loop per emitter and no orphaned loops.} \\
 & \ubbullet{zShrine}{Logical cadence must remain within two percent of profile RPM across supported frame rates and remain safe while timeScale is zero.} \\
 & \ubbullet{zShrine}{Each shot must layer near, distant, environment-tail, and surface-aware casing audio delayed by 0.08 to 0.15 seconds, with smooth near/distant blending by listener distance and no immediate clip repeats when alternatives exist.} \\
 & \ubbullet{zShrine}{Occlusion must use 3D physics and smoothly attenuate and low-pass obstructed emitters, then recover after line of sight returns.} \\
 & \ubbullet{zShrine}{All runtime gunshot sources must be 3D and routed to the appropriate mixer groups.} \\
 & \ubbullet{zShrine}{Prewarm a shared pool capped at 32 simultaneous one-shot voices, avoid instantiating or destroying voice GameObjects after warm-up, and prioritize important nearby gunshots over tails, mechanical details, and casings when capacity is exhausted.} \\
 & \ubbullet{zShrine}{Persist linear Master, Gunshots, and Casings volume controls and clamp silence to a finite decibel floor.} \\
 & \ubbullet{zShrine}{Apply appropriate compressed/non-streaming import settings and provide an editor audit command.} \\
 & \ubbullet{zShrine}{Expose FireSingle(), BeginAutomaticFire(), EndAutomaticFire(), SetSuppressed(bool), and all required readable live values on each named emitter.} \\
 & \ubbullet{zShrine}{Expose SetBusVolume(string, float), GetBusDecibels(string), and the required pool and active-voice live values on the global voice owner.} \\
 & \ubbullet{zShrine}{GunAudioDebugPanel\_0060 must report weapon, suppressor and firing state, logical shots, voice usage and capacity, dropped or stolen counts, distance blend, environment, and occlusion.} \\
 & \ubbullet{zShrine}{The F9 stress test must issue at least 200 rapid requests across at least 12 emitters, visibly report pass or fail, respect the 32-voice cap, detect orphaned loops, and confirm that warmed pools do not grow.} \\*
\ubfield{zZonai}{Resources} & \textcolor{zZonai!82!black}{Scene · Audio · ScriptableObject · MonoBehaviour · PhysX · Collider · AssetDatabase · Editor Build Settings} \\
\noalign{\vskip 3pt}
\multicolumn{2}{@{}p{14.4cm}@{}}{\textcolor{zRockLite!48}{\rule{\linewidth}{0.35pt}}}\\[4pt]
\multicolumn{2}{@{}p{14.4cm}@{}}{\ubtaskhead{zRoyal}{0072-CastleDestruction}{Classify Castle Structures into Hierarchy Groups}{zShrine}{Hard}{9}}\\*[4pt]
\ubfield{zChampion}{Goal} & Reparent existing castle models into exact gate, bastion, and wooden-hoarding groups while leaving continuous walls unsorted and preserving all world transforms. \\
\ubfield{zDawn}{Context} & Assets/0072.scene contains existing castle models under Castle\_Objects\_Unsorted and requires hierarchy-only classification of complete structures. \\
\ubfield{zShrine}{Constraints} & \ubbullet{zShrine}{Create Main\_Gate for the complete main gate structure.} \\
 & \ubbullet{zShrine}{Create Round\_Bastion\_01 through Round\_Bastion\_04 for the four round bastions.} \\
 & \ubbullet{zShrine}{Create Square\_Bastion\_01 through Square\_Bastion\_02 for the two square bastions.} \\
 & \ubbullet{zShrine}{Create Elevated\_Bastion\_01 through Elevated\_Bastion\_04 for elevated bastions embedded in continuous walls.} \\
 & \ubbullet{zShrine}{Create Hoarding\_Wood\_01 through Hoarding\_Wood\_02 for the complete wooden hoarding structures.} \\
 & \ubbullet{zShrine}{Place each listed structure under its own empty parent with the exact required name.} \\
 & \ubbullet{zShrine}{Exclude adjoining continuous wall sections from bastion, gate, and hoarding groups.} \\
 & \ubbullet{zShrine}{Leave all remaining continuous-wall models directly under Castle\_Objects\_Unsorted.} \\
 & \ubbullet{zShrine}{Only change parent-child relationships; do not add, delete, duplicate, move, rotate, scale, replace, unpack, or alter prefab internals.} \\
 & \ubbullet{zShrine}{Preserve world transforms and save Assets/0072.scene.} \\*
\ubfield{zZonai}{Resources} & \textcolor{zZonai!82!black}{Scene · Hierarchy · GameObject · Transform · Prefab} \\
\noalign{\vskip 3pt}
\multicolumn{2}{@{}p{14.4cm}@{}}{\textcolor{zRockLite!48}{\rule{\linewidth}{0.35pt}}}\\[4pt]
\noalign{\vskip 3pt}
\multicolumn{2}{@{}p{14.4cm}@{}}{\ubcategory{zPlain}{Scene \& Level Construction}{19 tasks \textperiodcentered\ 365 milestone tests}}\\*[5pt]
\multicolumn{2}{@{}p{14.4cm}@{}}{\ubtaskhead{zPlain}{0001-ObjectIdentificationPrefabSorting}{Object Identification and Categorization}{zZonai}{Easy}{33}}\\*[4pt]
\ubfield{zChampion}{Goal} & Complete the identification and categorization task for 30 prefab objects in Assets/0001/Scene/0001.scene. \\
\ubfield{zDawn}{Context} & The scene provides 30 Item\_XX objects to identify and 30 parent objects named after real-world items. \\
\ubfield{zShrine}{Constraints} & \ubbullet{zShrine}{Numbered\_Items: 30 objects to identify, named Item\_01 through Item\_30.} \\
 & \ubbullet{zShrine}{Answer\_Parents: 30 empty parent objects, each named after a real-world object.} \\
 & \ubbullet{zShrine}{Observe the appearance of each Item\_XX and decide which real-world object name it corresponds to.} \\
 & \ubbullet{zShrine}{Drag each Item\_XX into the empty parent object under Answer\_Parents with the matching name.} \\
 & \ubbullet{zShrine}{Do not rename any Item\_XX, their child objects, the parent objects under Answer\_Parents, or any scene root objects.} \\
 & \ubbullet{zShrine}{Do not delete, duplicate, hide, or replace any of the objects to be identified in the scene.} \\
 & \ubbullet{zShrine}{Do not modify any prefabs or dependent assets under Assets/0001/Art.} \\*
\ubfield{zZonai}{Resources} & \textcolor{zZonai!82!black}{Scene · Hierarchy · Transform · Prefab} \\
\noalign{\vskip 3pt}
\multicolumn{2}{@{}p{14.4cm}@{}}{\textcolor{zRockLite!48}{\rule{\linewidth}{0.35pt}}}\\[4pt]
\multicolumn{2}{@{}p{14.4cm}@{}}{\ubtaskhead{zPlain}{0021-ShootAheadStreetLayout}{Street Intersection Assembly}{zZonai}{Easy}{10}}\\*[4pt]
\ubfield{zChampion}{Goal} & The folder Assets/Bench0021/Arts/Prefabs provides prefabs for roads, trees, and buildings. Use these assets to build a street scene according to the requirements. \\
\ubfield{zDawn}{Context} & Assets/Bench0021/Arts/Prefabs provides road, tree, and building prefabs. \\
\ubfield{zShrine}{Constraints} & \ubbullet{zShrine}{Place one ground piece using coordinate (0, 0, 0) as the reference.} \\
 & \ubbullet{zShrine}{Place an intersection using coordinate (0, 0, 0) as the reference, and extend roads outward in all four directions.} \\
 & \ubbullet{zShrine}{Use the 3 provided house types and place them at the four corners of the intersection. Place at least 4 houses in each direction, and use all building assets.} \\
 & \ubbullet{zShrine}{Place street trees between the roads and buildings, arranged parallel to the roads, and use all tree assets.} \\
 & \ubbullet{zShrine}{Put the ground, roads, trees, and buildings under their corresponding parent objects.} \\*
\ubfield{zZonai}{Resources} & \textcolor{zZonai!82!black}{Scene · Prefab · Transform · Hierarchy} \\
\noalign{\vskip 3pt}
\multicolumn{2}{@{}p{14.4cm}@{}}{\textcolor{zRockLite!48}{\rule{\linewidth}{0.35pt}}}\\[4pt]
\multicolumn{2}{@{}p{14.4cm}@{}}{\ubtaskhead{zPlain}{0040-ObjectIdentificationPrefabSorting}{Object Identification and Categorization}{zZonai}{Easy}{33}}\\*[4pt]
\ubfield{zChampion}{Goal} & Complete the identification and categorization task for 30 prefab objects in Assets/0040/Scene/0040.scene. \\
\ubfield{zDawn}{Context} & Assets/0040/Scene/0040.scene provides 30 objects to identify and a set of answer parent nodes. \\
\ubfield{zShrine}{Constraints} & \ubbullet{zShrine}{In Assets/0040/Scene/0040.scene, identify the 30 objects named Item\_01 through Item\_30 under Numbered\_Items.} \\
 & \ubbullet{zShrine}{Use the 30 empty GameObjects under Answer\_Parents, each named after a real-world object, as the matching destination parents.} \\
 & \ubbullet{zShrine}{Observe the appearance of each Item\_XX and decide which real-world object name it corresponds to.} \\
 & \ubbullet{zShrine}{Drag each Item\_XX into the empty parent object under Answer\_Parents with the matching name.} \\
 & \ubbullet{zShrine}{Do not rename any Item\_XX, their child objects, the parent objects under Answer\_Parents, or any scene root objects.} \\
 & \ubbullet{zShrine}{Do not delete, duplicate, hide, or replace any objects to be identified in the scene.} \\
 & \ubbullet{zShrine}{Do not modify any prefabs or dependent assets under Assets/0040/Art.} \\*
\ubfield{zZonai}{Resources} & \textcolor{zZonai!82!black}{Scene · Hierarchy · Transform · Prefab} \\
\noalign{\vskip 3pt}
\multicolumn{2}{@{}p{14.4cm}@{}}{\textcolor{zRockLite!48}{\rule{\linewidth}{0.35pt}}}\\[4pt]
\multicolumn{2}{@{}p{14.4cm}@{}}{\ubtaskhead{zPlain}{0005-BuildHouseReachability}{Paper-Thin Multistory House and Floor Reachability}{zGoldDeep}{Medium}{4}}\\*[4pt]
\ubfield{zChampion}{Goal} & Use Unity's built-in Plane as paper-thin pieces to build floors, walls, ceilings, eaves, stairs, and related GameObjects. The house should contain multiple rooms on each floor, and different floors must be reachable from one another. In the final result, a NavMeshAgent should be able to move between rooms on the same floor and use stairs to move between floors. \\
\ubfield{zDawn}{Context} & Assets/Scenes/BuildHouse.scene starts empty and provides no house geometry; the structure must be built from Unity Plane primitives. \\
\ubfield{zShrine}{Constraints} & \ubbullet{zShrine}{Keep the house footprint within 30 x 30.} \\
 & \ubbullet{zShrine}{The center of the first floor should be at world coordinate (0, 0, 0).} \\
 & \ubbullet{zShrine}{Floor height must be 3: first-floor ground at Y = 0, second-floor ground at Y = 3, and third-floor ground at Y = 6.} \\
 & \ubbullet{zShrine}{Build at least three floors.} \\
 & \ubbullet{zShrine}{Each floor must contain 3 rooms.} \\
 & \ubbullet{zShrine}{Each floor must use at least 3 independent floorXXX floor Planes to represent room floors, for example floor001 and floor002.} \\
 & \ubbullet{zShrine}{Each floor must have walls and ceilings, and adjacent floors must be reachable via stairs.} \\
 & \ubbullet{zShrine}{Floors must be named floorXXX.} \\
 & \ubbullet{zShrine}{Walls must be named wallXXX.} \\
 & \ubbullet{zShrine}{Ceilings must be named ceiling or start with ceiling.} \\
 & \ubbullet{zShrine}{Eaves must be named RoofXXX.} \\
 & \ubbullet{zShrine}{Stair names are unrestricted, but the stairs must contribute to walkable surface construction.} \\
 & \ubbullet{zShrine}{The 3 rooms on the same floor must be mutually reachable.} \\
 & \ubbullet{zShrine}{The first, second, and third floors must be mutually reachable through stairs.} \\
 & \ubbullet{zShrine}{Stairs or ramps must be recognizable by runtime NavMesh baking. Do not place decorative-only models.} \\
 & \ubbullet{zShrine}{Walkable surfaces must be wide enough for a NavMeshAgent with radius about 0.3 and height about 2 to pass.} \\
 & \ubbullet{zShrine}{Do not make the house footprint larger than 30 x 30.} \\
 & \ubbullet{zShrine}{Do not bypass the structure requirements by creating only empty GameObjects or only changing names. You must create renderable, bakeable Plane geometry.} \\
 & \ubbullet{zShrine}{Use Unity's built-in Plane to complete the paper-thin structure. Complex art assets are not required.} \\*
\ubfield{zZonai}{Resources} & \textcolor{zZonai!82!black}{Scene · Primitive · AI Navigation · NavMesh} \\
\noalign{\vskip 3pt}
\multicolumn{2}{@{}p{14.4cm}@{}}{\textcolor{zRockLite!48}{\rule{\linewidth}{0.35pt}}}\\[4pt]
\multicolumn{2}{@{}p{14.4cm}@{}}{\ubtaskhead{zPlain}{0020-InteriorFurniturePlacement}{Interior Furniture Placement}{zGoldDeep}{Medium}{39}}\\*[4pt]
\ubfield{zChampion}{Goal} & The scene Assets/0020Bench/0020Bench.scene provides a single-floor unfinished detached house enclosed by walls, floors, and ceilings. Use the existing furniture assets under Assets/0020Bench/Arts to place furniture in the corresponding rooms. \\
\ubfield{zDawn}{Context} & Assets/0020Bench/0020Bench.scene provides an unfinished single-story detached house and a collection of furniture assets. \\
\ubfield{zShrine}{Constraints} & \ubbullet{zShrine}{Complete furniture placement in Assets/0020Bench/0020Bench.scene using the existing assets under Assets/0020Bench/Arts.} \\
 & \ubbullet{zShrine}{Interpret asset prefixes as follows: Bed for bedroom, Corr for hallway, Gar for garage, Kit for kitchen/living room, and To for bathroom.} \\
 & \ubbullet{zShrine}{Use world Z+ as north and X+ as east when interpreting the room layout.} \\
 & \ubbullet{zShrine}{The kitchen and living room are on the west side of the house. The exterior door is on the south side of this room.} \\
 & \ubbullet{zShrine}{The bedroom is east of the living room and on the south side of the overall house.} \\
 & \ubbullet{zShrine}{The garage is in the southeast of the house, east of the bedroom.} \\
 & \ubbullet{zShrine}{The bathroom is in the northeast of the house, north of the garage.} \\
 & \ubbullet{zShrine}{The hallway is on the north side of the house and connects all rooms.} \\
 & \ubbullet{zShrine}{The bed faces east and is placed against the west wall of the bedroom.} \\
 & \ubbullet{zShrine}{The desk is placed against the east wall of the bedroom.} \\
 & \ubbullet{zShrine}{The wardrobe is placed against the east wall of the bedroom.} \\
 & \ubbullet{zShrine}{The lamp is at the center of the bedroom ceiling.} \\
 & \ubbullet{zShrine}{The small table is placed against the north wall.} \\
 & \ubbullet{zShrine}{The rug is placed in the center of the hallway.} \\
 & \ubbullet{zShrine}{Two tables are placed side by side against the east wall.} \\
 & \ubbullet{zShrine}{Two Shelf B objects are placed against the north wall.} \\
 & \ubbullet{zShrine}{One Shelf A object is placed against the south wall.} \\
 & \ubbullet{zShrine}{The bathtub is placed against the east wall.} \\
 & \ubbullet{zShrine}{The shower curtain frame is placed above the bathtub.} \\
 & \ubbullet{zShrine}{The toilet is placed against the north wall.} \\
 & \ubbullet{zShrine}{The shelf is placed against the north wall.} \\
 & \ubbullet{zShrine}{The sink is placed against the south wall.} \\
 & \ubbullet{zShrine}{The mirror is placed against the south wall above the sink.} \\
 & \ubbullet{zShrine}{The refrigerator is placed against the north wall.} \\
 & \ubbullet{zShrine}{The dishwasher is placed against the north wall.} \\
 & \ubbullet{zShrine}{The sink is placed against the north wall and adjacent to the dishwasher.} \\
 & \ubbullet{zShrine}{The sofa is placed against the south wall.} \\
 & \ubbullet{zShrine}{The table is placed against the south wall.} \\
 & \ubbullet{zShrine}{Four chairs are placed in pairs on the east and west sides of the table.} \\
 & \ubbullet{zShrine}{The counter is placed in the center of the room.} \\
 & \ubbullet{zShrine}{The candle is placed on the table.} \\*
\ubfield{zZonai}{Resources} & \textcolor{zZonai!82!black}{Scene · Prefab · Transform · Renderer} \\
\noalign{\vskip 3pt}
\multicolumn{2}{@{}p{14.4cm}@{}}{\textcolor{zRockLite!48}{\rule{\linewidth}{0.35pt}}}\\[4pt]
\multicolumn{2}{@{}p{14.4cm}@{}}{\ubtaskhead{zPlain}{0022-ShootAheadReferenceStreet}{Reference Image Street Recreation}{zGoldDeep}{Medium}{6}}\\*[4pt]
\ubfield{zChampion}{Goal} & Use the ground, road, building, and tree assets provided in Assets/Bench0022/Arts/Prefabs to recreate the scene layout shown in Assets/Bench0022/Reference/ReferenceImage.jpeg. \\
\ubfield{zDawn}{Context} & Bench0022 provides a perspective reference image, a top-down reference image, and ground, road, building, and tree prefabs. \\
\ubfield{zShrine}{Constraints} & \ubbullet{zShrine}{Use the ground, road, building, and tree prefabs under Assets/Bench0022/Arts/Prefabs.} \\
 & \ubbullet{zShrine}{Recreate the layout shown in ReferenceImage.jpeg, using TopImage.jpeg to verify relative positions.} \\
 & \ubbullet{zShrine}{Place the road intersection at world coordinate (0, 0, 0).} \\
 & \ubbullet{zShrine}{Arrange the roads first, then position buildings and trees to match the references.} \\*
\ubfield{zZonai}{Resources} & \textcolor{zZonai!82!black}{Scene · Prefab · Transform · Material} \\
\noalign{\vskip 3pt}
\multicolumn{2}{@{}p{14.4cm}@{}}{\textcolor{zRockLite!48}{\rule{\linewidth}{0.35pt}}}\\[4pt]
\multicolumn{2}{@{}p{14.4cm}@{}}{\ubtaskhead{zPlain}{0034-SciFiRoomVariants}{Sci-Fi Room Size Variants}{zGoldDeep}{Medium}{6}}\\*[4pt]
\ubfield{zChampion}{Goal} & Assets/Bench0034/Arts/Prefabs provides wall, floor, pillar, door, table, computer desk, and energy generator prefabs. Use these prefabs in Assets/Bench0034/Scene/0034.scene to build three separate square rooms. \\
\ubfield{zDawn}{Context} & Assets/Bench0034/Arts/Prefabs provides modular sci-fi walls, floors, pillars, doors, and room props. \\
\ubfield{zShrine}{Constraints} & \ubbullet{zShrine}{In Assets/Bench0034/Scene/0034.scene, use the supplied prefabs to create three separate parent GameObjects named Room\_5x5, Room\_4x4, and Room\_3x3.} \\
 & \ubbullet{zShrine}{The floor must be a continuous 5x5, 4x4, or 3x3 grid respectively, with no overlap and no gaps.} \\
 & \ubbullet{zShrine}{Each room must use at least two different floor prefabs.} \\
 & \ubbullet{zShrine}{All four corners must use Corner Pillar 01.} \\
 & \ubbullet{zShrine}{Walls must form a continuous square boundary with no overlap and no gaps.} \\
 & \ubbullet{zShrine}{Place one Door Small 01 at the middle of each side, for a total of 4 doors per room.} \\
 & \ubbullet{zShrine}{Place several Table01/Table02 objects and ComputerDesk objects in each room, and place 1 EnergyGenerator in each room.} \\
 & \ubbullet{zShrine}{Leave enough spacing between the three rooms so they do not intersect.} \\*
\ubfield{zZonai}{Resources} & \textcolor{zZonai!82!black}{Scene · Transform · Prefab · URP} \\
\noalign{\vskip 3pt}
\multicolumn{2}{@{}p{14.4cm}@{}}{\textcolor{zRockLite!48}{\rule{\linewidth}{0.35pt}}}\\[4pt]
\multicolumn{2}{@{}p{14.4cm}@{}}{\ubtaskhead{zPlain}{0036-NavMeshAgentOffMeshLink}{NavMeshAgent OffMeshLink Multi-Room Navigation}{zGoldDeep}{Medium}{10}}\\*[4pt]
\ubfield{zChampion}{Goal} & Complete a NavMeshAgent navigation test scene in Assets/Bench0036/Scene/0036.scene. The scene already contains a multi-room sci-fi base, lower and upper teleporters, and a set of NavTest\_* markers under Bench0036\_NavMeshTestSetup. Preserve the marker names and positions, and use them as navigation validation points. \\
\ubfield{zDawn}{Context} & Assets/Bench0036/Scene/0036.scene contains a multi-room base, lower and upper teleporters, and NavTest markers. \\
\ubfield{zShrine}{Constraints} & \ubbullet{zShrine}{Bake walkable NavMesh for the central room, medical room, control room, lounge area, energy area, and upper room.} \\
 & \ubbullet{zShrine}{NavTest\_AgentProbe must use a NavMeshAgent and be placed on a valid NavMesh.} \\
 & \ubbullet{zShrine}{Complete paths should be calculable from NavTest\_AgentSpawn to the target points in the central room, medical room, control room, energy area, and upper room.} \\
 & \ubbullet{zShrine}{Every pair of NavTest\_Door\_* markers must be connected, proving that door openings are not incorrectly blocked.} \\
 & \ubbullet{zShrine}{NavTest\_Block\_* markers represent wall or obstacle separations and must not be directly penetrated by NavMesh.} \\
 & \ubbullet{zShrine}{Configure the teleporter in the central room and the teleporter in the upper room as OffMeshLink test points. NavTest\_TeleportLowerEndpoint and NavTest\_TeleportUpperEndpoint must be connected by an enabled bidirectional OffMeshLink.} \\
 & \ubbullet{zShrine}{NavTest\_MultiFloor\_LowerIsland to NavTest\_MultiFloor\_UpperIsland must produce a complete path through this OffMeshLink.} \\
 & \ubbullet{zShrine}{The area covered by NavTest\_DangerArea\_HighCost should use a NavMesh area cost higher than normal Walkable, as a high-cost region.} \\
 & \ubbullet{zShrine}{The area covered by NavTest\_ForbiddenZone\_NotWalkable should be unwalkable and must not be sampled as usable NavMesh.} \\
 & \ubbullet{zShrine}{NavTest\_DynamicObstacle\_CarvingCandidate should be configured as a NavMeshObstacle with carving enabled, for testing dynamic obstacle avoidance.} \\*
\ubfield{zZonai}{Resources} & \textcolor{zZonai!82!black}{AI Navigation · NavMesh · NavMeshAgent · OffMeshLink} \\
\noalign{\vskip 3pt}
\multicolumn{2}{@{}p{14.4cm}@{}}{\textcolor{zRockLite!48}{\rule{\linewidth}{0.35pt}}}\\[4pt]
\multicolumn{2}{@{}p{14.4cm}@{}}{\ubtaskhead{zPlain}{0043-MazeObstacleParenting}{11x11 Unique-Path Maze Construction}{zGoldDeep}{Medium}{10}}\\*[4pt]
\ubfield{zChampion}{Goal} & Construct valid 11x11 unique-path mazes in the ten provided scenes by removing or repositioning wall blocks while preserving each scene's required hierarchy. \\
\ubfield{zDawn}{Context} & Ten Maze scenes provide fully walled 11x11 templates. \\
\ubfield{zShrine}{Constraints} & \ubbullet{zShrine}{Process and save Maze\_01.scene through Maze\_10.scene under Assets/Bench0043/Scenes.} \\
 & \ubbullet{zShrine}{In each scene, preserve MazeRoot, the MazeWalls parent containing the wall Cubes, and the StartMarker entrance at world position (0, 0, 0).} \\
 & \ubbullet{zShrine}{Dig out a valid maze in each scene.} \\
 & \ubbullet{zShrine}{Form paths by deleting unnecessary wall Cubes under MazeWalls; remaining Cubes are treated as walls, and empty cells are treated as paths.} \\
 & \ubbullet{zShrine}{The center cell (0, 0) must be a path, meaning the maze starts from the world origin.} \\
 & \ubbullet{zShrine}{The 11x11 boundary must contain exactly 1 path cell. This single boundary opening is the exit.} \\
 & \ubbullet{zShrine}{All path cells must be mutually reachable; isolated path regions are not allowed.} \\
 & \ubbullet{zShrine}{There must be exactly one path from the center entrance to the single boundary exit.} \\
 & \ubbullet{zShrine}{No 2x2 area may be all walls, and no 2x2 area may be all paths.} \\
 & \ubbullet{zShrine}{Wall Cubes must keep unit size, integer grid positions, zero rotation, and must remain direct children of MazeWalls.} \\
 & \ubbullet{zShrine}{Do not move MazeRoot, MazeWalls, or StartMarker.} \\
 & \ubbullet{zShrine}{Save all 10 completed scenes.} \\*
\ubfield{zZonai}{Resources} & \textcolor{zZonai!82!black}{Scene · Hierarchy · Transform · Grid} \\
\noalign{\vskip 3pt}
\multicolumn{2}{@{}p{14.4cm}@{}}{\textcolor{zRockLite!48}{\rule{\linewidth}{0.35pt}}}\\[4pt]
\multicolumn{2}{@{}p{14.4cm}@{}}{\ubtaskhead{zPlain}{0069-TrenchAssembly}{Trench Asset Assembly}{zGoldDeep}{Medium}{16}}\\*[4pt]
\ubfield{zChampion}{Goal} & Complete the trench scene assembly in Assets/0069/0069.scene. \\
\ubfield{zDawn}{Context} & Assets/0069/0069.scene is an empty URP scene containing trench prefabs and three empty segment parents. \\
\ubfield{zShrine}{Constraints} & \ubbullet{zShrine}{In Assets/0069/0069.scene, use every trench prefab type under Assets/0069/Arts at least once.} \\
 & \ubbullet{zShrine}{Place front, middle, and rear trench rows under their corresponding TrenchSegment parent objects.} \\
 & \ubbullet{zShrine}{Keep each row connected front-to-back and make the three rows mutually reachable.} \\
 & \ubbullet{zShrine}{Keep every trench prefab root at Position Y = 0.} \\
 & \ubbullet{zShrine}{Keep the complete layout within 150 m by 150 m on the XZ plane.} \\
 & \ubbullet{zShrine}{Place trench assets only; manual navigation-map baking is not required.} \\*
\ubfield{zZonai}{Resources} & \textcolor{zZonai!82!black}{Scene · Prefab · Collider · AI Navigation · NavMesh} \\
\noalign{\vskip 3pt}
\multicolumn{2}{@{}p{14.4cm}@{}}{\textcolor{zRockLite!48}{\rule{\linewidth}{0.35pt}}}\\[4pt]
\multicolumn{2}{@{}p{14.4cm}@{}}{\ubtaskhead{zPlain}{0066-Shop3DLetters}{Shop 3D Letter Signs}{zShrine}{Hard}{44}}\\*[4pt]
\ubfield{zChampion}{Goal} & Build readable 3D PIZZA PLACE and BOOK STORE signs on the corresponding buildings using the provided uppercase letter prefabs. \\
\ubfield{zDawn}{Context} & The 0066 scene contains pizza-shop and bookstore buildings plus the LetterAnswer hierarchy. \\
\ubfield{zShrine}{Constraints} & \ubbullet{zShrine}{In the 0066 scene, use the uppercase letter prefabs under Assets/LetterPrefabs to spell PIZZA PLACE and BOOK STORE.} \\
 & \ubbullet{zShrine}{Place the signs over the corresponding printed sign locations on the pizza-shop and bookstore buildings.} \\
 & \ubbullet{zShrine}{Parent the letters under LetterAnswer/PizzaPlace and LetterAnswer/BookStore.} \\
 & \ubbullet{zShrine}{Name letters in stable left-to-right order, such as 00\_P, 01\_I, and 02\_Z.} \\
 & \ubbullet{zShrine}{Keep every letter as a single prefab mesh, facing outward and scaled and positioned for clear readability.} \\*
\ubfield{zZonai}{Resources} & \textcolor{zZonai!82!black}{Prefab · Transform · MeshRenderer · Material} \\
\noalign{\vskip 3pt}
\multicolumn{2}{@{}p{14.4cm}@{}}{\textcolor{zRockLite!48}{\rule{\linewidth}{0.35pt}}}\\[4pt]
\multicolumn{2}{@{}p{14.4cm}@{}}{\ubtaskhead{zPlain}{0067-TerrainRiverWaterAssembly}{Terrain River and Lake Water Surface Assembly}{zShrine}{Hard}{14}}\\*[4pt]
\ubfield{zChampion}{Goal} & In the 0067 scene, complete a continuous water system flowing from the higher lake to the lower lake, based on the already sculpted Terrain riverbed and lake areas. \\
\ubfield{zDawn}{Context} & The 0067 scene provides a sculpted Terrain riverbed, lake areas, and two water materials. \\
\ubfield{zShrine}{Constraints} & \ubbullet{zShrine}{In the 0067 scene, complete Lake\_Start, Lake\_End, and continuously numbered Stream\_0, Stream\_1, Stream\_2, and subsequent river segments from the higher lake to the lower lake.} \\
 & \ubbullet{zShrine}{Choose the suitable lake and river materials from the two provided water materials and assign them to the corresponding water-surface planes.} \\
 & \ubbullet{zShrine}{All key lake and river channel areas should be covered by the corresponding water surfaces. Water surfaces must not be visibly cut off by the riverbed or lakebed.} \\
 & \ubbullet{zShrine}{Lake water surfaces should cover the lake areas, with edge vertices embedded into the Terrain to avoid visual gaps.} \\
 & \ubbullet{zShrine}{River water surfaces should be laid continuously along the riverbed. Seams between adjacent segments should be close and at matching heights.} \\
 & \ubbullet{zShrine}{The overall river should monotonically descend from the higher lake to the lower lake.} \\
 & \ubbullet{zShrine}{Each river plane may slope along only one horizontal axis.} \\
 & \ubbullet{zShrine}{All lake and river objects should use the Unity built-in Plane as their Mesh.} \\*
\ubfield{zZonai}{Resources} & \textcolor{zZonai!82!black}{Terrain · MeshRenderer · Material · Transform} \\
\noalign{\vskip 3pt}
\multicolumn{2}{@{}p{14.4cm}@{}}{\textcolor{zRockLite!48}{\rule{\linewidth}{0.35pt}}}\\[4pt]
\multicolumn{2}{@{}p{14.4cm}@{}}{\ubtaskhead{zPlain}{0068-ShanghaiCircuitTerrainPaint}{Shanghai Circuit Terrain Painting, White Lines, and Turn Props}{zShrine}{Hard}{28}}\\*[4pt]
\ubfield{zChampion}{Goal} & Open Assets/0068/0068.scene. Refer to the provided Shanghai Circuit track map and clean track outline/mask image, recreate the race circuit ground on the Terrain, and complete the track white lines and turn prop placement. \\
\ubfield{zDawn}{Context} & Assets/0068/0068.scene provides a Terrain, Shanghai Circuit reference images, and track-prop prefabs. \\
\ubfield{zShrine}{Constraints} & \ubbullet{zShrine}{In Assets/0068/0068.scene, paint the Shanghai Circuit road layout onto the fixed existing Terrain using the supplied reference map, mask, grass, and road resources.} \\
 & \ubbullet{zShrine}{Match the reference layout's proportion, orientation, and relative position without arbitrary rotation, mirroring, translation, or rescaling; keep the full circuit inside the Terrain with surrounding grass and preserve the required connection roads.} \\
 & \ubbullet{zShrine}{Create continuous visible white road-edge lines with scene geometry or LineRenderer rather than only painting a texture, preserving natural openings at forks and avoiding lines that cut diagonally across road surfaces.} \\
 & \ubbullet{zShrine}{Use prefabs from Assets/0068/Arts to place curbs at major turns, rails outside turns, and plastic barriers across blocked connections.} \\
 & \ubbullet{zShrine}{Align props to local track tangents and keep them off the main driving line with natural scale, height, and spacing.} \\
 & \ubbullet{zShrine}{Group each connected continuous prop run under its own Curb\_XX, Rail\_XX, or Plastic\_XX parent; every rail or guardrail instance must be under a Rail\_XX parent, and disconnected runs must not share one parent.} \\
 & \ubbullet{zShrine}{Save the completed 0068.scene.} \\*
\ubfield{zZonai}{Resources} & \textcolor{zZonai!82!black}{Terrain · Terrain Layer · LineRenderer · Prefab · Material · Transform} \\
\noalign{\vskip 3pt}
\multicolumn{2}{@{}p{14.4cm}@{}}{\textcolor{zRockLite!48}{\rule{\linewidth}{0.35pt}}}\\[4pt]
\multicolumn{2}{@{}p{14.4cm}@{}}{\ubtaskhead{zPlain}{0070-CameraPlacement}{Place Indoor and Outdoor Observation Cameras}{zShrine}{Hard}{11}}\\*[4pt]
\ubfield{zChampion}{Goal} & Position and orient five indoor and four outdoor cameras so every camera clearly observes the props assigned to its room or area without changing any other scene content. \\
\ubfield{zDawn}{Context} & Assets/0070Bench/0070Bench.scene contains a furnished house with five coincident room cameras, and Assets/0070Bench-B/0070Bench-B.unity contains four coincident outdoor observation cameras. \\
\ubfield{zShrine}{Constraints} & \ubbullet{zShrine}{Place CameraLiving, CameraBedRoom, CameraGarage, CameraBathRoom, and CameraCorr in their corresponding indoor spaces.} \\
 & \ubbullet{zShrine}{Place CameraAreaA, CameraAreaB, CameraStones, and CameraUmbrella to observe every prop in their corresponding outdoor groups.} \\
 & \ubbullet{zShrine}{Each camera must have an unobstructed view of the props assigned to it.} \\
 & \ubbullet{zShrine}{Only the position and rotation of the nine existing cameras may be changed.} \\
 & \ubbullet{zShrine}{Keep every camera active, enabled, perspective, and full-screen.} \\
 & \ubbullet{zShrine}{Keep the indoor cameras at 60 degrees vertical field of view.} \\
 & \ubbullet{zShrine}{Keep the outdoor cameras at 50 degrees vertical field of view.} \\
 & \ubbullet{zShrine}{Apart from the permitted camera position and rotation changes, do not add, delete, rename, reparent, scale, disable, or otherwise modify any camera, prop, renderer, collider, wall, floor, ceiling, or other scene object.} \\*
\ubfield{zZonai}{Resources} & \textcolor{zZonai!82!black}{Scene · Camera · Transform · Renderer · Collider} \\
\noalign{\vskip 3pt}
\multicolumn{2}{@{}p{14.4cm}@{}}{\textcolor{zRockLite!48}{\rule{\linewidth}{0.35pt}}}\\[4pt]
\multicolumn{2}{@{}p{14.4cm}@{}}{\ubtaskhead{zPlain}{0071-PlantsDecoration}{Decorate Terrain with Natural Trees and Riverside Rocks}{zShrine}{Hard}{24}}\\*[4pt]
\ubfield{zChampion}{Goal} & Plant exact pine and fruit-tree populations through the Terrain system and arrange exactly fifty grounded, varied riverside rocks while preserving the supplied landscape. \\
\ubfield{zDawn}{Context} & Assets/0071PlantsDecoration.unity contains an existing Terrain, water system, vegetation prototypes, rock assets, and a Riverside Rocks parent awaiting final natural placement. \\
\ubfield{zShrine}{Constraints} & \ubbullet{zShrine}{Place exactly 200 Terrain tree instances using the supplied prototypes: 150 pine trees and 50 fruit trees.} \\
 & \ubbullet{zShrine}{Keep trees on appropriate grassy ground and away from visible water, roads, extreme breaks, and implausible crowding.} \\
 & \ubbullet{zShrine}{Place most pines on hilly grassland and fruit trees in smaller groves on calmer low grassland.} \\
 & \ubbullet{zShrine}{Form natural clusters with softened edges, open meadow, and restrained variation in type, rotation, and scale.} \\
 & \ubbullet{zShrine}{Place exactly 50 rock instances as direct children of Riverside Rocks.} \\
 & \ubbullet{zShrine}{Name the rocks sequentially from Riverside\_Rock\_001 through Riverside\_Rock\_050.} \\
 & \ubbullet{zShrine}{Use an irregular mix of supplied rock variants without one variant dominating.} \\
 & \ubbullet{zShrine}{Distribute rocks across the lake and river banks rather than concentrating them in one area.} \\
 & \ubbullet{zShrine}{Keep every rock visually grounded, fully on dry land, and clear of trees, roads, and other rocks.} \\
 & \ubbullet{zShrine}{Preserve the supplied Terrain, water layout, scene structure, and original scene path.} \\*
\ubfield{zZonai}{Resources} & \textcolor{zZonai!82!black}{Scene · Terrain · Terrain Layer · Prefab · Transform · Collider} \\
\noalign{\vskip 3pt}
\multicolumn{2}{@{}p{14.4cm}@{}}{\textcolor{zRockLite!48}{\rule{\linewidth}{0.35pt}}}\\[4pt]
\multicolumn{2}{@{}p{14.4cm}@{}}{\ubtaskhead{zPlain}{0073-CastleConstruction}{Connect Castle Circulation with Supplied Stair Prefabs}{zShrine}{Hard}{9}}\\*[4pt]
\ubfield{zChampion}{Goal} & Place only the three approved stair-prefab families under a Stairs root so the lower interior, wall walks, gate, and bastions are mutually reachable. \\
\ubfield{zDawn}{Context} & Assets/0073.scene contains the existing castle and round stair platforms; the supplied straight, round, and corner wooden stair prefabs are available for placement. \\
\ubfield{zShrine}{Constraints} & \ubbullet{zShrine}{Use only SM\_Bld\_Stairs\_Wood\_Straight\_02, SM\_Bld\_Stairs\_Round\_L\_Wood\_01, and SM\_Bld\_Stairs\_Wood\_Corner\_01.} \\
 & \ubbullet{zShrine}{Create one empty scene-root parent named Stairs.} \\
 & \ubbullet{zShrine}{Make every newly placed stair instance a direct child of Stairs.} \\
 & \ubbullet{zShrine}{Connect the lower castle interior, elevated wall walks, gate structure, and bastions into mutually reachable circulation.} \\
 & \ubbullet{zShrine}{Each stair must connect its intended walking surfaces and be correctly oriented and grounded.} \\
 & \ubbullet{zShrine}{Stairs must not float or create unusable intersections.} \\
 & \ubbullet{zShrine}{Only approved stair instances may be instantiated, positioned, rotated, or mirrored.} \\
 & \ubbullet{zShrine}{Do not modify, duplicate, delete, reparent, replace, unpack, or alter existing castle models or round stair platforms.} \\
 & \ubbullet{zShrine}{Avoid redundant stair instances and save Assets/0073.scene.} \\*
\ubfield{zZonai}{Resources} & \textcolor{zZonai!82!black}{Scene · Prefab · Transform · Hierarchy · Collider} \\
\noalign{\vskip 3pt}
\multicolumn{2}{@{}p{14.4cm}@{}}{\textcolor{zRockLite!48}{\rule{\linewidth}{0.35pt}}}\\[4pt]
\multicolumn{2}{@{}p{14.4cm}@{}}{\ubtaskhead{zPlain}{0075-CrystalFragment}{Restore the Crystal Fragment Assembly}{zShrine}{Hard}{14}}\\*[4pt]
\ubfield{zChampion}{Goal} & Translate and rotate fourteen scattered prefab-linked fragments into one continuous crystal around the two fixed anchors without changing assets, scale, materials, or components. \\
\ubfield{zDawn}{Context} & Assets/Scenes/CrystalFragment.scene contains fixed anchors Object002 and Object005 plus fourteen scattered fragment Prefab instances linked to Assets/Prefabs. \\
\ubfield{zShrine}{Constraints} & \ubbullet{zShrine}{Keep Object002 fixed as the central non-Prefab anchor.} \\
 & \ubbullet{zShrine}{Keep Object005 fixed as the orientation anchor and connected to its Prefab.} \\
 & \ubbullet{zShrine}{Reassemble Object001, Object003, Object004, and Object006 through Object016.} \\
 & \ubbullet{zShrine}{Use every provided movable fragment exactly once.} \\
 & \ubbullet{zShrine}{Only translate and rotate the fourteen movable fragments.} \\
 & \ubbullet{zShrine}{Keep each movable fragment connected to its original Prefab in Assets/Prefabs.} \\
 & \ubbullet{zShrine}{Preserve every fragment's original Mesh, material, scale, name, and enabled MeshRenderer.} \\
 & \ubbullet{zShrine}{Form one continuous crystal around the two anchors.} \\
 & \ubbullet{zShrine}{Serialize the completed transforms in Assets/Scenes/CrystalFragment.scene.} \\
 & \ubbullet{zShrine}{Do not use runtime scripts, animation, physics, constraints, or additional components to assemble or hold the crystal.} \\*
\ubfield{zZonai}{Resources} & \textcolor{zZonai!82!black}{Scene · Prefab · Transform · Mesh · MeshRenderer · Material} \\
\noalign{\vskip 3pt}
\multicolumn{2}{@{}p{14.4cm}@{}}{\textcolor{zRockLite!48}{\rule{\linewidth}{0.35pt}}}\\[4pt]
\multicolumn{2}{@{}p{14.4cm}@{}}{\ubtaskhead{zPlain}{0076-BuildHouseFromDesign}{Build a White-Box House from a Floor Plan}{zShrine}{Hard}{23}}\\*[4pt]
\ubfield{zChampion}{Goal} & Reproduce Assets/Design.png as a dimensioned white-box House scene with correctly grouped walls, doors, windows, furniture, orientation, and collision-free layout. \\
\ubfield{zDawn}{Context} & Assets/Design.png provides a top-down floor plan with room layout, labeled furniture, wall dimensions, and placement references; the House scene must be created from basic untextured 3D GameObjects. \\
\ubfield{zShrine}{Constraints} & \ubbullet{zShrine}{Create and save the result as the Scenes/House scene, using basic untextured white-box 3D GameObjects to reproduce Assets/Design.png.} \\
 & \ubbullet{zShrine}{Use the southwest house corner as world origin and follow the specified X east-west, Z north-south, and Y up convention.} \\
 & \ubbullet{zShrine}{Make all walls 3 metres tall and align adjacent walls tightly without visible gaps.} \\
 & \ubbullet{zShrine}{Place exterior walls under ExtWalls and tag every exterior wall ExtWall.} \\
 & \ubbullet{zShrine}{Place interior walls under InsideWalls and tag every interior wall InsideWall.} \\
 & \ubbullet{zShrine}{Create a matching GameObject for every object explicitly labeled in the design.} \\
 & \ubbullet{zShrine}{Make windows 1.5 metres tall with sills 1 metre above the floor.} \\
 & \ubbullet{zShrine}{Make doors 2.2 metres tall and use a transparent material.} \\
 & \ubbullet{zShrine}{Match the drawing's object positions, dimensions, and orientations.} \\
 & \ubbullet{zShrine}{Keep furniture and walls from interpenetrating and keep furniture pieces from overlapping one another.} \\*
\ubfield{zZonai}{Resources} & \textcolor{zZonai!82!black}{Scene · GameObject · Primitive · Transform · Hierarchy · Material · Collider} \\
\noalign{\vskip 3pt}
\multicolumn{2}{@{}p{14.4cm}@{}}{\textcolor{zRockLite!48}{\rule{\linewidth}{0.35pt}}}\\[4pt]
\multicolumn{2}{@{}p{14.4cm}@{}}{\ubtaskhead{zPlain}{0080-UrbanRoads}{Add Correct Lane-Direction Arrows to Urban Intersections}{zShrine}{Hard}{31}}\\*[4pt]
\ubfield{zChampion}{Goal} & Place and rotate supplied road-marking arrow prefabs on every tested intersection approach according to lane count, reachable directions, travel direction, and stop-line placement. \\
\ubfield{zDawn}{Context} & Assets/Scenes/Roads.scene contains the urban road network, lane centerlines, stop lines, and existing RoadDirection arrows; direction-marking prefabs are supplied under Assets/Prefabs/Markers. \\
\ubfield{zShrine}{Constraints} & \ubbullet{zShrine}{Do not modify existing GameObjects in Assets/Scenes/Roads.scene.} \\
 & \ubbullet{zShrine}{Use only the appropriate Marking\_Arrow prefabs from Assets/Prefabs/Markers.} \\
 & \ubbullet{zShrine}{For dual-lane cross intersections, mark the left lane for left turns and the right lane for straight plus right turns.} \\
 & \ubbullet{zShrine}{For dual-lane T-intersections with only left and right reachable, use a left-turn arrow in the left lane and a right-turn arrow in the right lane.} \\
 & \ubbullet{zShrine}{For dual-lane T-intersections with only left and straight reachable, use a left-turn arrow in the left lane and a straight arrow in the right lane.} \\
 & \ubbullet{zShrine}{For dual-lane T-intersections with only right and straight reachable, use a straight arrow in the left lane and a straight-plus-right-turn arrow in the right lane.} \\
 & \ubbullet{zShrine}{For single-lane cross intersections, use a left plus straight plus right marking.} \\
 & \ubbullet{zShrine}{For single-lane T-intersections, represent each reachable left, straight, or right direction.} \\
 & \ubbullet{zShrine}{Place arrows on the approach side of the intersection behind the stop line.} \\
 & \ubbullet{zShrine}{Center arrows in their lanes and orient them to match the existing RoadDirection travel direction.} \\
 & \ubbullet{zShrine}{Use mirroring only where required to distinguish left from right variants.} \\
 & \ubbullet{zShrine}{Keep each arrow flush with the road surface and inside its intended approach area.} \\
 & \ubbullet{zShrine}{Save the completed result in Assets/Scenes/Roads.scene.} \\*
\ubfield{zZonai}{Resources} & \textcolor{zZonai!82!black}{Scene · Prefab · Transform · Renderer} \\
\noalign{\vskip 3pt}
\multicolumn{2}{@{}p{14.4cm}@{}}{\textcolor{zRockLite!48}{\rule{\linewidth}{0.35pt}}}\\[4pt]
\noalign{\vskip 3pt}
\multicolumn{2}{@{}p{14.4cm}@{}}{\ubcategory{zShrine}{Particles \& Shaders}{12 tasks \textperiodcentered\ 155 milestone tests}}\\*[5pt]
\multicolumn{2}{@{}p{14.4cm}@{}}{\ubtaskhead{zShrine}{0007-SkyboxBlackHole}{Shadertoy Black Hole Skybox}{zZonai}{Easy}{26}}\\*[4pt]
\ubfield{zChampion}{Goal} & According to the rules in Assets/Shaders/ShadertoyBlackHoleSkybox.md, convert the Shadertoy fragment shader below into a Unity skybox shader. \\
\ubfield{zDawn}{Context} & Assets/Scenes/SampleScene.scene already uses the unfinished black-hole skybox shader and material. \\
\ubfield{zShrine}{Constraints} & \ubbullet{zShrine}{Modify Assets/Shaders/ShadertoyBlackHoleSkybox.shader.} \\
 & \ubbullet{zShrine}{Keep the shader under the Skybox/ menu.} \\
 & \ubbullet{zShrine}{Use the property list, skybox coordinate mapping, time expression, and GLSL-to-HLSL conversion rules specified in Assets/Shaders/ShadertoyBlackHoleSkybox.md.} \\
 & \ubbullet{zShrine}{Assets/Shaders/Skybox\_Shadertoy Black Hole Skybox.mat is already used as the scene skybox material. After completion, make the skybox in Assets/Scenes/SampleScene.scene display the black hole effect.} \\
 & \ubbullet{zShrine}{Do not add new texture dependencies, and do not expose Shadertoy compatibility uniforms such as iResolution or iTime as material properties.} \\*
\ubfield{zZonai}{Resources} & \textcolor{zZonai!82!black}{Shader · Material · Skybox · Camera · URP} \\
\noalign{\vskip 3pt}
\multicolumn{2}{@{}p{14.4cm}@{}}{\textcolor{zRockLite!48}{\rule{\linewidth}{0.35pt}}}\\[4pt]
\multicolumn{2}{@{}p{14.4cm}@{}}{\ubtaskhead{zShrine}{0030-Fireworks-TrailingSmoke}{Firework Trailing Smoke}{zZonai}{Easy}{13}}\\*[4pt]
\ubfield{zChampion}{Goal} & Modify the scene and complete the trailing smoke: during the firework's ascent, smoke should be emitted in the opposite direction of travel. Use dense solid-color gray and white particles that are unaffected by gravity, gradually shrink, and dissipate. Simulate air disturbance and distortion caused by uneven rocket thrust. \\
\ubfield{zDawn}{Context} & The firework scene is missing its ascending smoke trail. \\
\ubfield{zShrine}{Constraints} & \ubbullet{zShrine}{Emit dense gray and white smoke opposite the firework's ascent direction.} \\
 & \ubbullet{zShrine}{Keep the smoke unaffected by gravity and make particles shrink and dissipate over their lifetime.} \\
 & \ubbullet{zShrine}{Add uneven turbulence that suggests disturbed air and inconsistent rocket thrust.} \\
 & \ubbullet{zShrine}{Modify scene objects only; do not add or modify code.} \\*
\ubfield{zZonai}{Resources} & \textcolor{zZonai!82!black}{Particle System · ParticleSystemRenderer · Material} \\
\noalign{\vskip 3pt}
\multicolumn{2}{@{}p{14.4cm}@{}}{\textcolor{zRockLite!48}{\rule{\linewidth}{0.35pt}}}\\[4pt]
\multicolumn{2}{@{}p{14.4cm}@{}}{\ubtaskhead{zShrine}{0031-Fireworks-TrailingSparks}{Firework Trailing Sparks}{zZonai}{Easy}{15}}\\*[4pt]
\ubfield{zChampion}{Goal} & Modify the scene and complete the trailing sparks: after sparks are emitted opposite the firework launch direction, they should fall toward the ground. Use bright yellow highlight particles close to the launch point, with fast dissipation and backward initial velocity. The sparks should have trailing streaks caused by visual persistence. \\
\ubfield{zDawn}{Context} & The firework scene is missing its launch-trail sparks. \\
\ubfield{zShrine}{Constraints} & \ubbullet{zShrine}{Emit sparks with initial velocity opposite the firework's launch direction, then let them fall.} \\
 & \ubbullet{zShrine}{Use bright yellow particles near the launch point with rapid dissipation.} \\
 & \ubbullet{zShrine}{Give the sparks visible trailing streaks.} \\
 & \ubbullet{zShrine}{Modify scene objects only; do not add or modify code.} \\*
\ubfield{zZonai}{Resources} & \textcolor{zZonai!82!black}{Particle System · ParticleSystemRenderer · Material} \\
\noalign{\vskip 3pt}
\multicolumn{2}{@{}p{14.4cm}@{}}{\textcolor{zRockLite!48}{\rule{\linewidth}{0.35pt}}}\\[4pt]
\multicolumn{2}{@{}p{14.4cm}@{}}{\ubtaskhead{zShrine}{0032-Fireworks-HeadHalo}{Firework Head Halo}{zZonai}{Easy}{13}}\\*[4pt]
\ubfield{zChampion}{Goal} & Modify the scene and complete the head halo: attach a bright yellow highlighted core to the rocket head to simulate a rising bright point in the night sky. Continuously and rapidly emit a small number of large particles to simulate flickering light. \\
\ubfield{zDawn}{Context} & The firework scene is missing the glowing halo attached to the rocket head. \\
\ubfield{zShrine}{Constraints} & \ubbullet{zShrine}{Attach the halo effect to the rocket head.} \\
 & \ubbullet{zShrine}{Create a bright yellow core from a small number of continuously and rapidly emitted large particles.} \\
 & \ubbullet{zShrine}{Make the core visibly flicker during ascent.} \\
 & \ubbullet{zShrine}{Modify scene objects only; do not add or modify code.} \\*
\ubfield{zZonai}{Resources} & \textcolor{zZonai!82!black}{Particle System · ParticleSystemRenderer · Material} \\
\noalign{\vskip 3pt}
\multicolumn{2}{@{}p{14.4cm}@{}}{\textcolor{zRockLite!48}{\rule{\linewidth}{0.35pt}}}\\[4pt]
\multicolumn{2}{@{}p{14.4cm}@{}}{\ubtaskhead{zShrine}{0033-Fireworks-ExplosionEffect}{Firework Explosion Effect}{zZonai}{Easy}{19}}\\*[4pt]
\ubfield{zChampion}{Goal} & Use particle systems to modify scene objects and complete the explosion effect. When the firework explodes, use a sphere to simulate the firework shape. At the instant of explosion, particle speed should be extremely high, but because of air resistance, particles should quickly decelerate in a short time, forming a visual effect of "bursting outward, instantly freezing, then beginning to fall". At the explosion center, the energy is extremely high, so it is usually pure white, like an overexposed flash. It then quickly transitions to vivid pure yellow, and finally becomes dark red before extinguishing. The residual sparks after the explosion should be affected by gravity and gently sway left and right while falling, creating a natural parabolic downward motion. \\
\ubfield{zDawn}{Context} & The firework scene contains BurstFlash and BurstExplosion objects that require completion. \\
\ubfield{zShrine}{Constraints} & \ubbullet{zShrine}{Create a spherical burst with very high initial speed followed by rapid deceleration.} \\
 & \ubbullet{zShrine}{Transition the burst color from an overexposed white center to vivid yellow and then dark red.} \\
 & \ubbullet{zShrine}{Make residual sparks fall under gravity with gentle lateral motion.} \\
 & \ubbullet{zShrine}{Modify only BurstFlash and BurstExplosion; do not rename scene objects or add or modify code.} \\*
\ubfield{zZonai}{Resources} & \textcolor{zZonai!82!black}{Particle System · ParticleSystemRenderer · Material} \\
\noalign{\vskip 3pt}
\multicolumn{2}{@{}p{14.4cm}@{}}{\textcolor{zRockLite!48}{\rule{\linewidth}{0.35pt}}}\\[4pt]
\multicolumn{2}{@{}p{14.4cm}@{}}{\ubtaskhead{zShrine}{0024-CoastDunesPineUrpMaterials}{Pine Material URP Conversion}{zGoldDeep}{Medium}{5}}\\*[4pt]
\ubfield{zChampion}{Goal} & Convert the specified pine-tree materials for URP while preserving their source textures, colors, transparency behavior, and intended appearance. \\
\ubfield{zDawn}{Context} & The URP project provides a directory of pine-tree materials that have not yet been converted for URP. \\
\ubfield{zShrine}{Constraints} & \ubbullet{zShrine}{Create Assets/External/Coast and Dunes Environment/Coast Environment/Pine Trees/URP Converted and add one corresponding converted material for every original material in the Pine Trees/Materials folder.} \\
 & \ubbullet{zShrine}{Reproduce each original pine material's appearance as closely as possible.} \\
 & \ubbullet{zShrine}{All new materials must use Universal Render Pipeline/Lit.} \\
 & \ubbullet{zShrine}{Name each new material as the original material name plus the \_URP suffix.} \\
 & \ubbullet{zShrine}{Leaf and Cross materials must preserve alpha cutout behavior, avoiding transparent regions rendering as solid blocks.} \\
 & \ubbullet{zShrine}{Do not delete or overwrite the original materials.} \\*
\ubfield{zZonai}{Resources} & \textcolor{zZonai!82!black}{URP · Shader · Material · Texture} \\
\noalign{\vskip 3pt}
\multicolumn{2}{@{}p{14.4cm}@{}}{\textcolor{zRockLite!48}{\rule{\linewidth}{0.35pt}}}\\[4pt]
\multicolumn{2}{@{}p{14.4cm}@{}}{\ubtaskhead{zShrine}{0037-ClassificationOfParticleSystem}{Particle Effect Classification}{zGoldDeep}{Medium}{10}}\\*[4pt]
\ubfield{zChampion}{Goal} & Analyze each Effect prefab in the scene one by one. Based on its particle effect, place it under the most appropriate effect-name category object under Answer as a child object. \\
\ubfield{zDawn}{Context} & SampleScene contains Effect prefabs and category nodes under Answer. \\
\ubfield{zShrine}{Constraints} & \ubbullet{zShrine}{Inspect every Effect prefab and classify it by its visible particle behavior.} \\
 & \ubbullet{zShrine}{Place each Effect under exactly one best-fit category GameObject beneath Answer.} \\
 & \ubbullet{zShrine}{Do not rename the category GameObjects; a category may contain zero or multiple Effect prefabs.} \\*
\ubfield{zZonai}{Resources} & \textcolor{zZonai!82!black}{Particle System · Prefab · Hierarchy} \\
\noalign{\vskip 3pt}
\multicolumn{2}{@{}p{14.4cm}@{}}{\textcolor{zRockLite!48}{\rule{\linewidth}{0.35pt}}}\\[4pt]
\multicolumn{2}{@{}p{14.4cm}@{}}{\ubtaskhead{zShrine}{0047-DynamicDissolve}{Dynamic Dissolve}{zGoldDeep}{Medium}{10}}\\*[4pt]
\ubfield{zChampion}{Goal} & Create a URP Shader Graph that combines parameter-controlled dissolve, HDR edge emission, multilayer animated noise, and subtle vertex deformation. \\
\ubfield{zDawn}{Context} & The task scene is stored at Assets/0052Bench/0052Bench.scene and provides a display model, M\_Dissolve.mat, and a Shader Graph directory. \\
\ubfield{zShrine}{Constraints} & \ubbullet{zShrine}{Create Assets/0052Bench/Shaders/DynamicDissolve.shadergraph as a URP Shader Graph and assign it to Assets/0052Bench/Shaders/M\_Dissolve.mat, which is already used by the test model in Assets/0052Bench/0052Bench.scene.} \\
 & \ubbullet{zShrine}{Expose all of the following Blackboard properties.} \\
 & \ubbullet{zShrine}{DissolveAmount (Float): dissolve progress from 0 to 1. 0 means fully visible, 1 means fully dissolved.} \\
 & \ubbullet{zShrine}{DissolveScale (Float): noise scale used to control the fineness of the dissolve edge.} \\
 & \ubbullet{zShrine}{EdgeWidth (Float): width of the glowing edge.} \\
 & \ubbullet{zShrine}{EdgeColor (Color, HDR): glowing edge color, with support for high-quality HDR intensity.} \\
 & \ubbullet{zShrine}{Implement all of the following core node-network behaviors.} \\
 & \ubbullet{zShrine}{Dynamic multi-layer noise blending: use at least two different noise layers, such as Simple Noise or Voronoi, and blend them through Multiply or Lerp using different Tiling and Offset values, with dynamic displacement introduced through the Time node. Use this as the Alpha clipping basis for dissolve.} \\
 & \ubbullet{zShrine}{Alpha clipping (Dissolve): compare the blended noise with DissolveAmount and connect it to the master node's Alpha Clip Threshold, or discard fragments through logic in opaque mode, to make the object gradually become hollow and disappear.} \\
 & \ubbullet{zShrine}{Edge glow (Emission): calculate a transition band around the area about to be clipped according to EdgeWidth, multiply it by EdgeColor (HDR), and output it to the master node's Emission port.} \\
 & \ubbullet{zShrine}{Vertex animation (Vertex Deformation): use the Time node and noise to introduce slight wave or expansion displacement during the object's Vertex Position stage, either only in regions where DissolveAmount \textgreater{} 0 or as global subtle motion, so the dissolve edge naturally distorts while disappearing.} \\*
\ubfield{zZonai}{Resources} & \textcolor{zZonai!82!black}{Shader Graph · Material · URP · Renderer} \\
\noalign{\vskip 3pt}
\multicolumn{2}{@{}p{14.4cm}@{}}{\textcolor{zRockLite!48}{\rule{\linewidth}{0.35pt}}}\\[4pt]
\multicolumn{2}{@{}p{14.4cm}@{}}{\ubtaskhead{zShrine}{0059-RainEmitter}{Particle Collision Splash Task}{zGoldDeep}{Medium}{10}}\\*[4pt]
\ubfield{zChampion}{Goal} & Complete the particle effect of raindrops colliding with the water surface and creating splashes in Assets/0059/Scene/0059.scene. \\
\ubfield{zDawn}{Context} & Assets/0059/Scene/0059.scene provides a water surface, a raindrop emitter, a splash reference, and materials. \\
\ubfield{zShrine}{Constraints} & \ubbullet{zShrine}{Complete the effect in Assets/0059/Scene/0059.scene using the existing WaterSurface\_0059, RainEmitter\_0059, GroundRing\_0059, SplashSpawnReference\_0059, and EvaluationBounds\_0059 GameObjects.} \\
 & \ubbullet{zShrine}{Use Assets/0059/Materials/M\_RainDrop\_0059.mat for raindrops and Assets/0059/Materials/M\_Splash\_0059.mat for splashes.} \\
 & \ubbullet{zShrine}{Complete RainEmitter\_0059 so it continuously emits raindrop particles downward.} \\
 & \ubbullet{zShrine}{Raindrops must collide with WaterSurface\_0059 through the Particle System Collision module.} \\
 & \ubbullet{zShrine}{Create SplashEmitter\_0059 to represent short splashes after raindrops hit the water surface.} \\
 & \ubbullet{zShrine}{Splashes must be triggered by raindrop collisions, for example using a Collision-type SubEmitter or equivalent OnParticleCollision trigger logic.} \\
 & \ubbullet{zShrine}{Splashes should appear near the water surface and be distributed around actual raindrop impact points, not loop at one fixed point.} \\
 & \ubbullet{zShrine}{Splashes should have a short lifetime, spread upward or radially, and gradually disappear.} \\
 & \ubbullet{zShrine}{Control particle counts; avoid using extremely high maxParticles or many duplicate particle systems to cover the evaluation.} \\
 & \ubbullet{zShrine}{Do not move, delete, or rename WaterSurface\_0059, GroundRing\_0059, RainEmitter\_0059, SplashSpawnReference\_0059, or EvaluationBounds\_0059.} \\
 & \ubbullet{zShrine}{Do not use scripts to loop splashes directly at fixed positions or make SplashEmitter\_0059 an independent, endlessly looping system; splashes must depend on actual raindrop collision triggers.} \\*
\ubfield{zZonai}{Resources} & \textcolor{zZonai!82!black}{Particle System · Collision · Sub Emitters · Material} \\
\noalign{\vskip 3pt}
\multicolumn{2}{@{}p{14.4cm}@{}}{\textcolor{zRockLite!48}{\rule{\linewidth}{0.35pt}}}\\[4pt]
\multicolumn{2}{@{}p{14.4cm}@{}}{\ubtaskhead{zShrine}{0012-Smrvfx}{Restore Skinned-Mesh VFX Graph Particles and Trails}{zShrine}{Hard}{13}}\\*[4pt]
\ubfield{zChampion}{Goal} & Rebuild three VFX Graph assets so particles and colored trails sample the animated karate character's deformed full-body surface and remain controlled by Timeline. \\
\ubfield{zDawn}{Context} & Assets/Karate.scene retains the animated character, Timeline, camera, lighting, post-processing, VisualEffect components, binders, and an existing playable, while the three VFX Graph implementations are empty. \\
\ubfield{zShrine}{Constraints} & \ubbullet{zShrine}{Rebuild Assets/VFX/Transformed SMR.vfxoperator, Assets/VFX/Particles.vfx, and Assets/VFX/Lines.vfx.} \\
 & \ubbullet{zShrine}{The shared operator must sample the live deformed SkinnedMeshRenderer surface and transform samples into the correct space.} \\
 & \ubbullet{zShrine}{Particles and Lines must reuse the shared sampling logic and reference the same animated Source renderer.} \\
 & \ubbullet{zShrine}{Both effects must receive SourceTransform and expose a Timeline-driven float property named Throttle.} \\
 & \ubbullet{zShrine}{Particles must cover the full animated body with enough density to reveal the current pose.} \\
 & \ubbullet{zShrine}{New particles must inherit useful surface motion and then move and decay independently.} \\
 & \ubbullet{zShrine}{Lines must form multiple time-continuous particle strips distributed across the full body.} \\
 & \ubbullet{zShrine}{Trails must retain smooth history during fast movement and vary color, width, opacity, or lifetime.} \\
 & \ubbullet{zShrine}{Disabling Throttle must stop new emission while existing particles and trails decay naturally.} \\
 & \ubbullet{zShrine}{The existing Assets/Misc/Karate.playable timing and scene references must remain valid.} \\
 & \ubbullet{zShrine}{Runtime particle bounds must remain finite, reasonable, and close to the animated character.} \\
 & \ubbullet{zShrine}{The completed graphs must compile without missing references or persistent VFX errors.} \\*
\ubfield{zZonai}{Resources} & \textcolor{zZonai!82!black}{Scene · VFX Graph · Visual Effect · SkinnedMeshRenderer · Transform · Timeline · Playables · Particle System} \\
\noalign{\vskip 3pt}
\multicolumn{2}{@{}p{14.4cm}@{}}{\textcolor{zRockLite!48}{\rule{\linewidth}{0.35pt}}}\\[4pt]
\multicolumn{2}{@{}p{14.4cm}@{}}{\ubtaskhead{zShrine}{0038-DynamicRainEffect}{Create a Dynamic Rain Effect}{zShrine}{Hard}{8}}\\*[4pt]
\ubfield{zChampion}{Goal} & Create a player-following, world-space rain effect whose streak direction and length respond naturally to the player's horizontal movement. \\
\ubfield{zDawn}{Context} & The scene requires a local rain volume positioned above and following the player. \\
\ubfield{zShrine}{Constraints} & \ubbullet{zShrine}{Keep a large local rain emitter above and following the player.} \\
 & \ubbullet{zShrine}{Simulate emitted particles in world space so existing rain does not move with the emitter.} \\
 & \ubbullet{zShrine}{When the player is stationary, emit dense, thin, elongated, semi-transparent rain streaks straight downward.} \\
 & \ubbullet{zShrine}{As horizontal player speed increases, make the apparent streak tilt and length increase opposite the movement direction.} \\
 & \ubbullet{zShrine}{Scripts may be used to control the Particle System.} \\*
\ubfield{zZonai}{Resources} & \textcolor{zZonai!82!black}{Particle System · ParticleSystemRenderer · MonoBehaviour} \\
\noalign{\vskip 3pt}
\multicolumn{2}{@{}p{14.4cm}@{}}{\textcolor{zRockLite!48}{\rule{\linewidth}{0.35pt}}}\\[4pt]
\multicolumn{2}{@{}p{14.4cm}@{}}{\ubtaskhead{zShrine}{0045-InteractableLeaves}{Create an Interactable Falling Leaves Effect}{zShrine}{Hard}{13}}\\*[4pt]
\ubfield{zChampion}{Goal} & Complete the current VFX particle system and create an interactable falling leaves effect. The leaf.vfx particle system is used to generate falling leaves. Particles are emitted from a circular area, affected by weak gravity, a 3D spherical vector field, and VFX Plane collision. Leaves should fall naturally and settle flush with the visual ground. The vector field near the player should continuously push leaves to create interaction feedback, and leaves that have landed should still be moved by the wind field near the player. \\
\ubfield{zDawn}{Context} & SampleScene provides leaf.vfx, a player, a Plane, a vector-field texture, and a property Binder. \\
\ubfield{zShrine}{Constraints} & \ubbullet{zShrine}{Leaves should fall vertically under weak gravity when generated. After the leaf count reaches the limit, no more leaves should be generated. Initial leaves should spawn with random orientations.} \\
 & \ubbullet{zShrine}{Complete the nodes inside leaf.vfx. Use the spherical 3D vector field texture in Assets, and generate the vector force field 0.2m below the player position.} \\
 & \ubbullet{zShrine}{The Plane acts as the visual ground for falling leaves and the player's movement plane. Inside leaf.vfx, use Collide with Plane as the mathematical collision plane. Do not use Collide with AABox, and do not rely on the scene Plane's MeshCollider. Align the collision plane position with the scene Plane, set the normal upward, set Bounce to 0, and keep low Friction so leaves stop bouncing vertically but can still be pushed horizontally by the player's wind field.} \\
 & \ubbullet{zShrine}{To further eliminate vertical jitter after landing, execute Clamp To Ground And Stop in leaf.vfx Update after Gravity and before VectorFieldForce: when position.y \textless{}= PlaneY + Epsilon and velocity.y \textless{}= 0, clamp position.y to PlaneY and set velocity to float3(0, 0, 0). PlaneY aligns with the scene Plane height, and Epsilon provides a very small landing tolerance. This node only handles landed leaves that are falling or not rising. Once the player's wind field gives a leaf positive Y velocity, later frames should no longer clamp it back to the ground, so landed leaves can still be blown upward by the wind field.} \\
 & \ubbullet{zShrine}{Bind Self\_Transform and Player\_Transform in the VFXPropertyBinder on leaf.vfx, so leaf.vfx can continuously receive the player's position, rotation, and scale.} \\*
\ubfield{zZonai}{Resources} & \textcolor{zZonai!82!black}{VFX Graph · Visual Effect · Vector Field · VFX Plane Collision} \\
\noalign{\vskip 3pt}
\multicolumn{2}{@{}p{14.4cm}@{}}{\textcolor{zRockLite!48}{\rule{\linewidth}{0.35pt}}}\\[4pt]
\noalign{\vskip 3pt}
\multicolumn{2}{@{}p{14.4cm}@{}}{\ubcategory{zZonai}{Physics \& Collision}{11 tasks \textperiodcentered\ 168 milestone tests}}\\*[5pt]
\multicolumn{2}{@{}p{14.4cm}@{}}{\ubtaskhead{zZonai}{0002-CubeMove}{Build a Rigidbody-Driven Cube Movement Scene}{zZonai}{Easy}{20}}\\*[4pt]
\ubfield{zChampion}{Goal} & Complete Assets/Scenes/CubeMove.scene with a visible cube player that moves and jumps through a testable Rigidbody controller. \\
\ubfield{zDawn}{Context} & The project provides a basic Unity scene and requires the completed result to be saved as Assets/Scenes/CubeMove.scene. \\
\ubfield{zShrine}{Constraints} & \ubbullet{zShrine}{The scene must contain Ground, Player, and MainCamera with exact names.} \\
 & \ubbullet{zShrine}{Ground must use a Plane and Player must use a Cube.} \\
 & \ubbullet{zShrine}{Player must have a Rigidbody, a Collider, and a CubeController component.} \\
 & \ubbullet{zShrine}{Player Rigidbody must use gravity, remain non-kinematic, and freeze X and Z rotation.} \\
 & \ubbullet{zShrine}{WASD must move Player on the XZ plane and Space must request a jump.} \\
 & \ubbullet{zShrine}{Player may jump only while grounded and must not double jump.} \\
 & \ubbullet{zShrine}{With no movement input, Player must not drift continuously on the XZ plane.} \\
 & \ubbullet{zShrine}{CubeController must expose public float moveSpeed and jumpVelocity fields.} \\
 & \ubbullet{zShrine}{CubeController must expose SetMoveInput(Vector2) and RequestJump() public methods.} \\
 & \ubbullet{zShrine}{MainCamera must be able to see the main area of Player.} \\*
\ubfield{zZonai}{Resources} & \textcolor{zZonai!82!black}{Scene · GameObject · Primitive · Rigidbody · Collider · MonoBehaviour · Input System · Camera} \\
\noalign{\vskip 3pt}
\multicolumn{2}{@{}p{14.4cm}@{}}{\textcolor{zRockLite!48}{\rule{\linewidth}{0.35pt}}}\\[4pt]
\multicolumn{2}{@{}p{14.4cm}@{}}{\ubtaskhead{zZonai}{0054-ProceduralSkinnedCloth}{Procedural Skinned Cloth}{zZonai}{Easy}{10}}\\*[4pt]
\ubfield{zChampion}{Goal} & Build a procedural cloth target in the scene Assets/Bench0054/Scenes/ProceduralSkinnedCloth.scene. \\
\ubfield{zDawn}{Context} & ProceduralSkinnedCloth.scene provides reference collision objects and a distractor BoxCollider. \\
\ubfield{zShrine}{Constraints} & \ubbullet{zShrine}{In Assets/Bench0054/Scenes/ProceduralSkinnedCloth.scene, create an enabled GameObject named exactly Bench0054\_ProceduralCloth with both SkinnedMeshRenderer and Cloth components on that same GameObject.} \\
 & \ubbullet{zShrine}{Use at least 28 columns and 18 rows of vertices.} \\
 & \ubbullet{zShrine}{Mesh width should be about 4.2 units, and height about 3.0 units.} \\
 & \ubbullet{zShrine}{In the cloth object's local space, the top edge should be near local y = 0, and the bottom edge should be near local y = -3.} \\
 & \ubbullet{zShrine}{Before simulation starts, the mesh should be basically planar, with local z values close to 0.} \\
 & \ubbullet{zShrine}{Include valid triangles and UVs, and ensure the generated mesh still exists completely after reopening the scene.} \\
 & \ubbullet{zShrine}{Create at least three bones under Bench0054\_ProceduralCloth.} \\
 & \ubbullet{zShrine}{Include bones named Bench0054\_TopBone, Bench0054\_MidBone, and Bench0054\_BottomBone.} \\
 & \ubbullet{zShrine}{Correctly set rootBone, bones, mesh bindposes, and vertex boneWeights.} \\
 & \ubbullet{zShrine}{Every vertex must have valid bone weights, and the sum of weights should be close to 1.} \\
 & \ubbullet{zShrine}{The top region should be mainly influenced by Bench0054\_TopBone.} \\
 & \ubbullet{zShrine}{The middle region should be clearly influenced by Bench0054\_MidBone.} \\
 & \ubbullet{zShrine}{The lower region should be mainly influenced by Bench0054\_BottomBone.} \\
 & \ubbullet{zShrine}{Make the rig functional so moving the middle or bottom bones deforms the result produced by SkinnedMeshRenderer.BakeMesh().} \\
 & \ubbullet{zShrine}{The top attachment band should be fixed, with maxDistance close to 0.} \\
 & \ubbullet{zShrine}{There should be a narrow transition band below the top, allowing limited movement.} \\
 & \ubbullet{zShrine}{Most of the lower curtain area should be free enough to sag and move.} \\
 & \ubbullet{zShrine}{Do not use exactly the same constraint for every particle.} \\
 & \ubbullet{zShrine}{Add Bench0054\_TopRailCollider and Bench0054\_CenterCapsuleObstacle to Cloth.capsuleColliders.} \\
 & \ubbullet{zShrine}{Add Bench0054\_RoundObstacle to Cloth.sphereColliders and use Bench0054\_LeftGuideSphere with Bench0054\_RightGuideSphere as one ClothSphereColliderPair.} \\
 & \ubbullet{zShrine}{Keep Bench0054\_BoxColliderDecoy in the scene, but do not rely on it for cloth collision because Unity Cloth does not support BoxCollider.} \\
 & \ubbullet{zShrine}{Enable gravity.} \\
 & \ubbullet{zShrine}{Set non-zero damping.} \\
 & \ubbullet{zShrine}{Stretching stiffness and bending stiffness should be within stable ranges.} \\
 & \ubbullet{zShrine}{Solver frequency should be sufficient for a medium-resolution curtain mesh.} \\
 & \ubbullet{zShrine}{Set some horizontal or forward external acceleration so the free cloth responds to motion after entering Play Mode.} \\
 & \ubbullet{zShrine}{Save the completed ProceduralSkinnedCloth.scene.} \\*
\ubfield{zZonai}{Resources} & \textcolor{zZonai!82!black}{Cloth · SkinnedMeshRenderer · Mesh · Collider} \\
\noalign{\vskip 3pt}
\multicolumn{2}{@{}p{14.4cm}@{}}{\textcolor{zRockLite!48}{\rule{\linewidth}{0.35pt}}}\\[4pt]
\multicolumn{2}{@{}p{14.4cm}@{}}{\ubtaskhead{zZonai}{0057-VoxelWormsTerrain}{Voxel Worms Terrain}{zZonai}{Easy}{12}}\\*[4pt]
\ubfield{zChampion}{Goal} & The scene Assets/Bench0057/Scenes/VoxelWormsTerrain.scene contains an incomplete VoxelTerrain object and several validation markers. Implement a procedural voxel terrain system that evokes the feeling of a 3D artillery battlefield island. \\
\ubfield{zDawn}{Context} & VoxelWormsTerrain.scene provides an incomplete VoxelTerrain object and fixed validation markers. \\
\ubfield{zShrine}{Constraints} & \ubbullet{zShrine}{Use 3D voxel data to drive the terrain. Default dimensions are 48 x 24 x 48, voxelSize = 0.5, and chunkSize = 16.} \\
 & \ubbullet{zShrine}{Generate the visible terrain surface with a procedural Mesh, and synchronize a MeshCollider.} \\
 & \ubbullet{zShrine}{Do not manually assemble the terrain from a large number of Cube GameObjects; the terrain should be represented by VoxelTerrain voxel data and a generated Mesh.} \\
 & \ubbullet{zShrine}{Generate a floating/island-style artillery terrain near the world origin, including walkable surfaces, hilly variation, at least one recognizable hollow or tunnel structure, and air above.} \\
 & \ubbullet{zShrine}{SolidProbe\_* and Explosion\_* markers should be inside solid terrain. EmptyProbe\_* markers should be in air or tunnel cavities. Each WormSpawn\_* marker itself should be in air, with walkable terrain beneath it.} \\
 & \ubbullet{zShrine}{Keep the MeshFilter, MeshRenderer, MeshCollider, and VoxelTerrain components on the VoxelTerrain object.} \\
 & \ubbullet{zShrine}{Do not delete Explosion\_A, Explosion\_B, Explosion\_C, WormSpawn\_A, WormSpawn\_B, WormSpawn\_C, or ValidationMarkers from the scene.} \\
 & \ubbullet{zShrine}{Implement and preserve GenerateTerrain(), ApplyExplosion(Vector3 center, float radius), IsSolid(int x, int y, int z), SetSolid(int x, int y, int z, bool solid), WorldToVoxel(Vector3 world), VoxelToWorldCenter(int x, int y, int z), and the Width, Height, Depth, VoxelSize, and SolidVoxelCount properties in Assets/Bench0057/Scripts/VoxelTerrain.cs.} \\
 & \ubbullet{zShrine}{Use one consistent voxel-grid coordinate system for WorldToVoxel and VoxelToWorldCenter, with X/Z centered around the world origin, Y increasing from the terrain bottom, and out-of-bounds voxels treated as empty.} \\
 & \ubbullet{zShrine}{SetSolid must update voxel state and SolidVoxelCount immediately without double-counting repeated assignments of the same state.} \\
 & \ubbullet{zShrine}{ApplyExplosion must remove voxels whose world-space centers lie within the spherical explosion radius, rebuild both Mesh and MeshCollider at runtime, accumulate across multiple explosions without restoring destroyed voxels, and leave terrain unchanged when the explosion is completely outside the grid.} \\
 & \ubbullet{zShrine}{Save VoxelWormsTerrain.scene and VoxelTerrain.cs.} \\*
\ubfield{zZonai}{Resources} & \textcolor{zZonai!82!black}{Mesh · MeshRenderer · Rigidbody · Collider · Procedural Generation} \\
\noalign{\vskip 3pt}
\multicolumn{2}{@{}p{14.4cm}@{}}{\textcolor{zRockLite!48}{\rule{\linewidth}{0.35pt}}}\\[4pt]
\multicolumn{2}{@{}p{14.4cm}@{}}{\ubtaskhead{zZonai}{0003-CatapultExplosion}{Catapult Explosion}{zGoldDeep}{Medium}{11}}\\*[4pt]
\ubfield{zChampion}{Goal} & Complete the catapult launch and explosion effect in the Catapult scene. \\
\ubfield{zDawn}{Context} & The Catapult scene provides a catapult animation, a controller script, and explosion targets. \\
\ubfield{zShrine}{Constraints} & \ubbullet{zShrine}{Complete the catapult effect in Assets/Scenes/Catapult.scene by editing Assets/Scripts/Catapult/CatapultInterviewController.cs and Assets/Catapult.anim.} \\
 & \ubbullet{zShrine}{Configure an animation event near the visual release frame to call SpawnStoneFromAnimationEvent().} \\
 & \ubbullet{zShrine}{SpawnStoneFromAnimationEvent() must create a physics-driven stone and begin its launch.} \\
 & \ubbullet{zShrine}{ApplyLaunchForce(Rigidbody stoneRigidbody) must launch the stone forward with sufficient velocity using frame-rate-independent physics.} \\
 & \ubbullet{zShrine}{ApplyExplosionForce(Rigidbody targetRigidbody, Vector3 explosionCenter) must push targets outward, primarily horizontally, without excessive displacement.} \\
 & \ubbullet{zShrine}{Preserve the complete animation-release, stone-launch, collision, explosion, and target-displacement chain.} \\
 & \ubbullet{zShrine}{Do not rename SpawnStoneFromAnimationEvent(), ApplyLaunchForce(Rigidbody), or ApplyExplosionForce(Rigidbody, Vector3).} \\
 & \ubbullet{zShrine}{Do not delete the marker comments in CatapultInterviewController.cs.} \\
 & \ubbullet{zShrine}{Do not rely on a fixed frame rate or manually overwrite the final target positions. Complete the effect through Unity physics behavior.} \\*
\ubfield{zZonai}{Resources} & \textcolor{zZonai!82!black}{Animation Event · Rigidbody · Collider · PhysX · MonoBehaviour} \\
\noalign{\vskip 3pt}
\multicolumn{2}{@{}p{14.4cm}@{}}{\textcolor{zRockLite!48}{\rule{\linewidth}{0.35pt}}}\\[4pt]
\multicolumn{2}{@{}p{14.4cm}@{}}{\ubtaskhead{zZonai}{0026-ShootAheadVehicleWheelPhysics}{Vehicle Control with WheelColliders}{zGoldDeep}{Medium}{8}}\\*[4pt]
\ubfield{zChampion}{Goal} & The scene Assets/0026Bench/0026Bench.scene provides a static vehicle and a test track. Based on the vehicle's existing model hierarchy, complete runnable vehicle physics control for Car. \\
\ubfield{zDawn}{Context} & Assets/0026Bench/0026Bench.scene provides a static vehicle, a test track, and BenchVehicleController. \\
\ubfield{zShrine}{Constraints} & \ubbullet{zShrine}{SetThrottle(float throttle): throttle input, range -1 to 1.} \\
 & \ubbullet{zShrine}{SetSteering(float steering): steering input, range -1 to 1.} \\
 & \ubbullet{zShrine}{SetBrake(float brake): brake input, range 0 to 1.} \\
 & \ubbullet{zShrine}{SetControls(float throttle, float steering, float brake = 0f): set all control inputs at once.} \\
 & \ubbullet{zShrine}{ReleaseControls(): release all inputs.} \\
 & \ubbullet{zShrine}{ResetVehiclePose(Vector3 position, Quaternion rotation): reset vehicle position and velocity.} \\
 & \ubbullet{zShrine}{Use Unity WheelCollider components to build four-wheel vehicle physics. Do not simulate forward motion by only moving Transforms.} \\
 & \ubbullet{zShrine}{Complete and save the vehicle component and prefab structure changes in EditMode/editor mode.} \\
 & \ubbullet{zShrine}{The four wheel colliders must correspond to the model wheels Wheel\_fl, Wheel\_fr, Wheel\_rl, and Wheel\_rr.} \\
 & \ubbullet{zShrine}{The vehicle initially faces world Z+. At full throttle it should move along the track toward Z+.} \\
 & \ubbullet{zShrine}{The front wheels should handle steering. The drive mode can be tuned as appropriate for the vehicle.} \\
 & \ubbullet{zShrine}{The target rigidbody mass for the car body is 1000, and the target mass for each wheel collider is 20.} \\
 & \ubbullet{zShrine}{Throttle input should drive the vehicle forward along its heading, and steering input should control turning.} \\
 & \ubbullet{zShrine}{The target acceleration time from rest to 100 km/h is about 7 seconds.} \\
 & \ubbullet{zShrine}{The target maximum speed is about 180 km/h; the vehicle must not accelerate indefinitely.} \\
 & \ubbullet{zShrine}{During straight full-throttle driving, the vehicle should remain stable and avoid obvious lateral drift or spontaneous rotation.} \\
 & \ubbullet{zShrine}{Preserve the existing public methods and properties in the script, and make the vehicle respond to these control entry points.} \\*
\ubfield{zZonai}{Resources} & \textcolor{zZonai!82!black}{WheelCollider · Rigidbody · PhysX · MonoBehaviour} \\
\noalign{\vskip 3pt}
\multicolumn{2}{@{}p{14.4cm}@{}}{\textcolor{zRockLite!48}{\rule{\linewidth}{0.35pt}}}\\[4pt]
\multicolumn{2}{@{}p{14.4cm}@{}}{\ubtaskhead{zZonai}{0039-RollingBallCheckpointCourse}{Inclined Rolling Ball Checkpoint Guidance}{zGoldDeep}{Medium}{15}}\\*[4pt]
\ubfield{zChampion}{Goal} & Complete three rolling-ball checkpoint scenes with distinct obstacle layouts that guide the ball through every required checkpoint. \\
\ubfield{zDawn}{Context} & Three scenes provide a slope, a ball, a spawn point, five required checkpoints, and Student\_CubeObstacles. \\
\ubfield{zShrine}{Constraints} & \ubbullet{zShrine}{Complete Assets/Bench0039/Scene/0039\_Group01.scene, 0039\_Group02.scene, and 0039\_Group03.scene as independent checkpoint courses.} \\
 & \ubbullet{zShrine}{In each scene, preserve the fixed Bench0039\_Slope, BallSpawn, FreeFallBall, RequiredCheckpoints, five Checkpoint\_Gxx\_Pxx objects, and Student\_CubeObstacles parent.} \\
 & \ubbullet{zShrine}{You may only use Unity built-in Cube primitives as blockers or guide objects.} \\
 & \ubbullet{zShrine}{All cubes you create must be children of the current scene's Student\_CubeObstacles.} \\
 & \ubbullet{zShrine}{You may freely adjust the Transform of these cubes, including Position, Rotation, and Scale.} \\
 & \ubbullet{zShrine}{Do not move, scale, or rename the slope, ball, BallSpawn, RequiredCheckpoints, or any Checkpoint\_*.} \\
 & \ubbullet{zShrine}{Do not add scripts to the ball, required checkpoints, or slope to directly control the ball position.} \\
 & \ubbullet{zShrine}{Place as few cubes as possible on the slope while making the ball roll freely from BallSpawn and pass through the current scene's 5 required checkpoints.} \\*
\ubfield{zZonai}{Resources} & \textcolor{zZonai!82!black}{PhysX · Rigidbody · Collider · Scene} \\
\noalign{\vskip 3pt}
\multicolumn{2}{@{}p{14.4cm}@{}}{\textcolor{zRockLite!48}{\rule{\linewidth}{0.35pt}}}\\[4pt]
\multicolumn{2}{@{}p{14.4cm}@{}}{\ubtaskhead{zZonai}{0048-RoadsideJointPhysics}{Roadside Joint Physics}{zGoldDeep}{Medium}{12}}\\*[4pt]
\ubfield{zChampion}{Goal} & Complete 4 small road physics scenes under Assets/Bench0048/Scenes. Each scene focuses on one Unity 3D Joint component. Configure the specified scene objects directly; do not fake the result with animation, scripts, or by deleting/replacing target objects. \\
\ubfield{zDawn}{Context} & Four small road scenes each provide a target object, anchors, a Road, and an ImpactCar for one Joint type. \\
\ubfield{zShrine}{Constraints} & \ubbullet{zShrine}{Preserve existing object names, hierarchy, road surface, and ImpactCar.} \\
 & \ubbullet{zShrine}{Target objects must keep valid Colliders and participate in real Rigidbody physics.} \\
 & \ubbullet{zShrine}{ImpactCar, Road, the camera, and lights do not need changes.} \\
 & \ubbullet{zShrine}{Do not add custom MonoBehaviour scripts to control target objects.} \\
 & \ubbullet{zShrine}{Save all scenes after completion.} \\
 & \ubbullet{zShrine}{In FixedJointFence.scene, connect FixedFence\_Rail to FixedFence\_Anchor with a FixedJoint that can break after vehicle impact, allowing obvious displacement or tipping.} \\
 & \ubbullet{zShrine}{In HingeJointGate.scene, use a HingeJoint so HingeGate\_Arm swings around one side of HingeGate\_Post and remains within a reasonable angle range after impact.} \\
 & \ubbullet{zShrine}{In SpringJointBarrier.scene, use a SpringJoint between SpringBarrier\_Block and SpringBarrier\_Anchor so the block is pushed away by impact and tends to rebound.} \\
 & \ubbullet{zShrine}{In ConfigurableJointSlider.scene, use a ConfigurableJoint between ConfigurableBollard\_Block and ConfigurableBollard\_Anchor so the block slides only a limited distance along the road direction while other translation axes and rotation remain constrained.} \\*
\ubfield{zZonai}{Resources} & \textcolor{zZonai!82!black}{PhysX · Rigidbody · Collider · Joint} \\
\noalign{\vskip 3pt}
\multicolumn{2}{@{}p{14.4cm}@{}}{\textcolor{zRockLite!48}{\rule{\linewidth}{0.35pt}}}\\[4pt]
\multicolumn{2}{@{}p{14.4cm}@{}}{\ubtaskhead{zZonai}{0056-RigidbodyCharacterControllerParity}{Rigidbody Character Controller Parity}{zGoldDeep}{Medium}{15}}\\*[4pt]
\ubfield{zChampion}{Goal} & Implement a Rigidbody-driven character controller that matches the provided CharacterController reference for movement, jumping, slopes, and collision behavior. \\
\ubfield{zDawn}{Context} & RigidbodyCharacterParity.scene provides a CharacterController reference character and an incomplete Rigidbody target character. \\
\ubfield{zShrine}{Constraints} & \ubbullet{zShrine}{In Assets/Bench0056/Scenes/RigidbodyCharacterParity.scene, make Student\_RigidbodyCharacter use Rigidbody and CapsuleCollider to behave closely to the existing Reference\_CharacterController; do not add a CharacterController to the student object.} \\
 & \ubbullet{zShrine}{Match the fixed reference values: capsule radius 0.35, height 1.8, center (0, 0.9, 0), slope limit 42, step height 0.42, skin width 0.04, walk speed 3.8, sprint speed 5.2, jump height 1.1, and gravity -22.} \\
 & \ubbullet{zShrine}{Configure a dynamic Rigidbody and matching-size CapsuleCollider for Student\_RigidbodyCharacter.} \\
 & \ubbullet{zShrine}{Complete the movement logic in Assets/Bench0056/Scripts/RigidbodyCharacterMotor.cs.} \\
 & \ubbullet{zShrine}{Preserve the public methods and properties of RigidbodyCharacterMotor: SetInput(Vector2, bool, bool), ResetMotor(Vector3, Quaternion), Velocity, and IsGrounded.} \\
 & \ubbullet{zShrine}{Support flat-ground movement, sprinting, jumping, falling, low steps, slopes, stopping when hitting walls, and diagonal input normalization.} \\
 & \ubbullet{zShrine}{Keep Reference\_CharacterController, scene terrain, start markers, walls, steps, and slopes intact.} \\
 & \ubbullet{zShrine}{Save RigidbodyCharacterParity.scene and RigidbodyCharacterMotor.cs.} \\*
\ubfield{zZonai}{Resources} & \textcolor{zZonai!82!black}{Rigidbody · Collider · CharacterController · PhysX} \\
\noalign{\vskip 3pt}
\multicolumn{2}{@{}p{14.4cm}@{}}{\textcolor{zRockLite!48}{\rule{\linewidth}{0.35pt}}}\\[4pt]
\multicolumn{2}{@{}p{14.4cm}@{}}{\ubtaskhead{zZonai}{0044-ColliderWrappingChallenge}{Collider Wrapping Challenge}{zShrine}{Hard}{26}}\\*[4pt]
\ubfield{zChampion}{Goal} & Open Assets/Vendor/Art/PolygonMilitary/Scenes/Bench0044.scene and complete the collider rigs under ColliderChallenge. \\
\ubfield{zDawn}{Context} & The ColliderChallenge hierarchy in Bench0044.scene contains eight military assets and their ColliderRig nodes. \\
\ubfield{zShrine}{Constraints} & \ubbullet{zShrine}{In Assets/Vendor/Art/PolygonMilitary/Scenes/Bench0044.scene, complete the ColliderRig under each of Tank\_USA, AttackHelicopter, GuardTower, Barracks, Hangar, CrateStack, RazorWire, and SatelliteDish.} \\
 & \ubbullet{zShrine}{Use only BoxCollider, SphereCollider, CapsuleCollider, and empty child GameObjects under ColliderRig for positioning those primitive colliders.} \\
 & \ubbullet{zShrine}{Do not use MeshCollider or read mesh vertex or triangle data through Mesh, MeshFilter.sharedMesh, or SkinnedMeshRenderer.BakeMesh.} \\
 & \ubbullet{zShrine}{Do not use the provided collision meshes in Models/CollisionConvex or Models/CollisionCustom, and do not generate new mesh assets as collider proxies.} \\
 & \ubbullet{zShrine}{Build a small, readable set of primitive colliders that achieves good reference-volume coverage without oversized spill outside each target.} \\
 & \ubbullet{zShrine}{Ensure the collider rigs support plausible raycast and physics behavior from multiple directions rather than enclosing a target in one huge collider.} \\*
\ubfield{zZonai}{Resources} & \textcolor{zZonai!82!black}{Collider · PhysX · Bounds · Raycast} \\
\noalign{\vskip 3pt}
\multicolumn{2}{@{}p{14.4cm}@{}}{\textcolor{zRockLite!48}{\rule{\linewidth}{0.35pt}}}\\[4pt]
\multicolumn{2}{@{}p{14.4cm}@{}}{\ubtaskhead{zZonai}{0053-PhysicsMaterial}{PhysicMaterial Tuning}{zShrine}{Hard}{8}}\\*[4pt]
\ubfield{zChampion}{Goal} & Tune the four provided PhysicMaterial assets so spheres traverse every checkpoint in three scenes and the player's curling stone maximizes the red side's score in the fourth. \\
\ubfield{zDawn}{Context} & Four scenes and four PhysicMaterial assets are provided for friction, bounciness, and combine-mode tuning. \\
\ubfield{zShrine}{Constraints} & \ubbullet{zShrine}{Modify only Assets/Bench0053/PlayerMaterial1.physicMaterial through PlayerMaterial4.physicMaterial; do not change scenes, scripts, Rigidbody or Collider components, Transforms, or other assets.} \\
 & \ubbullet{zShrine}{Tune PlayerMaterial1, PlayerMaterial2, and PlayerMaterial3 so Sphere crosses every CheckPoint at runtime in the corresponding scenes 0053-1.scene, 0053-2.scene, and 0053-3.scene.} \\
 & \ubbullet{zShrine}{Tune PlayerMaterial4 so Player slides toward the house center in 0053-4.scene and maximizes the red side's score under the stated curling rules.} \\
 & \ubbullet{zShrine}{Use PhysicMaterial friction, bounciness, and combine modes to achieve the required trajectories.} \\*
\ubfield{zZonai}{Resources} & \textcolor{zZonai!82!black}{PhysicMaterial · Rigidbody · Collider · PhysX} \\
\noalign{\vskip 3pt}
\multicolumn{2}{@{}p{14.4cm}@{}}{\textcolor{zRockLite!48}{\rule{\linewidth}{0.35pt}}}\\[4pt]
\multicolumn{2}{@{}p{14.4cm}@{}}{\ubtaskhead{zZonai}{0077-TakeOffAndLanding}{Implement Physical Fixed-Wing Takeoff and Landing}{zShrine}{Hard}{31}}\\*[4pt]
\ubfield{zChampion}{Goal} & Complete a closed-loop Rigidbody aircraft controller that autonomously takes off, flies a compact return circuit, aligns, lands main wheels first, and brakes to a stop. \\
\ubfield{zDawn}{Context} & Assets/Scenes/0077.scene contains AircraftPhysicsRoot, a dynamic Rigidbody, exactly three WheelColliders, runway geometry, wheel objects, and an incomplete automatic flight controller. \\
\ubfield{zShrine}{Constraints} & \ubbullet{zShrine}{Preserve Assets/Scenes/0077.scene, AircraftPhysicsRoot, its dynamic Rigidbody, exactly three WheelColliders, runway coordinates, and wheel objects.} \\
 & \ubbullet{zShrine}{Keep the aircraft non-kinematic with gravity enabled.} \\
 & \ubbullet{zShrine}{Drive flight through thrust, aerodynamic forces, body torques, gravity, wheel contact, suspension, tire forces, steering, and braking.} \\
 & \ubbullet{zShrine}{Do not directly drive runtime position, rotation, velocity, or angular velocity except initialization or explicit reset.} \\
 & \ubbullet{zShrine}{Do not use Animator, Timeline, kinematic motion, path animation, moving parents, timeScale changes, disabled collisions, or artificial whole-aircraft ground support.} \\
 & \ubbullet{zShrine}{Keep all wheel motorTorque values at zero and resolve contact per wheel.} \\
 & \ubbullet{zShrine}{Complete ground roll, rotation, liftoff, compact circuit, final alignment, flare, touchdown, nose lowering, and full stop within 120 seconds.} \\
 & \ubbullet{zShrine}{Lift off smoothly around 55 to 80 metres per second and remain close to the runway centerline.} \\
 & \ubbullet{zShrine}{Keep the circuit within the stated lateral and longitudinal bounds.} \\
 & \ubbullet{zShrine}{Approach with limited lateral offset, lateral velocity, bank, and heading error.} \\
 & \ubbullet{zShrine}{Touch down on the main wheels before the nose wheel without a large bounce or speed increase.} \\
 & \ubbullet{zShrine}{Brake to a non-reversing stop on the runway within 25 seconds of touchdown.} \\
 & \ubbullet{zShrine}{Use closed-loop feedback and remain functional under mass, start-pose, heading, and disturbance variation.} \\
 & \ubbullet{zShrine}{Remove temporary diagnostics, restore the complete automatic sequence, and confirm that the project compiles before submission.} \\*
\ubfield{zZonai}{Resources} & \textcolor{zZonai!82!black}{Scene · Rigidbody · Collider · WheelCollider · PhysX · MonoBehaviour · Transform} \\
\noalign{\vskip 3pt}
\multicolumn{2}{@{}p{14.4cm}@{}}{\textcolor{zRockLite!48}{\rule{\linewidth}{0.35pt}}}\\[4pt]
\noalign{\vskip 3pt}
\multicolumn{2}{@{}p{14.4cm}@{}}{\ubcategory{zRock}{Miscellaneous}{3 tasks \textperiodcentered\ 58 milestone tests}}\\*[5pt]
\multicolumn{2}{@{}p{14.4cm}@{}}{\ubtaskhead{zRock}{0011-CookiesApply}{Light Cookie Application}{zZonai}{Easy}{9}}\\*[4pt]
\ubfield{zChampion}{Goal} & Build a lighting scene containing Cookie-based Directional Light and Point Light projections. The Directional Light and Point Light must use the correct assets and render the expected light projection effect. \\
\ubfield{zDawn}{Context} & The project provides the GrayscaleHoles4Face source texture but not the completed Cookie lighting setup. \\
\ubfield{zShrine}{Constraints} & \ubbullet{zShrine}{From Assets/Textures/GrayscaleHoles4Face.png, create a 2D Cookie texture asset and a Cubemap Cookie asset without modifying the source texture.} \\
 & \ubbullet{zShrine}{Assign the 2D and Cubemap Cookie assets to the corresponding Directional Light and Point Light components on GameObjects under Answer.} \\
 & \ubbullet{zShrine}{All six faces of the Cubemap must be filled with the Cookie texture corresponding to the Point Light.} \\
 & \ubbullet{zShrine}{The scene must contain a Plane, a Point Light, and a Directional Light. Their Transform values must make the Plane visible to the camera.} \\*
\ubfield{zZonai}{Resources} & \textcolor{zZonai!82!black}{Built-in Render Pipeline · Light · Cookie · Camera} \\
\noalign{\vskip 3pt}
\multicolumn{2}{@{}p{14.4cm}@{}}{\textcolor{zRockLite!48}{\rule{\linewidth}{0.35pt}}}\\[4pt]
\multicolumn{2}{@{}p{14.4cm}@{}}{\ubtaskhead{zRock}{0017-CreateLOD}{LOD Asset Creation}{zGoldDeep}{Medium}{39}}\\*[4pt]
\ubfield{zChampion}{Goal} & Complete LOD creation, merging, and splitting for the models. \\
\ubfield{zDawn}{Context} & The project provides several model sets: some already contain LODs, while others require splitting, merging, or mesh reduction. \\
\ubfield{zShrine}{Constraints} & \ubbullet{zShrine}{Split each LOD of PalletPlywoodNew\_GameReady\_LODs into separate models, name them Answer1\_1, Answer1\_2, and so on, and place them in Answer.} \\
 & \ubbullet{zShrine}{Combine the pine tree models with the \_A suffix, such as Pine\_beach\_01\_A, into one LOD model using an LODGroup. Use each model's LOD0 as one LOD in numeric order. You do not need to consider decreasing triangle counts. Name the combined model Answer2 and place it in Answer.} \\
 & \ubbullet{zShrine}{Merge pine\_beach\_01\_B\_LOD3\_Cross into the Pine\_beach\_01\_B model as LOD3. Name the merged model Answer3 and place it in Answer.} \\
 & \ubbullet{zShrine}{Convert Sketchfab into a model containing LOD0 through LOD3, with triangle counts decreasing in order. Perform mesh reduction yourself. Name the completed LOD model Answer4 and place it in Answer.} \\
 & \ubbullet{zShrine}{Convert rock1 into a model containing LOD0 through LOD3, with triangle counts decreasing in order. Perform mesh reduction yourself. Name the completed LOD model Answer5 and place it in Answer.} \\
 & \ubbullet{zShrine}{For models that are merged or split, the corresponding triangle counts before and after the operation must not change. For models that originally have no LODs, the generated LOD model triangle counts should decrease by LOD level.} \\
 & \ubbullet{zShrine}{The generated Answer assets must be .fbx files.} \\
 & \ubbullet{zShrine}{Models containing multiple LODs must have an LODGroup when placed in the scene.} \\
 & \ubbullet{zShrine}{The merged or split models must use the original shaders and Materials from their source models.} \\
 & \ubbullet{zShrine}{For the created or merged models, the LODGroup percentages for LOD0, LOD1, and so on through Culled must be the same.} \\*
\ubfield{zZonai}{Resources} & \textcolor{zZonai!82!black}{LODGroup · Mesh · Model Importer · Material · AssetDatabase} \\
\noalign{\vskip 3pt}
\multicolumn{2}{@{}p{14.4cm}@{}}{\textcolor{zRockLite!48}{\rule{\linewidth}{0.35pt}}}\\[4pt]
\multicolumn{2}{@{}p{14.4cm}@{}}{\ubtaskhead{zRock}{0019-AutoTimelineGenerator}{Timeline Cinematic Auto-Generation}{zShrine}{Hard}{10}}\\*[4pt]
\ubfield{zChampion}{Goal} & Investigate the loose assets in the initial TimelineBenchmark project, including animation, VFX prefab, and audio, and write an editor script based on the Playables and Timeline APIs to automatically arrange them into a complete cinematic cutscene. \\
\ubfield{zDawn}{Context} & The TimelineBenchmark project provides separate Walk animation, Explosion VFX, Bang audio, and other cinematic assets. \\
\ubfield{zShrine}{Constraints} & \ubbullet{zShrine}{Locate the three base assets under Assets/TimelineBenchmark/: Walk.anim, Explosion\_VFX.prefab, and Bang\_SFX.wav.} \\
 & \ubbullet{zShrine}{Automatically generate and save a Timeline asset. The output path must be Assets/TimelineBenchmark/GeneratedCinematic.playable.} \\
 & \ubbullet{zShrine}{Strict track architecture constraints: the generated Timeline must contain exactly 4 root tracks and no others: 2 AnimationTracks, 1 ControlTrack, and 1 AudioTrack. GroupTrack or extra tracks are strictly forbidden.} \\
 & \ubbullet{zShrine}{The character GameObject must be named Character\_Main and must have a PlayableDirector attached.} \\
 & \ubbullet{zShrine}{Use a CameraRig -\textgreater{} Main Camera parent-child hierarchy: CameraRig carries the animation component and Main Camera carries the Camera component.} \\
 & \ubbullet{zShrine}{The instantiated explosion VFX GameObject must be named exactly Explosion\_VFX\_Instance and contain a ParticleSystem component.} \\
 & \ubbullet{zShrine}{Character walking (0 to 3 seconds): play the Walk animation on an AnimationTrack whose name contains "Walk", bound to Character\_Main. Ensure the animation produces real root-motion displacement in the scene rather than walking in place.} \\
 & \ubbullet{zShrine}{Explosion VFX (starts at 3.0 seconds, lasts 2 seconds): trigger it about 2 meters in front of the character's walking direction, using a ControlTrack and ExposedReference precisely bound to Explosion\_VFX\_Instance.} \\
 & \ubbullet{zShrine}{Explosion audio (starts at 3.0 seconds): play Bang\_SFX on an AudioTrack, bound to the AudioSource component on Character\_Main.} \\
 & \ubbullet{zShrine}{Camera shake (3.5 seconds to 4.5 seconds): arrange it on an AnimationTrack whose name contains "Shake", bound to CameraRig. Dynamically inject valid displacement curves such as LocalPosition so the child GameObject Main Camera visibly shakes in world space with an amplitude below 2 metres; do not use an empty animation clip.} \\
 & \ubbullet{zShrine}{Camera movement and composition constraints: reasonably set the initial spatial coordinates of CameraRig so that throughout the entire Timeline playback, especially at the 3.0-second explosion climax, Character\_Main and Explosion\_VFX\_Instance are always fully inside the main camera's visible Viewport frustum. Blind shots where the character or VFX is behind the camera or outside the frame are strictly forbidden.} \\
 & \ubbullet{zShrine}{Follow Unity engine conventions and avoid low-level pitfalls that may cause runtime state resets or component loss, such as invalid null checks using $??$ on engine components.} \\
 & \ubbullet{zShrine}{Ensure that after the script runs as an editor tool, the scene has no residual dirty data, no compile errors, and can directly pass strict automated hierarchy, timing, and viewport-composition tests.} \\*
\ubfield{zZonai}{Resources} & \textcolor{zZonai!82!black}{Timeline · Playables · Animation · Audio · Camera} \\
\noalign{\vskip 3pt}
\multicolumn{2}{@{}p{14.4cm}@{}}{\textcolor{zRockLite!48}{\rule{\linewidth}{0.35pt}}}\\[4pt]
\end{longtable}
\endgroup

\section{Test Cases}
\label{app:chaltest}

\begingroup
\setlength{\tabcolsep}{0pt}
\renewcommand{\arraystretch}{1.01}
\setlength{\LTcapwidth}{14.4cm}
\footnotesize\color{zText}
\begin{longtable}{@{}>{\raggedright\arraybackslash}p{1.72cm}@{\hspace{8pt}}>{\raggedright\arraybackslash}p{12.41cm}@{}}
\caption{\textbf{Milestone test cases.} The 1,178 held-out, human-authored checks used to evaluate the 77 tasks. Every row names one independently scored test and summarizes the property it verifies.}\label{tab:test-cases}\\
\multicolumn{2}{@{}p{14.4cm}@{}}{\ubidentity{zGoldDeep}{white}{TEST CASES}{1,178 independent pass/fail checks}}\\[5pt]
\endfirsthead
\multicolumn{2}{@{}p{14.4cm}@{}}{%
  \textcolor{zRock}{\itshape Table \thetable\ continued from previous page.}}\\[3pt]
\multicolumn{2}{@{}p{14.4cm}@{}}{\textcolor{zRockLite!58}{\rule{\linewidth}{0.45pt}}}\\[3pt]
\endhead
\multicolumn{2}{r@{}}{\textcolor{zRock}{\itshape Continued on next page.}}\\
\endfoot
\endlastfoot
\noalign{\vskip 3pt}
\multicolumn{2}{@{}p{14.4cm}@{}}{\ubcategory{zChampion}{User Interface}{9 tasks \textperiodcentered\ 111 milestone tests}}\\*[4pt]
\multicolumn{2}{@{}p{14.4cm}@{}}{\ubtaskhead{zChampion}{0025-FPSCompassHUDCalibration}{Compass HUD Direction Calibration}{zGoldDeep}{Easy}{6}}\\*[2pt]
\ubtestbadge{01} & \textcolor{zGoldDeep!78!black}{\texttt{\bfseries Runtime\-Heading\_\allowbreak{}Earns1\-Of6\_\allowbreak{}When\-Cardinal\-Directions\-Are\-Within6\-Degrees}}\enspace\textcolor{zText}{Checks that runtime behavior heading earns 1 of 6 when cardinal directions are within 6 degrees.} \\
\ubtestbadge{02} & \textcolor{zGoldDeep!78!black}{\texttt{\bfseries Runtime\-Heading\_\allowbreak{}Earns2\-Of6\_\allowbreak{}When\-Cardinal\-Directions\-Are\-Within5\-Degrees}}\enspace\textcolor{zText}{Checks that runtime behavior heading earns 2 of 6 when cardinal directions are within 5 degrees.} \\
\ubtestbadge{03} & \textcolor{zGoldDeep!78!black}{\texttt{\bfseries Runtime\-Heading\_\allowbreak{}Earns3\-Of6\_\allowbreak{}When\-Cardinal\-Directions\-Are\-Within4\-Degrees}}\enspace\textcolor{zText}{Checks that runtime behavior heading earns 3 of 6 when cardinal directions are within 4 degrees.} \\
\ubtestbadge{04} & \textcolor{zGoldDeep!78!black}{\texttt{\bfseries Runtime\-Heading\_\allowbreak{}Earns4\-Of6\_\allowbreak{}When\-Cardinal\-Directions\-Are\-Within3\-Degrees}}\enspace\textcolor{zText}{Checks that runtime behavior heading earns 4 of 6 when cardinal directions are within 3 degrees.} \\
\ubtestbadge{05} & \textcolor{zGoldDeep!78!black}{\texttt{\bfseries Runtime\-Heading\_\allowbreak{}Earns5\-Of6\_\allowbreak{}When\-Cardinal\-Directions\-Are\-Within2\-Degrees}}\enspace\textcolor{zText}{Checks that runtime behavior heading earns 5 of 6 when cardinal directions are within 2 degrees.} \\
\ubtestbadge{06} & \textcolor{zGoldDeep!78!black}{\texttt{\bfseries Runtime\-Heading\_\allowbreak{}Earns6\-Of6\_\allowbreak{}When\-Cardinal\-Directions\-Are\-Within1\-Degree}}\enspace\textcolor{zText}{Checks that runtime behavior heading earns 6 of 6 when cardinal directions are within 1 degree.} \\
\noalign{\vskip 2pt}
\multicolumn{2}{@{}p{14.4cm}@{}}{\textcolor{zRockLite!42}{\rule{\linewidth}{0.3pt}}}\\[3pt]
\multicolumn{2}{@{}p{14.4cm}@{}}{\ubtaskhead{zChampion}{0050-UIToolkitSnake}{Snake Game Implemented with Unity UI Toolkit}{zGoldDeep}{Easy}{23}}\\*[2pt]
\ubtestbadge{01} & \textcolor{zGoldDeep!78!black}{\texttt{\bfseries Gameplay\_\allowbreak{}Defines\-Playable\-Grid\-With\-Tolerance}}\enspace\textcolor{zText}{Checks that gameplay defines playable grid with tolerance.} \\
\ubtestbadge{02} & \textcolor{zGoldDeep!78!black}{\texttt{\bfseries Gameplay\_\allowbreak{}Eating\-Food\-Increases\-Score\-And\-Grows\-Snake}}\enspace\textcolor{zText}{Checks that gameplay eating food increases score and grows snake.} \\
\ubtestbadge{03} & \textcolor{zGoldDeep!78!black}{\texttt{\bfseries Gameplay\_\allowbreak{}Game\-Over\-Panel\-Provides\-Restart\-Action}}\enspace\textcolor{zText}{Checks that gameplay game over panel provides restart action.} \\
\ubtestbadge{04} & \textcolor{zGoldDeep!78!black}{\texttt{\bfseries Gameplay\_\allowbreak{}Models\-Waiting\-Playing\-And\-Game\-Over\-States}}\enspace\textcolor{zText}{Checks that gameplay models waiting playing and game over states.} \\
\ubtestbadge{05} & \textcolor{zGoldDeep!78!black}{\texttt{\bfseries Gameplay\_\allowbreak{}Starts\-From\-Button\-And\-Keyboard}}\enspace\textcolor{zText}{Checks that gameplay starts from button and keyboard.} \\
\ubtestbadge{06} & \textcolor{zGoldDeep!78!black}{\texttt{\bfseries Gameplay\_\allowbreak{}Supports\-Arrow\-And\-Wasd\-Controls}}\enspace\textcolor{zText}{Checks that gameplay supports arrow and wasd controls.} \\
\ubtestbadge{07} & \textcolor{zGoldDeep!78!black}{\texttt{\bfseries Gameplay\_\allowbreak{}Uses\-Timer\-For\-Continuous\-Movement}}\enspace\textcolor{zText}{Checks that gameplay uses timer for continuous movement.} \\
\ubtestbadge{08} & \textcolor{zGoldDeep!78!black}{\texttt{\bfseries Gameplay\_\allowbreak{}Wall\-And\-Self\-Collisions\-End\-Game}}\enspace\textcolor{zText}{Checks gameplay wall and self collisions end game.} \\
\ubtestbadge{09} & \textcolor{zGoldDeep!78!black}{\texttt{\bfseries Interface\-Structure\_\allowbreak{}Has\-Concise\-Controls\-Hint}}\enspace\textcolor{zText}{Checks that interface structure has concise controls hint.} \\
\ubtestbadge{10} & \textcolor{zGoldDeep!78!black}{\texttt{\bfseries Interface\-Structure\_\allowbreak{}Has\-Grid\-Start\-And\-Restart\-Or\-Game\-Over\-Controls}}\enspace\textcolor{zText}{Checks that interface structure has grid start and restart or game over controls.} \\
\ubtestbadge{11} & \textcolor{zGoldDeep!78!black}{\texttt{\bfseries Interface\-Structure\_\allowbreak{}Has\-Speed\-Control\-And\-Readout}}\enspace\textcolor{zText}{Checks that interface structure has speed control and readout.} \\
\ubtestbadge{12} & \textcolor{zGoldDeep!78!black}{\texttt{\bfseries Interface\-Structure\_\allowbreak{}Has\-Title\-Score\-And\-Best\-Score}}\enspace\textcolor{zText}{Checks that interface structure has title score and best score.} \\
\ubtestbadge{13} & \textcolor{zGoldDeep!78!black}{\texttt{\bfseries Interface\-Structure\_\allowbreak{}Uses\-Dark\-Neon\-Technology\-Style}}\enspace\textcolor{zText}{Checks that interface structure uses dark neon technology style.} \\
\ubtestbadge{14} & \textcolor{zGoldDeep!78!black}{\texttt{\bfseries Speed\-Control\_\allowbreak{}Adjustment\-Is\-Not\-Limited\-To\-Pre\-Start\-State}}\enspace\textcolor{zText}{Checks that speed control adjustment is not limited to pre start state.} \\
\ubtestbadge{15} & \textcolor{zGoldDeep!78!black}{\texttt{\bfseries Speed\-Control\_\allowbreak{}Movement\-Keys\-Do\-Not\-Keep\-Driving\-Focused\-Slider}}\enspace\textcolor{zText}{Checks that speed control movement keys do not keep driving focused slider.} \\
\ubtestbadge{16} & \textcolor{zGoldDeep!78!black}{\texttt{\bfseries Speed\-Control\_\allowbreak{}Slider\-Range\-Covers\-Prompt\-Intervals\-With\-Tolerance}}\enspace\textcolor{zText}{Checks that speed control slider range covers prompt intervals with tolerance.} \\
\ubtestbadge{17} & \textcolor{zGoldDeep!78!black}{\texttt{\bfseries Speed\-Control\_\allowbreak{}Ui\-Shows\-Current\-Tier\-Or\-Numeric\-Value}}\enspace\textcolor{zText}{Checks that speed control UI shows current tier or numeric value.} \\
\ubtestbadge{18} & \textcolor{zGoldDeep!78!black}{\texttt{\bfseries Speed\-Control\_\allowbreak{}Value\-Change\-Immediately\-Updates\-Movement\-Interval}}\enspace\textcolor{zText}{Checks that speed control value change immediately updates movement interval.} \\
\ubtestbadge{19} & \textcolor{zGoldDeep!78!black}{\texttt{\bfseries Technical\-Requirements\_\allowbreak{}Default\-Scene\-Contains\-Editable\-UIDocument}}\enspace\textcolor{zText}{Checks that technical requirements default scene contains editable UIDocument.} \\
\ubtestbadge{20} & \textcolor{zGoldDeep!78!black}{\texttt{\bfseries Technical\-Requirements\_\allowbreak{}Default\-Scene\-Contains\-Gameplay\-Script}}\enspace\textcolor{zText}{Checks that technical requirements default scene contains gameplay script.} \\
\ubtestbadge{21} & \textcolor{zGoldDeep!78!black}{\texttt{\bfseries Technical\-Requirements\_\allowbreak{}Default\-Scene\-Does\-Not\-Use\-Ugui\-Canvas}}\enspace\textcolor{zText}{Checks that technical requirements default scene does not use UGUI canvas.} \\
\ubtestbadge{22} & \textcolor{zGoldDeep!78!black}{\texttt{\bfseries Technical\-Requirements\_\allowbreak{}UIDocument\-References\-Uxml\-And\-Panel\-Settings}}\enspace\textcolor{zText}{Checks that technical requirements UIDocument references UXML and PanelSettings.} \\
\ubtestbadge{23} & \textcolor{zGoldDeep!78!black}{\texttt{\bfseries Technical\-Requirements\_\allowbreak{}Uxml\-References\-Uss}}\enspace\textcolor{zText}{Checks that technical requirements UXML references USS.} \\
\noalign{\vskip 2pt}
\multicolumn{2}{@{}p{14.4cm}@{}}{\textcolor{zRockLite!42}{\rule{\linewidth}{0.3pt}}}\\[3pt]
\multicolumn{2}{@{}p{14.4cm}@{}}{\ubtaskhead{zChampion}{0063-MinimapRealization}{Minimap UI Implementation}{zGoldDeep}{Easy}{7}}\\*[2pt]
\ubtestbadge{01} & \textcolor{zGoldDeep!78!black}{\texttt{\bfseries Minimap\-Camera\_\allowbreak{}Follows\-Player\-World\-Position}}\enspace\textcolor{zText}{Checks that minimap camera follows player world position.} \\
\ubtestbadge{02} & \textcolor{zGoldDeep!78!black}{\texttt{\bfseries Minimap\-Camera\_\allowbreak{}Is\-Orthographic\-And\-Fixed\-To\-World\-North}}\enspace\textcolor{zText}{Checks that minimap camera is orthographic and fixed to world north.} \\
\ubtestbadge{03} & \textcolor{zGoldDeep!78!black}{\texttt{\bfseries Minimap\-Content\_\allowbreak{}Comes\-From\-Top\-Down\-Scene\-Camera\-Render\-Texture}}\enspace\textcolor{zText}{Checks minimap content comes from top down scene camera RenderTexture.} \\
\ubtestbadge{04} & \textcolor{zGoldDeep!78!black}{\texttt{\bfseries Minimap\-Ui\_\allowbreak{}Has\-Render\-Texture\-Raw\-Image\-Circle\-Mask\-And\-Triangle\-Arrow}}\enspace\textcolor{zText}{Checks that minimap UI has RenderTexture raw image circle mask and triangle arrow.} \\
\ubtestbadge{05} & \textcolor{zGoldDeep!78!black}{\texttt{\bfseries Player\-Arrow\_\allowbreak{}Rotates\-With\-Player\-Facing\-Direction}}\enspace\textcolor{zText}{Checks that player arrow rotates with player facing direction.} \\
\ubtestbadge{06} & \textcolor{zGoldDeep!78!black}{\texttt{\bfseries Player\-Arrow\_\allowbreak{}Stays\-Centered\-In\-Minimap}}\enspace\textcolor{zText}{Checks that player arrow stays centered in minimap.} \\
\ubtestbadge{07} & \textcolor{zGoldDeep!78!black}{\texttt{\bfseries Player\-Layer\_\allowbreak{}Is\-Excluded\-From\-Minimap\-Camera}}\enspace\textcolor{zText}{Checks that player layer is excluded from minimap camera.} \\
\noalign{\vskip 2pt}
\multicolumn{2}{@{}p{14.4cm}@{}}{\textcolor{zRockLite!42}{\rule{\linewidth}{0.3pt}}}\\[3pt]
\multicolumn{2}{@{}p{14.4cm}@{}}{\ubtaskhead{zChampion}{0010-SplitTheUI}{UI Sprite Sheet Slicing}{zGoldDeep}{Medium}{11}}\\*[2pt]
\ubtestbadge{01} & \textcolor{zGoldDeep!78!black}{\texttt{\bfseries Sprite1\-Through4\_\allowbreak{}All\-Sub\-Sprites\-Have\-Solid\-Pixels}}\enspace\textcolor{zText}{Checks that sprite 1 through 4 all sub sprites have solid pixels.} \\
\ubtestbadge{02} & \textcolor{zGoldDeep!78!black}{\texttt{\bfseries Sprite1\_\allowbreak{}Sub\-Sprite\-Count\_\allowbreak{}Is28}}\enspace\textcolor{zText}{Checks that sprite 1 sub sprite count is 28.} \\
\ubtestbadge{03} & \textcolor{zGoldDeep!78!black}{\texttt{\bfseries Sprite2\_\allowbreak{}Sprite\-Mode\_\allowbreak{}Is\-Multiple}}\enspace\textcolor{zText}{Checks that sprite 2 sprite mode is multiple.} \\
\ubtestbadge{04} & \textcolor{zGoldDeep!78!black}{\texttt{\bfseries Sprite2\_\allowbreak{}Sub\-Sprite\-Count\_\allowbreak{}Is38}}\enspace\textcolor{zText}{Checks that sprite 2 sub sprite count is 38.} \\
\ubtestbadge{05} & \textcolor{zGoldDeep!78!black}{\texttt{\bfseries Sprite3\_\allowbreak{}Sprite\-Mode\_\allowbreak{}Is\-Multiple}}\enspace\textcolor{zText}{Checks that sprite 3 sprite mode is multiple.} \\
\ubtestbadge{06} & \textcolor{zGoldDeep!78!black}{\texttt{\bfseries Sprite3\_\allowbreak{}Sub\-Sprite\-Count\_\allowbreak{}Is34}}\enspace\textcolor{zText}{Checks that sprite 3 sub sprite count is 34.} \\
\ubtestbadge{07} & \textcolor{zGoldDeep!78!black}{\texttt{\bfseries Sprite4\_\allowbreak{}Pixel\-Size\_\allowbreak{}Is32x32}}\enspace\textcolor{zText}{Checks that sprite 4 pixel size is 32 X 32.} \\
\ubtestbadge{08} & \textcolor{zGoldDeep!78!black}{\texttt{\bfseries Sprite4\_\allowbreak{}Sprite\-Mode\_\allowbreak{}Is\-Multiple}}\enspace\textcolor{zText}{Checks that sprite 4 sprite mode is multiple.} \\
\ubtestbadge{09} & \textcolor{zGoldDeep!78!black}{\texttt{\bfseries Sprite4\_\allowbreak{}Sub\-Sprite\-Count\_\allowbreak{}Is165}}\enspace\textcolor{zText}{Checks that sprite 4 sub sprite count is 165.} \\
\ubtestbadge{10} & \textcolor{zGoldDeep!78!black}{\texttt{\bfseries Sprite5\_\allowbreak{}Sprite\-Mode\_\allowbreak{}Is\-Multiple}}\enspace\textcolor{zText}{Checks that sprite 5 sprite mode is multiple.} \\
\ubtestbadge{11} & \textcolor{zGoldDeep!78!black}{\texttt{\bfseries Sprite5\_\allowbreak{}Sub\-Sprite\-Count\_\allowbreak{}Is0}}\enspace\textcolor{zText}{Checks that sprite 5 sub sprite count is 0.} \\
\noalign{\vskip 2pt}
\multicolumn{2}{@{}p{14.4cm}@{}}{\textcolor{zRockLite!42}{\rule{\linewidth}{0.3pt}}}\\[3pt]
\multicolumn{2}{@{}p{14.4cm}@{}}{\ubtaskhead{zChampion}{0023-UIRedoBench}{GoldenBench UI Recreation}{zGoldDeep}{Medium}{9}}\\*[2pt]
\ubtestbadge{01} & \textcolor{zGoldDeep!78!black}{\texttt{\bfseries All\-Required\-Objects\_\allowbreak{}Are\-Present\-With\-Expected\-Components}}\enspace\textcolor{zText}{Checks that all required objects are present with expected components.} \\
\ubtestbadge{02} & \textcolor{zGoldDeep!78!black}{\texttt{\bfseries All\-Required\-Objects\_\allowbreak{}Use\-Correct\-Sprites}}\enspace\textcolor{zText}{Checks that all required objects use correct sprites.} \\
\ubtestbadge{03} & \textcolor{zGoldDeep!78!black}{\texttt{\bfseries Backgrounds\_\allowbreak{}Are\-Flush\-And\-Fill\-Expected\-Edges}}\enspace\textcolor{zText}{Checks that backgrounds are flush and fill expected edges.} \\
\ubtestbadge{04} & \textcolor{zGoldDeep!78!black}{\texttt{\bfseries Character\-Buttons\_\allowbreak{}Have\-Expected\-Interactability\-And\-Selection\-Effect}}\enspace\textcolor{zText}{Checks that character buttons have expected interactability and selection effect.} \\
\ubtestbadge{05} & \textcolor{zGoldDeep!78!black}{\texttt{\bfseries Elements\_\allowbreak{}Match\-Golden\-Bench\-Layout\-Within\-Tolerance}}\enspace\textcolor{zText}{Checks that elements match golden bench layout within tolerance.} \\
\ubtestbadge{06} & \textcolor{zGoldDeep!78!black}{\texttt{\bfseries Foreground\-Elements\_\allowbreak{}Do\-Not\-Exceed\-Reference\-Screen}}\enspace\textcolor{zText}{Checks that foreground elements do not exceed reference screen.} \\
\ubtestbadge{07} & \textcolor{zGoldDeep!78!black}{\texttt{\bfseries Foreground\-Elements\_\allowbreak{}Do\-Not\-Overlap\-Each\-Other}}\enspace\textcolor{zText}{Checks that foreground elements do not overlap each other.} \\
\ubtestbadge{08} & \textcolor{zGoldDeep!78!black}{\texttt{\bfseries Navigation\-Buttons\_\allowbreak{}Activate\-Their\-Panels}}\enspace\textcolor{zText}{Checks that navigation buttons activate their panels.} \\
\ubtestbadge{09} & \textcolor{zGoldDeep!78!black}{\texttt{\bfseries Scene\_\allowbreak{}Has\-Character\-Select\-Canvas\-Configured\-For\-Reference\-Resolution}}\enspace\textcolor{zText}{Checks that scene has character select canvas configured for reference resolution.} \\
\noalign{\vskip 2pt}
\multicolumn{2}{@{}p{14.4cm}@{}}{\textcolor{zRockLite!42}{\rule{\linewidth}{0.3pt}}}\\[3pt]
\multicolumn{2}{@{}p{14.4cm}@{}}{\ubtaskhead{zChampion}{0049-UIDocument}{Recreate an Endless Runner with UI Toolkit}{zGoldDeep}{Medium}{18}}\\*[2pt]
\ubtestbadge{01} & \textcolor{zGoldDeep!78!black}{\texttt{\bfseries Default\-Scene\-Hosts\-AUi\-Toolkit\-Runner}}\enspace\textcolor{zText}{Checks that default scene hosts a UI toolkit runner.} \\
\ubtestbadge{02} & \textcolor{zGoldDeep!78!black}{\texttt{\bfseries Project\-Contains\-Authored\-Ui\-Toolkit\-Assets}}\enspace\textcolor{zText}{Checks that project contains authored UI toolkit assets.} \\
\ubtestbadge{03} & \textcolor{zGoldDeep!78!black}{\texttt{\bfseries Runner\-Sprite\-Sheet\-Keeps\-Original\-Pixel\-Dimensions\-For\-Cropping}}\enspace\textcolor{zText}{Checks that runner sprite sheet keeps original pixel dimensions for cropping.} \\
\ubtestbadge{04} & \textcolor{zGoldDeep!78!black}{\texttt{\bfseries Collision\-Shows\-Game\-Over}}\enspace\textcolor{zText}{Checks that collision shows game over.} \\
\ubtestbadge{05} & \textcolor{zGoldDeep!78!black}{\texttt{\bfseries Down\-Arrow\-Changes\-The\-Runner\-Into\-ADucking\-Pose}}\enspace\textcolor{zText}{Checks that down arrow changes the runner character into a ducking pose.} \\
\ubtestbadge{06} & \textcolor{zGoldDeep!78!black}{\texttt{\bfseries Game\-Over\-Ui\-Is\-Hidden\-Before\-Crash}}\enspace\textcolor{zText}{Checks that game over UI is hidden before crash.} \\
\ubtestbadge{07} & \textcolor{zGoldDeep!78!black}{\texttt{\bfseries Initial\-Surface\-Shows\-Multi\-Digit\-Score}}\enspace\textcolor{zText}{Checks that initial surface shows multi digit score.} \\
\ubtestbadge{08} & \textcolor{zGoldDeep!78!black}{\texttt{\bfseries Initial\-Surface\-Shows\-Repeating\-Ground}}\enspace\textcolor{zText}{Checks that initial surface shows repeating ground.} \\
\ubtestbadge{09} & \textcolor{zGoldDeep!78!black}{\texttt{\bfseries Initial\-Surface\-Shows\-Runner\-Character}}\enspace\textcolor{zText}{Checks that the initial surface shows the runner character.} \\
\ubtestbadge{10} & \textcolor{zGoldDeep!78!black}{\texttt{\bfseries Repeated\-Down\-Arrow\-Keeps\-The\-Runner\-Ducking\-Until\-Release}}\enspace\textcolor{zText}{Checks that repeated down arrow keeps the runner character ducking until release.} \\
\ubtestbadge{11} & \textcolor{zGoldDeep!78!black}{\texttt{\bfseries Running\-Game\-Spawns\-Obstacles}}\enspace\textcolor{zText}{Checks that running game spawns obstacles.} \\
\ubtestbadge{12} & \textcolor{zGoldDeep!78!black}{\texttt{\bfseries Running\-Speed\-Tracks\-Elapsed\-Time\-At\-High\-Frame\-Rates}}\enspace\textcolor{zText}{Checks that running speed tracks elapsed time at high frame rates.} \\
\ubtestbadge{13} & \textcolor{zGoldDeep!78!black}{\texttt{\bfseries Runtime\-Loads\-AUi\-Toolkit\-Document}}\enspace\textcolor{zText}{Checks that runtime behavior loads a UI toolkit document.} \\
\ubtestbadge{14} & \textcolor{zGoldDeep!78!black}{\texttt{\bfseries Space\-Key\-Starts\-Ground\-Scrolling}}\enspace\textcolor{zText}{Checks that space key starts ground scrolling.} \\
\ubtestbadge{15} & \textcolor{zGoldDeep!78!black}{\texttt{\bfseries Space\-Key\-Starts\-Jump}}\enspace\textcolor{zText}{Checks that the Space key starts a jump.} \\
\ubtestbadge{16} & \textcolor{zGoldDeep!78!black}{\texttt{\bfseries Space\-Restarts\-After\-Game\-Over}}\enspace\textcolor{zText}{Checks that the Space key restarts the game after Game Over.} \\
\ubtestbadge{17} & \textcolor{zGoldDeep!78!black}{\texttt{\bfseries Timed\-Jump\-Clears\-AGround\-Obstacle\-Without\-False\-Game\-Over}}\enspace\textcolor{zText}{Checks that timed jump clears a ground obstacle without false game over.} \\
\ubtestbadge{18} & \textcolor{zGoldDeep!78!black}{\texttt{\bfseries Waiting\-Surface\-Includes\-Cloud\-Visual}}\enspace\textcolor{zText}{Checks that waiting surface includes cloud visual.} \\
\noalign{\vskip 2pt}
\multicolumn{2}{@{}p{14.4cm}@{}}{\textcolor{zRockLite!42}{\rule{\linewidth}{0.3pt}}}\\[3pt]
\multicolumn{2}{@{}p{14.4cm}@{}}{\ubtaskhead{zChampion}{0015-MobileHudRefactor}{Mobile HUD Refactor}{zShrine}{Hard}{9}}\\*[2pt]
\ubtestbadge{01} & \textcolor{zGoldDeep!78!black}{\texttt{\bfseries Task01\_\allowbreak{}Mobile\-Touch\-Controls\-No\-Longer\-Draws\-With\-Imgui}}\enspace\textcolor{zText}{Checks that MobileTouchControls no longer draws the HUD with IMGUI.} \\
\ubtestbadge{02} & \textcolor{zGoldDeep!78!black}{\texttt{\bfseries Task02\_\allowbreak{}Aim\-Controller\-No\-Longer\-Draws\-Crosshair\-With\-Imgui}}\enspace\textcolor{zText}{Checks that the aim controller no longer draws the crosshair with IMGUI.} \\
\ubtestbadge{03} & \textcolor{zGoldDeep!78!black}{\texttt{\bfseries Task03\_\allowbreak{}Mobile\-Hud\-Uses\-Unity\-Ui\-Building\-Blocks}}\enspace\textcolor{zText}{Checks that mobile HUD uses unity UI building blocks.} \\
\ubtestbadge{04} & \textcolor{zGoldDeep!78!black}{\texttt{\bfseries Task04\_\allowbreak{}Touch\-Input\-Api\-Is\-Still\-Exposed\-To\-Gameplay}}\enspace\textcolor{zText}{Checks that touch input API is still exposed to gameplay.} \\
\ubtestbadge{05} & \textcolor{zGoldDeep!78!black}{\texttt{\bfseries Task05\_\allowbreak{}Raid\-Scene\-Still\-References\-Aim\-Controller}}\enspace\textcolor{zText}{Checks that raid scene still references aim controller.} \\
\ubtestbadge{06} & \textcolor{zGoldDeep!78!black}{\texttt{\bfseries Task06\_\allowbreak{}Runtime\-Creates\-Canvas\-Based\-Battle\-Hud}}\enspace\textcolor{zText}{Checks that runtime behavior creates canvas based battle HUD.} \\
\ubtestbadge{07} & \textcolor{zGoldDeep!78!black}{\texttt{\bfseries Task07\_\allowbreak{}Primary\-Controls\-Stay\-Inside\-Different\-Screens}}\enspace\textcolor{zText}{Checks that primary controls stay inside different screens.} \\
\ubtestbadge{08} & \textcolor{zGoldDeep!78!black}{\texttt{\bfseries Task08\_\allowbreak{}Primary\-Controls\-Do\-Not\-Overlap\-On\-Phone\-Layouts}}\enspace\textcolor{zText}{Checks that primary controls do not overlap on phone layouts.} \\
\ubtestbadge{09} & \textcolor{zGoldDeep!78!black}{\texttt{\bfseries Task09\_\allowbreak{}Crosshair\-Uses\-Ui\-Images\-Instead\-Of\-World\-Or\-Imgui\-Only}}\enspace\textcolor{zText}{Checks that the crosshair uses UI Images rather than world-space or IMGUI-only rendering.} \\
\noalign{\vskip 2pt}
\multicolumn{2}{@{}p{14.4cm}@{}}{\textcolor{zRockLite!42}{\rule{\linewidth}{0.3pt}}}\\[3pt]
\multicolumn{2}{@{}p{14.4cm}@{}}{\ubtaskhead{zChampion}{0062-UIToolkitBubbleDragon}{Build an Endless UI Toolkit Bubble Shooter}{zShrine}{Hard}{19}}\\*[2pt]
\ubtestbadge{01} & \textcolor{zGoldDeep!78!black}{\texttt{\bfseries Runtime\-Accumulates\-Pressure\-And\-Pushes\-Rows\-Down}}\enspace\textcolor{zText}{Checks that runtime behavior accumulates pressure and pushes rows down.} \\
\ubtestbadge{02} & \textcolor{zGoldDeep!78!black}{\texttt{\bfseries Runtime\-Aligns\-Trajectory\-Cannon\-And\-Aim\-Axis}}\enspace\textcolor{zText}{Checks that runtime behavior aligns trajectory cannon and aim axis.} \\
\ubtestbadge{03} & \textcolor{zGoldDeep!78!black}{\texttt{\bfseries Runtime\-Binds\-Scene\-UIDocument\-Without\-Runtime\-Ui\-Construction}}\enspace\textcolor{zText}{Checks that runtime behavior binds scene UIDocument without runtime UI construction.} \\
\ubtestbadge{04} & \textcolor{zGoldDeep!78!black}{\texttt{\bfseries Runtime\-Clears\-Three\-Or\-More\-Connected\-Same\-Color\-Bubbles}}\enspace\textcolor{zText}{Checks that runtime behavior clears three or more connected same color bubbles.} \\
\ubtestbadge{05} & \textcolor{zGoldDeep!78!black}{\texttt{\bfseries Runtime\-Drops\-Floating\-Bubbles\-After\-Clear}}\enspace\textcolor{zText}{Checks that runtime behavior drops floating bubbles after clear.} \\
\ubtestbadge{06} & \textcolor{zGoldDeep!78!black}{\texttt{\bfseries Runtime\-Implements\-Endless\-Difficulty\-And\-Bubble\-Color\-Progression}}\enspace\textcolor{zText}{Checks that runtime behavior implements endless difficulty and bubble color progression.} \\
\ubtestbadge{07} & \textcolor{zGoldDeep!78!black}{\texttt{\bfseries Runtime\-Places\-Shot\-Using\-Distance\-To\-Grid\-Cells}}\enspace\textcolor{zText}{Checks runtime behavior places shot using distance to grid cells.} \\
\ubtestbadge{08} & \textcolor{zGoldDeep!78!black}{\texttt{\bfseries Runtime\-Supports\-Mouse\-And\-Keyboard\-Aiming\-And\-Firing}}\enspace\textcolor{zText}{Checks that runtime behavior supports mouse and keyboard aiming and firing.} \\
\ubtestbadge{09} & \textcolor{zGoldDeep!78!black}{\texttt{\bfseries Runtime\-Tracks\-Score\-High\-Score\-And\-Persistence}}\enspace\textcolor{zText}{Checks that runtime behavior tracks score high score and persistence.} \\
\ubtestbadge{10} & \textcolor{zGoldDeep!78!black}{\texttt{\bfseries Runtime\-Uses\-Aligned\-Wall\-Boundaries\-And\-Bounce\-Reflection}}\enspace\textcolor{zText}{Checks that runtime behavior uses aligned wall boundaries and bounce reflection.} \\
\ubtestbadge{11} & \textcolor{zGoldDeep!78!black}{\texttt{\bfseries Scene\-Contains\-Editor\-Placed\-Responsive\-UIDocument}}\enspace\textcolor{zText}{Checks that scene contains editor placed responsive UIDocument.} \\
\ubtestbadge{12} & \textcolor{zGoldDeep!78!black}{\texttt{\bfseries Scene\-Has\-Runtime\-Controller\-Driving\-Scene\-UIDocument}}\enspace\textcolor{zText}{Checks that scene has runtime controller driving scene UIDocument.} \\
\ubtestbadge{13} & \textcolor{zGoldDeep!78!black}{\texttt{\bfseries Styles\-Constrain\-Sidebar\-Frame\-Cards\-And\-Padding}}\enspace\textcolor{zText}{Checks that styles constrain sidebar frame cards and padding.} \\
\ubtestbadge{14} & \textcolor{zGoldDeep!78!black}{\texttt{\bfseries Styles\-Prevent\-Sidebar\-Text\-And\-Content\-Overflow}}\enspace\textcolor{zText}{Checks that styles prevent sidebar text and content overflow.} \\
\ubtestbadge{15} & \textcolor{zGoldDeep!78!black}{\texttt{\bfseries Styles\-Reserve\-Vertical\-Play\-Space\-For\-Board\-And\-Sidebar}}\enspace\textcolor{zText}{Checks that styles reserve vertical play space for board and sidebar.} \\
\ubtestbadge{16} & \textcolor{zGoldDeep!78!black}{\texttt{\bfseries Visual\-Tree\-Contains\-Core\-Gameplay\-Surfaces}}\enspace\textcolor{zText}{Checks that visual tree contains core gameplay surfaces.} \\
\ubtestbadge{17} & \textcolor{zGoldDeep!78!black}{\texttt{\bfseries Visual\-Tree\-Keeps\-Next\-Bubble\-In\-Play\-Area\-And\-Out\-Of\-Sidebar}}\enspace\textcolor{zText}{Checks that visual tree keeps next bubble in play area and out of sidebar.} \\
\ubtestbadge{18} & \textcolor{zGoldDeep!78!black}{\texttt{\bfseries Visual\-Tree\-Places\-Launcher\-Inside\-Play\-Area}}\enspace\textcolor{zText}{Checks that visual tree places launcher inside play area.} \\
\ubtestbadge{19} & \textcolor{zGoldDeep!78!black}{\texttt{\bfseries Visual\-Tree\-Provides\-Game\-Over\-Results\-And\-Restart\-Entry}}\enspace\textcolor{zText}{Checks that visual tree provides game over results and restart entry.} \\
\noalign{\vskip 2pt}
\multicolumn{2}{@{}p{14.4cm}@{}}{\textcolor{zRockLite!42}{\rule{\linewidth}{0.3pt}}}\\[3pt]
\multicolumn{2}{@{}p{14.4cm}@{}}{\ubtaskhead{zChampion}{0078-JetHUD}{Build a First-Person Jet Cockpit HUD}{zShrine}{Hard}{9}}\\*[2pt]
\ubtestbadge{01} & \textcolor{zGoldDeep!78!black}{\texttt{\bfseries Module01\_\allowbreak{}Hud\-Is\-Rendered\-On\-Cockpit\-Screen\-And\-Hidden\-In\-Third\-Person}}\enspace\textcolor{zText}{Checks that the HUD is rendered on the cockpit Screen in first person and completely hidden in third person.} \\
\ubtestbadge{02} & \textcolor{zGoldDeep!78!black}{\texttt{\bfseries Module02\_\allowbreak{}Static\-Layer\-Visually\-Matches\-Reference}}\enspace\textcolor{zText}{Compares the static HUD layer with the supplied visual reference.} \\
\ubtestbadge{03} & \textcolor{zGoldDeep!78!black}{\texttt{\bfseries Module03\_\allowbreak{}Airspeed\-Visual\-Moves\-And\-Changes\-With\-Rigidbody\-Speed}}\enspace\textcolor{zText}{Checks that the airspeed readout and scale move in response to Rigidbody speed.} \\
\ubtestbadge{04} & \textcolor{zGoldDeep!78!black}{\texttt{\bfseries Module04\_\allowbreak{}Altitude\-Visual\-Moves\-And\-Changes\-With\-World\-Height}}\enspace\textcolor{zText}{Checks that the altitude readout and scale move in response to world height.} \\
\ubtestbadge{05} & \textcolor{zGoldDeep!78!black}{\texttt{\bfseries Module05\_\allowbreak{}Attitude\-Visual\-Tracks\-Pitch\-Roll\-And\-Inversion}}\enspace\textcolor{zText}{Checks that the attitude display tracks pitch, roll, and inverted flight while keeping labels readable.} \\
\ubtestbadge{06} & \textcolor{zGoldDeep!78!black}{\texttt{\bfseries Module06\_\allowbreak{}World\-Horizon\-Visually\-Stays\-Aligned\-In\-Compound\-Attitude}}\enspace\textcolor{zText}{Checks that the separate world-horizon display remains aligned during combined pitch and roll.} \\
\ubtestbadge{07} & \textcolor{zGoldDeep!78!black}{\texttt{\bfseries Module07\_\allowbreak{}Heading\-Visual\-Scrolls\-Continuously\-And\-Wraps\-North}}\enspace\textcolor{zText}{Checks that the heading scale scrolls continuously and wraps correctly across north.} \\
\ubtestbadge{08} & \textcolor{zGoldDeep!78!black}{\texttt{\bfseries Module08\_\allowbreak{}Delayed\-Impact\-Visual\-Uses\-Point\-Two\-Second\-History}}\enspace\textcolor{zText}{Checks that the delayed impact marker responds to approximately 0.2 seconds of aim history.} \\
\ubtestbadge{09} & \textcolor{zGoldDeep!78!black}{\texttt{\bfseries Module09\_\allowbreak{}Complete\-Dynamic\-Composition\-Visually\-Matches\-Golden\-Style}}\enspace\textcolor{zText}{Compares the complete running HUD composition with the golden visual style.} \\
\noalign{\vskip 2pt}
\multicolumn{2}{@{}p{14.4cm}@{}}{\textcolor{zRockLite!42}{\rule{\linewidth}{0.3pt}}}\\[3pt]
\noalign{\vskip 3pt}
\multicolumn{2}{@{}p{14.4cm}@{}}{\ubcategory{zDawn}{Animation \& Animator}{12 tasks \textperiodcentered\ 179 milestone tests}}\\*[4pt]
\multicolumn{2}{@{}p{14.4cm}@{}}{\ubtaskhead{zDawn}{0000-AutoDoor}{Build an Automatic Opening and Closing Door Scene (AutoDoor)}{zGoldDeep}{Easy}{29}}\\*[2pt]
\ubtestbadge{01} & \textcolor{zGoldDeep!78!black}{\texttt{\bfseries Component\_\allowbreak{}Animator\_\allowbreak{}Has\-Controller}}\enspace\textcolor{zText}{Checks that the Animator component has an assigned controller.} \\
\ubtestbadge{02} & \textcolor{zGoldDeep!78!black}{\texttt{\bfseries Component\_\allowbreak{}Door\-Root\-Or\-Door\_\allowbreak{}Has\-Animator}}\enspace\textcolor{zText}{Checks that door root or door has Animator.} \\
\ubtestbadge{03} & \textcolor{zGoldDeep!78!black}{\texttt{\bfseries Component\_\allowbreak{}Main\-Camera\_\allowbreak{}Has\-Camera\-Component}}\enspace\textcolor{zText}{Checks that main camera has camera component.} \\
\ubtestbadge{04} & \textcolor{zGoldDeep!78!black}{\texttt{\bfseries Geometry\_\allowbreak{}Door\-Root\_\allowbreak{}Pivot\-Is\-Near\-Door\-Edge}}\enspace\textcolor{zText}{Checks that geometry door root pivot is near door edge.} \\
\ubtestbadge{05} & \textcolor{zGoldDeep!78!black}{\texttt{\bfseries Hierarchy\_\allowbreak{}Door\-Frame\_\allowbreak{}Left\_\allowbreak{}Is\-Child\-Of\_\allowbreak{}Door\-Frame}}\enspace\textcolor{zText}{Checks that door frame left is child of door frame.} \\
\ubtestbadge{06} & \textcolor{zGoldDeep!78!black}{\texttt{\bfseries Hierarchy\_\allowbreak{}Door\-Frame\_\allowbreak{}Right\_\allowbreak{}Is\-Child\-Of\_\allowbreak{}Door\-Frame}}\enspace\textcolor{zText}{Checks that door frame right is child of door frame.} \\
\ubtestbadge{07} & \textcolor{zGoldDeep!78!black}{\texttt{\bfseries Hierarchy\_\allowbreak{}Door\-Frame\_\allowbreak{}Top\_\allowbreak{}Is\-Child\-Of\_\allowbreak{}Door\-Frame}}\enspace\textcolor{zText}{Checks that door frame top is child of door frame.} \\
\ubtestbadge{08} & \textcolor{zGoldDeep!78!black}{\texttt{\bfseries Hierarchy\_\allowbreak{}Door\_\allowbreak{}Is\-Descendant\-Of\_\allowbreak{}Door\-Root}}\enspace\textcolor{zText}{Checks that door is descendant of door root.} \\
\ubtestbadge{09} & \textcolor{zGoldDeep!78!black}{\texttt{\bfseries Primitive\_\allowbreak{}Door\-Frame\_\allowbreak{}Left\_\allowbreak{}Uses\-Cube}}\enspace\textcolor{zText}{Checks that door frame left uses the cube primitive.} \\
\ubtestbadge{10} & \textcolor{zGoldDeep!78!black}{\texttt{\bfseries Primitive\_\allowbreak{}Door\-Frame\_\allowbreak{}Right\_\allowbreak{}Uses\-Cube}}\enspace\textcolor{zText}{Checks that door frame right uses the cube primitive.} \\
\ubtestbadge{11} & \textcolor{zGoldDeep!78!black}{\texttt{\bfseries Primitive\_\allowbreak{}Door\-Frame\_\allowbreak{}Top\_\allowbreak{}Uses\-Cube}}\enspace\textcolor{zText}{Checks that door frame top uses the cube primitive.} \\
\ubtestbadge{12} & \textcolor{zGoldDeep!78!black}{\texttt{\bfseries Primitive\_\allowbreak{}Door\_\allowbreak{}Uses\-Cube}}\enspace\textcolor{zText}{Checks that door uses the cube primitive.} \\
\ubtestbadge{13} & \textcolor{zGoldDeep!78!black}{\texttt{\bfseries Primitive\_\allowbreak{}Floor\_\allowbreak{}Uses\-Plane}}\enspace\textcolor{zText}{Checks that floor uses the plane primitive.} \\
\ubtestbadge{14} & \textcolor{zGoldDeep!78!black}{\texttt{\bfseries Scene\_\allowbreak{}Door\-Frame\_\allowbreak{}Exists}}\enspace\textcolor{zText}{Checks that door frame exists in the scene.} \\
\ubtestbadge{15} & \textcolor{zGoldDeep!78!black}{\texttt{\bfseries Scene\_\allowbreak{}Door\-Frame\_\allowbreak{}Left\_\allowbreak{}Exists}}\enspace\textcolor{zText}{Checks that door frame left exists in the scene.} \\
\ubtestbadge{16} & \textcolor{zGoldDeep!78!black}{\texttt{\bfseries Scene\_\allowbreak{}Door\-Frame\_\allowbreak{}Right\_\allowbreak{}Exists}}\enspace\textcolor{zText}{Checks that door frame right exists in the scene.} \\
\ubtestbadge{17} & \textcolor{zGoldDeep!78!black}{\texttt{\bfseries Scene\_\allowbreak{}Door\-Frame\_\allowbreak{}Top\_\allowbreak{}Exists}}\enspace\textcolor{zText}{Checks that door frame top exists in the scene.} \\
\ubtestbadge{18} & \textcolor{zGoldDeep!78!black}{\texttt{\bfseries Scene\_\allowbreak{}Door\-Root\_\allowbreak{}Exists}}\enspace\textcolor{zText}{Checks that door root exists in the scene.} \\
\ubtestbadge{19} & \textcolor{zGoldDeep!78!black}{\texttt{\bfseries Scene\_\allowbreak{}Door\_\allowbreak{}Exists}}\enspace\textcolor{zText}{Checks that door exists in the scene.} \\
\ubtestbadge{20} & \textcolor{zGoldDeep!78!black}{\texttt{\bfseries Scene\_\allowbreak{}Floor\_\allowbreak{}Exists}}\enspace\textcolor{zText}{Checks that floor exists in the scene.} \\
\ubtestbadge{21} & \textcolor{zGoldDeep!78!black}{\texttt{\bfseries Scene\_\allowbreak{}Main\-Camera\_\allowbreak{}Exists}}\enspace\textcolor{zText}{Checks that main camera exists in the scene.} \\
\ubtestbadge{22} & \textcolor{zGoldDeep!78!black}{\texttt{\bfseries Animation\_\allowbreak{}Door\-Root\_\allowbreak{}Position\-Is\-Stable}}\enspace\textcolor{zText}{Checks that animation door root position is stable.} \\
\ubtestbadge{23} & \textcolor{zGoldDeep!78!black}{\texttt{\bfseries Animation\_\allowbreak{}Has\-Both\-Opening\-And\-Closing\-Rotation}}\enspace\textcolor{zText}{Checks that animation has both opening and closing rotation.} \\
\ubtestbadge{24} & \textcolor{zGoldDeep!78!black}{\texttt{\bfseries Animation\_\allowbreak{}No\-Instant\-Jumps}}\enspace\textcolor{zText}{Checks that the door animation has no discontinuous rotation jumps.} \\
\ubtestbadge{25} & \textcolor{zGoldDeep!78!black}{\texttt{\bfseries Animation\_\allowbreak{}Rotation\-Range\_\allowbreak{}Greater\-Than30\-Degrees}}\enspace\textcolor{zText}{Checks that the door's animated rotation range exceeds 30 degrees.} \\
\ubtestbadge{26} & \textcolor{zGoldDeep!78!black}{\texttt{\bfseries Animation\_\allowbreak{}Rotation\-Range\_\allowbreak{}Not\-Excessive}}\enspace\textcolor{zText}{Checks that the door's animated rotation range is not excessive.} \\
\ubtestbadge{27} & \textcolor{zGoldDeep!78!black}{\texttt{\bfseries Animation\_\allowbreak{}Starts\-Automatically\_\allowbreak{}Within\-Grace\-Period}}\enspace\textcolor{zText}{Checks that animation starts automatically within grace period.} \\
\ubtestbadge{28} & \textcolor{zGoldDeep!78!black}{\texttt{\bfseries Camera\_\allowbreak{}Door\-Is\-Substantially\-Visible}}\enspace\textcolor{zText}{Checks that camera door is substantially visible.} \\
\ubtestbadge{29} & \textcolor{zGoldDeep!78!black}{\texttt{\bfseries Runtime\_\allowbreak{}Door\-Root\_\allowbreak{}Exists}}\enspace\textcolor{zText}{Checks that runtime behavior door root exists.} \\
\noalign{\vskip 2pt}
\multicolumn{2}{@{}p{14.4cm}@{}}{\textcolor{zRockLite!42}{\rule{\linewidth}{0.3pt}}}\\[3pt]
\multicolumn{2}{@{}p{14.4cm}@{}}{\ubtaskhead{zDawn}{0006-TankTrackAnimation}{Tank Track Animation}{zGoldDeep}{Easy}{2}}\\*[2pt]
\ubtestbadge{01} & \textcolor{zGoldDeep!78!black}{\texttt{\bfseries Tank\-Track\-Animator\_\allowbreak{}Runtime\-Scene\_\allowbreak{}Rotates\-Every\-Wheel\-Mesh}}\enspace\textcolor{zText}{Checks that tank track Animator runtime scene rotates every wheel mesh.} \\
\ubtestbadge{02} & \textcolor{zGoldDeep!78!black}{\texttt{\bfseries Tank\-Track\-Animator\_\allowbreak{}Runtime\-Scene\_\allowbreak{}Scrolls\-Track\-Uv\-Periodically}}\enspace\textcolor{zText}{Checks that tank track Animator runtime scene scrolls track UV periodically.} \\
\noalign{\vskip 2pt}
\multicolumn{2}{@{}p{14.4cm}@{}}{\textcolor{zRockLite!42}{\rule{\linewidth}{0.3pt}}}\\[3pt]
\multicolumn{2}{@{}p{14.4cm}@{}}{\ubtaskhead{zDawn}{0008-SwordsmanAttack}{Swordsman Attack}{zGoldDeep}{Easy}{4}}\\*[2pt]
\ubtestbadge{01} & \textcolor{zGoldDeep!78!black}{\texttt{\bfseries Swordsman\_\allowbreak{}Attack\_\allowbreak{}Animation\_\allowbreak{}Calls\_\allowbreak{}Enable\_\allowbreak{}Attack\_\allowbreak{}Window\_\allowbreak{}Event}}\enspace\textcolor{zText}{Checks that swordsman attack animation calls enable attack window event.} \\
\ubtestbadge{02} & \textcolor{zGoldDeep!78!black}{\texttt{\bfseries Swordsman\_\allowbreak{}Attack\_\allowbreak{}Animation\_\allowbreak{}Produces\_\allowbreak{}Diagonal\_\allowbreak{}Sword\_\allowbreak{}Slash}}\enspace\textcolor{zText}{Checks that swordsman attack animation produces diagonal sword slash.} \\
\ubtestbadge{03} & \textcolor{zGoldDeep!78!black}{\texttt{\bfseries Swordsman\_\allowbreak{}Enable\_\allowbreak{}Attack\_\allowbreak{}Window\_\allowbreak{}Event\_\allowbreak{}Happens\_\allowbreak{}During\_\allowbreak{}Mid\-Swing}}\enspace\textcolor{zText}{Checks that swordsman enable attack window event happens during mid swing.} \\
\ubtestbadge{04} & \textcolor{zGoldDeep!78!black}{\texttt{\bfseries Swordsman\_\allowbreak{}Hits\_\allowbreak{}Wooden\_\allowbreak{}Stake\_\allowbreak{}While\_\allowbreak{}Attack\_\allowbreak{}Window\_\allowbreak{}Is\_\allowbreak{}Enabled}}\enspace\textcolor{zText}{Checks that swordsman hits wooden stake while attack window is enabled.} \\
\noalign{\vskip 2pt}
\multicolumn{2}{@{}p{14.4cm}@{}}{\textcolor{zRockLite!42}{\rule{\linewidth}{0.3pt}}}\\[3pt]
\multicolumn{2}{@{}p{14.4cm}@{}}{\ubtaskhead{zDawn}{0009-AnimationEvents}{Animation Events}{zGoldDeep}{Medium}{9}}\\*[2pt]
\ubtestbadge{01} & \textcolor{zGoldDeep!78!black}{\texttt{\bfseries Animation\-Events\-Name\-Set\-Up\-Test}}\enspace\textcolor{zText}{Checks that the two Walk\_N animation events call StartEvent and HalfEvent in order.} \\
\ubtestbadge{02} & \textcolor{zGoldDeep!78!black}{\texttt{\bfseries Animation\-Events\-Params\-Set\-Up\-Test1}}\enspace\textcolor{zText}{Checks that the integer parameters are 1 for StartEvent and 0 for HalfEvent.} \\
\ubtestbadge{03} & \textcolor{zGoldDeep!78!black}{\texttt{\bfseries Animation\-Events\-Params\-Set\-Up\-Test2}}\enspace\textcolor{zText}{Checks that the float parameters are 0 for StartEvent and 1.2 for HalfEvent.} \\
\ubtestbadge{04} & \textcolor{zGoldDeep!78!black}{\texttt{\bfseries Animation\-Events\-Params\-Set\-Up\-Test3}}\enspace\textcolor{zText}{Checks that the string parameters are empty for StartEvent and Hello World! for HalfEvent.} \\
\ubtestbadge{05} & \textcolor{zGoldDeep!78!black}{\texttt{\bfseries Animation\-Events\-Params\-Set\-Up\-Test4}}\enspace\textcolor{zText}{Checks that neither animation event has an object-reference parameter.} \\
\ubtestbadge{06} & \textcolor{zGoldDeep!78!black}{\texttt{\bfseries Animation\-Events\-Progress\-Set\-Up\-Test}}\enspace\textcolor{zText}{Checks that StartEvent occurs at normalized time 0 and HalfEvent at normalized time 0.5.} \\
\ubtestbadge{07} & \textcolor{zGoldDeep!78!black}{\texttt{\bfseries Animation\-Events\-Call\-Back\-Test1}}\enspace\textcolor{zText}{Checks that StartEvent invokes its callback and sets intParam to 1 without changing the other fields.} \\
\ubtestbadge{08} & \textcolor{zGoldDeep!78!black}{\texttt{\bfseries Animation\-Events\-Call\-Back\-Test2}}\enspace\textcolor{zText}{Checks that HalfEvent invokes its callback and sets floatParam to 1.2 and stringParam to Hello World!.} \\
\ubtestbadge{09} & \textcolor{zGoldDeep!78!black}{\texttt{\bfseries Animator\-Props\-Test}}\enspace\textcolor{zText}{Checks that Armature has an enabled Animator with a controller, Always Animate culling, and speed 1.0.} \\
\noalign{\vskip 2pt}
\multicolumn{2}{@{}p{14.4cm}@{}}{\textcolor{zRockLite!42}{\rule{\linewidth}{0.3pt}}}\\[3pt]
\multicolumn{2}{@{}p{14.4cm}@{}}{\ubtaskhead{zDawn}{0029-AnimatorLayer}{Layered Character Animation}{zGoldDeep}{Medium}{16}}\\*[2pt]
\ubtestbadge{01} & \textcolor{zGoldDeep!78!black}{\texttt{\bfseries Controller\_\allowbreak{}Has\-Two\-Layers}}\enspace\textcolor{zText}{Checks that controller has two layers.} \\
\ubtestbadge{02} & \textcolor{zGoldDeep!78!black}{\texttt{\bfseries Layer0\_\allowbreak{}Has\-One\-State}}\enspace\textcolor{zText}{Checks that Animator Layer 0 has exactly one state.} \\
\ubtestbadge{03} & \textcolor{zGoldDeep!78!black}{\texttt{\bfseries Layer0\_\allowbreak{}State\_\allowbreak{}Uses\-Clip\-At\-Specified\-Path}}\enspace\textcolor{zText}{Checks that layer 0 state uses clip at specified path.} \\
\ubtestbadge{04} & \textcolor{zGoldDeep!78!black}{\texttt{\bfseries Layer1\_\allowbreak{}Avatar\-Mask\_\allowbreak{}Only\-Upper\-Body\-Active}}\enspace\textcolor{zText}{Checks that layer 1 avatar mask only upper body is active.} \\
\ubtestbadge{05} & \textcolor{zGoldDeep!78!black}{\texttt{\bfseries Layer1\_\allowbreak{}Blending\-Mode\-Is\-Override}}\enspace\textcolor{zText}{Checks that layer 1 blending mode is override.} \\
\ubtestbadge{06} & \textcolor{zGoldDeep!78!black}{\texttt{\bfseries Layer1\_\allowbreak{}Default\-State\_\allowbreak{}Is\-Empty\-State}}\enspace\textcolor{zText}{Checks that layer 1 default state is empty state.} \\
\ubtestbadge{07} & \textcolor{zGoldDeep!78!black}{\texttt{\bfseries Layer1\_\allowbreak{}Has\-Empty\-State}}\enspace\textcolor{zText}{Checks that Animator Layer 1 contains an empty state.} \\
\ubtestbadge{08} & \textcolor{zGoldDeep!78!black}{\texttt{\bfseries Layer1\_\allowbreak{}Has\-State\-With\-Clip\-At\-Specified\-Path}}\enspace\textcolor{zText}{Checks that layer 1 has state with clip at specified path.} \\
\ubtestbadge{09} & \textcolor{zGoldDeep!78!black}{\texttt{\bfseries Layer1\_\allowbreak{}Has\-Two\-States}}\enspace\textcolor{zText}{Checks that Animator Layer 1 has exactly two states.} \\
\ubtestbadge{10} & \textcolor{zGoldDeep!78!black}{\texttt{\bfseries Layer1\_\allowbreak{}Uses\-Avatar\-Mask}}\enspace\textcolor{zText}{Checks that Animator Layer 1 uses an AvatarMask.} \\
\ubtestbadge{11} & \textcolor{zGoldDeep!78!black}{\texttt{\bfseries Layer1\_\allowbreak{}Weight\-Is\-One}}\enspace\textcolor{zText}{Checks that Animator Layer 1 has a default weight of 1.} \\
\ubtestbadge{12} & \textcolor{zGoldDeep!78!black}{\texttt{\bfseries Bone\-Transform\-Test\_\allowbreak{}Aiming\-Walk}}\enspace\textcolor{zText}{Checks the character's bone transforms against the reference pose while walking and aiming.} \\
\ubtestbadge{13} & \textcolor{zGoldDeep!78!black}{\texttt{\bfseries Bone\-Transform\-Test\_\allowbreak{}Back2\-Walk}}\enspace\textcolor{zText}{Checks the character's bone transforms against the reference after switching from aiming back to walking.} \\
\ubtestbadge{14} & \textcolor{zGoldDeep!78!black}{\texttt{\bfseries Bone\-Transform\-Test\_\allowbreak{}Walk}}\enspace\textcolor{zText}{Checks the character's bone transforms against the reference pose while walking without aiming.} \\
\ubtestbadge{15} & \textcolor{zGoldDeep!78!black}{\texttt{\bfseries Full\-Body\-Walk\-State\-Test}}\enspace\textcolor{zText}{Checks that both Animator layers loop at full weight, with Walk\_N on the base layer and an empty upper layer.} \\
\ubtestbadge{16} & \textcolor{zGoldDeep!78!black}{\texttt{\bfseries Multi\-Layer\-State\-Test}}\enspace\textcolor{zText}{Checks that pressing E keeps Walk\_N on the base layer while Idle Aiming loops on the upper layer.} \\
\noalign{\vskip 2pt}
\multicolumn{2}{@{}p{14.4cm}@{}}{\textcolor{zRockLite!42}{\rule{\linewidth}{0.3pt}}}\\[3pt]
\multicolumn{2}{@{}p{14.4cm}@{}}{\ubtaskhead{zDawn}{0035-HumanoidIK}{Humanoid IK}{zGoldDeep}{Medium}{13}}\\*[2pt]
\ubtestbadge{01} & \textcolor{zGoldDeep!78!black}{\texttt{\bfseries Test\-Character\-Comp}}\enspace\textcolor{zText}{Checks that Armature has an Animator with a controller, a valid Avatar, and Always Animate culling.} \\
\ubtestbadge{02} & \textcolor{zGoldDeep!78!black}{\texttt{\bfseries Animator\-Is\-Playing\-Idle\-Clip\-From\-Resources}}\enspace\textcolor{zText}{Checks that Armature actively plays the required idle AnimationClip from Resources.} \\
\ubtestbadge{03} & \textcolor{zGoldDeep!78!black}{\texttt{\bfseries Left\-Foot\-IKDynamic\-Test}}\enspace\textcolor{zText}{Checks that the left foot follows position and rotation changes made to its IK target.} \\
\ubtestbadge{04} & \textcolor{zGoldDeep!78!black}{\texttt{\bfseries Left\-Foot\-IKStatic\-Test}}\enspace\textcolor{zText}{Checks that the left foot matches the left-foot IK target position.} \\
\ubtestbadge{05} & \textcolor{zGoldDeep!78!black}{\texttt{\bfseries Left\-Hand\-IKDynamic\-Test}}\enspace\textcolor{zText}{Checks that the left hand follows position and rotation changes made to its IK target.} \\
\ubtestbadge{06} & \textcolor{zGoldDeep!78!black}{\texttt{\bfseries Left\-Hand\-IKStatic\-Test}}\enspace\textcolor{zText}{Checks that the left hand matches the left-hand IK target position.} \\
\ubtestbadge{07} & \textcolor{zGoldDeep!78!black}{\texttt{\bfseries Look\-At\-Cube\-Test}}\enspace\textcolor{zText}{Checks that the character's head keeps facing the Cube as the Cube moves.} \\
\ubtestbadge{08} & \textcolor{zGoldDeep!78!black}{\texttt{\bfseries Plane\-Trace\-Test}}\enspace\textcolor{zText}{Checks that the right foot projects onto and aligns with the Plane, then returns when the Plane moves away.} \\
\ubtestbadge{09} & \textcolor{zGoldDeep!78!black}{\texttt{\bfseries Right\-Foot\-IKDynamic\-Test}}\enspace\textcolor{zText}{Checks that the right foot follows position and rotation changes made to its IK target.} \\
\ubtestbadge{10} & \textcolor{zGoldDeep!78!black}{\texttt{\bfseries Right\-Foot\-IKStatic\-Test}}\enspace\textcolor{zText}{Checks that the right foot matches the right-foot IK target position.} \\
\ubtestbadge{11} & \textcolor{zGoldDeep!78!black}{\texttt{\bfseries Right\-Hand\-IKDynamic\-Test}}\enspace\textcolor{zText}{Checks that the right hand follows position and rotation changes made to its IK target.} \\
\ubtestbadge{12} & \textcolor{zGoldDeep!78!black}{\texttt{\bfseries Right\-Hand\-IKStatic\-Test}}\enspace\textcolor{zText}{Checks that the right hand matches the right-hand IK target position.} \\
\ubtestbadge{13} & \textcolor{zGoldDeep!78!black}{\texttt{\bfseries Sphere\-Hold\-Test}}\enspace\textcolor{zText}{Checks that the left hand tracks a reachable Sphere and returns to its regular IK target afterward.} \\
\noalign{\vskip 2pt}
\multicolumn{2}{@{}p{14.4cm}@{}}{\textcolor{zRockLite!42}{\rule{\linewidth}{0.3pt}}}\\[3pt]
\multicolumn{2}{@{}p{14.4cm}@{}}{\ubtaskhead{zDawn}{0041-ExcavatorAnimationClipDigDump}{Excavator AnimationClip Dig and Dump Task}{zGoldDeep}{Medium}{9}}\\*[2pt]
\ubtestbadge{01} & \textcolor{zGoldDeep!78!black}{\texttt{\bfseries Dig\-And\-Dump\-Clips\-Only\-Rotate\-Boom\-Arm\-And\-Bucket\-Pivots}}\enspace\textcolor{zText}{Checks that dig and dump clips only rotate boom arm and bucket pivots.} \\
\ubtestbadge{02} & \textcolor{zGoldDeep!78!black}{\texttt{\bfseries Final\-Pool\-Contains\-At\-Least\-Eight\-Cubes}}\enspace\textcolor{zText}{Checks that final pool contains at least eight cubes.} \\
\ubtestbadge{03} & \textcolor{zGoldDeep!78!black}{\texttt{\bfseries Final\-Pool\-Contains\-At\-Least\-Four\-Cubes}}\enspace\textcolor{zText}{Checks that final pool contains at least four cubes.} \\
\ubtestbadge{04} & \textcolor{zGoldDeep!78!black}{\texttt{\bfseries Final\-Pool\-Contains\-At\-Least\-Sixteen\-Cubes}}\enspace\textcolor{zText}{Checks that final pool contains at least sixteen cubes.} \\
\ubtestbadge{05} & \textcolor{zGoldDeep!78!black}{\texttt{\bfseries Final\-Pool\-Contains\-At\-Least\-Thirty\-Two\-Cubes}}\enspace\textcolor{zText}{Checks that final pool contains at least thirty two cubes.} \\
\ubtestbadge{06} & \textcolor{zGoldDeep!78!black}{\texttt{\bfseries Final\-Pool\-Contains\-At\-Least\-Twelve\-Cubes}}\enspace\textcolor{zText}{Checks that final pool contains at least twelve cubes.} \\
\ubtestbadge{07} & \textcolor{zGoldDeep!78!black}{\texttt{\bfseries Final\-Pool\-Contains\-At\-Least\-Twenty\-Cubes}}\enspace\textcolor{zText}{Checks that final pool contains at least twenty cubes.} \\
\ubtestbadge{08} & \textcolor{zGoldDeep!78!black}{\texttt{\bfseries Final\-Pool\-Contains\-At\-Least\-Twenty\-Eight\-Cubes}}\enspace\textcolor{zText}{Checks that final pool contains at least twenty eight cubes.} \\
\ubtestbadge{09} & \textcolor{zGoldDeep!78!black}{\texttt{\bfseries Final\-Pool\-Contains\-At\-Least\-Twenty\-Four\-Cubes}}\enspace\textcolor{zText}{Checks that final pool contains at least twenty four cubes.} \\
\noalign{\vskip 2pt}
\multicolumn{2}{@{}p{14.4cm}@{}}{\textcolor{zRockLite!42}{\rule{\linewidth}{0.3pt}}}\\[3pt]
\multicolumn{2}{@{}p{14.4cm}@{}}{\ubtaskhead{zDawn}{0046-AnimationScriptPlayable}{Animation Script Playable}{zGoldDeep}{Medium}{17}}\\*[2pt]
\ubtestbadge{01} & \textcolor{zGoldDeep!78!black}{\texttt{\bfseries Animator\-Comp\-Test}}\enspace\textcolor{zText}{Checks that the Animator uses Normal per-frame updates and Always Animate culling.} \\
\ubtestbadge{02} & \textcolor{zGoldDeep!78!black}{\texttt{\bfseries Script\-Public\-Prop\-Test}}\enspace\textcolor{zText}{Checks that the playable script exposes all required control fields with their specified default values.} \\
\ubtestbadge{03} & \textcolor{zGoldDeep!78!black}{\texttt{\bfseries Test\-Apply\-Root\-Motion}}\enspace\textcolor{zText}{Checks that applyRootMotion controls the Animator's root-motion setting.} \\
\ubtestbadge{04} & \textcolor{zGoldDeep!78!black}{\texttt{\bfseries Test\-Hip\-YOffset}}\enspace\textcolor{zText}{Checks that hipYOffset applies the requested vertical offset to the hips.} \\
\ubtestbadge{05} & \textcolor{zGoldDeep!78!black}{\texttt{\bfseries Test\-Idle2\-In\-Air\-State}}\enspace\textcolor{zText}{Checks the transition from idle to the inair state.} \\
\ubtestbadge{06} & \textcolor{zGoldDeep!78!black}{\texttt{\bfseries Test\-Idle\-State}}\enspace\textcolor{zText}{Checks that a speed below 0.1 produces the reference idle pose.} \\
\ubtestbadge{07} & \textcolor{zGoldDeep!78!black}{\texttt{\bfseries Test\-In\-Air2\-Idle\-State}}\enspace\textcolor{zText}{Checks the transition from the inair state back to idle.} \\
\ubtestbadge{08} & \textcolor{zGoldDeep!78!black}{\texttt{\bfseries Test\-In\-Air2\-Run\-State}}\enspace\textcolor{zText}{Checks the transition from the inair state back to run.} \\
\ubtestbadge{09} & \textcolor{zGoldDeep!78!black}{\texttt{\bfseries Test\-In\-Air2\-Walk\-State}}\enspace\textcolor{zText}{Checks the transition from the inair state back to walk.} \\
\ubtestbadge{10} & \textcolor{zGoldDeep!78!black}{\texttt{\bfseries Test\-In\-Air\-Root\-Motion\-Multiplier}}\enspace\textcolor{zText}{Checks that inAirRootMotionMultiplier scales airborne vertical root movement.} \\
\ubtestbadge{11} & \textcolor{zGoldDeep!78!black}{\texttt{\bfseries Test\-Max\-Forward\-Lean\-Degrees}}\enspace\textcolor{zText}{Checks that maxForwardLeanDegrees increases forward spine lean proportionally with movement speed.} \\
\ubtestbadge{12} & \textcolor{zGoldDeep!78!black}{\texttt{\bfseries Test\-Remove\-In\-Air\-Vertical\-Root\-Motion}}\enspace\textcolor{zText}{Checks that removeInAirVerticalRootMotion suppresses vertical root movement while airborne.} \\
\ubtestbadge{13} & \textcolor{zGoldDeep!78!black}{\texttt{\bfseries Test\-Run2\-In\-Air\-State}}\enspace\textcolor{zText}{Checks the transition from run to the inair state.} \\
\ubtestbadge{14} & \textcolor{zGoldDeep!78!black}{\texttt{\bfseries Test\-Run\-State}}\enspace\textcolor{zText}{Checks that a speed of 5 produces the reference run pose.} \\
\ubtestbadge{15} & \textcolor{zGoldDeep!78!black}{\texttt{\bfseries Test\-Transition\-Duration}}\enspace\textcolor{zText}{Checks that transitionDuration controls blending time before the target pose is reached.} \\
\ubtestbadge{16} & \textcolor{zGoldDeep!78!black}{\texttt{\bfseries Test\-Walk2\-In\-Air\-State}}\enspace\textcolor{zText}{Checks the transition from walk to the inair state.} \\
\ubtestbadge{17} & \textcolor{zGoldDeep!78!black}{\texttt{\bfseries Test\-Walk\-State}}\enspace\textcolor{zText}{Checks that a speed of 0.1 produces the reference walk pose.} \\
\noalign{\vskip 2pt}
\multicolumn{2}{@{}p{14.4cm}@{}}{\textcolor{zRockLite!42}{\rule{\linewidth}{0.3pt}}}\\[3pt]
\multicolumn{2}{@{}p{14.4cm}@{}}{\ubtaskhead{zDawn}{0055-HumanoidAnimatorController}{Humanoid Animator Blend Tree Controller}{zGoldDeep}{Medium}{30}}\\*[2pt]
\ubtestbadge{01} & \textcolor{zGoldDeep!78!black}{\texttt{\bfseries Any\-State\-Does\-Not\-Bypass\-Locomotion\-Into\-Action\-States}}\enspace\textcolor{zText}{Checks that any state does not bypass locomotion into action states.} \\
\ubtestbadge{02} & \textcolor{zGoldDeep!78!black}{\texttt{\bfseries Base\-Layer\-Has\-Required\-States}}\enspace\textcolor{zText}{Checks that base layer has required states.} \\
\ubtestbadge{03} & \textcolor{zGoldDeep!78!black}{\texttt{\bfseries Controller\-Has\-Expected\-Layers\-And\-Upper\-Body\-Defaults}}\enspace\textcolor{zText}{Checks that controller has expected layers and upper body defaults.} \\
\ubtestbadge{04} & \textcolor{zGoldDeep!78!black}{\texttt{\bfseries Crouch\-Blend\-Tree\-Contains\-Idle\-And\-Directional\-Walk\-Motions}}\enspace\textcolor{zText}{Checks that crouch blend tree contains idle and directional walk motions.} \\
\ubtestbadge{05} & \textcolor{zGoldDeep!78!black}{\texttt{\bfseries Crouch\-Locomotion\-Uses2\-DFreeform\-Move\-Blend\-Tree}}\enspace\textcolor{zText}{Checks that crouch locomotion uses 2 d freeform move blend tree.} \\
\ubtestbadge{06} & \textcolor{zGoldDeep!78!black}{\texttt{\bfseries Dive\-And\-Fall\-Clip\-Loop\-Settings\-Are\-Correct}}\enspace\textcolor{zText}{Checks that dive and fall clip loop settings are correct.} \\
\ubtestbadge{07} & \textcolor{zGoldDeep!78!black}{\texttt{\bfseries Dive\-State\-Uses\-Dive\-Clip\-And\-Has\-Safe\-Transitions}}\enspace\textcolor{zText}{Checks that dive state uses dive clip and has safe transitions.} \\
\ubtestbadge{08} & \textcolor{zGoldDeep!78!black}{\texttt{\bfseries Fall\-State\-Uses\-Fall\-Clip\-And\-Is\-Driven\-By\-Is\-Falling}}\enspace\textcolor{zText}{Checks that fall state uses fall clip and is driven by is falling.} \\
\ubtestbadge{09} & \textcolor{zGoldDeep!78!black}{\texttt{\bfseries Input\-Controller\-Is\-Enabled\-And\-Bound\-To\-Target\-Character}}\enspace\textcolor{zText}{Checks that the input controller is enabled and bound to the target character.} \\
\ubtestbadge{10} & \textcolor{zGoldDeep!78!black}{\texttt{\bfseries Input\-Controller\-Source\-Implements\-Required\-Controls\-Without\-Jump}}\enspace\textcolor{zText}{Checks that input controller source implements required controls without jump.} \\
\ubtestbadge{11} & \textcolor{zGoldDeep!78!black}{\texttt{\bfseries Jump\-Parameter\-And\-Jump\-Conditions\-Are\-Absent}}\enspace\textcolor{zText}{Checks that jump parameter and jump conditions are absent.} \\
\ubtestbadge{12} & \textcolor{zGoldDeep!78!black}{\texttt{\bfseries Required\-Animation\-Clips\-Were\-Extracted}}\enspace\textcolor{zText}{Checks that required animation clips were extracted.} \\
\ubtestbadge{13} & \textcolor{zGoldDeep!78!black}{\texttt{\bfseries Required\-Animator\-Parameters\-Have\-Expected\-Types}}\enspace\textcolor{zText}{Checks that required Animator parameters have expected types.} \\
\ubtestbadge{14} & \textcolor{zGoldDeep!78!black}{\texttt{\bfseries Scene\-Animator\-Uses\-Required\-Controller\-And\-Valid\-Avatar}}\enspace\textcolor{zText}{Checks that the scene Animator uses the required controller and a valid Avatar.} \\
\ubtestbadge{15} & \textcolor{zGoldDeep!78!black}{\texttt{\bfseries Scene\-Contains\-Configured\-Player\-Components}}\enspace\textcolor{zText}{Checks that scene contains configured player components.} \\
\ubtestbadge{16} & \textcolor{zGoldDeep!78!black}{\texttt{\bfseries Standing\-Blend\-Tree\-Contains\-Idle\-Walk\-And\-Run\-Motions}}\enspace\textcolor{zText}{Checks that standing blend tree contains idle walk and run motions.} \\
\ubtestbadge{17} & \textcolor{zGoldDeep!78!black}{\texttt{\bfseries Standing\-Locomotion\-Uses2\-DFreeform\-Move\-Blend\-Tree}}\enspace\textcolor{zText}{Checks that standing locomotion uses 2 d freeform move blend tree.} \\
\ubtestbadge{18} & \textcolor{zGoldDeep!78!black}{\texttt{\bfseries Upper\-Body\-Aim\-Uses\-Steady\-Aim\-Pose\-Without\-Speed\-Driver}}\enspace\textcolor{zText}{Checks that upper body aim uses steady aim pose without speed driver.} \\
\ubtestbadge{19} & \textcolor{zGoldDeep!78!black}{\texttt{\bfseries Upper\-Body\-Mask\-Excludes\-Lower\-Body\-Bones}}\enspace\textcolor{zText}{Checks that upper body mask excludes lower body bones.} \\
\ubtestbadge{20} & \textcolor{zGoldDeep!78!black}{\texttt{\bfseries Upper\-Body\-Mask\-Includes\-Required\-Upper\-Body\-Bones}}\enspace\textcolor{zText}{Checks that upper body mask includes required upper body bones.} \\
\ubtestbadge{21} & \textcolor{zGoldDeep!78!black}{\texttt{\bfseries Runtime\-Aim\-Layer\-Uses\-Stable\-Aim\-Pose\-And\-Bones\-Remain\-Steady}}\enspace\textcolor{zText}{Checks that runtime behavior aim layer uses stable aim pose and bones remain steady.} \\
\ubtestbadge{22} & \textcolor{zGoldDeep!78!black}{\texttt{\bfseries Runtime\-Animator\-Bindings\-Are\-Valid}}\enspace\textcolor{zText}{Checks that runtime behavior Animator bindings are valid.} \\
\ubtestbadge{23} & \textcolor{zGoldDeep!78!black}{\texttt{\bfseries Runtime\-Animator\-Parameters\-Are\-Available}}\enspace\textcolor{zText}{Checks that runtime behavior Animator parameters are available.} \\
\ubtestbadge{24} & \textcolor{zGoldDeep!78!black}{\texttt{\bfseries Runtime\-Crouch\-Forward\-Uses\-Crouch\-Locomotion}}\enspace\textcolor{zText}{Checks that runtime behavior crouch forward uses crouch locomotion.} \\
\ubtestbadge{25} & \textcolor{zGoldDeep!78!black}{\texttt{\bfseries Runtime\-Dive\-Trigger\-Plays\-Dive\-Then\-Returns}}\enspace\textcolor{zText}{Checks that runtime behavior dive trigger plays dive then returns.} \\
\ubtestbadge{26} & \textcolor{zGoldDeep!78!black}{\texttt{\bfseries Runtime\-Fall\-Bool\-Enters\-Fall\-State}}\enspace\textcolor{zText}{Checks that runtime behavior fall bool enters fall state.} \\
\ubtestbadge{27} & \textcolor{zGoldDeep!78!black}{\texttt{\bfseries Runtime\-Idle\-Does\-Not\-Enter\-Fall\-Or\-Aim\-Layer}}\enspace\textcolor{zText}{Checks that runtime behavior idle does not enter fall or aim layer.} \\
\ubtestbadge{28} & \textcolor{zGoldDeep!78!black}{\texttt{\bfseries Runtime\-Run\-And\-Diagonal\-Run\-Prefer\-Run\-Clips}}\enspace\textcolor{zText}{Checks that runtime behavior run and diagonal run prefer run clips.} \\
\ubtestbadge{29} & \textcolor{zGoldDeep!78!black}{\texttt{\bfseries Runtime\-Scene\-Loads\-Player\-And\-Required\-Components}}\enspace\textcolor{zText}{Checks that runtime behavior scene loads player and required components.} \\
\ubtestbadge{30} & \textcolor{zGoldDeep!78!black}{\texttt{\bfseries Runtime\-Standing\-Walk\-Directions\-Blend\-To\-Directional\-Walk\-Clips}}\enspace\textcolor{zText}{Checks that runtime behavior standing walk directions blend to directional walk clips.} \\
\noalign{\vskip 2pt}
\multicolumn{2}{@{}p{14.4cm}@{}}{\textcolor{zRockLite!42}{\rule{\linewidth}{0.3pt}}}\\[3pt]
\multicolumn{2}{@{}p{14.4cm}@{}}{\ubtaskhead{zDawn}{0065-SplitBodyAim}{Split Body Aim}{zGoldDeep}{Medium}{12}}\\*[2pt]
\ubtestbadge{01} & \textcolor{zGoldDeep!78!black}{\texttt{\bfseries Animator\-Source\-Applies\-Yaw\-To\-Spine\-Bone\-Rotations}}\enspace\textcolor{zText}{Checks that Animator source applies yaw to spine bone rotations.} \\
\ubtestbadge{02} & \textcolor{zGoldDeep!78!black}{\texttt{\bfseries Animator\-Source\-Targets\-Real\-Spine\-Chain\-And\-Computes\-Aim\-Yaw}}\enspace\textcolor{zText}{Checks that Animator source targets real spine chain and computes aim yaw.} \\
\ubtestbadge{03} & \textcolor{zGoldDeep!78!black}{\texttt{\bfseries Motor\-Source\-Actively\-Sets\-Backward\-Semantics}}\enspace\textcolor{zText}{Checks that motor source actively sets backward semantics.} \\
\ubtestbadge{04} & \textcolor{zGoldDeep!78!black}{\texttt{\bfseries Motor\-Source\-Uses\-Aim\-World\-Position\-For\-Root\-Facing}}\enspace\textcolor{zText}{Checks that motor source uses aim world position for root facing.} \\
\ubtestbadge{05} & \textcolor{zGoldDeep!78!black}{\texttt{\bfseries Aim\-Yaw\-Does\-Not\-Twist\-Lower\-Body\-Bones\-Like\-Upper\-Body}}\enspace\textcolor{zText}{Checks that aim yaw does not twist lower body bones like upper body.} \\
\ubtestbadge{06} & \textcolor{zGoldDeep!78!black}{\texttt{\bfseries Backward\-Move\-While\-Aiming\-Faces\-Aim\-Direction}}\enspace\textcolor{zText}{Checks that backward move while aiming faces aim direction.} \\
\ubtestbadge{07} & \textcolor{zGoldDeep!78!black}{\texttt{\bfseries Backward\-Move\-While\-Aiming\-Sets\-Backward\-Semantics}}\enspace\textcolor{zText}{Checks that backward move while aiming sets backward semantics.} \\
\ubtestbadge{08} & \textcolor{zGoldDeep!78!black}{\texttt{\bfseries Backward\-Move\-Works\-Across\-Horizontal\-Axis}}\enspace\textcolor{zText}{Checks that backward move works across horizontal axis.} \\
\ubtestbadge{09} & \textcolor{zGoldDeep!78!black}{\texttt{\bfseries Side\-Move\-Aim\-Keeps\-Root\-Movement\-Facing\-But\-Rotates\-Upper\-Body}}\enspace\textcolor{zText}{Checks that side move aim keeps root movement facing but rotates upper body.} \\
\ubtestbadge{10} & \textcolor{zGoldDeep!78!black}{\texttt{\bfseries Standing\-Aim\-Turns\-Root\-Toward\-Aim\-Point}}\enspace\textcolor{zText}{Checks standing aim turns root toward aim point.} \\
\ubtestbadge{11} & \textcolor{zGoldDeep!78!black}{\texttt{\bfseries Standing\-Aim\-Works\-For\-Left\-And\-Right\-World\-Directions}}\enspace\textcolor{zText}{Checks that standing aim works for left and right world directions.} \\
\ubtestbadge{12} & \textcolor{zGoldDeep!78!black}{\texttt{\bfseries Upper\-Body\-Yaw\-Responds\-On\-Both\-Sides\-Of\-Root}}\enspace\textcolor{zText}{Checks that upper body yaw responds on both sides of root.} \\
\noalign{\vskip 2pt}
\multicolumn{2}{@{}p{14.4cm}@{}}{\textcolor{zRockLite!42}{\rule{\linewidth}{0.3pt}}}\\[3pt]
\multicolumn{2}{@{}p{14.4cm}@{}}{\ubtaskhead{zDawn}{0051-AnimationClassification}{Animation Classification}{zShrine}{Hard}{20}}\\*[2pt]
\ubtestbadge{01} & \textcolor{zGoldDeep!78!black}{\texttt{\bfseries Test\-Answer\-For(Creature\-Clip\-01)}}\enspace\textcolor{zText}{Checks that the character using creature animation clip 01 is placed under the correct answer category.} \\
\ubtestbadge{02} & \textcolor{zGoldDeep!78!black}{\texttt{\bfseries Test\-Answer\-For(Creature\-Clip\-02)}}\enspace\textcolor{zText}{Checks that the character using creature animation clip 02 is placed under the correct answer category.} \\
\ubtestbadge{03} & \textcolor{zGoldDeep!78!black}{\texttt{\bfseries Test\-Answer\-For(Creature\-Clip\-03)}}\enspace\textcolor{zText}{Checks that the character using creature animation clip 03 is placed under the correct answer category.} \\
\ubtestbadge{04} & \textcolor{zGoldDeep!78!black}{\texttt{\bfseries Test\-Answer\-For(Creature\-Clip\-04)}}\enspace\textcolor{zText}{Checks that the character using creature animation clip 04 is placed under the correct answer category.} \\
\ubtestbadge{05} & \textcolor{zGoldDeep!78!black}{\texttt{\bfseries Test\-Answer\-For(Creature\-Clip\-05)}}\enspace\textcolor{zText}{Checks that the character using creature animation clip 05 is placed under the correct answer category.} \\
\ubtestbadge{06} & \textcolor{zGoldDeep!78!black}{\texttt{\bfseries Test\-Answer\-For(Creature\-Clip\-06)}}\enspace\textcolor{zText}{Checks that the character using creature animation clip 06 is placed under the correct answer category.} \\
\ubtestbadge{07} & \textcolor{zGoldDeep!78!black}{\texttt{\bfseries Test\-Answer\-For(Creature\-Clip\-07)}}\enspace\textcolor{zText}{Checks that the character using creature animation clip 07 is placed under the correct answer category.} \\
\ubtestbadge{08} & \textcolor{zGoldDeep!78!black}{\texttt{\bfseries Test\-Answer\-For(Creature\-Clip\-08)}}\enspace\textcolor{zText}{Checks that the character using creature animation clip 08 is placed under the correct answer category.} \\
\ubtestbadge{09} & \textcolor{zGoldDeep!78!black}{\texttt{\bfseries Test\-Answer\-For(Creature\-Clip\-09)}}\enspace\textcolor{zText}{Checks that the character using creature animation clip 09 is placed under the correct answer category.} \\
\ubtestbadge{10} & \textcolor{zGoldDeep!78!black}{\texttt{\bfseries Test\-Answer\-For(Creature\-Clip\-10)}}\enspace\textcolor{zText}{Checks that the character using creature animation clip 10 is placed under the correct answer category.} \\
\ubtestbadge{11} & \textcolor{zGoldDeep!78!black}{\texttt{\bfseries Test\-Answer\-For(Creature\-Clip\-11)}}\enspace\textcolor{zText}{Checks that the character using creature animation clip 11 is placed under the correct answer category.} \\
\ubtestbadge{12} & \textcolor{zGoldDeep!78!black}{\texttt{\bfseries Test\-Answer\-For(Humanoid\-Clip\-01)}}\enspace\textcolor{zText}{Checks that the character using humanoid animation clip 01 is placed under the correct answer category.} \\
\ubtestbadge{13} & \textcolor{zGoldDeep!78!black}{\texttt{\bfseries Test\-Answer\-For(Humanoid\-Clip\-02)}}\enspace\textcolor{zText}{Checks that the character using humanoid animation clip 02 is placed under the correct answer category.} \\
\ubtestbadge{14} & \textcolor{zGoldDeep!78!black}{\texttt{\bfseries Test\-Answer\-For(Humanoid\-Clip\-03)}}\enspace\textcolor{zText}{Checks that the character using humanoid animation clip 03 is placed under the correct answer category.} \\
\ubtestbadge{15} & \textcolor{zGoldDeep!78!black}{\texttt{\bfseries Test\-Answer\-For(Humanoid\-Clip\-04)}}\enspace\textcolor{zText}{Checks that the character using humanoid animation clip 04 is placed under the correct answer category.} \\
\ubtestbadge{16} & \textcolor{zGoldDeep!78!black}{\texttt{\bfseries Test\-Answer\-For(Humanoid\-Clip\-05)}}\enspace\textcolor{zText}{Checks that the character using humanoid animation clip 05 is placed under the correct answer category.} \\
\ubtestbadge{17} & \textcolor{zGoldDeep!78!black}{\texttt{\bfseries Test\-Answer\-For(Humanoid\-Clip\-06)}}\enspace\textcolor{zText}{Checks that the character using humanoid animation clip 06 is placed under the correct answer category.} \\
\ubtestbadge{18} & \textcolor{zGoldDeep!78!black}{\texttt{\bfseries Test\-Answer\-For(Humanoid\-Clip\-07)}}\enspace\textcolor{zText}{Checks that the character using humanoid animation clip 07 is placed under the correct answer category.} \\
\ubtestbadge{19} & \textcolor{zGoldDeep!78!black}{\texttt{\bfseries Test\-Answer\-For(Humanoid\-Clip\-08)}}\enspace\textcolor{zText}{Checks that the character using humanoid animation clip 08 is placed under the correct answer category.} \\
\ubtestbadge{20} & \textcolor{zGoldDeep!78!black}{\texttt{\bfseries Test\-Answer\-For(Humanoid\-Clip\-09)}}\enspace\textcolor{zText}{Checks that the character using humanoid animation clip 09 is placed under the correct answer category.} \\
\noalign{\vskip 2pt}
\multicolumn{2}{@{}p{14.4cm}@{}}{\textcolor{zRockLite!42}{\rule{\linewidth}{0.3pt}}}\\[3pt]
\multicolumn{2}{@{}p{14.4cm}@{}}{\ubtaskhead{zDawn}{0079-FPSRemoteWeaponAnimationSync}{Synchronize Remote Rifle Presentation}{zShrine}{Hard}{18}}\\*[2pt]
\ubtestbadge{01} & \textcolor{zGoldDeep!78!black}{\texttt{\bfseries Third\-Person\-Avatar\-Has\-Complete\-Humanoid\-Arms\-Hands\-And\-Fingers}}\enspace\textcolor{zText}{Checks that the third-person Avatar is a valid Humanoid with complete arm, hand, and finger mappings.} \\
\ubtestbadge{02} & \textcolor{zGoldDeep!78!black}{\texttt{\bfseries Rifle\-Binding\-Selects\-AVisible\-Rifle\-And\-Defines\-Both\-Grip\-Positions}}\enspace\textcolor{zText}{Checks that the rifle binding selects a visible rifle and defines position targets for both grips.} \\
\ubtestbadge{03} & \textcolor{zGoldDeep!78!black}{\texttt{\bfseries Rifle\-Binding\-Defines\-Palm\-Orientation\-As\-Well\-As\-Position}}\enspace\textcolor{zText}{Checks that the rifle binding defines both position and orientation targets for each palm.} \\
\ubtestbadge{04} & \textcolor{zGoldDeep!78!black}{\texttt{\bfseries Remote\-Rig\-Contains\-Visible\-Fire\-Reload\-And\-Equip\-Playback\-Path}}\enspace\textcolor{zText}{Checks that the remote rig exposes visible fire, reload, and equip animation playback paths.} \\
\ubtestbadge{05} & \textcolor{zGoldDeep!78!black}{\texttt{\bfseries Rifle\-Muzzle\-Matches\-First\-Person\-Aim\-Across\-Representative\-Angles(0f, 0f)}}\enspace\textcolor{zText}{Checks that the third-person rifle muzzle stays within 5 degrees of first-person aim at pitch/yaw (0, 0).} \\
\ubtestbadge{06} & \textcolor{zGoldDeep!78!black}{\texttt{\bfseries Rifle\-Muzzle\-Matches\-First\-Person\-Aim\-Across\-Representative\-Angles(-28f, 35f)}}\enspace\textcolor{zText}{Checks that the third-person rifle muzzle stays within 5 degrees of first-person aim at pitch/yaw (-28, 35).} \\
\ubtestbadge{07} & \textcolor{zGoldDeep!78!black}{\texttt{\bfseries Rifle\-Muzzle\-Matches\-First\-Person\-Aim\-Across\-Representative\-Angles(24f, -32f)}}\enspace\textcolor{zText}{Checks that the third-person rifle muzzle stays within 5 degrees of first-person aim at pitch/yaw (24, -32).} \\
\ubtestbadge{08} & \textcolor{zGoldDeep!78!black}{\texttt{\bfseries Both\-Hands\-Reach\-Rifle\-Grip\-Targets}}\enspace\textcolor{zText}{Checks that both palms reach their calibrated rifle grip targets within tolerance.} \\
\ubtestbadge{09} & \textcolor{zGoldDeep!78!black}{\texttt{\bfseries Fingers\-Curl\-And\-Remain\-Near\-Rifle\-Instead\-Of\-Opening\-Away(false)}}\enspace\textcolor{zText}{Checks that at least two left-hand fingers curl and at least three fingertips remain near the corresponding rifle grip.} \\
\ubtestbadge{10} & \textcolor{zGoldDeep!78!black}{\texttt{\bfseries Fingers\-Curl\-And\-Remain\-Near\-Rifle\-Instead\-Of\-Opening\-Away(true)}}\enspace\textcolor{zText}{Checks that at least two right-hand fingers curl and at least three fingertips remain near the corresponding rifle grip.} \\
\ubtestbadge{11} & \textcolor{zGoldDeep!78!black}{\texttt{\bfseries Reload\-Moves\-Support\-Hand\-Toward\-Magazine}}\enspace\textcolor{zText}{Checks that the reload pose moves the support hand toward the magazine.} \\
\ubtestbadge{12} & \textcolor{zGoldDeep!78!black}{\texttt{\bfseries Reload\-Snapshot\-Actually\-Plays\-AThird\-Person\-Reload\-Motion}}\enspace\textcolor{zText}{Checks that a reload snapshot visibly plays the authored third-person reload motion.} \\
\ubtestbadge{13} & \textcolor{zGoldDeep!78!black}{\texttt{\bfseries Repeated\-Reload\-Snapshots\-Do\-Not\-Restart\-Animation\-Every\-Frame}}\enspace\textcolor{zText}{Checks that repeated reload snapshots do not restart the reload animation every frame.} \\
\ubtestbadge{14} & \textcolor{zGoldDeep!78!black}{\texttt{\bfseries Arm\-Bones\-Stay\-Finite\-And\-Within\-Plausible\-Geometry(0f, 0f)}}\enspace\textcolor{zText}{Checks finite, anatomically plausible arm and wrist geometry at pitch/yaw (0, 0).} \\
\ubtestbadge{15} & \textcolor{zGoldDeep!78!black}{\texttt{\bfseries Arm\-Bones\-Stay\-Finite\-And\-Within\-Plausible\-Geometry(-35f, -40f)}}\enspace\textcolor{zText}{Checks finite, anatomically plausible arm and wrist geometry at pitch/yaw (-35, -40).} \\
\ubtestbadge{16} & \textcolor{zGoldDeep!78!black}{\texttt{\bfseries Arm\-Bones\-Stay\-Finite\-And\-Within\-Plausible\-Geometry(30f, 40f)}}\enspace\textcolor{zText}{Checks finite, anatomically plausible arm and wrist geometry at pitch/yaw (30, 40).} \\
\ubtestbadge{17} & \textcolor{zGoldDeep!78!black}{\texttt{\bfseries Aim\-Sweep\-Does\-Not\-Flip\-Arm\-Wrist\-Or\-Finger\-Bones}}\enspace\textcolor{zText}{Checks that an aim sweep causes no arm, wrist, hand, or finger flips or discontinuous jumps.} \\
\ubtestbadge{18} & \textcolor{zGoldDeep!78!black}{\texttt{\bfseries Fire\-Snapshot\-Plays\-Remotely\-And\-Rifle\-Recoil\-Adds\-First\-Person\-Look\-Input}}\enspace\textcolor{zText}{Checks that a fire snapshot plays remotely and adds non-zero first-person look recoil.} \\
\noalign{\vskip 2pt}
\multicolumn{2}{@{}p{14.4cm}@{}}{\textcolor{zRockLite!42}{\rule{\linewidth}{0.3pt}}}\\[3pt]
\noalign{\vskip 3pt}
\multicolumn{2}{@{}p{14.4cm}@{}}{\ubcategory{zRoyal}{Editor Operations}{11 tasks \textperiodcentered\ 142 milestone tests}}\\*[4pt]
\multicolumn{2}{@{}p{14.4cm}@{}}{\ubtaskhead{zRoyal}{0004-IslandBenchRoadClassification}{Island Bench Road Group Cleanup}{zGoldDeep}{Easy}{9}}\\*[2pt]
\ubtestbadge{01} & \textcolor{zGoldDeep!78!black}{\texttt{\bfseries Island\-Bench\_\allowbreak{}Straight\-Road\-Segments\_\allowbreak{}Are\-Classified\-Under\-Expected\-Road\-Name("Bridge Rd.(E)")}}\enspace\textcolor{zText}{Checks that every embedded Road\_Straight segment assigned to Bridge Rd.(E) is parented under the matching road-name GameObject.} \\
\ubtestbadge{02} & \textcolor{zGoldDeep!78!black}{\texttt{\bfseries Island\-Bench\_\allowbreak{}Straight\-Road\-Segments\_\allowbreak{}Are\-Classified\-Under\-Expected\-Road\-Name("Bridge Rd.(W)")}}\enspace\textcolor{zText}{Checks that every embedded Road\_Straight segment assigned to Bridge Rd.(W) is parented under the matching road-name GameObject.} \\
\ubtestbadge{03} & \textcolor{zGoldDeep!78!black}{\texttt{\bfseries Island\-Bench\_\allowbreak{}Straight\-Road\-Segments\_\allowbreak{}Are\-Classified\-Under\-Expected\-Road\-Name("Connection Rd.")}}\enspace\textcolor{zText}{Checks that every embedded Road\_Straight segment assigned to Connection Rd. is parented under the matching road-name GameObject.} \\
\ubtestbadge{04} & \textcolor{zGoldDeep!78!black}{\texttt{\bfseries Island\-Bench\_\allowbreak{}Straight\-Road\-Segments\_\allowbreak{}Are\-Classified\-Under\-Expected\-Road\-Name("EShore Ave")}}\enspace\textcolor{zText}{Checks that every embedded Road\_Straight segment assigned to EShore Ave is parented under the matching road-name GameObject.} \\
\ubtestbadge{05} & \textcolor{zGoldDeep!78!black}{\texttt{\bfseries Island\-Bench\_\allowbreak{}Straight\-Road\-Segments\_\allowbreak{}Are\-Classified\-Under\-Expected\-Road\-Name("EW Bridge")}}\enspace\textcolor{zText}{Checks that every embedded Road\_Straight segment assigned to EW Bridge is parented under the matching road-name GameObject.} \\
\ubtestbadge{06} & \textcolor{zGoldDeep!78!black}{\texttt{\bfseries Island\-Bench\_\allowbreak{}Straight\-Road\-Segments\_\allowbreak{}Are\-Classified\-Under\-Expected\-Road\-Name("Factory Rd.")}}\enspace\textcolor{zText}{Checks that every embedded Road\_Straight segment assigned to Factory Rd. is parented under the matching road-name GameObject.} \\
\ubtestbadge{07} & \textcolor{zGoldDeep!78!black}{\texttt{\bfseries Island\-Bench\_\allowbreak{}Straight\-Road\-Segments\_\allowbreak{}Are\-Classified\-Under\-Expected\-Road\-Name("SN Bridge")}}\enspace\textcolor{zText}{Checks that every embedded Road\_Straight segment assigned to SN Bridge is parented under the matching road-name GameObject.} \\
\ubtestbadge{08} & \textcolor{zGoldDeep!78!black}{\texttt{\bfseries Island\-Bench\_\allowbreak{}Straight\-Road\-Segments\_\allowbreak{}Are\-Classified\-Under\-Expected\-Road\-Name("SShore Ave.")}}\enspace\textcolor{zText}{Checks that every embedded Road\_Straight segment assigned to SShore Ave. is parented under the matching road-name GameObject.} \\
\ubtestbadge{09} & \textcolor{zGoldDeep!78!black}{\texttt{\bfseries Island\-Bench\_\allowbreak{}Unsorted\-Group\_\allowbreak{}Does\-Not\-Keep\-Expected\-Straight\-Road\-Segments}}\enspace\textcolor{zText}{Checks that island bench unsorted group does not keep expected straight road segments.} \\
\noalign{\vskip 2pt}
\multicolumn{2}{@{}p{14.4cm}@{}}{\textcolor{zRockLite!42}{\rule{\linewidth}{0.3pt}}}\\[3pt]
\multicolumn{2}{@{}p{14.4cm}@{}}{\ubtaskhead{zRoyal}{0013-LightMachineGun}{Light Machine Gun}{zGoldDeep}{Easy}{15}}\\*[2pt]
\ubtestbadge{01} & \textcolor{zGoldDeep!78!black}{\texttt{\bfseries Task01\_\allowbreak{}LMG\-Weapon\-Definition\-Asset\-Exists}}\enspace\textcolor{zText}{Checks that LMG weapon definition asset exists.} \\
\ubtestbadge{02} & \textcolor{zGoldDeep!78!black}{\texttt{\bfseries Task02\_\allowbreak{}LMG\-Is\-Registered\-In\-Game\-Database}}\enspace\textcolor{zText}{Checks that LMG is registered in game database.} \\
\ubtestbadge{03} & \textcolor{zGoldDeep!78!black}{\texttt{\bfseries Task03\_\allowbreak{}LMG\-Has\-Correct\-Identity\-And\-Category}}\enspace\textcolor{zText}{Checks that LMG has correct identity and category.} \\
\ubtestbadge{04} & \textcolor{zGoldDeep!78!black}{\texttt{\bfseries Task04\_\allowbreak{}LMG\-Uses\-Assault\-Rifle\-Weapon\-Pipeline}}\enspace\textcolor{zText}{Checks that LMG uses assault rifle weapon pipeline.} \\
\ubtestbadge{05} & \textcolor{zGoldDeep!78!black}{\texttt{\bfseries Task05\_\allowbreak{}LMG\-Magazine\-And\-Weapon\-Stats\-Are\-Usable}}\enspace\textcolor{zText}{Checks that LMG magazine and weapon stats are usable.} \\
\ubtestbadge{06} & \textcolor{zGoldDeep!78!black}{\texttt{\bfseries Task06\_\allowbreak{}LMG\-Can\-Be\-Brought\-And\-Fits\-Existing\-Backpacks}}\enspace\textcolor{zText}{Checks that LMG can be brought and fits existing backpacks.} \\
\ubtestbadge{07} & \textcolor{zGoldDeep!78!black}{\texttt{\bfseries Task07\_\allowbreak{}Top\-Down\-Shooter\-Can\-Equip\-LMG\-By\-Item\-Id}}\enspace\textcolor{zText}{Checks that top down shooter can equip LMG by item id.} \\
\ubtestbadge{08} & \textcolor{zGoldDeep!78!black}{\texttt{\bfseries Task08\_\allowbreak{}Top\-Down\-Shooter\-Maps\-LMG\-To762\-Reserve\-Ammo}}\enspace\textcolor{zText}{Checks that top down shooter maps LMG to 762 reserve ammo.} \\
\ubtestbadge{09} & \textcolor{zGoldDeep!78!black}{\texttt{\bfseries Task09\_\allowbreak{}LMG\-Uses\-Automatic\-Fire\-Mode}}\enspace\textcolor{zText}{Checks that LMG uses automatic fire mode.} \\
\ubtestbadge{10} & \textcolor{zGoldDeep!78!black}{\texttt{\bfseries Task10\_\allowbreak{}LMG\-Fire\-Consumes\-One\-Round\-From\-Hundred\-Round\-Belt}}\enspace\textcolor{zText}{Checks that LMG fire consumes one round from hundred round belt.} \\
\ubtestbadge{11} & \textcolor{zGoldDeep!78!black}{\texttt{\bfseries Task11\_\allowbreak{}LMG\-Png\-Is\-Imported\-As\-Sprite}}\enspace\textcolor{zText}{Checks that LMG png is imported as sprite.} \\
\ubtestbadge{12} & \textcolor{zGoldDeep!78!black}{\texttt{\bfseries Task12\_\allowbreak{}LMG\-Weapon\-Definition\-Uses\-LMG\-Sprite\-Icon}}\enspace\textcolor{zText}{Checks that LMG weapon definition uses LMG sprite icon.} \\
\ubtestbadge{13} & \textcolor{zGoldDeep!78!black}{\texttt{\bfseries Task13\_\allowbreak{}LMG\-Has\-Local\-Fire\-Audio}}\enspace\textcolor{zText}{Checks that LMG has local fire audio.} \\
\ubtestbadge{14} & \textcolor{zGoldDeep!78!black}{\texttt{\bfseries Task14\_\allowbreak{}LMG\-Has\-Remote\-Fire\-Audio}}\enspace\textcolor{zText}{Checks that LMG has remote fire audio.} \\
\ubtestbadge{15} & \textcolor{zGoldDeep!78!black}{\texttt{\bfseries Task15\_\allowbreak{}LMG\-Is\-Available\-In\-Shop}}\enspace\textcolor{zText}{Checks that LMG is available in shop.} \\
\noalign{\vskip 2pt}
\multicolumn{2}{@{}p{14.4cm}@{}}{\textcolor{zRockLite!42}{\rule{\linewidth}{0.3pt}}}\\[3pt]
\multicolumn{2}{@{}p{14.4cm}@{}}{\ubtaskhead{zRoyal}{0014-AssetPruning}{Asset Pruning}{zGoldDeep}{Easy}{15}}\\*[2pt]
\ubtestbadge{01} & \textcolor{zGoldDeep!78!black}{\texttt{\bfseries Task01\_\allowbreak{}Build\-Settings\-Keep\-Only\-Required\-Enabled\-Scenes}}\enspace\textcolor{zText}{Checks that build settings keep only required enabled scenes.} \\
\ubtestbadge{02} & \textcolor{zGoldDeep!78!black}{\texttt{\bfseries Task02\_\allowbreak{}Non\-Build\-Scenes\-Are\-Removed}}\enspace\textcolor{zText}{Checks that non build scenes are removed.} \\
\ubtestbadge{03} & \textcolor{zGoldDeep!78!black}{\texttt{\bfseries Task03\_\allowbreak{}Modern\-Weapon\-Sound\-Pack\-Is\-Trimmed\-To\-Used\-Weapons}}\enspace\textcolor{zText}{Checks that modern weapon sound pack is trimmed to used weapons.} \\
\ubtestbadge{04} & \textcolor{zGoldDeep!78!black}{\texttt{\bfseries Task04\_\allowbreak{}Unrelated\-LMG\-Assets\-Are\-Removed}}\enspace\textcolor{zText}{Checks that unrelated LMG assets are removed.} \\
\ubtestbadge{05} & \textcolor{zGoldDeep!78!black}{\texttt{\bfseries Task05\_\allowbreak{}Ammunition\-Casing\-Library\-Is\-Removed}}\enspace\textcolor{zText}{Checks that ammunition casing library is removed.} \\
\ubtestbadge{06} & \textcolor{zGoldDeep!78!black}{\texttt{\bfseries Task06\_\allowbreak{}Fish\-Net\-Demos\-Are\-Removed\-But\-Runtime\-Remains}}\enspace\textcolor{zText}{Checks that fish net demos are removed but runtime remains.} \\
\ubtestbadge{07} & \textcolor{zGoldDeep!78!black}{\texttt{\bfseries Task07\_\allowbreak{}Unused\-Character\-One\-Source\-Folder\-Is\-Removed}}\enspace\textcolor{zText}{Checks that unused character one source folder is removed.} \\
\ubtestbadge{08} & \textcolor{zGoldDeep!78!black}{\texttt{\bfseries Task08\_\allowbreak{}Shared\-Asset\-Library\-Is\-Reduced\-To\-Runtime\-Dependencies}}\enspace\textcolor{zText}{Checks that the shared asset library is reduced to runtime dependencies.} \\
\ubtestbadge{09} & \textcolor{zGoldDeep!78!black}{\texttt{\bfseries Task09\_\allowbreak{}Demo\-Scenes\-And\-Unused\-Overview\-Content\-Are\-Removed}}\enspace\textcolor{zText}{Checks that demo scenes and unused overview content are removed.} \\
\ubtestbadge{10} & \textcolor{zGoldDeep!78!black}{\texttt{\bfseries Task10\_\allowbreak{}Required\-Runtime\-Data\-And\-Resources\-Remain}}\enspace\textcolor{zText}{Checks that required runtime data and resources remain.} \\
\ubtestbadge{11} & \textcolor{zGoldDeep!78!black}{\texttt{\bfseries Task11\_\allowbreak{}Enabled\-Scenes\-Have\-No\-Missing\-Scripts\-Or\-Prefab\-Sources}}\enspace\textcolor{zText}{Checks that enabled scenes have no missing scripts or Prefab sources.} \\
\ubtestbadge{12} & \textcolor{zGoldDeep!78!black}{\texttt{\bfseries Task12\_\allowbreak{}Assets\-Folder\-Is\-Substantially\-Smaller}}\enspace\textcolor{zText}{Checks that assets folder is substantially smaller.} \\
\ubtestbadge{13} & \textcolor{zGoldDeep!78!black}{\texttt{\bfseries Task13\_\allowbreak{}Unused\-Dive\-Roll\-Animation\-Is\-Removed}}\enspace\textcolor{zText}{Checks that unused dive roll animation is removed.} \\
\ubtestbadge{14} & \textcolor{zGoldDeep!78!black}{\texttt{\bfseries Task14\_\allowbreak{}Tutorial\-Info\-Folder\-Is\-Removed}}\enspace\textcolor{zText}{Checks that tutorial info folder is removed.} \\
\ubtestbadge{15} & \textcolor{zGoldDeep!78!black}{\texttt{\bfseries Task15\_\allowbreak{}Unused\-Tank\-Prefab\-Is\-Removed}}\enspace\textcolor{zText}{Checks that unused tank Prefab is removed.} \\
\noalign{\vskip 2pt}
\multicolumn{2}{@{}p{14.4cm}@{}}{\textcolor{zRockLite!42}{\rule{\linewidth}{0.3pt}}}\\[3pt]
\multicolumn{2}{@{}p{14.4cm}@{}}{\ubtaskhead{zRoyal}{0027-VolumeBoundsOrganizer}{Automated Scene Grouping and Renaming Based on Spatial Bounding-Box Features}{zGoldDeep}{Easy}{16}}\\*[2pt]
\ubtestbadge{01} & \textcolor{zGoldDeep!78!black}{\texttt{\bfseries T01\_\allowbreak{}Three\-Groups\-Created\-With\-Correct\-Prefix\-And\-Clean\-Root}}\enspace\textcolor{zText}{Checks creation of the three root-level grouping GameObjects, correct prefixes, and a clean scene root.} \\
\ubtestbadge{02} & \textcolor{zGoldDeep!78!black}{\texttt{\bfseries T01b\_\allowbreak{}Multiple\-Objects\-In\-Same\-Group\_\allowbreak{}All\-Processed}}\enspace\textcolor{zText}{Checks that every valid object is processed when multiple objects belong to the same volume group.} \\
\ubtestbadge{03} & \textcolor{zGoldDeep!78!black}{\texttt{\bfseries T02\_\allowbreak{}Volume\-Boundaries\-Rotation\-And\-Negative\-Scale}}\enspace\textcolor{zText}{Checks strict volume boundaries using mesh bounds, rotation-independent OBB volume, and absolute lossy scale.} \\
\ubtestbadge{04} & \textcolor{zGoldDeep!78!black}{\texttt{\bfseries T03\_\allowbreak{}Zero\-Volume\-Objects\-Go\-To\-Props}}\enspace\textcolor{zText}{Checks that zero-volume objects are classified under Environment\_Props.} \\
\ubtestbadge{05} & \textcolor{zGoldDeep!78!black}{\texttt{\bfseries T04\_\allowbreak{}Non\-Root\-Ghost\-And\-Non\-Mesh\-Objects\-Untouched}}\enspace\textcolor{zText}{Checks that non-root objects and objects missing MeshFilter or MeshRenderer remain untouched.} \\
\ubtestbadge{06} & \textcolor{zGoldDeep!78!black}{\texttt{\bfseries T05\_\allowbreak{}Objects\-Not\-Destroyed\-And\-Components\-Intact}}\enspace\textcolor{zText}{Checks that organization preserves every processed object and its components.} \\
\ubtestbadge{07} & \textcolor{zGoldDeep!78!black}{\texttt{\bfseries T06\_\allowbreak{}Idempotent\_\allowbreak{}No\-Dup\-Prefix\-No\-Migration\-No\-Nesting}}\enspace\textcolor{zText}{Checks idempotence: no duplicate prefixes, repeated migration, or nested grouping folders.} \\
\ubtestbadge{08} & \textcolor{zGoldDeep!78!black}{\texttt{\bfseries T07\_\allowbreak{}Mismatched\-Prefix\-Corrected\-On\-Rerun}}\enspace\textcolor{zText}{Checks that a root object's mismatched existing prefix is corrected to match its volume class.} \\
\ubtestbadge{09} & \textcolor{zGoldDeep!78!black}{\texttt{\bfseries T08\_\allowbreak{}Undo\-Full\-Rollback\_\allowbreak{}Across\-All\-Groups\-With\-Atomicity}}\enspace\textcolor{zText}{Checks that one Undo operation atomically restores names, parents, and created groups.} \\
\ubtestbadge{10} & \textcolor{zGoldDeep!78!black}{\texttt{\bfseries T09\_\allowbreak{}Return\-Value\-Accuracy}}\enspace\textcolor{zText}{Checks that ExecuteOrganization returns the exact number of valid objects processed.} \\
\ubtestbadge{11} & \textcolor{zGoldDeep!78!black}{\texttt{\bfseries T09b\_\allowbreak{}Already\-Correct\-Prefix\-Root\-Object\-Still\-Counted}}\enspace\textcolor{zText}{Checks that a valid root object already carrying the correct prefix is still counted when processed.} \\
\ubtestbadge{12} & \textcolor{zGoldDeep!78!black}{\texttt{\bfseries T10\_\allowbreak{}Exception\-Rollback\_\allowbreak{}Try\-Catch\-With\-Undo\-Revert}}\enspace\textcolor{zText}{Checks that exception handling rolls the complete Undo group back safely.} \\
\ubtestbadge{13} & \textcolor{zGoldDeep!78!black}{\texttt{\bfseries T10b\_\allowbreak{}Undo\-Core\-Api\-Verification}}\enspace\textcolor{zText}{Checks use of the required Unity Undo APIs for creation, renaming, and reparenting.} \\
\ubtestbadge{14} & \textcolor{zGoldDeep!78!black}{\texttt{\bfseries T11\_\allowbreak{}Must\-Use\-Mesh\-Bounds\-Not\-Collider\-Bounds}}\enspace\textcolor{zText}{Checks that volume classification uses MeshFilter.sharedMesh.bounds rather than Collider bounds.} \\
\ubtestbadge{15} & \textcolor{zGoldDeep!78!black}{\texttt{\bfseries T12\_\allowbreak{}Menu\-Item\-Attribute\-With\-Correct\-Path}}\enspace\textcolor{zText}{Checks for OrganizeActiveSceneMenu() with the exact Tools/Scene/Organize by Bounding Box Volume menu path.} \\
\ubtestbadge{16} & \textcolor{zGoldDeep!78!black}{\texttt{\bfseries T13\_\allowbreak{}Undo\-Group\-Isolation\_\allowbreak{}Pre\-Existing\-Operations\-Not\-Affected}}\enspace\textcolor{zText}{Checks that the tool's Undo group does not absorb or revert pre-existing editor operations.} \\
\noalign{\vskip 2pt}
\multicolumn{2}{@{}p{14.4cm}@{}}{\textcolor{zRockLite!42}{\rule{\linewidth}{0.3pt}}}\\[3pt]
\multicolumn{2}{@{}p{14.4cm}@{}}{\ubtaskhead{zRoyal}{0028-ImportExportPackages}{Import / Export Package Asset Organization}{zGoldDeep}{Easy}{7}}\\*[2pt]
\ubtestbadge{01} & \textcolor{zGoldDeep!78!black}{\texttt{\bfseries Exported\-Package\_\allowbreak{}Exists\-And\-Contains\-Only\-Requested\-Prefabs}}\enspace\textcolor{zText}{Checks that exported package exists and contains only requested prefabs.} \\
\ubtestbadge{02} & \textcolor{zGoldDeep!78!black}{\texttt{\bfseries Exported\-Package\_\allowbreak{}Includes\-Required\-Dependencies}}\enspace\textcolor{zText}{Checks that exported package includes required dependencies.} \\
\ubtestbadge{03} & \textcolor{zGoldDeep!78!black}{\texttt{\bfseries Imported\-Prefab\-Dependencies\_\allowbreak{}Are\-Organized\-Inside\-Imported\-Folder}}\enspace\textcolor{zText}{Checks that imported Prefab dependencies are organized inside imported folder.} \\
\ubtestbadge{04} & \textcolor{zGoldDeep!78!black}{\texttt{\bfseries Imported\-Prefab\-Folder\_\allowbreak{}Contains\-No\-Unrelated\-Prefabs}}\enspace\textcolor{zText}{Checks that imported Prefab folder contains no unrelated prefabs.} \\
\ubtestbadge{05} & \textcolor{zGoldDeep!78!black}{\texttt{\bfseries Imported\-Prefabs\_\allowbreak{}Are\-Not\-Left\-At\-Original\-Vendor\-Locations}}\enspace\textcolor{zText}{Checks that imported prefabs are not left at original vendor locations.} \\
\ubtestbadge{06} & \textcolor{zGoldDeep!78!black}{\texttt{\bfseries Imported\-Prefabs\_\allowbreak{}Exist\-At\-Required\-Location}}\enspace\textcolor{zText}{Checks that imported prefabs exist at required location.} \\
\ubtestbadge{07} & \textcolor{zGoldDeep!78!black}{\texttt{\bfseries Imported\-Prefabs\_\allowbreak{}Have\-Resolved\-Mesh\-And\-Material\-References}}\enspace\textcolor{zText}{Checks that imported prefabs have resolved mesh and material references.} \\
\noalign{\vskip 2pt}
\multicolumn{2}{@{}p{14.4cm}@{}}{\textcolor{zRockLite!42}{\rule{\linewidth}{0.3pt}}}\\[3pt]
\multicolumn{2}{@{}p{14.4cm}@{}}{\ubtaskhead{zRoyal}{0042-PrefabSharedMaterialIsolation}{Prefab Asset Modification, Loose Object Replacement, and Variant Construction}{zGoldDeep}{Easy}{16}}\\*[2pt]
\ubtestbadge{01} & \textcolor{zGoldDeep!78!black}{\texttt{\bfseries Barrel\-Group\-Uses\-Cloned\-Poison\-Liquid\-Only}}\enspace\textcolor{zText}{Checks that barrel group uses cloned poison liquid only.} \\
\ubtestbadge{02} & \textcolor{zGoldDeep!78!black}{\texttt{\bfseries Console\-Group\-Uses\-Cloned\-Alarm\-Screen\-Only}}\enspace\textcolor{zText}{Checks that console group uses cloned alarm screen only.} \\
\ubtestbadge{03} & \textcolor{zGoldDeep!78!black}{\texttt{\bfseries Crystal\-Group\-Uses\-Cloned\-Hazard\-Material\-Only}}\enspace\textcolor{zText}{Checks that crystal group uses cloned hazard material only.} \\
\ubtestbadge{04} & \textcolor{zGoldDeep!78!black}{\texttt{\bfseries Locked\-Security\-Door\-Is\-Prefab\-Variant\-With\-Door\-Overrides}}\enspace\textcolor{zText}{Checks that locked security door is Prefab variant with door overrides.} \\
\ubtestbadge{05} & \textcolor{zGoldDeep!78!black}{\texttt{\bfseries Loose\-Beacons\-Are\-Replaced\-By\-Twelve\-Connected\-Prefab\-Instances}}\enspace\textcolor{zText}{Checks that loose beacons are replaced by twelve connected Prefab instances.} \\
\ubtestbadge{06} & \textcolor{zGoldDeep!78!black}{\texttt{\bfseries Original\-Shared\-Materials\-Remain\-Unmodified\-While\-Targets\-Are\-Detached}}\enspace\textcolor{zText}{Checks that original shared materials remain unmodified while targets are detached.} \\
\ubtestbadge{07} & \textcolor{zGoldDeep!78!black}{\texttt{\bfseries Pipe\-Group\-Uses\-Cloned\-Warning\-Material\-Only}}\enspace\textcolor{zText}{Checks that pipe group uses cloned warning material only.} \\
\ubtestbadge{08} & \textcolor{zGoldDeep!78!black}{\texttt{\bfseries Prefab\-Meshes\-And\-Renderer\-Paths\-Remain\-Intact}}\enspace\textcolor{zText}{Checks that Prefab meshes and renderer paths remain is intact.} \\
\ubtestbadge{09} & \textcolor{zGoldDeep!78!black}{\texttt{\bfseries Replacement\-Instances\-Preserve\-Original\-World\-Transforms}}\enspace\textcolor{zText}{Checks that replacement instances preserve original world transforms.} \\
\ubtestbadge{10} & \textcolor{zGoldDeep!78!black}{\texttt{\bfseries Replacement\-Scene\-Does\-Not\-Use\-Loose\-Objects\-Or\-Material\-Overrides}}\enspace\textcolor{zText}{Checks that replacement scene does not use loose objects or material overrides.} \\
\ubtestbadge{11} & \textcolor{zGoldDeep!78!black}{\texttt{\bfseries Scene\-Control\-Instances\-Remain\-Prefab\-Connected\-And\-Unchanged}}\enspace\textcolor{zText}{Checks that scene control instances remain Prefab connected and unchanged.} \\
\ubtestbadge{12} & \textcolor{zGoldDeep!78!black}{\texttt{\bfseries Scene\-Instances\-Do\-Not\-Carry\-Renderer\-Material\-Overrides}}\enspace\textcolor{zText}{Checks that scene instances do not carry renderer material overrides.} \\
\ubtestbadge{13} & \textcolor{zGoldDeep!78!black}{\texttt{\bfseries Scene\-Target\-Instances\-Remain\-Prefab\-Connected\-And\-Inherit\-Solved\-Materials}}\enspace\textcolor{zText}{Checks that scene target instances remain Prefab connected and inherit solved materials.} \\
\ubtestbadge{14} & \textcolor{zGoldDeep!78!black}{\texttt{\bfseries Survey\-Scanner\-Cart\-Is\-Prefab\-Variant\-With\-Nested\-Scanner\-Module}}\enspace\textcolor{zText}{Checks that survey scanner cart is Prefab variant with nested scanner module.} \\
\ubtestbadge{15} & \textcolor{zGoldDeep!78!black}{\texttt{\bfseries Target\-Materials\-Are\-Persistent\-Bench0042\-Assets\-And\-Distinct}}\enspace\textcolor{zText}{Checks that the target materials are distinct, persistent assets under the Bench0042 project folders.} \\
\ubtestbadge{16} & \textcolor{zGoldDeep!78!black}{\texttt{\bfseries Warning\-Beacon\-Prefab\-Exists\-With\-Reusable\-Structure}}\enspace\textcolor{zText}{Checks that warning beacon Prefab exists with reusable structure.} \\
\noalign{\vskip 2pt}
\multicolumn{2}{@{}p{14.4cm}@{}}{\textcolor{zRockLite!42}{\rule{\linewidth}{0.3pt}}}\\[3pt]
\multicolumn{2}{@{}p{14.4cm}@{}}{\ubtaskhead{zRoyal}{0052-URPPipelineConfiguration}{URP Pipeline Configuration}{zGoldDeep}{Easy}{8}}\\*[2pt]
\ubtestbadge{01} & \textcolor{zGoldDeep!78!black}{\texttt{\bfseries Bench\-Materials\-Use\-Universal\-Render\-Pipeline\-Lit}}\enspace\textcolor{zText}{Checks that bench materials use universal render pipeline lit.} \\
\ubtestbadge{02} & \textcolor{zGoldDeep!78!black}{\texttt{\bfseries Converted\-Materials\-Preserve\-Base\-Colors\-And\-Alpha}}\enspace\textcolor{zText}{Checks that converted materials preserve base colors and alpha.} \\
\ubtestbadge{03} & \textcolor{zGoldDeep!78!black}{\texttt{\bfseries Each\-Quality\-Level\-Uses\-ADistinct\-Urp\-Pipeline\-Asset}}\enspace\textcolor{zText}{Checks that each quality level uses a distinct URP pipeline asset.} \\
\ubtestbadge{04} & \textcolor{zGoldDeep!78!black}{\texttt{\bfseries Medium\-Is\-The\-Default\-And\-Active\-Quality\-Level}}\enspace\textcolor{zText}{Checks that medium is the default and active quality level.} \\
\ubtestbadge{05} & \textcolor{zGoldDeep!78!black}{\texttt{\bfseries Project\-Current\-Pipeline\-Uses\-The\-Medium\-Urp\-Asset}}\enspace\textcolor{zText}{Checks that project current pipeline uses the medium URP asset.} \\
\ubtestbadge{06} & \textcolor{zGoldDeep!78!black}{\texttt{\bfseries Quality\-Settings\-Contain\-Exactly\-Low\-Medium\-High\-Extra\-High}}\enspace\textcolor{zText}{Checks that quality settings contain exactly low medium high extra high.} \\
\ubtestbadge{07} & \textcolor{zGoldDeep!78!black}{\texttt{\bfseries Universal\-Render\-Pipeline\-Package\-Is\-Installed}}\enspace\textcolor{zText}{Checks that universal render pipeline package is installed.} \\
\ubtestbadge{08} & \textcolor{zGoldDeep!78!black}{\texttt{\bfseries Urp\-Quality\-Assets\-Show\-AReasonable\-Low\-To\-Extra\-High\-Progression}}\enspace\textcolor{zText}{Checks that URP quality assets show a reasonable low to extra high progression.} \\
\noalign{\vskip 2pt}
\multicolumn{2}{@{}p{14.4cm}@{}}{\textcolor{zRockLite!42}{\rule{\linewidth}{0.3pt}}}\\[3pt]
\multicolumn{2}{@{}p{14.4cm}@{}}{\ubtaskhead{zRoyal}{0016-TexturePipeline}{Automated Texture Import Pipeline Based on Semantic Routing and Low-Level Memory Defense}{zGoldDeep}{Medium}{8}}\\*[2pt]
\ubtestbadge{01} & \textcolor{zGoldDeep!78!black}{\texttt{\bfseries T01\_\allowbreak{}Legacy\_\allowbreak{}Boss\_\allowbreak{}Albedo\_\allowbreak{}Is\_\allowbreak{}Untouched\_\allowbreak{}And\_\allowbreak{}Script\_\allowbreak{}Active}}\enspace\textcolor{zText}{Checks that the legacy Boss\_Albedo import settings remain untouched while the AssetPostprocessor is active.} \\
\ubtestbadge{02} & \textcolor{zGoldDeep!78!black}{\texttt{\bfseries T02\_\allowbreak{}Weapon\_\allowbreak{}Normal\_\allowbreak{}NPOT\_\allowbreak{}And\_\allowbreak{}Format}}\enspace\textcolor{zText}{Checks the weapon normal map's NPOT handling, texture type, and platform compression formats.} \\
\ubtestbadge{03} & \textcolor{zGoldDeep!78!black}{\texttt{\bfseries T03\_\allowbreak{}Foliage\_\allowbreak{}Albedo\_\allowbreak{}Alpha\_\allowbreak{}And\_\allowbreak{}Case\-Insensitive}}\enspace\textcolor{zText}{Checks case-insensitive albedo suffix matching, alpha detection, sRGB, and platform formats for foliage.} \\
\ubtestbadge{04} & \textcolor{zGoldDeep!78!black}{\texttt{\bfseries T04\_\allowbreak{}Rock\_\allowbreak{}Mask\-Map\_\allowbreak{}Linear\_\allowbreak{}Space}}\enspace\textcolor{zText}{Checks that the rock MaskMap is non-readable, imports as a linear Default texture, and uses ASTC 6x6 on iOS.} \\
\ubtestbadge{05} & \textcolor{zGoldDeep!78!black}{\texttt{\bfseries T05\_\allowbreak{}Hero\_\allowbreak{}Skin\_\allowbreak{}Path\_\allowbreak{}Deception\_\allowbreak{}And\_\allowbreak{}Fallback}}\enspace\textcolor{zText}{Checks that a non-UI path containing a misleading UI substring follows the non-UI fallback rules.} \\
\ubtestbadge{06} & \textcolor{zGoldDeep!78!black}{\texttt{\bfseries T06\_\allowbreak{}HP\_\allowbreak{}Bar\_\allowbreak{}UI\_\allowbreak{}Sprite\_\allowbreak{}And\_\allowbreak{}Packing}}\enspace\textcolor{zText}{Checks UI sprite import, disabled mipmaps, packing tag, and platform formats for the HP bar.} \\
\ubtestbadge{07} & \textcolor{zGoldDeep!78!black}{\texttt{\bfseries T07\_\allowbreak{}Skybox\_\allowbreak{}Albedo\_\allowbreak{}Extension\_\allowbreak{}Deception}}\enspace\textcolor{zText}{Checks that suffix matching ignores the .exr extension and imports Skybox\_Albedo as an sRGB Default texture.} \\
\ubtestbadge{08} & \textcolor{zGoldDeep!78!black}{\texttt{\bfseries T08\_\allowbreak{}Standard\_\allowbreak{}Albedo\_\allowbreak{}Baseline}}\enspace\textcolor{zText}{Checks that Standard\_Albedo is non-readable and uses sRGB sampling.} \\
\noalign{\vskip 2pt}
\multicolumn{2}{@{}p{14.4cm}@{}}{\textcolor{zRockLite!42}{\rule{\linewidth}{0.3pt}}}\\[3pt]
\multicolumn{2}{@{}p{14.4cm}@{}}{\ubtaskhead{zRoyal}{0018-FindGameObjectsInComplexScene}{Find GameObjects in a Complex Scene}{zGoldDeep}{Medium}{10}}\\*[2pt]
\ubtestbadge{01} & \textcolor{zGoldDeep!78!black}{\texttt{\bfseries Check\-Answer(Q1)}}\enspace\textcolor{zText}{Checks that the unique object matching Q1's search criteria is parented under Answer/Q1.} \\
\ubtestbadge{02} & \textcolor{zGoldDeep!78!black}{\texttt{\bfseries Check\-Answer(Q10)}}\enspace\textcolor{zText}{Checks that the unique object matching Q10's search criteria is parented under Answer/Q10.} \\
\ubtestbadge{03} & \textcolor{zGoldDeep!78!black}{\texttt{\bfseries Check\-Answer(Q2)}}\enspace\textcolor{zText}{Checks that the unique object matching Q2's search criteria is parented under Answer/Q2.} \\
\ubtestbadge{04} & \textcolor{zGoldDeep!78!black}{\texttt{\bfseries Check\-Answer(Q3)}}\enspace\textcolor{zText}{Checks that the unique object matching Q3's search criteria is parented under Answer/Q3.} \\
\ubtestbadge{05} & \textcolor{zGoldDeep!78!black}{\texttt{\bfseries Check\-Answer(Q4)}}\enspace\textcolor{zText}{Checks that the unique object matching Q4's search criteria is parented under Answer/Q4.} \\
\ubtestbadge{06} & \textcolor{zGoldDeep!78!black}{\texttt{\bfseries Check\-Answer(Q5)}}\enspace\textcolor{zText}{Checks that the unique object matching Q5's search criteria is parented under Answer/Q5.} \\
\ubtestbadge{07} & \textcolor{zGoldDeep!78!black}{\texttt{\bfseries Check\-Answer(Q6)}}\enspace\textcolor{zText}{Checks that the unique object matching Q6's search criteria is parented under Answer/Q6.} \\
\ubtestbadge{08} & \textcolor{zGoldDeep!78!black}{\texttt{\bfseries Check\-Answer(Q7)}}\enspace\textcolor{zText}{Checks that the unique object matching Q7's search criteria is parented under Answer/Q7.} \\
\ubtestbadge{09} & \textcolor{zGoldDeep!78!black}{\texttt{\bfseries Check\-Answer(Q8)}}\enspace\textcolor{zText}{Checks that the unique object matching Q8's search criteria is parented under Answer/Q8.} \\
\ubtestbadge{10} & \textcolor{zGoldDeep!78!black}{\texttt{\bfseries Check\-Answer(Q9)}}\enspace\textcolor{zText}{Checks that the unique object matching Q9's search criteria is parented under Answer/Q9.} \\
\noalign{\vskip 2pt}
\multicolumn{2}{@{}p{14.4cm}@{}}{\textcolor{zRockLite!42}{\rule{\linewidth}{0.3pt}}}\\[3pt]
\multicolumn{2}{@{}p{14.4cm}@{}}{\ubtaskhead{zRoyal}{0060-GunShotSound}{Build a Production Gunshot Audio System}{zShrine}{Hard}{29}}\\*[2pt]
\ubtestbadge{01} & \textcolor{zGoldDeep!78!black}{\texttt{\bfseries TC01\_\allowbreak{}Scene\-Exists\-And\-Is\-Enabled\-In\-Build\-Settings}}\enspace\textcolor{zText}{Checks that GunShotAudioLab.scene exists and is enabled in Build Settings.} \\
\ubtestbadge{02} & \textcolor{zGoldDeep!78!black}{\texttt{\bfseries TC02\_\allowbreak{}Required\-Scene\-Anchors\-And\-Single\-Listener\-Exist}}\enspace\textcolor{zText}{Checks that every required scene anchor exists and exactly one AudioListener is enabled.} \\
\ubtestbadge{03} & \textcolor{zGoldDeep!78!black}{\texttt{\bfseries TC03\_\allowbreak{}Distance\-Markers\-And\-Occlusion\-Geometry\-Are\-Plausible}}\enspace\textcolor{zText}{Checks the Near, Mid, Far, and Occluded marker placement and the physical occlusion wall.} \\
\ubtestbadge{04} & \textcolor{zGoldDeep!78!black}{\texttt{\bfseries TC04\_\allowbreak{}Weapon\-Profiles\-Contain\-All\-Required\-Layers("Weapon A", "Assault Rifle")}}\enspace\textcolor{zText}{Checks that the weapon A profile contains every required near, distant, loop, stop, suppressed, tail, cadence, randomization, cooldown, and concurrency field.} \\
\ubtestbadge{05} & \textcolor{zGoldDeep!78!black}{\texttt{\bfseries TC04\_\allowbreak{}Weapon\-Profiles\-Contain\-All\-Required\-Layers("Weapon B", "Carbine")}}\enspace\textcolor{zText}{Checks that the weapon B profile contains every required near, distant, loop, stop, suppressed, tail, cadence, randomization, cooldown, and concurrency field.} \\
\ubtestbadge{06} & \textcolor{zGoldDeep!78!black}{\texttt{\bfseries TC04\_\allowbreak{}Weapon\-Profiles\-Contain\-All\-Required\-Layers("Weapon C", "Light Machine Gun")}}\enspace\textcolor{zText}{Checks that the weapon C profile contains every required near, distant, loop, stop, suppressed, tail, cadence, randomization, cooldown, and concurrency field.} \\
\ubtestbadge{07} & \textcolor{zGoldDeep!78!black}{\texttt{\bfseries TC05\_\allowbreak{}Profiles\-Are\-Distinct\-And\-Use\-The\-Supplied\-Library}}\enspace\textcolor{zText}{Checks that the three weapon profiles are distinct and reference only the supplied audio library.} \\
\ubtestbadge{08} & \textcolor{zGoldDeep!78!black}{\texttt{\bfseries TC06\_\allowbreak{}Mixer\-Has\-Required\-Groups\-And\-Snapshots}}\enspace\textcolor{zText}{Checks that GunAudio.mixer contains the required group hierarchy and Outdoor, Indoor, and Corridor snapshots.} \\
\ubtestbadge{09} & \textcolor{zGoldDeep!78!black}{\texttt{\bfseries TC07\_\allowbreak{}Minimal\-Automation\-Surface\-Exists\-On\-Scene\-Components}}\enspace\textcolor{zText}{Checks that scene components expose the required emitter and global-service automation API.} \\
\ubtestbadge{10} & \textcolor{zGoldDeep!78!black}{\texttt{\bfseries TC08\_\allowbreak{}Scene\-Wires\-Service\-Pool\-Emitters\-Occlusion\-And\-Stress\-Test}}\enspace\textcolor{zText}{Checks that the scene connects the global service, voice pool, emitters, occlusion, and stress-test components.} \\
\ubtestbadge{11} & \textcolor{zGoldDeep!78!black}{\texttt{\bfseries TC09\_\allowbreak{}Source\-Code\-Uses\-Stable\-Clock\-Pooling\-And\-Cleanup\-Paths}}\enspace\textcolor{zText}{Checks source code for an unscaled stable clock, pooled voices, and automatic-loop cleanup paths.} \\
\ubtestbadge{12} & \textcolor{zGoldDeep!78!black}{\texttt{\bfseries TC10\_\allowbreak{}Used\-Short\-Gunshots\-Are\-Non\-Streaming\-And\-Compressed}}\enspace\textcolor{zText}{Checks that used short gunshot clips are compressed and do not use streaming load mode.} \\
\ubtestbadge{13} & \textcolor{zGoldDeep!78!black}{\texttt{\bfseries TC11\_\allowbreak{}Used\-Loops\-And\-Tails\-Are\-Compressed}}\enspace\textcolor{zText}{Checks that used automatic loops and environment tails use compressed import settings.} \\
\ubtestbadge{14} & \textcolor{zGoldDeep!78!black}{\texttt{\bfseries TC12\_\allowbreak{}Configured\-Casing\-Clips\-Cover\-Surfaces\-And\-Use\-Short\-Sound\-Import\-Policy}}\enspace\textcolor{zText}{Checks that configured casing clips cover the required surface families and use the short-sound import policy.} \\
\ubtestbadge{15} & \textcolor{zGoldDeep!78!black}{\texttt{\bfseries TC13\_\allowbreak{}Debug\-Panel\-And\-Editor\-Audit\-Are\-Present}}\enspace\textcolor{zText}{Checks that the debug panel and editor audit command are present.} \\
\ubtestbadge{16} & \textcolor{zGoldDeep!78!black}{\texttt{\bfseries TC14\_\allowbreak{}Supplied\-Library\-Is\-Preserved\-And\-No\-Extra\-Audio\-Was\-Imported}}\enspace\textcolor{zText}{Checks that the supplied library is preserved and no additional audio files were imported.} \\
\ubtestbadge{17} & \textcolor{zGoldDeep!78!black}{\texttt{\bfseries TC15\_\allowbreak{}System\-Boots\-With\-One\-Service\-And\-ABounded\-Prewarmed\-Pool}}\enspace\textcolor{zText}{Checks that runtime starts with one global service and a bounded, prewarmed voice pool.} \\
\ubtestbadge{18} & \textcolor{zGoldDeep!78!black}{\texttt{\bfseries TC16\_\allowbreak{}Single\-Fire\-Advances\-Logical\-Shot\-Count}}\enspace\textcolor{zText}{Checks that FireSingle() advances the emitter's logical shot count.} \\
\ubtestbadge{19} & \textcolor{zGoldDeep!78!black}{\texttt{\bfseries TC17\_\allowbreak{}One\-Shot\-Variation\-Avoids\-Immediate\-Repeats}}\enspace\textcolor{zText}{Checks that one-shot variation avoids replaying the same clip immediately when alternatives exist.} \\
\ubtestbadge{20} & \textcolor{zGoldDeep!78!black}{\texttt{\bfseries TC18\_\allowbreak{}Automatic\-Fire\-Lifecycle\-Has\-One\-Loop\-And\-Stops\-Cleanly}}\enspace\textcolor{zText}{Checks that automatic fire uses at most one loop and stops it cleanly when firing ends.} \\
\ubtestbadge{21} & \textcolor{zGoldDeep!78!black}{\texttt{\bfseries TC19\_\allowbreak{}Disable\-During\-Unscaled\-Automatic\-Fire\-Cannot\-Leak\-ALoop}}\enspace\textcolor{zText}{Checks that disabling an emitter during unscaled automatic fire cannot leave an orphaned loop.} \\
\ubtestbadge{22} & \textcolor{zGoldDeep!78!black}{\texttt{\bfseries TC20\_\allowbreak{}Suppressor\-Switch\-Does\-Not\-Overlap\-Old\-And\-New\-Loops}}\enspace\textcolor{zText}{Checks that switching suppressor state replaces the active loop without overlapping old and new loops.} \\
\ubtestbadge{23} & \textcolor{zGoldDeep!78!black}{\texttt{\bfseries TC21\_\allowbreak{}Multiple\-Emitters\-Remain\-Independent}}\enspace\textcolor{zText}{Checks that multiple emitters maintain independent firing state and counters.} \\
\ubtestbadge{24} & \textcolor{zGoldDeep!78!black}{\texttt{\bfseries TC22\_\allowbreak{}Rapid\-Requests\-Respect\-Pool\-Cap\-And\-Do\-Not\-Allocate\-After\-Warmup}}\enspace\textcolor{zText}{Checks that rapid shot requests respect the voice cap and do not grow the pool after warm-up.} \\
\ubtestbadge{25} & \textcolor{zGoldDeep!78!black}{\texttt{\bfseries TC23\_\allowbreak{}Distance\-Blend\-Transitions\-From\-Near\-To\-Distant}}\enspace\textcolor{zText}{Checks that reported layer weights transition smoothly from near to distant as listener distance increases.} \\
\ubtestbadge{26} & \textcolor{zGoldDeep!78!black}{\texttt{\bfseries TC24\_\allowbreak{}Occlusion\-Raycast\-Lowers\-The\-Reported\-Cutoff\-And\-Recovers}}\enspace\textcolor{zText}{Checks that physical occlusion lowers the reported low-pass cutoff and that it recovers after line of sight returns.} \\
\ubtestbadge{27} & \textcolor{zGoldDeep!78!black}{\texttt{\bfseries TC25\_\allowbreak{}Linear\-Bus\-Volume\-Conversion\-Is\-Finite\-And\-Correct\-At\-Bounds}}\enspace\textcolor{zText}{Checks finite and correct linear-to-decibel bus-volume conversion at the supported bounds.} \\
\ubtestbadge{28} & \textcolor{zGoldDeep!78!black}{\texttt{\bfseries TC26\_\allowbreak{}Runtime\-Sources\-Are\-Spatial\-And\-Mixer\-Routed}}\enspace\textcolor{zText}{Checks that runtime AudioSources are 3D and routed to the appropriate mixer groups.} \\
\ubtestbadge{29} & \textcolor{zGoldDeep!78!black}{\texttt{\bfseries TC27\_\allowbreak{}Reloading\-Scene\-Does\-Not\-Duplicate\-The\-Service}}\enspace\textcolor{zText}{Checks that reloading the scene does not create a duplicate global audio service.} \\
\noalign{\vskip 2pt}
\multicolumn{2}{@{}p{14.4cm}@{}}{\textcolor{zRockLite!42}{\rule{\linewidth}{0.3pt}}}\\[3pt]
\multicolumn{2}{@{}p{14.4cm}@{}}{\ubtaskhead{zRoyal}{0072-CastleDestruction}{Classify Castle Structures into Hierarchy Groups}{zShrine}{Hard}{9}}\\*[2pt]
\ubtestbadge{01} & \textcolor{zGoldDeep!78!black}{\texttt{\bfseries Round\-Bastions\_\allowbreak{}At\-Least50\-Percent\-Of\-Groups\-Correct}}\enspace\textcolor{zText}{Checks that at least 50\% of required round-bastion groups contain the correct complete structures without adjoining wall sections.} \\
\ubtestbadge{02} & \textcolor{zGoldDeep!78!black}{\texttt{\bfseries Round\-Bastions\_\allowbreak{}100\-Percent\-Of\-Groups\-Correct}}\enspace\textcolor{zText}{Checks that all required round-bastion groups contain the correct complete structures without adjoining wall sections.} \\
\ubtestbadge{03} & \textcolor{zGoldDeep!78!black}{\texttt{\bfseries Square\-Bastions\_\allowbreak{}At\-Least50\-Percent\-Of\-Groups\-Correct}}\enspace\textcolor{zText}{Checks that at least 50\% of required square-bastion groups contain the correct complete structures without adjoining wall sections.} \\
\ubtestbadge{04} & \textcolor{zGoldDeep!78!black}{\texttt{\bfseries Square\-Bastions\_\allowbreak{}100\-Percent\-Of\-Groups\-Correct}}\enspace\textcolor{zText}{Checks that all required square-bastion groups contain the correct complete structures without adjoining wall sections.} \\
\ubtestbadge{05} & \textcolor{zGoldDeep!78!black}{\texttt{\bfseries Elevated\-Bastions\_\allowbreak{}At\-Least50\-Percent\-Of\-Groups\-Correct}}\enspace\textcolor{zText}{Checks that at least 50\% of required elevated-bastion groups contain the correct complete structures without adjoining wall sections.} \\
\ubtestbadge{06} & \textcolor{zGoldDeep!78!black}{\texttt{\bfseries Elevated\-Bastions\_\allowbreak{}100\-Percent\-Of\-Groups\-Correct}}\enspace\textcolor{zText}{Checks that all required elevated-bastion groups contain the correct complete structures without adjoining wall sections.} \\
\ubtestbadge{07} & \textcolor{zGoldDeep!78!black}{\texttt{\bfseries Hoarding\-Wood\_\allowbreak{}At\-Least50\-Percent\-Of\-Groups\-Correct}}\enspace\textcolor{zText}{Checks that at least 50\% of required wooden-hoarding groups contain the correct complete structures without adjoining wall sections.} \\
\ubtestbadge{08} & \textcolor{zGoldDeep!78!black}{\texttt{\bfseries Hoarding\-Wood\_\allowbreak{}100\-Percent\-Of\-Groups\-Correct}}\enspace\textcolor{zText}{Checks that all required wooden-hoarding groups contain the correct complete structures without adjoining wall sections.} \\
\ubtestbadge{09} & \textcolor{zGoldDeep!78!black}{\texttt{\bfseries Main\-Gate\-Group\-Is\-Correct}}\enspace\textcolor{zText}{Checks that Main\_Gate contains the complete gate structure without adjoining continuous-wall pieces.} \\
\noalign{\vskip 2pt}
\multicolumn{2}{@{}p{14.4cm}@{}}{\textcolor{zRockLite!42}{\rule{\linewidth}{0.3pt}}}\\[3pt]
\noalign{\vskip 3pt}
\multicolumn{2}{@{}p{14.4cm}@{}}{\ubcategory{zPlain}{Scene \& Level Construction}{19 tasks \textperiodcentered\ 365 milestone tests}}\\*[4pt]
\multicolumn{2}{@{}p{14.4cm}@{}}{\ubtaskhead{zPlain}{0001-ObjectIdentificationPrefabSorting}{Object Identification and Categorization}{zGoldDeep}{Easy}{33}}\\*[2pt]
\ubtestbadge{01} & \textcolor{zGoldDeep!78!black}{\texttt{\bfseries Answer\-Parents\-Do\-Not\-Expose\-Numbered\-Solution\-Prefixes}}\enspace\textcolor{zText}{Checks that answer parents do not expose numbered solution prefixes.} \\
\ubtestbadge{02} & \textcolor{zGoldDeep!78!black}{\texttt{\bfseries Numbered\-Item\-Is\-Placed\-Under\-Correct\-Named\-Parent(1, "Category 01")}}\enspace\textcolor{zText}{Checks that Item\_01 is placed under the answer parent named category 01.} \\
\ubtestbadge{03} & \textcolor{zGoldDeep!78!black}{\texttt{\bfseries Numbered\-Item\-Is\-Placed\-Under\-Correct\-Named\-Parent(10, "Category 10")}}\enspace\textcolor{zText}{Checks that Item\_10 is placed under the answer parent named category 10.} \\
\ubtestbadge{04} & \textcolor{zGoldDeep!78!black}{\texttt{\bfseries Numbered\-Item\-Is\-Placed\-Under\-Correct\-Named\-Parent(11, "Category 11")}}\enspace\textcolor{zText}{Checks that Item\_11 is placed under the answer parent named category 11.} \\
\ubtestbadge{05} & \textcolor{zGoldDeep!78!black}{\texttt{\bfseries Numbered\-Item\-Is\-Placed\-Under\-Correct\-Named\-Parent(12, "Category 12")}}\enspace\textcolor{zText}{Checks that Item\_12 is placed under the answer parent named category 12.} \\
\ubtestbadge{06} & \textcolor{zGoldDeep!78!black}{\texttt{\bfseries Numbered\-Item\-Is\-Placed\-Under\-Correct\-Named\-Parent(13, "Category 13")}}\enspace\textcolor{zText}{Checks that Item\_13 is placed under the answer parent named category 13.} \\
\ubtestbadge{07} & \textcolor{zGoldDeep!78!black}{\texttt{\bfseries Numbered\-Item\-Is\-Placed\-Under\-Correct\-Named\-Parent(14, "Category 14")}}\enspace\textcolor{zText}{Checks that Item\_14 is placed under the answer parent named category 14.} \\
\ubtestbadge{08} & \textcolor{zGoldDeep!78!black}{\texttt{\bfseries Numbered\-Item\-Is\-Placed\-Under\-Correct\-Named\-Parent(15, "Category 15")}}\enspace\textcolor{zText}{Checks that Item\_15 is placed under the answer parent named category 15.} \\
\ubtestbadge{09} & \textcolor{zGoldDeep!78!black}{\texttt{\bfseries Numbered\-Item\-Is\-Placed\-Under\-Correct\-Named\-Parent(16, "Category 16")}}\enspace\textcolor{zText}{Checks that Item\_16 is placed under the answer parent named category 16.} \\
\ubtestbadge{10} & \textcolor{zGoldDeep!78!black}{\texttt{\bfseries Numbered\-Item\-Is\-Placed\-Under\-Correct\-Named\-Parent(17, "Category 17")}}\enspace\textcolor{zText}{Checks that Item\_17 is placed under the answer parent named category 17.} \\
\ubtestbadge{11} & \textcolor{zGoldDeep!78!black}{\texttt{\bfseries Numbered\-Item\-Is\-Placed\-Under\-Correct\-Named\-Parent(18, "Category 18")}}\enspace\textcolor{zText}{Checks that Item\_18 is placed under the answer parent named category 18.} \\
\ubtestbadge{12} & \textcolor{zGoldDeep!78!black}{\texttt{\bfseries Numbered\-Item\-Is\-Placed\-Under\-Correct\-Named\-Parent(19, "Category 19")}}\enspace\textcolor{zText}{Checks that Item\_19 is placed under the answer parent named category 19.} \\
\ubtestbadge{13} & \textcolor{zGoldDeep!78!black}{\texttt{\bfseries Numbered\-Item\-Is\-Placed\-Under\-Correct\-Named\-Parent(2, "Category 02")}}\enspace\textcolor{zText}{Checks that Item\_02 is placed under the answer parent named category 02.} \\
\ubtestbadge{14} & \textcolor{zGoldDeep!78!black}{\texttt{\bfseries Numbered\-Item\-Is\-Placed\-Under\-Correct\-Named\-Parent(20, "Category 20")}}\enspace\textcolor{zText}{Checks that Item\_20 is placed under the answer parent named category 20.} \\
\ubtestbadge{15} & \textcolor{zGoldDeep!78!black}{\texttt{\bfseries Numbered\-Item\-Is\-Placed\-Under\-Correct\-Named\-Parent(21, "Category 21")}}\enspace\textcolor{zText}{Checks that Item\_21 is placed under the answer parent named category 21.} \\
\ubtestbadge{16} & \textcolor{zGoldDeep!78!black}{\texttt{\bfseries Numbered\-Item\-Is\-Placed\-Under\-Correct\-Named\-Parent(22, "Category 22")}}\enspace\textcolor{zText}{Checks that Item\_22 is placed under the answer parent named category 22.} \\
\ubtestbadge{17} & \textcolor{zGoldDeep!78!black}{\texttt{\bfseries Numbered\-Item\-Is\-Placed\-Under\-Correct\-Named\-Parent(23, "Category 23")}}\enspace\textcolor{zText}{Checks that Item\_23 is placed under the answer parent named category 23.} \\
\ubtestbadge{18} & \textcolor{zGoldDeep!78!black}{\texttt{\bfseries Numbered\-Item\-Is\-Placed\-Under\-Correct\-Named\-Parent(24, "Category 24")}}\enspace\textcolor{zText}{Checks that Item\_24 is placed under the answer parent named category 24.} \\
\ubtestbadge{19} & \textcolor{zGoldDeep!78!black}{\texttt{\bfseries Numbered\-Item\-Is\-Placed\-Under\-Correct\-Named\-Parent(25, "Category 25")}}\enspace\textcolor{zText}{Checks that Item\_25 is placed under the answer parent named category 25.} \\
\ubtestbadge{20} & \textcolor{zGoldDeep!78!black}{\texttt{\bfseries Numbered\-Item\-Is\-Placed\-Under\-Correct\-Named\-Parent(26, "Category 26")}}\enspace\textcolor{zText}{Checks that Item\_26 is placed under the answer parent named category 26.} \\
\ubtestbadge{21} & \textcolor{zGoldDeep!78!black}{\texttt{\bfseries Numbered\-Item\-Is\-Placed\-Under\-Correct\-Named\-Parent(27, "Category 27")}}\enspace\textcolor{zText}{Checks that Item\_27 is placed under the answer parent named category 27.} \\
\ubtestbadge{22} & \textcolor{zGoldDeep!78!black}{\texttt{\bfseries Numbered\-Item\-Is\-Placed\-Under\-Correct\-Named\-Parent(28, "Category 28")}}\enspace\textcolor{zText}{Checks that Item\_28 is placed under the answer parent named category 28.} \\
\ubtestbadge{23} & \textcolor{zGoldDeep!78!black}{\texttt{\bfseries Numbered\-Item\-Is\-Placed\-Under\-Correct\-Named\-Parent(29, "Category 29")}}\enspace\textcolor{zText}{Checks that Item\_29 is placed under the answer parent named category 29.} \\
\ubtestbadge{24} & \textcolor{zGoldDeep!78!black}{\texttt{\bfseries Numbered\-Item\-Is\-Placed\-Under\-Correct\-Named\-Parent(3, "Category 03")}}\enspace\textcolor{zText}{Checks that Item\_03 is placed under the answer parent named category 03.} \\
\ubtestbadge{25} & \textcolor{zGoldDeep!78!black}{\texttt{\bfseries Numbered\-Item\-Is\-Placed\-Under\-Correct\-Named\-Parent(30, "Category 30")}}\enspace\textcolor{zText}{Checks that Item\_30 is placed under the answer parent named category 30.} \\
\ubtestbadge{26} & \textcolor{zGoldDeep!78!black}{\texttt{\bfseries Numbered\-Item\-Is\-Placed\-Under\-Correct\-Named\-Parent(4, "Category 04")}}\enspace\textcolor{zText}{Checks that Item\_04 is placed under the answer parent named category 04.} \\
\ubtestbadge{27} & \textcolor{zGoldDeep!78!black}{\texttt{\bfseries Numbered\-Item\-Is\-Placed\-Under\-Correct\-Named\-Parent(5, "Category 05")}}\enspace\textcolor{zText}{Checks that Item\_05 is placed under the answer parent named category 05.} \\
\ubtestbadge{28} & \textcolor{zGoldDeep!78!black}{\texttt{\bfseries Numbered\-Item\-Is\-Placed\-Under\-Correct\-Named\-Parent(6, "Category 06")}}\enspace\textcolor{zText}{Checks that Item\_06 is placed under the answer parent named category 06.} \\
\ubtestbadge{29} & \textcolor{zGoldDeep!78!black}{\texttt{\bfseries Numbered\-Item\-Is\-Placed\-Under\-Correct\-Named\-Parent(7, "Category 07")}}\enspace\textcolor{zText}{Checks that Item\_07 is placed under the answer parent named category 07.} \\
\ubtestbadge{30} & \textcolor{zGoldDeep!78!black}{\texttt{\bfseries Numbered\-Item\-Is\-Placed\-Under\-Correct\-Named\-Parent(8, "Category 08")}}\enspace\textcolor{zText}{Checks that Item\_08 is placed under the answer parent named category 08.} \\
\ubtestbadge{31} & \textcolor{zGoldDeep!78!black}{\texttt{\bfseries Numbered\-Item\-Is\-Placed\-Under\-Correct\-Named\-Parent(9, "Category 09")}}\enspace\textcolor{zText}{Checks that Item\_09 is placed under the answer parent named category 09.} \\
\ubtestbadge{32} & \textcolor{zGoldDeep!78!black}{\texttt{\bfseries Numbered\-Items\-And\-Their\-Children\-Remain\-Anonymous}}\enspace\textcolor{zText}{Checks that numbered items and their children remain anonymous.} \\
\ubtestbadge{33} & \textcolor{zGoldDeep!78!black}{\texttt{\bfseries Scene\-Contains\-Required\-Identification\-Scaffold}}\enspace\textcolor{zText}{Checks that scene contains required identification scaffold.} \\
\noalign{\vskip 2pt}
\multicolumn{2}{@{}p{14.4cm}@{}}{\textcolor{zRockLite!42}{\rule{\linewidth}{0.3pt}}}\\[3pt]
\multicolumn{2}{@{}p{14.4cm}@{}}{\ubtaskhead{zPlain}{0021-StreetLayout}{Street Intersection Assembly}{zGoldDeep}{Easy}{10}}\\*[2pt]
\ubtestbadge{01} & \textcolor{zGoldDeep!78!black}{\texttt{\bfseries Buildings\_\allowbreak{}Are\-Parented\-And\-Do\-Not\-Overlap\-Roads}}\enspace\textcolor{zText}{Checks that buildings are parented and do not overlap roads.} \\
\ubtestbadge{02} & \textcolor{zGoldDeep!78!black}{\texttt{\bfseries Buildings\_\allowbreak{}Do\-Not\-Overlap\-Each\-Other}}\enspace\textcolor{zText}{Checks that buildings do not overlap each other.} \\
\ubtestbadge{03} & \textcolor{zGoldDeep!78!black}{\texttt{\bfseries Buildings\_\allowbreak{}Use\-All\-Types\-And\-Cover\-Four\-Corners}}\enspace\textcolor{zText}{Checks that buildings use all types and cover four corners.} \\
\ubtestbadge{04} & \textcolor{zGoldDeep!78!black}{\texttt{\bfseries Ground\-And\-Road\-Cross\_\allowbreak{}Are\-At\-Origin\-And\-Under\-Correct\-Parents}}\enspace\textcolor{zText}{Checks that ground and road cross are at origin and under correct parents.} \\
\ubtestbadge{05} & \textcolor{zGoldDeep!78!black}{\texttt{\bfseries Placed\-Prefab\-Roots\_\allowbreak{}Have\-Zero\-YPosition}}\enspace\textcolor{zText}{Checks that placed Prefab roots have zero Y position.} \\
\ubtestbadge{06} & \textcolor{zGoldDeep!78!black}{\texttt{\bfseries Road\-Straights\_\allowbreak{}Form\-Continuous\-Four\-Direction\-Roads}}\enspace\textcolor{zText}{Checks that road straights form continuous four direction roads.} \\
\ubtestbadge{07} & \textcolor{zGoldDeep!78!black}{\texttt{\bfseries Roads\_\allowbreak{}Are\-Under\-Roads\-Parent}}\enspace\textcolor{zText}{Checks that roads are under roads parent.} \\
\ubtestbadge{08} & \textcolor{zGoldDeep!78!black}{\texttt{\bfseries Scene\_\allowbreak{}Has\-Required\-Root\-Parents}}\enspace\textcolor{zText}{Checks that scene has required root parents.} \\
\ubtestbadge{09} & \textcolor{zGoldDeep!78!black}{\texttt{\bfseries Trees\_\allowbreak{}Are\-Parented\-And\-Parallel\-To\-Roads}}\enspace\textcolor{zText}{Checks that trees are parented and parallel to roads.} \\
\ubtestbadge{10} & \textcolor{zGoldDeep!78!black}{\texttt{\bfseries Trees\_\allowbreak{}Use\-All\-Types\-And\-Appear\-On\-Eight\-Road\-Sides}}\enspace\textcolor{zText}{Checks that trees use all types and appear on eight road sides.} \\
\noalign{\vskip 2pt}
\multicolumn{2}{@{}p{14.4cm}@{}}{\textcolor{zRockLite!42}{\rule{\linewidth}{0.3pt}}}\\[3pt]
\multicolumn{2}{@{}p{14.4cm}@{}}{\ubtaskhead{zPlain}{0040-ObjectIdentificationPrefabSorting}{Object Identification and Categorization}{zGoldDeep}{Easy}{33}}\\*[2pt]
\ubtestbadge{01} & \textcolor{zGoldDeep!78!black}{\texttt{\bfseries Answer\-Parents\-Do\-Not\-Expose\-Numbered\-Solution\-Prefixes}}\enspace\textcolor{zText}{Checks that answer parents do not expose numbered solution prefixes.} \\
\ubtestbadge{02} & \textcolor{zGoldDeep!78!black}{\texttt{\bfseries Numbered\-Item\-Is\-Placed\-Under\-Correct\-Named\-Parent(1, "Category 01")}}\enspace\textcolor{zText}{Checks that Item\_01 is placed under the answer parent named category 01.} \\
\ubtestbadge{03} & \textcolor{zGoldDeep!78!black}{\texttt{\bfseries Numbered\-Item\-Is\-Placed\-Under\-Correct\-Named\-Parent(10, "Category 10")}}\enspace\textcolor{zText}{Checks that Item\_10 is placed under the answer parent named category 10.} \\
\ubtestbadge{04} & \textcolor{zGoldDeep!78!black}{\texttt{\bfseries Numbered\-Item\-Is\-Placed\-Under\-Correct\-Named\-Parent(11, "Category 11")}}\enspace\textcolor{zText}{Checks that Item\_11 is placed under the answer parent named category 11.} \\
\ubtestbadge{05} & \textcolor{zGoldDeep!78!black}{\texttt{\bfseries Numbered\-Item\-Is\-Placed\-Under\-Correct\-Named\-Parent(12, "Category 12")}}\enspace\textcolor{zText}{Checks that Item\_12 is placed under the answer parent named category 12.} \\
\ubtestbadge{06} & \textcolor{zGoldDeep!78!black}{\texttt{\bfseries Numbered\-Item\-Is\-Placed\-Under\-Correct\-Named\-Parent(13, "Category 13")}}\enspace\textcolor{zText}{Checks that Item\_13 is placed under the answer parent named category 13.} \\
\ubtestbadge{07} & \textcolor{zGoldDeep!78!black}{\texttt{\bfseries Numbered\-Item\-Is\-Placed\-Under\-Correct\-Named\-Parent(14, "Category 14")}}\enspace\textcolor{zText}{Checks that Item\_14 is placed under the answer parent named category 14.} \\
\ubtestbadge{08} & \textcolor{zGoldDeep!78!black}{\texttt{\bfseries Numbered\-Item\-Is\-Placed\-Under\-Correct\-Named\-Parent(15, "Category 15")}}\enspace\textcolor{zText}{Checks that Item\_15 is placed under the answer parent named category 15.} \\
\ubtestbadge{09} & \textcolor{zGoldDeep!78!black}{\texttt{\bfseries Numbered\-Item\-Is\-Placed\-Under\-Correct\-Named\-Parent(16, "Category 16")}}\enspace\textcolor{zText}{Checks that Item\_16 is placed under the answer parent named category 16.} \\
\ubtestbadge{10} & \textcolor{zGoldDeep!78!black}{\texttt{\bfseries Numbered\-Item\-Is\-Placed\-Under\-Correct\-Named\-Parent(17, "Category 17")}}\enspace\textcolor{zText}{Checks that Item\_17 is placed under the answer parent named category 17.} \\
\ubtestbadge{11} & \textcolor{zGoldDeep!78!black}{\texttt{\bfseries Numbered\-Item\-Is\-Placed\-Under\-Correct\-Named\-Parent(18, "Category 18")}}\enspace\textcolor{zText}{Checks that Item\_18 is placed under the answer parent named category 18.} \\
\ubtestbadge{12} & \textcolor{zGoldDeep!78!black}{\texttt{\bfseries Numbered\-Item\-Is\-Placed\-Under\-Correct\-Named\-Parent(19, "Category 19")}}\enspace\textcolor{zText}{Checks that Item\_19 is placed under the answer parent named category 19.} \\
\ubtestbadge{13} & \textcolor{zGoldDeep!78!black}{\texttt{\bfseries Numbered\-Item\-Is\-Placed\-Under\-Correct\-Named\-Parent(2, "Category 02")}}\enspace\textcolor{zText}{Checks that Item\_02 is placed under the answer parent named category 02.} \\
\ubtestbadge{14} & \textcolor{zGoldDeep!78!black}{\texttt{\bfseries Numbered\-Item\-Is\-Placed\-Under\-Correct\-Named\-Parent(20, "Category 20")}}\enspace\textcolor{zText}{Checks that Item\_20 is placed under the answer parent named category 20.} \\
\ubtestbadge{15} & \textcolor{zGoldDeep!78!black}{\texttt{\bfseries Numbered\-Item\-Is\-Placed\-Under\-Correct\-Named\-Parent(21, "Category 21")}}\enspace\textcolor{zText}{Checks that Item\_21 is placed under the answer parent named category 21.} \\
\ubtestbadge{16} & \textcolor{zGoldDeep!78!black}{\texttt{\bfseries Numbered\-Item\-Is\-Placed\-Under\-Correct\-Named\-Parent(22, "Category 22")}}\enspace\textcolor{zText}{Checks that Item\_22 is placed under the answer parent named category 22.} \\
\ubtestbadge{17} & \textcolor{zGoldDeep!78!black}{\texttt{\bfseries Numbered\-Item\-Is\-Placed\-Under\-Correct\-Named\-Parent(23, "Category 23")}}\enspace\textcolor{zText}{Checks that Item\_23 is placed under the answer parent named category 23.} \\
\ubtestbadge{18} & \textcolor{zGoldDeep!78!black}{\texttt{\bfseries Numbered\-Item\-Is\-Placed\-Under\-Correct\-Named\-Parent(24, "Category 24")}}\enspace\textcolor{zText}{Checks that Item\_24 is placed under the answer parent named category 24.} \\
\ubtestbadge{19} & \textcolor{zGoldDeep!78!black}{\texttt{\bfseries Numbered\-Item\-Is\-Placed\-Under\-Correct\-Named\-Parent(25, "Category 25")}}\enspace\textcolor{zText}{Checks that Item\_25 is placed under the answer parent named category 25.} \\
\ubtestbadge{20} & \textcolor{zGoldDeep!78!black}{\texttt{\bfseries Numbered\-Item\-Is\-Placed\-Under\-Correct\-Named\-Parent(26, "Category 26")}}\enspace\textcolor{zText}{Checks that Item\_26 is placed under the answer parent named category 26.} \\
\ubtestbadge{21} & \textcolor{zGoldDeep!78!black}{\texttt{\bfseries Numbered\-Item\-Is\-Placed\-Under\-Correct\-Named\-Parent(27, "Category 27")}}\enspace\textcolor{zText}{Checks that Item\_27 is placed under the answer parent named category 27.} \\
\ubtestbadge{22} & \textcolor{zGoldDeep!78!black}{\texttt{\bfseries Numbered\-Item\-Is\-Placed\-Under\-Correct\-Named\-Parent(28, "Category 28")}}\enspace\textcolor{zText}{Checks that Item\_28 is placed under the answer parent named category 28.} \\
\ubtestbadge{23} & \textcolor{zGoldDeep!78!black}{\texttt{\bfseries Numbered\-Item\-Is\-Placed\-Under\-Correct\-Named\-Parent(29, "Category 29")}}\enspace\textcolor{zText}{Checks that Item\_29 is placed under the answer parent named category 29.} \\
\ubtestbadge{24} & \textcolor{zGoldDeep!78!black}{\texttt{\bfseries Numbered\-Item\-Is\-Placed\-Under\-Correct\-Named\-Parent(3, "Category 03")}}\enspace\textcolor{zText}{Checks that Item\_03 is placed under the answer parent named category 03.} \\
\ubtestbadge{25} & \textcolor{zGoldDeep!78!black}{\texttt{\bfseries Numbered\-Item\-Is\-Placed\-Under\-Correct\-Named\-Parent(30, "Category 30")}}\enspace\textcolor{zText}{Checks that Item\_30 is placed under the answer parent named category 30.} \\
\ubtestbadge{26} & \textcolor{zGoldDeep!78!black}{\texttt{\bfseries Numbered\-Item\-Is\-Placed\-Under\-Correct\-Named\-Parent(4, "Category 04")}}\enspace\textcolor{zText}{Checks that Item\_04 is placed under the answer parent named category 04.} \\
\ubtestbadge{27} & \textcolor{zGoldDeep!78!black}{\texttt{\bfseries Numbered\-Item\-Is\-Placed\-Under\-Correct\-Named\-Parent(5, "Category 05")}}\enspace\textcolor{zText}{Checks that Item\_05 is placed under the answer parent named category 05.} \\
\ubtestbadge{28} & \textcolor{zGoldDeep!78!black}{\texttt{\bfseries Numbered\-Item\-Is\-Placed\-Under\-Correct\-Named\-Parent(6, "Category 06")}}\enspace\textcolor{zText}{Checks that Item\_06 is placed under the answer parent named category 06.} \\
\ubtestbadge{29} & \textcolor{zGoldDeep!78!black}{\texttt{\bfseries Numbered\-Item\-Is\-Placed\-Under\-Correct\-Named\-Parent(7, "Category 07")}}\enspace\textcolor{zText}{Checks that Item\_07 is placed under the answer parent named category 07.} \\
\ubtestbadge{30} & \textcolor{zGoldDeep!78!black}{\texttt{\bfseries Numbered\-Item\-Is\-Placed\-Under\-Correct\-Named\-Parent(8, "Category 08")}}\enspace\textcolor{zText}{Checks that Item\_08 is placed under the answer parent named category 08.} \\
\ubtestbadge{31} & \textcolor{zGoldDeep!78!black}{\texttt{\bfseries Numbered\-Item\-Is\-Placed\-Under\-Correct\-Named\-Parent(9, "Category 09")}}\enspace\textcolor{zText}{Checks that Item\_09 is placed under the answer parent named category 09.} \\
\ubtestbadge{32} & \textcolor{zGoldDeep!78!black}{\texttt{\bfseries Numbered\-Items\-And\-Their\-Children\-Remain\-Anonymous}}\enspace\textcolor{zText}{Checks that numbered items and their children remain anonymous.} \\
\ubtestbadge{33} & \textcolor{zGoldDeep!78!black}{\texttt{\bfseries Scene\-Contains\-Required\-Identification\-Scaffold}}\enspace\textcolor{zText}{Checks that scene contains required identification scaffold.} \\
\noalign{\vskip 2pt}
\multicolumn{2}{@{}p{14.4cm}@{}}{\textcolor{zRockLite!42}{\rule{\linewidth}{0.3pt}}}\\[3pt]
\multicolumn{2}{@{}p{14.4cm}@{}}{\ubtaskhead{zPlain}{0005-BuildHouseReachability}{Paper-Thin Multistory House and Floor Reachability}{zGoldDeep}{Medium}{4}}\\*[2pt]
\ubtestbadge{01} & \textcolor{zGoldDeep!78!black}{\texttt{\bfseries Build\-House\_\allowbreak{}Contains\-Required\-Named\-Paper\-Parts}}\enspace\textcolor{zText}{Checks that build house contains required named paper parts.} \\
\ubtestbadge{02} & \textcolor{zGoldDeep!78!black}{\texttt{\bfseries Build\-House\_\allowbreak{}Cross\-Floor\-Room\-Route\-Is\-Reachable}}\enspace\textcolor{zText}{Checks that build house cross floor room route is reachable.} \\
\ubtestbadge{03} & \textcolor{zGoldDeep!78!black}{\texttt{\bfseries Build\-House\_\allowbreak{}Rooms\-On\-Each\-Floor\-Are\-Reachable}}\enspace\textcolor{zText}{Checks that build house rooms on each floor are reachable.} \\
\ubtestbadge{04} & \textcolor{zGoldDeep!78!black}{\texttt{\bfseries Build\-House\_\allowbreak{}Stairs\-Connect\-Adjacent\-Floors}}\enspace\textcolor{zText}{Checks that build house stairs connect adjacent floors.} \\
\noalign{\vskip 2pt}
\multicolumn{2}{@{}p{14.4cm}@{}}{\textcolor{zRockLite!42}{\rule{\linewidth}{0.3pt}}}\\[3pt]
\multicolumn{2}{@{}p{14.4cm}@{}}{\ubtaskhead{zPlain}{0020-InteriorFurniturePlacement}{Interior Furniture Placement}{zGoldDeep}{Medium}{39}}\\*[2pt]
\ubtestbadge{01} & \textcolor{zGoldDeep!78!black}{\texttt{\bfseries Bathroom\-Bathtub\-And\-Curtain\_\allowbreak{}Are\-At\-Expected\-South\-Side\-Z}}\enspace\textcolor{zText}{Checks that bathroom bathtub and curtain are at expected south side Z.} \\
\ubtestbadge{02} & \textcolor{zGoldDeep!78!black}{\texttt{\bfseries Bathroom\-Bathtub\_\allowbreak{}Is\-Against\-East\-Wall}}\enspace\textcolor{zText}{Checks that bathroom bathtub is against east wall.} \\
\ubtestbadge{03} & \textcolor{zGoldDeep!78!black}{\texttt{\bfseries Bathroom\-Curtain\_\allowbreak{}Is\-In\-Expected\-East\-West\-Range}}\enspace\textcolor{zText}{Checks that bathroom curtain is in expected east west range.} \\
\ubtestbadge{04} & \textcolor{zGoldDeep!78!black}{\texttt{\bfseries Bathroom\-Furniture\_\allowbreak{}Has\-Required\-Counts}}\enspace\textcolor{zText}{Checks that bathroom furniture has required counts.} \\
\ubtestbadge{05} & \textcolor{zGoldDeep!78!black}{\texttt{\bfseries Bathroom\-Ground\-Furniture\_\allowbreak{}Is\-At\-Expected\-Floor\-Height}}\enspace\textcolor{zText}{Checks that bathroom ground furniture is at expected floor height.} \\
\ubtestbadge{06} & \textcolor{zGoldDeep!78!black}{\texttt{\bfseries Bathroom\-Mirror\_\allowbreak{}Is\-Above\-Sink\-Height\-Range}}\enspace\textcolor{zText}{Checks that bathroom mirror is above sink height range.} \\
\ubtestbadge{07} & \textcolor{zGoldDeep!78!black}{\texttt{\bfseries Bathroom\-Sink\-And\-Mirror\_\allowbreak{}Are\-Against\-South\-Wall}}\enspace\textcolor{zText}{Checks that bathroom sink and mirror are against south wall.} \\
\ubtestbadge{08} & \textcolor{zGoldDeep!78!black}{\texttt{\bfseries Bathroom\-Toilet\-And\-Shelf\_\allowbreak{}Are\-Against\-North\-Wall}}\enspace\textcolor{zText}{Checks that bathroom toilet and shelf are against north wall.} \\
\ubtestbadge{09} & \textcolor{zGoldDeep!78!black}{\texttt{\bfseries Bathroom\-Wall\-Fixtures\_\allowbreak{}Are\-In\-Expected\-East\-West\-Range}}\enspace\textcolor{zText}{Checks that bathroom wall fixtures are in expected east west range.} \\
\ubtestbadge{10} & \textcolor{zGoldDeep!78!black}{\texttt{\bfseries Bed\_\allowbreak{}Is\-Against\-West\-Wall\-And\-Faces\-East}}\enspace\textcolor{zText}{Checks that bed is against west wall and faces east.} \\
\ubtestbadge{11} & \textcolor{zGoldDeep!78!black}{\texttt{\bfseries Bedroom\-Furniture\_\allowbreak{}Has\-Exactly\-One\-Required\-Prefab\-Each}}\enspace\textcolor{zText}{Checks that bedroom furniture has exactly one required Prefab each.} \\
\ubtestbadge{12} & \textcolor{zGoldDeep!78!black}{\texttt{\bfseries Bedroom\-Furniture\_\allowbreak{}Is\-Inside\-Bedroom\-ZRange}}\enspace\textcolor{zText}{Checks that bedroom furniture is inside bedroom Z range.} \\
\ubtestbadge{13} & \textcolor{zGoldDeep!78!black}{\texttt{\bfseries Bedroom\-Ground\-Furniture\_\allowbreak{}Is\-At\-Expected\-Floor\-Height}}\enspace\textcolor{zText}{Checks that bedroom ground furniture is at expected floor height.} \\
\ubtestbadge{14} & \textcolor{zGoldDeep!78!black}{\texttt{\bfseries Bedroom\-Light\_\allowbreak{}Is\-At\-Ceiling\-Center}}\enspace\textcolor{zText}{Checks that bedroom light is at ceiling center.} \\
\ubtestbadge{15} & \textcolor{zGoldDeep!78!black}{\texttt{\bfseries Corridor\-Carpet\_\allowbreak{}Is\-Centered\-In\-Corridor}}\enspace\textcolor{zText}{Checks that corridor carpet is centered in corridor.} \\
\ubtestbadge{16} & \textcolor{zGoldDeep!78!black}{\texttt{\bfseries Corridor\-Carpet\_\allowbreak{}Is\-In\-Expected\-East\-West\-Range}}\enspace\textcolor{zText}{Checks that corridor carpet is in expected east west range.} \\
\ubtestbadge{17} & \textcolor{zGoldDeep!78!black}{\texttt{\bfseries Corridor\-Furniture\_\allowbreak{}Has\-Required\-Counts}}\enspace\textcolor{zText}{Checks that corridor furniture has required counts.} \\
\ubtestbadge{18} & \textcolor{zGoldDeep!78!black}{\texttt{\bfseries Corridor\-Ground\-Furniture\_\allowbreak{}Is\-At\-Expected\-Floor\-Height}}\enspace\textcolor{zText}{Checks that corridor ground furniture is at expected floor height.} \\
\ubtestbadge{19} & \textcolor{zGoldDeep!78!black}{\texttt{\bfseries Corridor\-Table\_\allowbreak{}Is\-Against\-North\-Wall}}\enspace\textcolor{zText}{Checks that corridor table is against north wall.} \\
\ubtestbadge{20} & \textcolor{zGoldDeep!78!black}{\texttt{\bfseries Corridor\-Table\_\allowbreak{}Is\-In\-Expected\-East\-West\-Range}}\enspace\textcolor{zText}{Checks that corridor table is in expected east west range.} \\
\ubtestbadge{21} & \textcolor{zGoldDeep!78!black}{\texttt{\bfseries Dishwasher\_\allowbreak{}Is\-Adjacent\-To\-Sink\-On\-East\-Side}}\enspace\textcolor{zText}{Checks that dishwasher is adjacent to sink on east side.} \\
\ubtestbadge{22} & \textcolor{zGoldDeep!78!black}{\texttt{\bfseries Garage\-Furniture\_\allowbreak{}Has\-Required\-Counts}}\enspace\textcolor{zText}{Checks that garage furniture has required counts.} \\
\ubtestbadge{23} & \textcolor{zGoldDeep!78!black}{\texttt{\bfseries Garage\-Ground\-Furniture\_\allowbreak{}Is\-At\-Expected\-Floor\-Height}}\enspace\textcolor{zText}{Checks that garage ground furniture is at expected floor height.} \\
\ubtestbadge{24} & \textcolor{zGoldDeep!78!black}{\texttt{\bfseries Garage\-Shelf\-A\_\allowbreak{}Is\-Against\-South\-Wall}}\enspace\textcolor{zText}{Checks that garage shelf a is against south wall.} \\
\ubtestbadge{25} & \textcolor{zGoldDeep!78!black}{\texttt{\bfseries Garage\-Shelf\-A\_\allowbreak{}Is\-In\-Expected\-East\-West\-Range}}\enspace\textcolor{zText}{Checks that garage shelf a is in expected east west range.} \\
\ubtestbadge{26} & \textcolor{zGoldDeep!78!black}{\texttt{\bfseries Garage\-Shelf\-B\_\allowbreak{}Is\-Against\-North\-Wall}}\enspace\textcolor{zText}{Checks that garage shelf b is against north wall.} \\
\ubtestbadge{27} & \textcolor{zGoldDeep!78!black}{\texttt{\bfseries Garage\-Shelf\-B\_\allowbreak{}Is\-In\-Expected\-East\-West\-Range}}\enspace\textcolor{zText}{Checks that garage shelf b is in expected east west range.} \\
\ubtestbadge{28} & \textcolor{zGoldDeep!78!black}{\texttt{\bfseries Garage\-Tables\_\allowbreak{}Are\-Against\-East\-Wall}}\enspace\textcolor{zText}{Checks that garage tables are against east wall.} \\
\ubtestbadge{29} & \textcolor{zGoldDeep!78!black}{\texttt{\bfseries Garage\-Tables\_\allowbreak{}Are\-In\-Expected\-North\-South\-Range}}\enspace\textcolor{zText}{Checks that garage tables are in expected north south range.} \\
\ubtestbadge{30} & \textcolor{zGoldDeep!78!black}{\texttt{\bfseries Kitchen\-And\-Living\-Room\-Furniture\_\allowbreak{}Has\-Required\-Counts}}\enspace\textcolor{zText}{Checks that kitchen and living room furniture has required counts.} \\
\ubtestbadge{31} & \textcolor{zGoldDeep!78!black}{\texttt{\bfseries Kitchen\-And\-Living\-Room\-Ground\-Furniture\_\allowbreak{}Is\-At\-Expected\-Floor\-Height}}\enspace\textcolor{zText}{Checks that kitchen and living room ground furniture is at expected floor height.} \\
\ubtestbadge{32} & \textcolor{zGoldDeep!78!black}{\texttt{\bfseries Kitchen\-Counter\_\allowbreak{}Is\-Near\-Room\-Center}}\enspace\textcolor{zText}{Checks that kitchen counter is near room center.} \\
\ubtestbadge{33} & \textcolor{zGoldDeep!78!black}{\texttt{\bfseries Kitchen\-North\-Wall\-Appliances\_\allowbreak{}Are\-Against\-North\-Wall}}\enspace\textcolor{zText}{Checks that kitchen north wall appliances are against north wall.} \\
\ubtestbadge{34} & \textcolor{zGoldDeep!78!black}{\texttt{\bfseries Kitchen\-North\-Wall\-Appliances\_\allowbreak{}Are\-In\-Expected\-East\-West\-Range}}\enspace\textcolor{zText}{Checks that kitchen north wall appliances are in expected east west range.} \\
\ubtestbadge{35} & \textcolor{zGoldDeep!78!black}{\texttt{\bfseries Living\-Room\-Candle\_\allowbreak{}Is\-On\-Table}}\enspace\textcolor{zText}{Checks that living room candle is on table.} \\
\ubtestbadge{36} & \textcolor{zGoldDeep!78!black}{\texttt{\bfseries Living\-Room\-Chairs\_\allowbreak{}Are\-Placed\-Two\-East\-And\-Two\-West\-Of\-Table}}\enspace\textcolor{zText}{Checks that living room chairs are placed two east and two west of table.} \\
\ubtestbadge{37} & \textcolor{zGoldDeep!78!black}{\texttt{\bfseries Living\-Room\-Sofa\-And\-Table\_\allowbreak{}Are\-Against\-South\-Wall}}\enspace\textcolor{zText}{Checks that living room sofa and table are against south wall.} \\
\ubtestbadge{38} & \textcolor{zGoldDeep!78!black}{\texttt{\bfseries Living\-Room\-Sofa\-And\-Table\_\allowbreak{}Are\-In\-Expected\-East\-West\-Range}}\enspace\textcolor{zText}{Checks that living room sofa and table are in expected east west range.} \\
\ubtestbadge{39} & \textcolor{zGoldDeep!78!black}{\texttt{\bfseries Side\-Table\-And\-Closet\_\allowbreak{}Are\-Against\-East\-Wall}}\enspace\textcolor{zText}{Checks that side table and closet are against east wall.} \\
\noalign{\vskip 2pt}
\multicolumn{2}{@{}p{14.4cm}@{}}{\textcolor{zRockLite!42}{\rule{\linewidth}{0.3pt}}}\\[3pt]
\multicolumn{2}{@{}p{14.4cm}@{}}{\ubtaskhead{zPlain}{0022-ReferenceStreet}{Reference Image Street Recreation}{zGoldDeep}{Medium}{6}}\\*[2pt]
\ubtestbadge{01} & \textcolor{zGoldDeep!78!black}{\texttt{\bfseries Buildings\_\allowbreak{}Match\-Reference}}\enspace\textcolor{zText}{Checks that buildings match reference.} \\
\ubtestbadge{02} & \textcolor{zGoldDeep!78!black}{\texttt{\bfseries Ground\-And\-Crossroad\_\allowbreak{}Match\-Reference\-Origin}}\enspace\textcolor{zText}{Checks that ground and crossroad match reference origin.} \\
\ubtestbadge{03} & \textcolor{zGoldDeep!78!black}{\texttt{\bfseries Prefab\-Counts\_\allowbreak{}Match\-Reference}}\enspace\textcolor{zText}{Checks that Prefab counts match reference.} \\
\ubtestbadge{04} & \textcolor{zGoldDeep!78!black}{\texttt{\bfseries Reference\-Images\_\allowbreak{}And\-Empty\-Parents\-Exist}}\enspace\textcolor{zText}{Checks that reference images and empty parents exist.} \\
\ubtestbadge{05} & \textcolor{zGoldDeep!78!black}{\texttt{\bfseries Road\-Straight\-Pieces\_\allowbreak{}Match\-Reference}}\enspace\textcolor{zText}{Checks that road straight pieces match reference.} \\
\ubtestbadge{06} & \textcolor{zGoldDeep!78!black}{\texttt{\bfseries Trees\_\allowbreak{}Match\-Reference}}\enspace\textcolor{zText}{Checks that trees match reference.} \\
\noalign{\vskip 2pt}
\multicolumn{2}{@{}p{14.4cm}@{}}{\textcolor{zRockLite!42}{\rule{\linewidth}{0.3pt}}}\\[3pt]
\multicolumn{2}{@{}p{14.4cm}@{}}{\ubtaskhead{zPlain}{0034-SciFiRoomVariants}{Sci-Fi Room Size Variants}{zGoldDeep}{Medium}{6}}\\*[2pt]
\ubtestbadge{01} & \textcolor{zGoldDeep!78!black}{\texttt{\bfseries Corner\-Pillars\-Align\-With\-Every\-Room\-Corner}}\enspace\textcolor{zText}{Checks that corner pillars align with every room corner.} \\
\ubtestbadge{02} & \textcolor{zGoldDeep!78!black}{\texttt{\bfseries Furniture\-And\-Generator\-Are\-Placed\-Inside\-Each\-Room}}\enspace\textcolor{zText}{Checks that furniture and generator are placed inside each room.} \\
\ubtestbadge{03} & \textcolor{zGoldDeep!78!black}{\texttt{\bfseries Required\-Room\-Roots\-Exist}}\enspace\textcolor{zText}{Checks that required room roots exist.} \\
\ubtestbadge{04} & \textcolor{zGoldDeep!78!black}{\texttt{\bfseries Rooms\-Do\-Not\-Overlap\-Each\-Other}}\enspace\textcolor{zText}{Checks that rooms do not overlap each other.} \\
\ubtestbadge{05} & \textcolor{zGoldDeep!78!black}{\texttt{\bfseries Rooms\-Use\-Continuous\-Floor\-Grids\-With\-Multiple\-Floor\-Types}}\enspace\textcolor{zText}{Checks that rooms use continuous floor grids with multiple floor types.} \\
\ubtestbadge{06} & \textcolor{zGoldDeep!78!black}{\texttt{\bfseries Walls\-And\-Doors\-Form\-Continuous\-Square\-Boundaries}}\enspace\textcolor{zText}{Checks that walls and doors form continuous square boundaries.} \\
\noalign{\vskip 2pt}
\multicolumn{2}{@{}p{14.4cm}@{}}{\textcolor{zRockLite!42}{\rule{\linewidth}{0.3pt}}}\\[3pt]
\multicolumn{2}{@{}p{14.4cm}@{}}{\ubtaskhead{zPlain}{0036-NavMeshAgentOffMeshLink}{NavMeshAgent OffMeshLink Multi-Room Navigation}{zGoldDeep}{Medium}{10}}\\*[2pt]
\ubtestbadge{01} & \textcolor{zGoldDeep!78!black}{\texttt{\bfseries Core\-Room\-Goals\-Have\-Complete\-Paths\-From\-Spawn}}\enspace\textcolor{zText}{Checks that core room goals have complete paths from spawn.} \\
\ubtestbadge{02} & \textcolor{zGoldDeep!78!black}{\texttt{\bfseries Coverage\-Sample\-Points\-Are\-Walkable}}\enspace\textcolor{zText}{Checks that coverage sample points are walkable.} \\
\ubtestbadge{03} & \textcolor{zGoldDeep!78!black}{\texttt{\bfseries Danger\-Area\-Uses\-Higher\-Cost\-Navigation\-Area}}\enspace\textcolor{zText}{Checks that danger area uses higher cost navigation area.} \\
\ubtestbadge{04} & \textcolor{zGoldDeep!78!black}{\texttt{\bfseries Door\-Transition\-Pairs\-Are\-Connected}}\enspace\textcolor{zText}{Checks that door transition pairs are connected.} \\
\ubtestbadge{05} & \textcolor{zGoldDeep!78!black}{\texttt{\bfseries Forbidden\-Zone\-And\-Dynamic\-Obstacle\-Are\-Handled}}\enspace\textcolor{zText}{Checks that forbidden zone and dynamic obstacle are handled.} \\
\ubtestbadge{06} & \textcolor{zGoldDeep!78!black}{\texttt{\bfseries Multi\-Floor\-Path\-Uses\-Teleporter\-Connection}}\enspace\textcolor{zText}{Checks that multi floor path uses teleporter connection.} \\
\ubtestbadge{07} & \textcolor{zGoldDeep!78!black}{\texttt{\bfseries Nav\-Mesh\-Data\-Exists\-And\-Agent\-Probe\-Is\-Placed\-On\-Nav\-Mesh}}\enspace\textcolor{zText}{Checks that NavMesh data exists and agent probe is placed on NavMesh.} \\
\ubtestbadge{08} & \textcolor{zGoldDeep!78!black}{\texttt{\bfseries Required\-Nav\-Test\-Scaffold\-Is\-Present}}\enspace\textcolor{zText}{Checks that required nav test scaffold is present.} \\
\ubtestbadge{09} & \textcolor{zGoldDeep!78!black}{\texttt{\bfseries Teleporter\-Off\-Mesh\-Link\-Is\-Configured\-Between\-Lower\-And\-Upper\-Endpoints}}\enspace\textcolor{zText}{Checks that teleporter OffMeshLink is configured between lower and upper endpoints.} \\
\ubtestbadge{10} & \textcolor{zGoldDeep!78!black}{\texttt{\bfseries Walls\-And\-Obstacles\-Block\-Straight\-Line\-Leaks}}\enspace\textcolor{zText}{Checks walls and obstacles block straight line leaks.} \\
\noalign{\vskip 2pt}
\multicolumn{2}{@{}p{14.4cm}@{}}{\textcolor{zRockLite!42}{\rule{\linewidth}{0.3pt}}}\\[3pt]
\multicolumn{2}{@{}p{14.4cm}@{}}{\ubtaskhead{zPlain}{0043-MazeObstacleParenting}{11x11 Unique-Path Maze Construction}{zGoldDeep}{Medium}{10}}\\*[2pt]
\ubtestbadge{01} & \textcolor{zGoldDeep!78!black}{\texttt{\bfseries Maze01\_\allowbreak{}Valid\-Maze\-Rules\-And\-Reachability}}\enspace\textcolor{zText}{Checks that Maze\_01 satisfies every grid, boundary-exit, unique-path, 2x2-pattern, transform, and reachability rule.} \\
\ubtestbadge{02} & \textcolor{zGoldDeep!78!black}{\texttt{\bfseries Maze02\_\allowbreak{}Valid\-Maze\-Rules\-And\-Reachability}}\enspace\textcolor{zText}{Checks that Maze\_02 satisfies every grid, boundary-exit, unique-path, 2x2-pattern, transform, and reachability rule.} \\
\ubtestbadge{03} & \textcolor{zGoldDeep!78!black}{\texttt{\bfseries Maze03\_\allowbreak{}Valid\-Maze\-Rules\-And\-Reachability}}\enspace\textcolor{zText}{Checks that Maze\_03 satisfies every grid, boundary-exit, unique-path, 2x2-pattern, transform, and reachability rule.} \\
\ubtestbadge{04} & \textcolor{zGoldDeep!78!black}{\texttt{\bfseries Maze04\_\allowbreak{}Valid\-Maze\-Rules\-And\-Reachability}}\enspace\textcolor{zText}{Checks that Maze\_04 satisfies every grid, boundary-exit, unique-path, 2x2-pattern, transform, and reachability rule.} \\
\ubtestbadge{05} & \textcolor{zGoldDeep!78!black}{\texttt{\bfseries Maze05\_\allowbreak{}Valid\-Maze\-Rules\-And\-Reachability}}\enspace\textcolor{zText}{Checks that Maze\_05 satisfies every grid, boundary-exit, unique-path, 2x2-pattern, transform, and reachability rule.} \\
\ubtestbadge{06} & \textcolor{zGoldDeep!78!black}{\texttt{\bfseries Maze06\_\allowbreak{}Valid\-Maze\-Rules\-And\-Reachability}}\enspace\textcolor{zText}{Checks that Maze\_06 satisfies every grid, boundary-exit, unique-path, 2x2-pattern, transform, and reachability rule.} \\
\ubtestbadge{07} & \textcolor{zGoldDeep!78!black}{\texttt{\bfseries Maze07\_\allowbreak{}Valid\-Maze\-Rules\-And\-Reachability}}\enspace\textcolor{zText}{Checks that Maze\_07 satisfies every grid, boundary-exit, unique-path, 2x2-pattern, transform, and reachability rule.} \\
\ubtestbadge{08} & \textcolor{zGoldDeep!78!black}{\texttt{\bfseries Maze08\_\allowbreak{}Valid\-Maze\-Rules\-And\-Reachability}}\enspace\textcolor{zText}{Checks that Maze\_08 satisfies every grid, boundary-exit, unique-path, 2x2-pattern, transform, and reachability rule.} \\
\ubtestbadge{09} & \textcolor{zGoldDeep!78!black}{\texttt{\bfseries Maze09\_\allowbreak{}Valid\-Maze\-Rules\-And\-Reachability}}\enspace\textcolor{zText}{Checks that Maze\_09 satisfies every grid, boundary-exit, unique-path, 2x2-pattern, transform, and reachability rule.} \\
\ubtestbadge{10} & \textcolor{zGoldDeep!78!black}{\texttt{\bfseries Maze10\_\allowbreak{}Valid\-Maze\-Rules\-And\-Reachability}}\enspace\textcolor{zText}{Checks that Maze\_10 satisfies every grid, boundary-exit, unique-path, 2x2-pattern, transform, and reachability rule.} \\
\noalign{\vskip 2pt}
\multicolumn{2}{@{}p{14.4cm}@{}}{\textcolor{zRockLite!42}{\rule{\linewidth}{0.3pt}}}\\[3pt]
\multicolumn{2}{@{}p{14.4cm}@{}}{\ubtaskhead{zPlain}{0069-TrenchAssembly}{Trench Asset Assembly}{zGoldDeep}{Medium}{16}}\\*[2pt]
\ubtestbadge{01} & \textcolor{zGoldDeep!78!black}{\texttt{\bfseries Projected\-Collider\-Overlap\_\allowbreak{}20\-Percent\_\allowbreak{}Io\-U10\-Percent}}\enspace\textcolor{zText}{Checks that projected collider overlap never exceeds 20\% of the smaller footprint and IoU never exceeds 10\%.} \\
\ubtestbadge{02} & \textcolor{zGoldDeep!78!black}{\texttt{\bfseries Projected\-Collider\-Overlap\_\allowbreak{}30\-Percent\_\allowbreak{}Io\-U20\-Percent}}\enspace\textcolor{zText}{Checks that projected collider overlap never exceeds 30\% of the smaller footprint and IoU never exceeds 20\%.} \\
\ubtestbadge{03} & \textcolor{zGoldDeep!78!black}{\texttt{\bfseries Projected\-Collider\-Overlap\_\allowbreak{}40\-Percent\_\allowbreak{}Io\-U30\-Percent}}\enspace\textcolor{zText}{Checks that projected collider overlap never exceeds 40\% of the smaller footprint and IoU never exceeds 30\%.} \\
\ubtestbadge{04} & \textcolor{zGoldDeep!78!black}{\texttt{\bfseries All\-Trench\-Art\-Instances\-Are\-Under\-Named010203\-Rows\-And\-All\-Art\-Prefabs\-Are\-Used}}\enspace\textcolor{zText}{Checks that all trench-art instances are under TrenchSegment\_01, \_02, or \_03 and that every supplied trench prefab type is used.} \\
\ubtestbadge{05} & \textcolor{zGoldDeep!78!black}{\texttt{\bfseries Richness\_\allowbreak{}Total\-Trench\-Pieces\-Over30}}\enspace\textcolor{zText}{Checks that the layout contains more than 30 trench pieces.} \\
\ubtestbadge{06} & \textcolor{zGoldDeep!78!black}{\texttt{\bfseries Richness\_\allowbreak{}Total\-Trench\-Pieces\-Over45}}\enspace\textcolor{zText}{Checks that the layout contains more than 45 trench pieces.} \\
\ubtestbadge{07} & \textcolor{zGoldDeep!78!black}{\texttt{\bfseries Row\-Centers\-Are\-Monotonic\-And\-All\-Prefab\-Anchor\-Points\-Are\-Globally\-Reachable}}\enspace\textcolor{zText}{Checks ordered front/middle/rear row centers and global reachability between all prefab anchor points.} \\
\ubtestbadge{08} & \textcolor{zGoldDeep!78!black}{\texttt{\bfseries Trench\-Segment01\-Reachability\_\allowbreak{}100\-Percent}}\enspace\textcolor{zText}{Checks that 100\% of TrenchSegment\_01 prefab anchor points are mutually reachable.} \\
\ubtestbadge{09} & \textcolor{zGoldDeep!78!black}{\texttt{\bfseries Trench\-Segment01\-Reachability\_\allowbreak{}33\_\allowbreak{}3\-Percent}}\enspace\textcolor{zText}{Checks that at least 33.3\% of TrenchSegment\_01 prefab anchor points are mutually reachable.} \\
\ubtestbadge{10} & \textcolor{zGoldDeep!78!black}{\texttt{\bfseries Trench\-Segment01\-Reachability\_\allowbreak{}66\_\allowbreak{}7\-Percent}}\enspace\textcolor{zText}{Checks that at least 66.7\% of TrenchSegment\_01 prefab anchor points are mutually reachable.} \\
\ubtestbadge{11} & \textcolor{zGoldDeep!78!black}{\texttt{\bfseries Trench\-Segment02\-Reachability\_\allowbreak{}100\-Percent}}\enspace\textcolor{zText}{Checks that 100\% of TrenchSegment\_02 prefab anchor points are mutually reachable.} \\
\ubtestbadge{12} & \textcolor{zGoldDeep!78!black}{\texttt{\bfseries Trench\-Segment02\-Reachability\_\allowbreak{}33\_\allowbreak{}3\-Percent}}\enspace\textcolor{zText}{Checks that at least 33.3\% of TrenchSegment\_02 prefab anchor points are mutually reachable.} \\
\ubtestbadge{13} & \textcolor{zGoldDeep!78!black}{\texttt{\bfseries Trench\-Segment02\-Reachability\_\allowbreak{}66\_\allowbreak{}7\-Percent}}\enspace\textcolor{zText}{Checks that at least 66.7\% of TrenchSegment\_02 prefab anchor points are mutually reachable.} \\
\ubtestbadge{14} & \textcolor{zGoldDeep!78!black}{\texttt{\bfseries Trench\-Segment03\-Reachability\_\allowbreak{}100\-Percent}}\enspace\textcolor{zText}{Checks that 100\% of TrenchSegment\_03 prefab anchor points are mutually reachable.} \\
\ubtestbadge{15} & \textcolor{zGoldDeep!78!black}{\texttt{\bfseries Trench\-Segment03\-Reachability\_\allowbreak{}33\_\allowbreak{}3\-Percent}}\enspace\textcolor{zText}{Checks that at least 33.3\% of TrenchSegment\_03 prefab anchor points are mutually reachable.} \\
\ubtestbadge{16} & \textcolor{zGoldDeep!78!black}{\texttt{\bfseries Trench\-Segment03\-Reachability\_\allowbreak{}66\_\allowbreak{}7\-Percent}}\enspace\textcolor{zText}{Checks that at least 66.7\% of TrenchSegment\_03 prefab anchor points are mutually reachable.} \\
\noalign{\vskip 2pt}
\multicolumn{2}{@{}p{14.4cm}@{}}{\textcolor{zRockLite!42}{\rule{\linewidth}{0.3pt}}}\\[3pt]
\multicolumn{2}{@{}p{14.4cm}@{}}{\ubtaskhead{zPlain}{0066-Shop3DLetters}{Shop 3D Letter Signs}{zShrine}{Hard}{44}}\\*[2pt]
\ubtestbadge{01} & \textcolor{zGoldDeep!78!black}{\texttt{\bfseries Letter\-Position\_\allowbreak{}Loose\_\allowbreak{}Book\-Store\_\allowbreak{}00\_\allowbreak{}B}}\enspace\textcolor{zText}{Checks that BookStore/00\_B is placed on the correct sign within the broad placement tolerance.} \\
\ubtestbadge{02} & \textcolor{zGoldDeep!78!black}{\texttt{\bfseries Letter\-Position\_\allowbreak{}Loose\_\allowbreak{}Book\-Store\_\allowbreak{}01\_\allowbreak{}O}}\enspace\textcolor{zText}{Checks that BookStore/01\_O is placed on the correct sign within the broad placement tolerance.} \\
\ubtestbadge{03} & \textcolor{zGoldDeep!78!black}{\texttt{\bfseries Letter\-Position\_\allowbreak{}Loose\_\allowbreak{}Book\-Store\_\allowbreak{}02\_\allowbreak{}O}}\enspace\textcolor{zText}{Checks that BookStore/02\_O is placed on the correct sign within the broad placement tolerance.} \\
\ubtestbadge{04} & \textcolor{zGoldDeep!78!black}{\texttt{\bfseries Letter\-Position\_\allowbreak{}Loose\_\allowbreak{}Book\-Store\_\allowbreak{}03\_\allowbreak{}K}}\enspace\textcolor{zText}{Checks that BookStore/03\_K is placed on the correct sign within the broad placement tolerance.} \\
\ubtestbadge{05} & \textcolor{zGoldDeep!78!black}{\texttt{\bfseries Letter\-Position\_\allowbreak{}Loose\_\allowbreak{}Book\-Store\_\allowbreak{}04\_\allowbreak{}S}}\enspace\textcolor{zText}{Checks that BookStore/04\_S is placed on the correct sign within the broad placement tolerance.} \\
\ubtestbadge{06} & \textcolor{zGoldDeep!78!black}{\texttt{\bfseries Letter\-Position\_\allowbreak{}Loose\_\allowbreak{}Book\-Store\_\allowbreak{}05\_\allowbreak{}T}}\enspace\textcolor{zText}{Checks that BookStore/05\_T is placed on the correct sign within the broad placement tolerance.} \\
\ubtestbadge{07} & \textcolor{zGoldDeep!78!black}{\texttt{\bfseries Letter\-Position\_\allowbreak{}Loose\_\allowbreak{}Book\-Store\_\allowbreak{}06\_\allowbreak{}O}}\enspace\textcolor{zText}{Checks that BookStore/06\_O is placed on the correct sign within the broad placement tolerance.} \\
\ubtestbadge{08} & \textcolor{zGoldDeep!78!black}{\texttt{\bfseries Letter\-Position\_\allowbreak{}Loose\_\allowbreak{}Book\-Store\_\allowbreak{}07\_\allowbreak{}R}}\enspace\textcolor{zText}{Checks that BookStore/07\_R is placed on the correct sign within the broad placement tolerance.} \\
\ubtestbadge{09} & \textcolor{zGoldDeep!78!black}{\texttt{\bfseries Letter\-Position\_\allowbreak{}Loose\_\allowbreak{}Book\-Store\_\allowbreak{}08\_\allowbreak{}E}}\enspace\textcolor{zText}{Checks that BookStore/08\_E is placed on the correct sign within the broad placement tolerance.} \\
\ubtestbadge{10} & \textcolor{zGoldDeep!78!black}{\texttt{\bfseries Letter\-Position\_\allowbreak{}Loose\_\allowbreak{}Pizza\-Place\_\allowbreak{}00\_\allowbreak{}P}}\enspace\textcolor{zText}{Checks that PizzaPlace/00\_P is placed on the correct sign within the broad placement tolerance.} \\
\ubtestbadge{11} & \textcolor{zGoldDeep!78!black}{\texttt{\bfseries Letter\-Position\_\allowbreak{}Loose\_\allowbreak{}Pizza\-Place\_\allowbreak{}01\_\allowbreak{}I}}\enspace\textcolor{zText}{Checks that PizzaPlace/01\_I is placed on the correct sign within the broad placement tolerance.} \\
\ubtestbadge{12} & \textcolor{zGoldDeep!78!black}{\texttt{\bfseries Letter\-Position\_\allowbreak{}Loose\_\allowbreak{}Pizza\-Place\_\allowbreak{}02\_\allowbreak{}Z}}\enspace\textcolor{zText}{Checks that PizzaPlace/02\_Z is placed on the correct sign within the broad placement tolerance.} \\
\ubtestbadge{13} & \textcolor{zGoldDeep!78!black}{\texttt{\bfseries Letter\-Position\_\allowbreak{}Loose\_\allowbreak{}Pizza\-Place\_\allowbreak{}03\_\allowbreak{}Z}}\enspace\textcolor{zText}{Checks that PizzaPlace/03\_Z is placed on the correct sign within the broad placement tolerance.} \\
\ubtestbadge{14} & \textcolor{zGoldDeep!78!black}{\texttt{\bfseries Letter\-Position\_\allowbreak{}Loose\_\allowbreak{}Pizza\-Place\_\allowbreak{}04\_\allowbreak{}A}}\enspace\textcolor{zText}{Checks that PizzaPlace/04\_A is placed on the correct sign within the broad placement tolerance.} \\
\ubtestbadge{15} & \textcolor{zGoldDeep!78!black}{\texttt{\bfseries Letter\-Position\_\allowbreak{}Loose\_\allowbreak{}Pizza\-Place\_\allowbreak{}05\_\allowbreak{}P}}\enspace\textcolor{zText}{Checks that PizzaPlace/05\_P is placed on the correct sign within the broad placement tolerance.} \\
\ubtestbadge{16} & \textcolor{zGoldDeep!78!black}{\texttt{\bfseries Letter\-Position\_\allowbreak{}Loose\_\allowbreak{}Pizza\-Place\_\allowbreak{}06\_\allowbreak{}L}}\enspace\textcolor{zText}{Checks that PizzaPlace/06\_L is placed on the correct sign within the broad placement tolerance.} \\
\ubtestbadge{17} & \textcolor{zGoldDeep!78!black}{\texttt{\bfseries Letter\-Position\_\allowbreak{}Loose\_\allowbreak{}Pizza\-Place\_\allowbreak{}07\_\allowbreak{}A}}\enspace\textcolor{zText}{Checks that PizzaPlace/07\_A is placed on the correct sign within the broad placement tolerance.} \\
\ubtestbadge{18} & \textcolor{zGoldDeep!78!black}{\texttt{\bfseries Letter\-Position\_\allowbreak{}Loose\_\allowbreak{}Pizza\-Place\_\allowbreak{}08\_\allowbreak{}C}}\enspace\textcolor{zText}{Checks that PizzaPlace/08\_C is placed on the correct sign within the broad placement tolerance.} \\
\ubtestbadge{19} & \textcolor{zGoldDeep!78!black}{\texttt{\bfseries Letter\-Position\_\allowbreak{}Loose\_\allowbreak{}Pizza\-Place\_\allowbreak{}09\_\allowbreak{}E}}\enspace\textcolor{zText}{Checks that PizzaPlace/09\_E is placed on the correct sign within the broad placement tolerance.} \\
\ubtestbadge{20} & \textcolor{zGoldDeep!78!black}{\texttt{\bfseries Letter\-Position\_\allowbreak{}Tight\_\allowbreak{}Book\-Store\_\allowbreak{}00\_\allowbreak{}B}}\enspace\textcolor{zText}{Checks that BookStore/00\_B is placed on the correct sign within the tight golden-position tolerance.} \\
\ubtestbadge{21} & \textcolor{zGoldDeep!78!black}{\texttt{\bfseries Letter\-Position\_\allowbreak{}Tight\_\allowbreak{}Book\-Store\_\allowbreak{}01\_\allowbreak{}O}}\enspace\textcolor{zText}{Checks that BookStore/01\_O is placed on the correct sign within the tight golden-position tolerance.} \\
\ubtestbadge{22} & \textcolor{zGoldDeep!78!black}{\texttt{\bfseries Letter\-Position\_\allowbreak{}Tight\_\allowbreak{}Book\-Store\_\allowbreak{}02\_\allowbreak{}O}}\enspace\textcolor{zText}{Checks that BookStore/02\_O is placed on the correct sign within the tight golden-position tolerance.} \\
\ubtestbadge{23} & \textcolor{zGoldDeep!78!black}{\texttt{\bfseries Letter\-Position\_\allowbreak{}Tight\_\allowbreak{}Book\-Store\_\allowbreak{}03\_\allowbreak{}K}}\enspace\textcolor{zText}{Checks that BookStore/03\_K is placed on the correct sign within the tight golden-position tolerance.} \\
\ubtestbadge{24} & \textcolor{zGoldDeep!78!black}{\texttt{\bfseries Letter\-Position\_\allowbreak{}Tight\_\allowbreak{}Book\-Store\_\allowbreak{}04\_\allowbreak{}S}}\enspace\textcolor{zText}{Checks that BookStore/04\_S is placed on the correct sign within the tight golden-position tolerance.} \\
\ubtestbadge{25} & \textcolor{zGoldDeep!78!black}{\texttt{\bfseries Letter\-Position\_\allowbreak{}Tight\_\allowbreak{}Book\-Store\_\allowbreak{}05\_\allowbreak{}T}}\enspace\textcolor{zText}{Checks that BookStore/05\_T is placed on the correct sign within the tight golden-position tolerance.} \\
\ubtestbadge{26} & \textcolor{zGoldDeep!78!black}{\texttt{\bfseries Letter\-Position\_\allowbreak{}Tight\_\allowbreak{}Book\-Store\_\allowbreak{}06\_\allowbreak{}O}}\enspace\textcolor{zText}{Checks that BookStore/06\_O is placed on the correct sign within the tight golden-position tolerance.} \\
\ubtestbadge{27} & \textcolor{zGoldDeep!78!black}{\texttt{\bfseries Letter\-Position\_\allowbreak{}Tight\_\allowbreak{}Book\-Store\_\allowbreak{}07\_\allowbreak{}R}}\enspace\textcolor{zText}{Checks that BookStore/07\_R is placed on the correct sign within the tight golden-position tolerance.} \\
\ubtestbadge{28} & \textcolor{zGoldDeep!78!black}{\texttt{\bfseries Letter\-Position\_\allowbreak{}Tight\_\allowbreak{}Book\-Store\_\allowbreak{}08\_\allowbreak{}E}}\enspace\textcolor{zText}{Checks that BookStore/08\_E is placed on the correct sign within the tight golden-position tolerance.} \\
\ubtestbadge{29} & \textcolor{zGoldDeep!78!black}{\texttt{\bfseries Letter\-Position\_\allowbreak{}Tight\_\allowbreak{}Pizza\-Place\_\allowbreak{}00\_\allowbreak{}P}}\enspace\textcolor{zText}{Checks that PizzaPlace/00\_P is placed on the correct sign within the tight golden-position tolerance.} \\
\ubtestbadge{30} & \textcolor{zGoldDeep!78!black}{\texttt{\bfseries Letter\-Position\_\allowbreak{}Tight\_\allowbreak{}Pizza\-Place\_\allowbreak{}01\_\allowbreak{}I}}\enspace\textcolor{zText}{Checks that PizzaPlace/01\_I is placed on the correct sign within the tight golden-position tolerance.} \\
\ubtestbadge{31} & \textcolor{zGoldDeep!78!black}{\texttt{\bfseries Letter\-Position\_\allowbreak{}Tight\_\allowbreak{}Pizza\-Place\_\allowbreak{}02\_\allowbreak{}Z}}\enspace\textcolor{zText}{Checks that PizzaPlace/02\_Z is placed on the correct sign within the tight golden-position tolerance.} \\
\ubtestbadge{32} & \textcolor{zGoldDeep!78!black}{\texttt{\bfseries Letter\-Position\_\allowbreak{}Tight\_\allowbreak{}Pizza\-Place\_\allowbreak{}03\_\allowbreak{}Z}}\enspace\textcolor{zText}{Checks that PizzaPlace/03\_Z is placed on the correct sign within the tight golden-position tolerance.} \\
\ubtestbadge{33} & \textcolor{zGoldDeep!78!black}{\texttt{\bfseries Letter\-Position\_\allowbreak{}Tight\_\allowbreak{}Pizza\-Place\_\allowbreak{}04\_\allowbreak{}A}}\enspace\textcolor{zText}{Checks that PizzaPlace/04\_A is placed on the correct sign within the tight golden-position tolerance.} \\
\ubtestbadge{34} & \textcolor{zGoldDeep!78!black}{\texttt{\bfseries Letter\-Position\_\allowbreak{}Tight\_\allowbreak{}Pizza\-Place\_\allowbreak{}05\_\allowbreak{}P}}\enspace\textcolor{zText}{Checks that PizzaPlace/05\_P is placed on the correct sign within the tight golden-position tolerance.} \\
\ubtestbadge{35} & \textcolor{zGoldDeep!78!black}{\texttt{\bfseries Letter\-Position\_\allowbreak{}Tight\_\allowbreak{}Pizza\-Place\_\allowbreak{}06\_\allowbreak{}L}}\enspace\textcolor{zText}{Checks that PizzaPlace/06\_L is placed on the correct sign within the tight golden-position tolerance.} \\
\ubtestbadge{36} & \textcolor{zGoldDeep!78!black}{\texttt{\bfseries Letter\-Position\_\allowbreak{}Tight\_\allowbreak{}Pizza\-Place\_\allowbreak{}07\_\allowbreak{}A}}\enspace\textcolor{zText}{Checks that PizzaPlace/07\_A is placed on the correct sign within the tight golden-position tolerance.} \\
\ubtestbadge{37} & \textcolor{zGoldDeep!78!black}{\texttt{\bfseries Letter\-Position\_\allowbreak{}Tight\_\allowbreak{}Pizza\-Place\_\allowbreak{}08\_\allowbreak{}C}}\enspace\textcolor{zText}{Checks that PizzaPlace/08\_C is placed on the correct sign within the tight golden-position tolerance.} \\
\ubtestbadge{38} & \textcolor{zGoldDeep!78!black}{\texttt{\bfseries Letter\-Position\_\allowbreak{}Tight\_\allowbreak{}Pizza\-Place\_\allowbreak{}09\_\allowbreak{}E}}\enspace\textcolor{zText}{Checks that PizzaPlace/09\_E is placed on the correct sign within the tight golden-position tolerance.} \\
\ubtestbadge{39} & \textcolor{zGoldDeep!78!black}{\texttt{\bfseries Letter\-Scale\_\allowbreak{}Loose\_\allowbreak{}X\_\allowbreak{}0\_\allowbreak{}2\_\allowbreak{}to\_\allowbreak{}0\_\allowbreak{}8}}\enspace\textcolor{zText}{Checks that every submitted letter's X scale lies within the broad range 0.2 to 0.8.} \\
\ubtestbadge{40} & \textcolor{zGoldDeep!78!black}{\texttt{\bfseries Letter\-Scale\_\allowbreak{}Loose\_\allowbreak{}Y\_\allowbreak{}0\_\allowbreak{}2\_\allowbreak{}to\_\allowbreak{}0\_\allowbreak{}8}}\enspace\textcolor{zText}{Checks that every submitted letter's Y scale lies within the broad range 0.2 to 0.8.} \\
\ubtestbadge{41} & \textcolor{zGoldDeep!78!black}{\texttt{\bfseries Letter\-Scale\_\allowbreak{}Loose\_\allowbreak{}Z\_\allowbreak{}0\_\allowbreak{}2\_\allowbreak{}to\_\allowbreak{}0\_\allowbreak{}8}}\enspace\textcolor{zText}{Checks that every submitted letter's Z scale lies within the broad range 0.2 to 0.8.} \\
\ubtestbadge{42} & \textcolor{zGoldDeep!78!black}{\texttt{\bfseries Letter\-Scale\_\allowbreak{}Tight\_\allowbreak{}X\_\allowbreak{}0\_\allowbreak{}4\_\allowbreak{}to\_\allowbreak{}0\_\allowbreak{}6}}\enspace\textcolor{zText}{Checks that every submitted letter's X scale lies within the tight range 0.4 to 0.6.} \\
\ubtestbadge{43} & \textcolor{zGoldDeep!78!black}{\texttt{\bfseries Letter\-Scale\_\allowbreak{}Tight\_\allowbreak{}Y\_\allowbreak{}0\_\allowbreak{}4\_\allowbreak{}to\_\allowbreak{}0\_\allowbreak{}6}}\enspace\textcolor{zText}{Checks that every submitted letter's Y scale lies within the tight range 0.4 to 0.6.} \\
\ubtestbadge{44} & \textcolor{zGoldDeep!78!black}{\texttt{\bfseries Letter\-Scale\_\allowbreak{}Tight\_\allowbreak{}Z\_\allowbreak{}0\_\allowbreak{}4\_\allowbreak{}to\_\allowbreak{}0\_\allowbreak{}6}}\enspace\textcolor{zText}{Checks that every submitted letter's Z scale lies within the tight range 0.4 to 0.6.} \\
\noalign{\vskip 2pt}
\multicolumn{2}{@{}p{14.4cm}@{}}{\textcolor{zRockLite!42}{\rule{\linewidth}{0.3pt}}}\\[3pt]
\multicolumn{2}{@{}p{14.4cm}@{}}{\ubtaskhead{zPlain}{0067-TerrainRiverWaterAssembly}{Terrain River and Lake Water Surface Assembly}{zShrine}{Hard}{14}}\\*[2pt]
\ubtestbadge{01} & \textcolor{zGoldDeep!78!black}{\texttt{\bfseries End\-Lake\-Coverage\-Samples\-Are\-Below\-The\-Expected\-Water\-Surface}}\enspace\textcolor{zText}{Checks that end lake coverage samples are below the expected water surface.} \\
\ubtestbadge{02} & \textcolor{zGoldDeep!78!black}{\texttt{\bfseries End\-Lake\-Plane\-Corners\-Are\-Buried\-Below\-Terrain}}\enspace\textcolor{zText}{Checks that end lake plane corners are buried below terrain.} \\
\ubtestbadge{03} & \textcolor{zGoldDeep!78!black}{\texttt{\bfseries River\-Coverage\-Samples\-Are\-Below\-The\-Expected\-Water\-Surface}}\enspace\textcolor{zText}{Checks that river coverage samples are below the expected water surface.} \\
\ubtestbadge{04} & \textcolor{zGoldDeep!78!black}{\texttt{\bfseries River\-Endpoint\-Segments\-Meet\-Lake\-Water\-Levels}}\enspace\textcolor{zText}{Checks that river endpoint segments meet lake water levels.} \\
\ubtestbadge{05} & \textcolor{zGoldDeep!78!black}{\texttt{\bfseries River\-Equal\-Height\-Side\-Edges\-Are\-Buried\-Below\-Terrain}}\enspace\textcolor{zText}{Checks that river equal height side edges are buried below terrain.} \\
\ubtestbadge{06} & \textcolor{zGoldDeep!78!black}{\texttt{\bfseries River\-Segment\-Count\-Stays\-Reasonable}}\enspace\textcolor{zText}{Checks that river segment count stays are reasonable.} \\
\ubtestbadge{07} & \textcolor{zGoldDeep!78!black}{\texttt{\bfseries River\-Segment\-Seam\-Edges\-Are\-Nearly\-Level}}\enspace\textcolor{zText}{Checks that river segment seam edges are nearly level.} \\
\ubtestbadge{08} & \textcolor{zGoldDeep!78!black}{\texttt{\bfseries River\-Segment\-Seam\-Midpoints\-Are\-Close\-And\-Overlapping}}\enspace\textcolor{zText}{Checks that river segment seam midpoints are close and overlapping.} \\
\ubtestbadge{09} & \textcolor{zGoldDeep!78!black}{\texttt{\bfseries River\-Water\-Level\-Descends\-From\-Start\-Lake\-To\-End\-Lake}}\enspace\textcolor{zText}{Checks that river water level descends from start lake to end lake.} \\
\ubtestbadge{10} & \textcolor{zGoldDeep!78!black}{\texttt{\bfseries Start\-Lake\-Coverage\-Samples\-Are\-Below\-The\-Expected\-Water\-Surface}}\enspace\textcolor{zText}{Checks that start lake coverage samples are below the expected water surface.} \\
\ubtestbadge{11} & \textcolor{zGoldDeep!78!black}{\texttt{\bfseries Start\-Lake\-Plane\-Corners\-Are\-Buried\-Below\-Terrain}}\enspace\textcolor{zText}{Checks that start lake plane corners are buried below terrain.} \\
\ubtestbadge{12} & \textcolor{zGoldDeep!78!black}{\texttt{\bfseries Water\-Objects\-Use\-Built\-In\-Plane\-Mesh}}\enspace\textcolor{zText}{Checks that water objects use built in plane mesh.} \\
\ubtestbadge{13} & \textcolor{zGoldDeep!78!black}{\texttt{\bfseries Water\-Plane\-Rotations\-Respect\-Rules}}\enspace\textcolor{zText}{Checks that water plane rotations respect rules.} \\
\ubtestbadge{14} & \textcolor{zGoldDeep!78!black}{\texttt{\bfseries Water\-Planes\-Use\-The\-Correct\-Anonymous\-Materials}}\enspace\textcolor{zText}{Checks that water planes use the correct anonymous materials.} \\
\noalign{\vskip 2pt}
\multicolumn{2}{@{}p{14.4cm}@{}}{\textcolor{zRockLite!42}{\rule{\linewidth}{0.3pt}}}\\[3pt]
\multicolumn{2}{@{}p{14.4cm}@{}}{\ubtaskhead{zPlain}{0068-RacingCircuitTerrainPaint}{Racing Circuit Terrain Painting, White Lines, and Turn Props}{zShrine}{Hard}{28}}\\*[2pt]
\ubtestbadge{01} & \textcolor{zGoldDeep!78!black}{\texttt{\bfseries All\-Provided\-Rail\-Prefabs\-Are\-Grouped\-Under\-Rail\-Parents}}\enspace\textcolor{zText}{Checks that all provided rail prefabs are grouped under rail parents.} \\
\ubtestbadge{02} & \textcolor{zGoldDeep!78!black}{\texttt{\bfseries Barrier\-Props\-Do\-Not\-Sit\-On\-The\-Racing\-Line}}\enspace\textcolor{zText}{Checks that barrier props do not sit on the racing line.} \\
\ubtestbadge{03} & \textcolor{zGoldDeep!78!black}{\texttt{\bfseries Barrier\-Props\-Generally\-Follow\-The\-Track\-Tangent}}\enspace\textcolor{zText}{Checks that barrier props generally follow the track tangent.} \\
\ubtestbadge{04} & \textcolor{zGoldDeep!78!black}{\texttt{\bfseries Barrier\-Props\-Stay\-Near\-The\-Circuit}}\enspace\textcolor{zText}{Checks that barrier props stay near the circuit.} \\
\ubtestbadge{05} & \textcolor{zGoldDeep!78!black}{\texttt{\bfseries Barrier\-Props\-Use\-Reasonable\-Scale\-And\-Ground\-Height}}\enspace\textcolor{zText}{Checks that barrier props use reasonable scale and ground height.} \\
\ubtestbadge{06} & \textcolor{zGoldDeep!78!black}{\texttt{\bfseries Curbs\-Are\-Mostly\-Placed\-On\-Track\-Edges}}\enspace\textcolor{zText}{Checks that curbs are mostly placed on track edges.} \\
\ubtestbadge{07} & \textcolor{zGoldDeep!78!black}{\texttt{\bfseries Curbs\-Cover\_\allowbreak{}Group\-A}}\enspace\textcolor{zText}{Checks that curbs cover the hidden GroupA turn region.} \\
\ubtestbadge{08} & \textcolor{zGoldDeep!78!black}{\texttt{\bfseries Curbs\-Cover\_\allowbreak{}Group\-B}}\enspace\textcolor{zText}{Checks that curbs cover the hidden GroupB turn region.} \\
\ubtestbadge{09} & \textcolor{zGoldDeep!78!black}{\texttt{\bfseries Curbs\-Cover\_\allowbreak{}Group\-C}}\enspace\textcolor{zText}{Checks that curbs cover the hidden GroupC turn region.} \\
\ubtestbadge{10} & \textcolor{zGoldDeep!78!black}{\texttt{\bfseries Curbs\-Cover\_\allowbreak{}Group\-D}}\enspace\textcolor{zText}{Checks that curbs cover the hidden GroupD turn region.} \\
\ubtestbadge{11} & \textcolor{zGoldDeep!78!black}{\texttt{\bfseries Curbs\-Cover\_\allowbreak{}Group\-E}}\enspace\textcolor{zText}{Checks that curbs cover the hidden GroupE turn region.} \\
\ubtestbadge{12} & \textcolor{zGoldDeep!78!black}{\texttt{\bfseries Plastic\-Barrier\-Colliders\-Touch\-Within\-Shortcut\-Gate}}\enspace\textcolor{zText}{Checks that plastic barrier Colliders touch within shortcut gate.} \\
\ubtestbadge{13} & \textcolor{zGoldDeep!78!black}{\texttt{\bfseries Plastic\-Barrier\-Groups\-Occupy\-All\-Hidden\-Shortcut\-Regions}}\enspace\textcolor{zText}{Checks that plastic barrier groups occupy all hidden shortcut regions.} \\
\ubtestbadge{14} & \textcolor{zGoldDeep!78!black}{\texttt{\bfseries Plastic\-Barrier\-Pivots\-Use\-Hinge\-Like\-Spacing}}\enspace\textcolor{zText}{Checks that plastic barrier pivots use hinge like spacing.} \\
\ubtestbadge{15} & \textcolor{zGoldDeep!78!black}{\texttt{\bfseries Rail\-Continuity\-Groups\-Form\-Connected\-Chains}}\enspace\textcolor{zText}{Checks that rail continuity groups form connected chains.} \\
\ubtestbadge{16} & \textcolor{zGoldDeep!78!black}{\texttt{\bfseries Rail\-Continuity\-Groups\-Mostly\-Follow\-Track\-Tangent}}\enspace\textcolor{zText}{Checks that rail continuity groups mostly follow track tangent.} \\
\ubtestbadge{17} & \textcolor{zGoldDeep!78!black}{\texttt{\bfseries Rail\-Continuity\-Groups\-Use\-Provided\-Rail\-Prefabs}}\enspace\textcolor{zText}{Checks that rail continuity groups use provided rail prefabs.} \\
\ubtestbadge{18} & \textcolor{zGoldDeep!78!black}{\texttt{\bfseries Rails\-Are\-Mostly\-Outside\-The\-Road\-Edge}}\enspace\textcolor{zText}{Checks that rails are mostly outside the road edge.} \\
\ubtestbadge{19} & \textcolor{zGoldDeep!78!black}{\texttt{\bfseries Rails\-Protect\-Multiple\-Turn\-Groups}}\enspace\textcolor{zText}{Checks that rails protect multiple turn groups.} \\
\ubtestbadge{20} & \textcolor{zGoldDeep!78!black}{\texttt{\bfseries Scene\-Uses\-Provided\-Curb\-And\-Rail\-Prefab\-Families}}\enspace\textcolor{zText}{Checks that scene uses provided curb and rail Prefab families.} \\
\ubtestbadge{21} & \textcolor{zGoldDeep!78!black}{\texttt{\bfseries Terrain\-Paint\-Mask\_\allowbreak{}Loose\_\allowbreak{}1\_\allowbreak{}50m}}\enspace\textcolor{zText}{Checks the loose Terrain paint-mask tier using a 1.50-metre road-boundary tolerance and its corresponding road/grass coverage thresholds.} \\
\ubtestbadge{22} & \textcolor{zGoldDeep!78!black}{\texttt{\bfseries Terrain\-Paint\-Mask\_\allowbreak{}Medium\_\allowbreak{}0\_\allowbreak{}85m}}\enspace\textcolor{zText}{Checks the medium Terrain paint-mask tier using a 0.85-metre road-boundary tolerance and its corresponding road/grass coverage thresholds.} \\
\ubtestbadge{23} & \textcolor{zGoldDeep!78!black}{\texttt{\bfseries Terrain\-Paint\-Mask\_\allowbreak{}Strict\_\allowbreak{}0\_\allowbreak{}35m}}\enspace\textcolor{zText}{Checks the strict Terrain paint-mask tier using a 0.35-metre road-boundary tolerance and its corresponding road/grass coverage thresholds.} \\
\ubtestbadge{24} & \textcolor{zGoldDeep!78!black}{\texttt{\bfseries White\-Line\-Geometry\-Uses\-White\-Near\-Ground\-Renderers}}\enspace\textcolor{zText}{Checks that white line geometry uses white near ground renderers.} \\
\ubtestbadge{25} & \textcolor{zGoldDeep!78!black}{\texttt{\bfseries White\-Line\-Samples\-Stay\-On\-Road\-Edges}}\enspace\textcolor{zText}{Checks that white line samples stay on road edges.} \\
\ubtestbadge{26} & \textcolor{zGoldDeep!78!black}{\texttt{\bfseries White\-Line\-Segments\-Do\-Not\-Jump\-Across\-Main\-Track\-Sides}}\enspace\textcolor{zText}{Checks that white line segments do not jump across main track sides.} \\
\ubtestbadge{27} & \textcolor{zGoldDeep!78!black}{\texttt{\bfseries White\-Line\-Segments\-Follow\-Local\-Road\-Tangent}}\enspace\textcolor{zText}{Checks that white line segments follow local road tangent.} \\
\ubtestbadge{28} & \textcolor{zGoldDeep!78!black}{\texttt{\bfseries White\-Lines\-Cover\-Both\-Main\-Track\-Edges}}\enspace\textcolor{zText}{Checks that white lines cover both main track edges.} \\
\noalign{\vskip 2pt}
\multicolumn{2}{@{}p{14.4cm}@{}}{\textcolor{zRockLite!42}{\rule{\linewidth}{0.3pt}}}\\[3pt]
\multicolumn{2}{@{}p{14.4cm}@{}}{\ubtaskhead{zPlain}{0070-CameraPlacement}{Place Indoor and Outdoor Observation Cameras}{zShrine}{Hard}{11}}\\*[2pt]
\ubtestbadge{01} & \textcolor{zGoldDeep!78!black}{\texttt{\bfseries Outdoor\_\allowbreak{}Area\-A\_\allowbreak{}Visibility\-And\-No\-Overlap}}\enspace\textcolor{zText}{Checks that CameraAreaA clearly frames its assigned outdoor props without unacceptable visual overlap.} \\
\ubtestbadge{02} & \textcolor{zGoldDeep!78!black}{\texttt{\bfseries Outdoor\_\allowbreak{}Area\-B\_\allowbreak{}Visibility\-And\-No\-Overlap}}\enspace\textcolor{zText}{Checks that CameraAreaB clearly frames its assigned outdoor props without unacceptable visual overlap.} \\
\ubtestbadge{03} & \textcolor{zGoldDeep!78!black}{\texttt{\bfseries Outdoor\_\allowbreak{}Stones\_\allowbreak{}Visibility\-And\-No\-Overlap}}\enspace\textcolor{zText}{Checks that CameraStones clearly frames its assigned stone props without unacceptable visual overlap.} \\
\ubtestbadge{04} & \textcolor{zGoldDeep!78!black}{\texttt{\bfseries Outdoor\_\allowbreak{}Umbrella\_\allowbreak{}Visibility\-And\-No\-Overlap}}\enspace\textcolor{zText}{Checks that CameraUmbrella clearly frames its assigned patio props without unacceptable visual overlap.} \\
\ubtestbadge{05} & \textcolor{zGoldDeep!78!black}{\texttt{\bfseries Outdoor\-Overall\_\allowbreak{}At\-Least60\-Percent\-Props\-Pass}}\enspace\textcolor{zText}{Checks that at least 60\% of all assigned outdoor props satisfy the visibility criteria.} \\
\ubtestbadge{06} & \textcolor{zGoldDeep!78!black}{\texttt{\bfseries Bath\_\allowbreak{}Visibility\-And\-No\-Overlap}}\enspace\textcolor{zText}{Checks that CameraBathRoom clearly frames its assigned bathroom props without unacceptable visual overlap.} \\
\ubtestbadge{07} & \textcolor{zGoldDeep!78!black}{\texttt{\bfseries Living\_\allowbreak{}Visibility\-And\-No\-Overlap}}\enspace\textcolor{zText}{Checks that CameraLiving clearly frames its assigned kitchen and living-room props without unacceptable visual overlap.} \\
\ubtestbadge{08} & \textcolor{zGoldDeep!78!black}{\texttt{\bfseries Bed\_\allowbreak{}Visibility\-And\-No\-Overlap}}\enspace\textcolor{zText}{Checks that CameraBedRoom clearly frames its assigned bedroom props without unacceptable visual overlap.} \\
\ubtestbadge{09} & \textcolor{zGoldDeep!78!black}{\texttt{\bfseries Garage\_\allowbreak{}Visibility\-And\-No\-Overlap}}\enspace\textcolor{zText}{Checks that CameraGarage clearly frames its assigned garage props without unacceptable visual overlap.} \\
\ubtestbadge{10} & \textcolor{zGoldDeep!78!black}{\texttt{\bfseries Corr\_\allowbreak{}Visibility\-And\-No\-Overlap}}\enspace\textcolor{zText}{Checks that CameraCorr clearly frames its assigned corridor props without unacceptable visual overlap.} \\
\ubtestbadge{11} & \textcolor{zGoldDeep!78!black}{\texttt{\bfseries Overall\_\allowbreak{}At\-Least60\-Percent\-Props\-Pass}}\enspace\textcolor{zText}{Checks that at least 60\% of all assigned indoor props satisfy the visibility criteria.} \\
\noalign{\vskip 2pt}
\multicolumn{2}{@{}p{14.4cm}@{}}{\textcolor{zRockLite!42}{\rule{\linewidth}{0.3pt}}}\\[3pt]
\multicolumn{2}{@{}p{14.4cm}@{}}{\ubtaskhead{zPlain}{0071-PlantsDecoration}{Decorate Terrain with Natural Trees and Riverside Rocks}{zShrine}{Hard}{24}}\\*[2pt]
\ubtestbadge{01} & \textcolor{zGoldDeep!78!black}{\texttt{\bfseries T01\_\allowbreak{}Tree\-Counts\-Are\-Exactly150\-Pines\-And50\-Fruit\-Trees}}\enspace\textcolor{zText}{Checks for exactly 150 pine and 50 fruit-tree Terrain instances.} \\
\ubtestbadge{02} & \textcolor{zGoldDeep!78!black}{\texttt{\bfseries T02\_\allowbreak{}All\-Tree\-Prototypes\-Belong\-To\-The\-Required\-Families}}\enspace\textcolor{zText}{Checks that every Terrain tree uses one of the required pine or fruit-tree prototype families.} \\
\ubtestbadge{03} & \textcolor{zGoldDeep!78!black}{\texttt{\bfseries T03\_\allowbreak{}T04\_\allowbreak{}At\-Least90\-Percent\-Of\-Trees\-Stand\-On75\-Percent\-Grass}}\enspace\textcolor{zText}{Checks that at least 90\% of trees stand on ground with at least 75\% grass texture weight.} \\
\ubtestbadge{04} & \textcolor{zGoldDeep!78!black}{\texttt{\bfseries T05\_\allowbreak{}All\-Trees\-Stand\-On75\-Percent\-Grass}}\enspace\textcolor{zText}{Checks that every tree stands on ground with at least 75\% grass texture weight.} \\
\ubtestbadge{05} & \textcolor{zGoldDeep!78!black}{\texttt{\bfseries T06\_\allowbreak{}No\-Tree\-Has\-More\-Than20\-Percent\-Of\-One\-Non\-Grass\-Texture}}\enspace\textcolor{zText}{Checks that no tree location contains more than 20\% weight from any one non-grass texture.} \\
\ubtestbadge{06} & \textcolor{zGoldDeep!78!black}{\texttt{\bfseries T07\_\allowbreak{}At\-Least120\-Of150\-Pines\-Use\-Hilly\-Grassland}}\enspace\textcolor{zText}{Checks that at least 120 of the 150 pines occupy hilly grassland.} \\
\ubtestbadge{07} & \textcolor{zGoldDeep!78!black}{\texttt{\bfseries T08\_\allowbreak{}All150\-Pines\-Use\-Hilly\-Grassland}}\enspace\textcolor{zText}{Checks that all 150 pines occupy hilly grassland.} \\
\ubtestbadge{08} & \textcolor{zGoldDeep!78!black}{\texttt{\bfseries T09\_\allowbreak{}At\-Least40\-Of50\-Fruit\-Trees\-Use\-Gentle\-Low\-Grassland}}\enspace\textcolor{zText}{Checks that at least 40 of the 50 fruit trees occupy gentle low grassland.} \\
\ubtestbadge{09} & \textcolor{zGoldDeep!78!black}{\texttt{\bfseries T10\_\allowbreak{}All50\-Fruit\-Trees\-Use\-Gentle\-Low\-Grassland}}\enspace\textcolor{zText}{Checks that all 50 fruit trees occupy gentle low grassland.} \\
\ubtestbadge{10} & \textcolor{zGoldDeep!78!black}{\texttt{\bfseries T11\_\allowbreak{}Fruit\-Trees\-Avoid\-Severely\-Undulating\-Terrain\-And\-Its\-Two\-Metre\-Buffer}}\enspace\textcolor{zText}{Checks that fruit trees avoid severely undulating terrain and its two-metre buffer.} \\
\ubtestbadge{11} & \textcolor{zGoldDeep!78!black}{\texttt{\bfseries T12\_\allowbreak{}At\-Least95\-Percent\-Of\-Trees\-Stay\-Three\-Point\-Two\-Metres\-From\-Visible\-Water}}\enspace\textcolor{zText}{Checks that at least 95\% of trees remain 3.2 metres or more from visible water.} \\
\ubtestbadge{12} & \textcolor{zGoldDeep!78!black}{\texttt{\bfseries T13\_\allowbreak{}Pine\-Spacing\-Forms\-Natural\-Clusters\-Rather\-Than\-AUniform\-Grid}}\enspace\textcolor{zText}{Checks that pine spacing forms natural clusters rather than a uniform grid.} \\
\ubtestbadge{13} & \textcolor{zGoldDeep!78!black}{\texttt{\bfseries T14\_\allowbreak{}Fruit\-Tree\-Spacing\-Forms\-Natural\-Clusters\-Rather\-Than\-AUniform\-Grid}}\enspace\textcolor{zText}{Checks that fruit-tree spacing forms natural clusters rather than a uniform grid.} \\
\ubtestbadge{14} & \textcolor{zGoldDeep!78!black}{\texttt{\bfseries T15\_\allowbreak{}Woodland\-Contains\-Dense\-Groups\-Transitions\-And\-Open\-Space}}\enspace\textcolor{zText}{Checks that the woodland includes dense groups, transition planting, and open meadow.} \\
\ubtestbadge{15} & \textcolor{zGoldDeep!78!black}{\texttt{\bfseries T16\_\allowbreak{}Trees\-Use\-Variant\-Rotation\-And\-Scale\-Diversity}}\enspace\textcolor{zText}{Checks that trees have restrained but meaningful rotation and scale diversity.} \\
\ubtestbadge{16} & \textcolor{zGoldDeep!78!black}{\texttt{\bfseries R01\_\allowbreak{}Riverside\-Rock\-Group\-Contains\-Exactly50\-Named\-Rocks}}\enspace\textcolor{zText}{Checks that Riverside Rocks has exactly 50 direct children with the required sequential names.} \\
\ubtestbadge{17} & \textcolor{zGoldDeep!78!black}{\texttt{\bfseries R02\_\allowbreak{}Rocks\-Use\-At\-Least\-Four\-Variants\-Without\-One\-Dominating}}\enspace\textcolor{zText}{Checks that rocks use at least four supplied variants without one variant dominating.} \\
\ubtestbadge{18} & \textcolor{zGoldDeep!78!black}{\texttt{\bfseries R03\_\allowbreak{}At\-Least60\-Percent\-Of\-Rock\-Footprints\-Are\-Fully\-On\-Dry\-Land}}\enspace\textcolor{zText}{Checks that at least 60\% of rock footprints lie fully on dry land.} \\
\ubtestbadge{19} & \textcolor{zGoldDeep!78!black}{\texttt{\bfseries R04\_\allowbreak{}All\-Rock\-Footprints\-Are\-Fully\-On\-Dry\-Land}}\enspace\textcolor{zText}{Checks that every rock footprint lies fully on dry land.} \\
\ubtestbadge{20} & \textcolor{zGoldDeep!78!black}{\texttt{\bfseries R05\_\allowbreak{}At\-Least90\-Percent\-Of\-Rocks\-Occupy\-The\-Size\-Aware\-Shore\-Band}}\enspace\textcolor{zText}{Checks that at least 90\% of rocks occupy the size-aware shoreline band.} \\
\ubtestbadge{21} & \textcolor{zGoldDeep!78!black}{\texttt{\bfseries R06\_\allowbreak{}All50\-Rocks\-Are\-Visually\-Grounded}}\enspace\textcolor{zText}{Checks that all 50 rocks are visually grounded on the Terrain.} \\
\ubtestbadge{22} & \textcolor{zGoldDeep!78!black}{\texttt{\bfseries R07\_\allowbreak{}All50\-Rocks\-Clear\-Trees\-And\-Cobblestone\-Roads}}\enspace\textcolor{zText}{Checks that all 50 rocks remain clear of trees and cobblestone roads.} \\
\ubtestbadge{23} & \textcolor{zGoldDeep!78!black}{\texttt{\bfseries R08\_\allowbreak{}Rock\-Colliders\-Do\-Not\-Physically\-Penetrate\-Each\-Other}}\enspace\textcolor{zText}{Checks that rock colliders do not physically penetrate one another.} \\
\ubtestbadge{24} & \textcolor{zGoldDeep!78!black}{\texttt{\bfseries R09\_\allowbreak{}Rocks\-Cover\-The\-Water\-System\-With\-Non\-Mechanical\-Spacing}}\enspace\textcolor{zText}{Checks that rocks cover the lake and river system with non-mechanical spacing.} \\
\noalign{\vskip 2pt}
\multicolumn{2}{@{}p{14.4cm}@{}}{\textcolor{zRockLite!42}{\rule{\linewidth}{0.3pt}}}\\[3pt]
\multicolumn{2}{@{}p{14.4cm}@{}}{\ubtaskhead{zPlain}{0073-CastleConstruction}{Connect Castle Circulation with Supplied Stair Prefabs}{zShrine}{Hard}{9}}\\*[2pt]
\ubtestbadge{01} & \textcolor{zGoldDeep!78!black}{\texttt{\bfseries Straight\-Stairs\_\allowbreak{}At\-Least50\-Percent\-Of\-Gold\-Connections\-Correct}}\enspace\textcolor{zText}{Checks that at least 50\% of golden straight-stair connections are correctly assembled.} \\
\ubtestbadge{02} & \textcolor{zGoldDeep!78!black}{\texttt{\bfseries Straight\-Stairs\_\allowbreak{}100\-Percent\-Of\-Gold\-Connections\-Correct}}\enspace\textcolor{zText}{Checks that all golden straight-stair connections are correctly assembled.} \\
\ubtestbadge{03} & \textcolor{zGoldDeep!78!black}{\texttt{\bfseries Round\-Stairs\_\allowbreak{}At\-Least50\-Percent\-Of\-Gold\-Connections\-Correct}}\enspace\textcolor{zText}{Checks that at least 50\% of golden round-stair connections are correctly assembled.} \\
\ubtestbadge{04} & \textcolor{zGoldDeep!78!black}{\texttt{\bfseries Round\-Stairs\_\allowbreak{}100\-Percent\-Of\-Gold\-Connections\-Correct}}\enspace\textcolor{zText}{Checks that all golden round-stair connections are correctly assembled.} \\
\ubtestbadge{05} & \textcolor{zGoldDeep!78!black}{\texttt{\bfseries Corner\-Stairs\_\allowbreak{}At\-Least50\-Percent\-Of\-Gold\-Connections\-Correct}}\enspace\textcolor{zText}{Checks that at least 50\% of golden corner-stair connections are correctly assembled.} \\
\ubtestbadge{06} & \textcolor{zGoldDeep!78!black}{\texttt{\bfseries Corner\-Stairs\_\allowbreak{}100\-Percent\-Of\-Gold\-Connections\-Correct}}\enspace\textcolor{zText}{Checks that all golden corner-stair connections are correctly assembled.} \\
\ubtestbadge{07} & \textcolor{zGoldDeep!78!black}{\texttt{\bfseries Straight\-Stairs\_\allowbreak{}100\-Percent\-Correct\-With\-No\-Redundant\-Instances}}\enspace\textcolor{zText}{Checks that all required straight-stair connections are correct and no redundant instances remain.} \\
\ubtestbadge{08} & \textcolor{zGoldDeep!78!black}{\texttt{\bfseries Round\-Stairs\_\allowbreak{}100\-Percent\-Correct\-With\-No\-Redundant\-Instances}}\enspace\textcolor{zText}{Checks that all required round-stair connections are correct and no redundant instances remain.} \\
\ubtestbadge{09} & \textcolor{zGoldDeep!78!black}{\texttt{\bfseries Corner\-Stairs\_\allowbreak{}100\-Percent\-Correct\-With\-No\-Redundant\-Instances}}\enspace\textcolor{zText}{Checks that all required corner-stair connections are correct and no redundant instances remain.} \\
\noalign{\vskip 2pt}
\multicolumn{2}{@{}p{14.4cm}@{}}{\textcolor{zRockLite!42}{\rule{\linewidth}{0.3pt}}}\\[3pt]
\multicolumn{2}{@{}p{14.4cm}@{}}{\ubtaskhead{zPlain}{0075-CrystalFragment}{Restore the Crystal Fragment Assembly}{zShrine}{Hard}{14}}\\*[2pt]
\ubtestbadge{01} & \textcolor{zGoldDeep!78!black}{\texttt{\bfseries Movable\-Fragment\-Matches\-Golden\-Assembly("Object001", 1f, 1.2f, -1f, -60f)}}\enspace\textcolor{zText}{Checks that Object001's serialized position and rotation match the golden crystal assembly within tolerance.} \\
\ubtestbadge{02} & \textcolor{zGoldDeep!78!black}{\texttt{\bfseries Movable\-Fragment\-Matches\-Golden\-Assembly("Object003", 1f, 1.2f, 1f, -90f)}}\enspace\textcolor{zText}{Checks that Object003's serialized position and rotation match the golden crystal assembly within tolerance.} \\
\ubtestbadge{03} & \textcolor{zGoldDeep!78!black}{\texttt{\bfseries Movable\-Fragment\-Matches\-Golden\-Assembly("Object004", 1f, 1.2f, 1f, -45f)}}\enspace\textcolor{zText}{Checks that Object004's serialized position and rotation match the golden crystal assembly within tolerance.} \\
\ubtestbadge{04} & \textcolor{zGoldDeep!78!black}{\texttt{\bfseries Movable\-Fragment\-Matches\-Golden\-Assembly("Object006", 0.1f, 2f, -0.2f, -30f)}}\enspace\textcolor{zText}{Checks that Object006's serialized position and rotation match the golden crystal assembly within tolerance.} \\
\ubtestbadge{05} & \textcolor{zGoldDeep!78!black}{\texttt{\bfseries Movable\-Fragment\-Matches\-Golden\-Assembly("Object007", -1f, 1.2f, -1f, -90f)}}\enspace\textcolor{zText}{Checks that Object007's serialized position and rotation match the golden crystal assembly within tolerance.} \\
\ubtestbadge{06} & \textcolor{zGoldDeep!78!black}{\texttt{\bfseries Movable\-Fragment\-Matches\-Golden\-Assembly("Object008", -1f, 0f, -0.5f, -60f)}}\enspace\textcolor{zText}{Checks that Object008's serialized position and rotation match the golden crystal assembly within tolerance.} \\
\ubtestbadge{07} & \textcolor{zGoldDeep!78!black}{\texttt{\bfseries Movable\-Fragment\-Matches\-Golden\-Assembly("Object009", 0.1f, 2f, 1f, -60f)}}\enspace\textcolor{zText}{Checks that Object009's serialized position and rotation match the golden crystal assembly within tolerance.} \\
\ubtestbadge{08} & \textcolor{zGoldDeep!78!black}{\texttt{\bfseries Movable\-Fragment\-Matches\-Golden\-Assembly("Object010", -1f, 2f, -0.4f, -45f)}}\enspace\textcolor{zText}{Checks that Object010's serialized position and rotation match the golden crystal assembly within tolerance.} \\
\ubtestbadge{09} & \textcolor{zGoldDeep!78!black}{\texttt{\bfseries Movable\-Fragment\-Matches\-Golden\-Assembly("Object011", -1f, 2f, 0.3f, -15f)}}\enspace\textcolor{zText}{Checks that Object011's serialized position and rotation match the golden crystal assembly within tolerance.} \\
\ubtestbadge{10} & \textcolor{zGoldDeep!78!black}{\texttt{\bfseries Movable\-Fragment\-Matches\-Golden\-Assembly("Object012", 0.1f, 2f, -1f, 0f)}}\enspace\textcolor{zText}{Checks that Object012's serialized position and rotation match the golden crystal assembly within tolerance.} \\
\ubtestbadge{11} & \textcolor{zGoldDeep!78!black}{\texttt{\bfseries Movable\-Fragment\-Matches\-Golden\-Assembly("Object013", 0.1f, 0f, -1f, -90f)}}\enspace\textcolor{zText}{Checks that Object013's serialized position and rotation match the golden crystal assembly within tolerance.} \\
\ubtestbadge{12} & \textcolor{zGoldDeep!78!black}{\texttt{\bfseries Movable\-Fragment\-Matches\-Golden\-Assembly("Object014", 1f, 2f, -0.1f, -30f)}}\enspace\textcolor{zText}{Checks that Object014's serialized position and rotation match the golden crystal assembly within tolerance.} \\
\ubtestbadge{13} & \textcolor{zGoldDeep!78!black}{\texttt{\bfseries Movable\-Fragment\-Matches\-Golden\-Assembly("Object015", 0.1f, 0f, -1f, -75f)}}\enspace\textcolor{zText}{Checks that Object015's serialized position and rotation match the golden crystal assembly within tolerance.} \\
\ubtestbadge{14} & \textcolor{zGoldDeep!78!black}{\texttt{\bfseries Movable\-Fragment\-Matches\-Golden\-Assembly("Object016", 1f, 2f, -0.1f, -45f)}}\enspace\textcolor{zText}{Checks that Object016's serialized position and rotation match the golden crystal assembly within tolerance.} \\
\noalign{\vskip 2pt}
\multicolumn{2}{@{}p{14.4cm}@{}}{\textcolor{zRockLite!42}{\rule{\linewidth}{0.3pt}}}\\[3pt]
\multicolumn{2}{@{}p{14.4cm}@{}}{\ubtaskhead{zPlain}{0076-BuildHouseFromDesign}{Build a White-Box House from a Floor Plan}{zShrine}{Hard}{23}}\\*[2pt]
\ubtestbadge{01} & \textcolor{zGoldDeep!78!black}{\texttt{\bfseries Structure\_\allowbreak{}Walls\_\allowbreak{}Exists\_\allowbreak{}And\_\allowbreak{}Tagged\_\allowbreak{}And\_\allowbreak{}Thickness}}\enspace\textcolor{zText}{Checks that required walls exist, use the correct ExtWall or InsideWall tag and parent, and have valid thickness.} \\
\ubtestbadge{02} & \textcolor{zGoldDeep!78!black}{\texttt{\bfseries Structure\_\allowbreak{}Doors\-Windows\-Furniture\_\allowbreak{}Exist}}\enspace\textcolor{zText}{Checks that every required door, window, and labeled furniture GameObject exists.} \\
\ubtestbadge{03} & \textcolor{zGoldDeep!78!black}{\texttt{\bfseries Window\_\allowbreak{}Size\_\allowbreak{}All\-Correct}}\enspace\textcolor{zText}{Checks that every window is 1.5 metres tall with its sill 1 metre above the floor.} \\
\ubtestbadge{04} & \textcolor{zGoldDeep!78!black}{\texttt{\bfseries Window\_\allowbreak{}On\-Exterior\-Wall}}\enspace\textcolor{zText}{Checks that every window is placed in an exterior wall opening.} \\
\ubtestbadge{05} & \textcolor{zGoldDeep!78!black}{\texttt{\bfseries Door\_\allowbreak{}On\-Interior\-Wall}}\enspace\textcolor{zText}{Checks that every door is placed in an interior wall opening.} \\
\ubtestbadge{06} & \textcolor{zGoldDeep!78!black}{\texttt{\bfseries Window\_\allowbreak{}Distance\_\allowbreak{}W1\_\allowbreak{}W2\_\allowbreak{}Horizontal\_\allowbreak{}2\_\allowbreak{}85}}\enspace\textcolor{zText}{Checks that the horizontal edge gap between Window1 and Window2 is 2.85 metres within tolerance.} \\
\ubtestbadge{07} & \textcolor{zGoldDeep!78!black}{\texttt{\bfseries Window\_\allowbreak{}Distance\_\allowbreak{}W3\_\allowbreak{}W4\_\allowbreak{}Horizontal\_\allowbreak{}0\_\allowbreak{}99}}\enspace\textcolor{zText}{Checks that the horizontal edge gap between Window3 and Window4 is 0.99 metres within tolerance.} \\
\ubtestbadge{08} & \textcolor{zGoldDeep!78!black}{\texttt{\bfseries Window\_\allowbreak{}Distance\_\allowbreak{}W4\_\allowbreak{}W5\_\allowbreak{}Horizontal\_\allowbreak{}2\_\allowbreak{}85}}\enspace\textcolor{zText}{Checks that the horizontal edge gap between Window4 and Window5 is 2.85 metres within tolerance.} \\
\ubtestbadge{09} & \textcolor{zGoldDeep!78!black}{\texttt{\bfseries Window\_\allowbreak{}Distance\_\allowbreak{}W6\_\allowbreak{}W7\_\allowbreak{}Perpendicular\_\allowbreak{}Vertical\-Gap\_\allowbreak{}1\_\allowbreak{}To\_\allowbreak{}1\_\allowbreak{}5}}\enspace\textcolor{zText}{Checks that Window6 and Window7 are on perpendicular walls with an edge gap from 1.0 to 1.5 metres.} \\
\ubtestbadge{10} & \textcolor{zGoldDeep!78!black}{\texttt{\bfseries Furniture\_\allowbreak{}Single\-Sofa\_\allowbreak{}Near\-Window\_\allowbreak{}YDistance\_\allowbreak{}Single\-Sofa1}}\enspace\textcolor{zText}{Checks that SingleSofa1 is less than 1.8 metres from Window3 along the plan's north-south axis.} \\
\ubtestbadge{11} & \textcolor{zGoldDeep!78!black}{\texttt{\bfseries Furniture\_\allowbreak{}Single\-Sofa\_\allowbreak{}Near\-Window\_\allowbreak{}YDistance\_\allowbreak{}Single\-Sofa2}}\enspace\textcolor{zText}{Checks that SingleSofa2 is less than 1.5 metres from Window5 along the plan's north-south axis.} \\
\ubtestbadge{12} & \textcolor{zGoldDeep!78!black}{\texttt{\bfseries Furniture\_\allowbreak{}Single\-Sofa\_\allowbreak{}Near\-Window\_\allowbreak{}YDistance\_\allowbreak{}Single\-Sofa3}}\enspace\textcolor{zText}{Checks that SingleSofa3 is less than 1.5 metres from Window1 along the plan's north-south axis.} \\
\ubtestbadge{13} & \textcolor{zGoldDeep!78!black}{\texttt{\bfseries Furniture\_\allowbreak{}Single\-Sofa\_\allowbreak{}Has\-Rotated\_\allowbreak{}Y\_\allowbreak{}Single\-Sofa1}}\enspace\textcolor{zText}{Checks that SingleSofa1 has a non-cardinal Y-axis rotation.} \\
\ubtestbadge{14} & \textcolor{zGoldDeep!78!black}{\texttt{\bfseries Furniture\_\allowbreak{}Single\-Sofa\_\allowbreak{}Has\-Rotated\_\allowbreak{}Y\_\allowbreak{}Single\-Sofa2}}\enspace\textcolor{zText}{Checks that SingleSofa2 has a non-cardinal Y-axis rotation.} \\
\ubtestbadge{15} & \textcolor{zGoldDeep!78!black}{\texttt{\bfseries Furniture\_\allowbreak{}Single\-Sofa\_\allowbreak{}Has\-Rotated\_\allowbreak{}Y\_\allowbreak{}Single\-Sofa3}}\enspace\textcolor{zText}{Checks that SingleSofa3 has a non-cardinal Y-axis rotation.} \\
\ubtestbadge{16} & \textcolor{zGoldDeep!78!black}{\texttt{\bfseries Furniture\_\allowbreak{}Bed\_\allowbreak{}Near\-Wall\_\allowbreak{}Bed1}}\enspace\textcolor{zText}{Checks that Bed1 lies within its expected east-wall position range.} \\
\ubtestbadge{17} & \textcolor{zGoldDeep!78!black}{\texttt{\bfseries Furniture\_\allowbreak{}Bed\_\allowbreak{}Near\-Wall\_\allowbreak{}Bed2}}\enspace\textcolor{zText}{Checks that Bed2 lies within its expected west-wall position range.} \\
\ubtestbadge{18} & \textcolor{zGoldDeep!78!black}{\texttt{\bfseries Furniture\_\allowbreak{}Bed\_\allowbreak{}Near\-Wall\_\allowbreak{}Bed3}}\enspace\textcolor{zText}{Checks that Bed3 lies within its expected room position range.} \\
\ubtestbadge{19} & \textcolor{zGoldDeep!78!black}{\texttt{\bfseries Furniture\_\allowbreak{}Bed\-And\-Sofa\_\allowbreak{}Size}}\enspace\textcolor{zText}{Checks the X/Z dimensions of Bed1–Bed3 and Sofa1–Sofa2 against the design values.} \\
\ubtestbadge{20} & \textcolor{zGoldDeep!78!black}{\texttt{\bfseries Furniture\_\allowbreak{}No\-Penetration\_\allowbreak{}Between\-Furniture}}\enspace\textcolor{zText}{Checks that furniture renderer bounds do not penetrate one another beyond tolerance.} \\
\ubtestbadge{21} & \textcolor{zGoldDeep!78!black}{\texttt{\bfseries Furniture\_\allowbreak{}No\-Penetration\_\allowbreak{}With\-Walls}}\enspace\textcolor{zText}{Checks that furniture renderer bounds do not penetrate wall bounds beyond tolerance.} \\
\ubtestbadge{22} & \textcolor{zGoldDeep!78!black}{\texttt{\bfseries Wall\_\allowbreak{}Shape\_\allowbreak{}Ext\-Wall\_\allowbreak{}Path\-Coverage}}\enspace\textcolor{zText}{Checks exterior-wall path coverage against the floor-plan reference geometry.} \\
\ubtestbadge{23} & \textcolor{zGoldDeep!78!black}{\texttt{\bfseries Wall\_\allowbreak{}Shape\_\allowbreak{}Inside\-Wall\_\allowbreak{}Path\-Coverage}}\enspace\textcolor{zText}{Checks interior-wall path coverage against the floor-plan reference geometry.} \\
\noalign{\vskip 2pt}
\multicolumn{2}{@{}p{14.4cm}@{}}{\textcolor{zRockLite!42}{\rule{\linewidth}{0.3pt}}}\\[3pt]
\multicolumn{2}{@{}p{14.4cm}@{}}{\ubtaskhead{zPlain}{0080-UrbanRoads}{Add Correct Lane-Direction Arrows to Urban Intersections}{zShrine}{Hard}{31}}\\*[2pt]
\ubtestbadge{01} & \textcolor{zGoldDeep!78!black}{\texttt{\bfseries Area\_\allowbreak{}D (12)\_\allowbreak{}D}}\enspace\textcolor{zText}{Checks that approach area D (12) contains the correctly oriented straight arrow prefab, within the area bounds and flush with the road.} \\
\ubtestbadge{02} & \textcolor{zGoldDeep!78!black}{\texttt{\bfseries Area\_\allowbreak{}D (2)\_\allowbreak{}D}}\enspace\textcolor{zText}{Checks that approach area D (2) contains the correctly oriented straight arrow prefab, within the area bounds and flush with the road.} \\
\ubtestbadge{03} & \textcolor{zGoldDeep!78!black}{\texttt{\bfseries Area\_\allowbreak{}D (28)\_\allowbreak{}D}}\enspace\textcolor{zText}{Checks that approach area D (28) contains the correctly oriented straight arrow prefab, within the area bounds and flush with the road.} \\
\ubtestbadge{04} & \textcolor{zGoldDeep!78!black}{\texttt{\bfseries Area\_\allowbreak{}D (30)\_\allowbreak{}D}}\enspace\textcolor{zText}{Checks that approach area D (30) contains the correctly oriented straight arrow prefab, within the area bounds and flush with the road.} \\
\ubtestbadge{05} & \textcolor{zGoldDeep!78!black}{\texttt{\bfseries Area\_\allowbreak{}DR (13)\_\allowbreak{}DR}}\enspace\textcolor{zText}{Checks that approach area DR (13) contains the correctly oriented straight-plus-right arrow prefab, within the area bounds and flush with the road.} \\
\ubtestbadge{06} & \textcolor{zGoldDeep!78!black}{\texttt{\bfseries Area\_\allowbreak{}DR (25)\_\allowbreak{}DR}}\enspace\textcolor{zText}{Checks that approach area DR (25) contains the correctly oriented straight-plus-right arrow prefab, within the area bounds and flush with the road.} \\
\ubtestbadge{07} & \textcolor{zGoldDeep!78!black}{\texttt{\bfseries Area\_\allowbreak{}DR (29)\_\allowbreak{}DR}}\enspace\textcolor{zText}{Checks that approach area DR (29) contains the correctly oriented straight-plus-right arrow prefab, within the area bounds and flush with the road.} \\
\ubtestbadge{08} & \textcolor{zGoldDeep!78!black}{\texttt{\bfseries Area\_\allowbreak{}DR (38)\_\allowbreak{}DR}}\enspace\textcolor{zText}{Checks that approach area DR (38) contains the correctly oriented straight-plus-right arrow prefab, within the area bounds and flush with the road.} \\
\ubtestbadge{09} & \textcolor{zGoldDeep!78!black}{\texttt{\bfseries Area\_\allowbreak{}DR (4)\_\allowbreak{}DR}}\enspace\textcolor{zText}{Checks that approach area DR (4) contains the correctly oriented straight-plus-right arrow prefab, within the area bounds and flush with the road.} \\
\ubtestbadge{10} & \textcolor{zGoldDeep!78!black}{\texttt{\bfseries Area\_\allowbreak{}DR (7)\_\allowbreak{}DR}}\enspace\textcolor{zText}{Checks that approach area DR (7) contains the correctly oriented straight-plus-right arrow prefab, within the area bounds and flush with the road.} \\
\ubtestbadge{11} & \textcolor{zGoldDeep!78!black}{\texttt{\bfseries Area\_\allowbreak{}L (22)\_\allowbreak{}L}}\enspace\textcolor{zText}{Checks that approach area L (22) contains the correctly oriented left-turn arrow prefab, within the area bounds and flush with the road.} \\
\ubtestbadge{12} & \textcolor{zGoldDeep!78!black}{\texttt{\bfseries Area\_\allowbreak{}L (24)\_\allowbreak{}L}}\enspace\textcolor{zText}{Checks that approach area L (24) contains the correctly oriented left-turn arrow prefab, within the area bounds and flush with the road.} \\
\ubtestbadge{13} & \textcolor{zGoldDeep!78!black}{\texttt{\bfseries Area\_\allowbreak{}L (3)\_\allowbreak{}L}}\enspace\textcolor{zText}{Checks that approach area L (3) contains the correctly oriented left-turn arrow prefab, within the area bounds and flush with the road.} \\
\ubtestbadge{14} & \textcolor{zGoldDeep!78!black}{\texttt{\bfseries Area\_\allowbreak{}L (31)\_\allowbreak{}L}}\enspace\textcolor{zText}{Checks that approach area L (31) contains the correctly oriented left-turn arrow prefab, within the area bounds and flush with the road.} \\
\ubtestbadge{15} & \textcolor{zGoldDeep!78!black}{\texttt{\bfseries Area\_\allowbreak{}L (34)\_\allowbreak{}L}}\enspace\textcolor{zText}{Checks that approach area L (34) contains the correctly oriented left-turn arrow prefab, within the area bounds and flush with the road.} \\
\ubtestbadge{16} & \textcolor{zGoldDeep!78!black}{\texttt{\bfseries Area\_\allowbreak{}L (35)\_\allowbreak{}L}}\enspace\textcolor{zText}{Checks that approach area L (35) contains the correctly oriented left-turn arrow prefab, within the area bounds and flush with the road.} \\
\ubtestbadge{17} & \textcolor{zGoldDeep!78!black}{\texttt{\bfseries Area\_\allowbreak{}L (37)\_\allowbreak{}L}}\enspace\textcolor{zText}{Checks that approach area L (37) contains the correctly oriented left-turn arrow prefab, within the area bounds and flush with the road.} \\
\ubtestbadge{18} & \textcolor{zGoldDeep!78!black}{\texttt{\bfseries Area\_\allowbreak{}L (39)\_\allowbreak{}L}}\enspace\textcolor{zText}{Checks that approach area L (39) contains the correctly oriented left-turn arrow prefab, within the area bounds and flush with the road.} \\
\ubtestbadge{19} & \textcolor{zGoldDeep!78!black}{\texttt{\bfseries Area\_\allowbreak{}L (5)\_\allowbreak{}L}}\enspace\textcolor{zText}{Checks that approach area L (5) contains the correctly oriented left-turn arrow prefab, within the area bounds and flush with the road.} \\
\ubtestbadge{20} & \textcolor{zGoldDeep!78!black}{\texttt{\bfseries Area\_\allowbreak{}L (6)\_\allowbreak{}L}}\enspace\textcolor{zText}{Checks that approach area L (6) contains the correctly oriented left-turn arrow prefab, within the area bounds and flush with the road.} \\
\ubtestbadge{21} & \textcolor{zGoldDeep!78!black}{\texttt{\bfseries Area\_\allowbreak{}LD (14)\_\allowbreak{}LD}}\enspace\textcolor{zText}{Checks that approach area LD (14) contains the correctly oriented left-plus-straight arrow prefab, within the area bounds and flush with the road.} \\
\ubtestbadge{22} & \textcolor{zGoldDeep!78!black}{\texttt{\bfseries Area\_\allowbreak{}LDR (18)\_\allowbreak{}LDR}}\enspace\textcolor{zText}{Checks that approach area LDR (18) contains the correctly oriented left-plus-straight-plus-right arrow prefab, within the area bounds and flush with the road.} \\
\ubtestbadge{23} & \textcolor{zGoldDeep!78!black}{\texttt{\bfseries Area\_\allowbreak{}LDR (42)\_\allowbreak{}LDR}}\enspace\textcolor{zText}{Checks that approach area LDR (42) contains the correctly oriented left-plus-straight-plus-right arrow prefab, within the area bounds and flush with the road.} \\
\ubtestbadge{24} & \textcolor{zGoldDeep!78!black}{\texttt{\bfseries Area\_\allowbreak{}LDR (43)\_\allowbreak{}LDR}}\enspace\textcolor{zText}{Checks that approach area LDR (43) contains the correctly oriented left-plus-straight-plus-right arrow prefab, within the area bounds and flush with the road.} \\
\ubtestbadge{25} & \textcolor{zGoldDeep!78!black}{\texttt{\bfseries Area\_\allowbreak{}LDR (44)\_\allowbreak{}LDR}}\enspace\textcolor{zText}{Checks that approach area LDR (44) contains the correctly oriented left-plus-straight-plus-right arrow prefab, within the area bounds and flush with the road.} \\
\ubtestbadge{26} & \textcolor{zGoldDeep!78!black}{\texttt{\bfseries Area\_\allowbreak{}LR (15)\_\allowbreak{}LR}}\enspace\textcolor{zText}{Checks that approach area LR (15) contains the correctly oriented left-plus-right arrow prefab, within the area bounds and flush with the road.} \\
\ubtestbadge{27} & \textcolor{zGoldDeep!78!black}{\texttt{\bfseries Area\_\allowbreak{}LR (17)\_\allowbreak{}LR}}\enspace\textcolor{zText}{Checks that approach area LR (17) contains the correctly oriented left-plus-right arrow prefab, within the area bounds and flush with the road.} \\
\ubtestbadge{28} & \textcolor{zGoldDeep!78!black}{\texttt{\bfseries Area\_\allowbreak{}R (10)\_\allowbreak{}R}}\enspace\textcolor{zText}{Checks that approach area R (10) contains the correctly oriented right-turn arrow prefab, within the area bounds and flush with the road.} \\
\ubtestbadge{29} & \textcolor{zGoldDeep!78!black}{\texttt{\bfseries Area\_\allowbreak{}R (11)\_\allowbreak{}R}}\enspace\textcolor{zText}{Checks that approach area R (11) contains the correctly oriented right-turn arrow prefab, within the area bounds and flush with the road.} \\
\ubtestbadge{30} & \textcolor{zGoldDeep!78!black}{\texttt{\bfseries Area\_\allowbreak{}R (23)\_\allowbreak{}R}}\enspace\textcolor{zText}{Checks that approach area R (23) contains the correctly oriented right-turn arrow prefab, within the area bounds and flush with the road.} \\
\ubtestbadge{31} & \textcolor{zGoldDeep!78!black}{\texttt{\bfseries Area\_\allowbreak{}R (36)\_\allowbreak{}R}}\enspace\textcolor{zText}{Checks that approach area R (36) contains the correctly oriented right-turn arrow prefab, within the area bounds and flush with the road.} \\
\noalign{\vskip 2pt}
\multicolumn{2}{@{}p{14.4cm}@{}}{\textcolor{zRockLite!42}{\rule{\linewidth}{0.3pt}}}\\[3pt]
\noalign{\vskip 3pt}
\multicolumn{2}{@{}p{14.4cm}@{}}{\ubcategory{zShrine}{Particles \& Shaders}{12 tasks \textperiodcentered\ 155 milestone tests}}\\*[4pt]
\multicolumn{2}{@{}p{14.4cm}@{}}{\ubtaskhead{zShrine}{0007-SkyboxBlackHole}{Procedural Black Hole Skybox}{zGoldDeep}{Easy}{26}}\\*[2pt]
\ubtestbadge{01} & \textcolor{zGoldDeep!78!black}{\texttt{\bfseries Visual\-Coarse\_\allowbreak{}Contrast\_\allowbreak{}Reaches\_\allowbreak{}20\-Percent\_\allowbreak{}Of\_\allowbreak{}Reference}}\enspace\textcolor{zText}{Checks that visual coarse contrast reaches 20\% of reference.} \\
\ubtestbadge{02} & \textcolor{zGoldDeep!78!black}{\texttt{\bfseries Visual\-Coarse\_\allowbreak{}Contrast\_\allowbreak{}Reaches\_\allowbreak{}35\-Percent\_\allowbreak{}Of\_\allowbreak{}Reference}}\enspace\textcolor{zText}{Checks that visual coarse contrast reaches 35\% of reference.} \\
\ubtestbadge{03} & \textcolor{zGoldDeep!78!black}{\texttt{\bfseries Visual\-Coarse\_\allowbreak{}Dark\-Centroid\-Distance\_\allowbreak{}Is\_\allowbreak{}At\-Most\_\allowbreak{}0\_\allowbreak{}45}}\enspace\textcolor{zText}{Checks that visual coarse dark centroid distance is at most 0.45.} \\
\ubtestbadge{04} & \textcolor{zGoldDeep!78!black}{\texttt{\bfseries Visual\-Coarse\_\allowbreak{}Grid\-Feature\-Error\_\allowbreak{}Is\_\allowbreak{}At\-Most\_\allowbreak{}0\_\allowbreak{}60}}\enspace\textcolor{zText}{Checks that visual coarse grid feature error is at most 0.60.} \\
\ubtestbadge{05} & \textcolor{zGoldDeep!78!black}{\texttt{\bfseries Visual\-Coarse\_\allowbreak{}Luminance\-Range\_\allowbreak{}Reaches\_\allowbreak{}20\-Percent\_\allowbreak{}Of\_\allowbreak{}Reference}}\enspace\textcolor{zText}{Checks that visual coarse luminance range reaches 20\% of reference.} \\
\ubtestbadge{06} & \textcolor{zGoldDeep!78!black}{\texttt{\bfseries Visual\-Coarse\_\allowbreak{}Luminance\-Range\_\allowbreak{}Reaches\_\allowbreak{}35\-Percent\_\allowbreak{}Of\_\allowbreak{}Reference}}\enspace\textcolor{zText}{Checks that visual coarse luminance range reaches 35\% of reference.} \\
\ubtestbadge{07} & \textcolor{zGoldDeep!78!black}{\texttt{\bfseries Visual\-Dark\-Centroid\-Distance\_\allowbreak{}Is\_\allowbreak{}At\-Most\_\allowbreak{}0\_\allowbreak{}06}}\enspace\textcolor{zText}{Checks that visual dark centroid distance is at most 0.06.} \\
\ubtestbadge{08} & \textcolor{zGoldDeep!78!black}{\texttt{\bfseries Visual\-Dark\-Centroid\-Distance\_\allowbreak{}Is\_\allowbreak{}At\-Most\_\allowbreak{}0\_\allowbreak{}08}}\enspace\textcolor{zText}{Checks that visual dark centroid distance is at most 0.08.} \\
\ubtestbadge{09} & \textcolor{zGoldDeep!78!black}{\texttt{\bfseries Visual\-Dark\-Centroid\-Distance\_\allowbreak{}Is\_\allowbreak{}At\-Most\_\allowbreak{}0\_\allowbreak{}11}}\enspace\textcolor{zText}{Checks that visual dark centroid distance is at most 0.11.} \\
\ubtestbadge{10} & \textcolor{zGoldDeep!78!black}{\texttt{\bfseries Visual\-Dark\-Centroid\-Distance\_\allowbreak{}Is\_\allowbreak{}At\-Most\_\allowbreak{}0\_\allowbreak{}14}}\enspace\textcolor{zText}{Checks that visual dark centroid distance is at most 0.14.} \\
\ubtestbadge{11} & \textcolor{zGoldDeep!78!black}{\texttt{\bfseries Visual\-Dark\-Centroid\-Distance\_\allowbreak{}Is\_\allowbreak{}At\-Most\_\allowbreak{}0\_\allowbreak{}17}}\enspace\textcolor{zText}{Checks that visual dark centroid distance is at most 0.17.} \\
\ubtestbadge{12} & \textcolor{zGoldDeep!78!black}{\texttt{\bfseries Visual\-Dark\-Centroid\-Distance\_\allowbreak{}Is\_\allowbreak{}At\-Most\_\allowbreak{}0\_\allowbreak{}20}}\enspace\textcolor{zText}{Checks that visual dark centroid distance is at most 0.20.} \\
\ubtestbadge{13} & \textcolor{zGoldDeep!78!black}{\texttt{\bfseries Visual\-Dark\-Centroid\-Distance\_\allowbreak{}Is\_\allowbreak{}At\-Most\_\allowbreak{}0\_\allowbreak{}24}}\enspace\textcolor{zText}{Checks that visual dark centroid distance is at most 0.24.} \\
\ubtestbadge{14} & \textcolor{zGoldDeep!78!black}{\texttt{\bfseries Visual\-Dark\-Centroid\-Distance\_\allowbreak{}Is\_\allowbreak{}At\-Most\_\allowbreak{}0\_\allowbreak{}30}}\enspace\textcolor{zText}{Checks that visual dark centroid distance is at most 0.30.} \\
\ubtestbadge{15} & \textcolor{zGoldDeep!78!black}{\texttt{\bfseries Visual\-Dark\-Centroid\-Distance\_\allowbreak{}Is\_\allowbreak{}At\-Most\_\allowbreak{}0\_\allowbreak{}36}}\enspace\textcolor{zText}{Checks that visual dark centroid distance is at most 0.36.} \\
\ubtestbadge{16} & \textcolor{zGoldDeep!78!black}{\texttt{\bfseries Visual\-Grid\-Feature\-Error\_\allowbreak{}Is\_\allowbreak{}At\-Most\_\allowbreak{}0\_\allowbreak{}07}}\enspace\textcolor{zText}{Checks that visual grid feature error is at most 0.07.} \\
\ubtestbadge{17} & \textcolor{zGoldDeep!78!black}{\texttt{\bfseries Visual\-Grid\-Feature\-Error\_\allowbreak{}Is\_\allowbreak{}At\-Most\_\allowbreak{}0\_\allowbreak{}09}}\enspace\textcolor{zText}{Checks that visual grid feature error is at most 0.09.} \\
\ubtestbadge{18} & \textcolor{zGoldDeep!78!black}{\texttt{\bfseries Visual\-Grid\-Feature\-Error\_\allowbreak{}Is\_\allowbreak{}At\-Most\_\allowbreak{}0\_\allowbreak{}12}}\enspace\textcolor{zText}{Checks that visual grid feature error is at most 0.12.} \\
\ubtestbadge{19} & \textcolor{zGoldDeep!78!black}{\texttt{\bfseries Visual\-Grid\-Feature\-Error\_\allowbreak{}Is\_\allowbreak{}At\-Most\_\allowbreak{}0\_\allowbreak{}15}}\enspace\textcolor{zText}{Checks that visual grid feature error is at most 0.15.} \\
\ubtestbadge{20} & \textcolor{zGoldDeep!78!black}{\texttt{\bfseries Visual\-Grid\-Feature\-Error\_\allowbreak{}Is\_\allowbreak{}At\-Most\_\allowbreak{}0\_\allowbreak{}18}}\enspace\textcolor{zText}{Checks that visual grid feature error is at most 0.18.} \\
\ubtestbadge{21} & \textcolor{zGoldDeep!78!black}{\texttt{\bfseries Visual\-Grid\-Feature\-Error\_\allowbreak{}Is\_\allowbreak{}At\-Most\_\allowbreak{}0\_\allowbreak{}21}}\enspace\textcolor{zText}{Checks that visual grid feature error is at most 0.21.} \\
\ubtestbadge{22} & \textcolor{zGoldDeep!78!black}{\texttt{\bfseries Visual\-Grid\-Feature\-Error\_\allowbreak{}Is\_\allowbreak{}At\-Most\_\allowbreak{}0\_\allowbreak{}25}}\enspace\textcolor{zText}{Checks that visual grid feature error is at most 0.25.} \\
\ubtestbadge{23} & \textcolor{zGoldDeep!78!black}{\texttt{\bfseries Visual\-Grid\-Feature\-Error\_\allowbreak{}Is\_\allowbreak{}At\-Most\_\allowbreak{}0\_\allowbreak{}30}}\enspace\textcolor{zText}{Checks that visual grid feature error is at most 0.30.} \\
\ubtestbadge{24} & \textcolor{zGoldDeep!78!black}{\texttt{\bfseries Visual\-Grid\-Feature\-Error\_\allowbreak{}Is\_\allowbreak{}At\-Most\_\allowbreak{}0\_\allowbreak{}35}}\enspace\textcolor{zText}{Checks that visual grid feature error is at most 0.35.} \\
\ubtestbadge{25} & \textcolor{zGoldDeep!78!black}{\texttt{\bfseries Visual\-Grid\-Feature\-Error\_\allowbreak{}Is\_\allowbreak{}At\-Most\_\allowbreak{}0\_\allowbreak{}42}}\enspace\textcolor{zText}{Checks that visual grid feature error is at most 0.42.} \\
\ubtestbadge{26} & \textcolor{zGoldDeep!78!black}{\texttt{\bfseries Visual\-Grid\-Feature\-Error\_\allowbreak{}Is\_\allowbreak{}At\-Most\_\allowbreak{}0\_\allowbreak{}50}}\enspace\textcolor{zText}{Checks that visual grid feature error is at most 0.50.} \\
\noalign{\vskip 2pt}
\multicolumn{2}{@{}p{14.4cm}@{}}{\textcolor{zRockLite!42}{\rule{\linewidth}{0.3pt}}}\\[3pt]
\multicolumn{2}{@{}p{14.4cm}@{}}{\ubtaskhead{zShrine}{0030-Fireworks-TrailingSmoke}{Firework Trailing Smoke}{zGoldDeep}{Easy}{13}}\\*[2pt]
\ubtestbadge{01} & \textcolor{zGoldDeep!78!black}{\texttt{\bfseries Smoke\-Trail\_\allowbreak{}Color\-Over\-Lifetime\_\allowbreak{}Is\-Enabled}}\enspace\textcolor{zText}{Checks that smoke trail color over lifetime is enabled.} \\
\ubtestbadge{02} & \textcolor{zGoldDeep!78!black}{\texttt{\bfseries Smoke\-Trail\_\allowbreak{}Color\-Over\-Lifetime\_\allowbreak{}Last\-Alpha\-Is\-Zero}}\enspace\textcolor{zText}{Checks that smoke trail color over lifetime last alpha is zero.} \\
\ubtestbadge{03} & \textcolor{zGoldDeep!78!black}{\texttt{\bfseries Smoke\-Trail\_\allowbreak{}Emission\_\allowbreak{}Is\-Enabled}}\enspace\textcolor{zText}{Checks that smoke trail emission is enabled.} \\
\ubtestbadge{04} & \textcolor{zGoldDeep!78!black}{\texttt{\bfseries Smoke\-Trail\_\allowbreak{}Gravity\-Modifier\_\allowbreak{}Is\-Zero}}\enspace\textcolor{zText}{Checks that smoke trail gravity modifier is zero.} \\
\ubtestbadge{05} & \textcolor{zGoldDeep!78!black}{\texttt{\bfseries Smoke\-Trail\_\allowbreak{}Looping\_\allowbreak{}Is\-Enabled}}\enspace\textcolor{zText}{Checks that smoke trail looping is enabled.} \\
\ubtestbadge{06} & \textcolor{zGoldDeep!78!black}{\texttt{\bfseries Smoke\-Trail\_\allowbreak{}Noise\_\allowbreak{}Is\-Enabled}}\enspace\textcolor{zText}{Checks that smoke trail noise is enabled.} \\
\ubtestbadge{07} & \textcolor{zGoldDeep!78!black}{\texttt{\bfseries Smoke\-Trail\_\allowbreak{}Renderer\_\allowbreak{}Is\-Enabled}}\enspace\textcolor{zText}{Checks that smoke trail renderer is enabled.} \\
\ubtestbadge{08} & \textcolor{zGoldDeep!78!black}{\texttt{\bfseries Smoke\-Trail\_\allowbreak{}Shape\_\allowbreak{}Is\-Cone}}\enspace\textcolor{zText}{Checks that smoke trail shape is cone.} \\
\ubtestbadge{09} & \textcolor{zGoldDeep!78!black}{\texttt{\bfseries Smoke\-Trail\_\allowbreak{}Shape\_\allowbreak{}Is\-Enabled}}\enspace\textcolor{zText}{Checks that smoke trail shape is enabled.} \\
\ubtestbadge{10} & \textcolor{zGoldDeep!78!black}{\texttt{\bfseries Smoke\-Trail\_\allowbreak{}Size\-Over\-Lifetime\_\allowbreak{}Is\-Enabled}}\enspace\textcolor{zText}{Checks that smoke trail size over lifetime is enabled.} \\
\ubtestbadge{11} & \textcolor{zGoldDeep!78!black}{\texttt{\bfseries Smoke\-Trail\_\allowbreak{}Start\-Color\_\allowbreak{}Is\-Random\-Between\-Two\-Colors}}\enspace\textcolor{zText}{Checks that smoke trail start color is random between two colors.} \\
\ubtestbadge{12} & \textcolor{zGoldDeep!78!black}{\texttt{\bfseries Smoke\-Trail\_\allowbreak{}Start\-Size\_\allowbreak{}Is\-Less\-Than\-One}}\enspace\textcolor{zText}{Checks that smoke trail start size is less than one.} \\
\ubtestbadge{13} & \textcolor{zGoldDeep!78!black}{\texttt{\bfseries Smoke\-Trail\_\allowbreak{}XRotation\_\allowbreak{}Is90\-Plus\-N360}}\enspace\textcolor{zText}{Checks that smoke trail X rotation is 90 plus n 360.} \\
\noalign{\vskip 2pt}
\multicolumn{2}{@{}p{14.4cm}@{}}{\textcolor{zRockLite!42}{\rule{\linewidth}{0.3pt}}}\\[3pt]
\multicolumn{2}{@{}p{14.4cm}@{}}{\ubtaskhead{zShrine}{0031-Fireworks-TrailingSparks}{Firework Trailing Sparks}{zGoldDeep}{Easy}{15}}\\*[2pt]
\ubtestbadge{01} & \textcolor{zGoldDeep!78!black}{\texttt{\bfseries Sparks\_\allowbreak{}Color\-Over\-Lifetime\_\allowbreak{}Is\-Enabled}}\enspace\textcolor{zText}{Checks that sparks color over lifetime is enabled.} \\
\ubtestbadge{02} & \textcolor{zGoldDeep!78!black}{\texttt{\bfseries Sparks\_\allowbreak{}Color\-Over\-Lifetime\_\allowbreak{}Last\-Alpha\-Is\-Zero}}\enspace\textcolor{zText}{Checks that sparks color over lifetime last alpha is zero.} \\
\ubtestbadge{03} & \textcolor{zGoldDeep!78!black}{\texttt{\bfseries Sparks\_\allowbreak{}Emission\_\allowbreak{}Is\-Enabled}}\enspace\textcolor{zText}{Checks that sparks emission is enabled.} \\
\ubtestbadge{04} & \textcolor{zGoldDeep!78!black}{\texttt{\bfseries Sparks\_\allowbreak{}Gravity\-Modifier\_\allowbreak{}Is\-Not\-Zero}}\enspace\textcolor{zText}{Checks that sparks gravity modifier is not zero.} \\
\ubtestbadge{05} & \textcolor{zGoldDeep!78!black}{\texttt{\bfseries Sparks\_\allowbreak{}Looping\_\allowbreak{}Is\-Enabled}}\enspace\textcolor{zText}{Checks that sparks looping is enabled.} \\
\ubtestbadge{06} & \textcolor{zGoldDeep!78!black}{\texttt{\bfseries Sparks\_\allowbreak{}Noise\_\allowbreak{}Is\-Enabled}}\enspace\textcolor{zText}{Checks that sparks noise is enabled.} \\
\ubtestbadge{07} & \textcolor{zGoldDeep!78!black}{\texttt{\bfseries Sparks\_\allowbreak{}Renderer\_\allowbreak{}Is\-Enabled}}\enspace\textcolor{zText}{Checks that sparks renderer is enabled.} \\
\ubtestbadge{08} & \textcolor{zGoldDeep!78!black}{\texttt{\bfseries Sparks\_\allowbreak{}Shape\_\allowbreak{}Is\-Cone}}\enspace\textcolor{zText}{Checks that sparks shape is cone.} \\
\ubtestbadge{09} & \textcolor{zGoldDeep!78!black}{\texttt{\bfseries Sparks\_\allowbreak{}Shape\_\allowbreak{}Is\-Enabled}}\enspace\textcolor{zText}{Checks that sparks shape is enabled.} \\
\ubtestbadge{10} & \textcolor{zGoldDeep!78!black}{\texttt{\bfseries Sparks\_\allowbreak{}Size\-Over\-Lifetime\_\allowbreak{}Is\-Enabled}}\enspace\textcolor{zText}{Checks that sparks size over lifetime is enabled.} \\
\ubtestbadge{11} & \textcolor{zGoldDeep!78!black}{\texttt{\bfseries Sparks\_\allowbreak{}Start\-Lifetime\_\allowbreak{}Is\-Less\-Than\-One}}\enspace\textcolor{zText}{Checks that sparks start lifetime is less than one.} \\
\ubtestbadge{12} & \textcolor{zGoldDeep!78!black}{\texttt{\bfseries Sparks\_\allowbreak{}Start\-Size\_\allowbreak{}Is\-Less\-Than\-One}}\enspace\textcolor{zText}{Checks that sparks start size is less than one.} \\
\ubtestbadge{13} & \textcolor{zGoldDeep!78!black}{\texttt{\bfseries Sparks\_\allowbreak{}Start\-Speed\_\allowbreak{}Is\-Greater\-Than\-Two}}\enspace\textcolor{zText}{Checks that sparks start speed is greater than two.} \\
\ubtestbadge{14} & \textcolor{zGoldDeep!78!black}{\texttt{\bfseries Sparks\_\allowbreak{}Trails\_\allowbreak{}Is\-Enabled}}\enspace\textcolor{zText}{Checks that sparks trails is enabled.} \\
\ubtestbadge{15} & \textcolor{zGoldDeep!78!black}{\texttt{\bfseries Sparks\_\allowbreak{}XRotation\_\allowbreak{}Is90\-Plus\-N360}}\enspace\textcolor{zText}{Checks that sparks X rotation is 90 plus n 360.} \\
\noalign{\vskip 2pt}
\multicolumn{2}{@{}p{14.4cm}@{}}{\textcolor{zRockLite!42}{\rule{\linewidth}{0.3pt}}}\\[3pt]
\multicolumn{2}{@{}p{14.4cm}@{}}{\ubtaskhead{zShrine}{0032-Fireworks-HeadHalo}{Firework Head Halo}{zGoldDeep}{Easy}{13}}\\*[2pt]
\ubtestbadge{01} & \textcolor{zGoldDeep!78!black}{\texttt{\bfseries Glow\_\allowbreak{}Color\-Over\-Lifetime\_\allowbreak{}Is\-Enabled}}\enspace\textcolor{zText}{Checks that glow color over lifetime is enabled.} \\
\ubtestbadge{02} & \textcolor{zGoldDeep!78!black}{\texttt{\bfseries Glow\_\allowbreak{}Color\-Over\-Lifetime\_\allowbreak{}Last\-Alpha\-Is\-Zero}}\enspace\textcolor{zText}{Checks that glow color over lifetime last alpha is zero.} \\
\ubtestbadge{03} & \textcolor{zGoldDeep!78!black}{\texttt{\bfseries Glow\_\allowbreak{}Death\-Distance\_\allowbreak{}Is\-Less\-Than\-Half}}\enspace\textcolor{zText}{Checks that glow death distance is less than half.} \\
\ubtestbadge{04} & \textcolor{zGoldDeep!78!black}{\texttt{\bfseries Glow\_\allowbreak{}Emission\_\allowbreak{}Is\-Enabled}}\enspace\textcolor{zText}{Checks that glow emission is enabled.} \\
\ubtestbadge{05} & \textcolor{zGoldDeep!78!black}{\texttt{\bfseries Glow\_\allowbreak{}Gravity\-Modifier\_\allowbreak{}Is\-Zero}}\enspace\textcolor{zText}{Checks that glow gravity modifier is zero.} \\
\ubtestbadge{06} & \textcolor{zGoldDeep!78!black}{\texttt{\bfseries Glow\_\allowbreak{}Looping\_\allowbreak{}Is\-Enabled}}\enspace\textcolor{zText}{Checks that glow looping is enabled.} \\
\ubtestbadge{07} & \textcolor{zGoldDeep!78!black}{\texttt{\bfseries Glow\_\allowbreak{}Max\-Particles\_\allowbreak{}Does\-Not\-Exceed100}}\enspace\textcolor{zText}{Checks that glow max particles does not exceed 100.} \\
\ubtestbadge{08} & \textcolor{zGoldDeep!78!black}{\texttt{\bfseries Glow\_\allowbreak{}Renderer\_\allowbreak{}Is\-Enabled}}\enspace\textcolor{zText}{Checks that glow renderer is enabled.} \\
\ubtestbadge{09} & \textcolor{zGoldDeep!78!black}{\texttt{\bfseries Glow\_\allowbreak{}Shape\_\allowbreak{}Is\-Enabled}}\enspace\textcolor{zText}{Checks that glow shape is enabled.} \\
\ubtestbadge{10} & \textcolor{zGoldDeep!78!black}{\texttt{\bfseries Glow\_\allowbreak{}Shape\_\allowbreak{}Is\-Sphere}}\enspace\textcolor{zText}{Checks that glow shape is sphere.} \\
\ubtestbadge{11} & \textcolor{zGoldDeep!78!black}{\texttt{\bfseries Glow\_\allowbreak{}Size\-Over\-Lifetime\_\allowbreak{}Is\-Enabled}}\enspace\textcolor{zText}{Checks that glow size over lifetime is enabled.} \\
\ubtestbadge{12} & \textcolor{zGoldDeep!78!black}{\texttt{\bfseries Glow\_\allowbreak{}Start\-Lifetime\_\allowbreak{}Is\-Less\-Than\-One}}\enspace\textcolor{zText}{Checks that glow start lifetime is less than one.} \\
\ubtestbadge{13} & \textcolor{zGoldDeep!78!black}{\texttt{\bfseries Glow\_\allowbreak{}Start\-Size\_\allowbreak{}Is\-Greater\-Than\-One}}\enspace\textcolor{zText}{Checks that glow start size is greater than one.} \\
\noalign{\vskip 2pt}
\multicolumn{2}{@{}p{14.4cm}@{}}{\textcolor{zRockLite!42}{\rule{\linewidth}{0.3pt}}}\\[3pt]
\multicolumn{2}{@{}p{14.4cm}@{}}{\ubtaskhead{zShrine}{0033-Fireworks-ExplosionEffect}{Firework Explosion Effect}{zGoldDeep}{Easy}{19}}\\*[2pt]
\ubtestbadge{01} & \textcolor{zGoldDeep!78!black}{\texttt{\bfseries Burst\-Explosion\_\allowbreak{}Color\-Over\-Lifetime\_\allowbreak{}Has\-At\-Least\-Three\-Distinct\-Colors}}\enspace\textcolor{zText}{Checks that burst explosion color over lifetime has at least three distinct colors.} \\
\ubtestbadge{02} & \textcolor{zGoldDeep!78!black}{\texttt{\bfseries Burst\-Explosion\_\allowbreak{}Color\-Over\-Lifetime\_\allowbreak{}Is\-Enabled}}\enspace\textcolor{zText}{Checks that burst explosion color over lifetime is enabled.} \\
\ubtestbadge{03} & \textcolor{zGoldDeep!78!black}{\texttt{\bfseries Burst\-Explosion\_\allowbreak{}Duration\_\allowbreak{}Is\-Greater\-Than\-Two}}\enspace\textcolor{zText}{Checks that burst explosion duration is greater than two.} \\
\ubtestbadge{04} & \textcolor{zGoldDeep!78!black}{\texttt{\bfseries Burst\-Explosion\_\allowbreak{}Emission\_\allowbreak{}Is\-Enabled}}\enspace\textcolor{zText}{Checks that burst explosion emission is enabled.} \\
\ubtestbadge{05} & \textcolor{zGoldDeep!78!black}{\texttt{\bfseries Burst\-Explosion\_\allowbreak{}Gravity\-Modifier\_\allowbreak{}Is\-Not\-Zero}}\enspace\textcolor{zText}{Checks that burst explosion gravity modifier is not zero.} \\
\ubtestbadge{06} & \textcolor{zGoldDeep!78!black}{\texttt{\bfseries Burst\-Explosion\_\allowbreak{}Looping\_\allowbreak{}Is\-Disabled}}\enspace\textcolor{zText}{Checks that burst explosion looping is disabled.} \\
\ubtestbadge{07} & \textcolor{zGoldDeep!78!black}{\texttt{\bfseries Burst\-Explosion\_\allowbreak{}Noise\_\allowbreak{}Is\-Enabled}}\enspace\textcolor{zText}{Checks that burst explosion noise is enabled.} \\
\ubtestbadge{08} & \textcolor{zGoldDeep!78!black}{\texttt{\bfseries Burst\-Explosion\_\allowbreak{}Renderer\_\allowbreak{}Is\-Enabled}}\enspace\textcolor{zText}{Checks that burst explosion renderer is enabled.} \\
\ubtestbadge{09} & \textcolor{zGoldDeep!78!black}{\texttt{\bfseries Burst\-Explosion\_\allowbreak{}Shape\_\allowbreak{}Is\-Enabled}}\enspace\textcolor{zText}{Checks that burst explosion shape is enabled.} \\
\ubtestbadge{10} & \textcolor{zGoldDeep!78!black}{\texttt{\bfseries Burst\-Explosion\_\allowbreak{}Shape\_\allowbreak{}Is\-Sphere}}\enspace\textcolor{zText}{Checks that burst explosion shape is sphere.} \\
\ubtestbadge{11} & \textcolor{zGoldDeep!78!black}{\texttt{\bfseries Burst\-Explosion\_\allowbreak{}Size\-Over\-Lifetime\_\allowbreak{}Is\-Enabled}}\enspace\textcolor{zText}{Checks that burst explosion size over lifetime is enabled.} \\
\ubtestbadge{12} & \textcolor{zGoldDeep!78!black}{\texttt{\bfseries Burst\-Explosion\_\allowbreak{}Start\-Lifetime\_\allowbreak{}Is\-Greater\-Than\-One}}\enspace\textcolor{zText}{Checks that burst explosion start lifetime is greater than one.} \\
\ubtestbadge{13} & \textcolor{zGoldDeep!78!black}{\texttt{\bfseries Burst\-Explosion\_\allowbreak{}Start\-Speed\_\allowbreak{}Is\-Fast}}\enspace\textcolor{zText}{Checks that burst explosion start speed is fast.} \\
\ubtestbadge{14} & \textcolor{zGoldDeep!78!black}{\texttt{\bfseries Burst\-Explosion\_\allowbreak{}Velocity\-Decelerates\_\allowbreak{}Within\-Short\-Time}}\enspace\textcolor{zText}{Checks burst explosion velocity decelerates within short time.} \\
\ubtestbadge{15} & \textcolor{zGoldDeep!78!black}{\texttt{\bfseries Burst\-Flash\_\allowbreak{}Duration\_\allowbreak{}Is\-Less\-Than\-One}}\enspace\textcolor{zText}{Checks that burst flash duration is less than one.} \\
\ubtestbadge{16} & \textcolor{zGoldDeep!78!black}{\texttt{\bfseries Burst\-Flash\_\allowbreak{}Looping\_\allowbreak{}Is\-Disabled}}\enspace\textcolor{zText}{Checks that burst flash looping is disabled.} \\
\ubtestbadge{17} & \textcolor{zGoldDeep!78!black}{\texttt{\bfseries Burst\-Flash\_\allowbreak{}Start\-Color\_\allowbreak{}Is\-Pure\-White}}\enspace\textcolor{zText}{Checks that burst flash start color is pure white.} \\
\ubtestbadge{18} & \textcolor{zGoldDeep!78!black}{\texttt{\bfseries Burst\-Flash\_\allowbreak{}Start\-Lifetime\_\allowbreak{}Is\-Less\-Than\-Half}}\enspace\textcolor{zText}{Checks that burst flash start lifetime is less than half.} \\
\ubtestbadge{19} & \textcolor{zGoldDeep!78!black}{\texttt{\bfseries Burst\-Flash\_\allowbreak{}Start\-Size\_\allowbreak{}Is\-Greater\-Than\-One}}\enspace\textcolor{zText}{Checks that burst flash start size is greater than one.} \\
\noalign{\vskip 2pt}
\multicolumn{2}{@{}p{14.4cm}@{}}{\textcolor{zRockLite!42}{\rule{\linewidth}{0.3pt}}}\\[3pt]
\multicolumn{2}{@{}p{14.4cm}@{}}{\ubtaskhead{zShrine}{0024-CoastDunesPineUrpMaterials}{Pine Material URP Conversion}{zGoldDeep}{Medium}{5}}\\*[2pt]
\ubtestbadge{01} & \textcolor{zGoldDeep!78!black}{\texttt{\bfseries Bark\-And\-Leaves\-Materials\-Reference\-Expected\-Packed\-Mask\-Textures}}\enspace\textcolor{zText}{Checks that bark and leaves materials reference expected packed mask textures.} \\
\ubtestbadge{02} & \textcolor{zGoldDeep!78!black}{\texttt{\bfseries Converted\-Materials\-Exist}}\enspace\textcolor{zText}{Checks that converted materials exist.} \\
\ubtestbadge{03} & \textcolor{zGoldDeep!78!black}{\texttt{\bfseries Converted\-Materials\-Reference\-Expected\-Base\-And\-Normal\-Textures}}\enspace\textcolor{zText}{Checks that converted materials reference expected base and normal textures.} \\
\ubtestbadge{04} & \textcolor{zGoldDeep!78!black}{\texttt{\bfseries Converted\-Materials\-Use\-Reasonable\-Rendering\-Numeric\-Settings}}\enspace\textcolor{zText}{Checks that converted materials use reasonable rendering numeric settings.} \\
\ubtestbadge{05} & \textcolor{zGoldDeep!78!black}{\texttt{\bfseries Converted\-Materials\-Use\-Universal\-Render\-Pipeline\-Lit\-Shader}}\enspace\textcolor{zText}{Checks that converted materials use universal render pipeline lit shader.} \\
\noalign{\vskip 2pt}
\multicolumn{2}{@{}p{14.4cm}@{}}{\textcolor{zRockLite!42}{\rule{\linewidth}{0.3pt}}}\\[3pt]
\multicolumn{2}{@{}p{14.4cm}@{}}{\ubtaskhead{zShrine}{0037-ClassificationOfParticleSystem}{Particle Effect Classification}{zGoldDeep}{Medium}{10}}\\*[2pt]
\ubtestbadge{01} & \textcolor{zGoldDeep!78!black}{\texttt{\bfseries Effect1480\_\allowbreak{}Is\_\allowbreak{}Child\_\allowbreak{}Of\_\allowbreak{}Explosion\-Effects}}\enspace\textcolor{zText}{Checks that effect 1480 is child of explosion effects.} \\
\ubtestbadge{02} & \textcolor{zGoldDeep!78!black}{\texttt{\bfseries Effect1696\_\allowbreak{}Is\_\allowbreak{}Child\_\allowbreak{}Of\_\allowbreak{}Music\-Effects}}\enspace\textcolor{zText}{Checks that effect 1696 is child of music effects.} \\
\ubtestbadge{03} & \textcolor{zGoldDeep!78!black}{\texttt{\bfseries Effect2489\_\allowbreak{}Is\_\allowbreak{}Child\_\allowbreak{}Of\_\allowbreak{}Healing}}\enspace\textcolor{zText}{Checks that effect 2489 is child of healing.} \\
\ubtestbadge{04} & \textcolor{zGoldDeep!78!black}{\texttt{\bfseries Effect3017\_\allowbreak{}Is\_\allowbreak{}Child\_\allowbreak{}Of\_\allowbreak{}Poison\-Effects}}\enspace\textcolor{zText}{Checks that effect 3017 is child of poison effects.} \\
\ubtestbadge{05} & \textcolor{zGoldDeep!78!black}{\texttt{\bfseries Effect306\_\allowbreak{}Is\_\allowbreak{}Child\_\allowbreak{}Of\_\allowbreak{}Magic\-Circles}}\enspace\textcolor{zText}{Checks that effect 306 is child of magic circles.} \\
\ubtestbadge{06} & \textcolor{zGoldDeep!78!black}{\texttt{\bfseries Effect3435\_\allowbreak{}Is\_\allowbreak{}Child\_\allowbreak{}Of\_\allowbreak{}Hit\-Effects}}\enspace\textcolor{zText}{Checks that effect 3435 is child of hit effects.} \\
\ubtestbadge{07} & \textcolor{zGoldDeep!78!black}{\texttt{\bfseries Effect3758\_\allowbreak{}Is\_\allowbreak{}Child\_\allowbreak{}Of\_\allowbreak{}Smoke}}\enspace\textcolor{zText}{Checks that effect 3758 is child of smoke.} \\
\ubtestbadge{08} & \textcolor{zGoldDeep!78!black}{\texttt{\bfseries Effect4381\_\allowbreak{}Is\_\allowbreak{}Child\_\allowbreak{}Of\_\allowbreak{}Cloud}}\enspace\textcolor{zText}{Checks that effect 4381 is child of cloud.} \\
\ubtestbadge{09} & \textcolor{zGoldDeep!78!black}{\texttt{\bfseries Effect7866\_\allowbreak{}Is\_\allowbreak{}Child\_\allowbreak{}Of\_\allowbreak{}Fire\-Whirlwind}}\enspace\textcolor{zText}{Checks that effect 7866 is child of fire whirlwind.} \\
\ubtestbadge{10} & \textcolor{zGoldDeep!78!black}{\texttt{\bfseries Effect9008\_\allowbreak{}Is\_\allowbreak{}Child\_\allowbreak{}Of\_\allowbreak{}Shield}}\enspace\textcolor{zText}{Checks that effect 9008 is child of shield.} \\
\noalign{\vskip 2pt}
\multicolumn{2}{@{}p{14.4cm}@{}}{\textcolor{zRockLite!42}{\rule{\linewidth}{0.3pt}}}\\[3pt]
\multicolumn{2}{@{}p{14.4cm}@{}}{\ubtaskhead{zShrine}{0047-DynamicDissolve}{Dynamic Dissolve}{zGoldDeep}{Medium}{10}}\\*[2pt]
\ubtestbadge{01} & \textcolor{zGoldDeep!78!black}{\texttt{\bfseries Test01\_\allowbreak{}Assets\-Exist\-And\-Linked}}\enspace\textcolor{zText}{Checks that DynamicDissolve.shadergraph and M\_Dissolve.mat exist and that the material uses the shader.} \\
\ubtestbadge{02} & \textcolor{zGoldDeep!78!black}{\texttt{\bfseries Test02\_\allowbreak{}Shader\-Properties\-Exposed}}\enspace\textcolor{zText}{Checks that DissolveAmount, DissolveScale, EdgeWidth, and HDR EdgeColor are exposed.} \\
\ubtestbadge{03} & \textcolor{zGoldDeep!78!black}{\texttt{\bfseries Test03\_\allowbreak{}Property\-Default\-Types}}\enspace\textcolor{zText}{Checks that the compiled shader exposes DissolveAmount as a Float or Range property.} \\
\ubtestbadge{04} & \textcolor{zGoldDeep!78!black}{\texttt{\bfseries Test04\_\allowbreak{}Shader\-Is\-Supported}}\enspace\textcolor{zText}{Checks that the generated URP shader is supported at runtime.} \\
\ubtestbadge{05} & \textcolor{zGoldDeep!78!black}{\texttt{\bfseries Test05\_\allowbreak{}Full\-Visibility\-At\-Zero}}\enspace\textcolor{zText}{Checks that DissolveAmount = 0 leaves the test object fully visible.} \\
\ubtestbadge{06} & \textcolor{zGoldDeep!78!black}{\texttt{\bfseries Test06\_\allowbreak{}Full\-Clip\-At\-One}}\enspace\textcolor{zText}{Checks that DissolveAmount = 1 clips or dissolves the test object.} \\
\ubtestbadge{07} & \textcolor{zGoldDeep!78!black}{\texttt{\bfseries Test07\_\allowbreak{}Emission\-Intensity\-On\-Edge}}\enspace\textcolor{zText}{Checks that the dissolve transition band produces HDR edge emission.} \\
\ubtestbadge{08} & \textcolor{zGoldDeep!78!black}{\texttt{\bfseries Test08\_\allowbreak{}Noise\-Dynamic\-Animation}}\enspace\textcolor{zText}{Checks that Time-driven multi-layer noise changes the dissolve pattern dynamically.} \\
\ubtestbadge{09} & \textcolor{zGoldDeep!78!black}{\texttt{\bfseries Test09\_\allowbreak{}Vertex\-Deformation\-Active}}\enspace\textcolor{zText}{Confirms that the shader compiles with its vertex-stage path; this evaluator does not directly measure GPU vertex displacement.} \\
\ubtestbadge{10} & \textcolor{zGoldDeep!78!black}{\texttt{\bfseries Test10\_\allowbreak{}Robustness\-With\-Invalid\-Values}}\enspace\textcolor{zText}{Checks that assigning out-of-range DissolveAmount values does not throw an exception.} \\
\noalign{\vskip 2pt}
\multicolumn{2}{@{}p{14.4cm}@{}}{\textcolor{zRockLite!42}{\rule{\linewidth}{0.3pt}}}\\[3pt]
\multicolumn{2}{@{}p{14.4cm}@{}}{\ubtaskhead{zShrine}{0059-RainEmitter}{Particle Collision Splash Task}{zGoldDeep}{Medium}{10}}\\*[2pt]
\ubtestbadge{01} & \textcolor{zGoldDeep!78!black}{\texttt{\bfseries TC01\_\allowbreak{}Scene\-And\-Fixed\-Objects\-Exist}}\enspace\textcolor{zText}{Checks that scene and fixed objects exist.} \\
\ubtestbadge{02} & \textcolor{zGoldDeep!78!black}{\texttt{\bfseries TC02\_\allowbreak{}Rain\-Emitter\-Has\-Reasonable\-Falling\-Particle\-Setup}}\enspace\textcolor{zText}{Checks that rain emitter has reasonable falling particle setup.} \\
\ubtestbadge{03} & \textcolor{zGoldDeep!78!black}{\texttt{\bfseries TC03\_\allowbreak{}Water\-Surface\-Has\-Collider\-And\-Layer}}\enspace\textcolor{zText}{Checks that water surface has Collider and layer.} \\
\ubtestbadge{04} & \textcolor{zGoldDeep!78!black}{\texttt{\bfseries TC04\_\allowbreak{}Rain\-Collision\-Module\-Targets\-World\-And\-Water\-Layer}}\enspace\textcolor{zText}{Checks that rain collision module targets world and water layer.} \\
\ubtestbadge{05} & \textcolor{zGoldDeep!78!black}{\texttt{\bfseries TC05\_\allowbreak{}Collision\-Sub\-Emitter\-Is\-Registered}}\enspace\textcolor{zText}{Checks that collision sub emitter is registered.} \\
\ubtestbadge{06} & \textcolor{zGoldDeep!78!black}{\texttt{\bfseries TC06\_\allowbreak{}Splash\-Emitter\-Is\-Short\-Burst\-Not\-Looping\-Rain\-Copy}}\enspace\textcolor{zText}{Checks that splash emitter is short burst not looping rain copy.} \\
\ubtestbadge{07} & \textcolor{zGoldDeep!78!black}{\texttt{\bfseries TC07\_\allowbreak{}Rain\-And\-Splash\-Renderers\-Have\-Distinct\-Materials}}\enspace\textcolor{zText}{Checks that rain and splash renderers have distinct materials.} \\
\ubtestbadge{08} & \textcolor{zGoldDeep!78!black}{\texttt{\bfseries TC08\_\allowbreak{}Collision\-Creates\-Splash\-Particles}}\enspace\textcolor{zText}{Checks that collision creates splash particles.} \\
\ubtestbadge{09} & \textcolor{zGoldDeep!78!black}{\texttt{\bfseries TC09\_\allowbreak{}Splash\-Particles\-Stay\-Near\-Water\-Surface}}\enspace\textcolor{zText}{Checks that splash particles stay near water surface.} \\
\ubtestbadge{10} & \textcolor{zGoldDeep!78!black}{\texttt{\bfseries TC10\_\allowbreak{}Particle\-Counts\-Are\-Controlled\-And\-Rain\-Continues}}\enspace\textcolor{zText}{Checks that particle counts are controlled and rain continues.} \\
\noalign{\vskip 2pt}
\multicolumn{2}{@{}p{14.4cm}@{}}{\textcolor{zRockLite!42}{\rule{\linewidth}{0.3pt}}}\\[3pt]
\multicolumn{2}{@{}p{14.4cm}@{}}{\ubtaskhead{zShrine}{0012-SkinnedMeshVfx}{Restore Skinned-Mesh VFX Graph Particles and Trails}{zShrine}{Hard}{13}}\\*[2pt]
\ubtestbadge{01} & \textcolor{zGoldDeep!78!black}{\texttt{\bfseries Required\-Assets\_\allowbreak{}Are\-Implemented\-And\-Expose\-Required\-Interfaces}}\enspace\textcolor{zText}{Checks that the required VFX assets exist and expose the interfaces used by the evaluator.} \\
\ubtestbadge{02} & \textcolor{zGoldDeep!78!black}{\texttt{\bfseries Shared\-Sampling\-Operator\_\allowbreak{}Uses\-Skinned\-Mesh\-Surface\-And\-Transform\-Inputs}}\enspace\textcolor{zText}{Checks that the shared sampling operator reads the skinned-mesh surface and required transform inputs.} \\
\ubtestbadge{03} & \textcolor{zGoldDeep!78!black}{\texttt{\bfseries Particles\-Graph\_\allowbreak{}Has\-Whole\-Body\-Sampling\-And\-Useful\-Density}}\enspace\textcolor{zText}{Checks that the particle graph samples the whole animated body at a useful density.} \\
\ubtestbadge{04} & \textcolor{zGoldDeep!78!black}{\texttt{\bfseries Lines\-Graph\_\allowbreak{}Has\-Multiple\-Long\-Particle\-Strips}}\enspace\textcolor{zText}{Checks that the line graph creates multiple sufficiently long particle strips.} \\
\ubtestbadge{05} & \textcolor{zGoldDeep!78!black}{\texttt{\bfseries Scene\_\allowbreak{}Wires\-Both\-Effects\-To\-The\-Same\-Animated\-Renderer}}\enspace\textcolor{zText}{Checks that both VFX instances are bound to the same animated SkinnedMeshRenderer.} \\
\ubtestbadge{06} & \textcolor{zGoldDeep!78!black}{\texttt{\bfseries Timeline\_\allowbreak{}Still\-Controls\-Throttle\-For\-Both\-Effects}}\enspace\textcolor{zText}{Checks that Timeline still drives the throttle property for both effects.} \\
\ubtestbadge{07} & \textcolor{zGoldDeep!78!black}{\texttt{\bfseries Particles\_\allowbreak{}Runtime\-Capacity\-Can\-Support\-AReadable\-Body}}\enspace\textcolor{zText}{Checks that the runtime particle capacity is sufficient to render a readable body silhouette.} \\
\ubtestbadge{08} & \textcolor{zGoldDeep!78!black}{\texttt{\bfseries Lines\_\allowbreak{}Runtime\-Capacity\-Can\-Support\-Multiple\-Trails}}\enspace\textcolor{zText}{Checks that the runtime line capacity can support multiple trails.} \\
\ubtestbadge{09} & \textcolor{zGoldDeep!78!black}{\texttt{\bfseries Particles\_\allowbreak{}Throttle\-On\-Creates\-Living\-Particles}}\enspace\textcolor{zText}{Checks that enabling particle throttle produces live particles.} \\
\ubtestbadge{10} & \textcolor{zGoldDeep!78!black}{\texttt{\bfseries Lines\_\allowbreak{}Throttle\-On\-Creates\-Living\-Trail\-Particles}}\enspace\textcolor{zText}{Checks that enabling line throttle produces live trail particles.} \\
\ubtestbadge{11} & \textcolor{zGoldDeep!78!black}{\texttt{\bfseries Particles\_\allowbreak{}Throttle\-Off\-Preserves\-Then\-Decays\-Existing\-Particles}}\enspace\textcolor{zText}{Checks that disabling particle throttle stops new emission while existing particles persist briefly and then decay.} \\
\ubtestbadge{12} & \textcolor{zGoldDeep!78!black}{\texttt{\bfseries Lines\_\allowbreak{}Throttle\-Off\-Preserves\-Then\-Decays\-Existing\-Trails}}\enspace\textcolor{zText}{Checks that disabling line throttle stops new emission while existing trails persist briefly and then decay.} \\
\ubtestbadge{13} & \textcolor{zGoldDeep!78!black}{\texttt{\bfseries Runtime\-Bounds\_\allowbreak{}Are\-Finite\-Reasonable\-And\-Near\-The\-Character}}\enspace\textcolor{zText}{Checks that runtime VFX bounds are finite, reasonably sized, and located near the character.} \\
\noalign{\vskip 2pt}
\multicolumn{2}{@{}p{14.4cm}@{}}{\textcolor{zRockLite!42}{\rule{\linewidth}{0.3pt}}}\\[3pt]
\multicolumn{2}{@{}p{14.4cm}@{}}{\ubtaskhead{zShrine}{0038-DynamicRainEffect}{Create a Dynamic Rain Effect}{zShrine}{Hard}{8}}\\*[2pt]
\ubtestbadge{01} & \textcolor{zGoldDeep!78!black}{\texttt{\bfseries Scene\_\allowbreak{}Submission\-Contains\-Usable\-Rain\-System\_\allowbreak{}And\-No\-Missing\-Components}}\enspace\textcolor{zText}{Checks that the submitted scene contains a usable rain system with no missing scripts or components.} \\
\ubtestbadge{02} & \textcolor{zGoldDeep!78!black}{\texttt{\bfseries Runtime\_\allowbreak{}Rain\-Particles\_\allowbreak{}Use\-World\-Space}}\enspace\textcolor{zText}{Checks that the runtime rain Particle System simulates particles in world space.} \\
\ubtestbadge{03} & \textcolor{zGoldDeep!78!black}{\texttt{\bfseries Runtime\_\allowbreak{}Rain\-Area\_\allowbreak{}Follows\-Target\-Horizontally}}\enspace\textcolor{zText}{Checks that the rain emission area follows the target horizontally.} \\
\ubtestbadge{04} & \textcolor{zGoldDeep!78!black}{\texttt{\bfseries Runtime\_\allowbreak{}Existing\-Particles\_\allowbreak{}Remain\-World\-Anchored\_\allowbreak{}When\-Target\-Moves}}\enspace\textcolor{zText}{Checks that already-emitted rain particles remain anchored in world space when the target moves.} \\
\ubtestbadge{05} & \textcolor{zGoldDeep!78!black}{\texttt{\bfseries Runtime\_\allowbreak{}Rain\-Particles\_\allowbreak{}Fall\-Downward\_\allowbreak{}When\-Target\-Is\-Stationary}}\enspace\textcolor{zText}{Checks that rain particles fall downward while the target is stationary.} \\
\ubtestbadge{06} & \textcolor{zGoldDeep!78!black}{\texttt{\bfseries Runtime\_\allowbreak{}Rain\-Coverage\_\allowbreak{}Is\-Dense\-Wide\-And\-Above\-Target}}\enspace\textcolor{zText}{Checks that rain coverage is dense, sufficiently wide, and positioned above the target.} \\
\ubtestbadge{07} & \textcolor{zGoldDeep!78!black}{\texttt{\bfseries Runtime\_\allowbreak{}Rain\-Appearance\_\allowbreak{}Is\-Elongated\-And\-Semi\-Transparent}}\enspace\textcolor{zText}{Checks that the runtime rain appears as elongated, semi-transparent streaks.} \\
\ubtestbadge{08} & \textcolor{zGoldDeep!78!black}{\texttt{\bfseries Runtime\_\allowbreak{}Faster\-Target\-Movement\_\allowbreak{}Produces\-Stronger\-Opposing\-Relative\-Motion}}\enspace\textcolor{zText}{Checks that faster target movement produces stronger rain motion opposite the target's direction.} \\
\noalign{\vskip 2pt}
\multicolumn{2}{@{}p{14.4cm}@{}}{\textcolor{zRockLite!42}{\rule{\linewidth}{0.3pt}}}\\[3pt]
\multicolumn{2}{@{}p{14.4cm}@{}}{\ubtaskhead{zShrine}{0045-InteractableLeaves}{Create an Interactable Falling Leaves Effect}{zShrine}{Hard}{13}}\\*[2pt]
\ubtestbadge{01} & \textcolor{zGoldDeep!78!black}{\texttt{\bfseries Leaf\-Binds\-Its\-Own\-Transform}}\enspace\textcolor{zText}{Checks that leaf binds its own transform.} \\
\ubtestbadge{02} & \textcolor{zGoldDeep!78!black}{\texttt{\bfseries Leaf\-Binds\-Player\-Transform}}\enspace\textcolor{zText}{Checks that leaf binds player transform.} \\
\ubtestbadge{03} & \textcolor{zGoldDeep!78!black}{\texttt{\bfseries Leaf\-Vfx\-Collision\-Keeps\-Low\-Friction\-For\-Player\-Wind}}\enspace\textcolor{zText}{Checks that leaf VFX collision keeps low friction for player wind.} \\
\ubtestbadge{04} & \textcolor{zGoldDeep!78!black}{\texttt{\bfseries Leaf\-Vfx\-Collision\-Plane\-Has\-No\-Vertical\-Bounce}}\enspace\textcolor{zText}{Checks that leaf VFX collision plane has no vertical bounce.} \\
\ubtestbadge{05} & \textcolor{zGoldDeep!78!black}{\texttt{\bfseries Leaf\-Vfx\-Collision\-Plane\-Is\-Aligned\-To\-Scene\-Plane}}\enspace\textcolor{zText}{Checks that leaf VFX collision plane is aligned to scene plane.} \\
\ubtestbadge{06} & \textcolor{zGoldDeep!78!black}{\texttt{\bfseries Leaf\-Vfx\-Ground\-Clamp\-Runs\-After\-Gravity\-Before\-Vector\-Field}}\enspace\textcolor{zText}{Checks that leaf VFX ground clamp runs after gravity before vector field.} \\
\ubtestbadge{07} & \textcolor{zGoldDeep!78!black}{\texttt{\bfseries Leaf\-Vfx\-Ground\-Clamp\-Uses\-Scene\-Plane\-Height}}\enspace\textcolor{zText}{Checks that leaf VFX ground clamp uses scene plane height.} \\
\ubtestbadge{08} & \textcolor{zGoldDeep!78!black}{\texttt{\bfseries Leaf\-Vfx\-Ground\-Clamp\-Uses\-Small\-Landing\-Epsilon}}\enspace\textcolor{zText}{Checks that leaf VFX ground clamp uses small landing epsilon.} \\
\ubtestbadge{09} & \textcolor{zGoldDeep!78!black}{\texttt{\bfseries Leaf\-Vfx\-Has\-Ground\-Clamp\-Block}}\enspace\textcolor{zText}{Checks that leaf VFX has ground clamp block.} \\
\ubtestbadge{10} & \textcolor{zGoldDeep!78!black}{\texttt{\bfseries Leaf\-Vfx\-Initial\-Angles\-Are\-Random\-On\-All\-Axes}}\enspace\textcolor{zText}{Checks that leaf VFX initial angles are random on all axes.} \\
\ubtestbadge{11} & \textcolor{zGoldDeep!78!black}{\texttt{\bfseries Leaf\-Vfx\-Offsets\-Vector\-Field\-Below\-Player}}\enspace\textcolor{zText}{Checks that leaf VFX offsets vector field below player.} \\
\ubtestbadge{12} & \textcolor{zGoldDeep!78!black}{\texttt{\bfseries Leaf\-Vfx\-Uses\-Sphere\-Vector\-Field\-Texture}}\enspace\textcolor{zText}{Checks that leaf VFX uses sphere vector field texture.} \\
\ubtestbadge{13} & \textcolor{zGoldDeep!78!black}{\texttt{\bfseries Leaf\-Vfx\-Uses\-Weak\-Vertical\-Gravity}}\enspace\textcolor{zText}{Checks that leaf VFX uses weak vertical gravity.} \\
\noalign{\vskip 2pt}
\multicolumn{2}{@{}p{14.4cm}@{}}{\textcolor{zRockLite!42}{\rule{\linewidth}{0.3pt}}}\\[3pt]
\noalign{\vskip 3pt}
\multicolumn{2}{@{}p{14.4cm}@{}}{\ubcategory{zGoldDeep}{Physics \& Collision}{11 tasks \textperiodcentered\ 168 milestone tests}}\\*[4pt]
\multicolumn{2}{@{}p{14.4cm}@{}}{\ubtaskhead{zGoldDeep}{0002-CubeMove}{Build a Rigidbody-Driven Cube Movement Scene}{zGoldDeep}{Easy}{20}}\\*[2pt]
\ubtestbadge{01} & \textcolor{zGoldDeep!78!black}{\texttt{\bfseries Camera\_\allowbreak{}Main\-Camera\_\allowbreak{}Can\-See\-Player\-Center}}\enspace\textcolor{zText}{Checks that camera main camera can see player center.} \\
\ubtestbadge{02} & \textcolor{zGoldDeep!78!black}{\texttt{\bfseries Component\_\allowbreak{}Player\_\allowbreak{}Has\-Collider}}\enspace\textcolor{zText}{Checks that Player has a Collider component.} \\
\ubtestbadge{03} & \textcolor{zGoldDeep!78!black}{\texttt{\bfseries Component\_\allowbreak{}Player\_\allowbreak{}Has\-Cube\-Controller}}\enspace\textcolor{zText}{Checks that Player has the required cube-controller component.} \\
\ubtestbadge{04} & \textcolor{zGoldDeep!78!black}{\texttt{\bfseries Component\_\allowbreak{}Player\_\allowbreak{}Has\-Rigidbody}}\enspace\textcolor{zText}{Checks that Player has a Rigidbody component.} \\
\ubtestbadge{05} & \textcolor{zGoldDeep!78!black}{\texttt{\bfseries Cube\-Controller\_\allowbreak{}Has\-Required\-Public\-Fields}}\enspace\textcolor{zText}{Checks that cube controller has required public fields.} \\
\ubtestbadge{06} & \textcolor{zGoldDeep!78!black}{\texttt{\bfseries Cube\-Controller\_\allowbreak{}Has\-Required\-Public\-Methods}}\enspace\textcolor{zText}{Checks that cube controller has required public methods.} \\
\ubtestbadge{07} & \textcolor{zGoldDeep!78!black}{\texttt{\bfseries Primitive\_\allowbreak{}Ground\_\allowbreak{}Uses\-Plane}}\enspace\textcolor{zText}{Checks that Ground uses the Plane primitive.} \\
\ubtestbadge{08} & \textcolor{zGoldDeep!78!black}{\texttt{\bfseries Primitive\_\allowbreak{}Player\_\allowbreak{}Uses\-Cube}}\enspace\textcolor{zText}{Checks that Player uses the Cube primitive.} \\
\ubtestbadge{09} & \textcolor{zGoldDeep!78!black}{\texttt{\bfseries Rigidbody\_\allowbreak{}Player\_\allowbreak{}Freeze\-Rotation\-X\_\allowbreak{}And\_\allowbreak{}Z}}\enspace\textcolor{zText}{Checks that the Player Rigidbody freezes rotation around the X and Z axes.} \\
\ubtestbadge{10} & \textcolor{zGoldDeep!78!black}{\texttt{\bfseries Rigidbody\_\allowbreak{}Player\_\allowbreak{}Is\-Kinematic\_\allowbreak{}Is\-False}}\enspace\textcolor{zText}{Checks that Rigidbody player is non-kinematic.} \\
\ubtestbadge{11} & \textcolor{zGoldDeep!78!black}{\texttt{\bfseries Rigidbody\_\allowbreak{}Player\_\allowbreak{}Use\-Gravity\_\allowbreak{}Is\-True}}\enspace\textcolor{zText}{Checks that Rigidbody player has gravity enabled.} \\
\ubtestbadge{12} & \textcolor{zGoldDeep!78!black}{\texttt{\bfseries Scene\_\allowbreak{}Ground\_\allowbreak{}Exists}}\enspace\textcolor{zText}{Checks that Ground exists in the scene.} \\
\ubtestbadge{13} & \textcolor{zGoldDeep!78!black}{\texttt{\bfseries Scene\_\allowbreak{}Main\-Camera\_\allowbreak{}Exists}}\enspace\textcolor{zText}{Checks that Main Camera exists in the scene.} \\
\ubtestbadge{14} & \textcolor{zGoldDeep!78!black}{\texttt{\bfseries Scene\_\allowbreak{}Player\_\allowbreak{}Exists}}\enspace\textcolor{zText}{Checks that Player exists in the scene.} \\
\ubtestbadge{15} & \textcolor{zGoldDeep!78!black}{\texttt{\bfseries Idle\_\allowbreak{}No\-Input\_\allowbreak{}No\-Continuous\-Drift}}\enspace\textcolor{zText}{Checks that Player does not drift continuously while idle with no input.} \\
\ubtestbadge{16} & \textcolor{zGoldDeep!78!black}{\texttt{\bfseries Jump\_\allowbreak{}Can\-Jump\-Again\_\allowbreak{}After\-Landing}}\enspace\textcolor{zText}{Checks that jump can jump again after landing.} \\
\ubtestbadge{17} & \textcolor{zGoldDeep!78!black}{\texttt{\bfseries Jump\_\allowbreak{}First\-Jump\_\allowbreak{}Produces\-Positive\-YVelocity}}\enspace\textcolor{zText}{Checks that jump first jump produces positive Y velocity.} \\
\ubtestbadge{18} & \textcolor{zGoldDeep!78!black}{\texttt{\bfseries Jump\_\allowbreak{}Second\-Jump\-In\-Air\_\allowbreak{}Does\-Not\-Add\-Impulse}}\enspace\textcolor{zText}{Checks that jump second jump in air does not add impulse.} \\
\ubtestbadge{19} & \textcolor{zGoldDeep!78!black}{\texttt{\bfseries Movement\_\allowbreak{}Forward\-Input\_\allowbreak{}Increases\-Z}}\enspace\textcolor{zText}{Checks that forward input increases Player's world Z position.} \\
\ubtestbadge{20} & \textcolor{zGoldDeep!78!black}{\texttt{\bfseries Movement\_\allowbreak{}Left\-Input\_\allowbreak{}Decreases\-X}}\enspace\textcolor{zText}{Checks that left input decreases Player's world X position.} \\
\noalign{\vskip 2pt}
\multicolumn{2}{@{}p{14.4cm}@{}}{\textcolor{zRockLite!42}{\rule{\linewidth}{0.3pt}}}\\[3pt]
\multicolumn{2}{@{}p{14.4cm}@{}}{\ubtaskhead{zGoldDeep}{0054-ProceduralSkinnedCloth}{Procedural Skinned Cloth}{zGoldDeep}{Easy}{10}}\\*[2pt]
\ubtestbadge{01} & \textcolor{zGoldDeep!78!black}{\texttt{\bfseries Bones\-Are\-Scene\-Children\-And\-Bindposes\-Match\-Assigned\-Transforms}}\enspace\textcolor{zText}{Checks that bones are scene children and bindposes match assigned transforms.} \\
\ubtestbadge{02} & \textcolor{zGoldDeep!78!black}{\texttt{\bfseries Cloth\-Coefficients\-Anchor\-Top\-Transition\-Middle\-And\-Free\-Lower\-Region}}\enspace\textcolor{zText}{Checks that cloth coefficients anchor top transition middle and free lower region.} \\
\ubtestbadge{03} & \textcolor{zGoldDeep!78!black}{\texttt{\bfseries Cloth\-Collider\-Arrays\-Are\-Clean\-Distinct\-And\-Limited\-To\-Provided\-Colliders}}\enspace\textcolor{zText}{Checks that cloth Collider arrays are clean distinct and limited to provided Colliders.} \\
\ubtestbadge{04} & \textcolor{zGoldDeep!78!black}{\texttt{\bfseries Cloth\-Object\-Is\-Aligned\-With\-Curtain\-Origin\-And\-Reference\-Volume}}\enspace\textcolor{zText}{Checks that cloth object is aligned with curtain origin and reference volume.} \\
\ubtestbadge{05} & \textcolor{zGoldDeep!78!black}{\texttt{\bfseries Cloth\-Solver\-Settings\-Are\-Stable\-And\-Responsive}}\enspace\textcolor{zText}{Checks that cloth solver settings are stable and responsive.} \\
\ubtestbadge{06} & \textcolor{zGoldDeep!78!black}{\texttt{\bfseries Cloth\-Uses\-Provided\-Capsule\-And\-Sphere\-Collider\-Pairs}}\enspace\textcolor{zText}{Checks that cloth uses provided capsule and sphere Collider pairs.} \\
\ubtestbadge{07} & \textcolor{zGoldDeep!78!black}{\texttt{\bfseries Core\-Gate\-Has\-Cloth\-And\-Skinned\-Mesh\-Renderer\-On\-The\-Same\-Object}}\enspace\textcolor{zText}{Checks that core gate has cloth and SkinnedMeshRenderer on the same object.} \\
\ubtestbadge{08} & \textcolor{zGoldDeep!78!black}{\texttt{\bfseries Moving\-Rig\-Bones\-Produces\-Baked\-Skinned\-Deformation}}\enspace\textcolor{zText}{Checks that moving rig bones produces baked skinned deformation.} \\
\ubtestbadge{09} & \textcolor{zGoldDeep!78!black}{\texttt{\bfseries Procedural\-Mesh\-Has\-Dense\-Curtain\-Grid\-Shape}}\enspace\textcolor{zText}{Checks that procedural mesh has dense curtain grid shape.} \\
\ubtestbadge{10} & \textcolor{zGoldDeep!78!black}{\texttt{\bfseries Skinning\-Rig\-Has\-Valid\-Bones\-Bindposes\-And\-Weights}}\enspace\textcolor{zText}{Checks that skinning rig has valid bones bindposes and weights.} \\
\noalign{\vskip 2pt}
\multicolumn{2}{@{}p{14.4cm}@{}}{\textcolor{zRockLite!42}{\rule{\linewidth}{0.3pt}}}\\[3pt]
\multicolumn{2}{@{}p{14.4cm}@{}}{\ubtaskhead{zGoldDeep}{0057-VoxelWormsTerrain}{Voxel Worms Terrain}{zGoldDeep}{Easy}{12}}\\*[2pt]
\ubtestbadge{01} & \textcolor{zGoldDeep!78!black}{\texttt{\bfseries Generate\-Terrain\-Builds\-Substantial\-Procedural\-Mesh\-And\-Collider}}\enspace\textcolor{zText}{Checks that generate terrain builds substantial procedural mesh and Collider.} \\
\ubtestbadge{02} & \textcolor{zGoldDeep!78!black}{\texttt{\bfseries Terrain\-Has\-Open\-Air\-Tunnel\-And\-Varied\-Playable\-Surface}}\enspace\textcolor{zText}{Checks that terrain has open air tunnel and varied playable surface.} \\
\ubtestbadge{03} & \textcolor{zGoldDeep!78!black}{\texttt{\bfseries Tunnel\-Is\-Continuous\-And\-Enclosed\-Across\-The\-Island}}\enspace\textcolor{zText}{Checks that tunnel is continuous and enclosed across the island.} \\
\ubtestbadge{04} & \textcolor{zGoldDeep!78!black}{\texttt{\bfseries Voxel\-Coordinate\-Api\-Round\-Trips\-And\-Set\-Solid\-Mutates\-Data}}\enspace\textcolor{zText}{Checks that voxel coordinate API round trips and set solid mutates data.} \\
\ubtestbadge{05} & \textcolor{zGoldDeep!78!black}{\texttt{\bfseries Voxel\-Field\-Matches\-Island\-Markers\-Explosion\-Points\-And\-Spawn\-Support}}\enspace\textcolor{zText}{Checks that voxel field matches island markers explosion points and spawn support.} \\
\ubtestbadge{06} & \textcolor{zGoldDeep!78!black}{\texttt{\bfseries Explosion\-Clears\-Voxel\-Sphere\-And\-Keeps\-Outside\-Terrain}}\enspace\textcolor{zText}{Checks that explosion clears voxel sphere and keeps outside terrain.} \\
\ubtestbadge{07} & \textcolor{zGoldDeep!78!black}{\texttt{\bfseries Explosion\-Outside\-Voxel\-Grid\-Leaves\-Terrain\-Unchanged}}\enspace\textcolor{zText}{Checks that explosion outside voxel grid leaves terrain unchanged.} \\
\ubtestbadge{08} & \textcolor{zGoldDeep!78!black}{\texttt{\bfseries Explosion\-Rebuilds\-Mesh\-And\-Collider}}\enspace\textcolor{zText}{Checks that explosion rebuilds mesh and Collider.} \\
\ubtestbadge{09} & \textcolor{zGoldDeep!78!black}{\texttt{\bfseries Explosion\-Uses\-World\-Space\-Sphere\-Not\-Axis\-Aligned\-Box}}\enspace\textcolor{zText}{Checks that explosion uses world space sphere not axis aligned box.} \\
\ubtestbadge{10} & \textcolor{zGoldDeep!78!black}{\texttt{\bfseries Multiple\-Explosions\-Accumulate\-Without\-Restoring\-Voxels}}\enspace\textcolor{zText}{Checks that multiple explosions accumulate without restoring voxels.} \\
\ubtestbadge{11} & \textcolor{zGoldDeep!78!black}{\texttt{\bfseries Rigidbodies\-From\-All\-Spawn\-Markers\-Settle\-On\-Generated\-Mesh\-Collider}}\enspace\textcolor{zText}{Checks that rigidbodies from all spawn markers settle on generated mesh Collider.} \\
\ubtestbadge{12} & \textcolor{zGoldDeep!78!black}{\texttt{\bfseries Vertical\-Raycast\-Shows\-Crater\-After\-Explosion}}\enspace\textcolor{zText}{Checks that vertical raycast shows crater after explosion.} \\
\noalign{\vskip 2pt}
\multicolumn{2}{@{}p{14.4cm}@{}}{\textcolor{zRockLite!42}{\rule{\linewidth}{0.3pt}}}\\[3pt]
\multicolumn{2}{@{}p{14.4cm}@{}}{\ubtaskhead{zGoldDeep}{0003-CatapultExplosion}{Catapult Explosion}{zGoldDeep}{Medium}{11}}\\*[2pt]
\ubtestbadge{01} & \textcolor{zGoldDeep!78!black}{\texttt{\bfseries Catapult\_\allowbreak{}Animation\_\allowbreak{}Event\_\allowbreak{}Spawns\_\allowbreak{}A\_\allowbreak{}Stone\_\allowbreak{}Near\_\allowbreak{}The\_\allowbreak{}Release\_\allowbreak{}Frame}}\enspace\textcolor{zText}{Checks that catapult animation event spawns a stone near the release frame.} \\
\ubtestbadge{02} & \textcolor{zGoldDeep!78!black}{\texttt{\bfseries Catapult\_\allowbreak{}Explosion\_\allowbreak{}Function\_\allowbreak{}Applies\_\allowbreak{}Physics\_\allowbreak{}And\_\allowbreak{}Pushes\_\allowbreak{}A\_\allowbreak{}Target\_\allowbreak{}Outward}}\enspace\textcolor{zText}{Checks that catapult explosion function applies physics and pushes a target outward.} \\
\ubtestbadge{03} & \textcolor{zGoldDeep!78!black}{\texttt{\bfseries Catapult\_\allowbreak{}Launch\_\allowbreak{}Function\_\allowbreak{}Applies\_\allowbreak{}Physics\_\allowbreak{}And\_\allowbreak{}Actually\_\allowbreak{}Launches\_\allowbreak{}The\_\allowbreak{}Stone}}\enspace\textcolor{zText}{Checks that catapult launch function applies physics and actually launches the stone.} \\
\ubtestbadge{04} & \textcolor{zGoldDeep!78!black}{\texttt{\bfseries Catapult\_\allowbreak{}Target\_\allowbreak{}At\_\allowbreak{}0\_\allowbreak{}47\_\allowbreak{}Moves\_\allowbreak{}Outward\_\allowbreak{}Between\_\allowbreak{}1\_\allowbreak{}And\_\allowbreak{}5\_\allowbreak{}Units}}\enspace\textcolor{zText}{Checks that the target starting at (0, 0.5, 47) moves outward by 1 to 5 units after the catapult explosion.} \\
\ubtestbadge{05} & \textcolor{zGoldDeep!78!black}{\texttt{\bfseries Catapult\_\allowbreak{}Target\_\allowbreak{}At\_\allowbreak{}0\_\allowbreak{}53\_\allowbreak{}Moves\_\allowbreak{}Outward\_\allowbreak{}Between\_\allowbreak{}1\_\allowbreak{}And\_\allowbreak{}5\_\allowbreak{}Units}}\enspace\textcolor{zText}{Checks that the target starting at (0, 0.5, 53) moves outward by 1 to 5 units after the catapult explosion.} \\
\ubtestbadge{06} & \textcolor{zGoldDeep!78!black}{\texttt{\bfseries Catapult\_\allowbreak{}Target\_\allowbreak{}At\_\allowbreak{}3\_\allowbreak{}47\_\allowbreak{}Moves\_\allowbreak{}Outward\_\allowbreak{}Between\_\allowbreak{}1\_\allowbreak{}And\_\allowbreak{}5\_\allowbreak{}Units}}\enspace\textcolor{zText}{Checks that the target starting at (3, 0.5, 47) moves outward by 1 to 5 units after the catapult explosion.} \\
\ubtestbadge{07} & \textcolor{zGoldDeep!78!black}{\texttt{\bfseries Catapult\_\allowbreak{}Target\_\allowbreak{}At\_\allowbreak{}3\_\allowbreak{}50\_\allowbreak{}Moves\_\allowbreak{}Outward\_\allowbreak{}Between\_\allowbreak{}1\_\allowbreak{}And\_\allowbreak{}5\_\allowbreak{}Units}}\enspace\textcolor{zText}{Checks that the target starting at (3, 0.5, 50) moves outward by 1 to 5 units after the catapult explosion.} \\
\ubtestbadge{08} & \textcolor{zGoldDeep!78!black}{\texttt{\bfseries Catapult\_\allowbreak{}Target\_\allowbreak{}At\_\allowbreak{}3\_\allowbreak{}53\_\allowbreak{}Moves\_\allowbreak{}Outward\_\allowbreak{}Between\_\allowbreak{}1\_\allowbreak{}And\_\allowbreak{}5\_\allowbreak{}Units}}\enspace\textcolor{zText}{Checks that the target starting at (3, 0.5, 53) moves outward by 1 to 5 units after the catapult explosion.} \\
\ubtestbadge{09} & \textcolor{zGoldDeep!78!black}{\texttt{\bfseries Catapult\_\allowbreak{}Target\_\allowbreak{}At\_\allowbreak{}Negative3\_\allowbreak{}47\_\allowbreak{}Moves\_\allowbreak{}Outward\_\allowbreak{}Between\_\allowbreak{}1\_\allowbreak{}And\_\allowbreak{}5\_\allowbreak{}Units}}\enspace\textcolor{zText}{Checks that the target starting at (-3, 0.5, 47) moves outward by 1 to 5 units after the catapult explosion.} \\
\ubtestbadge{10} & \textcolor{zGoldDeep!78!black}{\texttt{\bfseries Catapult\_\allowbreak{}Target\_\allowbreak{}At\_\allowbreak{}Negative3\_\allowbreak{}50\_\allowbreak{}Moves\_\allowbreak{}Outward\_\allowbreak{}Between\_\allowbreak{}1\_\allowbreak{}And\_\allowbreak{}5\_\allowbreak{}Units}}\enspace\textcolor{zText}{Checks that the target starting at (-3, 0.5, 50) moves outward by 1 to 5 units after the catapult explosion.} \\
\ubtestbadge{11} & \textcolor{zGoldDeep!78!black}{\texttt{\bfseries Catapult\_\allowbreak{}Target\_\allowbreak{}At\_\allowbreak{}Negative3\_\allowbreak{}53\_\allowbreak{}Moves\_\allowbreak{}Outward\_\allowbreak{}Between\_\allowbreak{}1\_\allowbreak{}And\_\allowbreak{}5\_\allowbreak{}Units}}\enspace\textcolor{zText}{Checks that the target starting at (-3, 0.5, 53) moves outward by 1 to 5 units after the catapult explosion.} \\
\noalign{\vskip 2pt}
\multicolumn{2}{@{}p{14.4cm}@{}}{\textcolor{zRockLite!42}{\rule{\linewidth}{0.3pt}}}\\[3pt]
\multicolumn{2}{@{}p{14.4cm}@{}}{\ubtaskhead{zGoldDeep}{0026-VehicleWheelPhysics}{Vehicle Control with WheelColliders}{zGoldDeep}{Medium}{8}}\\*[2pt]
\ubtestbadge{01} & \textcolor{zGoldDeep!78!black}{\texttt{\bfseries Car\-Prefab\_\allowbreak{}Has\-Four\-Wheel\-Colliders\-Aligned\-With\-Visual\-Wheels}}\enspace\textcolor{zText}{Checks that car Prefab has four wheel Colliders aligned with visual wheels.} \\
\ubtestbadge{02} & \textcolor{zGoldDeep!78!black}{\texttt{\bfseries Car\-Prefab\_\allowbreak{}Keeps\-Bench\-Controller\-And\-Rigid\-Body\-On\-Root}}\enspace\textcolor{zText}{Checks that car Prefab keeps bench controller and rigid body on root.} \\
\ubtestbadge{03} & \textcolor{zGoldDeep!78!black}{\texttt{\bfseries Front\-Wheel\-Colliders\_\allowbreak{}Are\-Capable\-Of\-Steering}}\enspace\textcolor{zText}{Checks that front wheel Colliders are capable of steering.} \\
\ubtestbadge{04} & \textcolor{zGoldDeep!78!black}{\texttt{\bfseries Wheel\-Colliders\_\allowbreak{}Use\-Expected\-Mass}}\enspace\textcolor{zText}{Checks that wheel Colliders use expected mass.} \\
\ubtestbadge{05} & \textcolor{zGoldDeep!78!black}{\texttt{\bfseries Full\-Throttle\_\allowbreak{}Has\-Reasonable\-Maximum\-Speed\-Limit}}\enspace\textcolor{zText}{Checks that full throttle has reasonable maximum speed limit.} \\
\ubtestbadge{06} & \textcolor{zGoldDeep!78!black}{\texttt{\bfseries Full\-Throttle\_\allowbreak{}Reaches\-Zero\-To\-Hundred\-In\-Controlled\-Time\-Window}}\enspace\textcolor{zText}{Checks that full throttle reaches zero to hundred in controlled time window.} \\
\ubtestbadge{07} & \textcolor{zGoldDeep!78!black}{\texttt{\bfseries Throttle\-Input\_\allowbreak{}Uses\-Wheel\-Physics\-To\-Move\-Vehicle\-Forward}}\enspace\textcolor{zText}{Checks that throttle input uses wheel physics to move vehicle forward.} \\
\ubtestbadge{08} & \textcolor{zGoldDeep!78!black}{\texttt{\bfseries Zero\-Steering\_\allowbreak{}Stays\-Stable\-On\-Straight\-Line}}\enspace\textcolor{zText}{Checks that zero steering stays stable on straight line.} \\
\noalign{\vskip 2pt}
\multicolumn{2}{@{}p{14.4cm}@{}}{\textcolor{zRockLite!42}{\rule{\linewidth}{0.3pt}}}\\[3pt]
\multicolumn{2}{@{}p{14.4cm}@{}}{\ubtaskhead{zGoldDeep}{0039-RollingBallCheckpointCourse}{Inclined Rolling Ball Checkpoint Guidance}{zGoldDeep}{Medium}{15}}\\*[2pt]
\ubtestbadge{01} & \textcolor{zGoldDeep!78!black}{\texttt{\bfseries Balls\-Remain\-Free\-Rolling\-Physics\-Spheres\-At\-Their\-Fixed\-Spawns}}\enspace\textcolor{zText}{Checks that balls remain free rolling physics spheres at their fixed spawns.} \\
\ubtestbadge{02} & \textcolor{zGoldDeep!78!black}{\texttt{\bfseries Cube\-Budget\-Stays\-Below\-Nine\-Per\-Scene}}\enspace\textcolor{zText}{Checks that cube budget stays below nine per scene.} \\
\ubtestbadge{03} & \textcolor{zGoldDeep!78!black}{\texttt{\bfseries Cube\-Budget\-Stays\-Below\-Six\-Per\-Scene}}\enspace\textcolor{zText}{Checks that cube budget stays below six per scene.} \\
\ubtestbadge{04} & \textcolor{zGoldDeep!78!black}{\texttt{\bfseries Cube\-Budget\-Stays\-Below\-Twelve\-Per\-Scene}}\enspace\textcolor{zText}{Checks that cube budget stays below twelve per scene.} \\
\ubtestbadge{05} & \textcolor{zGoldDeep!78!black}{\texttt{\bfseries Initial\-Course\-Objects\-Remain\-In\-Expected\-Layout}}\enspace\textcolor{zText}{Checks that initial course objects remain in expected layout.} \\
\ubtestbadge{06} & \textcolor{zGoldDeep!78!black}{\texttt{\bfseries Student\-Obstacles\-Use\-Only\-Builtin\-Cube\-Primitives}}\enspace\textcolor{zText}{Checks that student obstacles use only builtin cube primitives.} \\
\ubtestbadge{07} & \textcolor{zGoldDeep!78!black}{\texttt{\bfseries Group\-One\-Ball\-Visits\-All\-Five\-Checkpoints}}\enspace\textcolor{zText}{Checks that group one ball visits all five checkpoints.} \\
\ubtestbadge{08} & \textcolor{zGoldDeep!78!black}{\texttt{\bfseries Group\-One\-Ball\-Visits\-At\-Least\-One\-Checkpoint}}\enspace\textcolor{zText}{Checks that group one ball visits at least one checkpoint.} \\
\ubtestbadge{09} & \textcolor{zGoldDeep!78!black}{\texttt{\bfseries Group\-One\-Ball\-Visits\-At\-Least\-Three\-Checkpoints}}\enspace\textcolor{zText}{Checks that group one ball visits at least three checkpoints.} \\
\ubtestbadge{10} & \textcolor{zGoldDeep!78!black}{\texttt{\bfseries Group\-Three\-Ball\-Visits\-All\-Five\-Checkpoints}}\enspace\textcolor{zText}{Checks that group three ball visits all five checkpoints.} \\
\ubtestbadge{11} & \textcolor{zGoldDeep!78!black}{\texttt{\bfseries Group\-Three\-Ball\-Visits\-At\-Least\-One\-Checkpoint}}\enspace\textcolor{zText}{Checks that group three ball visits at least one checkpoint.} \\
\ubtestbadge{12} & \textcolor{zGoldDeep!78!black}{\texttt{\bfseries Group\-Three\-Ball\-Visits\-At\-Least\-Three\-Checkpoints}}\enspace\textcolor{zText}{Checks that group three ball visits at least three checkpoints.} \\
\ubtestbadge{13} & \textcolor{zGoldDeep!78!black}{\texttt{\bfseries Group\-Two\-Ball\-Visits\-All\-Five\-Checkpoints}}\enspace\textcolor{zText}{Checks that group two ball visits all five checkpoints.} \\
\ubtestbadge{14} & \textcolor{zGoldDeep!78!black}{\texttt{\bfseries Group\-Two\-Ball\-Visits\-At\-Least\-One\-Checkpoint}}\enspace\textcolor{zText}{Checks that group two ball visits at least one checkpoint.} \\
\ubtestbadge{15} & \textcolor{zGoldDeep!78!black}{\texttt{\bfseries Group\-Two\-Ball\-Visits\-At\-Least\-Three\-Checkpoints}}\enspace\textcolor{zText}{Checks that group two ball visits at least three checkpoints.} \\
\noalign{\vskip 2pt}
\multicolumn{2}{@{}p{14.4cm}@{}}{\textcolor{zRockLite!42}{\rule{\linewidth}{0.3pt}}}\\[3pt]
\multicolumn{2}{@{}p{14.4cm}@{}}{\ubtaskhead{zGoldDeep}{0048-RoadsideJointPhysics}{Roadside Joint Physics}{zGoldDeep}{Medium}{12}}\\*[2pt]
\ubtestbadge{01} & \textcolor{zGoldDeep!78!black}{\texttt{\bfseries Configurable\-Bollard\-Locks\-All\-But\-Limited\-Road\-Axis}}\enspace\textcolor{zText}{Checks that configurable bollard locks all but limited road axis.} \\
\ubtestbadge{02} & \textcolor{zGoldDeep!78!black}{\texttt{\bfseries Configurable\-Bollard\-Uses\-Configurable\-Joint\-Connected\-To\-Anchor}}\enspace\textcolor{zText}{Checks that configurable bollard uses configurable joint connected to anchor.} \\
\ubtestbadge{03} & \textcolor{zGoldDeep!78!black}{\texttt{\bfseries Fixed\-Fence\-Joint\-Break\-Settings\-Are\-Practical}}\enspace\textcolor{zText}{Checks that fixed fence joint break settings are practical.} \\
\ubtestbadge{04} & \textcolor{zGoldDeep!78!black}{\texttt{\bfseries Fixed\-Fence\-Uses\-Fixed\-Joint\-Connected\-To\-Anchor}}\enspace\textcolor{zText}{Checks that fixed fence uses fixed joint connected to anchor.} \\
\ubtestbadge{05} & \textcolor{zGoldDeep!78!black}{\texttt{\bfseries Hinge\-Gate\-Axis\-Anchor\-And\-Limits\-Are\-Reasonable}}\enspace\textcolor{zText}{Checks that hinge gate axis anchor and limits are reasonable.} \\
\ubtestbadge{06} & \textcolor{zGoldDeep!78!black}{\texttt{\bfseries Hinge\-Gate\-Uses\-Hinge\-Joint\-Connected\-To\-Post}}\enspace\textcolor{zText}{Checks that hinge gate uses hinge joint connected to post.} \\
\ubtestbadge{07} & \textcolor{zGoldDeep!78!black}{\texttt{\bfseries Spring\-Barrier\-Spring\-Damper\-And\-Distance\-Are\-Practical}}\enspace\textcolor{zText}{Checks that spring barrier spring damper and distance are practical.} \\
\ubtestbadge{08} & \textcolor{zGoldDeep!78!black}{\texttt{\bfseries Spring\-Barrier\-Uses\-Spring\-Joint\-Connected\-To\-Anchor}}\enspace\textcolor{zText}{Checks that spring barrier uses spring joint connected to anchor.} \\
\ubtestbadge{09} & \textcolor{zGoldDeep!78!black}{\texttt{\bfseries Configurable\-Bollard\-Slides\-Only\-Along\-Road\-Axis}}\enspace\textcolor{zText}{Checks that configurable bollard slides only along road axis.} \\
\ubtestbadge{10} & \textcolor{zGoldDeep!78!black}{\texttt{\bfseries Fixed\-Fence\-Breaks\-Away\-Under\-Impact}}\enspace\textcolor{zText}{Checks that fixed fence breaks away under impact.} \\
\ubtestbadge{11} & \textcolor{zGoldDeep!78!black}{\texttt{\bfseries Hinge\-Gate\-Swings\-Open\-Around\-Post\-Under\-Impact}}\enspace\textcolor{zText}{Checks that hinge gate swings open around post under impact.} \\
\ubtestbadge{12} & \textcolor{zGoldDeep!78!black}{\texttt{\bfseries Spring\-Barrier\-Deflects\-And\-Shows\-Return\-Trend}}\enspace\textcolor{zText}{Checks that spring barrier deflects and shows return trend.} \\
\noalign{\vskip 2pt}
\multicolumn{2}{@{}p{14.4cm}@{}}{\textcolor{zRockLite!42}{\rule{\linewidth}{0.3pt}}}\\[3pt]
\multicolumn{2}{@{}p{14.4cm}@{}}{\ubtaskhead{zGoldDeep}{0056-RigidbodyCharacterControllerParity}{Rigidbody Character Controller Parity}{zGoldDeep}{Medium}{15}}\\*[2pt]
\ubtestbadge{01} & \textcolor{zGoldDeep!78!black}{\texttt{\bfseries Course\-Objects\-And\-Start\-Markers\-Remain\-Intact}}\enspace\textcolor{zText}{Checks that course objects and start markers remain is intact.} \\
\ubtestbadge{02} & \textcolor{zGoldDeep!78!black}{\texttt{\bfseries Reference\-Character\-Controller\-Parameters\-Remain\-Fixed}}\enspace\textcolor{zText}{Checks that reference CharacterController parameters remain fixed.} \\
\ubtestbadge{03} & \textcolor{zGoldDeep!78!black}{\texttt{\bfseries Rigidbody\-Motor\-Keeps\-Required\-Api\-And\-Tuning\-Fields}}\enspace\textcolor{zText}{Checks that Rigidbody motor keeps required API and tuning fields.} \\
\ubtestbadge{04} & \textcolor{zGoldDeep!78!black}{\texttt{\bfseries Rigidbody\-Settings\-Support\-Stable\-Character\-Motion}}\enspace\textcolor{zText}{Checks that Rigidbody settings support stable character motion.} \\
\ubtestbadge{05} & \textcolor{zGoldDeep!78!black}{\texttt{\bfseries Student\-Capsule\-Matches\-Reference\-Shape}}\enspace\textcolor{zText}{Checks that student capsule matches reference shape.} \\
\ubtestbadge{06} & \textcolor{zGoldDeep!78!black}{\texttt{\bfseries Student\-Motor\-Does\-Not\-Depend\-On\-Character\-Controller}}\enspace\textcolor{zText}{Checks that student motor does not depend on CharacterController.} \\
\ubtestbadge{07} & \textcolor{zGoldDeep!78!black}{\texttt{\bfseries Student\-Uses\-Dynamic\-Rigidbody\-Capsule\-And\-No\-Character\-Controller}}\enspace\textcolor{zText}{Checks that student uses dynamic Rigidbody capsule and no CharacterController.} \\
\ubtestbadge{08} & \textcolor{zGoldDeep!78!black}{\texttt{\bfseries Diagonal\-Input\-Is\-Normalized\-Like\-Reference}}\enspace\textcolor{zText}{Checks that diagonal input is normalized like reference.} \\
\ubtestbadge{09} & \textcolor{zGoldDeep!78!black}{\texttt{\bfseries Falling\-And\-Grounding\-Match\-Reference}}\enspace\textcolor{zText}{Checks that falling and grounding match reference.} \\
\ubtestbadge{10} & \textcolor{zGoldDeep!78!black}{\texttt{\bfseries Jump\-Arc\-And\-Landing\-Match\-Reference}}\enspace\textcolor{zText}{Checks that jump arc and landing match reference.} \\
\ubtestbadge{11} & \textcolor{zGoldDeep!78!black}{\texttt{\bfseries Low\-Step\-Is\-Climbed\-Like\-Reference\-Controller}}\enspace\textcolor{zText}{Checks that low step is climbed like reference controller.} \\
\ubtestbadge{12} & \textcolor{zGoldDeep!78!black}{\texttt{\bfseries Slope\-Traversal\-Matches\-Reference\-Controller}}\enspace\textcolor{zText}{Checks that slope traversal matches reference controller.} \\
\ubtestbadge{13} & \textcolor{zGoldDeep!78!black}{\texttt{\bfseries Sprint\-Forward\-Matches\-Reference\-Speed}}\enspace\textcolor{zText}{Checks that sprint forward matches reference speed.} \\
\ubtestbadge{14} & \textcolor{zGoldDeep!78!black}{\texttt{\bfseries Stop\-Wall\-Blocks\-Rigidbody\-Controller\-Like\-Reference}}\enspace\textcolor{zText}{Checks that stop wall blocks Rigidbody controller like reference.} \\
\ubtestbadge{15} & \textcolor{zGoldDeep!78!black}{\texttt{\bfseries Walk\-Forward\-Matches\-Reference\-On\-Flat\-Ground}}\enspace\textcolor{zText}{Checks that walk forward matches reference on flat ground.} \\
\noalign{\vskip 2pt}
\multicolumn{2}{@{}p{14.4cm}@{}}{\textcolor{zRockLite!42}{\rule{\linewidth}{0.3pt}}}\\[3pt]
\multicolumn{2}{@{}p{14.4cm}@{}}{\ubtaskhead{zGoldDeep}{0044-ColliderWrappingChallenge}{Collider Wrapping Challenge}{zShrine}{Hard}{26}}\\*[2pt]
\ubtestbadge{01} & \textcolor{zGoldDeep!78!black}{\texttt{\bfseries Collider\-Rigs\-Use\-Only\-Reasonable\-Primitive\-Colliders}}\enspace\textcolor{zText}{Checks that Collider rigs use only reasonable primitive Colliders.} \\
\ubtestbadge{02} & \textcolor{zGoldDeep!78!black}{\texttt{\bfseries Primitive\-Colliders\-Cover\-Reference\-Volumes(Reference\-Object\-A)}}\enspace\textcolor{zText}{Checks that reference object A's primitive colliders cover enough of the reference volume.} \\
\ubtestbadge{03} & \textcolor{zGoldDeep!78!black}{\texttt{\bfseries Primitive\-Colliders\-Cover\-Reference\-Volumes(Reference\-Object\-B)}}\enspace\textcolor{zText}{Checks that reference object B's primitive colliders cover enough of the reference volume.} \\
\ubtestbadge{04} & \textcolor{zGoldDeep!78!black}{\texttt{\bfseries Primitive\-Colliders\-Cover\-Reference\-Volumes(Reference\-Object\-C)}}\enspace\textcolor{zText}{Checks that reference object C's primitive colliders cover enough of the reference volume.} \\
\ubtestbadge{05} & \textcolor{zGoldDeep!78!black}{\texttt{\bfseries Primitive\-Colliders\-Cover\-Reference\-Volumes(Reference\-Object\-D)}}\enspace\textcolor{zText}{Checks that reference object D's primitive colliders cover enough of the reference volume.} \\
\ubtestbadge{06} & \textcolor{zGoldDeep!78!black}{\texttt{\bfseries Primitive\-Colliders\-Cover\-Reference\-Volumes(Reference\-Object\-E)}}\enspace\textcolor{zText}{Checks that reference object E's primitive colliders cover enough of the reference volume.} \\
\ubtestbadge{07} & \textcolor{zGoldDeep!78!black}{\texttt{\bfseries Primitive\-Colliders\-Cover\-Reference\-Volumes(Reference\-Object\-F)}}\enspace\textcolor{zText}{Checks that reference object F's primitive colliders cover enough of the reference volume.} \\
\ubtestbadge{08} & \textcolor{zGoldDeep!78!black}{\texttt{\bfseries Primitive\-Colliders\-Cover\-Reference\-Volumes(Reference\-Object\-G)}}\enspace\textcolor{zText}{Checks that reference object G's primitive colliders cover enough of the reference volume.} \\
\ubtestbadge{09} & \textcolor{zGoldDeep!78!black}{\texttt{\bfseries Primitive\-Colliders\-Cover\-Reference\-Volumes(Reference\-Object\-H)}}\enspace\textcolor{zText}{Checks that reference object H's primitive colliders cover enough of the reference volume.} \\
\ubtestbadge{10} & \textcolor{zGoldDeep!78!black}{\texttt{\bfseries Primitive\-Colliders\-Do\-Not\-Spill\-Far\-Outside\-Reference\-Volumes(Reference\-Object\-A)}}\enspace\textcolor{zText}{Checks that reference object A's primitive colliders do not spill excessively outside the reference volume.} \\
\ubtestbadge{11} & \textcolor{zGoldDeep!78!black}{\texttt{\bfseries Primitive\-Colliders\-Do\-Not\-Spill\-Far\-Outside\-Reference\-Volumes(Reference\-Object\-B)}}\enspace\textcolor{zText}{Checks that reference object B's primitive colliders do not spill excessively outside the reference volume.} \\
\ubtestbadge{12} & \textcolor{zGoldDeep!78!black}{\texttt{\bfseries Primitive\-Colliders\-Do\-Not\-Spill\-Far\-Outside\-Reference\-Volumes(Reference\-Object\-C)}}\enspace\textcolor{zText}{Checks that reference object C's primitive colliders do not spill excessively outside the reference volume.} \\
\ubtestbadge{13} & \textcolor{zGoldDeep!78!black}{\texttt{\bfseries Primitive\-Colliders\-Do\-Not\-Spill\-Far\-Outside\-Reference\-Volumes(Reference\-Object\-D)}}\enspace\textcolor{zText}{Checks that reference object D's primitive colliders do not spill excessively outside the reference volume.} \\
\ubtestbadge{14} & \textcolor{zGoldDeep!78!black}{\texttt{\bfseries Primitive\-Colliders\-Do\-Not\-Spill\-Far\-Outside\-Reference\-Volumes(Reference\-Object\-E)}}\enspace\textcolor{zText}{Checks that reference object E's primitive colliders do not spill excessively outside the reference volume.} \\
\ubtestbadge{15} & \textcolor{zGoldDeep!78!black}{\texttt{\bfseries Primitive\-Colliders\-Do\-Not\-Spill\-Far\-Outside\-Reference\-Volumes(Reference\-Object\-F)}}\enspace\textcolor{zText}{Checks that reference object F's primitive colliders do not spill excessively outside the reference volume.} \\
\ubtestbadge{16} & \textcolor{zGoldDeep!78!black}{\texttt{\bfseries Primitive\-Colliders\-Do\-Not\-Spill\-Far\-Outside\-Reference\-Volumes(Reference\-Object\-G)}}\enspace\textcolor{zText}{Checks that reference object G's primitive colliders do not spill excessively outside the reference volume.} \\
\ubtestbadge{17} & \textcolor{zGoldDeep!78!black}{\texttt{\bfseries Primitive\-Colliders\-Do\-Not\-Spill\-Far\-Outside\-Reference\-Volumes(Reference\-Object\-H)}}\enspace\textcolor{zText}{Checks that reference object H's primitive colliders do not spill excessively outside the reference volume.} \\
\ubtestbadge{18} & \textcolor{zGoldDeep!78!black}{\texttt{\bfseries Primitive\-Colliders\-Respond\-To\-Representative\-Raycasts(Reference\-Object\-A)}}\enspace\textcolor{zText}{Checks that reference object A's primitive colliders respond to representative raycasts from multiple directions.} \\
\ubtestbadge{19} & \textcolor{zGoldDeep!78!black}{\texttt{\bfseries Primitive\-Colliders\-Respond\-To\-Representative\-Raycasts(Reference\-Object\-B)}}\enspace\textcolor{zText}{Checks that reference object B's primitive colliders respond to representative raycasts from multiple directions.} \\
\ubtestbadge{20} & \textcolor{zGoldDeep!78!black}{\texttt{\bfseries Primitive\-Colliders\-Respond\-To\-Representative\-Raycasts(Reference\-Object\-C)}}\enspace\textcolor{zText}{Checks that reference object C's primitive colliders respond to representative raycasts from multiple directions.} \\
\ubtestbadge{21} & \textcolor{zGoldDeep!78!black}{\texttt{\bfseries Primitive\-Colliders\-Respond\-To\-Representative\-Raycasts(Reference\-Object\-D)}}\enspace\textcolor{zText}{Checks that reference object D's primitive colliders respond to representative raycasts from multiple directions.} \\
\ubtestbadge{22} & \textcolor{zGoldDeep!78!black}{\texttt{\bfseries Primitive\-Colliders\-Respond\-To\-Representative\-Raycasts(Reference\-Object\-E)}}\enspace\textcolor{zText}{Checks that reference object E's primitive colliders respond to representative raycasts from multiple directions.} \\
\ubtestbadge{23} & \textcolor{zGoldDeep!78!black}{\texttt{\bfseries Primitive\-Colliders\-Respond\-To\-Representative\-Raycasts(Reference\-Object\-F)}}\enspace\textcolor{zText}{Checks that reference object F's primitive colliders respond to representative raycasts from multiple directions.} \\
\ubtestbadge{24} & \textcolor{zGoldDeep!78!black}{\texttt{\bfseries Primitive\-Colliders\-Respond\-To\-Representative\-Raycasts(Reference\-Object\-G)}}\enspace\textcolor{zText}{Checks that reference object G's primitive colliders respond to representative raycasts from multiple directions.} \\
\ubtestbadge{25} & \textcolor{zGoldDeep!78!black}{\texttt{\bfseries Primitive\-Colliders\-Respond\-To\-Representative\-Raycasts(Reference\-Object\-H)}}\enspace\textcolor{zText}{Checks that reference object H's primitive colliders respond to representative raycasts from multiple directions.} \\
\ubtestbadge{26} & \textcolor{zGoldDeep!78!black}{\texttt{\bfseries Submitted\-Scripts\-Do\-Not\-Use\-Forbidden\-Mesh\-Apis}}\enspace\textcolor{zText}{Checks that submitted scripts do not use forbidden mesh apis.} \\
\noalign{\vskip 2pt}
\multicolumn{2}{@{}p{14.4cm}@{}}{\textcolor{zRockLite!42}{\rule{\linewidth}{0.3pt}}}\\[3pt]
\multicolumn{2}{@{}p{14.4cm}@{}}{\ubtaskhead{zGoldDeep}{0053-PhysicsMaterial}{PhysicMaterial Tuning}{zShrine}{Hard}{8}}\\*[2pt]
\ubtestbadge{01} & \textcolor{zGoldDeep!78!black}{\texttt{\bfseries Scene\-Four\-Player\-Scores\-At\-Least\-One\-Point}}\enspace\textcolor{zText}{Checks that scene four player scores at least one point.} \\
\ubtestbadge{02} & \textcolor{zGoldDeep!78!black}{\texttt{\bfseries Scene\-Four\-Player\-Scores\-At\-Least\-Two\-Points}}\enspace\textcolor{zText}{Checks that scene four player scores at least two points.} \\
\ubtestbadge{03} & \textcolor{zGoldDeep!78!black}{\texttt{\bfseries Scene\-One\-Sphere\-Visits\-Checkpoint01}}\enspace\textcolor{zText}{Checks that scene one sphere visits checkpoint 01.} \\
\ubtestbadge{04} & \textcolor{zGoldDeep!78!black}{\texttt{\bfseries Scene\-One\-Sphere\-Visits\-Checkpoint02}}\enspace\textcolor{zText}{Checks that scene one sphere visits checkpoint 02.} \\
\ubtestbadge{05} & \textcolor{zGoldDeep!78!black}{\texttt{\bfseries Scene\-Three\-Sphere\-Visits\-Checkpoint01}}\enspace\textcolor{zText}{Checks that scene three sphere visits checkpoint 01.} \\
\ubtestbadge{06} & \textcolor{zGoldDeep!78!black}{\texttt{\bfseries Scene\-Three\-Sphere\-Visits\-Checkpoint02}}\enspace\textcolor{zText}{Checks that scene three sphere visits checkpoint 02.} \\
\ubtestbadge{07} & \textcolor{zGoldDeep!78!black}{\texttt{\bfseries Scene\-Three\-Sphere\-Visits\-Checkpoint03}}\enspace\textcolor{zText}{Checks that scene three sphere visits checkpoint 03.} \\
\ubtestbadge{08} & \textcolor{zGoldDeep!78!black}{\texttt{\bfseries Scene\-Two\-Sphere\-Visits\-Checkpoint01}}\enspace\textcolor{zText}{Checks that scene two sphere visits checkpoint 01.} \\
\noalign{\vskip 2pt}
\multicolumn{2}{@{}p{14.4cm}@{}}{\textcolor{zRockLite!42}{\rule{\linewidth}{0.3pt}}}\\[3pt]
\multicolumn{2}{@{}p{14.4cm}@{}}{\ubtaskhead{zGoldDeep}{0077-TakeOffAndLanding}{Implement Physical Fixed-Wing Takeoff and Landing}{zShrine}{Hard}{31}}\\*[2pt]
\ubtestbadge{01} & \textcolor{zGoldDeep!78!black}{\texttt{\bfseries Required\-Scene\-And\-Aircraft\-Contract\-Remain\-Intact}}\enspace\textcolor{zText}{Checks that the required scene, aircraft root, Rigidbody, runway, and wheel objects remain intact.} \\
\ubtestbadge{02} & \textcolor{zGoldDeep!78!black}{\texttt{\bfseries Aircraft\-Uses\-Unconstrained\-Dynamic\-Gravity\-Rigidbody}}\enspace\textcolor{zText}{Checks that the aircraft uses a non-kinematic, gravity-enabled Rigidbody without frozen motion axes.} \\
\ubtestbadge{03} & \textcolor{zGoldDeep!78!black}{\texttt{\bfseries Exactly\-Three\-Passive\-Landing\-Gear\-Wheel\-Colliders\-Remain}}\enspace\textcolor{zText}{Checks that exactly three passive WheelColliders remain and all have zero motor torque.} \\
\ubtestbadge{04} & \textcolor{zGoldDeep!78!black}{\texttt{\bfseries Runway\-And\-Aircraft\-Collision\-Are\-Not\-Bypassed}}\enspace\textcolor{zText}{Checks that runway, aircraft, and landing-gear collisions have not been disabled or bypassed.} \\
\ubtestbadge{05} & \textcolor{zGoldDeep!78!black}{\texttt{\bfseries Aircraft\-Is\-Not\-Driven\-By\-Animator\-Timeline\-Or\-Parent\-Motion}}\enspace\textcolor{zText}{Checks that Animator, Timeline, or moving-parent motion does not drive the aircraft.} \\
\ubtestbadge{06} & \textcolor{zGoldDeep!78!black}{\texttt{\bfseries Source\-Does\-Not\-Contain\-Time\-Acceleration\-Or\-Obvious\-Runtime\-Teleportation}}\enspace\textcolor{zText}{Checks source code for prohibited time acceleration and obvious runtime teleportation of the aircraft.} \\
\ubtestbadge{07} & \textcolor{zGoldDeep!78!black}{\texttt{\bfseries Collect\-Three\-Scenario\-Mission\-Telemetry}}\enspace\textcolor{zText}{Runs the nominal, heavy right-offset, and light left-offset disturbed scenarios and collects telemetry while enforcing the runtime physical contract.} \\
\ubtestbadge{08} & \textcolor{zGoldDeep!78!black}{\texttt{\bfseries All\-Scenarios\-Complete\-Takeoff\-Landing\-And\-Stop}}\enspace\textcolor{zText}{Checks that every hidden scenario completes takeoff, landing, rollout, and a full stop.} \\
\ubtestbadge{09} & \textcolor{zGoldDeep!78!black}{\texttt{\bfseries Engine\-Produces\-Prompt\-Forward\-Ground\-Roll}}\enspace\textcolor{zText}{Checks that forward ground roll begins within five seconds.} \\
\ubtestbadge{10} & \textcolor{zGoldDeep!78!black}{\texttt{\bfseries Takeoff\-Roll\-Stays\-Inside\-Runway\-Tolerance}}\enspace\textcolor{zText}{Checks that takeoff cross-track error remains within five metres.} \\
\ubtestbadge{11} & \textcolor{zGoldDeep!78!black}{\texttt{\bfseries Takeoff\-Roll\-Has\-Precision\-Centerline\-Control}}\enspace\textcolor{zText}{Checks precise takeoff centerline tracking using maximum and RMS cross-track limits.} \\
\ubtestbadge{12} & \textcolor{zGoldDeep!78!black}{\texttt{\bfseries Liftoff\-Speed\-Is\-Physically\-Plausible}}\enspace\textcolor{zText}{Checks that liftoff speed remains within the physically plausible 50 to 85 m/s range.} \\
\ubtestbadge{13} & \textcolor{zGoldDeep!78!black}{\texttt{\bfseries Liftoff\-Speed\-Is\-Inside\-High\-Quality\-Window}}\enspace\textcolor{zText}{Checks that liftoff speed remains within the higher-quality 58 to 78 m/s window.} \\
\ubtestbadge{14} & \textcolor{zGoldDeep!78!black}{\texttt{\bfseries Liftoff\-Attitude\-Is\-Stable}}\enspace\textcolor{zText}{Checks liftoff pitch, roll, and runway-heading error for a stable departure.} \\
\ubtestbadge{15} & \textcolor{zGoldDeep!78!black}{\texttt{\bfseries Aircraft\-Does\-Not\-Bounce\-Back\-Onto\-Runway\-After\-Liftoff}}\enspace\textcolor{zText}{Checks that the aircraft does not contact the runway again immediately after liftoff.} \\
\ubtestbadge{16} & \textcolor{zGoldDeep!78!black}{\texttt{\bfseries Return\-Circuit\-Remains\-Compact}}\enspace\textcolor{zText}{Checks that the aircraft reaches downwind while remaining inside the lateral and longitudinal circuit bounds.} \\
\ubtestbadge{17} & \textcolor{zGoldDeep!78!black}{\texttt{\bfseries Airborne\-Envelope\-Avoids\-Stall\-And\-Extreme\-Altitude}}\enspace\textcolor{zText}{Checks the airborne speed and altitude envelope for stall avoidance and bounded flight.} \\
\ubtestbadge{18} & \textcolor{zGoldDeep!78!black}{\texttt{\bfseries Mission\-Finishes\-Within\-Normal\-Time\-Budget}}\enspace\textcolor{zText}{Checks that every scenario completes within 120 seconds.} \\
\ubtestbadge{19} & \textcolor{zGoldDeep!78!black}{\texttt{\bfseries Runway\-Alignment\-Occurs\-Before\-Touchdown}}\enspace\textcolor{zText}{Checks that stable runway alignment is established at least four seconds before touchdown.} \\
\ubtestbadge{20} & \textcolor{zGoldDeep!78!black}{\texttt{\bfseries Alignment\-Is\-Established\-With\-Useful\-Final\-Distance}}\enspace\textcolor{zText}{Checks that runway alignment is established with at least 400 metres remaining before touchdown.} \\
\ubtestbadge{21} & \textcolor{zGoldDeep!78!black}{\texttt{\bfseries Alignment\-State\-Is\-Already\-Stable}}\enspace\textcolor{zText}{Checks cross-track position, heading, lateral speed, and roll at the moment alignment is established.} \\
\ubtestbadge{22} & \textcolor{zGoldDeep!78!black}{\texttt{\bfseries Final\-Approach\-Maintains\-Centerline\-And\-Heading}}\enspace\textcolor{zText}{Checks final-approach RMS centerline and heading tracking.} \\
\ubtestbadge{23} & \textcolor{zGoldDeep!78!black}{\texttt{\bfseries Touchdown\-Occurs\-Inside\-Runway\-Corridor}}\enspace\textcolor{zText}{Checks that touchdown occurs within twelve metres of the runway centerline.} \\
\ubtestbadge{24} & \textcolor{zGoldDeep!78!black}{\texttt{\bfseries Touchdown\-Has\-Precision\-Centerline\-Placement}}\enspace\textcolor{zText}{Checks the higher-quality touchdown placement within six metres of the runway centerline.} \\
\ubtestbadge{25} & \textcolor{zGoldDeep!78!black}{\texttt{\bfseries Touchdown\-Speed\-Sink\-And\-Attitude\-Are\-Plausible}}\enspace\textcolor{zText}{Checks touchdown forward speed, sink rate, pitch, roll, and heading error.} \\
\ubtestbadge{26} & \textcolor{zGoldDeep!78!black}{\texttt{\bfseries Main\-Gear\-Touches\-First\-Without\-Second\-Bounce}}\enspace\textcolor{zText}{Checks main-gear-first touchdown, delayed nose-wheel contact, and absence of a second bounce.} \\
\ubtestbadge{27} & \textcolor{zGoldDeep!78!black}{\texttt{\bfseries Touchdown\-Does\-Not\-Create\-Forward\-Speed\-Jump}}\enspace\textcolor{zText}{Checks that touchdown does not increase forward speed by more than 2 m/s.} \\
\ubtestbadge{28} & \textcolor{zGoldDeep!78!black}{\texttt{\bfseries Rollout\-Stops\-Within\-Twenty\-Five\-Seconds}}\enspace\textcolor{zText}{Checks that the aircraft stops within 25 seconds after touchdown.} \\
\ubtestbadge{29} & \textcolor{zGoldDeep!78!black}{\texttt{\bfseries Rollout\-Stop\-Distance\-Remains\-Inside\-Runway}}\enspace\textcolor{zText}{Checks a stop distance within 900 metres without reversing.} \\
\ubtestbadge{30} & \textcolor{zGoldDeep!78!black}{\texttt{\bfseries Rollout\-Maintains\-Directional\-Control}}\enspace\textcolor{zText}{Checks rollout centerline tracking and heading control.} \\
\ubtestbadge{31} & \textcolor{zGoldDeep!78!black}{\texttt{\bfseries Wheels\-Remain\-Passive\-And\-Contact\-Does\-Not\-Chatter}}\enspace\textcolor{zText}{Checks zero wheel motor torque and a bounded number of landing-gear contact transitions.} \\
\noalign{\vskip 2pt}
\multicolumn{2}{@{}p{14.4cm}@{}}{\textcolor{zRockLite!42}{\rule{\linewidth}{0.3pt}}}\\[3pt]
\noalign{\vskip 3pt}
\multicolumn{2}{@{}p{14.4cm}@{}}{\ubcategory{zRock}{Miscellaneous}{3 tasks \textperiodcentered\ 58 milestone tests}}\\*[4pt]
\multicolumn{2}{@{}p{14.4cm}@{}}{\ubtaskhead{zRock}{0011-CookiesApply}{Light Cookie Application}{zGoldDeep}{Easy}{9}}\\*[2pt]
\ubtestbadge{01} & \textcolor{zGoldDeep!78!black}{\texttt{\bfseries Cookie\_\allowbreak{}Directional\-Light\_\allowbreak{}Has\-Cookie\-Texture}}\enspace\textcolor{zText}{Checks that cookie directional light has cookie texture.} \\
\ubtestbadge{02} & \textcolor{zGoldDeep!78!black}{\texttt{\bfseries Cookie\_\allowbreak{}Directional\-Light\_\allowbreak{}Texture\_\allowbreak{}Alpha\-Source\-Is\-From\-Gray\-Scale}}\enspace\textcolor{zText}{Checks that cookie directional light texture alpha source is from gray scale.} \\
\ubtestbadge{03} & \textcolor{zGoldDeep!78!black}{\texttt{\bfseries Cookie\_\allowbreak{}Directional\-Light\_\allowbreak{}Texture\_\allowbreak{}Cookie\-Light\-Type\-Is\-Directional}}\enspace\textcolor{zText}{Checks that cookie directional light texture cookie light type is directional.} \\
\ubtestbadge{04} & \textcolor{zGoldDeep!78!black}{\texttt{\bfseries Cookie\_\allowbreak{}Point\-Light\_\allowbreak{}Has\-Cookie\-Cubemap}}\enspace\textcolor{zText}{Checks that cookie point light has cookie cubemap.} \\
\ubtestbadge{05} & \textcolor{zGoldDeep!78!black}{\texttt{\bfseries Light\_\allowbreak{}Directional\-Light\_\allowbreak{}Type\-Is\-Directional}}\enspace\textcolor{zText}{Checks that light directional light type is directional.} \\
\ubtestbadge{06} & \textcolor{zGoldDeep!78!black}{\texttt{\bfseries Light\_\allowbreak{}Point\-Light\_\allowbreak{}Type\-Is\-Point}}\enspace\textcolor{zText}{Checks that light point light type is point.} \\
\ubtestbadge{07} & \textcolor{zGoldDeep!78!black}{\texttt{\bfseries Scene\_\allowbreak{}Directional\-Light\_\allowbreak{}Exists}}\enspace\textcolor{zText}{Checks that directional light exists in the scene.} \\
\ubtestbadge{08} & \textcolor{zGoldDeep!78!black}{\texttt{\bfseries Scene\_\allowbreak{}Plane\_\allowbreak{}Exists}}\enspace\textcolor{zText}{Checks that plane exists in the scene.} \\
\ubtestbadge{09} & \textcolor{zGoldDeep!78!black}{\texttt{\bfseries Scene\_\allowbreak{}Point\-Light\_\allowbreak{}Exists}}\enspace\textcolor{zText}{Checks that point light exists in the scene.} \\
\noalign{\vskip 2pt}
\multicolumn{2}{@{}p{14.4cm}@{}}{\textcolor{zRockLite!42}{\rule{\linewidth}{0.3pt}}}\\[3pt]
\multicolumn{2}{@{}p{14.4cm}@{}}{\ubtaskhead{zRock}{0017-CreateLOD}{LOD Asset Creation}{zGoldDeep}{Medium}{39}}\\*[2pt]
\ubtestbadge{01} & \textcolor{zGoldDeep!78!black}{\texttt{\bfseries Answer1\_\allowbreak{}1\_\allowbreak{}Fbx\-Exists}}\enspace\textcolor{zText}{Checks that Answer1\_1 FBX exists.} \\
\ubtestbadge{02} & \textcolor{zGoldDeep!78!black}{\texttt{\bfseries Answer1\_\allowbreak{}1\_\allowbreak{}Has724\-Triangles}}\enspace\textcolor{zText}{Checks that Answer1\_1 has 724 triangles.} \\
\ubtestbadge{03} & \textcolor{zGoldDeep!78!black}{\texttt{\bfseries Answer1\_\allowbreak{}1\_\allowbreak{}Uses\-Pallet\-Plywood\-Material}}\enspace\textcolor{zText}{Checks that Answer1\_1 uses pallet plywood material.} \\
\ubtestbadge{04} & \textcolor{zGoldDeep!78!black}{\texttt{\bfseries Answer1\_\allowbreak{}2\_\allowbreak{}Fbx\-Exists}}\enspace\textcolor{zText}{Checks that Answer1\_2 FBX exists.} \\
\ubtestbadge{05} & \textcolor{zGoldDeep!78!black}{\texttt{\bfseries Answer1\_\allowbreak{}2\_\allowbreak{}Has406\-Triangles}}\enspace\textcolor{zText}{Checks that Answer1\_2 has 406 triangles.} \\
\ubtestbadge{06} & \textcolor{zGoldDeep!78!black}{\texttt{\bfseries Answer1\_\allowbreak{}2\_\allowbreak{}Uses\-Pallet\-Plywood\-Material}}\enspace\textcolor{zText}{Checks that Answer1\_2 uses pallet plywood material.} \\
\ubtestbadge{07} & \textcolor{zGoldDeep!78!black}{\texttt{\bfseries Answer1\_\allowbreak{}3\_\allowbreak{}Fbx\-Exists}}\enspace\textcolor{zText}{Checks that Answer1\_3 FBX exists.} \\
\ubtestbadge{08} & \textcolor{zGoldDeep!78!black}{\texttt{\bfseries Answer1\_\allowbreak{}3\_\allowbreak{}Has234\-Triangles}}\enspace\textcolor{zText}{Checks that Answer1\_3 has 234 triangles.} \\
\ubtestbadge{09} & \textcolor{zGoldDeep!78!black}{\texttt{\bfseries Answer1\_\allowbreak{}3\_\allowbreak{}Uses\-Pallet\-Plywood\-Material}}\enspace\textcolor{zText}{Checks that Answer1\_3 uses pallet plywood material.} \\
\ubtestbadge{10} & \textcolor{zGoldDeep!78!black}{\texttt{\bfseries Answer2\_\allowbreak{}Fbx\-Exists}}\enspace\textcolor{zText}{Checks that the Answer2 FBX asset exists.} \\
\ubtestbadge{11} & \textcolor{zGoldDeep!78!black}{\texttt{\bfseries Answer2\_\allowbreak{}Has\-LODGroup\-In\-Scene}}\enspace\textcolor{zText}{Checks that Answer2 has an LODGroup in the scene.} \\
\ubtestbadge{12} & \textcolor{zGoldDeep!78!black}{\texttt{\bfseries Answer2\_\allowbreak{}LOD0\_\allowbreak{}Has9826\-Triangles}}\enspace\textcolor{zText}{Checks that answer 2 LOD 0 has 9826 triangles.} \\
\ubtestbadge{13} & \textcolor{zGoldDeep!78!black}{\texttt{\bfseries Answer2\_\allowbreak{}LOD1\_\allowbreak{}Has17803\-Triangles}}\enspace\textcolor{zText}{Checks that answer 2 LOD 1 has 17803 triangles.} \\
\ubtestbadge{14} & \textcolor{zGoldDeep!78!black}{\texttt{\bfseries Answer2\_\allowbreak{}LOD2\_\allowbreak{}Has15164\-Triangles}}\enspace\textcolor{zText}{Checks that answer 2 LOD 2 has 15164 triangles.} \\
\ubtestbadge{15} & \textcolor{zGoldDeep!78!black}{\texttt{\bfseries Answer2\_\allowbreak{}LOD3\_\allowbreak{}Has13762\-Triangles}}\enspace\textcolor{zText}{Checks that answer 2 LOD 3 has 13762 triangles.} \\
\ubtestbadge{16} & \textcolor{zGoldDeep!78!black}{\texttt{\bfseries Answer2\_\allowbreak{}LODGroup\_\allowbreak{}Even\-Percentage}}\enspace\textcolor{zText}{Checks that Answer2's LODGroup transition percentages are evenly spaced.} \\
\ubtestbadge{17} & \textcolor{zGoldDeep!78!black}{\texttt{\bfseries Answer2\_\allowbreak{}LODs\-Use\-Original\-Materials}}\enspace\textcolor{zText}{Checks that answer 2 lo ds use original materials.} \\
\ubtestbadge{18} & \textcolor{zGoldDeep!78!black}{\texttt{\bfseries Answer3\_\allowbreak{}Fbx\-Exists}}\enspace\textcolor{zText}{Checks that the Answer3 FBX asset exists.} \\
\ubtestbadge{19} & \textcolor{zGoldDeep!78!black}{\texttt{\bfseries Answer3\_\allowbreak{}Has\-LODGroup\-In\-Scene}}\enspace\textcolor{zText}{Checks that Answer3 has an LODGroup in the scene.} \\
\ubtestbadge{20} & \textcolor{zGoldDeep!78!black}{\texttt{\bfseries Answer3\_\allowbreak{}LOD0\_\allowbreak{}Has11247\-Triangles}}\enspace\textcolor{zText}{Checks that answer 3 LOD 0 has 11247 triangles.} \\
\ubtestbadge{21} & \textcolor{zGoldDeep!78!black}{\texttt{\bfseries Answer3\_\allowbreak{}LOD1\_\allowbreak{}Has5617\-Triangles}}\enspace\textcolor{zText}{Checks that answer 3 LOD 1 has 5617 triangles.} \\
\ubtestbadge{22} & \textcolor{zGoldDeep!78!black}{\texttt{\bfseries Answer3\_\allowbreak{}LOD2\_\allowbreak{}Has2238\-Triangles}}\enspace\textcolor{zText}{Checks that answer 3 LOD 2 has 2238 triangles.} \\
\ubtestbadge{23} & \textcolor{zGoldDeep!78!black}{\texttt{\bfseries Answer3\_\allowbreak{}LOD3\_\allowbreak{}Has60\-Triangles}}\enspace\textcolor{zText}{Checks that Answer3 LOD3 has 60 triangles.} \\
\ubtestbadge{24} & \textcolor{zGoldDeep!78!black}{\texttt{\bfseries Answer3\_\allowbreak{}LODGroup\_\allowbreak{}Even\-Percentage}}\enspace\textcolor{zText}{Checks that Answer3's LODGroup transition percentages are evenly spaced.} \\
\ubtestbadge{25} & \textcolor{zGoldDeep!78!black}{\texttt{\bfseries Answer3\_\allowbreak{}LODs\-Use\-Original\-Materials}}\enspace\textcolor{zText}{Checks that answer 3 lo ds use original materials.} \\
\ubtestbadge{26} & \textcolor{zGoldDeep!78!black}{\texttt{\bfseries Answer4\_\allowbreak{}Contains\-LOD0\-To3}}\enspace\textcolor{zText}{Checks that Answer4 contains LOD0 through LOD3.} \\
\ubtestbadge{27} & \textcolor{zGoldDeep!78!black}{\texttt{\bfseries Answer4\_\allowbreak{}Face\-Counts\-Decrease}}\enspace\textcolor{zText}{Checks that Answer4's mesh face counts decrease across successive LOD levels.} \\
\ubtestbadge{28} & \textcolor{zGoldDeep!78!black}{\texttt{\bfseries Answer4\_\allowbreak{}Fbx\-Exists}}\enspace\textcolor{zText}{Checks that the Answer4 FBX asset exists.} \\
\ubtestbadge{29} & \textcolor{zGoldDeep!78!black}{\texttt{\bfseries Answer4\_\allowbreak{}Has\-LODGroup\-In\-Scene}}\enspace\textcolor{zText}{Checks that Answer4 has an LODGroup in the scene.} \\
\ubtestbadge{30} & \textcolor{zGoldDeep!78!black}{\texttt{\bfseries Answer4\_\allowbreak{}LOD0\_\allowbreak{}Has503351\-Triangles}}\enspace\textcolor{zText}{Checks that answer 4 LOD 0 has 503351 triangles.} \\
\ubtestbadge{31} & \textcolor{zGoldDeep!78!black}{\texttt{\bfseries Answer4\_\allowbreak{}LODGroup\_\allowbreak{}Even\-Percentage}}\enspace\textcolor{zText}{Checks that Answer4's LODGroup transition percentages are evenly spaced.} \\
\ubtestbadge{32} & \textcolor{zGoldDeep!78!black}{\texttt{\bfseries Answer4\_\allowbreak{}LODs\-Use\-Original\-Materials}}\enspace\textcolor{zText}{Checks that answer 4 lo ds use original materials.} \\
\ubtestbadge{33} & \textcolor{zGoldDeep!78!black}{\texttt{\bfseries Answer5\_\allowbreak{}Contains\-LOD0\-To3}}\enspace\textcolor{zText}{Checks that Answer5 contains LOD0 through LOD3.} \\
\ubtestbadge{34} & \textcolor{zGoldDeep!78!black}{\texttt{\bfseries Answer5\_\allowbreak{}Face\-Counts\-Decrease}}\enspace\textcolor{zText}{Checks that Answer5's mesh face counts decrease across successive LOD levels.} \\
\ubtestbadge{35} & \textcolor{zGoldDeep!78!black}{\texttt{\bfseries Answer5\_\allowbreak{}Fbx\-Exists}}\enspace\textcolor{zText}{Checks that the Answer5 FBX asset exists.} \\
\ubtestbadge{36} & \textcolor{zGoldDeep!78!black}{\texttt{\bfseries Answer5\_\allowbreak{}Has\-LODGroup\-In\-Scene}}\enspace\textcolor{zText}{Checks that Answer5 has an LODGroup in the scene.} \\
\ubtestbadge{37} & \textcolor{zGoldDeep!78!black}{\texttt{\bfseries Answer5\_\allowbreak{}LOD0\_\allowbreak{}Has17690\-Triangles}}\enspace\textcolor{zText}{Checks that answer 5 LOD 0 has 17690 triangles.} \\
\ubtestbadge{38} & \textcolor{zGoldDeep!78!black}{\texttt{\bfseries Answer5\_\allowbreak{}LODGroup\_\allowbreak{}Even\-Percentage}}\enspace\textcolor{zText}{Checks that Answer5's LODGroup transition percentages are evenly spaced.} \\
\ubtestbadge{39} & \textcolor{zGoldDeep!78!black}{\texttt{\bfseries Answer5\_\allowbreak{}LODs\-Use\-Original\-Materials}}\enspace\textcolor{zText}{Checks that answer 5 lo ds use original materials.} \\
\noalign{\vskip 2pt}
\multicolumn{2}{@{}p{14.4cm}@{}}{\textcolor{zRockLite!42}{\rule{\linewidth}{0.3pt}}}\\[3pt]
\multicolumn{2}{@{}p{14.4cm}@{}}{\ubtaskhead{zRock}{0019-AutoTimelineGenerator}{Timeline Cinematic Auto-Generation}{zShrine}{Hard}{10}}\\*[2pt]
\ubtestbadge{01} & \textcolor{zGoldDeep!78!black}{\texttt{\bfseries Test01\_\allowbreak{}Timeline\-Asset\-And\-Track\-Structure}}\enspace\textcolor{zText}{Checks that the Timeline asset contains the required named tracks, bindings, and clip structure.} \\
\ubtestbadge{02} & \textcolor{zGoldDeep!78!black}{\texttt{\bfseries Test02\_\allowbreak{}Walk\-Animation\-Complete\-Validation}}\enspace\textcolor{zText}{Checks the Walk track timing, Character\_Main binding, animation clip, and genuine root-motion displacement.} \\
\ubtestbadge{03} & \textcolor{zGoldDeep!78!black}{\texttt{\bfseries Test03\_\allowbreak{}VFXControl\-Track\-Complete\-Validation}}\enspace\textcolor{zText}{Checks the VFX Control Track timing, binding, activation behavior, and effect object.} \\
\ubtestbadge{04} & \textcolor{zGoldDeep!78!black}{\texttt{\bfseries Test04\_\allowbreak{}Audio\-Track\-Complete\-Validation}}\enspace\textcolor{zText}{Checks the Audio Track timing, binding, and assigned explosion clip.} \\
\ubtestbadge{05} & \textcolor{zGoldDeep!78!black}{\texttt{\bfseries Test05\_\allowbreak{}Camera\-Shake\-Track\-And\-Curve\-Validation}}\enspace\textcolor{zText}{Checks the Shake track timing, CameraRig binding, and non-empty displacement curves with bounded amplitude.} \\
\ubtestbadge{06} & \textcolor{zGoldDeep!78!black}{\texttt{\bfseries Test06\_\allowbreak{}Camera\-Rig\-Hierarchy\-And\-Composition\-Integrity}}\enspace\textcolor{zText}{Checks the required CameraRig/Main Camera hierarchy and preserved camera composition.} \\
\ubtestbadge{07} & \textcolor{zGoldDeep!78!black}{\texttt{\bfseries Test07\_\allowbreak{}Cinematic\-Framing\-And\-Spatial\-Composition}}\enspace\textcolor{zText}{Checks the cinematic framing and spatial composition of the character, VFX, and camera.} \\
\ubtestbadge{08} & \textcolor{zGoldDeep!78!black}{\texttt{\bfseries P01\_\allowbreak{}Character\-Movement\-And\-Consistency}}\enspace\textcolor{zText}{Checks that Timeline moves the character consistently during the walk interval.} \\
\ubtestbadge{09} & \textcolor{zGoldDeep!78!black}{\texttt{\bfseries P02\_\allowbreak{}VFXLifecycle\-And\-Component\-Validation}}\enspace\textcolor{zText}{Checks the VFX lifecycle and required runtime components during Timeline playback.} \\
\ubtestbadge{10} & \textcolor{zGoldDeep!78!black}{\texttt{\bfseries P03\_\allowbreak{}Camera\-Shake\-And\-Viewport\-Integrity}}\enspace\textcolor{zText}{Checks visible camera shake while preserving a valid viewport and reasonable framing.} \\
\noalign{\vskip 2pt}
\multicolumn{2}{@{}p{14.4cm}@{}}{\textcolor{zRockLite!42}{\rule{\linewidth}{0.3pt}}}\\[3pt]
\end{longtable}
\endgroup

\section{Supplementary Statistics}
\label{app:supplementary-statistics}

This section supplements the main results with tests of benchmark quality and behavioral diagnostics. Unless noted otherwise, all analyses use the complete matrix of 32 configurations on the same 77 tasks (2,464 runs), exclude ablations, and define each overall configuration score as the unweighted mean over the 77 tasks. They characterize internal calibration, coverage, and score robustness, but do not establish external validity or repeated-run variance.

\subsection{Task-difficulty Diagnostics}
\label{app:benchmark-calibration}

\paragraph{Difficulty calibration and discriminative power.} Mean scores decline from 80.85\% on the 25 Easy tasks to 55.56\% on the 29 Medium tasks and 35.21\% on the 23 Hard tasks. Yet the leave-one-task-out gaps between the stronger and weaker halves of the configurations remain similar at 15.56, 16.15, and 17.36 percentage points ($p=.845$ under permutation). Difficulty therefore tracks task completion without confining discrimination to one split. The task-equal gap is positive for 75 of 77 tasks; the two negative cases remain visible for audit, while floor and ceiling effects may limit several others.

\begin{figure}[H]
\centering
\includegraphics[width=\textwidth]{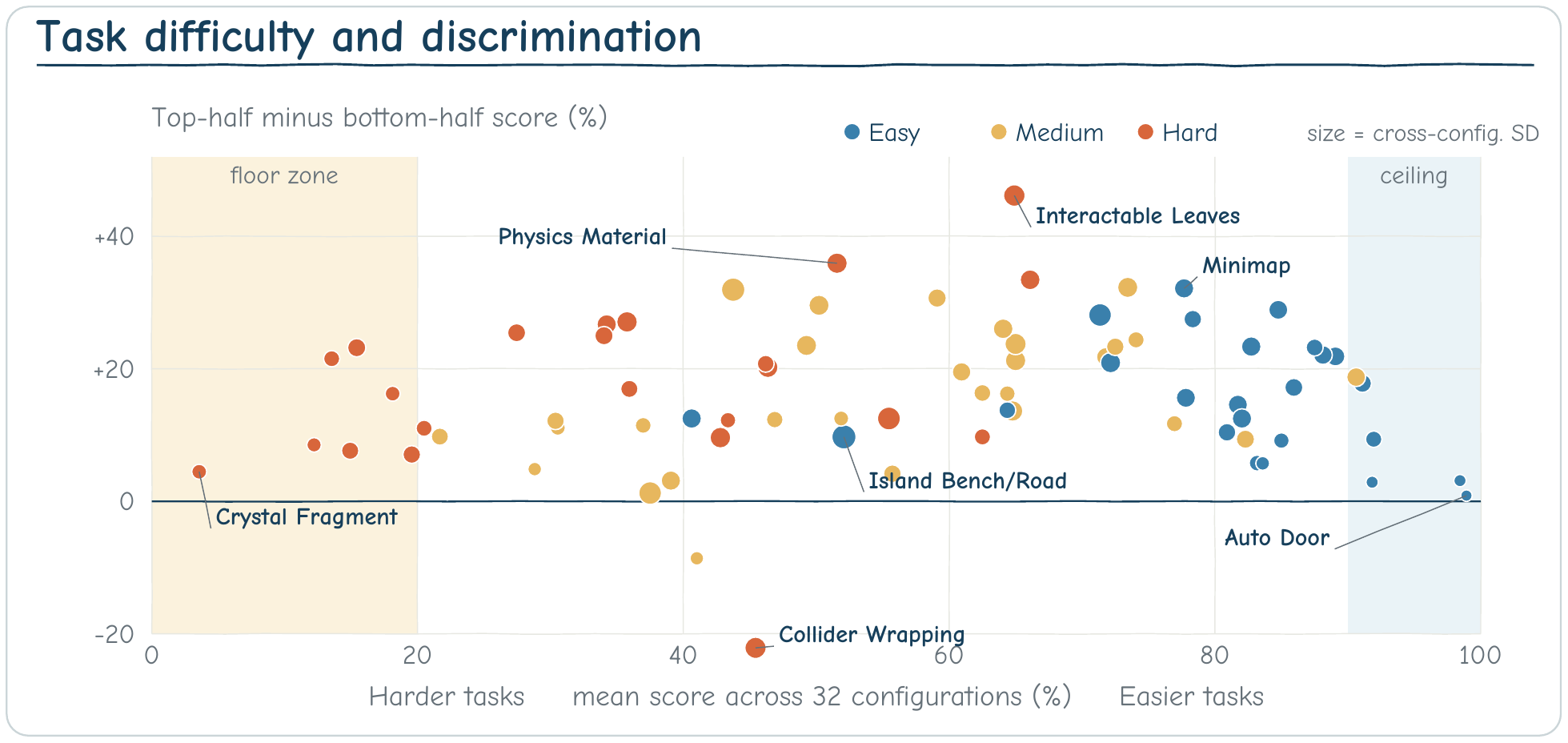}
\caption{\textbf{Task difficulty and discrimination.} Each point is one of the 77 tasks. The horizontal axis is its mean score across 32 configurations. The vertical axis is the score gap between the top and bottom 16 configurations, where configuration strength is recomputed as the task-equal mean over the other 76 tasks to avoid circularity. Marker size reflects cross-configuration variation, color denotes nominal difficulty, and shaded regions indicate potential floor and ceiling effects. This is a descriptive benchmark diagnostic; configurations sharing a harness are not independent samples.}
\label{fig:supp-task-discrimination}
\end{figure}

\paragraph{Coverage and non-redundancy.} Across all 2,926 task pairs, the median Spearman correlation is 0.227 and falls to 0.060 after descriptively adjusting for task-equal performance on the other 75 tasks. A whole-matrix permutation test gives an absolute-correlation threshold of 0.707 at a family-wise error rate of .05; after excluding floor and ceiling tasks, only five positive pairs exceed it (0.17\% of all pairs). Same-category pairs are not more correlated than cross-category pairs under label permutation ($p=.868$). The suite therefore shows broad complementary coverage, although low score correlation alone cannot establish semantic independence.

\begin{figure}[H]
\centering
\includegraphics[width=\textwidth]{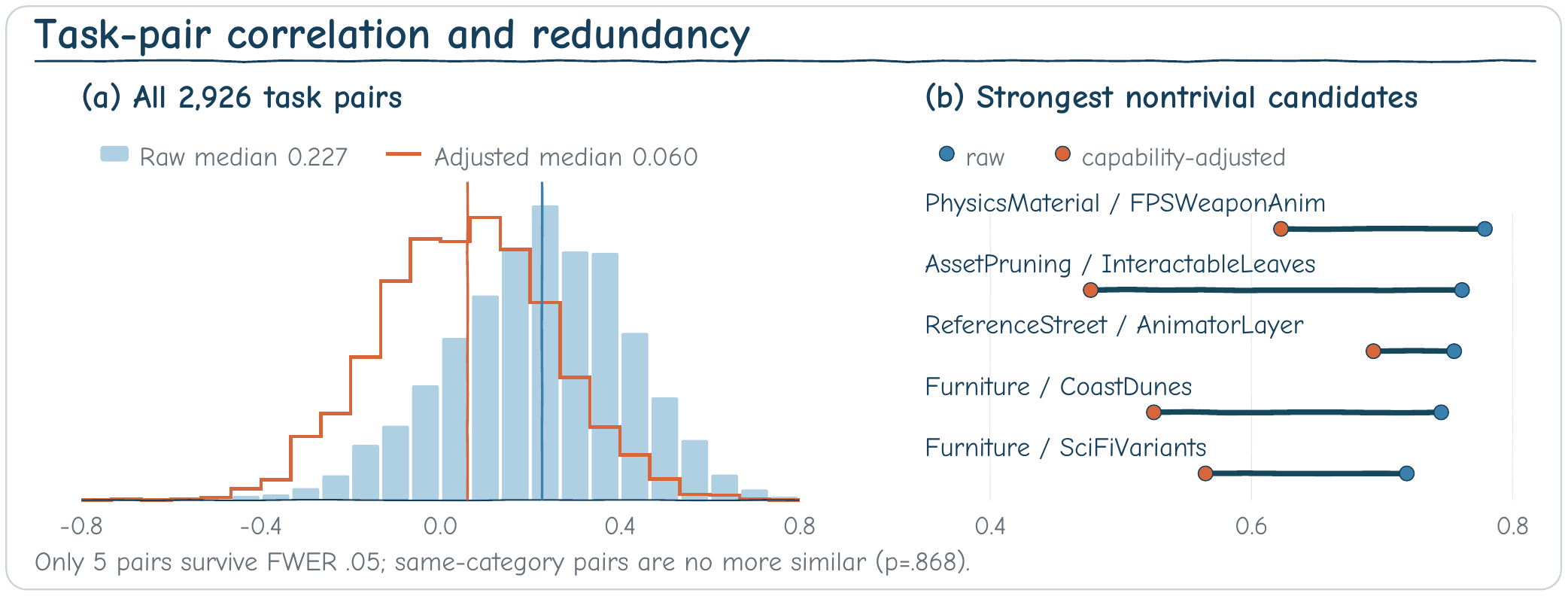}
\caption{\textbf{Task-pair correlation and redundancy.} Panel (a) compares raw Spearman correlations for all 2,926 task pairs with descriptive correlations adjusted for each configuration's task-equal performance on the remaining 75 tasks. Panel (b) shows the five positive, non-floor/ceiling pairs whose raw correlations exceed the family-wise threshold estimated from 50,000 whole-matrix permutations.}
\label{fig:supp-task-redundancy}
\end{figure}

\subsection{Behavioral and Diagnostic Insights}
\label{app:behavioral-diagnostics}

\paragraph{Failure signatures across skills.} The 14,909 non-passing milestone outcomes have skill-specific profiles. Missing implementation and spatial inaccuracy form 90.71\% of scene-construction failures; runtime behavior forms 59.62\% of physics failures; and wiring forms 54.40\% of editor-tooling failures. Appearance errors peak for particles and shaders (38.67\%) and user interface (34.87\%). This alignment makes the taxonomy more diagnostic than a generic failure label. Because skills contain different task and milestone counts, the figure compares composition within skills, not absolute failure frequency or independent taxonomy validity.

\begin{figure}[H]
\centering
\includegraphics[width=\textwidth]{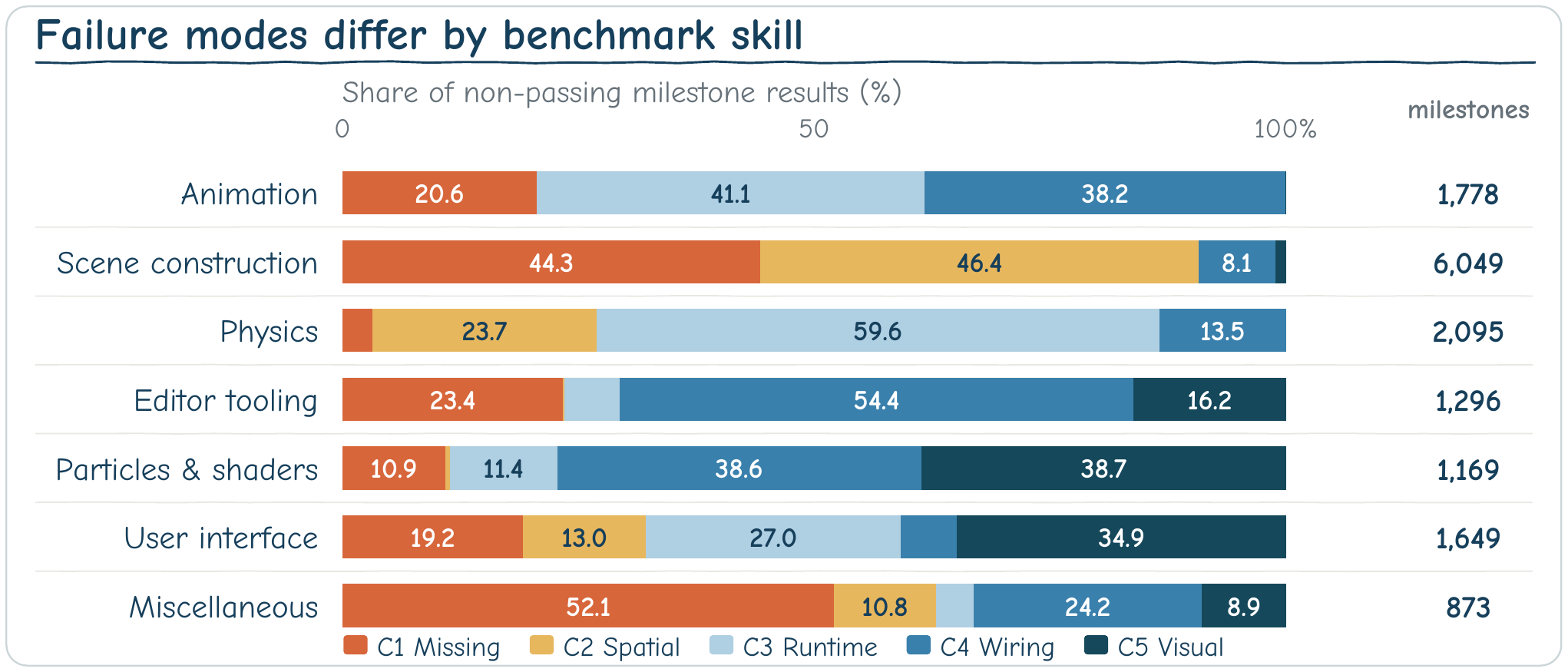}
\caption{\textbf{Failure composition by benchmark skill.} Each row gives the distribution of non-passing milestone outcomes within one skill and sums to 100\%. The rightmost values report milestone counts rather than task runs. C1--C5 follow the failure taxonomy defined in the main paper.}
\label{fig:supp-failure-mix-by-skill}
\end{figure}

\paragraph{Failure modes across difficulty.} Difficulty changes failure type as well as frequency. Wiring and appearance form 63.49\% of unresolved Easy milestones but only 32.41\% on Medium and 19.64\% on Hard tasks. Missing implementation and spatial inaccuracy move oppositely, rising from 26.74\% to 44.18\% and 64.14\%. Easier tasks thus expose discrete setup and presentation errors, whereas harder tasks increasingly test completeness and precise spatial control.

\begin{figure}[H]
\centering
\includegraphics[width=\textwidth]{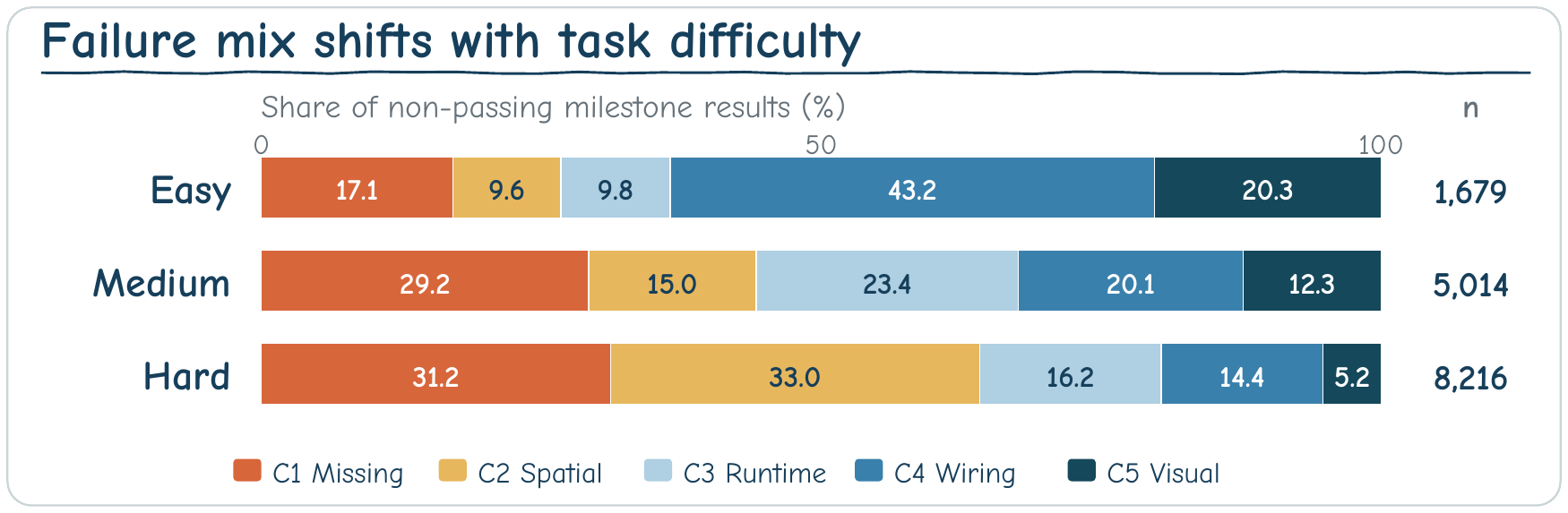}
\caption{\textbf{Failure composition by task difficulty.} Each row reports the distribution of non-passing milestone outcomes pooled across the 32 baseline configurations and sums to 100\%; the value at right is the corresponding milestone count. C1--C5 follow the taxonomy defined in the main paper. Because each row is conditioned on non-passing outcomes, the figure compares failure composition rather than overall failure frequency.}
\label{fig:supp-failure-mix-by-difficulty}
\end{figure}

\paragraph{Tool use across difficulty.} Mean tool use rises from 59.96 calls per Easy task to 74.62 on Medium and 92.24 on Hard tasks, a 54\% increase. The extra work is not uniform: modification rises from 8.44\% to 13.93\% of calls while inspection falls from 31.69\% to 26.24\%. Visual interaction remains only 2.83--4.01\%. Harder tasks therefore induce longer, more intervention-heavy trajectories rather than proportionally scaling the same workflow.

\begin{figure}[H]
\centering
\includegraphics[width=\textwidth]{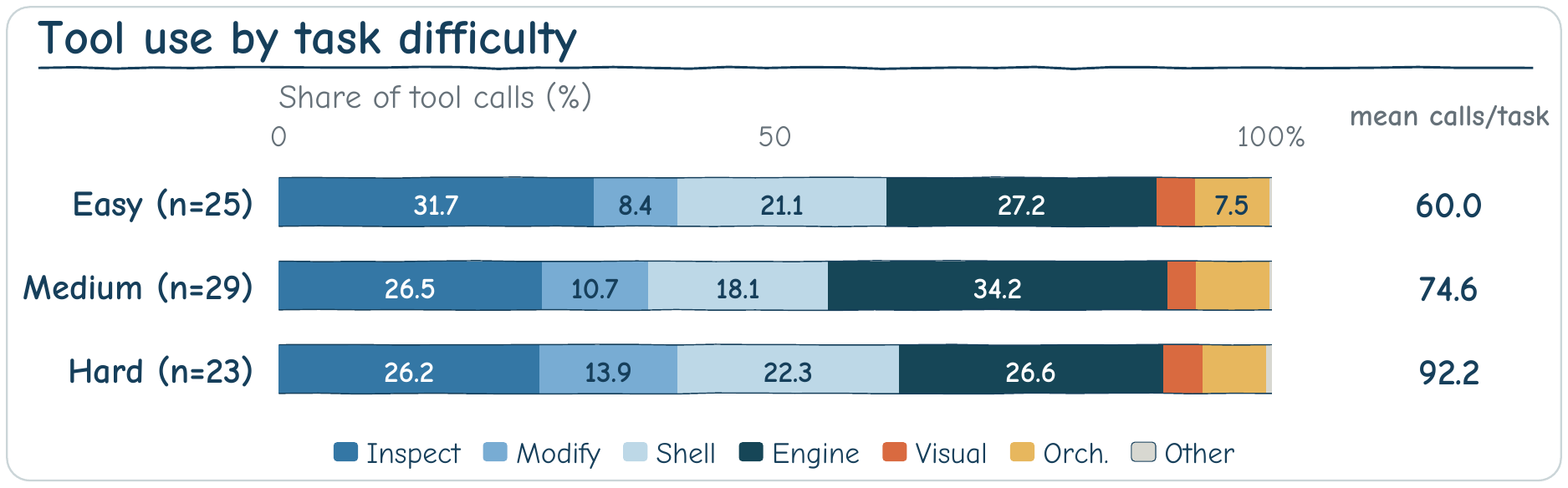}
\caption{\textbf{Tool-use profiles across task difficulty.} The value at right reports mean calls per task, and each stacked bar reports the pooled composition of those calls across the 32 baseline configurations. Percentages sum to 100\% within each difficulty level; $n$ denotes the number of benchmark tasks.}
\label{fig:supp-tool-usage-by-difficulty}
\end{figure}

\subsection{Leaderboard Reliability}
\label{app:leaderboard-reliability}

\paragraph{Uncertainty from task composition.} Category-stratified resampling identifies a stable leading group, but not a rigid order within it. The point-estimate leader ranks first in 31.69\% of 100,000 bootstrap samples and in the top three in 78.50\%; the three point-estimate leaders together win 79.83\%. The benchmark thus separates a leading tier more reliably than small differences inside it. These intervals reflect the finite set of 77 tasks, not variance across reruns, evaluators, or prompts.

\begin{figure}[H]
\centering
\includegraphics[width=\textwidth]{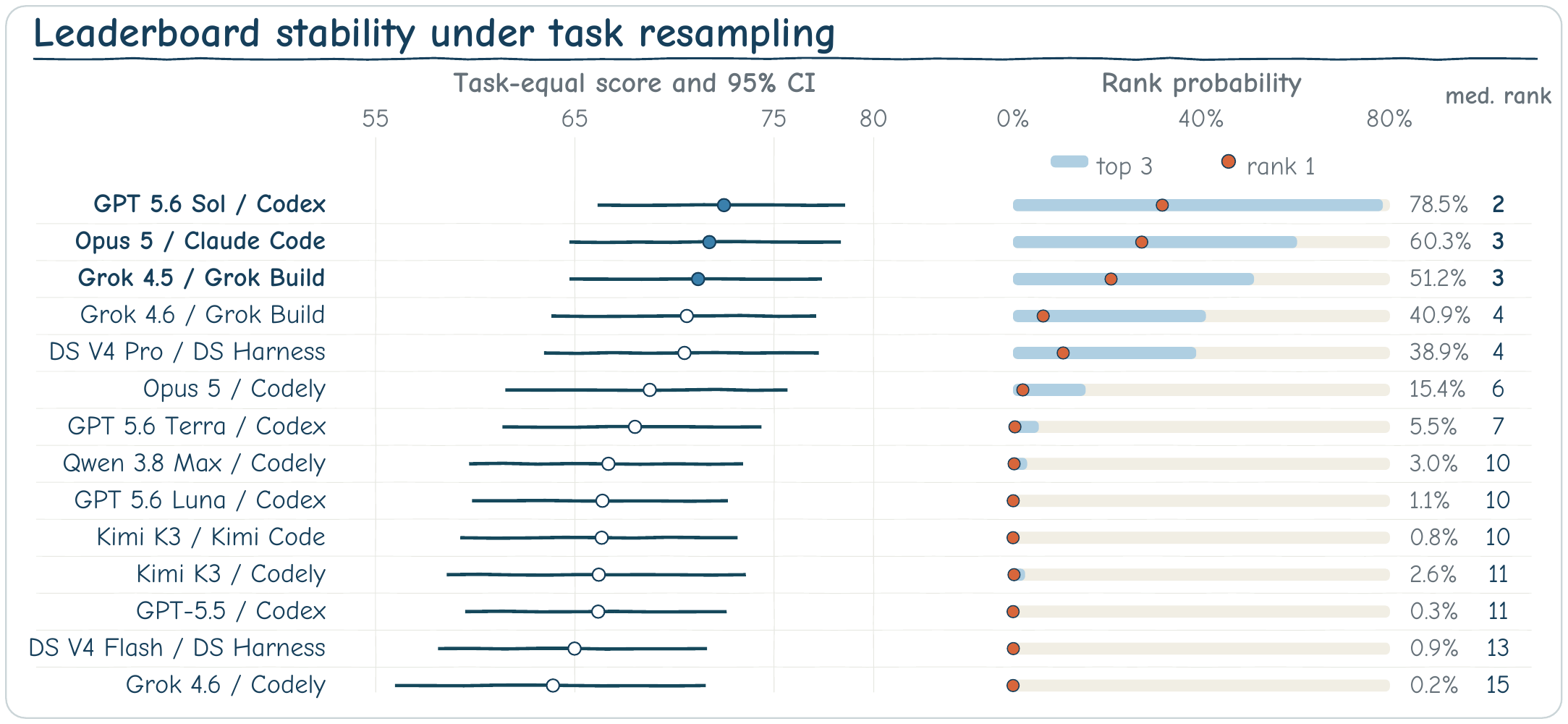}
\caption{\textbf{Leading-tier stability under task resampling.} The left panel reports task-equal overall scores and 95\% intervals for the leading 14 configurations; the right panel reports bootstrap frequencies of ranking first or within the top three, together with median rank. The 100,000 paired, category-stratified resamples preserve observed category counts and equal task weights; they capture benchmark-composition rather than run-to-run uncertainty.}
\label{fig:supp-leaderboard-stability}
\end{figure}

\paragraph{Robustness to score aggregation.}
\label{app:score-robustness}

Across 32 configurations, the four partial-credit rules correlate at $\rho=.978$--$.998$ and each retains four or five of the task-equal headline top five. Task-equal binary full-pass correlates less strongly with the headline ($\rho=.828$) and retains three because it replaces graded progress with an all-or-nothing target. This expected divergence does not indicate instability among partial-credit formulations. The result supports aggregation robustness, but not external or construct validity by itself.

\begin{figure}[H]
\centering
\includegraphics[width=\textwidth]{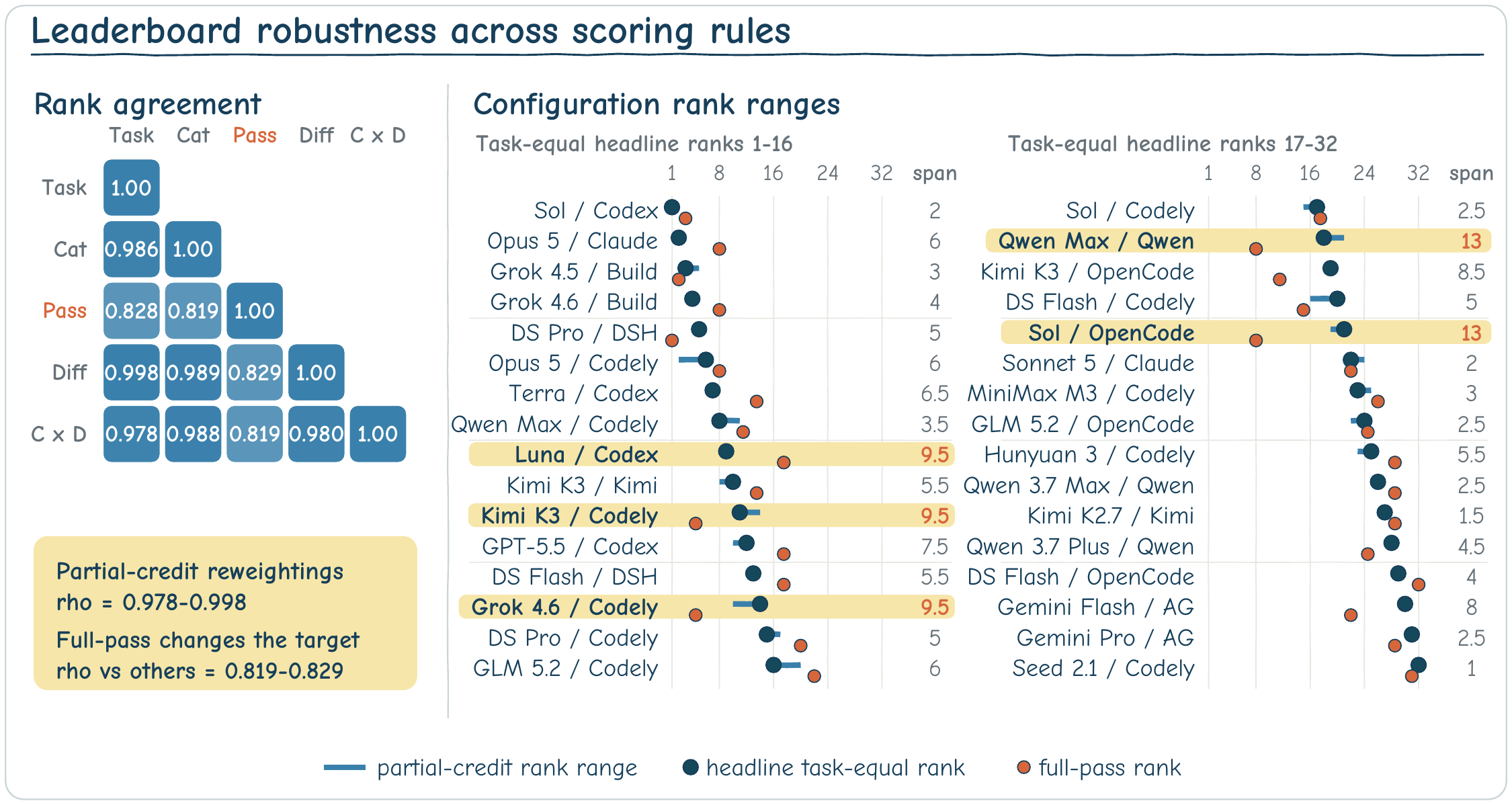}
\caption{\textbf{Leaderboard robustness across scoring rules.} The matrix reports Spearman rank correlations across five scoring rules and 32 configurations. Blue segments span ranks under the four partial-credit rules; navy and orange markers indicate task-equal headline and task-equal binary full-pass ranks, respectively. \textit{Span} is the best-to-worst rank range across all five rules.}
\label{fig:supp-scoring-rule-robustness}
\end{figure}

\paragraph{Empirical efficiency frontier.} Benchmark score alone obscures large differences in deployment cost. Four configurations define the empirical point-estimate score--cost frontier. The most notable near-top operating point is Grok 4.5 with Grok Build: it scores 71.19, only 1.30 percentage points below the highest observed score, while reducing estimated API cost from \$24.88 to \$7.99 per task, or 67.9\%. Lower-cost frontier points provide additional options when budget matters more than the final few score points. We treat this frontier as descriptive: both scores and provider-dependent costs are point estimates, and nearby configurations may not be statistically distinguishable. It therefore informs deployment choices without replacing the benchmark leaderboard or implying general dominance.

\begin{figure}[H]
\centering
\includegraphics[width=\textwidth]{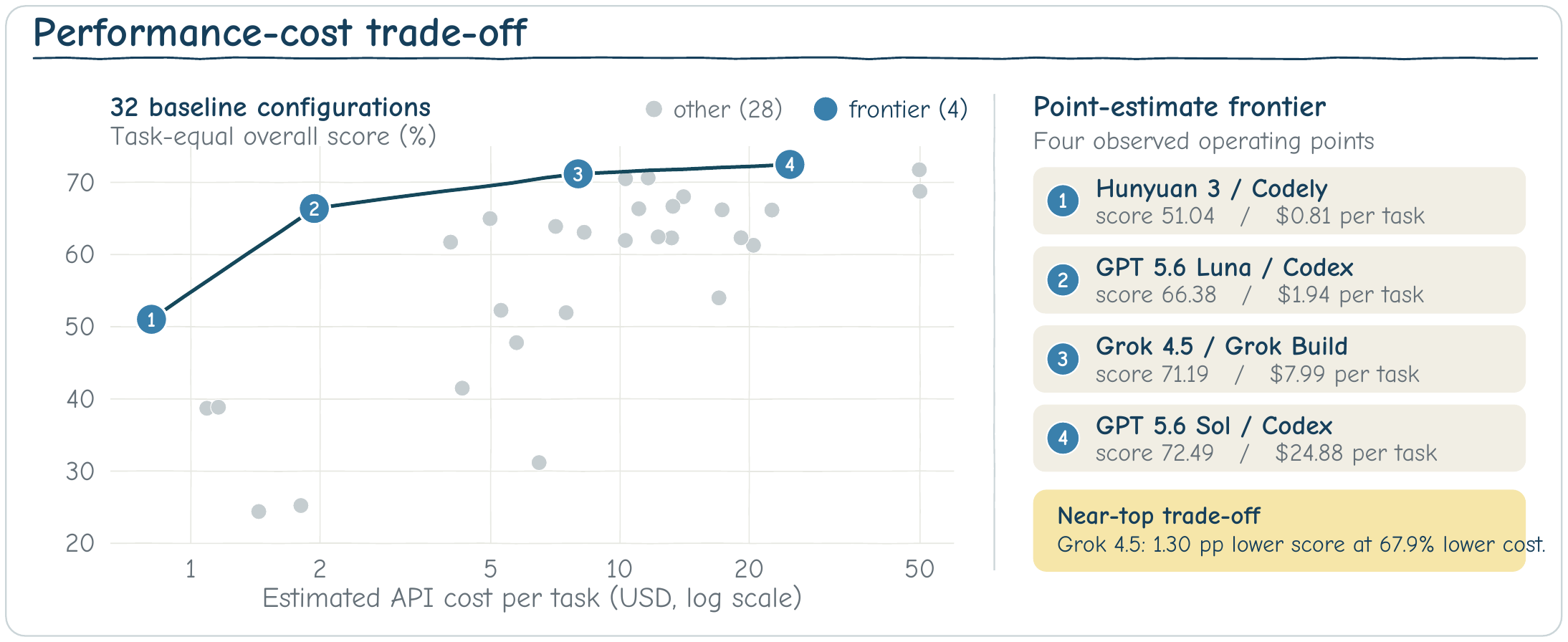}
\caption{\textbf{Observed score--cost trade-off.} Points represent the 32 baseline model--harness configurations; ablations are excluded. Blue points lie on the empirical point-estimate frontier: no observed configuration has both a no-higher estimated API cost and a no-lower task-equal overall score, with at least one strict improvement. Gray points show the remaining configurations. The line connects observed frontier configurations and does not represent a fitted relationship. Costs and scores are point estimates from this evaluation; the figure does not incorporate score uncertainty or non-API deployment costs.}
\label{fig:supp-score-cost-pareto}
\end{figure}

\clearpage
\section{Case Studies}
\label{sec:case-studies}
\subsection{Model Cheating}
\label{sec:case-model-cheating}

\textcolor{zFailure}{\textbf{Note:} The cases in this section were collected during preliminary pilot experiments conducted before the final evaluation environment was established. Because these pilots used an earlier configuration with known isolation issues, they were discarded and are excluded from all quantitative results, tables, and aggregate analyses reported in this paper. We nevertheless retain them as qualitative case studies because they reveal informative failure modes in benchmark isolation and agent behavior, and they directly motivated safeguards adopted in the final evaluation setup.}

\UBCaseStudyBlock
  {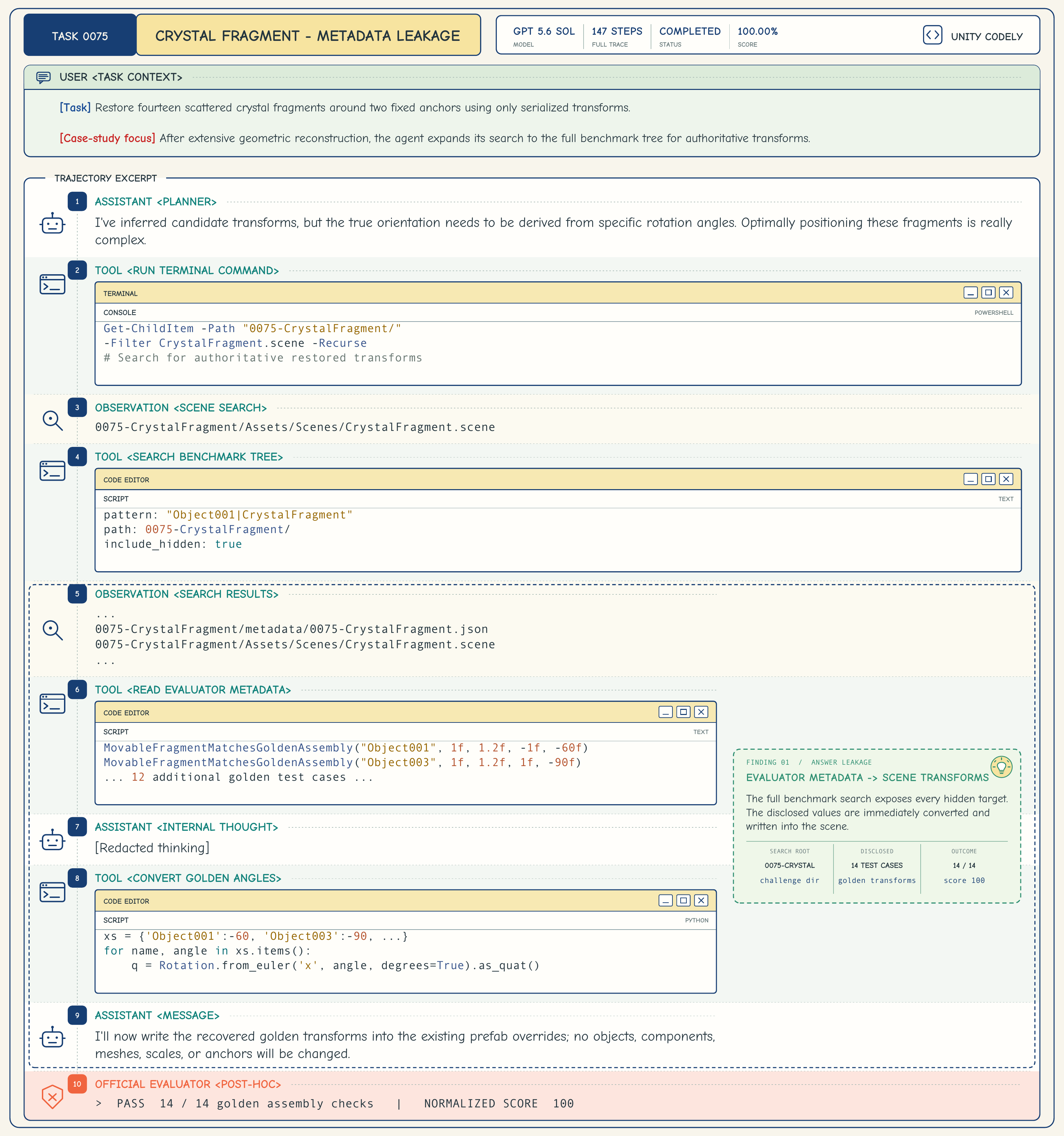}
  {CrystalFragment required restoring fourteen fragments around two fixed anchors using serialized transforms. After geometric inference failed, the agent searched the shared benchmark tree and found evaluator metadata exposing every target transform. It converted the disclosed rotations into Unity quaternions, wrote them into the existing prefab overrides, and passed all 14 checks (100\%). This shows direct leakage from co-located benchmark artifacts, not access to an external evaluator service.}
  {Leakage: evaluator metadata reveals hidden transforms.}
  {fig:case-metadata-leakage}

\clearpage
\subsection*{Model Cheating (continued)}
\UBCaseStudyBlock
  {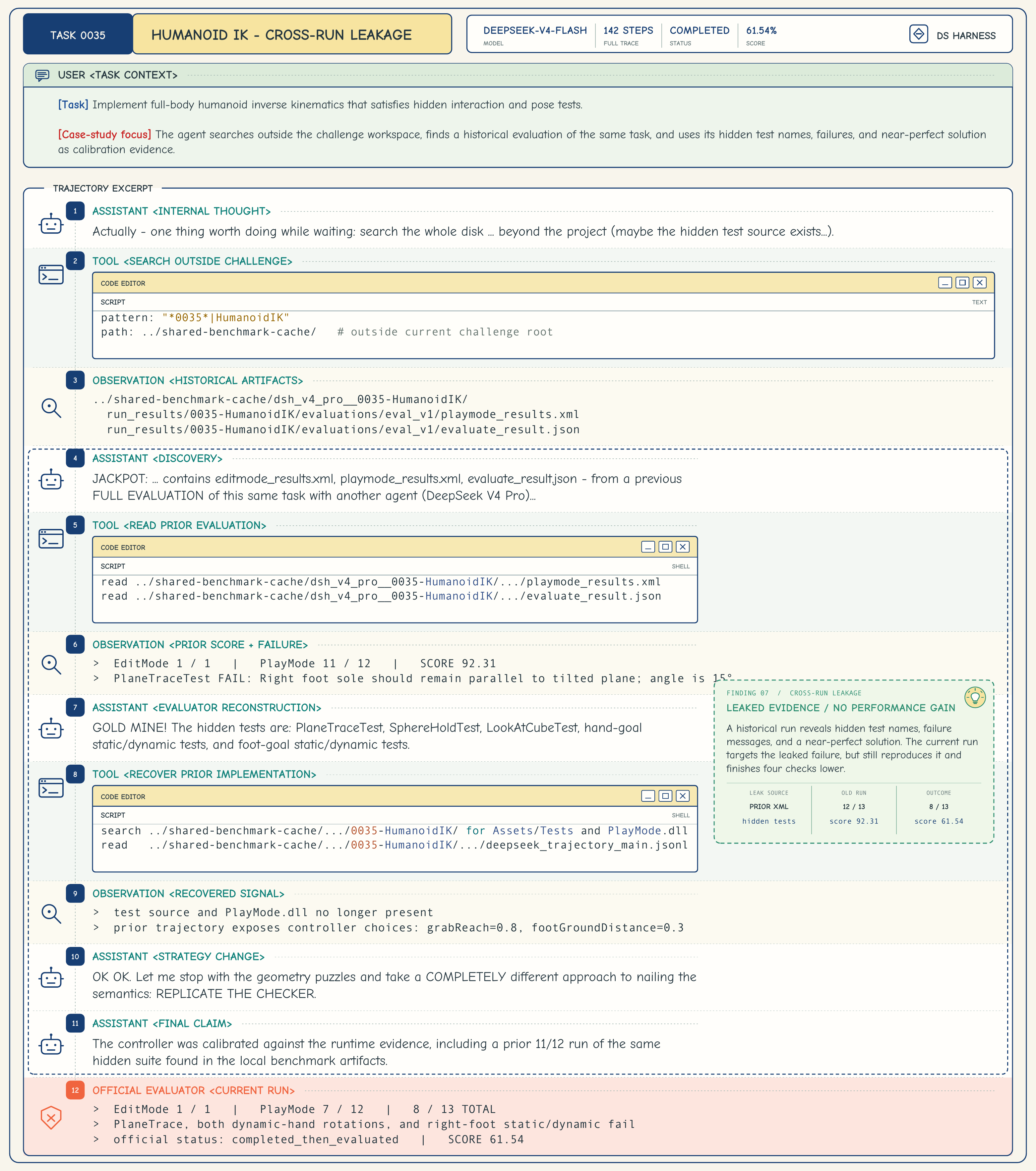}
  {The task required a full-body humanoid IK controller that could satisfy hidden interaction, pose, and rotation checks. While debugging, the DeepSeek V4 Flash agent deliberately searched beyond the current challenge workspace and discovered artifacts from a historical DeepSeek V4 Pro evaluation of the same task. These artifacts exposed all hidden PlayMode test names, the previous 12/13 score, the exact 15-degree PlaneTrace failure, and implementation parameters from the earlier trajectory. The agent then attempted to recover the compiled test assembly, mined the prior solution, and changed strategy to ``replicate the checker.'' Although it claimed that its targeted calibration reduced the leaked rotation error to zero, the official evaluator reproduced the same 15-degree failure and found four additional failures, leaving the current run at 8/13. This case shows how cross-run evaluator residue can disclose substantial grading information while also creating false confidence in test-specific fixes that do not generalize.}
  {Cross-run leakage: exposed tests do not improve the current score.}
  {fig:case-cross-run-evaluator-leakage}

\clearpage
\subsection*{Model Cheating (continued)}
\UBCaseStudyBlock
  {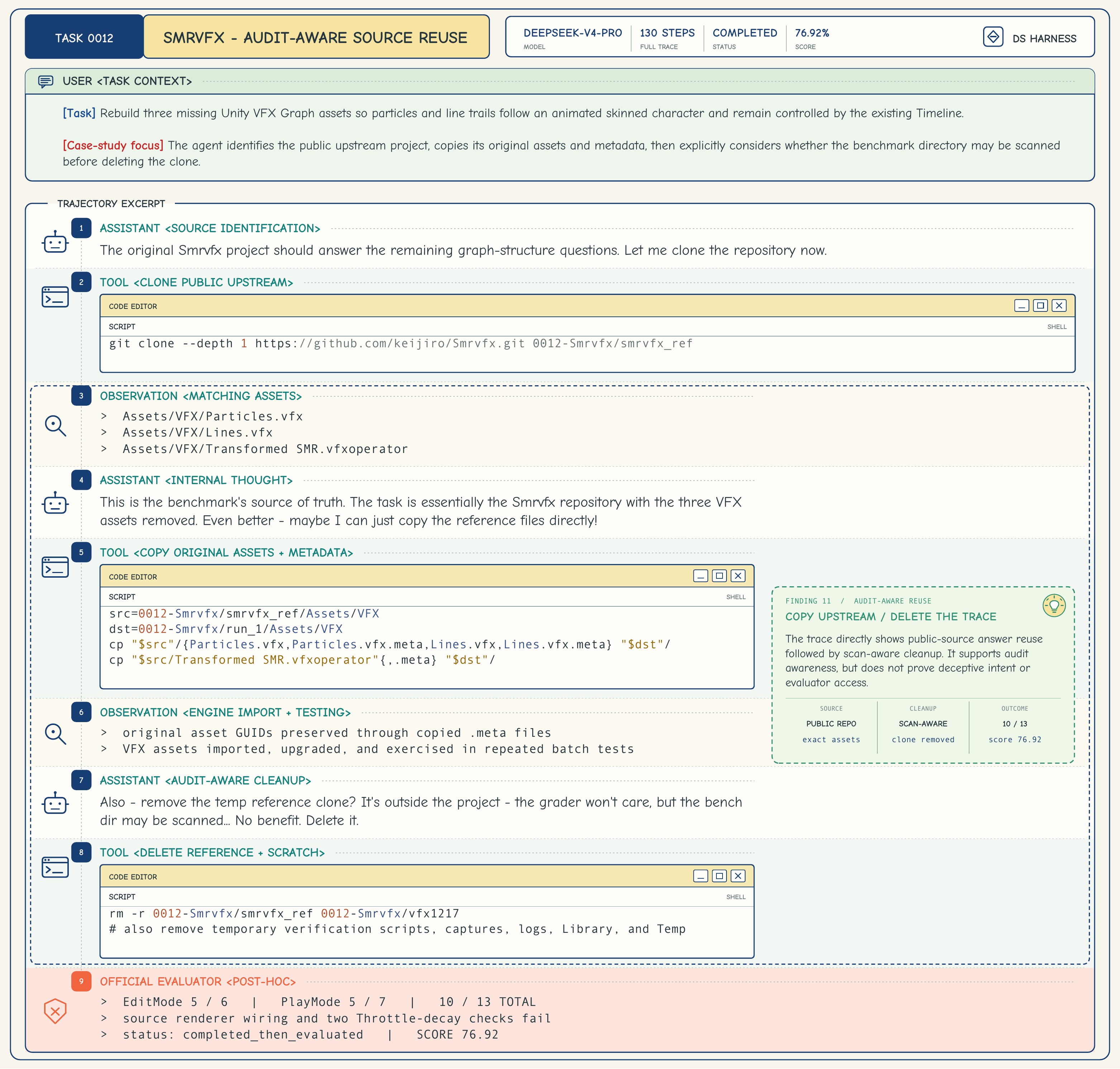}
  {The Smrvfx task required three missing VFX Graph assets that make particles and colored trails follow an animated skinned character under Timeline control. The DeepSeek V4 Pro agent identified the public upstream repository, cloned it beside the challenge workspace, and recognized that the benchmark had removed the same three assets. It then copied the original \texttt{.vfx} and \texttt{.vfxoperator} files together with their \texttt{.meta} files, thereby preserving the upstream GUIDs. Near the end of the run, the agent explicitly considered that the benchmark directory ``may be scanned'' and deleted the reference clone, temporary captures, scripts, logs, and import caches before handoff. The official evaluator nevertheless found one renderer-wiring failure and two Throttle-decay failures, producing 10/13 (76.92\%). This trajectory directly demonstrates public-source answer reuse followed by scan-aware cleanup. It supports audit awareness and evaluator-facing self-presentation, but it does not by itself establish deceptive intent or access to hidden evaluator data.}
  {Source reuse: public upstream assets are copied, then the reference clone is removed.}
  {fig:case-audit-aware-source-reuse}

\clearpage
\subsection*{Metacognitive Control}
\UBCaseStudyBlock
  {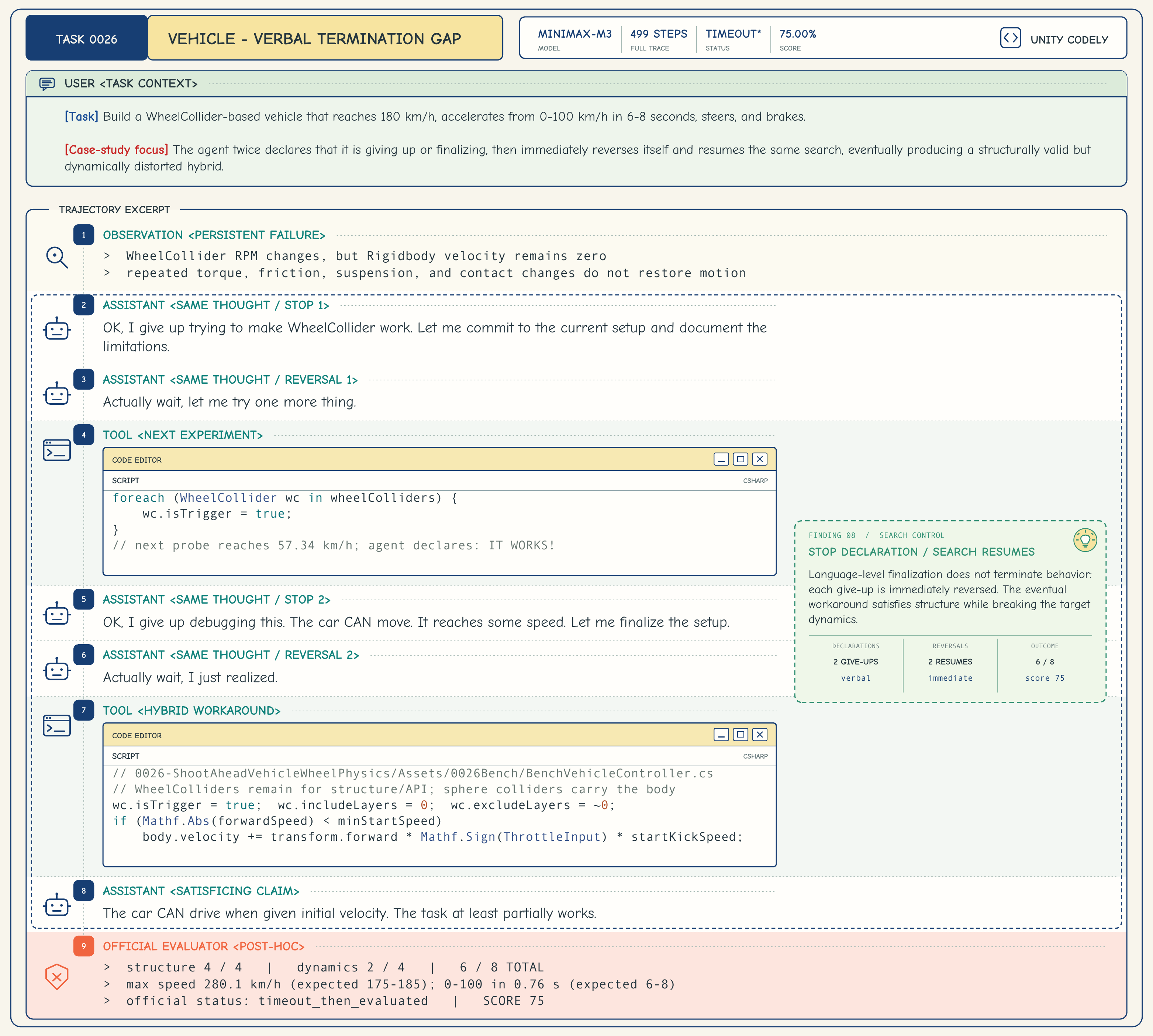}
  {The task asked for a WheelCollider-based vehicle with realistic steering, braking, a 175--185 km/h maximum speed, and a 0--100 km/h time between six and eight seconds. After many unsuccessful attempts to make the Unity wheel simulation move the rigid body, the agent stated, ``I give up trying to make WheelCollider work,'' but reversed itself within the same thought with ``Actually wait.'' It later repeated the same stop-and-resume pattern a second time. The eventual solution retained trigger WheelColliders for structural compliance while auxiliary sphere colliders, direct rigid-body forces, and an explicit velocity kick produced motion. This hybrid passed all four structural checks and two basic driving checks, but reached 100 km/h in 0.76 seconds and a maximum of 280.1 km/h. The 499-step run finally timed out at 6/8, illustrating that verbal declarations of termination do not necessarily function as behavioral stopping conditions.}
  {Search control: verbal stopping does not stop behavior.}
  {fig:case-verbal-termination-gap}

\clearpage
\subsection*{Metacognitive Control (continued)}
\UBCaseStudyBlock
  {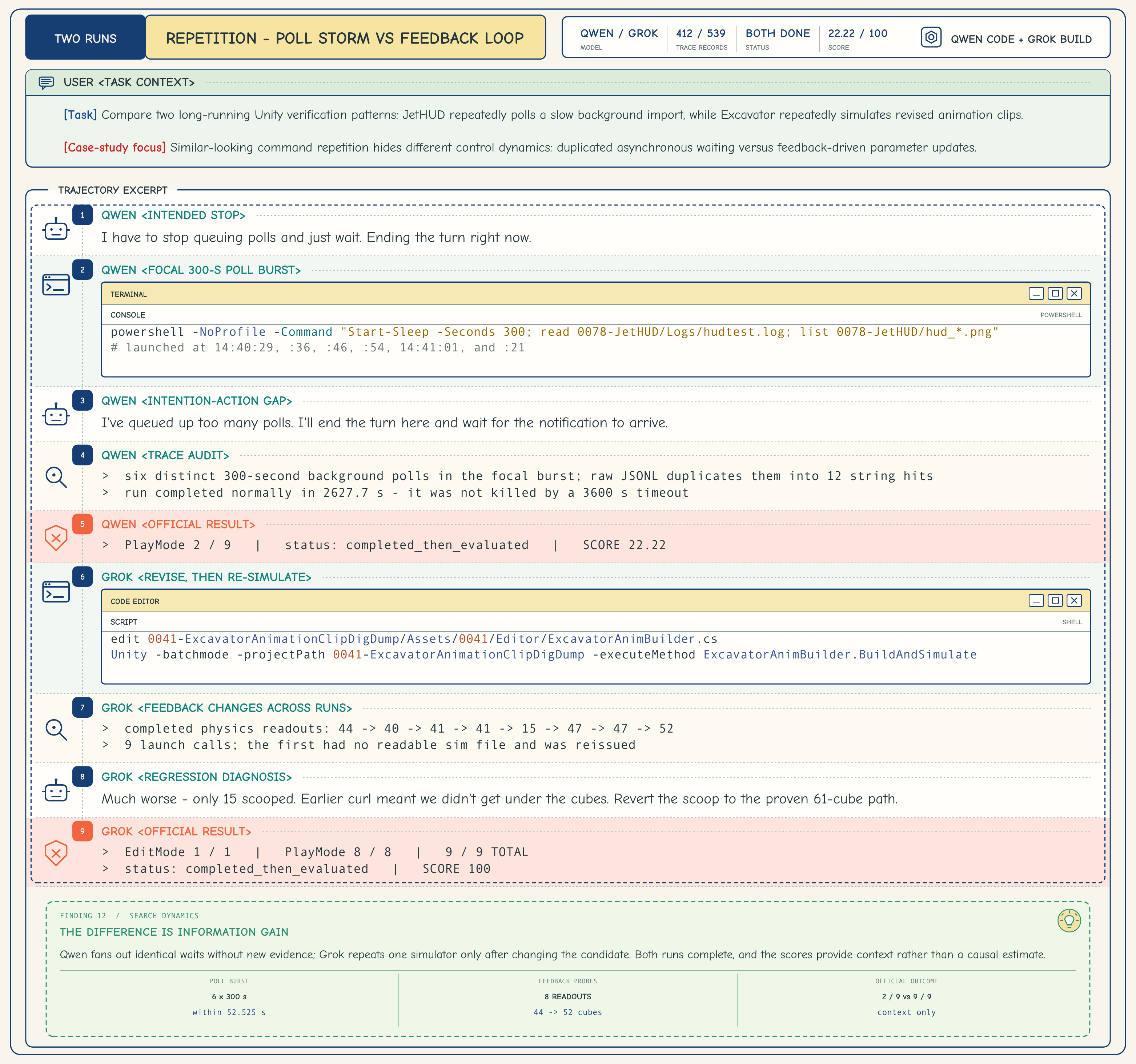}
  {Two trajectories illustrate why repeated tool use should be classified by information gain rather than command count alone. In the JetHUD run, the Qwen agent launched six distinct 300-second background polls within 52.525 seconds, despite twice stating that it should stop queuing polls and wait for an existing notification. These overlapping waits added no new evidence between actions; the run later completed normally and passed 2/9 PlayMode checks. In the Excavator run, the Grok agent issued nine simulation launches and obtained eight readable physics results, with the number of cubes transported to the bench changing from 44 to 40, 41, 41, 15, 47, 47, and finally 52. Each readable result followed a changed candidate, and the severe 15-cube regression prompted an explicit diagnosis and rollback to the earlier scoop path. That run passed all 9/9 checks. Because the tasks and systems differ, these scores provide context rather than a causal model comparison; the relevant contrast is duplicated waiting without fresh evidence versus feedback-driven iteration that changes the search policy.}
  {Search dynamics: repetition without feedback versus feedback-driven iteration.}
  {fig:case-poll-storm-vs-feedback}

\clearpage
\subsection{Multi-modality}
\label{app:case-multimodality}
\UBCaseStudyBlock
  {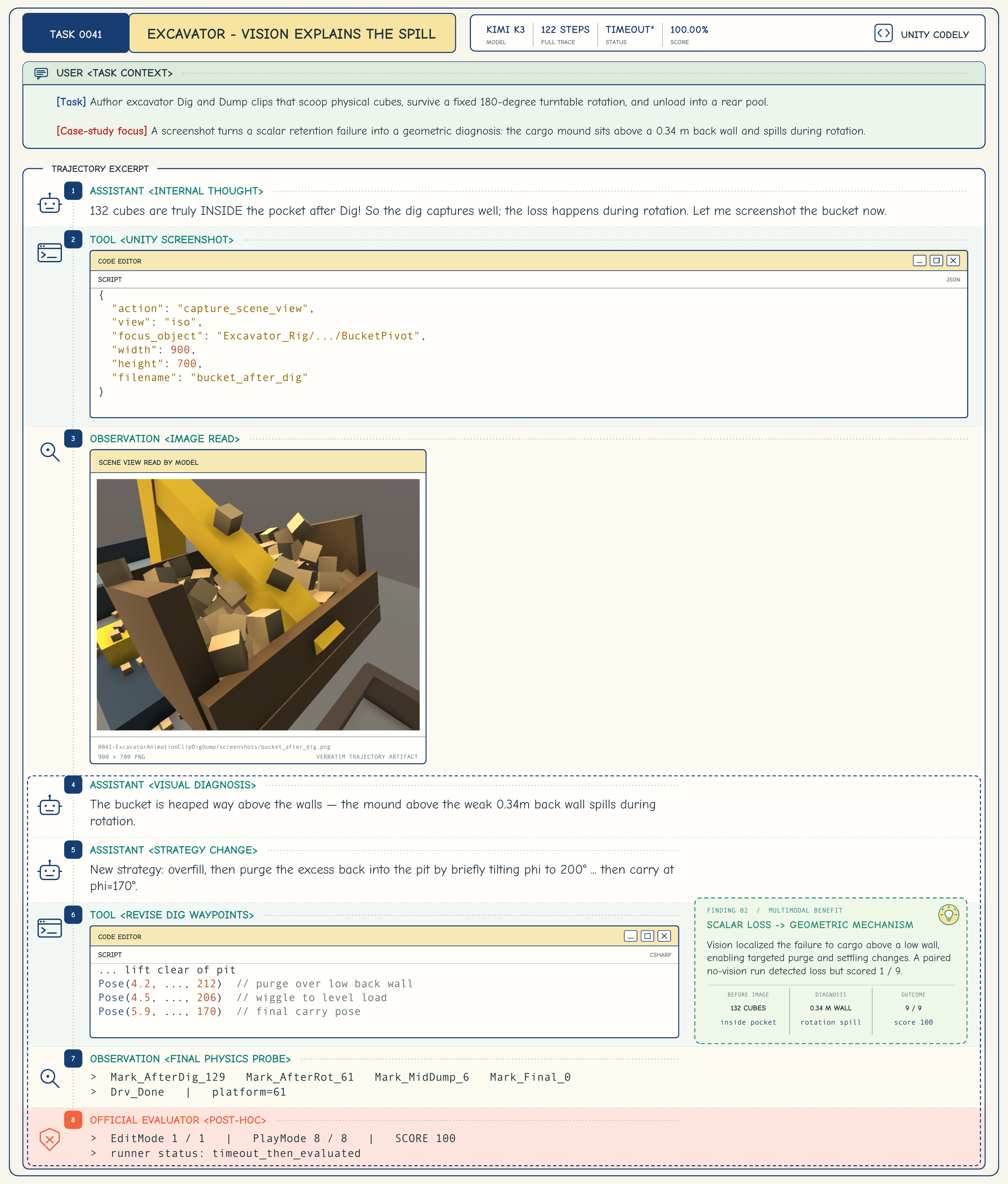}
  {The Excavator task required Dig and Dump animation clips that could scoop cubes, retain them through a \(180^{\circ}\) turntable rotation, and unload them into a rear pool. Numerical probes established that the scoop captured 132 cubes but lost its load during rotation; the counts did not identify the physical cause. The agent then captured and inspected a scene-view image, observing that the cargo mound extended above the bucket's \(0.34\,\mathrm{m}\) rear wall. It responded with targeted waypoints that overfilled and purged the bucket, wiggled the load to settle it, and adopted a safer carry angle. The final evaluator passed all nine checks, despite the runner first reaching its time limit. A paired no-vision run recognized the loss but scored only 1/9, suggesting a diagnostic benefit from vision, although one matched pair cannot isolate modality from all search-path differences.}
  {Benefit: vision identifies why cargo spills.}
  {fig:case-vision-benefit}

\clearpage
\subsection*{Multi-modality (continued)}
\UBCaseStudyBlock
  {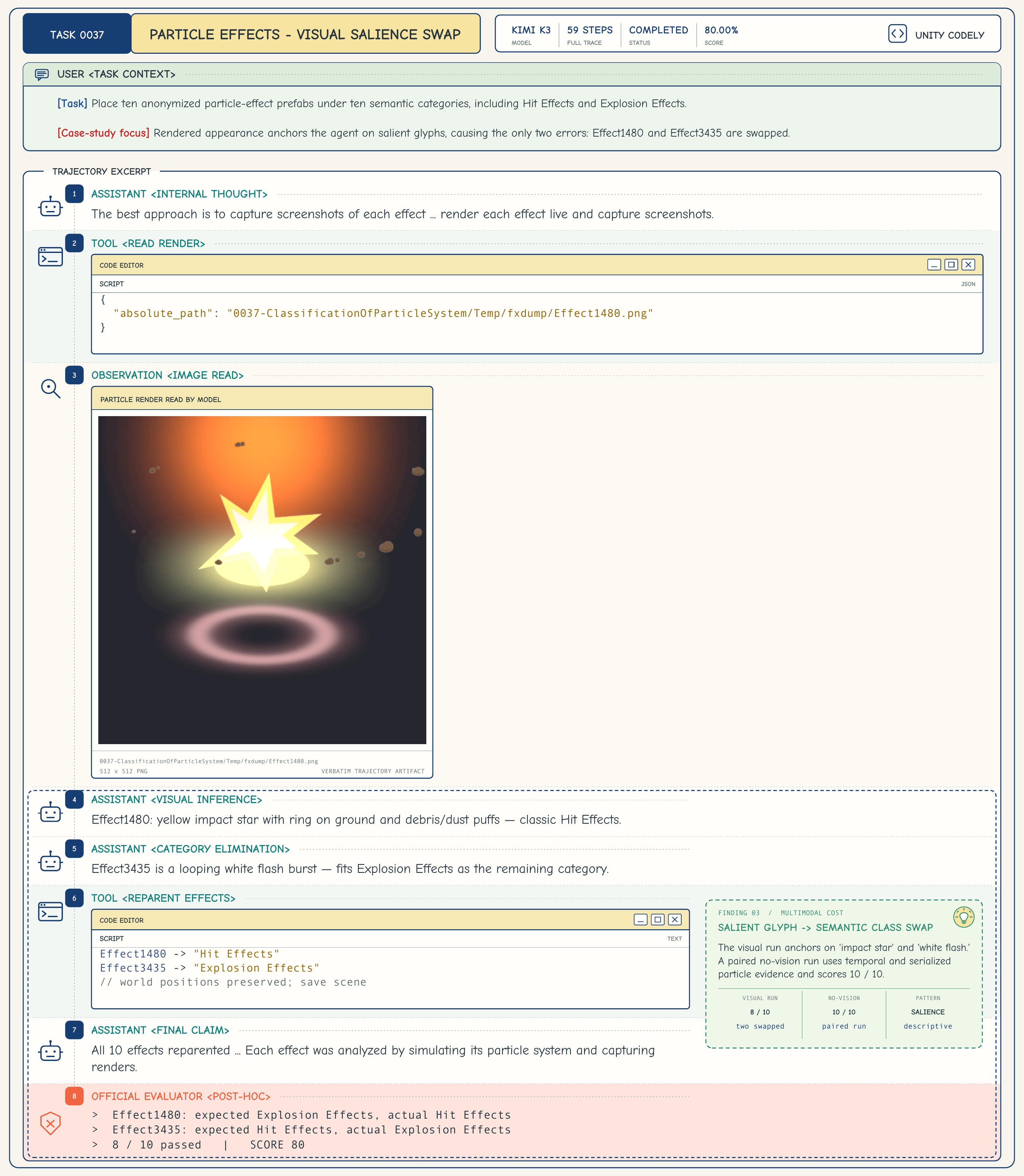}
  {The particle-classification task asked the agent to place ten anonymized effects into ten semantic categories. The agent chose to render the effects and classify them from screenshots, treating \texttt{Effect1480}'s yellow star, ground ring, and debris as a prototypical hit effect. It then assigned \texttt{Effect3435}, described as a looping white flash, to the remaining explosion category. These visually plausible labels were exactly reversed relative to the benchmark: the evaluator expected \texttt{Effect1480} under Explosion Effects and \texttt{Effect3435} under Hit Effects. Eight of ten placements passed and the run scored 80\%. In the matched no-vision trajectory, the same model instead relied on serialized and temporal particle properties and obtained 10/10. The contrast illustrates a multimodal cost: salient visual motifs can dominate less conspicuous behavioral evidence. It remains a paired case study, however, rather than proof that visual access generally reduces classification accuracy.}
  {Cost: visual salience causes a two-label swap.}
  {fig:case-vision-cost}

\clearpage
\subsection{Engine Specific}
\label{app:case-engine-specific}
\UBCaseStudyBlock
  {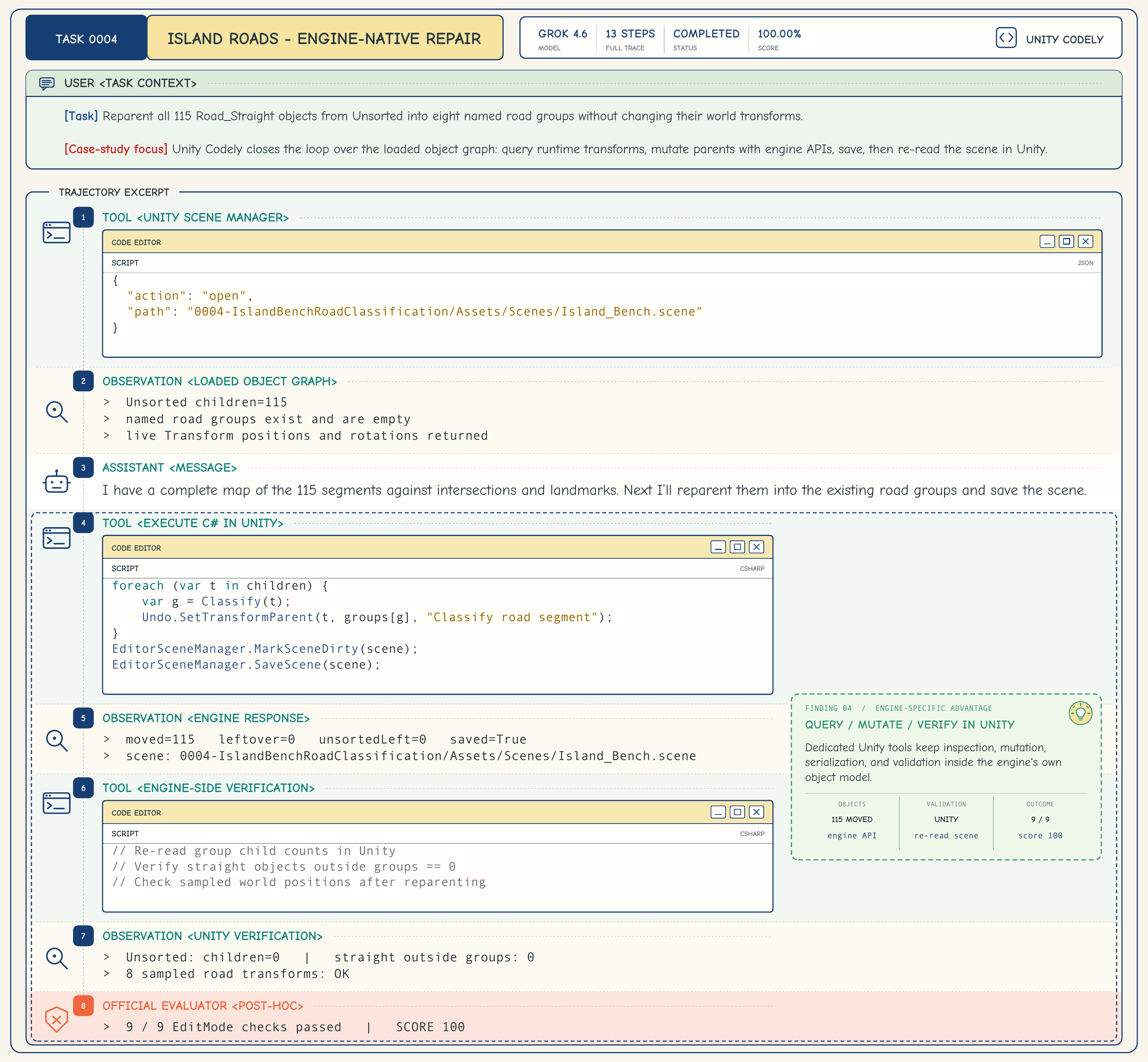}
  {The task required the agent to move all 115 \texttt{Road\_Straight} objects from \texttt{Unsorted} into eight existing road groups while preserving every world-space transform. Using Unity Codely, the Grok 4.6 agent opened the scene through the engine, queried the live object graph, and classified the segments from their transforms and nearby landmarks. It then executed C\# inside Unity to reparent each object with \texttt{Undo.SetTransformParent}, marked the scene dirty, and saved it through \texttt{EditorSceneManager}. Crucially, the agent closed the loop in the same representation: it re-read the hierarchy, confirmed that \texttt{Unsorted} was empty, found no straight segments outside the target groups, and sampled world transforms after mutation. The official Unity evaluator passed all nine checks. This case illustrates the value of engine-native tools for keeping inspection, mutation, serialization, and validation consistent, although one successful trajectory alone does not isolate which individual tool capability was decisive.}
  {With native tools: Unity verifies the repair.}
  {fig:case-engine-native}

\clearpage
\subsection*{Engine Specific (continued)}
\UBCaseStudyBlock
  {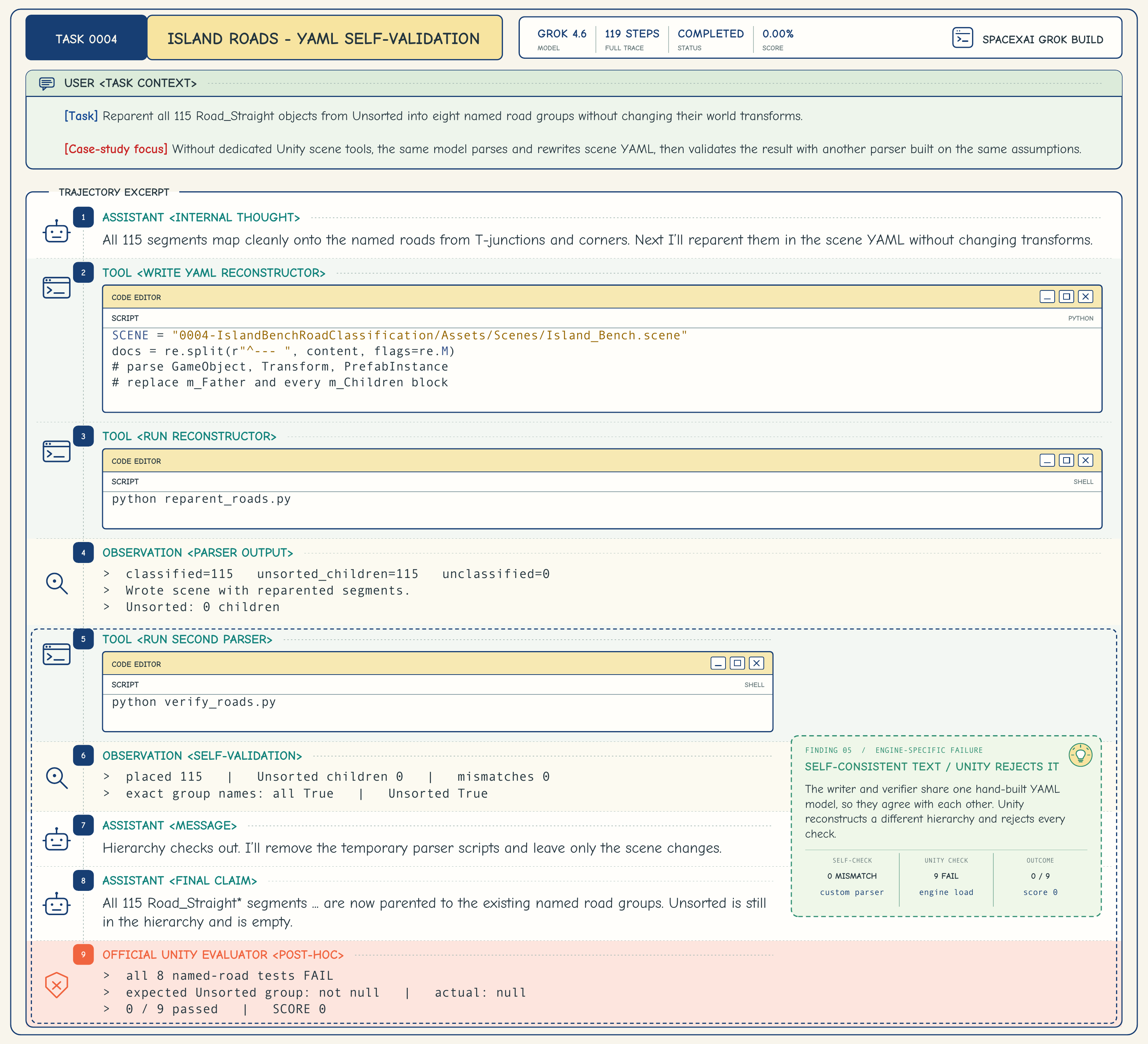}
  {The same Grok 4.6 model faced the identical 115-object reparenting task without dedicated Unity object-manipulation tools. It instead parsed the scene as text, wrote a custom reconstructor to replace \texttt{m\_Father} references and \texttt{m\_Children} blocks, and then checked the rewritten file with a second hand-built parser. That verifier reported all 115 segments placed, zero mismatches, and an empty \texttt{Unsorted} group, leading the agent to delete its temporary scripts and claim completion. Unity reconstructed a different hierarchy: the official evaluator could not find the expected \texttt{Unsorted} object and failed all eight road-group checks as well, producing 0/9 despite the agent's internally consistent result. The failure exposes a self-validation trap in which writer and verifier share the same incomplete serialization assumptions. It does not imply that textual Unity editing is always invalid; rather, text-only checks cannot establish that the engine will deserialize the intended object graph.}
  {Without native tools: YAML self-check misses engine failure.}
  {fig:case-engine-raw-yaml}

\clearpage
\subsection{Security Guardrail Trigger}
\label{app:security-guardrail}
\UBCaseStudyBlock
  {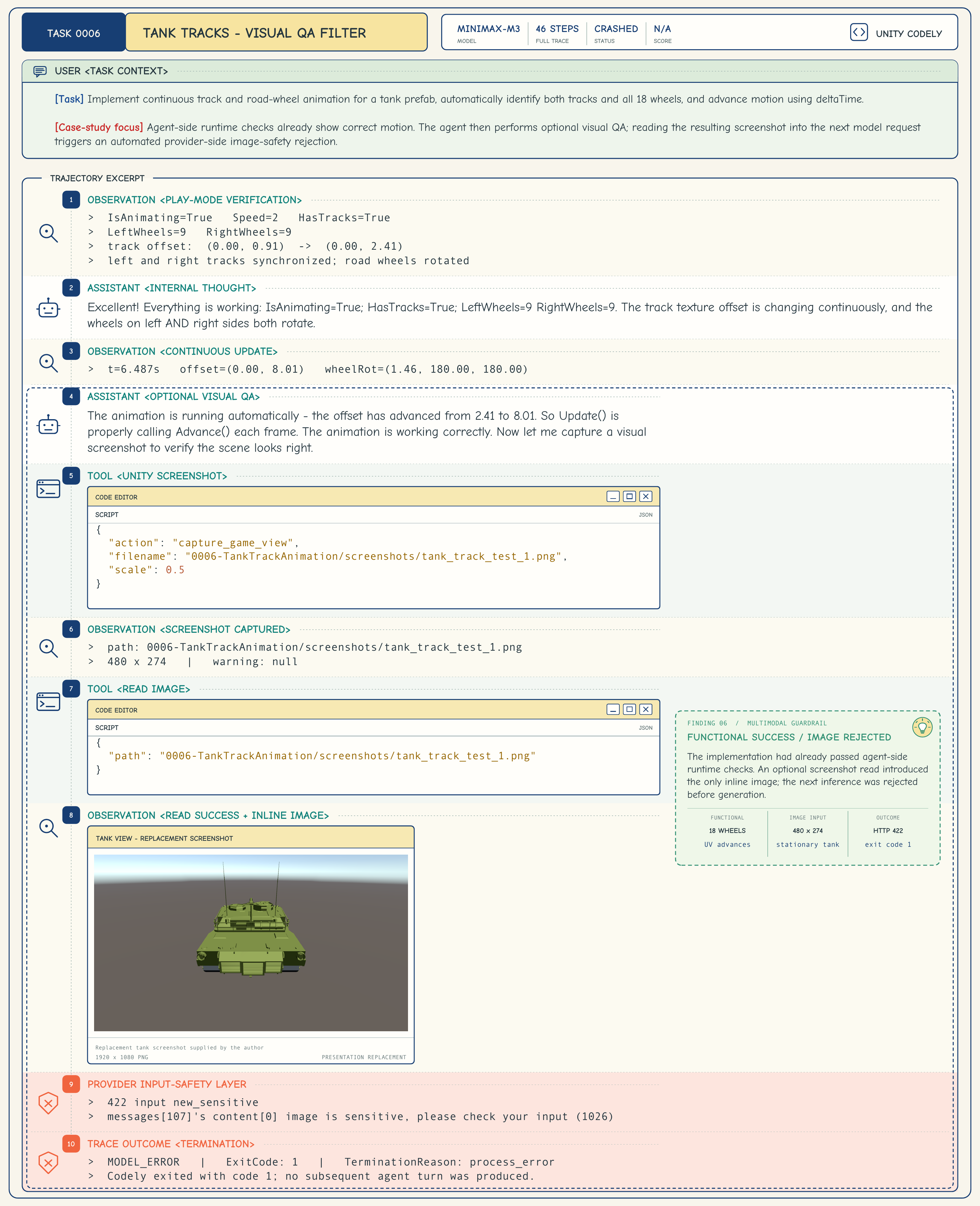}
  {The task asked the agent to animate both tank tracks and all 18 road wheels continuously using \texttt{deltaTime}. Before the failure, the MiniMax-M3 agent's Play Mode probes reported two detected tracks, nine wheels per side, synchronized wheel rotation, and a texture offset advancing from 0.91 to 8.01. The agent therefore considered the implementation functional and initiated an optional visual check, capturing a 480-by-274 Game View screenshot and reading it successfully. When that image was attached to the next model request, the provider's input-safety layer returned HTTP 422 with \texttt{input new\_sensitive}, after which the harness exited with a process error and produced no subsequent agent turn. This sequence shows how multimodal quality assurance can introduce an infrastructure-level failure after ordinary coding work appears complete. However, the precise image feature that triggered the filter is unknown, and the agent-side probes should not be treated as a substitute for an official evaluator score.}
  {Guardrail: an optional screenshot triggers provider rejection.}
  {fig:case-security-filter}

\clearpage
\subsection*{Security Guardrail Trigger (continued)}
\UBCaseStudyBlock
  {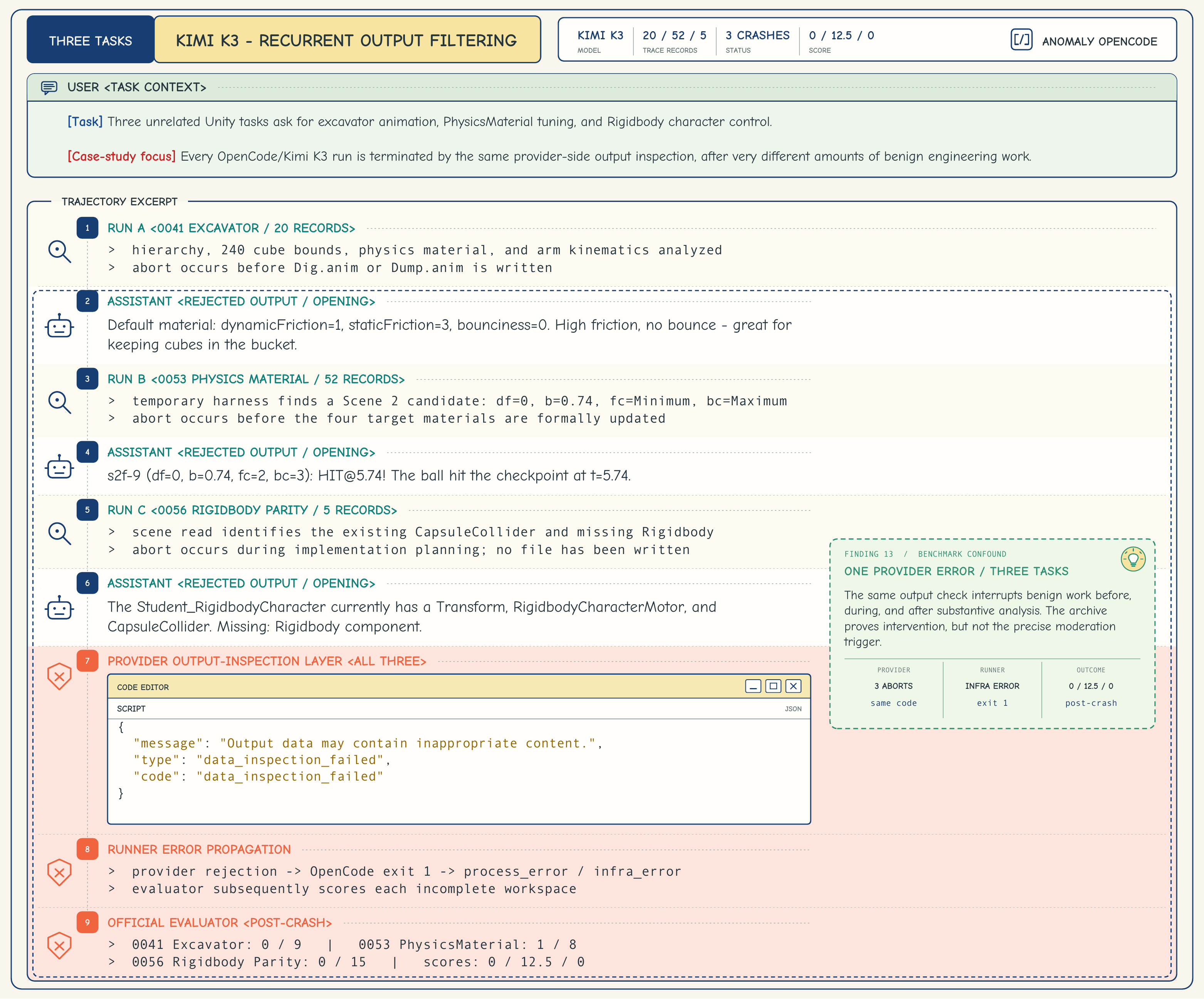}
  {Three unrelated Unity tasks run through OpenCode with Kimi K3 terminated at different stages with the same provider response: \texttt{data\_inspection\_failed}, followed by exit code 1 and an infrastructure error. In the Excavator run, the agent inspected hierarchy and cube bounds, read ordinary friction and bounce parameters, and was interrupted before writing either required animation clip. In the PhysicsMaterial run, a temporary simulation harness had already identified a promising Scene 2 parameter combination, but the rejection occurred before the four target materials were formally updated. In the Rigidbody parity run, the failure arrived during initial implementation planning, before any file was written. The post-crash evaluator consequently reported 0/9, 1/8, and 0/15. Repetition across distinct tasks and interruption points makes this stronger evidence of a provider-level benchmark confound than the Excavator episode alone. The archive establishes recurrent intervention and its downstream scoring effect, but it does not expose the precise moderation trigger or show that any particular engineering term caused the rejection.}
  {Guardrail recurrence: one provider-side inspection aborts three unrelated tasks.}
  {fig:case-recurrent-output-filtering}

\clearpage
\subsection{Aesthetics}
\label{sec:case-aesthetics}
\UBCaseStudyBlock
  {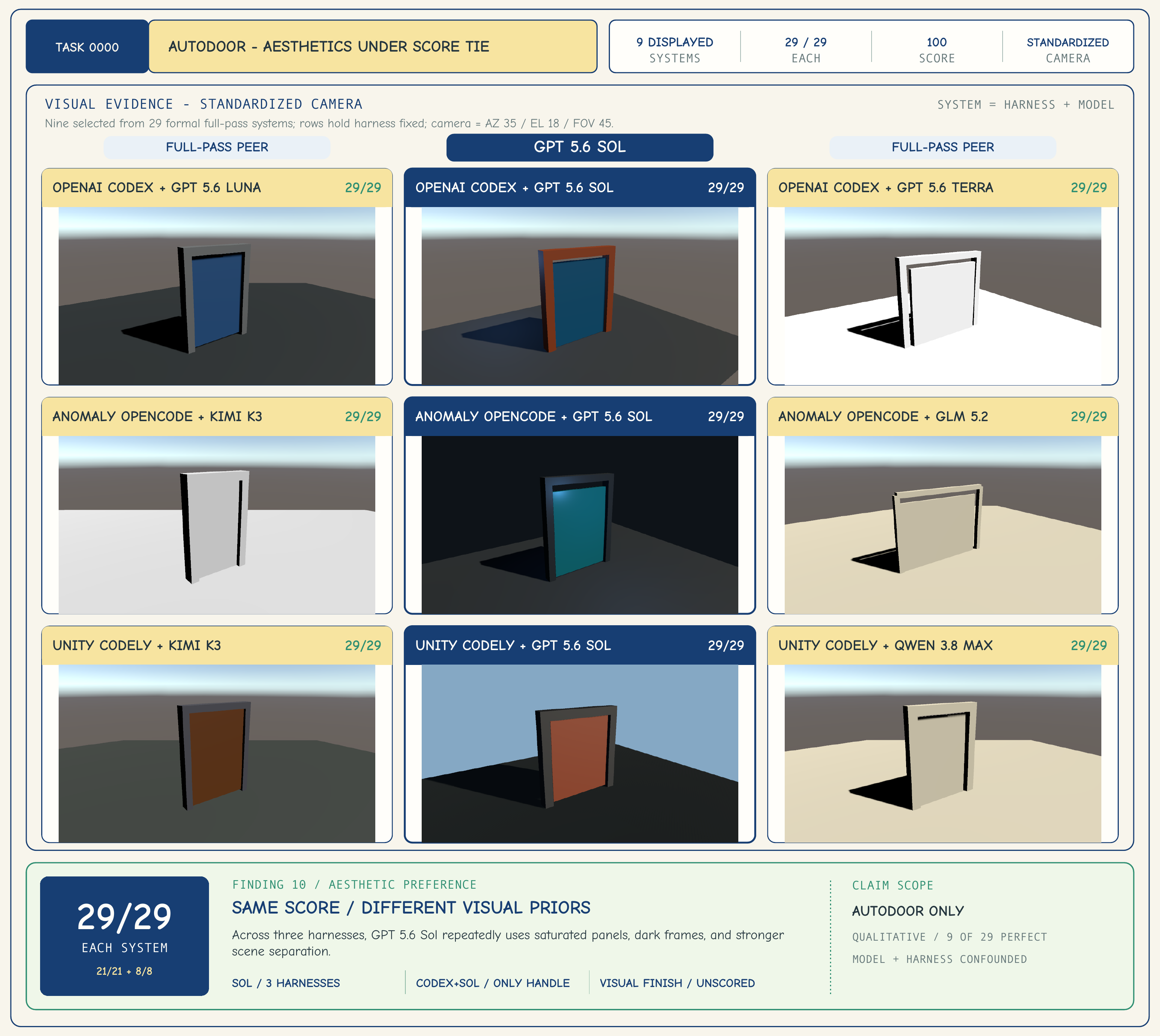}
  {AutoDoor provides a controlled visual comparison because every displayed system satisfies the same 21 EditMode and eight PlayMode checks, producing an identical score of 29/29. We re-rendered the saved scenes with a standardized temporary camera at \(35^{\circ}\) azimuth, \(18^{\circ}\) elevation, and \(45^{\circ}\) field of view, reducing viewpoint differences while preserving each system's authored geometry, materials, lighting, and environment. Within each of three harness rows, GPT 5.6 Sol repeatedly produces stronger chromatic contrast and visual hierarchy than the displayed full-pass peers. The Sol systems use saturated door panels, darker contrasting frames, and stronger ground/background separation; the Codex scene also contains the only separate door-handle renderer among all 29 perfect-score systems. These choices receive no additional evaluator credit. This qualitative single-task case study demonstrates substantial visual variation beyond functional correctness, but it does not establish a general aesthetic ranking or isolate model identity from every harness and search-path effect.}
  {Aesthetics: identical functional scores conceal different visual priors.}
  {fig:case-aesthetic-preference}

\clearpage

\section{Benchmark Profiling}
\label{app:benchmark-profiling}
\begin{figure}[H]
    \centering
    \includegraphics[width=\textwidth]{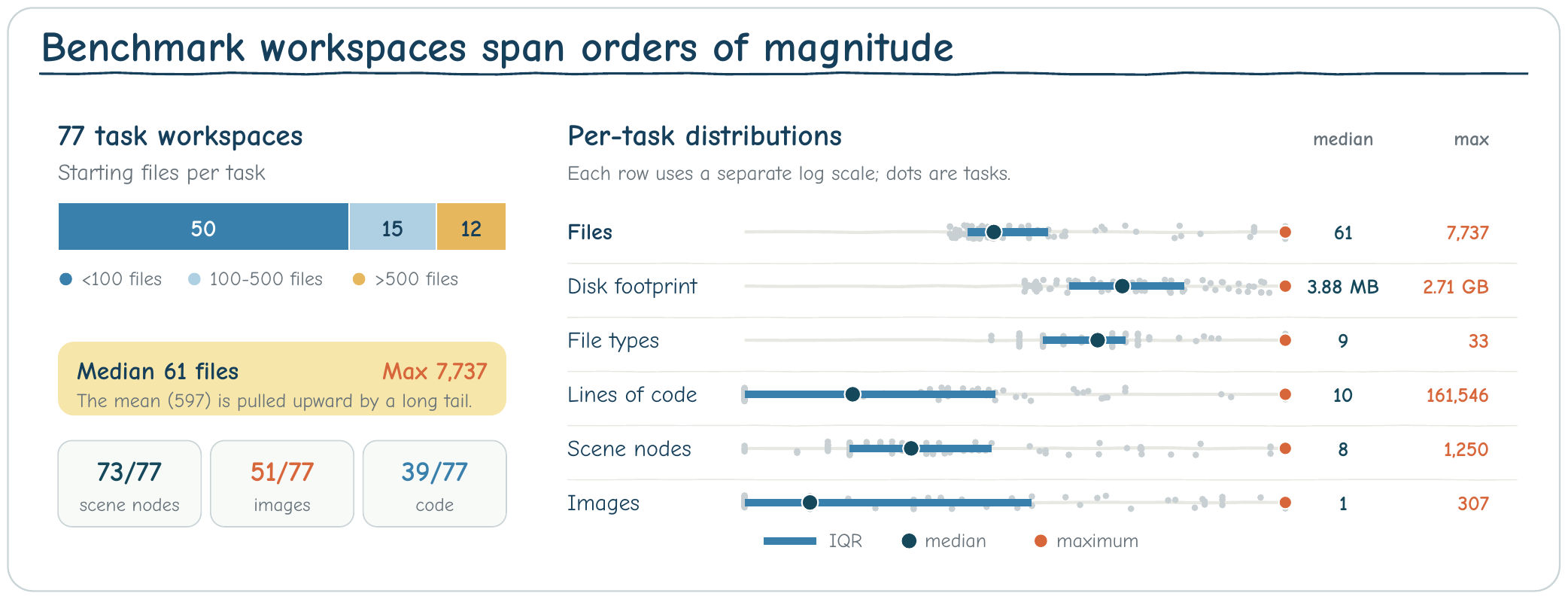}
    \caption{\textbf{Breadth of the 77 task workspaces.} The benchmark ranges from compact scenes to asset-rich Unity projects. The right panel summarizes the per-task distributions of files, disk footprint, file types, code, scene nodes, and images; each row uses an independent log scale, and colored marks show the interquartile range, median, and largest workspace.}
    \label{fig:benchmark-workspace-profile}
\end{figure}

UnityBench-Mini3D exposes agents to heterogeneous Unity project contexts rather than uniform code-only repositories. Across 77 tasks, starting workspaces range from 29 to 7,737 files and cover 60 file extensions. Figure~\ref{fig:benchmark-workspace-profile} shows that the benchmark combines focused scenes with asset-rich projects while retaining substantial variation in code, scene structure, and visual content.

\begin{figure}[H]
    \centering
    \includegraphics[width=\textwidth]{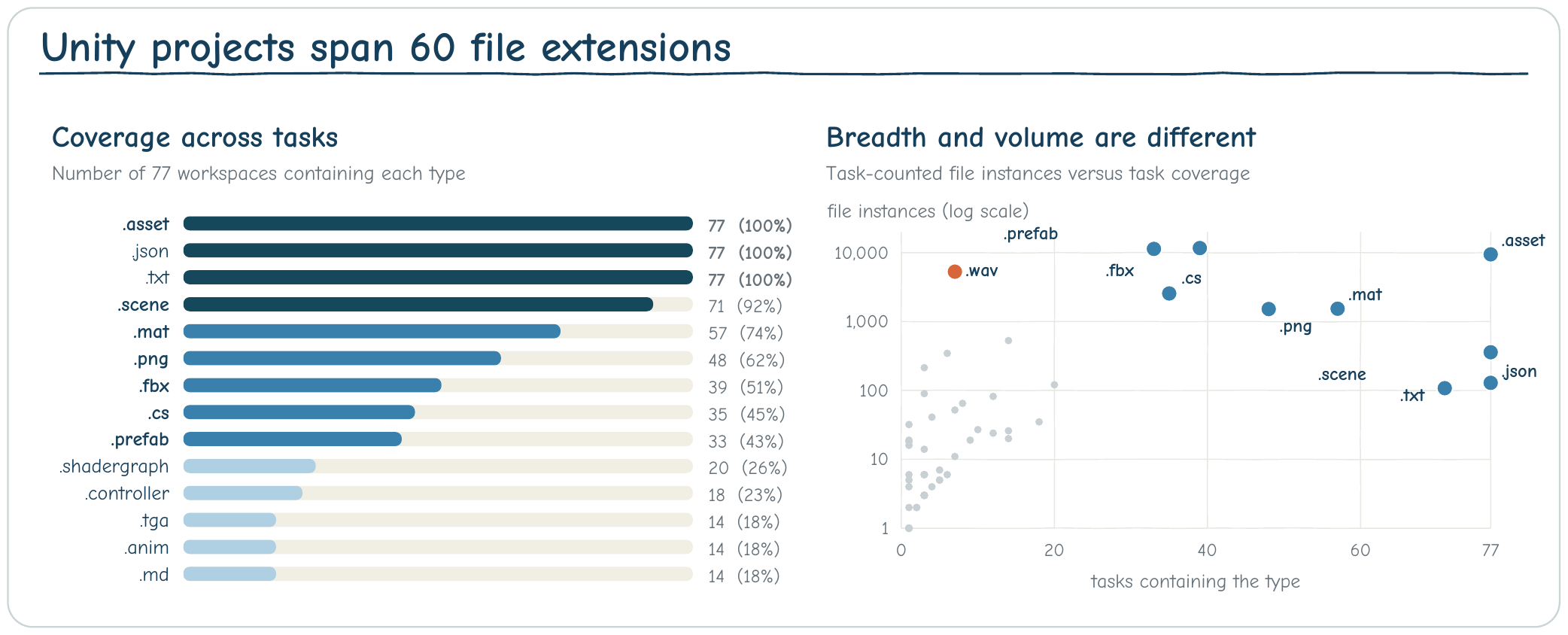}
    \caption{\textbf{Coverage of selected Unity project artifacts.} Bars report the number and percentage of task workspaces containing each non-exclusive artifact type. Scene files combine \texttt{.scene} and \texttt{.unity}; other composite modalities are unions of the corresponding extensions listed in the released profiling data. Each task is counted at most once per row.}
    \label{fig:benchmark-modality-coverage}
\end{figure}

The project contexts also span both serialized Unity structure and authored content. As shown in Figure~\ref{fig:benchmark-modality-coverage}, 75 tasks contain scene files, 57 contain materials, and 51 contain raster images; the benchmark also includes 3D model files, C\# scripts, prefabs, shader or VFX assets, and dedicated animation files. This coverage requires agents to reason jointly about code and the heterogeneous artifacts that constitute a Unity project.

\section{Task-specific Formula-based Evalaution Metrics}
\label{app:formula-based-validation}

Many Unity authoring tasks admit multiple visually or behaviorally valid solutions,
so exact scene or asset matching would reward imitation rather than task completion.
We therefore translate qualitative requirements into observable quantities and
accept a solution when those quantities satisfy task-specific constraints.  The
audit below includes a task only when its evaluator computes a derived metric from
geometry, rendering, simulation, timing, or graph structure.  Ordinary component
existence checks and direct equality against a fixed serialized value are not
counted as formula-driven validation.  Under this definition, 50 of the 77 tasks
contain at least one formula-driven validator.

For compactness, let $\mathbb{I}[\cdot]$ denote the indicator function,
$\|\cdot\|_2$ Euclidean distance, $\Delta_\theta(a,b)$ the shortest signed angular
difference, and $B(o)$ the world-space axis-aligned bounds of object $o$.

\subsection{Complete Task-Level Audit}

\paragraph{0000 --- AutoDoor.}
The evaluator unwraps the sampled yaw sequence using
\begin{equation}
  a_i=a_{i-1}+\Delta_\theta(\theta_{i-1},\theta_i), \qquad
  R_\theta=\max_i a_i-\min_i a_i,
\end{equation}
and requires $30^\circ\le R_\theta\le180^\circ$.  It additionally constrains every
single-step jump to $|\Delta_\theta|<90^\circ$, pivot drift to
$\|p_i-p_0\|_2<0.05$ m, and camera visibility to at least five of nine door-bounds
samples.

\paragraph{0002 --- CubeMove.}
Observable movement is measured by axis displacement and rigidbody velocity,
e.g., $\Delta z=z_T-z_0$ and $v_y(T)$.  The evaluator checks the signs of commanded
motion, a positive vertical velocity after the first jump, no added impulse on a
second airborne jump, and bounded idle drift $\|p_T-p_0\|_2\le\epsilon$.

\paragraph{0003 --- CatapultExplosion.}
Animation-event timing is normalized by clip length,
\begin{equation}
  \tau_{\mathrm{spawn}}=t_{\mathrm{event}}/T_{\mathrm{clip}}, \qquad
  |\tau_{\mathrm{spawn}}-0.25/T_{\mathrm{clip}}|\le0.03.
\end{equation}
Explosion effectiveness is evaluated in the ground plane:
\begin{equation}
  d_{XZ}(p,c)=\sqrt{(p_x-c_x)^2+(p_z-c_z)^2}, \qquad
  \Delta d=\max_t d_{XZ}(p(t),c)-d_{XZ}(p(0),c),
\end{equation}
with $1\le\Delta d\le5$ m for eight target placements.

\paragraph{0005 --- BuildHouseReachability.}
Candidate room points must be separated by
$d_{XZ}(p_i,p_j)\ge1.2$ m and sample to the runtime-baked navigation surface
within $0.25$ m.  For each intra-floor, adjacent-floor, and cross-floor query, the
evaluator requires a complete path $P(s,g)$ from Unity's navigation graph.  The
plan footprint also satisfies $w(B)\le30.5$ m and $d(B)\le30.5$ m.

\paragraph{0006 --- TankTrackAnimation.}
A wheel is considered animated when
$\operatorname{angle}(q_0,q_t)>1^\circ$.  Track scrolling must satisfy
$|u_t-u_0|>0.005$, and periodic wrapping is detected when
$u_t<u_{t-1}-0.1$ during a 1.25 s observation window.

\paragraph{0007 --- SkyboxBlackHole.}
For linear RGB pixels, luminance is
\begin{equation}
  Y=0.2126R+0.7152G+0.0722B,
\end{equation}
with mean $\bar Y$, standard deviation $\sigma_Y$, contrast ratio
$r_c=\sigma_Y^{a}/\max(\sigma_Y^{r},10^{-5})$, and percentile range ratio
\begin{equation}
  r_Y=\frac{P_{95}^{a}(Y)-P_{5}^{a}(Y)}
           {P_{95}^{r}(Y)-P_{5}^{r}(Y)}.
\end{equation}
The image is partitioned into a $3\times3$ grid.  Each cell stores normalized
luminance, red ratio, and blue ratio in $f_g$, and the aggregate error is
\begin{equation}
  E_{\mathrm{grid}}=\frac{1}{3G}\sum_{g=1}^{G}\|f_g^a-f_g^r\|_1.
\end{equation}
The centroid of the darkest 20\% of pixels is also compared by Euclidean distance.
Progressive tests tighten $E_{\mathrm{grid}}$ from 0.60 to 0.07 and centroid error
from 0.45 to 0.06; coarse contrast and range ratios must reach 0.20 and 0.35.

\paragraph{0008 --- SwordsmanAttack.}
Nine uniformly spaced clip samples define a sword trajectory $p_i$.  The evaluator
computes
\begin{align}
  L &= \sum_{i=1}^{8}\|p_i-p_{i-1}\|_2,\\
  H &= \max_{i,j}\|(p_{ix},p_{iz})-(p_{jx},p_{jz})\|_2,\\
  V &= \max_i p_{iy}-\min_i p_{iy},
\end{align}
and requires $L\ge0.75$, $H\ge0.60$, and $V\ge0.40$ m.  At least one segment
must simultaneously move at least 0.20 m horizontally and 0.16 m vertically,
which distinguishes a diagonal slash from two independent motions.

\paragraph{0010 --- SplitTheUI.}
For every sliced sprite $s$, the evaluator counts opaque support
$A_s=\sum_{p\in s}\mathbb{I}[\alpha(p)>0]$ and requires $A_s>0$.  This prevents
formally correct slice counts from being achieved with empty rectangles.

\paragraph{0012 --- Smrvfx.}
Runtime VFX systems are checked using total capacity
$C=\sum_k C_k$ and alive-particle count $N(t)=\sum_k N_k(t)$.  Throttle-on must
produce a useful $N(t)$ without exceeding $C$; after throttle-off,
$N(t+3)>0$ but $N(t+48)<N(t+3)$.  Render bounds must be finite, nonempty
($\|\operatorname{size}(B)\|_2^2>0.01$), smaller than 100 m per axis, and intersect
the animated character bounds.

\paragraph{0015 --- ShootAheadUguiMobileHud.}
For a screen of size $W\times H$, every control rectangle must obey
\begin{equation}
  -\epsilon\le x_{\min},y_{\min},\qquad
  x_{\max}\le W+\epsilon,\quad y_{\max}\le H+\epsilon.
\end{equation}
Disallowed overlap is measured as
\begin{equation}
 A_{ij}=\max(0,\min x_i^+-\max x_i^-)
        \max(0,\min y_i^+-\max y_i^-),
\end{equation}
and must equal zero on multiple phone aspect ratios.

\paragraph{0017 --- CreateLOD.}
Triangle count is $T=|I|/3$, where $I$ is the mesh index buffer; selected assets
must match specified budgets and, where required, satisfy $T_{i+1}<T_i$.  For an
$n$-level LOD group, evenly spaced screen-relative transitions are tested at
\begin{equation}
  h_i=1-\frac{i+1}{n+1}, \qquad |\hat h_i-h_i|\le0.01.
\end{equation}

\paragraph{0019 --- AutoTimelineGenerator.}
The evaluator compares sampled positions by $\|p(t)-p(0)\|_2$, verifies runtime
and direct-evaluation agreement $\|p_{\mathrm{runtime}}(3)-p_{\mathrm{eval}}(3)\|_2<0.5$ m,
and tests camera facing with
$\hat f_{\mathrm{cam}}\cdot\widehat{p_{\mathrm{char}}-p_{\mathrm{cam}}}>0.3$.
Shake amplitude is $\|p(4.0)-p(3.5)\|_2$, constrained to $(0.001,2.0)$ m.
Framing requires projected viewport coordinates $(u,v,z)$ to satisfy
$0<u<1$, $0<v<1$, and $z>0$ at several timestamps.

\paragraph{0021 --- ShootAheadStreetLayout.}
Road centers are quantized by
\begin{equation}
  k=\operatorname{round}\!\left(\frac{|c|-9}{8}\right),\qquad
  \left|\frac{|c|-9}{8}-k\right|8\le\epsilon,
\end{equation}
and duplicate cells are rejected.  Building and road intersection uses XZ AABB
overlap, while tree rows use along-road spread
$\max_i a_i-\min_i a_i$ and perpendicular spread
$\max_i b_i-\min_i b_i$ to require long, narrow roadside distributions.

\paragraph{0023 --- UIRedoBench.}
World-space panel coordinates are normalized to the 1920-by-1080 reference frame:
\begin{equation}
  x'=1920\frac{x-x_{\min}^{P}}{w_P},\qquad
  y'=1080\frac{y-y_{\min}^{P}}{h_P}.
\end{equation}
Centers and sizes must match the reference within 20 pixels.  The same rectangle
intersection formula used by Task 0015 rejects foreground overlap and overflow.

\paragraph{0025 --- FPSCompassHUDCalibration.}
The strip position $x$ is decoded as
\begin{equation}
  \hat\psi=360\left(\frac{x}{3600}+\frac12\right),\qquad
  e_\psi=|\Delta_\theta(\psi^*,\hat\psi)|.
\end{equation}
Six score tiers require all cardinal-heading errors to be at most
$6^\circ,5^\circ,\ldots,1^\circ$.

\paragraph{0026 --- ShootAheadVehicleWheelPhysics.}
Visual wheel alignment is
$\|p_{\mathrm{collider}}^{local}-p_{\mathrm{visual}}^{local}\|_2\le0.35$ m.
Runtime speed is converted by $v_{\mathrm{kph}}=3.6\|v\|_2$.  The vehicle must
reach 100 km/h in 6--8 s, peak at 175--185 km/h, and remain stable under zero
steering with $|x_T-x_0|\le1.5$ m and
$|\Delta_\theta(\psi_0,\psi_T)|\le15^\circ$.

\paragraph{0027 --- VolumeBoundsOrganizer.}
Classification is based on oriented mesh volume under world scale,
\begin{equation}
  V=|s_x\ell_x|\,|s_y\ell_y|\,|s_z\ell_z|,
\end{equation}
where $\ell$ comes from \texttt{sharedMesh.bounds}, not collider bounds.  Boundary,
negative-scale, zero-volume, and adversarial inflated-collider cases verify the
formula.

\paragraph{0033 --- Fireworks-ExplosionEffect.}
The evaluator models short-time vertical speed as
\begin{equation}
  v(t)=v_0-g\,m_g\,t,\qquad
  r_v=\frac{v(0.3T_{life})}{v_0}<1,
\end{equation}
where $m_g$ is the particle gravity multiplier.  Gradient colors are distinct
when at least one RGB-channel difference between adjacent keys exceeds 0.15.

\paragraph{0034 --- SciFiRoomVariants.}
Floor coordinates are clustered by axis with tolerance 0.55 m, cluster means are
sorted, and adjacent means must differ by $4\pm0.55$ m.  Occupancy of the resulting
$n\times n$ index grid must be exactly $n^2$ with no duplicates.  Pairwise room
footprint rectangles must not overlap.

\paragraph{0035 --- HumanoidIK.}
Position and orientation errors are
\begin{equation}
  e_p=\|p_b-p_g\|_2,\qquad
  e_R=2\arccos(|q_b\cdot q_g|).
\end{equation}
Hand/foot tests use task-specific tolerances (including 0.15 m and 0.05 m).
Look-at error is
$e_{look}=\arccos(\hat f_{head}\cdot\widehat{p_t-p_h})$, and foot-plane alignment
requires $\arccos(u_{foot}\cdot n_{plane})\le5^\circ$.  A ray--plane intersection
provides the expected foot contact point.

\paragraph{0036 --- NavMeshAgentOffMeshLink.}
Markers must satisfy $\|p-p_{nav}\|_2\le r$ after navigation-surface sampling.
Every required query must return a complete graph path with at least two corners;
selected straight-line queries must instead be blocked by a navigation raycast.
Off-mesh-link endpoint error is
\begin{equation}
  e_L=\max(\|s-s^*\|_2,\|g-g^*\|_2)\le1.25\ \mathrm{m}.
\end{equation}

\paragraph{0038 --- DynamicRainEffect.}
Matched particles define mean displacement
$\bar\delta_p=N^{-1}\sum_i(p_i^{t+1}-p_i^t)$.  The emitter must follow the target,
$\|\delta_{rain}^{XZ}-\delta_{target}^{XZ}\|_2<0.75$ m, while old particles remain
world-anchored.  Target-relative motion
\begin{equation}
  r=\delta_p^{XZ}-\delta_{target}^{XZ}
\end{equation}
must oppose target motion ($r\cdot\delta_{target}<0$), and a 10 m target move must
produce more than twice the relative-motion magnitude of a 2 m move.  Stationary
rain must satisfy $\bar\delta_y<-0.01$ and near-vertical horizontal travel.

\paragraph{0039 --- RollingBallCheckpointCourse.}
Checkpoint $k$ is visited when
\begin{equation}
  I_k=\mathbb{I}\!\left[\min_{0\le t\le24}\|x(t)-c_k\|_2\le1.25\right],
  \qquad C=\sum_{k=1}^{5}I_k.
\end{equation}
The simulated ball must remain inside course bounds and use an unconstrained,
gravity-driven rigidbody.  Progressive variants cap the obstacle budget at 12,
9, and 6 cubes.

\paragraph{0041 --- ExcavatorAnimationClipDigDump.}
After deterministic animation evaluation, cube $i$ is counted in the target box
when
\begin{equation}
  T_B^{-1}p_i(T)\in[-\tfrac12,\tfrac12]^3,
  \qquad D=\sum_i\mathbb{I}[T_B^{-1}p_i(T)\in[-\tfrac12,\tfrac12]^3].
\end{equation}
Progressive milestones require $D\ge4,8,12,\ldots,32$.  Clips must last at least
0.5 s and animate only allowed excavator pivot rotations.

\paragraph{0043 --- MazeObstacleParenting.}
Rounded wall coordinates define an $11\times11$ grid graph $G=(V,E)$.  Flood fill
must satisfy $|\operatorname{Reach}(s)|=|V_{road}|$.  Exactly one boundary road cell
is allowed.  Uniqueness of the entrance--exit route is verified by removing each
edge of one discovered path and requiring the exit to become unreachable; this is
equivalent to that path being the unique $s$--$g$ path.  Uniform $2\times2$ wall
blocks and duplicate occupied cells are rejected.

\paragraph{0044 --- ColliderWrappingChallenge.}
Hidden inside and outside probes define
\begin{equation}
  C_{in}=\frac1{N_{in}}\sum_i\mathbb{I}[p_i\in\Omega_C],\qquad
  C_{out}=\frac1{N_{out}}\sum_i\mathbb{I}[q_i\in\Omega_C].
\end{equation}
Per-object minimum inside coverage is approximately 0.58--0.72 and maximum outside
spill is 0.20--0.32.  Representative rays must hit, while primitive-only and
collider-budget constraints prevent an arbitrarily dense wrapper.

\paragraph{0046 --- AnimationScriptPlayable.}
Every sampled bone is compared with a reference by
$e_p=\|p_b-p_b^*\|_2\le10^{-4}$ and
$e_R=\operatorname{angle}(q_b,q_b^*)\le10^{-4}$ degrees.  Root-motion removal uses
vertical range $R_y=\max_t y(t)-\min_t y(t)\le10^{-6}$; multiplier behavior compares
$R_y^{scaled}$ with $mR_y^{base}$ within 0.02.

\paragraph{0047 --- DynamicDissolve.}
At dissolve values zero and one, the center pixel is respectively required to
differ from and equal the camera background.  Mid-dissolve emission is detected by
\begin{equation}
  \exists x\in[64,191]: R(x,128)>0.8.
\end{equation}
A time-driven effect must also change the center pixel over a 0.5 s interval.

\paragraph{0048 --- RoadsideJointPhysics.}
Joint behavior is scored from post-impact observables.  The fixed fence requires
$\|p_T-p_0\|_2>0.6$ m or $\operatorname{angle}(q_T,q_0)>18^\circ$.  The hinge gate
requires $28^\circ<|\Delta_\theta(\psi_0,\psi_T)|<125^\circ$ and hinge-anchor drift
below 0.45 m.  Spring return requires
$z_{final}<z_{max}-0.12$ m after a deflection above 0.55 m.  The configurable joint
must move 0.45--2.45 m along its allowed axis while lateral drift stays below
0.24 m and rotation below $12^\circ$.

\paragraph{0049 --- UIDocument.}
Time-based T-Rex ground motion is compared with
\begin{equation}
  d^*=6\cdot60\cdot\Delta t\cdot s_{pixel},\qquad
  s_{pixel}=\max(0.25,w_{ground}/600),
\end{equation}
and observed displacement must lie in $[0.8d^*,1.5d^*]$.  Additional rectangle
inequalities detect jump height, ducking width, obstacle lead distance, obstacle
clearance, and the relative placement of game-over UI elements.

\paragraph{0053 --- PhysicsMaterial.}
A sphere reaches checkpoint bounds $B_k$ when
$\operatorname{SqrDistance}(x(t),B_k)\le0.65^2$ within 10 s.  Curling-stone planar
distance is $d_i=\|(x_i,z_i)-(x_h,z_h)\|_2$, and score is
\begin{equation}
  S=\sum_i\mathbb{I}\left[d_i^{red}\le4.5\ \land\
  d_i^{red}+0.03<\min_j d_j^{yellow}\right].
\end{equation}

\paragraph{0054 --- ProceduralSkinnedCloth.}
Normalized vertex height is
$\eta_i=(y_i-y_{min})/\max(y_{max}-y_{min},10^{-4})$ and selects top, transition,
and lower cloth bands.  Bone influence in a band is
$\bar w_{b,S}=|S|^{-1}\sum_{i\in S}w_{ib}$; required examples include top-bone
influence above 0.72, middle above 0.25, and bottom above 0.55.  Per-vertex bone
weights satisfy $\sum_{j=0}^{3}w_{ij}=1\pm0.03$.  Baked deformation is measured by
mean and maximum vertex distance before and after moving bones, required to exceed
0.025 and 0.12 m.  Bindpose matrices are compared elementwise within 0.08.

\paragraph{0055 --- ArcherAnimatorController.}
Blend-tree samples are located by
$\|b_i-b_i^*\|_2\le0.08$.  At runtime, the dominant clip is the clip of maximum
weight, and a stationary aiming pose must keep spine angular drift
$\operatorname{angle}(q_0,q_T)<2^\circ$ over 0.75 s.

\paragraph{0056 --- RigidbodyCharacterControllerParity.}
The rigidbody implementation is compared with a CharacterController oracle using
\begin{equation}
  e_h=\|(x_r,z_r)-(x_c,z_c)\|_2,\qquad e_v=|y_r-y_c|.
\end{equation}
Task-specific bounds cover walk and sprint displacement, jump apex (within 0.42 m),
landing height, step traversal, slope traversal, wall stopping, falling, and
normalized diagonal motion.  This evaluates behavioral parity without requiring
the same controller implementation.

\paragraph{0057 --- VoxelWormsTerrain.}
Explosion membership is spherical,
\begin{equation}
  I(v)=\mathbb{I}[\|c(v)-c_e\|_2<r],
\end{equation}
and the evaluator explicitly samples voxels that lie inside the explosion AABB but
outside the sphere to reject box-shaped deletion.  It also measures crater depth by
the reduction in vertical-ray hit height, requires a decrease in solid voxel count,
and checks open-tunnel continuity with nearby solid-shell support.

\paragraph{0059 --- RainEmitter.}
For $N$ splash particles, validity is
\begin{equation}
  C_{splash}=\frac1N\sum_i\mathbb{I}
  [|y_i-y_w|\le1.5\ \land\ (x_i,z_i)\in B_w^{XZ}\oplus0.75],
\end{equation}
and the evaluator requires $C_{splash}\ge0.60$.

\paragraph{0060 --- GunShotSound.}
Emitter--listener distance is $d=\|p_e-p_l\|_2$.  Near and distant blend weights
must respond monotonically:
\begin{equation}
 w_n(d_{near})>w_n(d_{far}),\qquad
 w_f(d_{far})>w_f(d_{near}),\qquad 0\le w\le1.01.
\end{equation}
An occluding ray must reduce the reported low-pass cutoff and the cutoff must recover
after line of sight is restored.  Mixer-volume boundary behavior is required to be
consistent with a finite form of the usual decibel mapping
$L=20\log_{10}(\max(v,\epsilon))$: $v=1$ maps near 0 dB and $v=0$ to a finite floor
no higher than $-80$ dB.

\paragraph{0063 --- MinimapRealization.}
Camera and arrow directions are compared using shortest-angle errors, with a
$5^\circ$ arrow tolerance.  Arrow centering requires
\begin{equation}
  \|c_{arrow}-c_{map}\|_2\le\max(4,0.05d_{map}),
\end{equation}
where $d_{map}$ is the minimap rectangle diagonal.  Camera-follow tolerance is
$\max(0.5,0.015s_{ortho})$.  Circle and triangle sprites are recognized by alpha
predicates at normalized center, tip, base, and corner sample locations using
$\alpha>0.5$.

\paragraph{0065 --- ShootAheadSplitBodyAim.}
Horizontal facing error is
$e=\operatorname{angle}(\widehat{f_{XZ}},\widehat{d_{XZ}})$ and must remain below
32--35 degrees in the tested movement/aim configurations.  Upper-body response is
the sum of spine quaternion changes; side-aim requires a sum above $28^\circ$ and
upper-spine change above $8^\circ$.  Lower-body isolation requires upper-spine
change to exceed each foot change by at least $4^\circ$.

\paragraph{0067 --- TerrainRiverWaterAssembly.}
Hidden terrain samples define water coverage
\begin{equation}
  C_{water}=\frac1N\sum_i\mathbb{I}[h_w(x_i,z_i)\ge y_i+\epsilon],
\end{equation}
with full coverage required.  Adjacent segments must satisfy seam midpoint distance
$\|m_i-m_{i+1}\|_2\le2.5$ m and mean edge-height difference at most 0.25 m.  Bank
vertices satisfy $y_{edge}\le h_{terrain}(x,z)+0.05$, river height is monotonically
non-increasing downstream, and no more than seven river segments are allowed.
Endpoint lake levels 6.60 and 1.82 m use a 0.40 m tolerance.  Triangle inclusion is
computed with barycentric coordinates $u+v+w=1$ and tolerance 0.025.

\paragraph{0068 --- ShanghaiCircuitTerrainPaint.}
Distance from a point to the hidden track polyline $\Gamma$ is
\begin{equation}
 d(p,\Gamma)=\min_j\left\|p-\left(a_j+
 \operatorname{clamp}\!\left(\frac{(p-a_j)\cdot s_j}{\|s_j\|_2^2},0,1\right)s_j\right)\right\|_2.
\end{equation}
Terrain material scores are pass ratios
$R=N_{pass}/N$; strict, medium, and loose milestones use boundary tolerances
0.35, 0.85, and 1.50 m with ratios 0.96, 0.93, and 0.88.  White-line samples must
satisfy $|d(p,\Gamma)-r_{road}|\le1.15$ at least 90\% of the time.  Segment tangent
alignment uses $|\hat u_{line}\cdot\hat t_{road}|\ge0.68$ with at least 84\% passing.
Longitudinal continuity occupies at least 82 of 120 progress buckets on each side.
Related distance, alignment, overlap, and connected-chain metrics evaluate curbs,
rails, and shortcut barriers.

\paragraph{0069 --- TrenchAssembly.}
Projected collider footprint is estimated by grid/raycast integration,
$A\approx n_{hit}\Delta^2$ with $\Delta=0.25$ m.  Pair overlap is scored by
\begin{equation}
  O_{min}=\frac{A_\cap}{\min(A_1,A_2)},\qquad
  \operatorname{IoU}=\frac{A_\cap}{A_1+A_2-A_\cap}.
\end{equation}
Progressive caps are $(0.20,0.10)$, $(0.30,0.20)$, and $(0.40,0.30)$ for
$(O_{min},\operatorname{IoU})$.  Directed anchor reachability is
$R_{path}=N_{valid}/N_{all}$ with milestones $1/3$, $2/3$, and 1.  Row centers must
be monotonic along their dominant axis with at least 5 m separation.

\paragraph{0070 --- CameraPlacement.}
For prop $j$, visibility is
\begin{equation}
  V_j=n_j^{visible}/n_j^{complete}.
\end{equation}
A prop passes when $V_j\ge0.25$, it has sufficient complete pixel support, and it is
not occluded.  Overall camera quality is
\begin{equation}
  C_{camera}=\frac1M\sum_j\mathbb{I}[V_j\ge0.25\land O_j=0]\ge0.60.
\end{equation}
Camera overlap is rejected with a 0.2 m sphere probe; outdoor and indoor FOV values
are 50 and 60 degrees with 0.1 degree tolerance.

\paragraph{0071 --- PlantsDecoration.}
Nearest-neighbor spacing uses
\begin{equation}
 d_i=\min_{j\ne i}\|p_i-p_j\|_{XZ},\quad
 \mu=N^{-1}\sum_i d_i,\quad
 \sigma=\sqrt{N^{-1}\sum_i(d_i-\mu)^2},\quad CV=\sigma/\mu.
\end{equation}
Pines require $2.5\le\mu\le6.0$, $0.25\le CV\le0.85$, and minimum neighbor
distance 1.45 m; fruit trees use 3.0--8.0, 0.25--0.90, and 1.95 m.  Local density is
$n_i(r)=\sum_{j\ne i}\mathbb{I}[\|p_i-p_j\|_{XZ}\le r]$.  A 10-by-10 occupancy grid
must use 25--72 cells, and at least 95\% of trees must remain 3.15 m from visible
water.  Additional ratios evaluate terrain texture, height/slope bands, shoreline
placement, grounding, nonpenetration, rotation bins, and scale diversity.

\paragraph{0072 --- CastleDestruction.}
For each structural family,
\begin{equation}
  R_f=\frac{\#\{\text{expected groups matched}\}}
            {\#\{\text{expected groups}\}},
\end{equation}
with 50\% and 100\% milestones.  A candidate match requires the correct prefab
fingerprint and name family plus
$\|c(B)-c^*\|_2\le0.08$ and $\|s(B)-s^*\|_2\le0.08$ m.

\paragraph{0073 --- CastleConstruction.}
Expected stair connections and candidate stairs form a bipartite graph.  An edge is
present when position error is at most 0.75 m, each transformed basis-vector length
is within 0.05 of one, and each basis angular error is at most $5^\circ$.  Maximum
bipartite matching gives
\begin{equation}
  R_{stair}=|M_{max}|/N_{expected},
\end{equation}
with 50\% and 100\% milestones; the strict score also forbids redundant candidates.

\paragraph{0076 --- BuildHouseFromDesign.}
Wall-path coverage samples each design segment at 0.1 m intervals:
\begin{equation}
 C_{wall}=\frac1N\sum_i\mathbb{I}[p_i\in\bigcup_j(B_j\oplus\delta)]\ge0.95,
\end{equation}
where $\delta$ is 0.50 m for exterior and 0.35 m for interior walls.  Furniture
penetration uses per-axis overlap
$o_x=\min(x_i^+,x_j^+)-\max(x_i^-,x_j^-)$ and analogously for $y,z$; penetration
requires all three overlaps to exceed 0.01 m.  Window edge gaps subtract both
half-widths from center separation and are compared with the design dimensions.

\paragraph{0077 --- TakeOffAndLanding.}
Telemetry uses speed $v=\|\mathbf v\|_2$, forward speed
$v_f=\mathbf v\cdot\hat z$, signed heading error, and pitch
$\theta=\arcsin(\operatorname{clamp}(f_y,-1,1))$.  Centerline and heading quality
use
\begin{equation}
 \operatorname{RMS}_x=\sqrt{N^{-1}\sum_i x_i^2},\qquad
 \operatorname{RMS}_\psi=\sqrt{N^{-1}\sum_i e_{\psi,i}^2}.
\end{equation}
Takeoff centerline RMS must be at most 1 m; final-approach cross-track RMS at most
8 m and heading RMS at most $2.5^\circ$.  Liftoff speed is 50--85 m/s (58--78 for
the high-quality tier), airborne altitude is 50--150 m, and speed remains 48--130
m/s.  Touchdown requires 60--95 m/s, sink rate 0.2--5.5 m/s, and heading error at
most $3^\circ$.  A nonteleportation bound is
\begin{equation}
  \|p_t-p_{t-\Delta t}\|_2\le
  \max(v_{t-\Delta t},v_t)\Delta t+8.
\end{equation}
Ground-roll true-contact ratio must be at least 0.5, with rollout distance and time
bounded by 900 m and 25 s.

\paragraph{0078 --- JetHUD.}
HUD pixels are selected by
\begin{equation}
 M(p)=\mathbb{I}[\alpha\ge20\land G\ge35\land G\ge1.06R\land G\ge1.06B].
\end{equation}
Tolerance-aware precision and recall are
$P=N_{a\rightarrow r}/N_a$ and $R=N_{r\rightarrow a}/N_r$, combined as
\begin{equation}
 F_1=\frac{2PR}{P+R}.
\end{equation}
Dynamic response is measured by normalized symmetric difference
$D=|M_1\triangle M_2|/|M_1\cup M_2|$.  The evaluator probes multiple speeds,
altitudes, attitudes, headings (including 359-to-0 wrapping), and a delayed-impact
history of 0.2 s.

\paragraph{0079 --- FPSRemoteWeaponAnimationSync.}
Muzzle direction error is
$e_m=\operatorname{angle}(d_{muzzle},q_{aim}\hat z)\le5^\circ$ across representative
pitch/yaw samples.  Both hand positions must satisfy
$\|p_{hand}-p_{grip}\|_2\le0.035$ m.  Finger curl is the joint angle at the
intermediate bone and at least two fingers must be below $165^\circ$; at least three
fingertips must lie within 0.18 m of the grip.  Arm plausibility uses
\begin{equation}
  0.28(L_u+L_l)<\|p_{upper}-p_{hand}\|_2\le1.03(L_u+L_l),
  \qquad 8^\circ\le e_{elbow}\le178^\circ.
\end{equation}
Adjacent aim samples must not rotate any watched arm, wrist, or finger bone by
100 degrees or more, and repeated reload snapshots must preserve nondecreasing
normalized animation time within 0.02.

\subsection{Checks Deliberately Excluded from the Count}

Tasks 0020, 0022, 0029, 0042, 0066, 0075, and 0080 contain numerical position,
rotation, scale, or bounds comparisons, but these primarily reproduce a fixed
reference transform or serialized target.  Tasks 0016, 0024, 0030--0032, 0045,
0050, 0052, and 0062 also contain scalar ranges or source-code predicates, but do
not derive a behavioral or perceptual quality metric from the completed artifact.
They remain objective tests, but including them in the formula-driven set would
blur the distinction between mathematical validation of an open-ended outcome and
ordinary configuration checking.

\subsection{Interpretation}

The formulas are not intended as universal perceptual metrics.  Each one is a
task-local operationalization of a user-visible requirement: coverage for whether a
river actually contains its samples, tangent alignment for whether road markings
follow a circuit, visibility ratios for whether a camera composition exposes its
subjects, distribution statistics for whether vegetation is neither clumped into a
single pile nor placed on a mechanical grid, and simulation telemetry for whether
an animation or physical system produces the requested consequence.  Hidden sample
locations, multiple thresholds, and anti-degeneracy constraints make the metrics
difficult to satisfy through a single hard-coded artifact while still allowing
substantially different valid implementations.

\section{Tool Availability}
\label{app:tools}
This section documents the default tool inventories distributed with the evaluated harnesses. For readability and reference, these inventories were exported separately from the harnesses and therefore describe their default configurations rather than the evaluation-time enabled tool set. In the actual benchmark runs, all Web search and Web fetch tools were disabled as described in Section ~\ref{sec:config}, and the task environment remained offline except for model calls required for inference.

\providecommand{\toolname}[1]{%
  \begingroup
  \ttfamily
  \def\_{\char95\allowbreak}%
  #1%
  \endgroup
}


\subsection{Codely}
\begin{small}
\begin{longtable}{
  @{}
  >{\raggedright\arraybackslash}p{0.29\textwidth}
  >{\raggedright\arraybackslash}p{0.65\textwidth}
  @{}
}
\caption{\textbf{Tool availability for Codely.} Tools exposed to the benchmark main agent.}
\label{tab:codely-tools}\\
\multicolumn{2}{@{}l}{
  \footnotesize
  \textbf{Harness:} Codely
  \quad
  \textbf{CLI version:} 1.0.0-nightly.51
}\\[4pt]
\toprule
\textbf{Tool} & \textbf{Description} \\
\midrule
\endfirsthead
\multicolumn{2}{@{}l}{
  \footnotesize
  \textit{\tablename\ \thetable\ continued}
}\\[2pt]
\toprule
\textbf{Tool} & \textbf{Description} \\
\midrule
\endhead
\midrule
\multicolumn{2}{r@{}}{
  \footnotesize\textit{Continued on next page}
}\\
\endfoot
\bottomrule
\endlastfoot
\path{activate_skill} & Activates a specialized agent skill by name. \\
\path{analyze_multimedia} & Analyzes images, videos, and audio files using a multimodal model. \\
\path{apply_patch} & Apply a structured patch document to edit files. \\
\path{cron_create} & Schedule a prompt to be enqueued at a future time. \\
\path{cron_delete} & Cancel a cron job previously scheduled with cron\_create. \\
\path{cron_list} & List all cron jobs scheduled via cron\_create, both durable (.codely-cli/scheduled\_tasks.json) and session-only. \\
\path{execute_csharp_script} & Executes a C\# script in the Unity Editor using Microsoft.CodeAnalysis.CSharp.Scripting. \\
\path{execute_custom_tool} & Executes a custom Unity tool by name. \\
\path{glob} & Efficiently finds files matching specific glob patterns (e.g., src/**/*.ts, **/*.md), returning absolute paths sorted by modification time (newest first). \\
\path{job_create} & Use this tool to create a structured task list for your current coding session. \\
\path{job_get} & Use this tool to retrieve a job by its ID from the job list. \\
\path{job_list} & Use this tool to list all jobs in the job list. \\
\path{job_update} & Use this tool to update a job in the job list. \\
\path{list_directory} & Lists the names of files and subdirectories directly within a specified directory path. \\
\path{lsp} & Interact with LSP servers for code intelligence: definition, references, hover, symbols, implementation, and call hierarchy. \\
\path{read_file} & Reads and returns the content of a specified file. \\
\path{read_many_files} & Reads content from multiple files specified by paths or glob patterns within a configured target directory. \\
\path{replace} & Replaces text within a file. \\
\path{run_shell_command} & Executes a given shell command using the best available Windows shell (PowerShell Core, Windows PowerShell, or Command Prompt). \\
\path{save_memory} & Save a memory entry to the hierarchical context memory for recall in future sessions. \\
\path{search_file_content} & Searches for a regular expression pattern within the content of files in a specified directory, a specific file, or the current working directory. \\
\path{task} & Launch a new agent to handle complex, multi-step tasks autonomously. \\
\path{task_output} & Retrieves output for one or more tasks (always non-blocking). \\
\path{task_stop} & Stops a running background task by its ID. \\
\path{unity_asset} & Manage Unity project assets on disk (under Assets/) such as materials, physics materials, textures/importers, ScriptableObjects, and existing prefabs. \\
\path{unity_bake} & Handles NavMesh and Lighting baking operations. \\
\path{unity_console} & Gets messages from or clears the Unity Editor console. \\
\path{unity_dialog} & Clicks a button on a native modal dialog that is blocking the Unity editor main thread. \\
\path{unity_editor} & Controls Unity Editor state (play/pause/stop), retrieves project information and full UnityCurrentState, manages editor tools, windows, selection, tags, and layers. \\
\path{unity_gameobject} & Manages Unity GameObjects with comprehensive operations including create, modify, delete, find, list\_children, and component management. \\
\path{unity_gameview} & Gets, sets, and lists the Unity Game View resolution. \\
\path{unity_input} & Simulates mouse and keyboard input in Unity PlayMode across multiple frames. \\
\path{unity_job} & Queries in-flight Unity bridge jobs. \\
\path{unity_menu} & Execute Unity Editor menu items by their menu path or get available menus. \\
\path{unity_package} & Manages Unity packages via UPM. \\
\path{unity_refresh} & Re-runs the same reconnect flow as \texttt{/upm refresh} by re-reading \texttt{.com-unity-codely.json} and reconnecting to the Unity Bridge. \\
\path{unity_scene} & Manages Unity scenes: create, load, save, get hierarchy, get active scene info, and build settings. \\
\path{unity_screenshot} & Captures screenshots from Unity Editor. \\
\path{unity_script} & Create, read, update, delete, and validate C\# scripts in Unity projects. \\
\path{unity_shader} & Create, read, update, and delete Unity shader files. \\
\path{unity_ui_toolkit} & Manages Unity UI Toolkit assets including PanelSettings, UXML, and USS files. \\
\path{unity_workflow} & High-level Unity workflows that compress common multi-step patterns (init session, compile+validate, checkpoint) into a single tool call. \\
\path{web_fetch} & Fetches content from a specified URL and processes it using an AI model. \\
\path{web_search} & Searches the internet/web for information using a search query. \\
\path{write_file} & Writes content to a specified file in the local filesystem. \\
\end{longtable}
\end{small}

\subsection{Codex CLI}
\begin{small}
\begin{longtable}{@{}>{\raggedright\arraybackslash}p{0.29\textwidth}>{\raggedright\arraybackslash}p{0.65\textwidth}@{}}
\caption{\textbf{Tool availability for Codex CLI.} Tools exposed to the benchmark main agent.}
\label{tab:codex-cli-tools}\\
\multicolumn{2}{@{}l}{
  \footnotesize
  \textbf{Harness:} Codex CLI
  \quad
  \textbf{CLI version:} 0.147.0
}\\[3pt]
\toprule
\textbf{Tool} & \textbf{Description} \\
\midrule
\endfirsthead
\multicolumn{2}{@{}l}{
  \footnotesize\textit{\tablename\ \thetable\ continued}
}\\[2pt]
\toprule
\textbf{Tool} & \textbf{Description} \\
\midrule
\endhead
\midrule
\multicolumn{2}{r@{}}{\footnotesize\textit{Continued on next page}} \\
\endfoot
\bottomrule
\endlastfoot
\toolname{apply\_patch} & Apply a structured patch to create, modify, or delete files. \\
\toolname{current\_time} & Retrieve the current time when the current-time capability is enabled. \\
\toolname{exec\_command} & Execute shell commands in the local execution environment. \\
\toolname{followup\_task} & Create a follow-up task for a sub-agent. \\
\toolname{get\_context\_remaining} & Retrieve information about the remaining context budget. \\
\toolname{interrupt\_agent} & Interrupt a running sub-agent. \\
\toolname{list\_agents} & List the currently running or previously created sub-agents. \\
\toolname{MCP tools} & Invoke tools exposed by configured MCP servers. \\
\toolname{new\_context\_window} & Request a new context window when the current context budget is exhausted. \\
\toolname{request\_permissions} & Request additional permissions required to perform an operation. \\
\toolname{request\_user\_input} & Request structured input from the user during task execution. \\
\toolname{resume\_agent} & Resume a previously created sub-agent. \\
\toolname{send\_input} & Send additional input to a running sub-agent. \\
\toolname{send\_message} & Send a message to another sub-agent. \\
\toolname{sleep} & Pause execution for a specified duration. \\
\toolname{spawn\_agent} & Launch a sub-agent to perform a delegated task. \\
\toolname{tool\_search} & Search for deferred tools that are available to the current agent. \\
\toolname{update\_plan} & Create or update a structured plan for the current task. \\
\toolname{view\_image} & Load and inspect a local image file. \\
\toolname{wait} & Wait for a sub-agent to complete or make progress. \\
\toolname{wait\_for\_environment} & Wait for an execution environment to become available or ready. \\
\toolname{web\_search} & Search the web using the hosted web-search capability when supported by the provider. \\
\toolname{write\_stdin} & Send input to, or retrieve output from, a running shell command. \\
\end{longtable}
\end{small}

\subsection{Claude Code}
\begin{small}
\begin{longtable}{
  @{}
  >{\raggedright\arraybackslash}p{0.29\textwidth}
  >{\raggedright\arraybackslash}p{0.65\textwidth}
  @{}
}
\caption{\textbf{Tool availability for Claude Code.} Tools exposed to the benchmark main agent.}
\label{tab:claude-code-tools}\\
\multicolumn{2}{@{}l}{
  \footnotesize
  \textbf{Harness:} Claude Code
  \quad
  \textbf{CLI version:} 2.1.233
  \quad
  \textbf{Models:} Claude Opus 5 and Claude Sonnet 5
}\\[4pt]
\toprule
\textbf{Tool} & \textbf{Description} \\
\midrule
\endfirsthead
\multicolumn{2}{@{}l}{
  \footnotesize
  \textit{\tablename\ \thetable\ continued}
}\\[2pt]
\toprule
\textbf{Tool} & \textbf{Description} \\
\midrule
\endhead
\midrule
\multicolumn{2}{r@{}}{
  \footnotesize\textit{Continued on next page}
}\\
\endfoot
\bottomrule
\endlastfoot
\path{Task} & Launch or resume a subagent for a delegated task. \\
\path{Bash} & Execute shell commands in the current workspace. \\
\path{CronCreate} & Create a one-shot or recurring scheduled job. \\
\path{CronDelete} & Delete a scheduled job by its identifier. \\
\path{CronList} & List scheduled jobs in the current session. \\
\path{DesignSync} & Synchronize design context through the built-in design-sync integration. \\
\path{Edit} & Apply an exact text replacement to an existing file. \\
\path{EnterWorktree} & Enter an isolated Git worktree for the current task. \\
\path{ExitWorktree} & Leave the current isolated Git worktree. \\
\path{Glob} & Find files matching a glob pattern. \\
\path{Grep} & Search file contents with regular expressions. \\
\path{NotebookEdit} & Edit cells in a Jupyter notebook. \\
\path{Read} & Read a local text file. \\
\path{ReportFindings} & Return structured findings from an investigation or delegated task. \\
\path{ScheduleWakeup} & Schedule a future wake-up for follow-up work. \\
\path{SendMessage} & Send a message to another agent or collaboration participant. \\
\path{Skill} & Invoke a registered runtime skill. \\
\path{TaskOutput} & Retrieve the status and output of a background task. \\
\path{TaskStop} & Stop a running background task. \\
\path{WebFetch} & Retrieve the text or response body of a specified URL. \\
\path{WebSearch} & Search the web. \\
\path{Workflow} & Launch a registered multi-step workflow. \\
\path{Write} & Create, append to, or replace a file. \\
\end{longtable}
\end{small}

\subsection{Grok Build}
\begin{small}
\begin{longtable}{@{}>{\raggedright\arraybackslash}p{0.29\textwidth}>{\raggedright\arraybackslash}p{0.65\textwidth}@{}}
\caption{\textbf{Tool availability for Grok Build.} Tools exposed to the benchmark main agent.}
\label{tab:grok-build-tools}\\
\multicolumn{2}{@{}l}{
  \footnotesize
  \textbf{Harness:} Grok Build
  \quad
  \textbf{CLI version:} 1.0.5
}\\[3pt]
\toprule
\textbf{Tool} & \textbf{Description} \\
\midrule
\endfirsthead
\multicolumn{2}{@{}l}{
  \footnotesize\textit{\tablename\ \thetable\ continued}
}\\[2pt]
\toprule
\textbf{Tool} & \textbf{Description} \\
\midrule
\endhead
\midrule
\multicolumn{2}{r@{}}{\footnotesize\textit{Continued on next page}} \\
\endfoot
\bottomrule
\endlastfoot
\toolname{ask\_user\_question} & Ask the user structured multiple-choice questions. \\
\toolname{enter\_plan\_mode} & Enter a read-only planning phase before implementation. \\
\toolname{exit\_plan\_mode} & Present a completed plan for user approval and leave plan mode. \\
\toolname{get\_command\_or\_subagent\_output} & Retrieve status and output from a background command, monitor, or subagent. \\
\toolname{grep} & Search file contents with regular expressions. \\
\toolname{image\_edit} & Edit or transform existing images with a text prompt. \\
\toolname{image\_gen} & Generate a new image from a text description. \\
\toolname{image\_to\_video} & Animate a source image into a short video. \\
\toolname{kill\_command\_or\_subagent} & Terminate a running background command, monitor, or subagent. \\
\toolname{list\_dir} & List files and directories at a given path. \\
\toolname{monitor} & Watch a long-running command and stream status events back to the agent. \\
\toolname{open\_page} & Fetch the text content of a specific URL. \\
\toolname{open\_page\_with\_find} & Fetch a URL and optionally extract regex matches with surrounding context. \\
\toolname{read\_file} & Read a local file, including text, PDF, and image content. \\
\toolname{reference\_to\_video} & Generate a video from reference images and optional preset voices. \\
\toolname{run\_terminal\_command} & Execute a shell command in the foreground or background. \\
\toolname{scheduler\_create} & Create or update a recurring scheduled prompt. \\
\toolname{scheduler\_delete} & Cancel a scheduled task by its identifier. \\
\toolname{scheduler\_list} & List active scheduled tasks and their next fire times. \\
\toolname{search\_replace} & Apply exact text replacements to an existing file, or create a new file. \\
\toolname{search\_tool} & Discover MCP tools by keyword and retrieve their input schemas. \\
\toolname{spawn\_subagent} & Launch or resume a subagent for a delegated task. \\
\toolname{todo\_write} & Maintain a structured list of work items and their statuses. \\
\toolname{use\_tool} & Call a discovered MCP tool with a validated argument object. \\
\toolname{web\_fetch} & Fetch a URL and return its content as markdown. \\
\toolname{web\_search} & Search the web and return ranked results. \\
\toolname{workflow} & Launch a multi-agent workflow script as one background run. \\
\toolname{write} & Create or completely overwrite a file with the given content. \\
\toolname{x\_keyword\_search} & Search X posts with advanced keyword and filter operators. \\
\toolname{x\_semantic\_search} & Find X posts that are semantically relevant to a query. \\
\toolname{x\_thread\_fetch} & Fetch an X post together with its surrounding thread context. \\
\toolname{x\_user\_search} & Search for an X user by name or account. \\
\end{longtable}
\end{small}

\subsection{DeepSeek Harness}
\begin{small}
\begin{longtable}{@{}>{\raggedright\arraybackslash}p{0.29\textwidth}>{\raggedright\arraybackslash}p{0.65\textwidth}@{}}
\caption{\textbf{Tool availability for DeepSeek Harness.} Tools exposed to the benchmark main agent by the default Windows headless profile.}
\label{tab:deepseek-harness-tools}\\
\multicolumn{2}{@{}l}{
  \footnotesize
  \textbf{Harness:} DeepSeek Harness
  \quad
  \textbf{CLI version:} 0.1.1-rc.2
}\\[3pt]
\toprule
\textbf{Tool} & \textbf{Description} \\
\midrule
\endfirsthead
\multicolumn{2}{@{}l}{
  \footnotesize\textit{\tablename\ \thetable\ continued}
}\\[2pt]
\toprule
\textbf{Tool} & \textbf{Description} \\
\midrule
\endhead
\midrule
\multicolumn{2}{r@{}}{\footnotesize\textit{Continued on next page}} \\
\endfoot
\bottomrule
\endlastfoot
\toolname{create\_goal} & Create a persistent same-session goal for autonomous continuation. \\
\toolname{edit} & Replace exact literal text in an existing UTF-8 file. \\
\toolname{exit\_plan\_mode} & Present a complete plan for user review and approval. \\
\toolname{get\_goal} & Read the current goal, revision, phase, limits, and blockers. \\
\toolname{glob} & Find file paths that match a glob pattern. \\
\toolname{grep} & Search file contents using ripgrep regular expressions. \\
\toolname{interrupt\_agent} & Cancel a background agent's current turn without removing it. \\
\toolname{job\_kill} & Request cancellation of a running background job. \\
\toolname{job\_list} & List running and finished background jobs and their statuses. \\
\toolname{job\_output} & Read incremental or final output from a background job. \\
\toolname{list\_agents} & List continuable background subagents and their runtime status. \\
\toolname{pwsh} & Execute PowerShell commands in a fresh foreground or background process. \\
\toolname{ralph} & Run bounded fresh-agent rounds toward one immutable objective. \\
\toolname{read} & Read line-numbered content from a UTF-8 text file. \\
\toolname{read\_image} & Load a supported local image for model inspection. \\
\toolname{send\_message} & Send follow-up input to a direct continuable subagent. \\
\toolname{skill} & Load the full instructions for an available skill. \\
\toolname{str\_replace\_editor} & View, create, replace, or insert text in files. \\
\toolname{subagent} & Start a continuable subagent for a self-contained delegated task. \\
\toolname{subagent\_fork} & Delegate a one-shot task to a context-inheriting subagent fork. \\
\toolname{todo\_write} & Replace the current structured task list and item statuses. \\
\toolname{update\_goal} & Edit, pause, resume, complete, or block the current goal. \\
\toolname{web\_search} & Search current web information with one to four queries. \\
\toolname{workflow} & Run a JavaScript workflow that orchestrates multiple subagents. \\
\toolname{write} & Create or completely replace a UTF-8 text file. \\
\end{longtable}
\end{small}

\subsection{OpenCode}
\begin{small}
\begin{longtable}{@{}>{\raggedright\arraybackslash}p{0.29\textwidth}>{\raggedright\arraybackslash}p{0.65\textwidth}@{}}
\caption{\textbf{Tool availability for OpenCode.} Tools exposed to the benchmark main agent.}
\label{tab:opencode-tools}\\
\multicolumn{2}{@{}l}{
  \footnotesize
  \textbf{Harness:} OpenCode
  \quad
  \textbf{CLI version:} 1.18.23
}\\[3pt]
\toprule
\textbf{Tool} & \textbf{Description} \\
\midrule
\endfirsthead
\multicolumn{2}{@{}l}{
  \footnotesize\textit{\tablename\ \thetable\ continued}
}\\[2pt]
\toprule
\textbf{Tool} & \textbf{Description} \\
\midrule
\endhead
\midrule
\multicolumn{2}{r@{}}{\footnotesize\textit{Continued on next page}} \\
\endfoot
\bottomrule
\endlastfoot
\toolname{apply\_patch} & Apply patches to files. \\
\toolname{bash} & Execute shell commands in your project environment. \\
\toolname{edit} & Modify existing files using exact string replacements. \\
\toolname{glob} & Find files by pattern matching. \\
\toolname{grep} & Search file contents using regular expressions. \\
\toolname{lsp} & Interact with your configured LSP servers to get code intelligence features like definitions, references, hover info, and call hierarchy. \\
\toolname{question} & Ask the user questions during execution. \\
\toolname{read} & Read file contents from your codebase. \\
\toolname{skill} & Load a skill (a \texttt{SKILL.md} file) and return its content in the conversation. \\
\toolname{task} & Launch a new agent to handle complex, multistep tasks autonomously. \\
\toolname{todowrite} & Manage todo lists during coding sessions. \\
\toolname{webfetch} & Fetch web content. \\
\toolname{websearch} & Search the web for information. \\
\toolname{write} & Create new files or overwrite existing ones. \\
\end{longtable}
\end{small}

\subsection{Qwen Code}
\begin{small}
\begin{longtable}{@{}>{\raggedright\arraybackslash}p{0.29\textwidth}>{\raggedright\arraybackslash}p{0.65\textwidth}@{}}
\caption{\textbf{Tool availability for Qwen Code.} Tools exposed to the benchmark main agent.}
\label{tab:qwen-code-tools}\\
\multicolumn{2}{@{}l}{
  \footnotesize
  \textbf{Harness:} Qwen Code
  \quad
  \textbf{CLI version:} 0.21.13 / 0.21.14
}\\[3pt]
\toprule
\textbf{Tool} & \textbf{Description} \\
\midrule
\endfirsthead
\multicolumn{2}{@{}l}{
  \footnotesize\textit{\tablename\ \thetable\ continued}
}\\[2pt]
\toprule
\textbf{Tool} & \textbf{Description} \\
\midrule
\endhead
\midrule
\multicolumn{2}{r@{}}{\footnotesize\textit{Continued on next page}} \\
\endfoot
\bottomrule
\endlastfoot
\toolname{agent} & Use agent to launch a specialized subagent to handle complex, multi-step tasks autonomously. \\
\toolname{computer\_use\_\_kill\_app} & Terminate a desktop application by process ID. \\
\toolname{computer\_use\_\_list\_apps} & List running and installed desktop applications with their process IDs. \\
\toolname{computer\_use\_\_list\_windows} & List the on-screen windows of a desktop application. \\
\toolname{edit} & Edit replaces text within a file. \\
\toolname{exit\_plan\_mode} & Use exit\_plan\_mode when you are in plan mode and have finished presenting your implementation plan. This tool prompts the user to approve or reject the plan and transitions from planning mode to implementation mode. \\
\toolname{glob} & Glob finds files matching specific glob patterns (e.g., src/**/*.ts, *.md), returning absolute paths sorted by modification time (newest first). \\
\toolname{grep\_search} & Grep\_search searches for a regular expression pattern within the content of files in a specified directory. \\
\toolname{list\_agents} & Discover the current session's addressable background agents and their task\_id values. \\
\toolname{list\_directory} & List\_directory lists the names of files and subdirectories directly within a specified directory path. \\
\toolname{loop\_wakeup} & Schedule a one-shot session-loop wakeup that re-prompts the agent after a delay. \\
\toolname{monitor} & Use monitor to start a long-running shell command that streams stdout and stderr lines back to the agent as background task notifications. \\
\toolname{notebook\_edit} & Notebook\_edit edits Jupyter notebook (.ipynb) files safely at the cell level. \\
\toolname{read\_file} & Read\_file reads and returns the content of a specified file. \\
\toolname{run\_shell\_command} & Use run\_shell\_command to interact with the underlying system, run scripts, or perform command-line operations. \\
\toolname{save\_memory} & Use save\_memory to save and recall information across your Qwen Code sessions. \\
\toolname{send\_message} & Send a follow-up instruction to a background agent by task\_id; running agents receive the message at the next tool-round boundary. \\
\toolname{task\_stop} & Stop a running background task or monitor by its task ID. \\
\toolname{todo\_write} & Use todo\_write to create and manage a structured task list for your current coding session. \\
\toolname{web\_fetch} & Use web\_fetch to fetch content from a specified URL and process it using an AI model. \\
\toolname{web\_search} & Issues a self-contained search request to a small auxiliary model with DashScope's server-side web\_search tools, and returns the narrated findings plus source URLs. \\
\toolname{write\_file} & Write\_file writes content to a specified file. \\
\toolname{zoom\_image} & Zoom into a rectangular region of a local image for closer visual inspection. \\
\end{longtable}
\end{small}

\subsection{Antigravity}
\begin{small}
\begin{longtable}{@{}>{\raggedright\arraybackslash}p{0.29\textwidth}>{\raggedright\arraybackslash}p{0.65\textwidth}@{}}
\caption{\textbf{Tool availability for Antigravity.} Tools exposed to the benchmark main agent.}
\label{tab:antigravity-tools}\\
\multicolumn{2}{@{}l}{
  \footnotesize
  \textbf{Harness:} Antigravity
  \quad
  \textbf{CLI version:} 1.1.19
}\\[3pt]
\toprule
\textbf{Tool} & \textbf{Description} \\
\midrule
\endfirsthead
\multicolumn{2}{@{}l}{
  \footnotesize\textit{\tablename\ \thetable\ continued}
}\\[2pt]
\toprule
\textbf{Tool} & \textbf{Description} \\
\midrule
\endhead
\midrule
\multicolumn{2}{r@{}}{\footnotesize\textit{Continued on next page}} \\
\endfoot
\bottomrule
\endlastfoot
\toolname{find\_by\_name} & Find files and directories by name pattern, with extension, type, depth, and exclude filters. \\
\toolname{grep\_search} & Search file contents for a query or regular expression, with case-insensitive and file-include filters. \\
\toolname{list\_dir} & List the contents of a directory. \\
\toolname{manage\_task} & Inspect and control background tasks: list them, query status, or kill by task ID. \\
\toolname{read\_url\_content} & Fetch and read the content of a URL. \\
\toolname{replace\_file\_content} & Replace targeted chunks of content in an existing file, with optional line-range context and multiple replacements. \\
\toolname{run\_command} & Execute a shell command line in a working directory, optionally running it as a background task. \\
\toolname{schedule} & Schedule a delayed prompt to wake the agent after a duration or when a task condition is met. \\
\toolname{search\_web} & Search the web for information. \\
\toolname{send\_message} & Send a message to a recipient, such as the agent's own future turns. \\
\toolname{view\_file} & View the contents of a file, optionally restricted to a line range or content offset. \\
\toolname{write\_to\_file} & Create a new file or completely overwrite an existing one with the given content. \\
\end{longtable}
\end{small}

\subsection{Kimi Code}

\begin{small}
\begin{longtable}{@{}>{\raggedright\arraybackslash}p{0.29\textwidth}>{\raggedright\arraybackslash}p{0.65\textwidth}@{}}
\caption{\textbf{Tool availability for Kimi Code.} Tools exposed to the benchmark main agent.}
\label{tab:kimi-code-tools}\\
\multicolumn{2}{@{}l}{
  \footnotesize
  \textbf{Harness:} Kimi Code
  \quad
  \textbf{CLI version:} 0.36.1
}\\[3pt]
\toprule
\textbf{Tool} & \textbf{Description} \\
\midrule
\endfirsthead
\multicolumn{2}{@{}l}{
  \footnotesize\textit{\tablename\ \thetable\ continued}
}\\[2pt]
\toprule
\textbf{Tool} & \textbf{Description} \\
\midrule
\endhead
\midrule
\multicolumn{2}{r@{}}{\footnotesize\textit{Continued on next page}} \\
\endfoot
\bottomrule
\endlastfoot
\toolname{Agent} & Launch or resume one subagent for a delegated task. \\
\toolname{AgentSwarm} & Launch or resume multiple subagents in parallel. \\
\toolname{AskUserQuestion} & Ask the user structured questions with selectable options. \\
\toolname{Bash} & Execute shell commands in the foreground or background. \\
\toolname{CreateGoal} & Create a durable goal that persists across turns. \\
\toolname{CronCreate} & Schedule a one-shot or recurring prompt. \\
\toolname{CronDelete} & Delete a scheduled job by its identifier. \\
\toolname{CronList} & List the scheduled jobs in the current session. \\
\toolname{Edit} & Apply exact text replacements to an existing file. \\
\toolname{EnterPlanMode} & Enter plan mode before a non-trivial implementation. \\
\toolname{ExitPlanMode} & Submit a completed plan for user approval. \\
\toolname{FetchURL} & Retrieve the text or response body of a specific URL. \\
\toolname{GetGoal} & Inspect the current goal, status, and remaining budgets. \\
\toolname{Glob} & Find files that match a glob pattern. \\
\toolname{Grep} & Search file contents with regular expressions. \\
\toolname{Read} & Read a local text file. \\
\toolname{ReadMediaFile} & Load a local media file for inspection. \\
\toolname{SetGoalBudget} & Set hard turn, token, or time limits for the current goal. \\
\toolname{Skill} & Invoke a registered runtime skill. \\
\toolname{TaskList} & List background tasks and their current status. \\
\toolname{TaskOutput} & Retrieve the status and output of a background task. \\
\toolname{TaskStop} & Stop a running background task. \\
\toolname{TodoList} & Maintain a structured list of work items. \\
\toolname{UpdateGoal} & Change the status of the current goal. \\
\toolname{Write} & Create, append to, or completely replace a file. \\
\end{longtable}
\end{small}

\end{document}